\documentclass[traditabstract]{aa} 
\usepackage{savesym} 
\usepackage{txfonts} 
\usepackage[english]{babel} 
\usepackage{t1enc} 
\usepackage{graphicx} 
\usepackage{subfig} 
\usepackage{pdflscape}
\usepackage{fixltx2e} 
\usepackage{natbib} 
\usepackage{footmisc}
\usepackage{url} 
\usepackage{longtable}
\usepackage{multirow}
\usepackage{setspace}
\usepackage[table]{xcolor}
\usepackage{caption}

\begin{document} 
 
\title{Temperature structures in Galactic Center clouds} 
\subtitle{Direct evidence for gas heating via turbulence}                                                   
\author{K. Immer\inst{1} \and J. Kauffmann\inst{2} \and T. Pillai\inst{2} \and A. Ginsburg\inst{1} \and K.~M. Menten\inst{2}}

\institute{European Southern Observatory, Karl-Schwarzschild-Strasse 2, 85748 Garching bei M\"unchen, Germany \and
Max-Planck-Institut f\"ur Radioastronomie, Auf dem H\"ugel 69, 53121 Bonn, Germany}                                                     
\date{Received xxxx; accepted xxxx} 
 
\abstract{
The Central Molecular Zone (CMZ) at the center of our Galaxy is the best template to study star formation processes under 
extreme conditions, similar to those in high-redshift galaxies. 
We observed on-the-fly maps of para-H$_{2}$CO transitions at 218 GHz and 291 GHz towards seven Galactic Center clouds. From the 
temperature-sensitive integrated intensity line ratios of H$_{2}$CO(3$_{2,1}-$2$_{2,0}$)/H$_{2}$CO(3$_{0,3}-$2$_{0,2}$) and
H$_{2}$CO(4$_{2,2}-$3$_{2,1}$)/H$_{2}$CO(4$_{0,4}-$3$_{0,3}$) in combination with radiative transfer models, we produce 
gas temperature maps of our targets. These transitions are sensitive to gas with densities of $\sim$10$^{5}$ cm$^{-3}$ and temperatures 
<150 K.  
 The measured gas temperatures in our sources are all higher (>40 K) than their dust temperatures ($\sim$25 K). Our targets have a complex 
 velocity structure that requires a careful disentanglement of the different components. We produce temperature maps for each of the velocity
 components and show that the temperatures of the components differ, revealing temperature gradients in the clouds. 
 Combining the temperature measurements with the integrated intensity line ratio of  
 H$_{2}$CO(4$_{0,4}-$3$_{0,3}$)/H$_{2}$CO(3$_{0,3}-$2$_{0,2}$), we constrain the density of this warm gas to 10$^{4}-$10$^{6}$ cm$^{-3}$.
 We find a positive correlation of the line width of the main H$_{2}$CO lines with the temperature of the gas, direct evidence for gas heating 
 via turbulence. Our data is consistent with a turbulence heating model with a density of n = 10$^5$ cm$^{-3}$.}
 
\keywords{Galaxy: center, ISM: molecules, ISM: structure, ISM: clouds, Submillimeter: ISM} 
 
\authorrunning{} 
\titlerunning{Temperature structures in Galactic Center clouds} 
 
\maketitle 
 
\section{Introduction}  
\label{Intro}

The central region of the Milky Way, the so-called Central Molecular Zone (CMZ), is an exceptional star-forming environment. 
This region contains $\sim$10\% of the Galaxy's total molecular gas and produces 5$-$10\% of the Galaxy's infrared and Lyman 
continuum luminosity \citep{Morris1996}. The conditions (pressure, magnetic field strength, turbulence, gas temperature, etc.) 
in this region are much more extreme than in Galactic plane clouds \citep{Morris1996}. The star formation rate in the CMZ is a 
factor of 10$-$100 lower than expected for the huge amount of dense gas and dust contained in this region 
\citep[e.g.][]{Yusef-Zadeh2009, Immer2012a, Longmore2013, Kruijssen2013}.

The high gas temperatures are one of the key properties of the CMZ clouds, influencing the chemistry of the gas as well as 
the star formation efficiency of the clouds. It determines the thermal Jeans mass as well as the sound speed which in turn sets the 
Mach number. Understanding the gas temperature structure of Galactic Center clouds is thus crucial for understanding 
the fragmentation and star forming mechanisms within them. 
The discrepancy between observed dust and gas temperatures is a long-known feature of CMZ clouds. While multi-wavelength 
observations of the dust emission in the CMZ yield dust temperatures of $\sim$20 K \citep{Lis1999,Molinari2011}, comparable 
to dust temperatures of Galactic plane clouds, the gas temperatures are much higher 
\citep[$>$50 K,][]{Guesten1981,Huettemeister1993,Ao2013,Mills2013,Ott2014,Ginsburg2016}. Many previous gas temperature measurements 
in the CMZ are based on observations of the NH$_{3}$ molecule which traces low-density gas 
\citep[n $\sim$ 10$^{3}$ cm$^{-3}$, e.g.][]{Guesten1981,Huettemeister1993, Ott2014}. \citet{Ao2013} 
mapped the inner $\sim$75 pc of the CMZ in the para-H$_{2}$CO transitions at 218 GHz,
sensitive to warmer (T $>$ 20 K) and denser (n $\sim$ 10$^{4}-$10$^{5}$ cm$^{-3}$) gas. Their results show 
high gas temperatures towards many CMZ clouds, comparable to those measured in prior studies.
This survey was extended to the whole CMZ ($-$0.4$^{\circ}$ $<$ l $<$ 1.6$^{\circ}$) by \citet{Ginsburg2016}. The inferred 
gas temperatures range from $\sim$ 60 K to $>$ 100 K, where the highest values are 
measured towards Sgr B2, the 20 and 50 km/s clouds, and G0.253+0.016 (``The Brick''). Comparing their results with 
dust temperature measurements of the whole CMZ, they show that the gas is uniformly hotter than the dust. The high 
gas temperatures are consistent with heating through turbulence, while uniform cosmic ray heating is excluded as a dominant 
heating mechanism.

H$_{2}$CO is a slightly asymmetric rotor molecule. It has two different species (i.e. ortho and para) for 
which the K$_{\rm a}$\footnote{K$_{\rm a}$ is the projection 
of J along the symmetry axis for the limiting case of a prolate (oblate) top.} quantum number is odd or even. These 
two species are not connected by radiative transitions. The differences in the population of levels separated by 
$\Delta$K$_{\rm a}$~=~2$\cdot$n are due to collisions.
\citet{Mangum1993} presented a detailed study of the usability of different H$_{2}$CO transitions for the determination
of the kinetic temperature and density of the gas in molecular clouds.  For a range of total 
angular momentum quantum numbers $J$ (here J = 2 and 3), modeling of the relative intensities of (in our case) 
para-H$_2$CO lines (K$_{\rm a}$ quantum number of 0 or 2), delivers estimates of the density and the 
temperature. 
The K$_{\rm a}$ ladders are 
close in frequency and thus can be observed with the same telescope, even very often in the same spectrum
which makes them calibration-independent. 

The J = 3$-$2 and 4$-$3 H$_{2}$CO K$_{\rm a}$ ladders at 218 and 291 GHz, respectively, can be observed with the 
Atacama Pathfinder Experiment (APEX) telescope, each group within one band. \citet{Mangum1993} showed that measuring 
several H$_{2}$CO intensity ratios of transitions with different J values yields better constraints of the kinetic temperature than can 
be obtained from just one H$_{2}$CO intensity ratio. This is clear since then a larger range of level energies is covered.

\begin{figure*}
	\caption{870 $\mu$m emission of the Central Molecular Zone from the ATLASGAL survey \citep{Schuller2009}. The targets of our 
	temperature study are marked. The boxes show the sizes of the observed OTF maps at 218 GHz for the 20 and 50 km/s clouds, G0.253+0.016, G0.411+0.050, and G0.480$-$0.006 and at 291 GHz for Sgr C and Sgr D.}
	\centering
       \includegraphics[height=0.95\textwidth,angle=270]{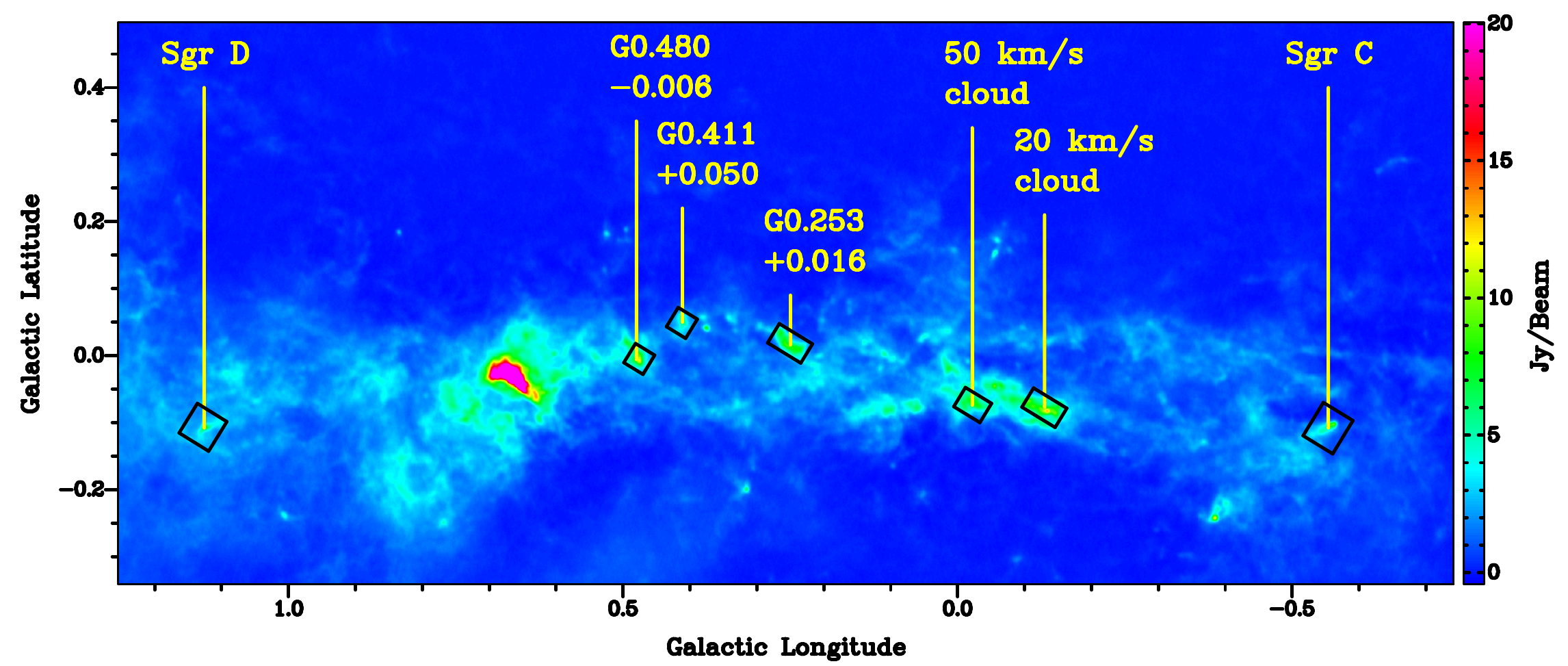}
	\label{Targets-CMZ}
\end{figure*}

In this paper, we report a detailed gas temperature study of seven molecular clouds in the CMZ, using the 
H$_{2}$CO thermometer at 218 and 291 GHz. The names of the observed sources and their coordinates are listed in Table 
\ref{SourceCoord}. We chose our targets to be high density clouds with previous warm gas temperature measurements. As shown
in Fig. \ref{Targets-CMZ}, they span the whole CMZ. None of these clouds are photon-dominated or X-ray-dominated regions. 
There is evidence of wide-spread shocks in the form of SiO emission in the CMZ including these clouds. 
Our sample contains potential star forming clouds (Sgr C, 20 km/s cloud, 50 km/s cloud, 
G0.480$-$0.006), quiescent clouds (G0.411+0.050) and shock heated clouds (G0.253+0.016).
There is an ongoing debate whether Sgr D is part of the CMZ or not \citep{Mehringer1998, Blum1999, Sawada2009}. 
However, this uncertainty does not influence our results or conclusions.

In Section \ref{Obs}, we describe the observations 
and the calibration of the data. In Section \ref{Analysis}, we present how the H$_{2}$CO spectra, ratio and 
uncertainty maps were produced. In Section \ref{Modeling}, the radiative transfer modeling is described. 
We discuss the different results in Section \ref{TempMeasurements} and give conclusions in Section \ref{Summary}.

\section{Observation and Data Reduction} 
\label{Obs}  

\subsection{Observations}

\begin{table*}
\setlength{\tabcolsep}{3pt}
\centering
\caption{Coordinates and sizes of observed OTF maps at 218 and 291 GHz as well as velocity ranges over which the H$_{2}$CO 
line emission was integrated (see Section \ref{VelRanges}).}
\begin{tabular}{cccccccccc}
Source & R.A.  & Dec & v$_{LSR}$  & \multicolumn{2}{c}{OTF Size} & \multicolumn{2}{c}{$\sigma_{median}(T_{mb})$} & \multicolumn{2}{c}{v$_{range}$}\\
& (hh mm ss.ss) & (dd $\arcmin$$\arcmin$ $\arcsec$$\arcsec$.$\arcsec$) & (km s$^{-1}$) & \multicolumn{2}{c}{($\arcsec$ x $\arcsec$)} & 
\multicolumn{2}{c}{(mK per pixel)} & \multicolumn{2}{c}{(km s$^{-1}$)}\\
& (EQ 2000) & (EQ 2000) & & 218 GHz & 291 GHz & 218 GHz & 291 GHz & Whole Source & Vel. Comp.\\ \hline
\multirow{3}{*}{20 km/s cloud} & \multirow{3}{*}{17 45 36.89} & \multirow{3}{*}{$-$29 05 12.0} & \multirow{3}{*}{20} & \multirow{3}{*}{115 x 
200} & \multirow{3}{*}{180 x 270} & \multirow{3}{*}{41} & \multirow{3}{*}{30} & \multirow{3}{*}{$-$15 $-$ 36} & 0 $-$ 6\\
& & & & & & & & & 8 $-$ 14\\
& & & & & & & & & 27 $-$ 33\\
\multirow{2}{*}{50 km/s cloud} & \multirow{2}{*}{17 45 51.18} & \multirow{2}{*}{$-$28 59 34.3} & \multirow{2}{*}{50} & \multirow{2}{*}{115 x 
150} & \multirow{2}{*}{180 x 180}  & \multirow{2}{*}{59} & \multirow{2}{*}{46} & \multirow{2}{*}{16 $-$ 80} & 41 $-$ 47\\
& & & & & & & & &57 $-$ 63\\
\multirow{4}{*}{G0.253+0.016} & \multirow{4}{*}{17 46 08.55} & \multirow{4}{*}{$-$28 42 44.0} & \multirow{4}{*}{25} & \multirow{4}{*}{115 x 
200} & \multirow{4}{*}{200 x 340} & \multirow{4}{*}{35} & \multirow{4}{*}{44} &\multirow{4}{*}{$-$6 $-$ 54} & $-$3 $-$ 3\\
& & & & & & & & &16 $-$ 22\\
& & & & & & & & & 36 $-$ 42\\
& & & & & & & & & 75 $-$ 81\\
G0.411+0.050 & 17 46 24.44 & $-$28 33 26.6 & 25 & 115 x 115 & 180 x 240 & 36 & 44 & 10 $-$ 30 & 19 $-$ 25\\
G0.480$-$0.006 & 17 46 46.00 & $-$28 31 51.0 & 25 & 115 x 115 & 180 x 240 & 49 & 43 & 19 $-$ 44 & 27 $-$ 33\\
Sgr C & 17 44 43.20 & $-$29 27 53.7 & $-$50 & & 200 x 180 & & 44 & $-$60 $-$ $-$46 & $-$55 $-$ $-$49 \\
Sgr D & 17 48 41.75 & $-$28 01 45.0 & $-$15 & & 180 x 180 & & 44 &$-$17 $-$ $-$14 & $-$19 $-$ $-$13\\
\end{tabular}
\tablefoot{Col. 6 and 7 give the median rms values of the H$_{2}$CO cubes of the whole 
velocity range of each source at 218 and 291 GHz.
}
\label{SourceCoord}
\end{table*}

\begin{table}
\centering
\caption{Properties of covered para-H$_{2}$CO transitions. }
\begin{tabular}{ccc}
Transition & $\nu$ & E$_{up}$ \\ 
&  (GHz) & (K) \\\hline
3$_{0,3}-$2$_{0,2}$ & 218.22219 & 20.96 \\
3$_{2,2}-$2$_{2,1}$ & 218.47563 & 68.09 \\
3$_{2,1}-$2$_{2,0}$ & 218.76007 & 68.11 \\ \hline
4$_{0,4}-$3$_{0,3}$ & 290.62341 & 34.90 \\
4$_{2,3}-$3$_{2,2}$& 291.23778  & 82.07 \\
4$_{2,2}-$3$_{2,1}$ & 291.94806 & 82.12\\
\end{tabular}
\tablefoot{Columns give quantum numbers, J$_{\rm{K_a},{K_c}}$, frequencies and 
level energies above the ground state. The critical densities of 3$_{0,3}-$2$_{0,2}$ and 4$_{0,4}-$3$_{0,3}$ are 
$\sim$6$\cdot$10$^{5}$ cm$^{-3}$ and $\sim$1$\cdot$10$^{6}$ cm$^{-3}$, respectively \citep{Shirley2015}.
}
\label{H2COTrans}
\end{table}

In 2012 and 2014, we observed five and seven molecular clouds in the Central Molecular Zone (Table \ref{SourceCoord}) with the 
APEX\footnote{This publication is based on data acquired with the Atacama Pathfinder Experiment (APEX). APEX is a collaboration between 
 the Max-Planck-Institut f\"ur Radioastronomie, the European Southern Observatory, and the Onsala Space Observatory.} 
telescope \citep{Guesten2006} at 218 and 291 GHz, respectively (project codes: M$-$089.F$-$0029$-$2012, M$-$093.F$-$0030$-
$2014). On-the-fly (OTF) maps were taken with the Swedish heterodyne facility instrument \citep[SHeFi,][]{Vassilev2008} 
as well as the First Light APEX Submillimeter Heterodyne (FLASH) receiver \citep{Heyminck2006, Klein2014}, 
using the eXtended bandwidth Fast Fourier Transform Spectrometer (XFFTS) backend \citep{Klein2012}. Table 
\ref{H2COTrans} shows the covered para-H$_{2}$CO transitions. 

The SHeFi observations were centered at 218.9 GHz and covered a bandwidth of 4 GHz. The FLASH observations were taken in the 
frequency ranges 278.5$-$282.5 GHz and 290.5$-$294.5 GHz. The XFFTS backend provides a fixed number of 32768 spectral channels, 
resulting in a resolution of 0.1 km s$^{-1}$ at 218 GHz and 0.04 km s$^{-1}$ at 291 GHz. To increase the signal-to-noise ratio of the spectra, 
we smoothed the two datasets by a factor of 9 and 24, respectively, yielding a velocity resolution of $\sim$ 0.9 km s$^{-1}$. Since we do not 
expect to detect lines in our spectra that are narrower than a few km s$^{-1}$, this velocity resolution is sufficient. The full widths at half 
maximum of the beams were 30$\arcsec$ at 218 GHz and 24 $\arcsec$ at 291 GHz.

A first calibration of the data was conducted already at the telescope, yielding spectra with calibrated fluxes at each position of the OTF 
maps. The calibration error of this step is about 15\% for each dataset
(SHeFI calibration plan; A. Belloche for FLASH, priv. comm.).
We then converted the antenna temperatures of the spectra into main beam temperatures, using Ruze's equation with the scaling 
factor being 0.69 and the width factor 19 micron. The two datasets were further edited and 
analyzed with the software CLASS from the GILDAS package \citep{Pety2005}. 

\subsection{Baseline subtraction}

 The 218 GHz spectra were strongly affected by bad baselines (Fig. \ref{Spectra-Example}). Since the baselines could not be cleanly 
 removed by fitting low-level polynomials to the data, we fitted the spectra with splines over the velocity ranges $-$90$-$250 km s$^{-1}$ 
 for the 20 km/s cloud, G0.253+0.016, G0.411+0.050, and G0.480$-$0.006, and $-$60$-$220 km s$^{-1}$ for the 50 km/s cloud. First, 
 the raw data were smoothed to a velocity resolution of $\sim$0.9 km s$^{-1}$, then lines in the spectra were fitted with Gaussians and 
 removed from the spectra if they were at least a 3$\sigma$ detection. Then, the data were down sampled to a resolution of 
 $\sim$36 km s$^{-1}$ to ensure that weak line emission that was not removed in the previous step would be smoothed out over the 
 adjacent channels. This spectrum was then resampled to the original resolution of 0.9 km s$^{-1}$, yielding the spline spectrum, and then 
 subtracted from the input spectrum. However, in some of the spectra, the baseline features are as narrow as the spectral lines and 
 could not be removed completely, resulting in negative features in the final spectra. 
 Figure \ref{Spectra-Example} in the Appendix shows one input spectrum, the fitted spline, and the final spectrum. The grey box marks one of 
 the narrow negative features in the spectrum.
 
 The 291 GHz dataset was much less affected by bad baselines. The baselines were removed by fitting each spectrum with a 1st-order 
 polynomial over the velocity ranges $-$120$-$120 km s$^{-1}$ (20 km/s cloud), $-$70$-$120 km s$^{-1}$ (50 km/s cloud), $-$90$-$120 
 km s$^{-1}$ (G0.253+0.016, G0.411+0.050, G0.480$-$0.006), and $-$200$-$100 km s$^{-1}$ (Sgr C, Sgr D). 
  
  \subsection {H$_{2}$CO spectra}
\label{H2COSpectra}

 \begin{figure*}
	\caption{Spectra of para-H$_{2}$CO transitions at 218 (left panels) and 291 GHz (right panels) of the 20 km/s cloud (upper panels) and 
	G0.411+0.050 (lower panels), averaged over the whole OTF maps. Transitions of H$_{2}$CO as well as other 
	molecules are marked in the spectra.}
	\centering
	\subfloat{\includegraphics[width=8cm]{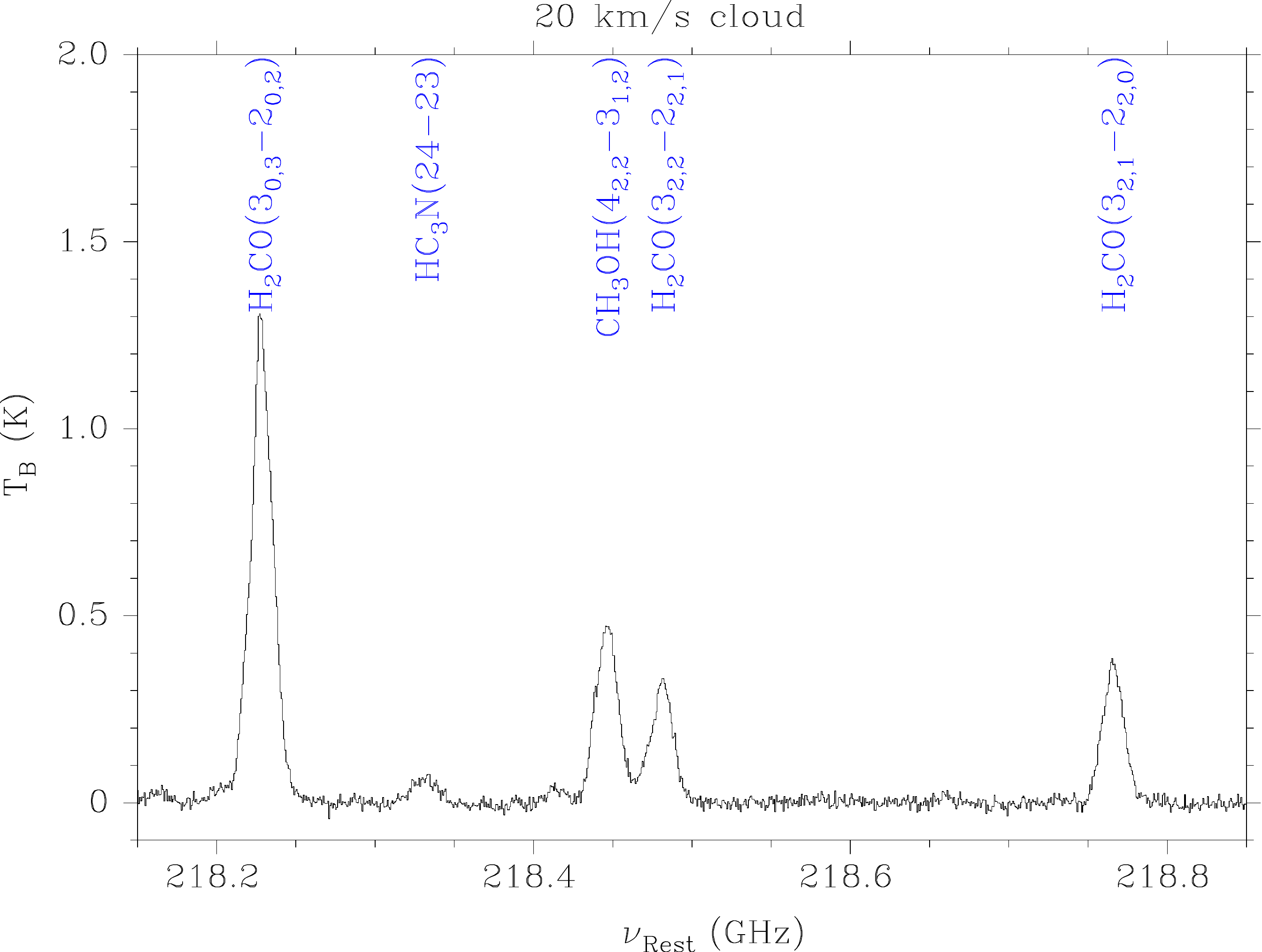}}
\hspace{0.1cm}
	\subfloat{\includegraphics[width=8cm]{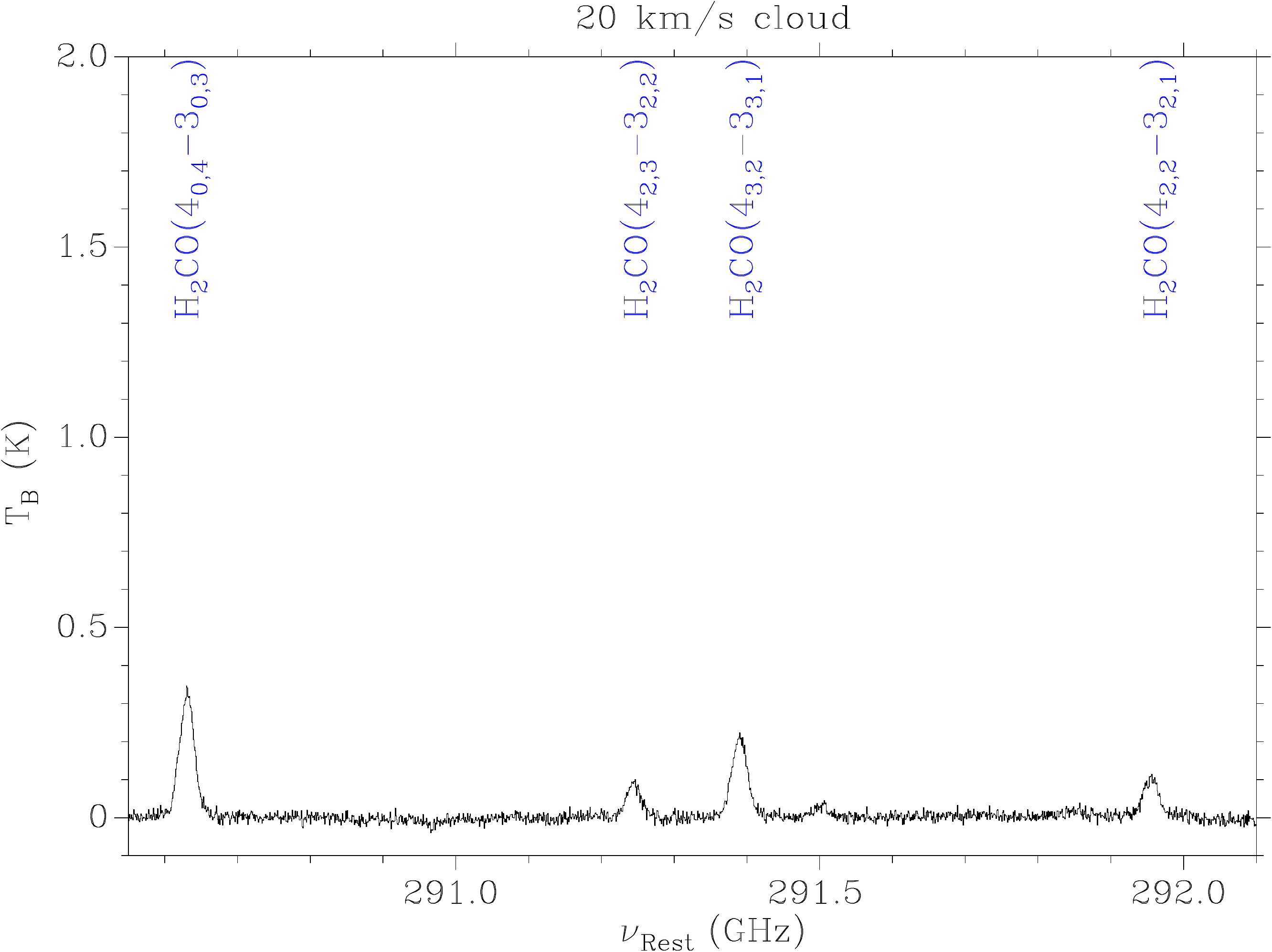}}\\
	\subfloat{\includegraphics[width=8cm]{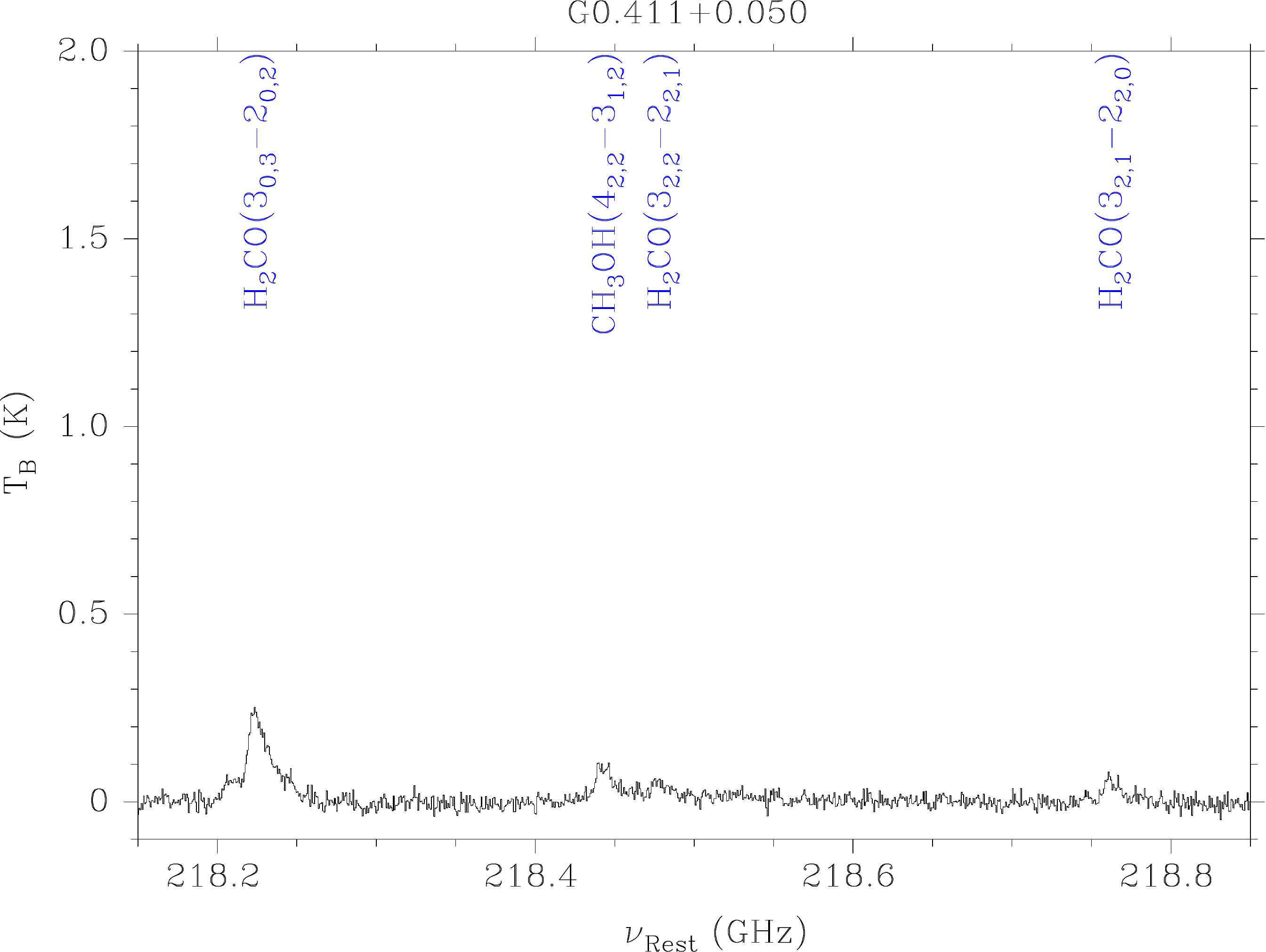}}
\hspace{0.1cm}
	\subfloat{\includegraphics[width=8cm]{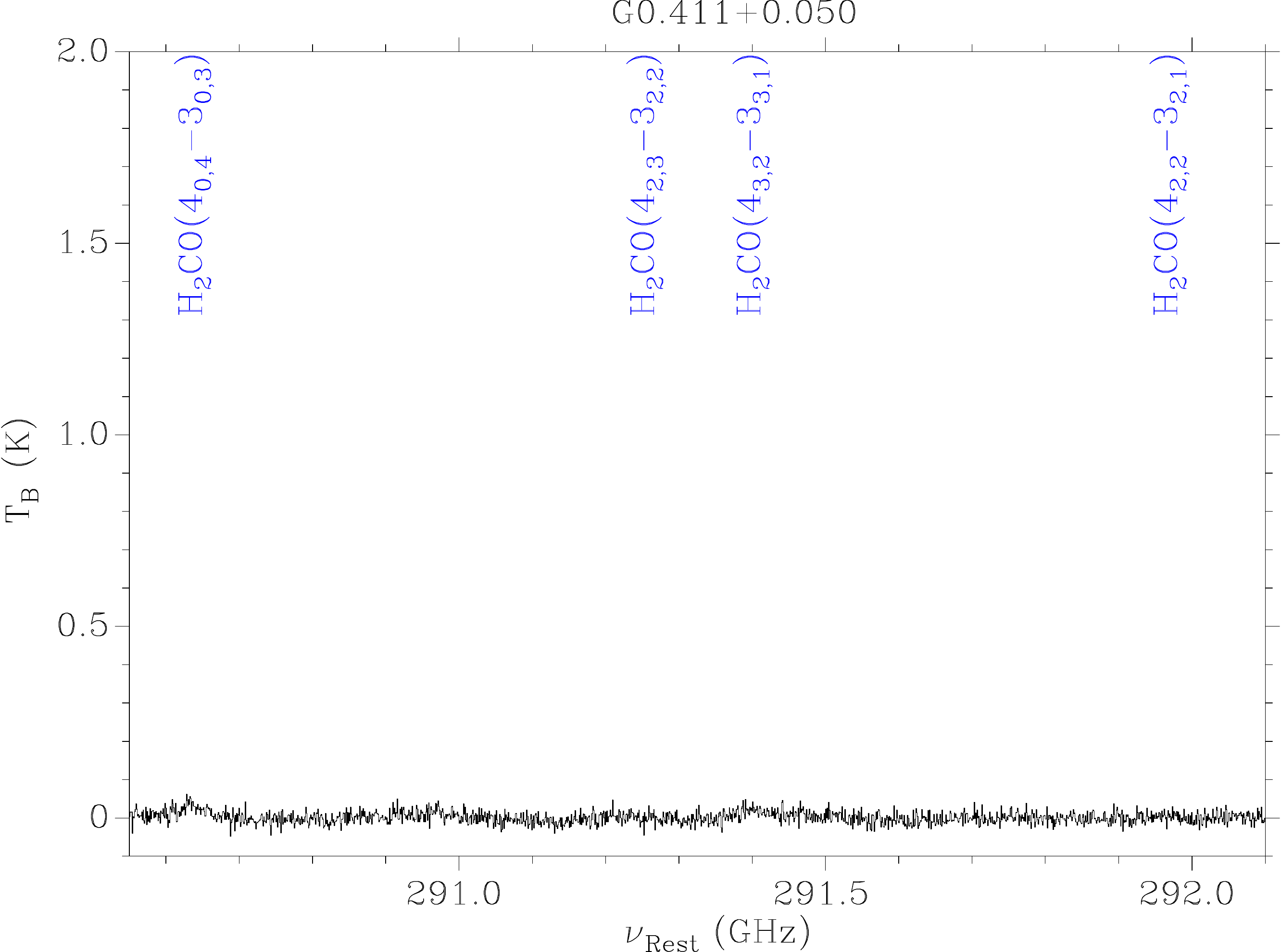}}\\
	\label{ExampleSpectra}
\end{figure*}

To obtain average spectra of the six H$_{2}$CO lines at 218 GHz and 291 GHz, we averaged the 
218.15$-$218.85 GHz and 290.55$-$292.1 GHz emission, respectively, over the OTF map of each source. 
As an example, the spectra of the 20 km/s cloud and G0.411+0.050 are shown in Fig. \ref{ExampleSpectra}.
In addition to the six p$-$H$_{2}$CO lines in the two frequency ranges, we marked the transitions HC$_{3}$N(24$-$23), 
CH$_{3}$OH(4$_{2,2}-$3$_{1,2}$), and o$-$H$_{2}$CO(4$_{3,2}-$3$_{3,1}$). The spectra of all sources are presented 
in Fig. \ref{Spectra} in the Appendix. To better compare the strength of the different lines, we chose the same intensity scale for all spectra.

The H$_{2}$CO emission at both frequencies is strongest in the 20 and 50 km/s clouds. Here, all six p$-$H$_{2}$CO
lines are well-detected. In G0.411+0.050 and Sgr D, the H$_{2}$CO emission at 291 GHz is almost not detected 
in the average spectra.

In Fig. \ref{Comp218290-Spectra} in the Appendix, we overplot the H$_{2}$CO(3$_{0,3}-$2$_{0,2}$) and the H$_{2}$CO(4$_{0,4}-$3$_{0,3}$) 
spectra for the 20 and 50 km/s clouds. The figure shows that the line width of the two transitions is comparable.

 \subsection{Methanol contamination}
 
 The H$_{2}$CO(3$_{2,2}-$2$_{2,1}$)  and the CH$_{3}$OH(4$_{2,2}-$3$_{1,2}$) transitions are separated by 35 MHz, corresponding to 
 25.5 km s$^{-1}$. These two lines are well-resolved and well-separated in our data. However, in sources for which the range of velocity 
 components extends over more than 26 km s$^{-1}$, the integrated intensity maps of 
 the H$_{2}$CO(3$_{2,2}-$2$_{2,1}$) line are contaminated by 
 methanol emission for certain velocity components (see Fig.  \ref{Methanol-Contamination} in the Appendix). We thus excluded this 
 line from the further analysis.

\section{Analysis}  
\label{Analysis}

 \subsection{Velocity ranges}
 \label{VelRanges}
 
 \captionsetup[subfloat]{labelformat=empty, labelsep=colon}

To determine the whole velocity range covered by H$_2$CO emission for each source, we used the 
218 GHz spectra (Fig. \ref{Spectra}, Sect. \ref{H2COSpectra}) and zoomed in 
on the H$_{2}$CO(3$_{0,3}-$2$_{0,2}$) line. We fitted a linear baseline 
to line-free channels close to the line to obtain the average rms level in each OTF map. 
For each source, we determined the velocity range where the 
line emission was above 3$\sigma$ (Col. 9 in Table \ref{SourceCoord}, Fig. \ref{SpectraWholeVelRange} in the Appendix). 
For Sgr C and Sgr D, we followed the same procedure but used the transition H$_{2}$CO(4$_{0,4}-$3$_{0,3}$).\\

To detect different velocity components, we subdivided the OTF maps into small tiles of size 33$\arcsec$~x~33$\arcsec$,
slightly larger than the beam, and averaged the H$_{2}$CO(3$_{0,3}-$2$_{0,2}$) spectra over 
these areas. We fitted the line with several Gaussians if more than one velocity component was present and 
determined the central velocity of each component in each tile. 

For each velocity component in each source, we averaged the H$_{2}$CO(3$_{0,3}-$2$_{0,2}$) spectra 
(H$_{2}$CO(4$_{0,4}-$3$_{0,3}$) in the case of Sgr C and Sgr D) over 
the area where this component dominated and again estimated the central velocity of the line in the averaged spectrum.
To separate the components well spatially and spectrally in the following analysis, we set the velocity ranges
over which the H$_{2}$CO emission is integrated as $\pm$~3 km s$^{-1}$ around the central velocity of each 
component (Col. 10 in Table \ref{SourceCoord}).

We detected velocity gradients in three of our clouds. In G0.253+0.016, the average velocities range from $\sim$40 km s$^{-1}$ 
in the south-west of the cloud to $\sim$0 km s$^{-1}$ in the North-East. We identified four spatially distinct velocity components 
at 0, 19, 39, and 78 km s$^{-1}$ in this source. While the emission of the first three velocity components peaks towards 
the cloud, the emission of the 78 km s$^{-1}$ component originates from the south-east border of the cloud and might not be 
associated with this cloud.

In the 20 km/s cloud, the velocity gradient spreads over $\sim$30 km s$^{-1}$ from the south-east end of the cloud to the 
north-west part. The three velocity components we identified in this source are at 3, 11, and 30 km s$^{-1}$. Most of the 50 
km/s cloud is at $\sim$45 km s$^{-1}$ but there are higher velocities detected at the north and lower velocities at the west
end of the cloud. The velocity gradient also spans about $\sim$15 km s$^{-1}$. We could distinguish between two 
velocity components at 44 and 60 km s$^{-1}$. 

We identified only one velocity component in the other clouds. The H$_{2}$CO emission peaks at 22 km s$^{-1}$ in G0.411+0.050, 
30 km s$^{-1}$ in G0.480$-$0.006, $-$52 km s$^{-1}$ in Sgr C and $-$16 km s$^{-1}$ in Sgr D.

 \subsection{Integrated intensity ratio and uncertainty maps}

For each H$_{2}$CO transition in Table \ref{H2COTrans}, we produced calibrated data cubes on a grid with a resolution of 1 km s$^{-1}$.
The cubes are smoothed to a resolution of 33$\arcsec$. The pixel size is 11$\arcsec$.
For each pixel of these cubes, we fitted linear baselines to line-free velocities at both sides of the lines. For each line and 
all velocity components of our targets, we produced maps of the uncertainty of the line intensity by assigning every pixel of the map the 
corresponding $\sigma(T_{mb})$ from the baseline fitting. 
The median of $\sigma(T_{mb})$ of the maps ranges between 35 and 59 mK per pixel at 218 GHz and 30 and 46 mK 
per pixel at 291 GHz (Cols. 7, 8 in Table \ref{SourceCoord}).

As described in Sect. \ref{Intro}, the relative intensities of para-H$_{2}$CO lines yield estimates of the temperature and density 
of the gas. We selected the following key line ratios 

\begin{gather*}
R_{321}~=~\frac{\int I_{H_{2}CO(3_{2,1}-2_{2,0})}~d\textnormal{v}}{\int I_{H_{2}CO(3_{0,3}-2_{0,2})}~d\textnormal{v}} \hspace{1cm} \Rightarrow  \rm{} Temperature\\
R_{422}~=~\frac{\int I_{H_{2}CO(4_{2,2}-3_{2,1})}~d\textnormal{v}}{\int I_{H_{2}CO(4_{0,4}-3_{0,3})}~d\textnormal{v}}   \hspace{1cm} \Rightarrow  \rm{} Temperature\\
R_{404}~=~\frac{\int I_{H_{2}CO(4_{0,4}-3_{0,3})}~d\textnormal{v}}{\int I_{H_{2}CO(3_{0,3}-2_{0,2})}~d\textnormal{v}} \hspace{1cm} \Rightarrow  \rm{} Density\\
\end{gather*}

For a better readability of the paper, we will abbreviate these ratios with R$_{321}$, R$_{404}$, and R$_{422}$, respectively, 
in the following text.

For each pixel in the data cubes, we integrated the emission of the para-H$_{2}$CO lines independently over the same velocity 
range v$_{range}$, yielding separate integrated intensity maps for the different lines. We produced these maps for all velocity components 
of our targets (v$_{range}$ in Cols. 9, 10 in Table \ref{SourceCoord}; Figs. \ref{20kms-Int-H2CO} $-$ \ref{SGRD-Int-H2CO} in the Appendix).

We identified the pixels in the OTF maps of each source where the integrated emission of the H$_{2}$CO(3$_{0,3}-$2$_{0,2}$) and 
the H$_{2}$CO(4$_{0,4}-$3$_{0,3}$) lines, respectively, was above the 
${5\cdot\sigma(T_{mb})\cdot\sqrt{\frac{\textnormal{v}_{range}}{\Delta \textnormal{v}_{res}}}\cdot\Delta \textnormal{v}_{res}}$ threshold. 
The parameter $\Delta$v$_{res}$ is the velocity resolution of our data. 
For these pixels, we determined the aforementioned ratios from the integrated intensity maps of these lines, yielding integrated 
intensity ratio maps (Figs. \ref{20kms-All-Ratio-H2CO} $-$ \ref{SGRD-All-Ratio-H2CO} in the Appendix).

For positions where the integrated emission of the line in the numerator was below the 5$\sigma$ threshold, the integrated intensity value was 
exchanged for the value ${5\cdot\sigma(T_{mb})\cdot\sqrt{\frac{\textnormal{v}_{range}}{\Delta \textnormal{v}_{res}}}\cdot\Delta \textnormal{v}_{res}}$, representing upper limits of the ratios (marked with Xs in the maps). 
For each pixel in the ratio maps (except upper limits), we then determined the uncertainty of the ratio via Gaussian error propagation, 
using the uncertainty maps of the intensity of the two corresponding lines. The remaining pixels (corresponding to the pixels with 
upper limits in the ratio maps) were blanked in grey. In two velocity components of two 
sources, all or almost all pixels are upper limits (G0.253+0.016: 16$-$22 km/s, 291 GHz; 75$-$81, 218 and 291 GHz; 
G0.411+0.050: both velocity components, 291 GHz).

\section{Modeling\label{Modeling}}
\subsection{Fundamental Approach}

\begin{figure*}
	\caption{Left: R$_{321}$ as a function of kinetic temperature T$_{kin}$ and density n. The plot shows that the ratio is mostly dependent 
	on temperature and independent of the density of the gas. Right: R$_{404}$ as a function of kinetic temperature T$_{kin}$ and density n. 
	The plot shows the strong dependence of the ratio on both temperature and density for densities $<$ 10$^{7}$ cm$^{-3}$. The red 
	plus signs indicate locations where the observed line ratios R$_{321}$ and R$_{404}$ are matched at the same time.}
	\centering
	\subfloat{\includegraphics[width=9.5cm]{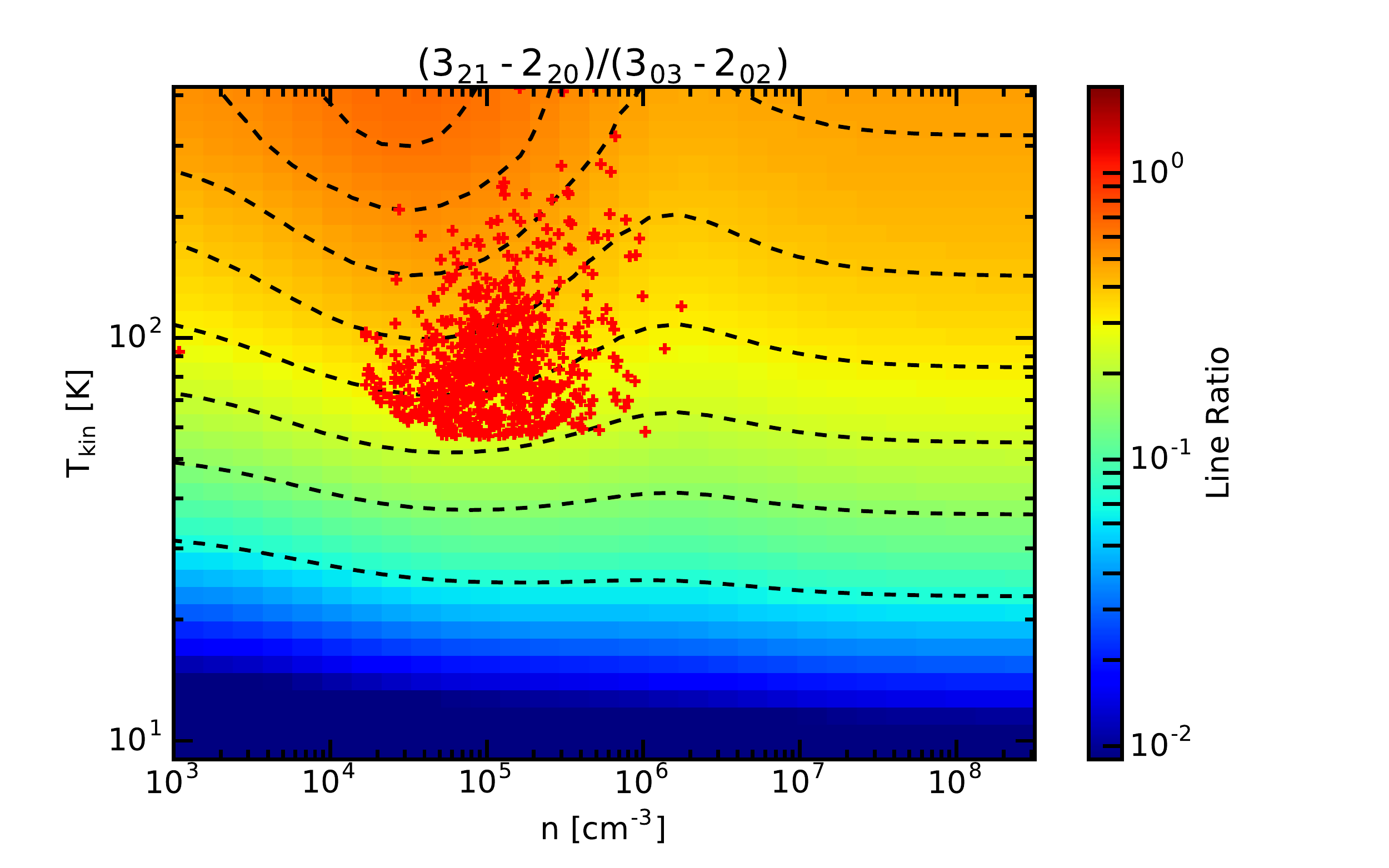}}
	\subfloat{\includegraphics[width=9.5cm]{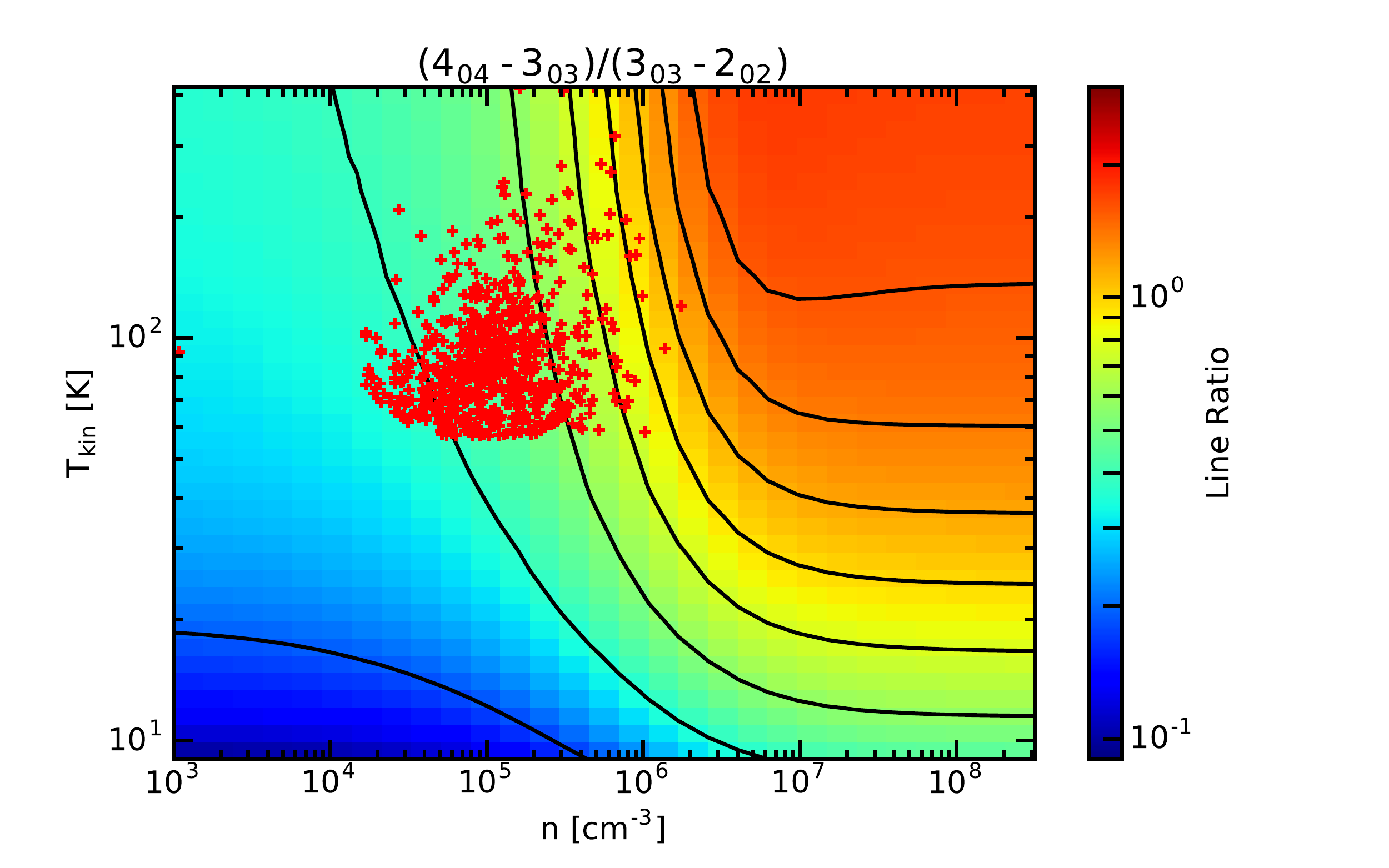}}
	\label{TempDensity}
\end{figure*}

We use RADEX \citep{vanderTak2007} and a related
solver (Fujun Du's myRadex; \url{https://github.com/fjdu/myRadex}) to
create model intensities for the para-H$_{2}$CO lines. The
calculations are executed using pyradex, a Python--based wrapper for
RADEX (developed by A.~Ginsburg;
\url{https://github.com/keflavich/pyradex}). The collision rates used
here \citep[from][]{Wiesenfeld2013} are retrieved
from the LAMDA data base \citep{Schoier2005}. This
approach is conceptually identical to the one employed by \citet{Ginsburg2016}, 
though we choose slightly different gas
properties to solve the radiative transfer problem.

We adopt a fiducial set of conditions for our calculations of emission
line intensities and ratios. Below we show that our temperature
estimates are robust with respect to the choice of values for these
parameters.  We assume $\rm{}H_2$ particle densities of
$10^5~\rm{}cm^{-3}$. This choice is motivated by the average densities of a
few $10^4~\rm{}cm^{-3}$ as estimated for our targets via the analysis
of dust emission maps (Kauffmann et al., subm.). The column density of
para-H$_{2}$CO is set to $5\times{}10^{13}~\rm{}cm^{-2}$. For a
representative $\rm{}H_2$ column density of
$5\times{}10^{22}~\rm{}cm^{-2}$ (Kauffmann et al., subm.) this
corresponds to an abundance of $10^{-9}$ with respect to $\rm{}H_2$, a value
that is suggested by previous work \citep[for a summary, see][]{Ginsburg2016}. We adopt a velocity
dispersion of $5~\rm{}km\,s^{-1}$. This value is appropriate for
structures with a radius of order 1~pc, which is about the diameter of
our beam. The calculations are executed assuming a spherical geometry.

We compute the intensities of para--$\rm{}H_2CO$ emission lines under
the aforementioned conditions. These intensities can be used to
calculate the R$_{321}$ and R$_{422}$ line ratios. These line ratios
are known to be highly sensitive to the temperature while being rather
insensitive to changes in other parameters \citep[e.g.][see left panel of Fig. \ref{TempDensity} for R$_{321}$]{Mangum1993}.  
This trend is confirmed in calculations
discussed below. For this reason, the R$_{321}$ and R$_{422}$ line
ratios can be used to estimate the kinetic gas
temperature.

In essence, we use the aforementioned RADEX calculations for kinetic
gas temperatures of 10 to 300~K to obtain functions that
yield estimates of the kinetic gas temperature for given R$_{321}$ or
R$_{422}$. The ratios range from 0 to values larger than 0.5 for this
temperature range. The resulting function for R$_{321}$ is
almost identical to the one used by \citet{Ginsburg2016}. This is in particular remarkable because
Ginsburg et al.\ use large velocity gradient (LVG) calculations where
we choose to adopt spherical model geometries. Here, we opt for the latter 
modeling geometry since it is close to the naive picture of a cloud with a well-defined 
boundary that is immersed in tenuous material. The comparison with the LVG case shows, however,
that our analysis is robust with respect to the adopted modeling geometry.

\subsection{Systematic Uncertainties}
\label{SystematicUncertainties}
We assess the robustness of this method by repeating the line ratio
calculations for different choices in gas properties. We vary the
velocity dispersion between 2 and $10~\rm{}km\,s^{-1}$, increase and
decrease in column density by a factor of 10, explore lower densities of
$10^4~\rm{}cm^{-3}$, and examine the results for the slab and LVG
models available within RADEX (Fig. \ref{TempUncertainties}). This shows that, for a given line ratio,
none of these modifications lead to temperatures that deviate by more
than $\pm{}30\%$ from the result obtained using our fiducial case (solid blue line in 
Fig. \ref{TempUncertainties}). For
reference, we note that spherical model geometries give the lowest
temperatures, followed by slightly elevated values for LVG
conditions. Slab geometries, however, give for a fixed line ratio
temperatures that are up to 30\% higher than those found for the fiducial
case. Note in particular that these calculations explore the density 
range between 10$^{4}$ and 10$^{5}$ cm$^{-3}$ that is suggested by the 
observations of dust emission discussed by Kauffmann et al. (subm.). 
Variations in density by a factor 10 have no significant impact on our results.

\medskip
\medskip
\noindent In the high column density Galactic center environment, the H$_{2}$CO(3$_{0,3}-$2$_{0,2}$) emission might be optically 
thick. This saturation would cause a falsely high 
R$_{321}$ line ratio in our integrated intensity ratio maps. A
 higher ratio implies higher temperatures. Thus, an optically thick H$_{2}$CO(3$_{0,3}-$2$_{0,2}$) line would mimic higher gas 
temperatures in the clouds.

We have executed more detailed experiments for selected positions in
order to understand the impact of the line optical depth. For example,
we modeled the emission from the $20~\rm{}km\,s^{-1}$ cloud in the
velocity slice of 8 to $14~\rm{}km\,s^{-1}$ (Fig. \ref{TempOpacity}, left panel). 
The 3$_{0,3}-$2$_{0,2}$
transition has a velocity--integrated peak intensity of
$15~\rm{}K\,km\,s^{-1}$ (blue line in top panel of Fig. \ref{TempOpacity}). 
At this peak position the 4$_{0,4}-$3$_{0,3}$
transition has an integrated intensity of $4~\rm{}K\,km\,s^{-1}$ (blue line in bottom panel of Fig. \ref{TempOpacity}), 
while we find
$R_{321}=0.25$ and $R_{422}=0.35$ (red line in top and bottom panel, respectively, 
of Fig. \ref{TempOpacity}). 
Further, the analysis of dust emission maps (Kauffmann et al., subm.) indicates 
H$_{2}$ densities below 10$^{5}$ cm$^{-3}$. This limits the parameter space 
explored in our radiative transfer experiments. In our modeling, 
we thus require that the
model intensities match the observed ones for $\rm{}H_2$ densities
below $10^5~\rm{}cm^{-3}$. For these constraints we find that the data
for the transitions near 218~GHz can only be matched if we increase
the column density to a multiple of the fiducial value (Fig. \ref{TempOpacity}, right panel,
for an increased column density of 5$\cdot$10$^{14}$ cm$^{-2}$). The optical depth of the 
3$_{0,3}-$2$_{0,2}$ line is larger than one under these conditions. One might think that the
high optical depth leads to massive biases in temperature
estimates. This is, however, not the case: given a fixed line ratio,
conditions with high optical depth can imply temperatures lower by
30\% for optical depths $\gtrsim{}5$ (i.e., our procedure, which
assumes low optical depth, \emph{overestimates} gas temperatures in
regions of high optical depth). 

\subsection{Observational Uncertainties}
Further uncertainties in the temperature result from observational
uncertainties in line ratios due to noise. To handle these
uncertainties we first obtain two additional temperature estimates per
map position. These two estimates are calculated for line ratios that
are equal to the observed value plus or minus the observed uncertainty
in the line ratio. Conceptually these values bracket the range of
possible temperatures for a given map position. The difference between
these values represents the change in kinetic gas temperatures when
changing the line ratio by \emph{twice} its uncertainty (i.e., from
$R_i-\sigma[R_i]$ to $R_i+\sigma[R_i]$, where $R_i$ and $\sigma[R_i]$
are the line ratio and its uncertainty). \emph{Half} of this
difference can thus be considered to represent the average temperature
uncertainty associated with a line ratio uncertainty
$\pm{}\sigma(R_i)$. In our analysis we thus calculate for every map
position the temperatures corresponding to $R_i\pm{}\sigma[R_i]$, and
we report half of the difference between these temperatures as the
observational uncertainty in kinetic gas temperatures. We are aware that 
the temperature uncertainties are asymmetric around the temperature value.
However, we decided that the chosen method is the best way to present the 
uncertainties in the paper.

For some locations, the observations only yield upper limits to the line ratios. 
In such cases we take this upper limit to $R_i$, we
calculate the associated temperature, and we report this resulting
value as an upper limit to the kinetic gas temperature.

\section{Discussion}
\label{TempMeasurements}

\begin{figure*}
	\caption{Summary of the produced maps for the 8$-$14 km s$^{-1}$ component of the 20 km/s cloud. The contours show the moment 0 map of the H$_{2}$CO(3$_{0,3}-$2$_{0,2}$) transition, 
	produced over the whole velocity range of the source (levels: 30\%$-$90\% of the maximum in steps of 10\%). 
	The circle in the lower left corner shows the 33$\arcsec$ beam.}
	\centering
	H$_{2}$CO integrated intensity maps (from 
	left to right: H$_{2}$CO(3$_{0,3}-$2$_{0,2}$), H$_{2}$CO(3$_{2,1}-$2$_{2,0}$), H$_{2}$CO(4$_{0,4}-$3$_{0,3}$), and 
	H$_{2}$CO(4$_{2,2}-$3$_{2,1}$)) \\
	\subfloat{\includegraphics[bb = 0 0 650 560, clip, height=4.3cm]{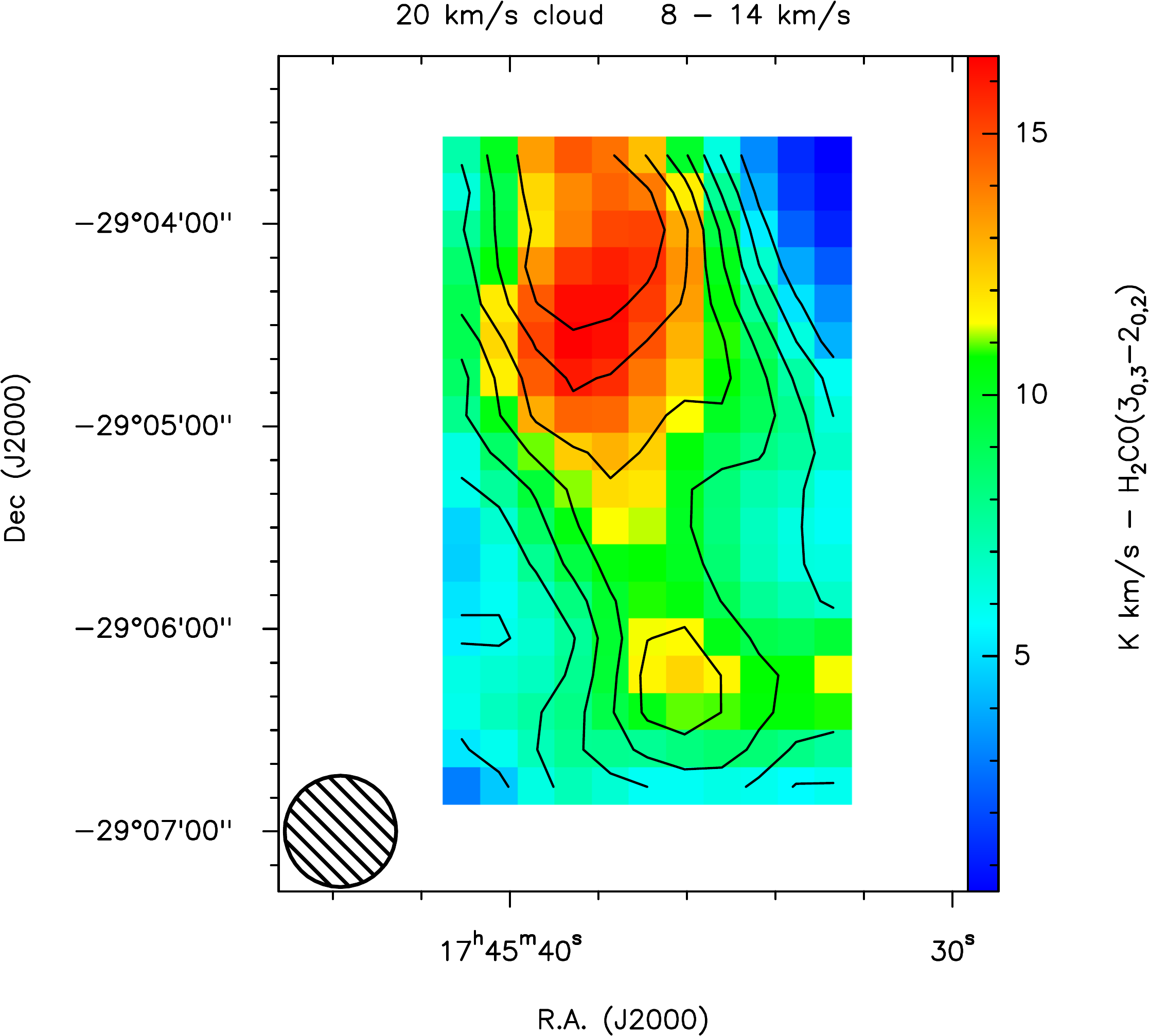}}
	\subfloat{\includegraphics[bb = 140 0 650 560, clip, height=4.3cm]{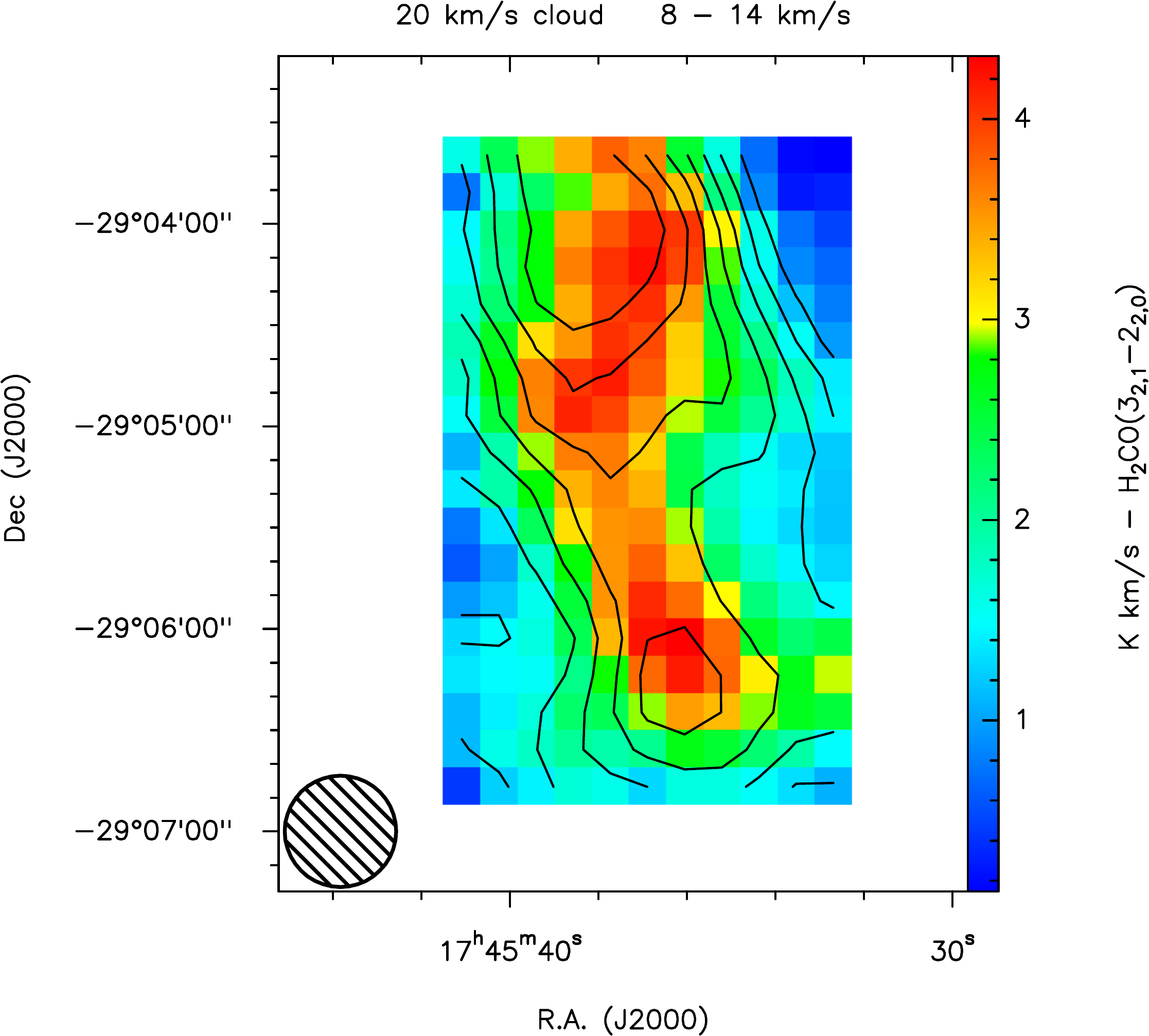}}
	\subfloat{\includegraphics[bb = 140 0 650 560, clip, height=4.3cm]{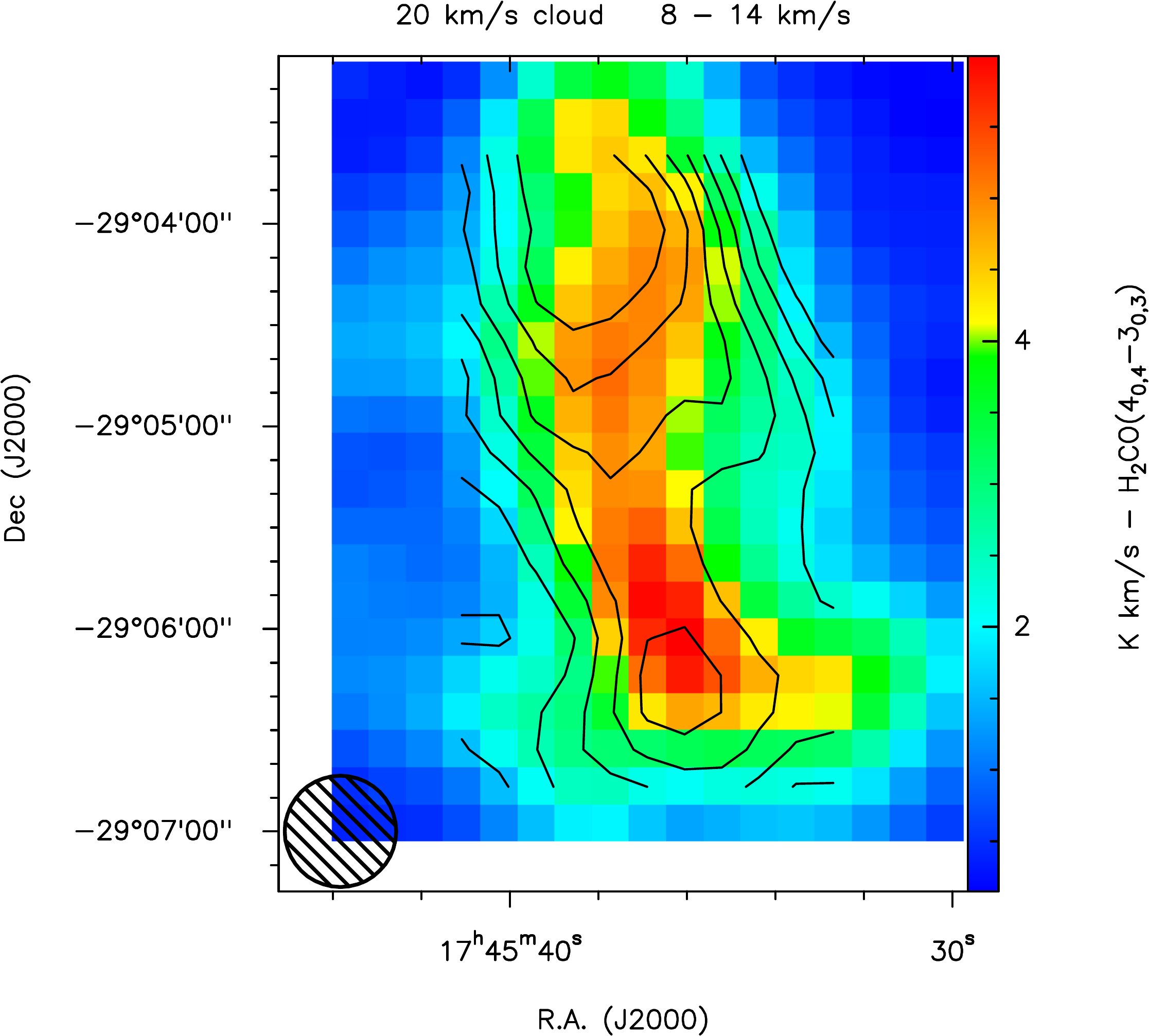}}
	\subfloat{\includegraphics[bb = 140 0 650 560, clip, height=4.3cm]{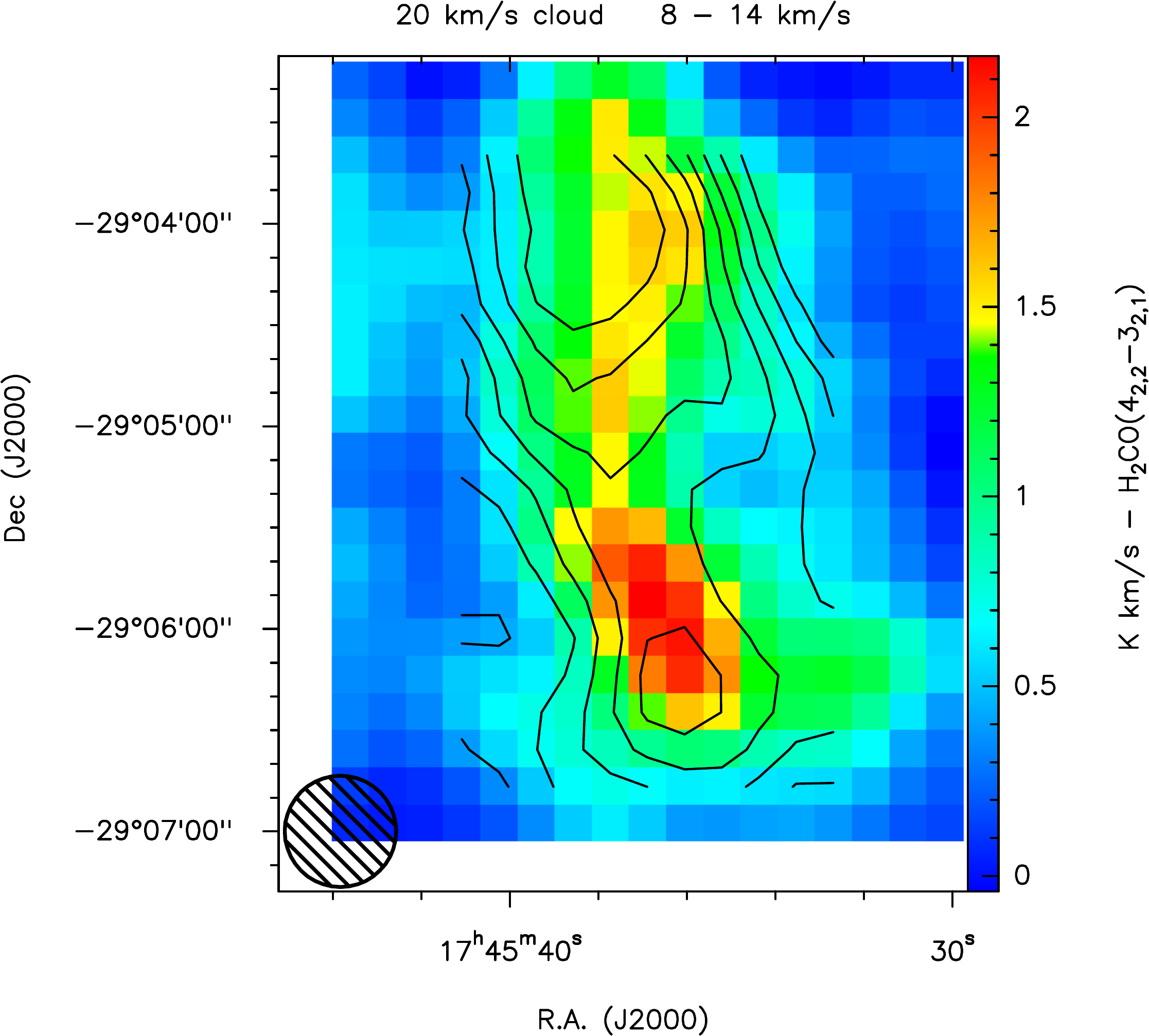}}\\
	H$_{2}$CO ratio (upper panels) and uncertainty maps (lower panels) (from left to right: 
	R$_{321}$, R$_{422}$, and R$_{404}$).\\
	\subfloat{\includegraphics[bb = 0 60 650 560, clip, height=3.84cm]{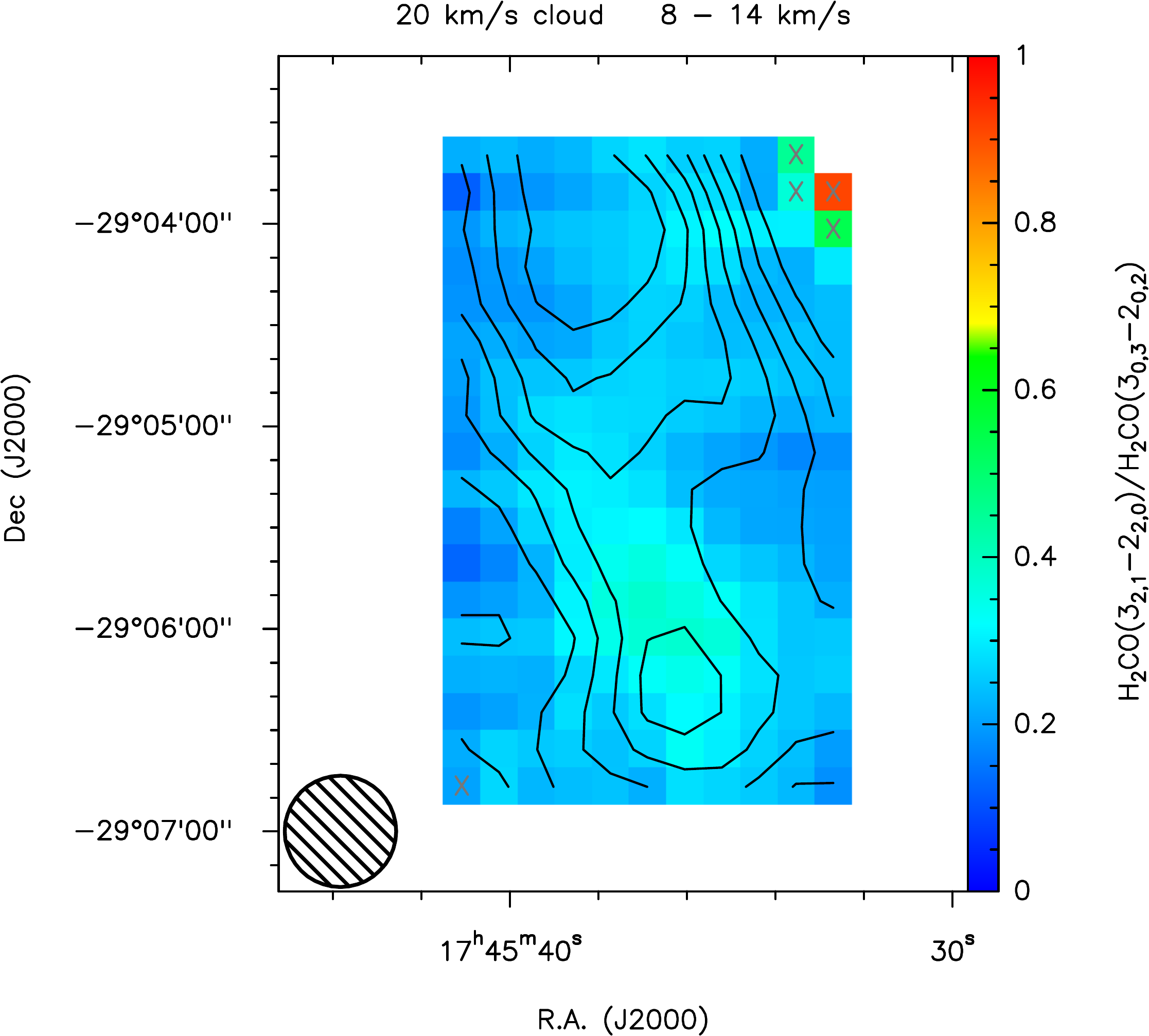}}
	\subfloat{\includegraphics[bb = 140 60 650 560, clip, height=3.84cm]{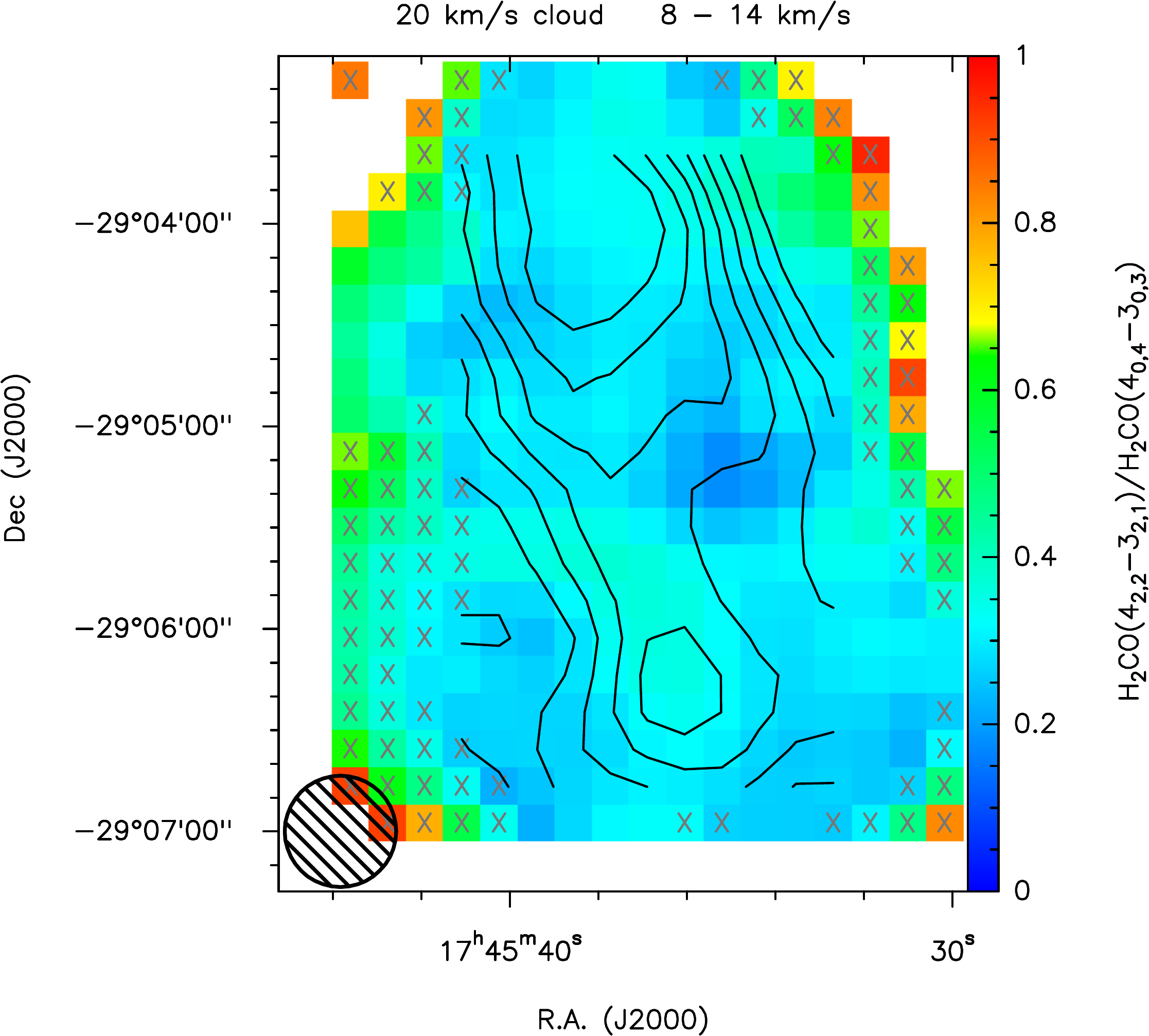}}
	\subfloat{\includegraphics[bb = 140 60 650 560, clip, height=3.84cm]{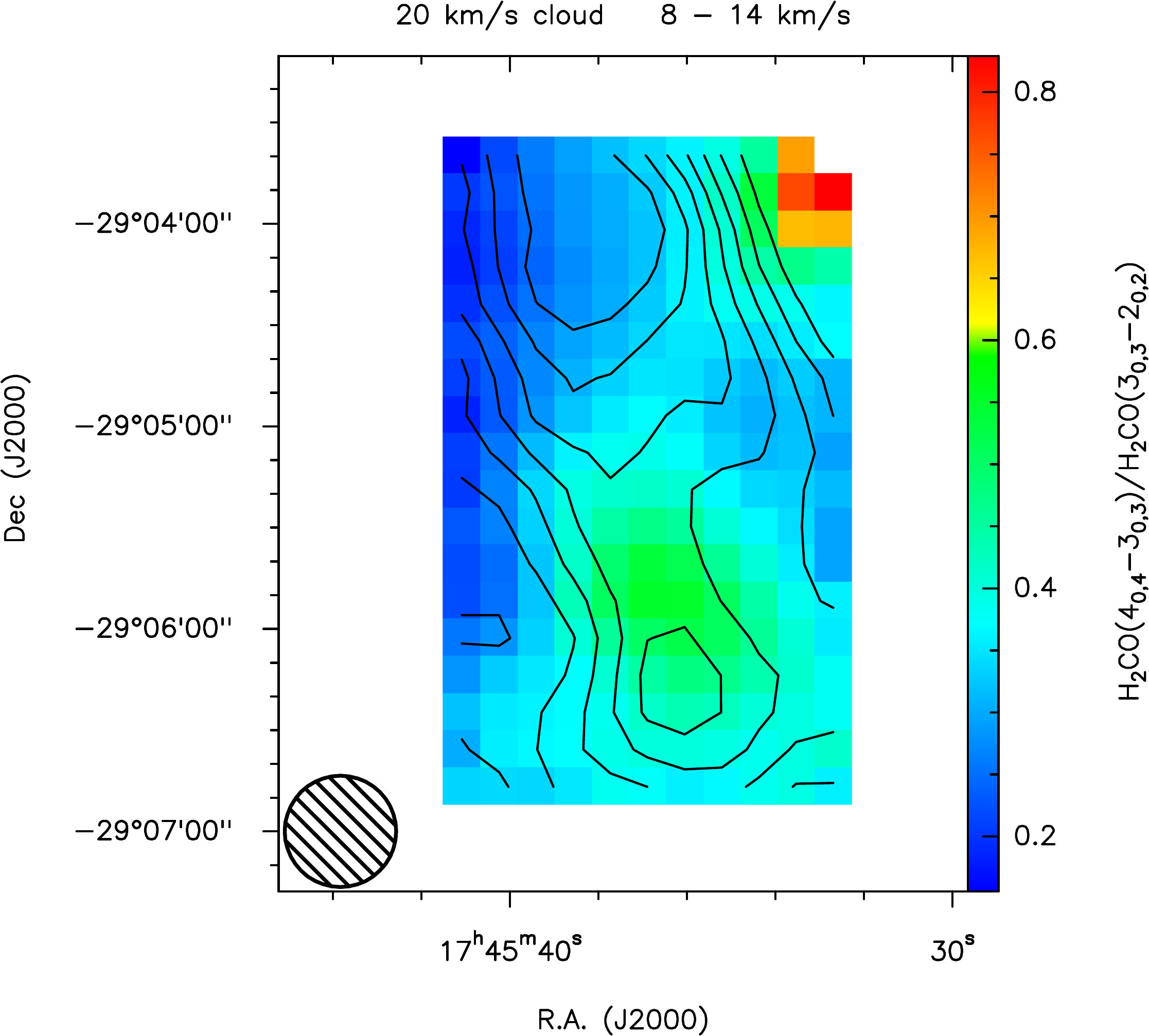}}\\
	\vspace{-0.5cm}
	\subfloat{\includegraphics[bb = 0 0 650 560, clip, height=4.3cm]{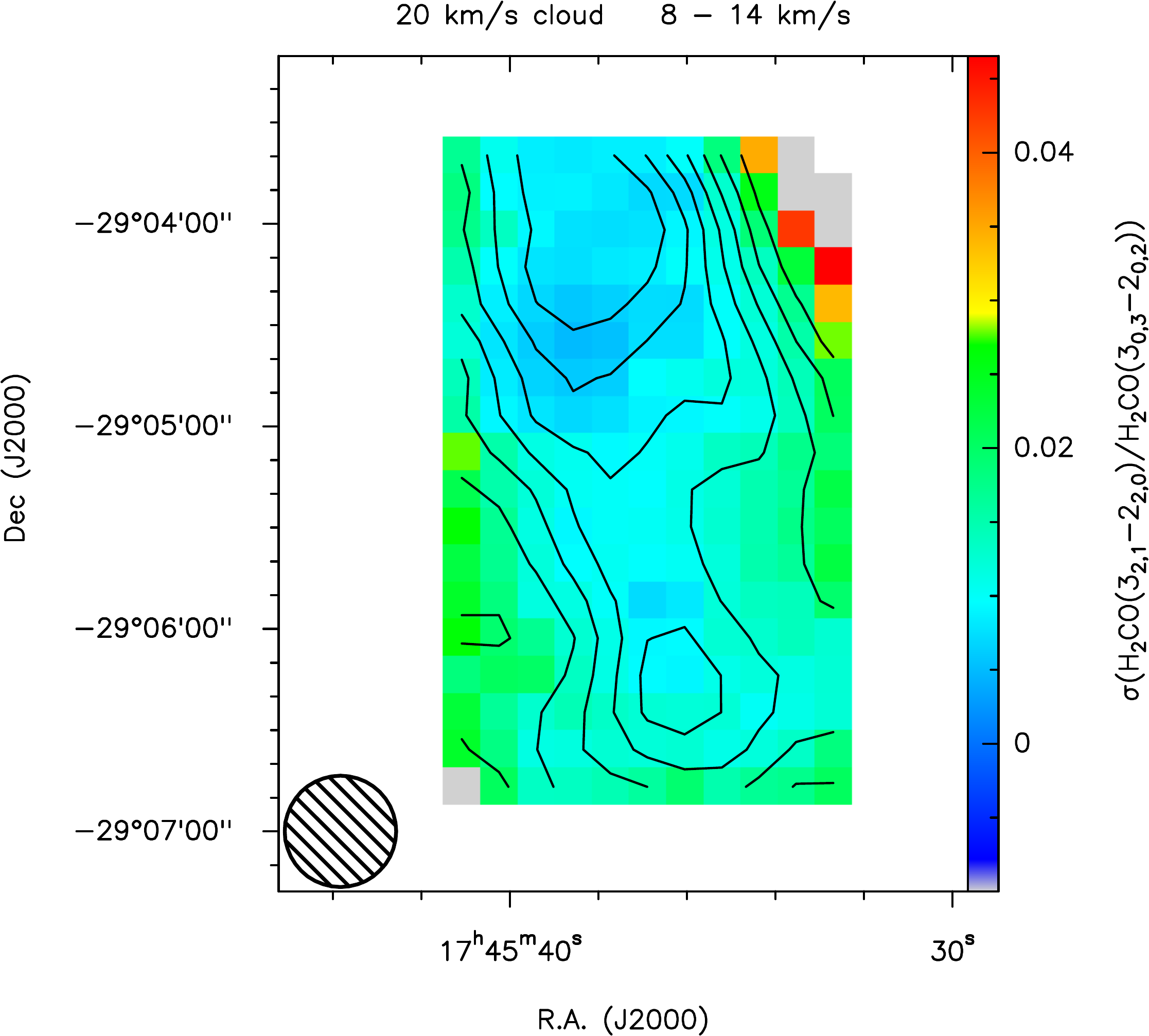}}
	\subfloat{\includegraphics[bb = 140 0 650 560, clip, height=4.3cm]{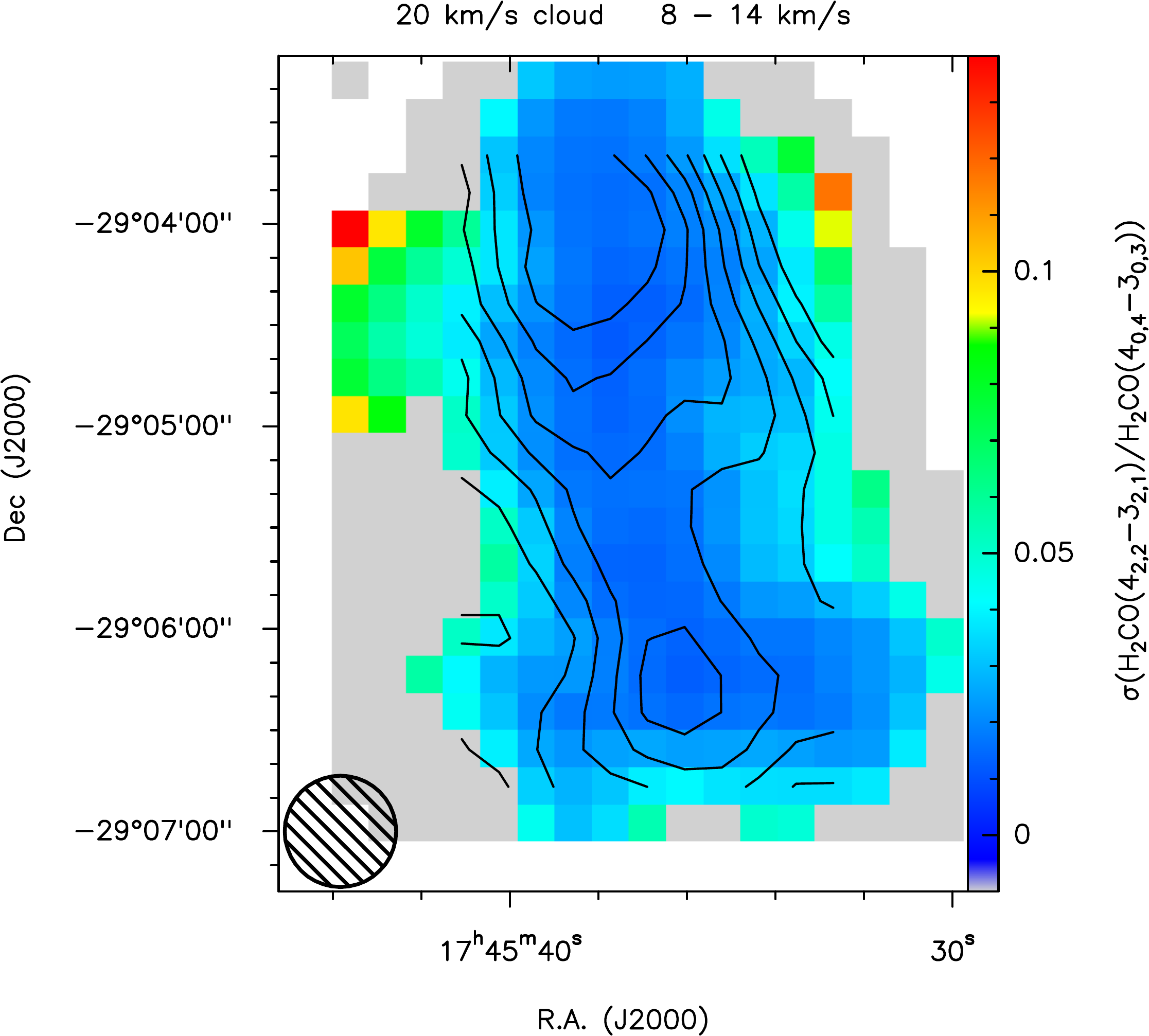}}
	\subfloat{\includegraphics[bb = 140 0 650 560, clip, height=4.3cm]{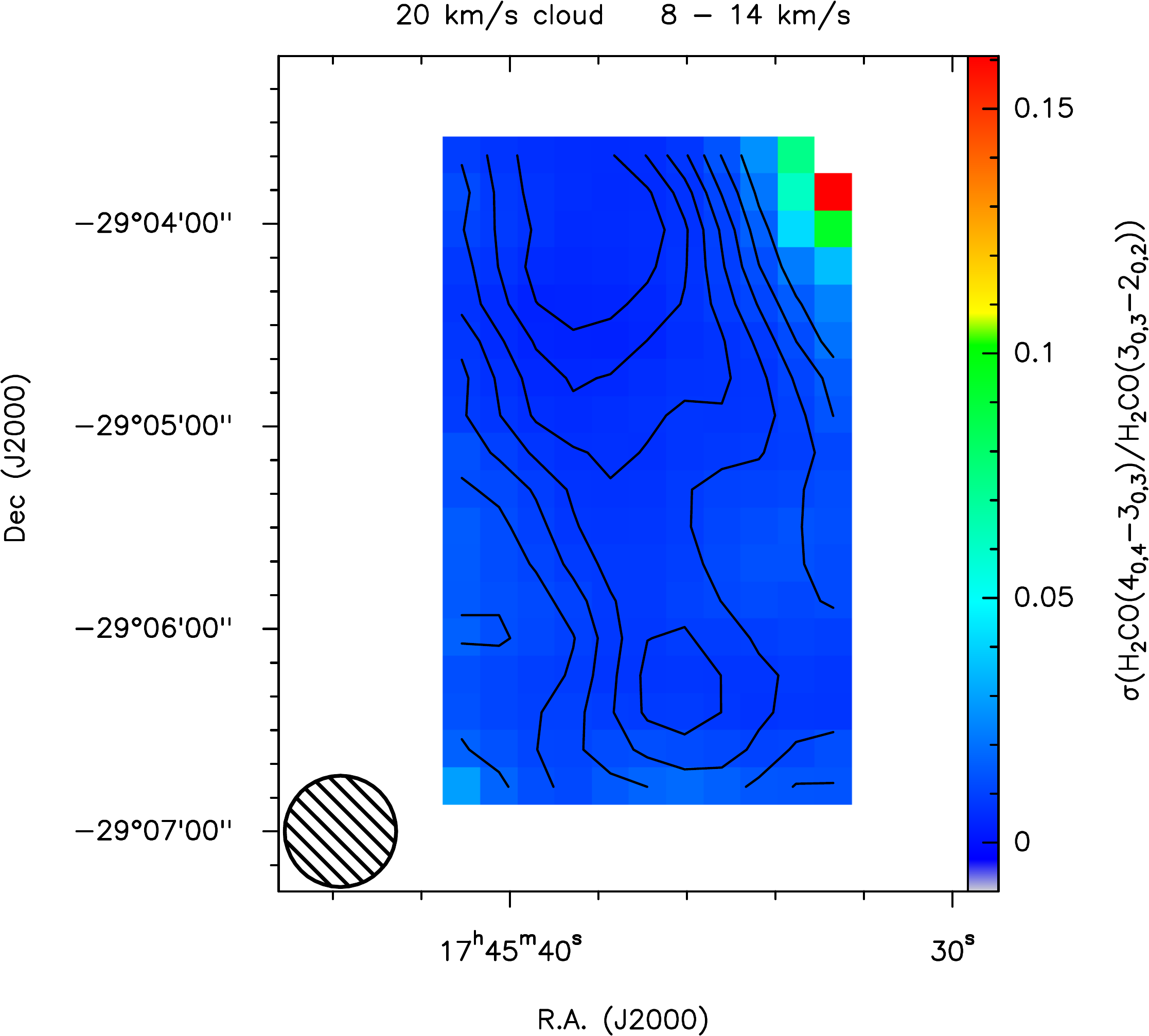}}\\
	\label{20kms-Ratio-H2CO}
	\begin{minipage}{0.6\textwidth}
	\centering
	H$_{2}$CO temperature (upper panels) and uncertainty maps (lower panels), derived from the 218 GHz (left panels) and 291 GHz data (right panels)\\
	\subfloat{\includegraphics[bb = 0 60 650 560, clip, height=3.84cm]{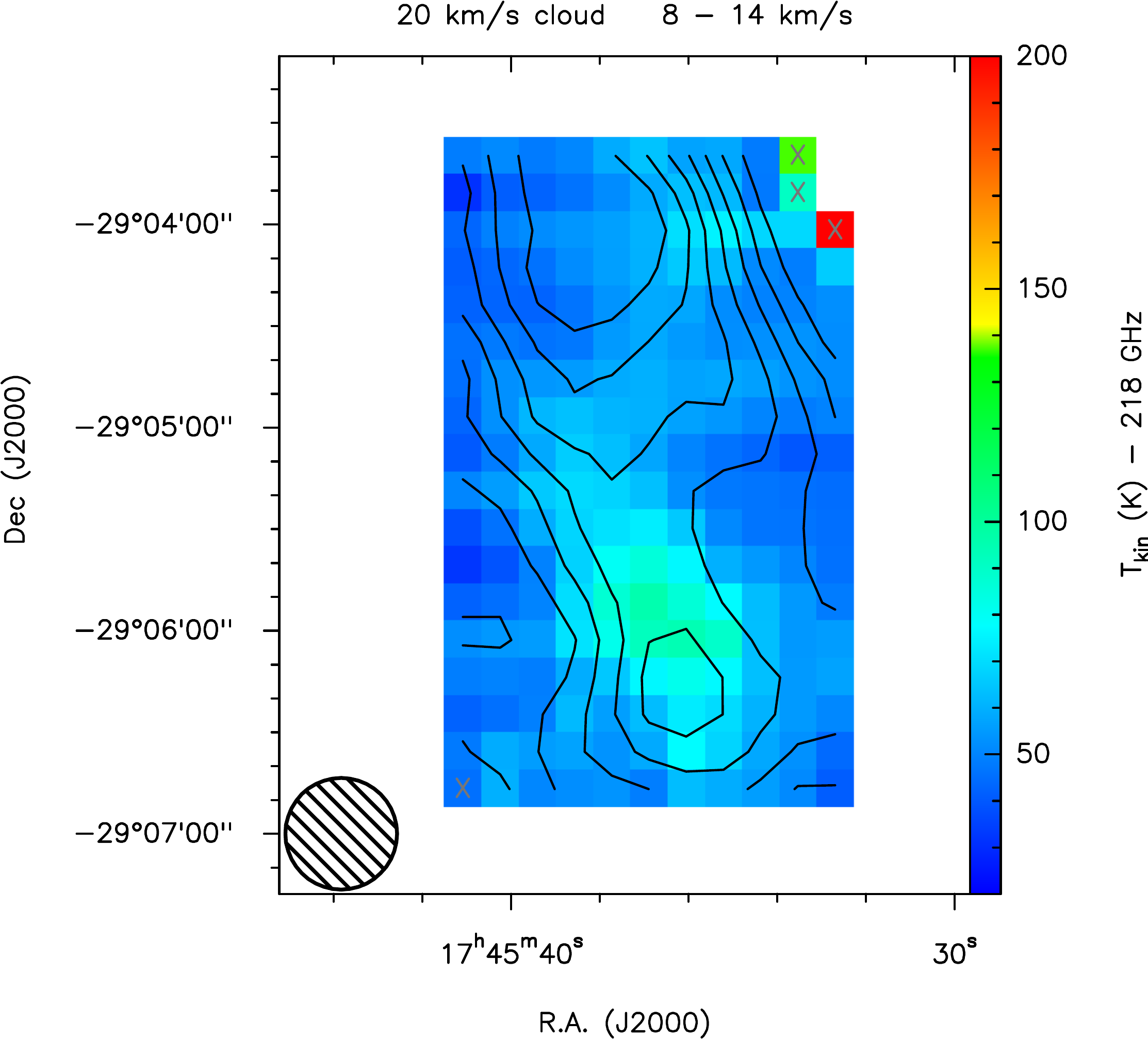}}
	\subfloat{\includegraphics[bb = 140 60 650 560, clip, height=3.84cm]{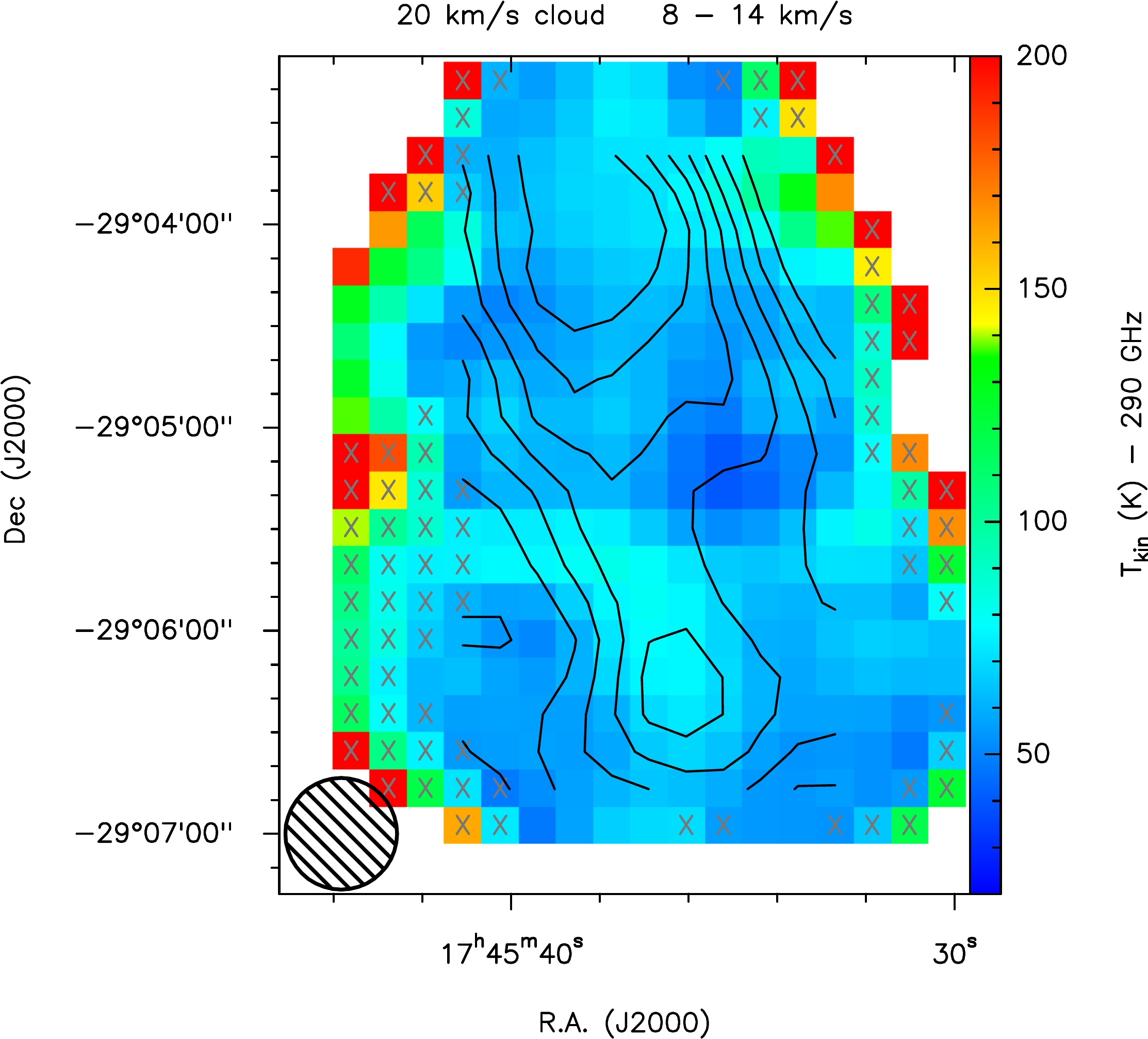}}\\
	\vspace{-0.5cm}
	\subfloat{\includegraphics[bb = 0 0 650 560, clip, height=4.3cm]{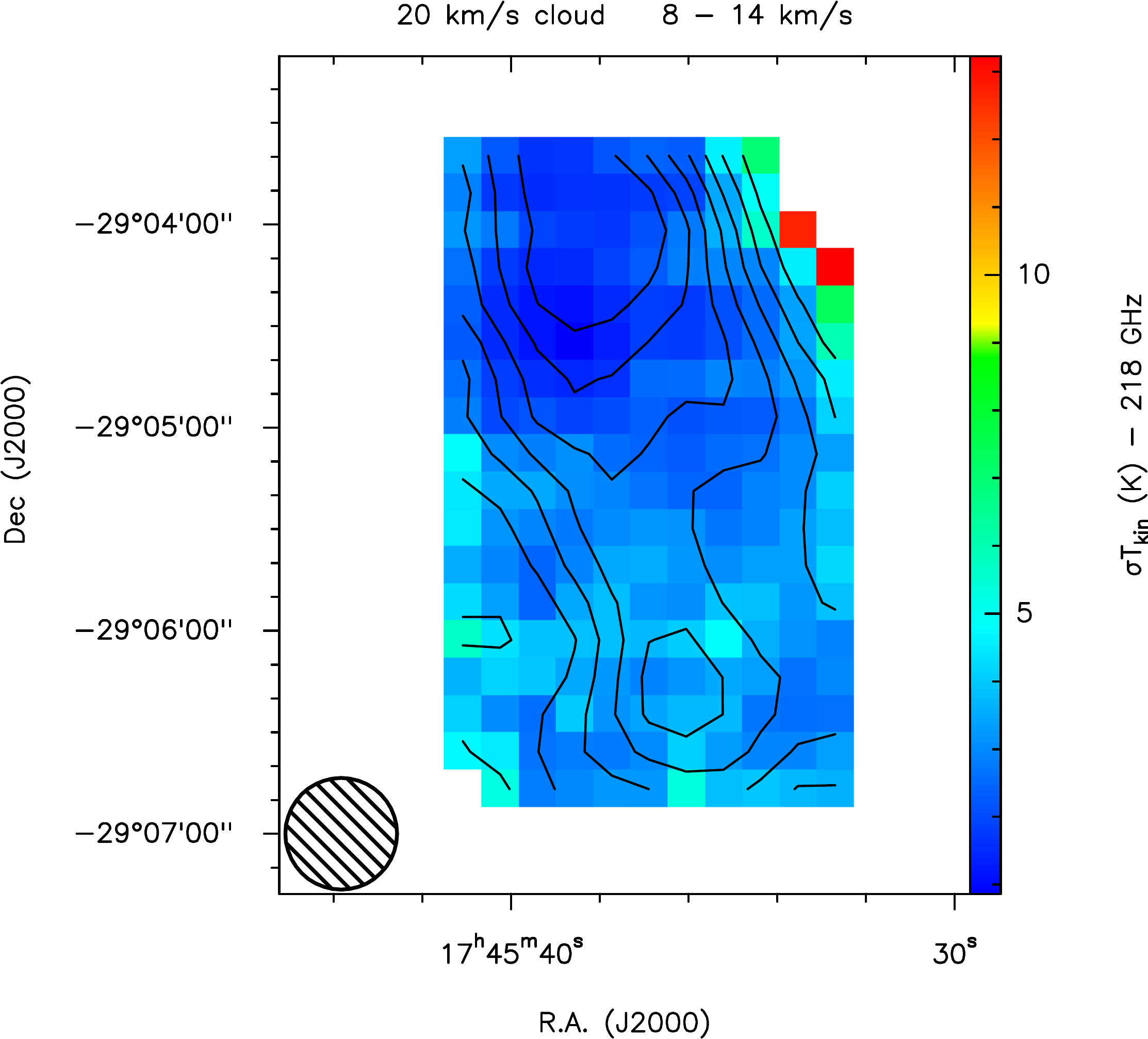}}
	\subfloat{\includegraphics[bb = 140 0 650 560, clip, height=4.3cm]{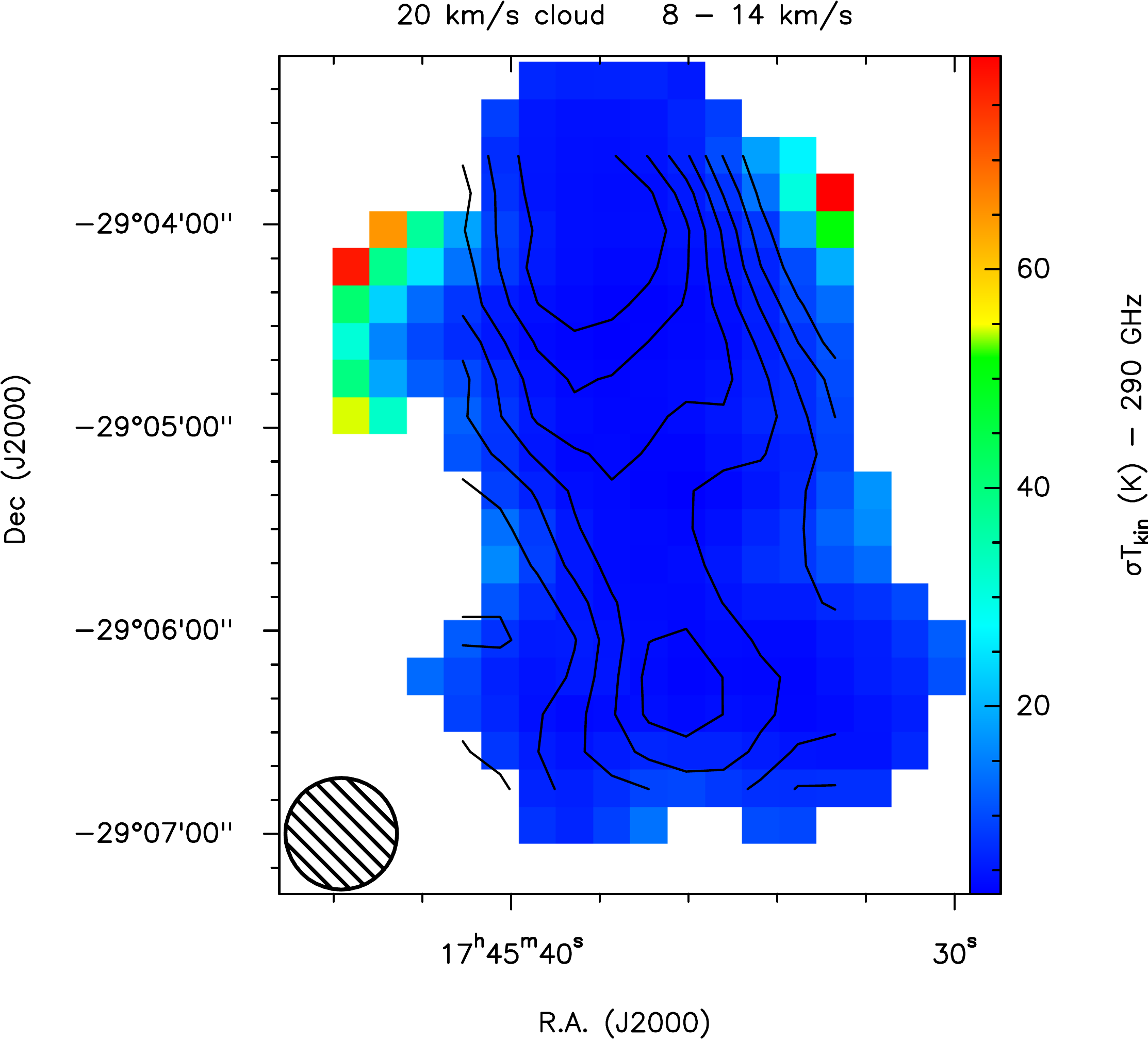}}
         \end{minipage}
	\begin{minipage}{0.35\textwidth}
	\centering
	Spitzer/GLIMPSE RGB image (blue = 3.6 $\mu$m, green = 4.5 $\mu$m, and red = 8.0 $\mu$m)
	\subfloat{\includegraphics[width=\textwidth]{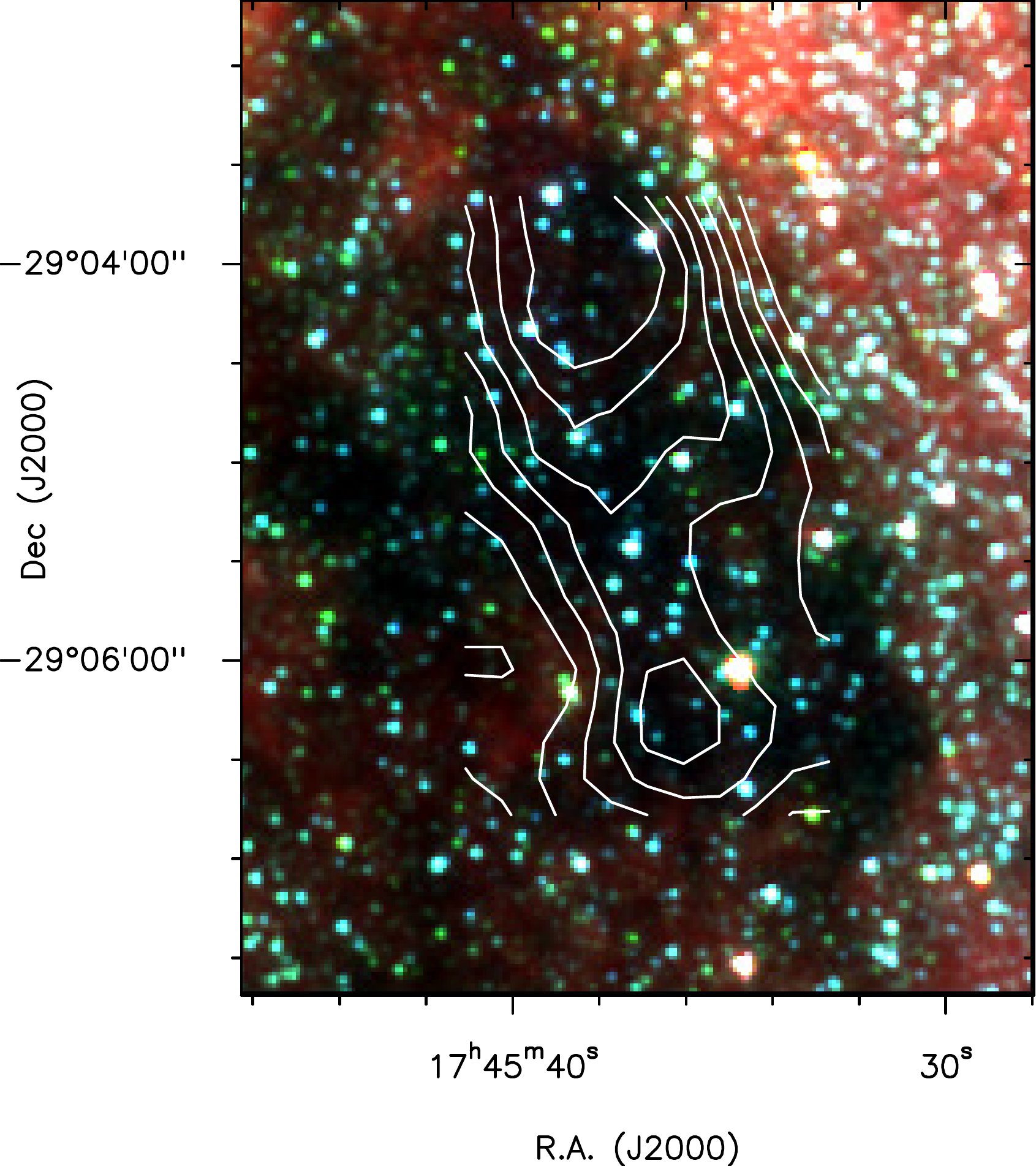}}\\
         \end{minipage}
         \label{20kms-cloud}
\end{figure*}

The temperature maps of all velocity components of all sources are shown in 
Figs. \ref{20kms-All-Temp-H2CO} $-$ \ref{SGRD-All-Temp-H2CO} in the Appendix. Upper limits 
in the temperature maps are marked again with Xs. The corresponding pixels in the uncertainty maps are blanked in white.
The measured temperatures range between 28 and 242 K at 218 GHz and between 37 and 252 K at 291 GHz (excluding 
upper limit values of the temperature). This shows that the gas temperatures are everywhere in the clouds much higher 
than the dust temperatures that are measured towards these clouds \citep[$\sim$25 K,][]{Molinari2011}.
However, temperatures above $\sim$150 K have to be considered as lower limits since the radiative transfer models 
start to diverge significantly at these temperatures and the para-H$_{2}$CO line ratios are 
intrinsically insensitive to higher temperatures \citep{Mangum1993, Ginsburg2016}.

We present the different plots we generated (integrated intensity maps, line ratio and uncertainty maps, temperature 
and uncertainty maps) for the 8$-$14 km s$^{-1}$ component of the 20 km/s cloud in Fig. \ref{20kms-cloud}. In the 
last panel, we show a three-color image of Spitzer/GLIMPSE data 
\citep[][blue = 3.6 $\mu$m, green = 4.5 $\mu$m, red = 8.0 $\mu$m]{Churchwell2009} of this source.

\subsection{Average Temperatures}

\begin{table*}
\centering
\caption{Minimal T$_{Min}$, maximal T$_{Max}$, and weighted average T$_{Average}$ temperatures as well as 1$\sigma$ uncertainties of the 
weighted average temperature for all temperature maps (i.e. each source, each velocity component, and each frequency).}
\begin{tabular}{ccccccccccc}
Source & Velocity Component & \multicolumn{4}{c}{218 GHz} &\multicolumn{4}{c}{291 GHz} & T$_{Ginsburg}$\\
& & T$_{Min}$ & T$_{Average}$ & $\sigma_{Average}$ & T$_{Max}$ & T$_{Min}$ & T$_{Average}$ & $\sigma_{Average}$  & T$_{Max}$ &  \\
& (km s$^{-1}$) & (K) & (K) & (K) & (K) & (K) & (K) & (K) & (K) & (K)\\ \hline
\multirow{4}{*}{20 km/s cloud} & $-$15 $-$ 36 & 28 & 58 & 16 & 133 & 37 & 62 & 11 & 163 & \multirow{4}{*}{64}\\
& 0 $-$ 6 & 36 & 70 & 16  & 183 & 43 & 71 & 9 & 115 & \\
& 8 $-$ 14 & 31 & 54 & 10 & 96 & 40 & 62 & 9 & 191 & \\
& 27 $-$ 33 & 31 & 48 & 11 & 93 & 45 & 62 & 13 & 199 & \\ 
\multirow{3}{*}{50 km/s cloud} &16 $-$ 80 & 30 & 82 & 29 & 242 & 41 & 82 & 14 & 223 & \multirow{3}{*}{91}\\
& 41 $-$ 47 & 44 & 87 & 18 & 142 & 53 & 91 & 15 & 150 & \\
& 57 $-$ 63 & 38 & 56 & 12 & 115 & 71 & 109 & 21 & 252 & \\ 
\multirow{5}{*}{G0.253+0.016} & $-$6~$-$~54 & 29 & 50 & 14 & 203 & 41 & 61 & 12 & 212 & \multirow{5}{*}{82} \\
&  $-$3 $-$ 3 & 61 & 81 & 16 & 180 & 56 & 77 & 17 & 140 & \\
& 16 $-$ 22 & 38 & 49 & 11 & 222 & 76 & 90 & 14 & 111 & \\
& 36 $-$ 42 & 42 & 72 & 21 & 189 & 47 & 69 & 16 & 210 & \\
& 75 $-$ 81 & 72 & 72 &  & 72 & 95 & 109 & 25 & 183 & \\ 
\multirow{2}{*}{G0.411+0.050} & 10 $-$ 30 & 41 & 49 & 8 & 132 & & & & & \multirow{2}{*}{57} \\
& 19 $-$ 25 & 49 & 59 & 11 & 197 & & & & &\\ 
\multirow{2}{*}{G0.480$-$0.006} & 19 $-$ 44 & 32 & 40 & 8 & 117 & 50 & 69 & 18 & 239 & \multirow{2}{*}{84}\\ 
& 27 $-$ 33 & 42 & 56 & 12 & 161 & 65 & 84 & 14 & 189 &\\ 
\multirow{2}{*}{Sgr C} & $-$60 $-$ $-$46 & & & & & 54 & 72 & 8 & 173 & \multirow{2}{*}{53} \\
& $-$55 $-$ $-$49 & & & & & 43 & 59 & 7 & 191 & \\ 
\multirow{2}{*}{Sgr D} & $-$17 $-$ $-$14 & & & & & 47 & 56 & 8 & 80 & \multirow{2}{*}{52}  \\
& $-$19 $-$ $-$13 & & & & & 39 & 50 & 7 & 66\\
\end{tabular}
\tablefoot{The last column gives the average temperature of each source from the survey of \citet{Ginsburg2016}.
}
\label{SourceTemp}
\end{table*}

To estimate and compare the overall temperatures of the clouds, we determined the weighted average of the 
temperature in each cloud for each velocity component, first, derived from the 218 GHz lines and, second, from the 291 GHz ladder. 
The squared reciprocal of the temperature uncertainties served as the weights. In this
calculation, temperature upper limits were not included. The results are listed in Table \ref{SourceTemp}.
In addition, we give the minimal and maximal temperature of each map in the table. 

In most cases, the average temperature derived from the 291 GHz data is higher than the average temperature derived from the 
218 GHz lines but the 
values agree within the 1$\sigma$ temperature uncertainty. In two cases, however, the temperature difference is 
rather large (50 km/s cloud: 57$-$63 km s$^{-1}$ component, G0.253+0.016: 16$-$22 km s$^{-1}$ component).
In Fig. \ref{Average-Temp-Boxplot} in the Appendix, we visualize these results in temperature box plots.

To check the validity of our results, we compared our temperature measurements to the H$_{2}$CO 218 GHz
survey of \citet{Ginsburg2016}. For each source, we averaged their temperature map (their Fig. 7c) over the sizes of our 
218 GHz OTF maps (the 291 GHz OTF maps for Sgr C and Sgr D). The obtained values are shown in the last column of 
Table \ref{SourceTemp}. These average temperatures are roughly consistent with our results except for G0.480$-$0.006. Here, 
the average temperature of \citet{Ginsburg2016} is much higher than our values at 218 GHz. However, our 291 GHz measurements do 
show a high temperature component. We conclude that on average our study yields similar temperatures for the observed 
clouds as the survey of \citet{Ginsburg2016}.

Comparing the average temperature of the whole velocity range of each source, we noticed a large spread of temperatures
between 40 K in G0.480$-$0.006 and 82 K in the 50 km/s cloud. There is no clear correlation of the average temperature 
with the location of the clouds in the CMZ.

\subsection{Temperature Gradients within clouds}

Table \ref{SourceTemp} shows that the average temperatures of the different velocity components in the sources are not the same. Differences of 
$\sim$30 K are observed. This indicates the presences of temperature gradients in the sources. These results show that averaging the 
temperatures over the whole velocity range of the clouds can yield misleading values of the temperature. It is thus important to look at the 
temperatures of the gas with different velocities to understand the underlying temperature structures of the clouds.

The 20 km/s cloud and G0.253+0.016 even show temperature gradients across the sources within one velocity component. 
In Fig. \ref{20kms-cloud}, the  temperature map of the 8$-$14 km s$^{-1}$ component of the 20 km/s cloud is plotted. 
There is a clear temperature gradient from $\sim$60$-$70 K 
at the north-east side of the cloud to $\sim$110$-$120 K in the southern part. A similar gradient is seen in the 36$-$42 km s$^{-1}$ component
of G0.253+0.016 (Fig. \ref{G0253-All-Temp-H2CO}, fourth panel), with the temperature increasing from $\sim$70 K in the north of the cloud 
to more than 150 K in the southern part.

In a forthcoming paper, we will compare the temperature structures of the clouds with tracers of shocks (e.g. observations of SiO, Class I 
methanol masers) or star formation (Class II methanol masers, water masers) to apprehend the cause for the different temperatures.

\subsection{Evidence for heating by turbulence}
\label{SectTempLineWidth}

To better understand the temperature structures in our clouds, we will investigate the energy balance in the gas 
following the analysis of \citet{Ginsburg2016}. We will not consider stellar heating or energy injection through supernovae 
since these processes can only explain local heating of the clouds but not their overall high gas temperature. 
\citet{Ao2013} already excluded diffuse X-rays as the main heating source, leaving only cosmic ray and mechanical 
heating (turbulence) as the energetically important heating processes.
\citet{Ginsburg2016} concluded that cosmic ray heating is either irrelevant in the CMZ or the heating is non-uniform since 
uniform heating cannot simultaneously explain the high (> 100 K) and low (< 50 K) temperatures measured in the CMZ.
The detected temperature gradients in our clouds also exclude the latter except if the cosmic ray heating is non-uniform
on very small scales ($\sim$1 pc).
 We will produce similar gas temperature vs line width plots (as their Fig. 9) for the 218 and 291 GHz temperatures 
 and compare our data to the same models as used in \citet{Ginsburg2016} to narrow down the parameter 
space of possible heating models. 

\begin{figure*}
	\caption{\textit{Left and Middle Panels:} Correlation plots of H$_{2}$CO integrated intensity ratio and kinetic temperature, respectively 
	vs line width. The error bars present 
	the standard deviation of the weighted average for the integrated intensity ratio, 20\% uncertainty in the line width and 
	30\% uncertainty in the temperature. The best linear fit to the temperature data using the total least square 
	method \citep{Vanderplas2012} is shown as the solid line. \textit{Right Panel:} The possible slope 
	and intercept values with the 1, 2 and 3$\sigma$ likelihood contours.   }
	\centering
	\subfloat{\includegraphics[height=8cm]{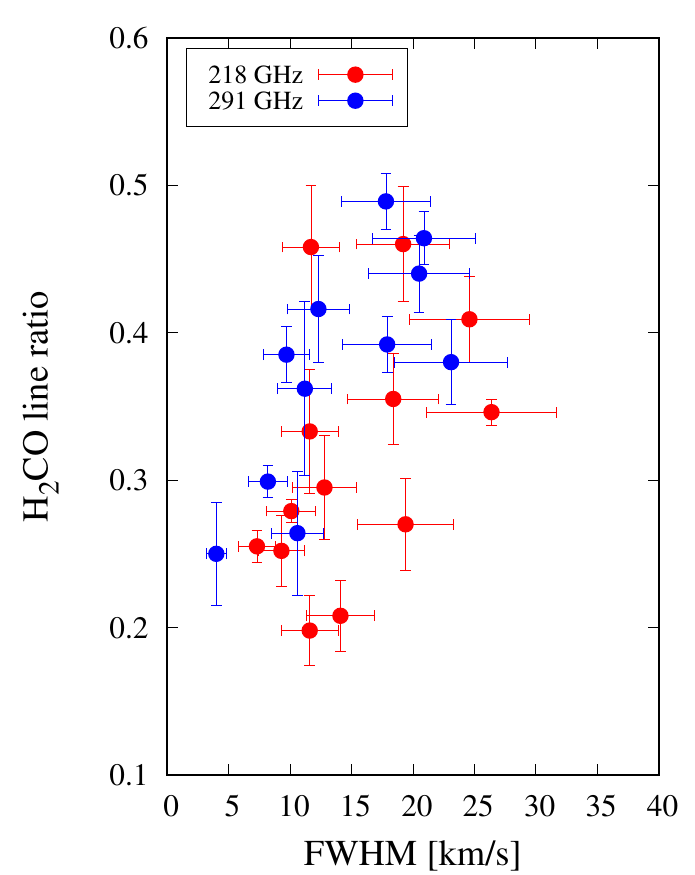}}	\hspace{-0.1cm}
        \subfloat{\includegraphics[bb = 10 30 685 480, clip,height=8cm]{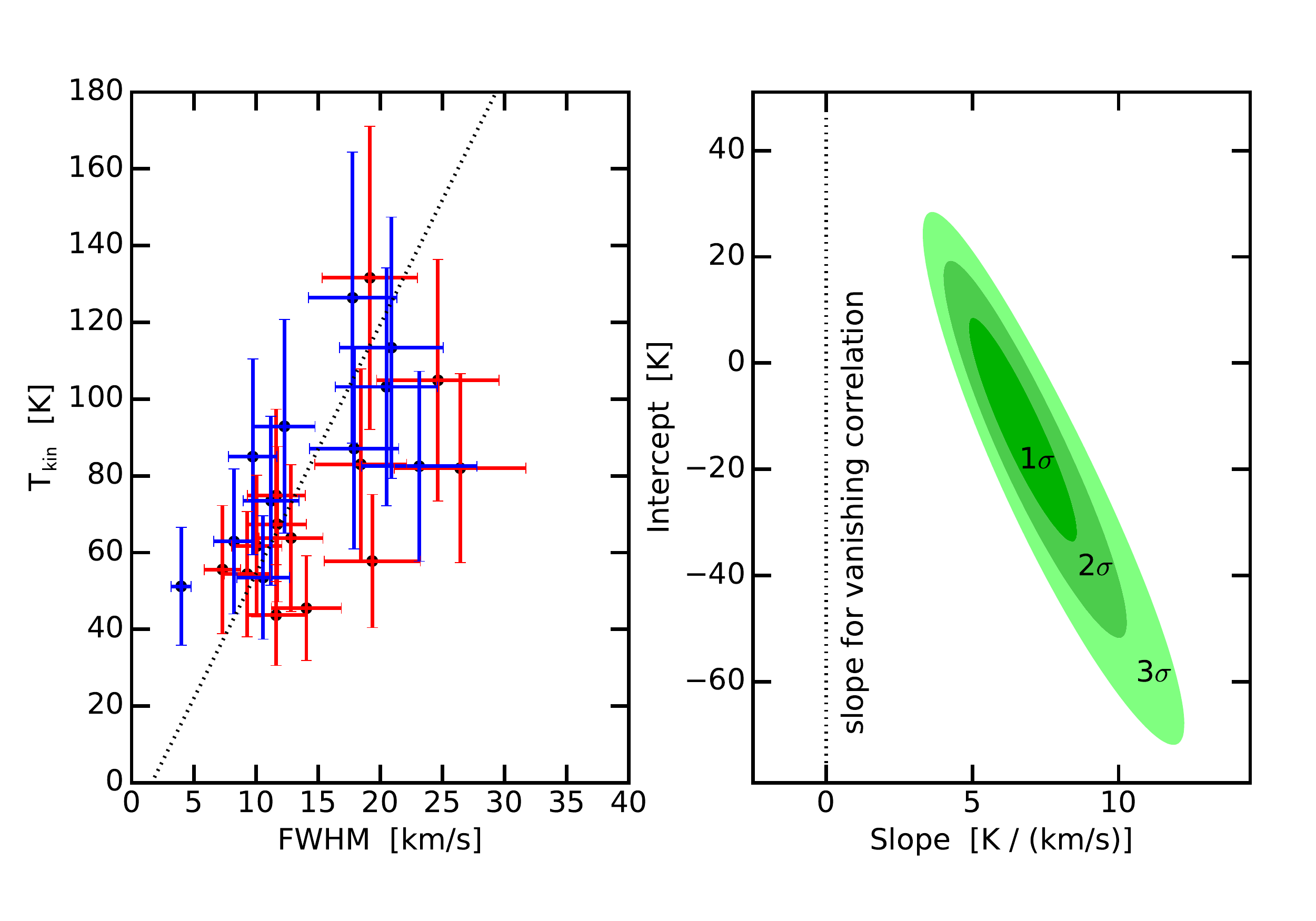}}
        \label{TempvsLinewidth}
\end{figure*}

To collect the data of these plots, we chose one or several positions in the 218 and 291 GHz temperature maps 
of the 6 km s$^{-1}$ broad velocity components of each source where none of the pixels in a 33$\arcsec$ x 33$\arcsec$ tile 
(roughly corresponding to the size of the beam) around these positions are upper limits. The boxes are overlayed on the 
temperature maps in Figs. \ref{20kms-All-Temp-H2CO} $-$ \ref{SGRD-All-Temp-H2CO} in the Appendix. 
We then determined the weighted average temperature within these boxes, using the squared reciprocal of the 
temperature uncertainties as weights. Averaging the H$_{2}$CO(3$_{0,3}-$2$_{0,2}$) spectra of the nine pixels in the boxes, 
we fitted the line with one or more velocity components depending on the source. We then assigned each average temperature value 
the line width of the corresponding velocity component. We excluded the 20 km/s cloud from this analysis because we could not 
fit the line width of the different components unambiguously. To make sure that any trend we see is real and not due to the radiative 
transfer modeling, we also obtained the average line ratios R$_{321}$ and R$_{422}$ in those tiles. The values for line ratio, temperature 
and line width are listed in Table \ref{TempLineWidth}. In Fig. \ref{TempvsLinewidth}, we show the relation of line ratio (left panel) 
and temperature (middle panel), respectively versus line width. 

Considering the statistical uncertainties associated with both parameters (20\% for the line width, 30\% for the temperature), 
we adopt the method of total least squares \citep{Vanderplas2012} to measure the slope of the dv-T$_{\rm kin}$ relation. 
Since we are integrating the emission of the H$_{2}$CO lines over a fixed velocity ranges, it is save to assume that the temperature 
and line width measurements are uncorrelated, so we set the covariance between them to be zero.  
Figure~\ref{TempvsLinewidth} shows a statistically significant positive correlation suggesting that regions of higher 
line widths in our clouds are expected to have higher gas temperature. This implies that turbulence might play a direct role in gas heating. 

\begin{figure*}
	\caption{Temperature versus line width plot, overlayed with thermal equilibrium models computed using the 
	DESPOTIC \citep{Krumholz2014} code \citep[see also Fig. 9 of][]{Ginsburg2016}. The data is consistent with 
	the n = 10$^5$ cm$^{-3}$ model.}
	\centering
        \includegraphics[width=12cm]{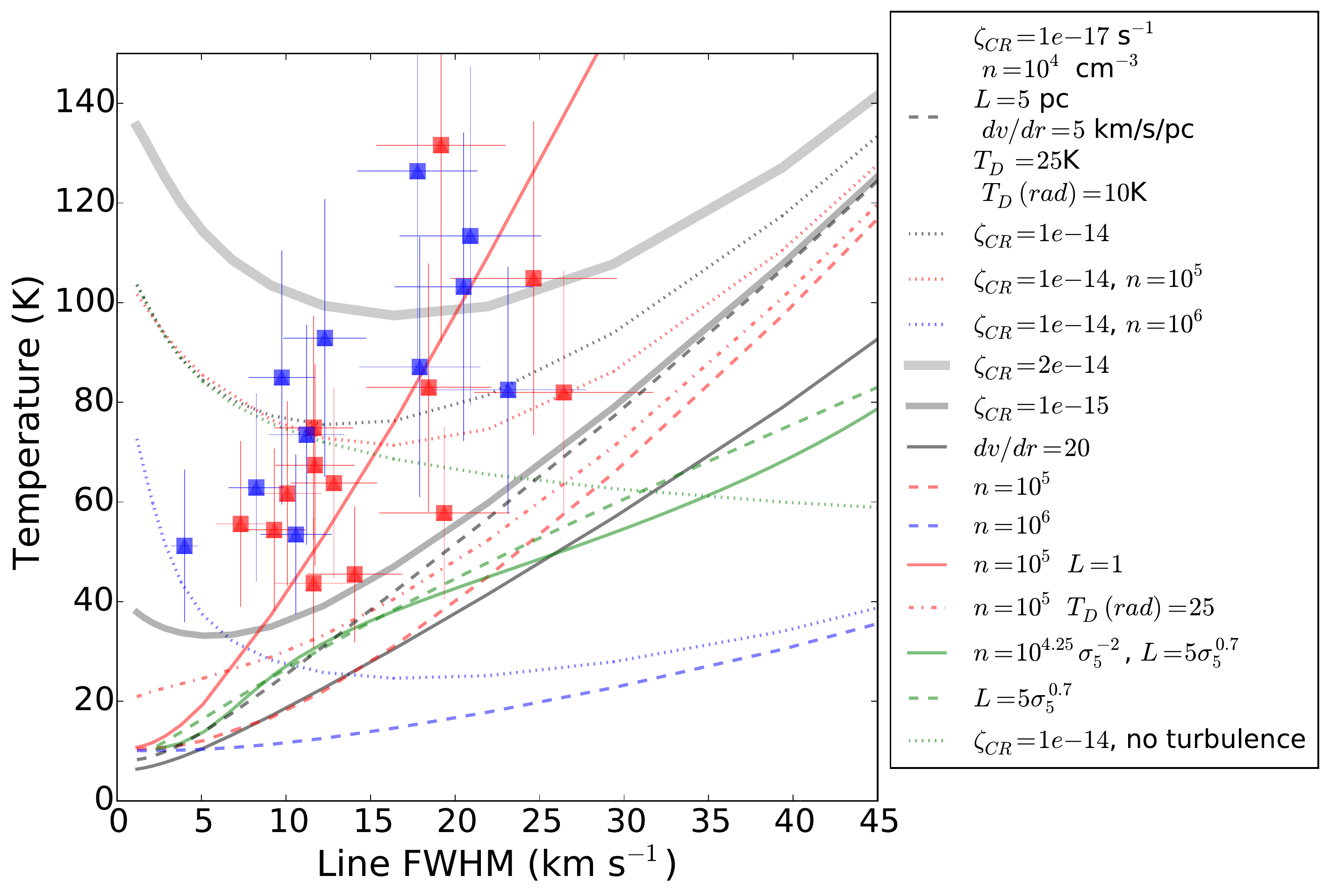}
	\label{DespoticModels}
\end{figure*}

Figure \ref{DespoticModels} shows a collection of thermal equilibrium models computed using the DESPOTIC \citep{Krumholz2014} code.  
Each model uses the heating terms specified in the legend and both line and dust cooling.  The models compute the equilibrium 
temperature achieved at the specified density.  The primary variable being varied is the line width, which is treated as an observational 
proxy for the 3D velocity dispersion following the equation $\sigma_{3D} = \sqrt{3} \sigma_{1D}$.  In contrast with \citet{Ginsburg2016}, 
who observed a wider range in densities and environments, our data show a strong trend between the observed temperature and the 
velocity dispersion.  They are readily explained by the n = 10$^5$ cm$^{-3}$ model which assumes that each cloud has a line-of-sight length of 
1 pc.  These observations therefore significantly strengthen the case put forth by \citet{Ao2013} and \citet{Ginsburg2016} that the densest 
Galactic center clouds are predominantly heated by turbulence.

\subsection{Density Constraints}

Following \citet{Mangum1993}, the integrated intensity ratio of H$_2$CO $\Delta$~J =  1 lines with different J, but the 
same K$_{\rm a}$ quantum numbers constrains the density of the gas. Figure \ref{TempDensity} (right panel) shows that the R$_{404}$
ratio depends strongly on both temperature and density for densities $<$ 10$^{7}$ cm$^{-3}$ while the R$_{321}$ ratio only 
depends on temperature (Fig. \ref{TempDensity}, left panel). 
Thus, combining the measurements of the R$_{321}$ and R$_{422}$
ratios which yield the temperature with the R$_{404}$ ratio we can estimate the density in the clouds. The red plus signs  
in Fig. \ref{TempDensity} mark the locations where the R$_{321}$ and R$_{404}$ ratios are matched at the same 
time. Due to the significant dependence of the density estimate on the temperature estimate and our temperature 
uncertainties of $\sim$30\%, we cannot compute density maps of our clouds. However, we constrain the density 
of the widespread warm gas in our clouds to 10$^{4}-$10$^{6}$ cm$^{-3}$. Efforts are ongoing to further constrain 
this range using the cm transitions of H$_{2}$CO (Ginsburg et al. in prep).

\section{Conclusion}
\label{Summary}

In this paper, we present H$_{2}$CO observations of five and seven CMZ clouds at 218 and 291 GHz, 
respectively. Combining integrated intensity H$_{2}$CO line ratios with radiative transfer models, we obtain gas temperature 
maps for our clouds. The two different sets of H$_{2}$CO lines (H$_{2}$CO(3$-$2) at 218 GHz and H$_{2}$CO(4$-$3) at 291 GHz)
yield two independent estimates of the gas temperature.

Our observations at 218 GHz are a factor of $\sim$1.5 deeper than previous H$_{2}$CO CMZ observations by \citet[][compare
our median rms values of $\sim$45 mK per pixel to their noise level of 70 mK per pixel]{Ginsburg2016}. While Ginsburg et al. 
focus on the global temperature differences in the CMZ, we disentangle the different velocity components of the gas in our 
sources and investigate their temperature structures. This is a significant step since the CMZ 
clouds exhibit the widest velocity components observed in our galaxy.

From a comparison of the H$_{2}$CO main lines at 218 and 291 GHz, we accessed that the H$_{2}$CO(3$_{0,3}-$2$_{0,2}$) 
line is optically thick in some parts of the clouds. Combining the observed line ratios R$_{321}$ and R$_{404}$, we 
constrain the density of the warm cloud gas to 10$^{4}-$10$^{6}$ cm$^{-3}$.

Our temperature maps at 218 and 291 GHz show clear temperature gradients in our clouds. This indicates that 
heating mechanisms that act on the bulk of the molecular gas cannot be the main heating sources. Cosmic ray 
heating is only possible if the heating is non-uniform on very small scales.
In a following paper, we 
will compare our results with complementary observations of shock and star formation tracers as well as 
supernova remnants in the clouds to understand if these gradients are caused by local heating through cloud collisions, 
feedback from new born stars or the explosion of stars.

Comparing the line widths of the main H$_{2}$CO lines at 218 and 291 GHz with the measured temperatures at selected 
positions in our clouds, we found a clear positive correlation between these two parameters. This indicates that turbulence 
plays an important role in the heating of the gas. Our data is consistent with a turbulence model with a density n = 10$^5$ cm$^{-3}$ 
which assumes that each cloud has a line-of-sight length of 1 pc.

\begin{acknowledgements}

We thank the APEX team and MPIfR observers for carrying out these service-mode observations.
The authors are thankful for the helpful comments of the anonymous referee. 
T.P. acknowledges support from the Deutsche Forschungsgemeinschaft, DFG via the SPP (priority program) 1573 ``Physics of the ISM''.
This research has made use of NASA's Astrophysics Data System Bibliographic Services.

\end{acknowledgements}

\bibliographystyle{aa}

\Online

\begin{appendix}

\section{Calibration}

\begin{figure*}
	\caption{Example spectra of the H$_{2}$CO(3$_{0,3}-$2$_{0,2}$) transition of target G0.253+0.016 to show the spline fitting of bad 
	baselines. Left: Input spectrum showing a bad baseline. The red line shows a linear baseline for comparison. Middle: Input spectrum with the 
	spline spectrum (red) overlayed. Right: Baseline-subtracted spectrum. The grey box shows a baseline feature that was not removed in the 
	spline fitting due to its narrow width.}
	\centering
	\subfloat[\label{BadSpectrum}]{\includegraphics[width=6cm]{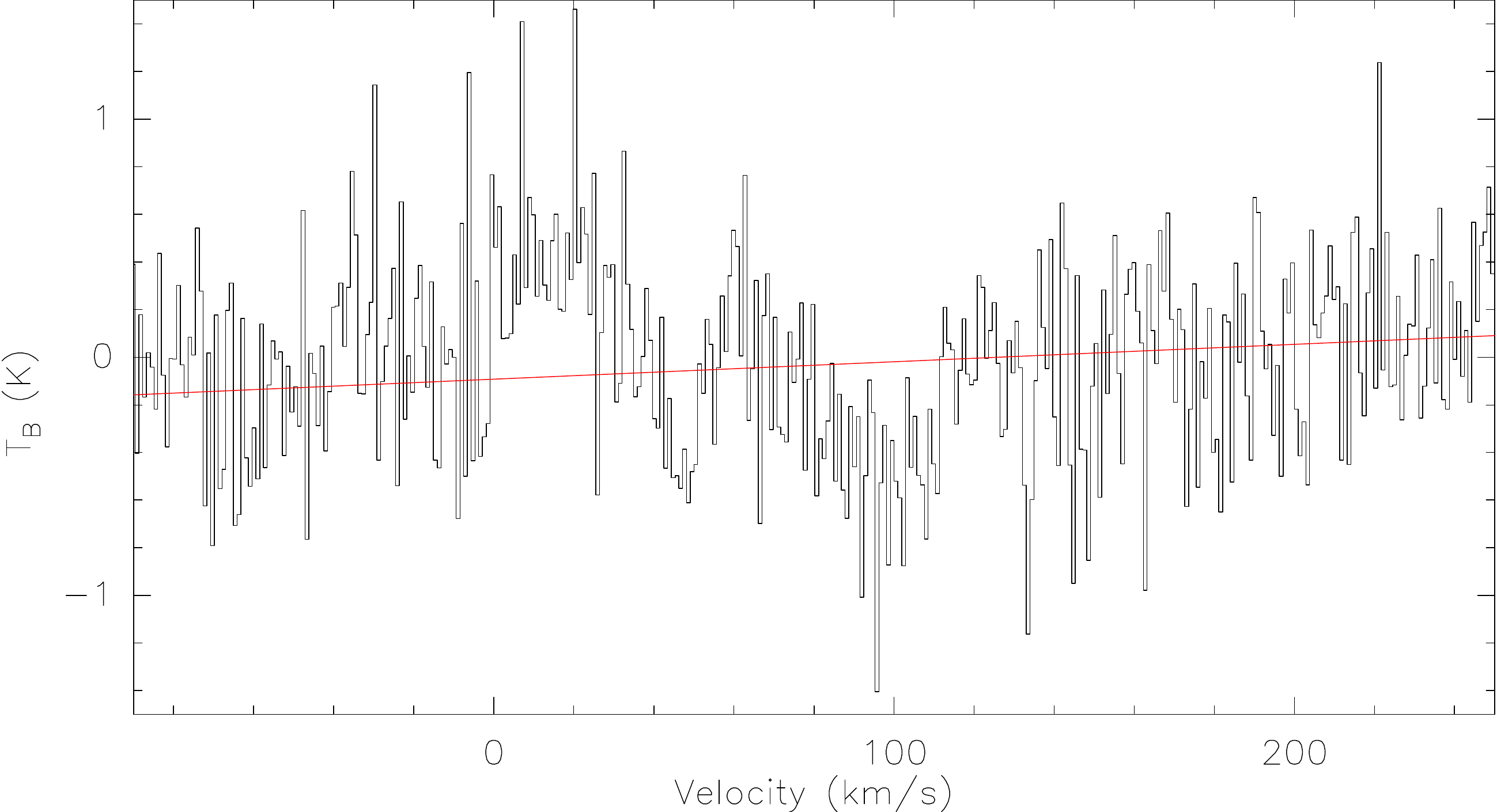}}\hspace{0.1cm}
	\subfloat[\label{Fitted Spline}]{\includegraphics[width=6cm]{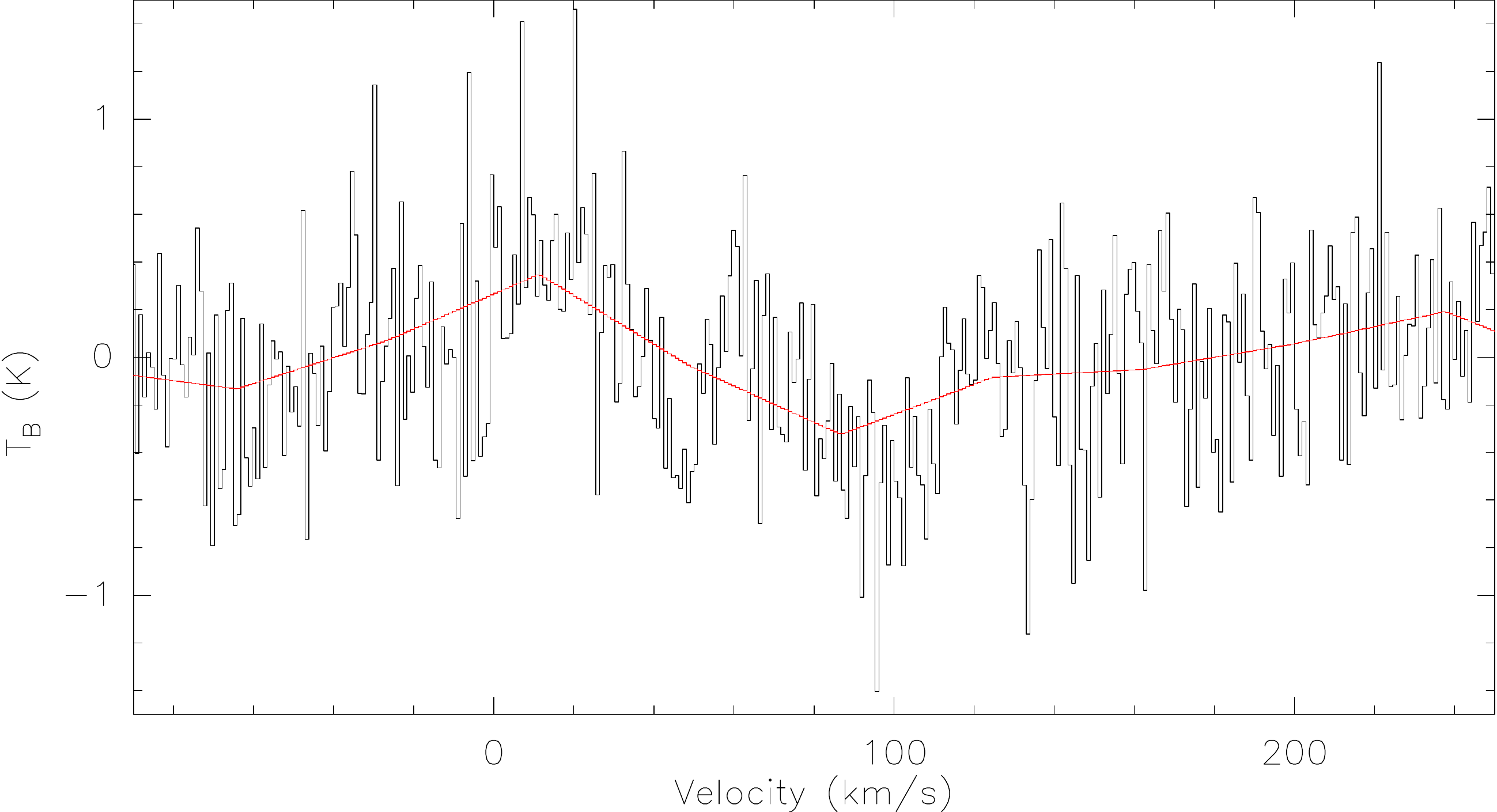}}\hspace{0.1cm}
	\subfloat[\label{CleanSpectrum}]{\includegraphics[width=6cm]{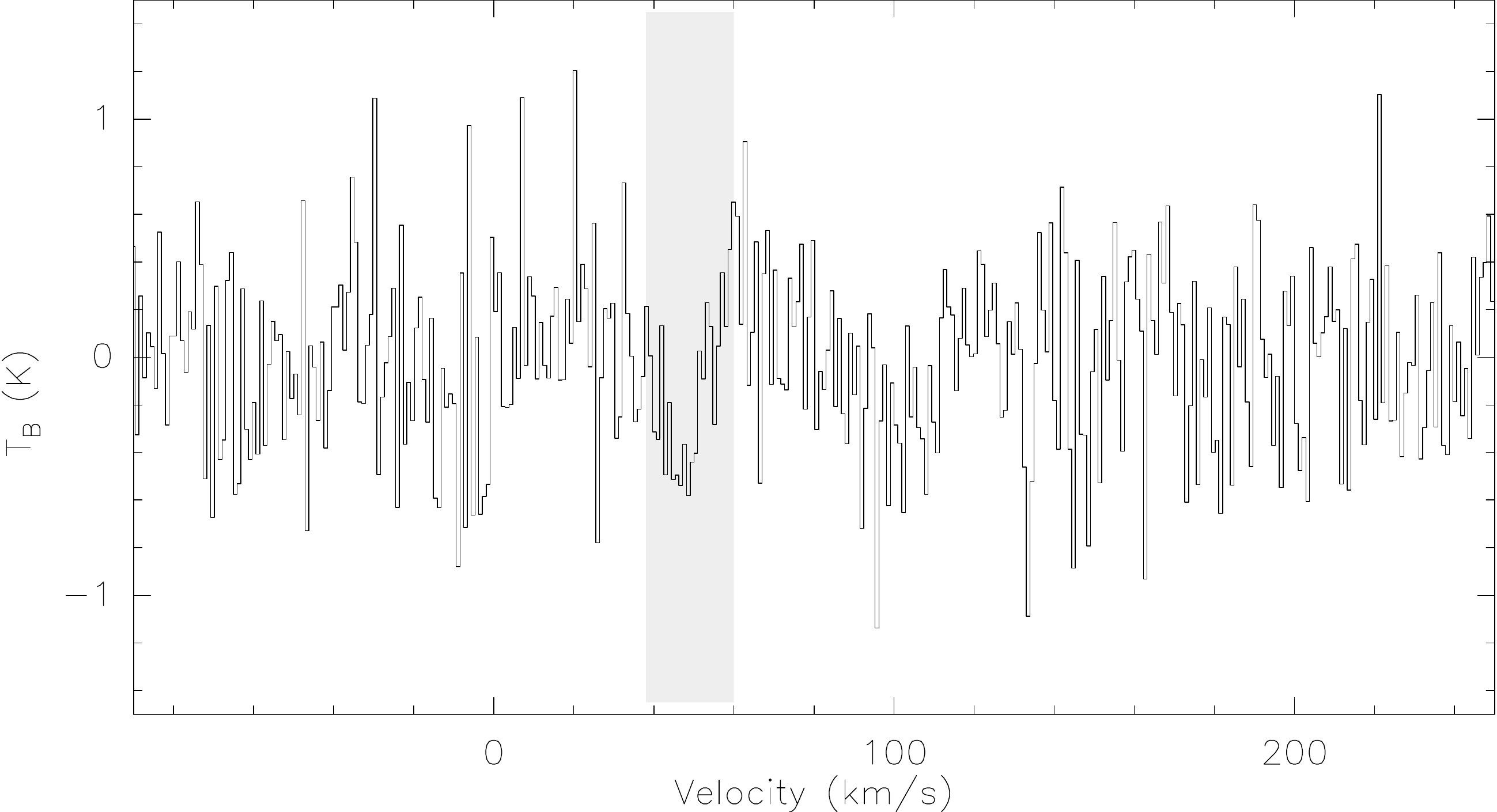}}	
	\label{Spectra-Example}
\end{figure*}

\clearpage

\section{Spectra}

 \begin{figure*}
	\caption{Spectra of para-H$_{2}$CO transitions at 218 (left panels) and 291 GHz (right panels), averaged over the whole OTF map of each source. H$_{2}$CO as well as other lines are marked in the spectra.}
	\centering
	20 km/s cloud\\
	\subfloat{\includegraphics[width=8cm]{20kms-cloud_H2CO-Spectrum_218GHz.pdf}}
\hspace{0.1cm}
	\subfloat{\includegraphics[width=8cm]{20kms-cloud_H2CO-Spectrum_290GHz.pdf}}\\
	50 km/s cloud\\
	\subfloat{\includegraphics[width=8cm]{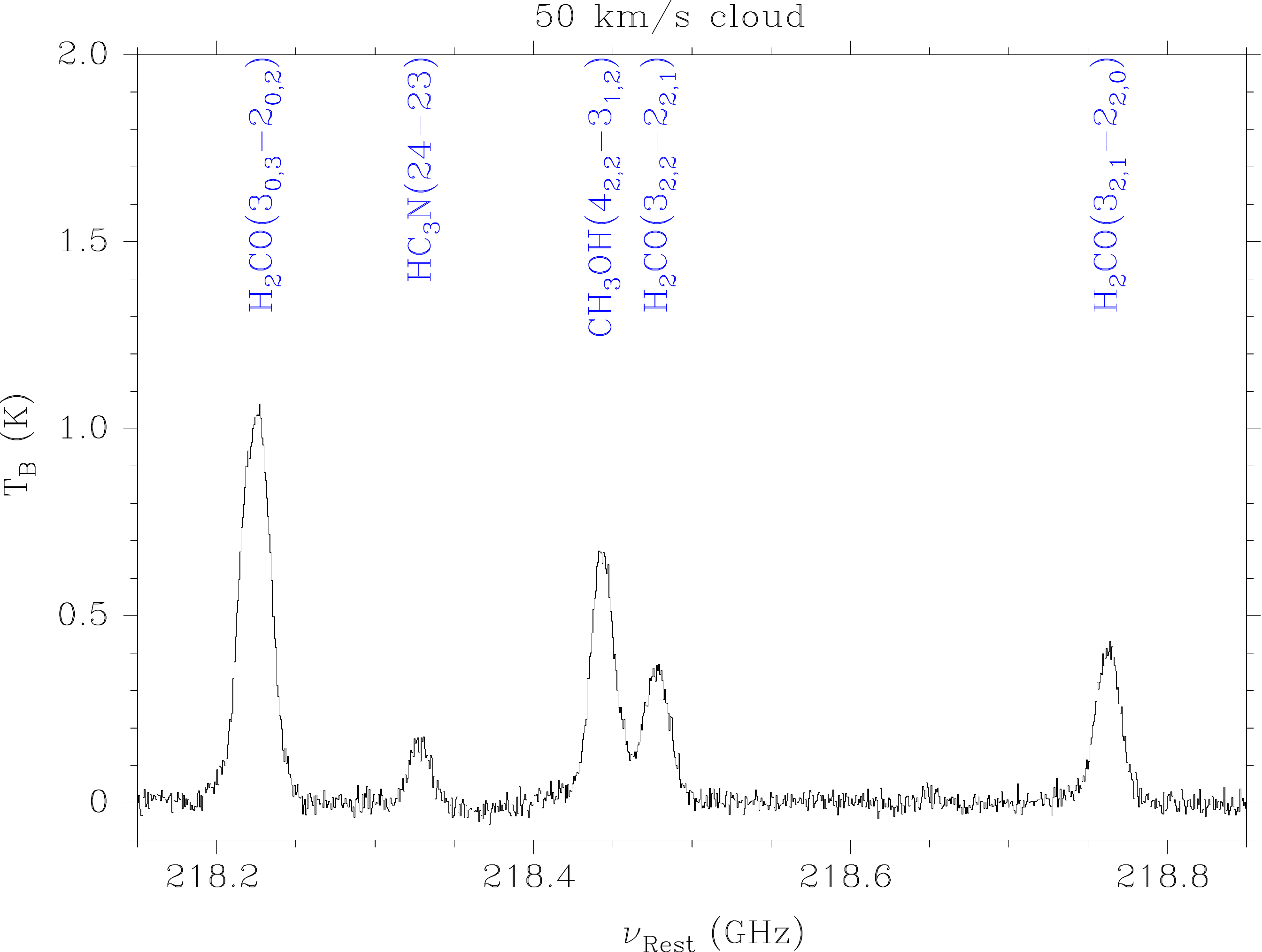}}
\hspace{0.1cm}
	\subfloat{\includegraphics[width=8cm]{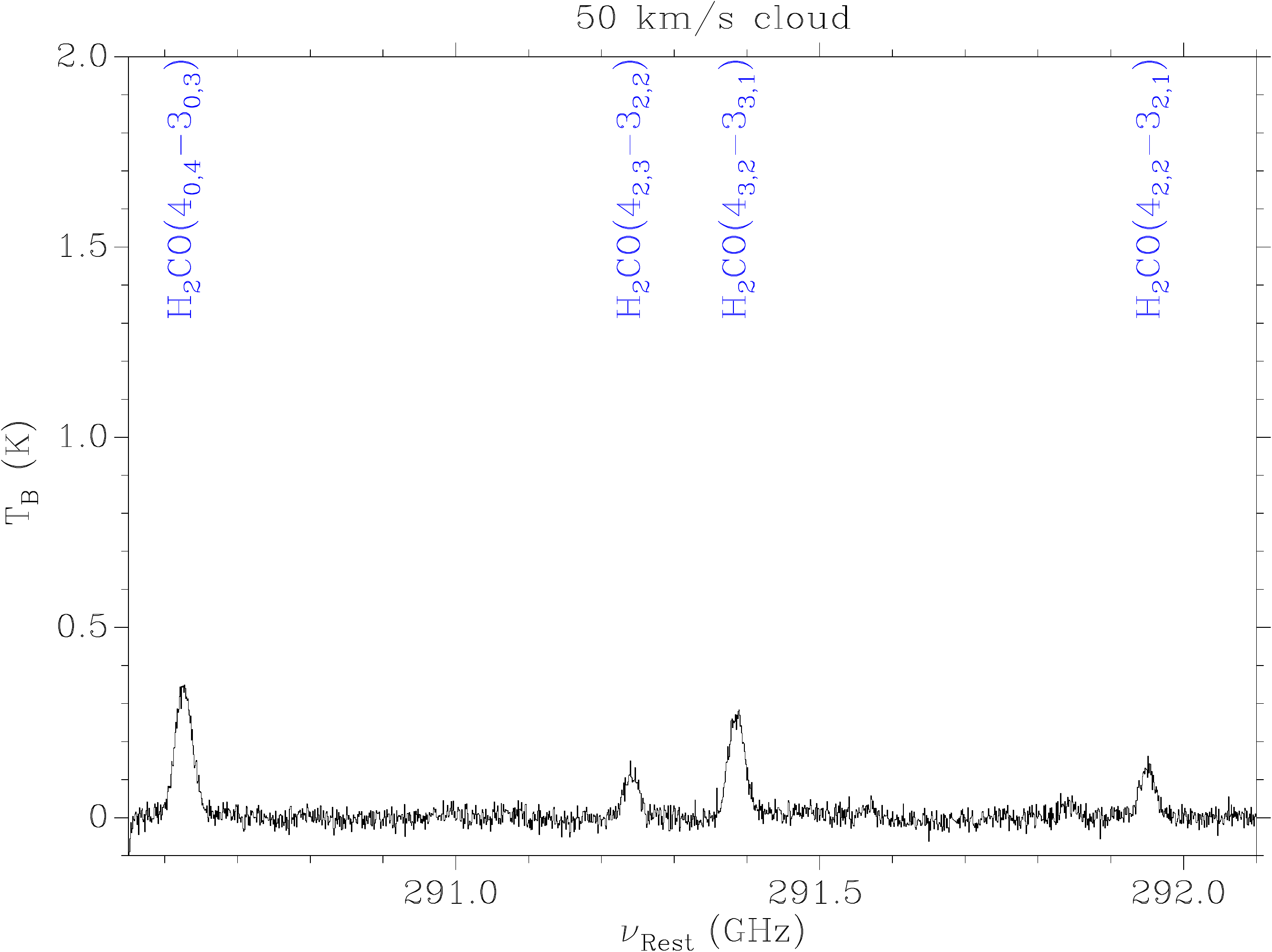}}\\
	G0.253+0.016\\
	\subfloat{\includegraphics[width=8cm]{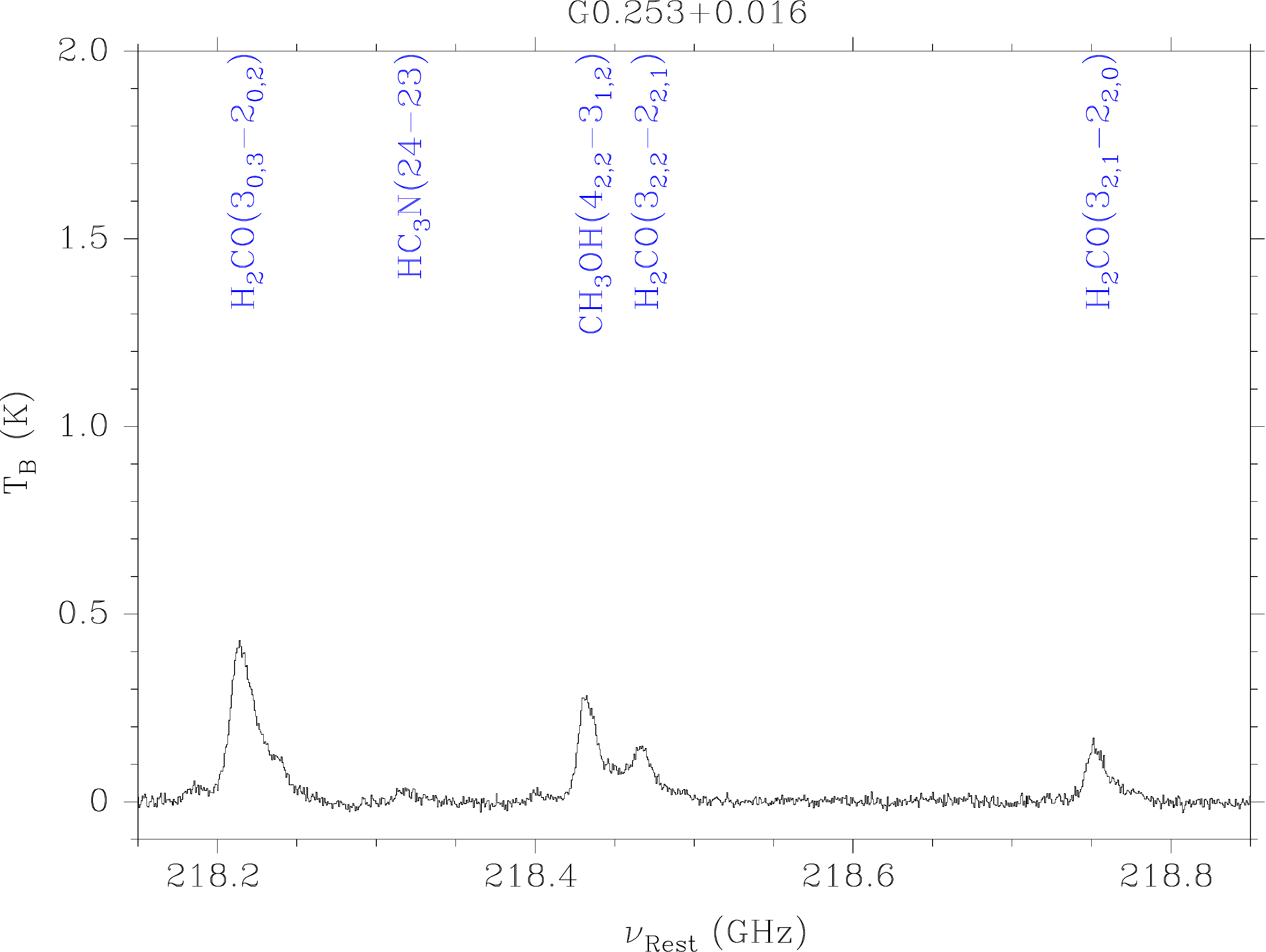}}
\hspace{0.1cm}
	\subfloat{\includegraphics[width=8cm]{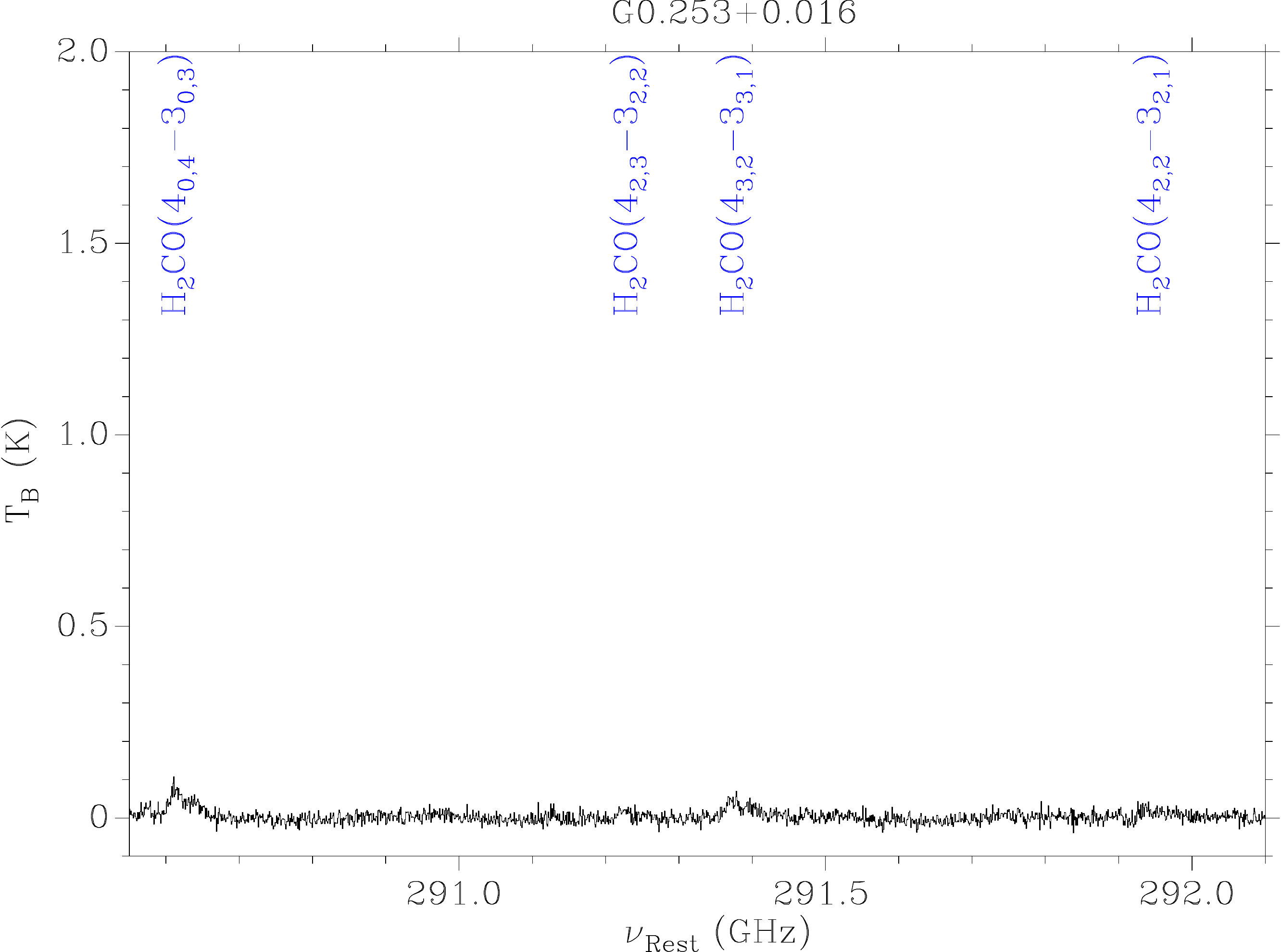}}\\
	\label{Spectra}
\end{figure*}

\addtocounter{figure}{-1}
 \begin{figure*}
	\caption{Continued.}
	\centering
	G0.411+0.050\\
	\subfloat{\includegraphics[width=8cm]{G0411_H2CO-Spectrum_218GHz.pdf}}
\hspace{0.1cm}
	\subfloat{\includegraphics[width=8cm]{G0411_H2CO-Spectrum_290GHz.pdf}}\\
	G0.480$-$0.006\\
	\subfloat{\includegraphics[width=8cm]{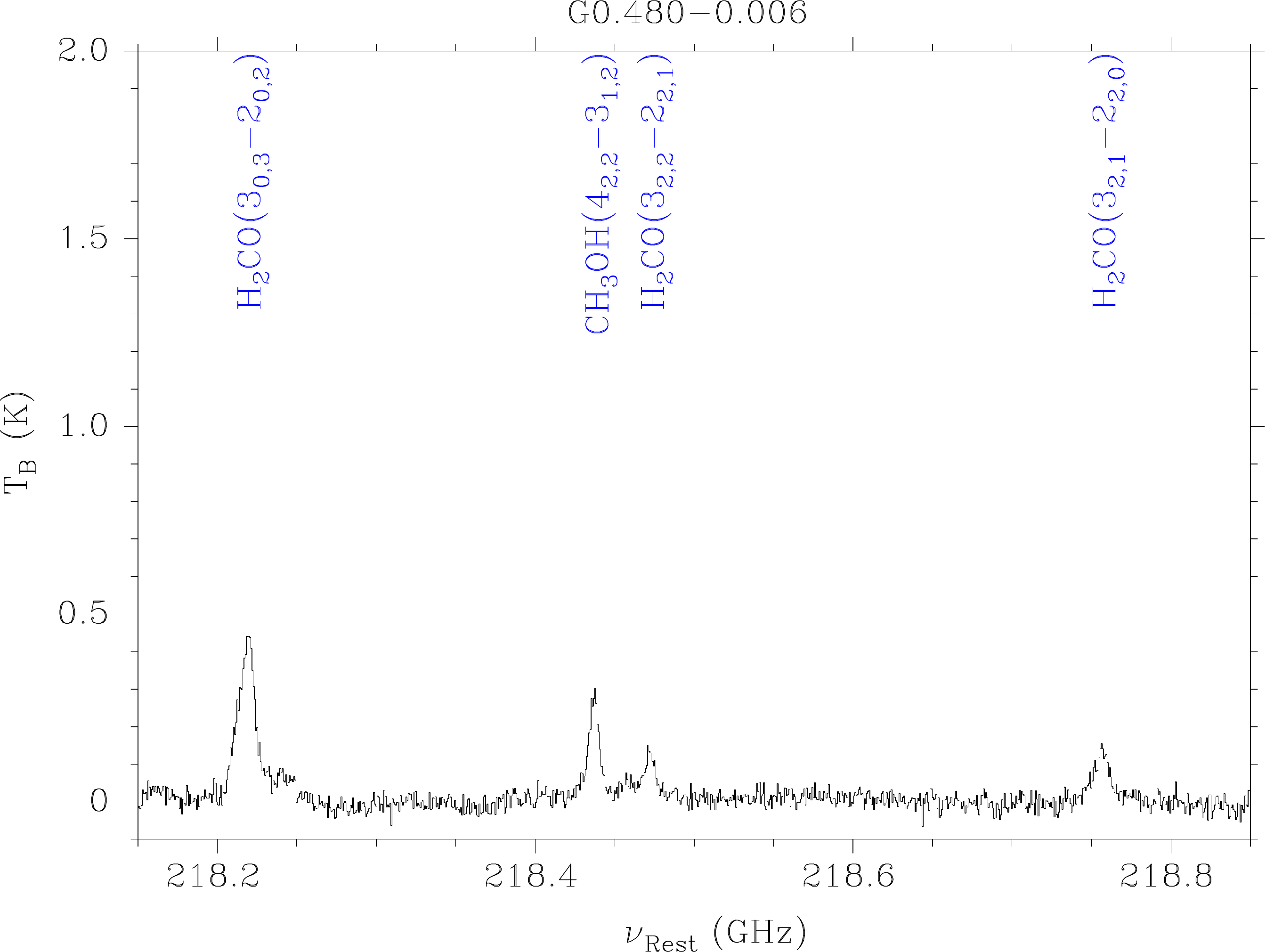}}
\hspace{0.1cm}
	\subfloat{\includegraphics[width=8cm]{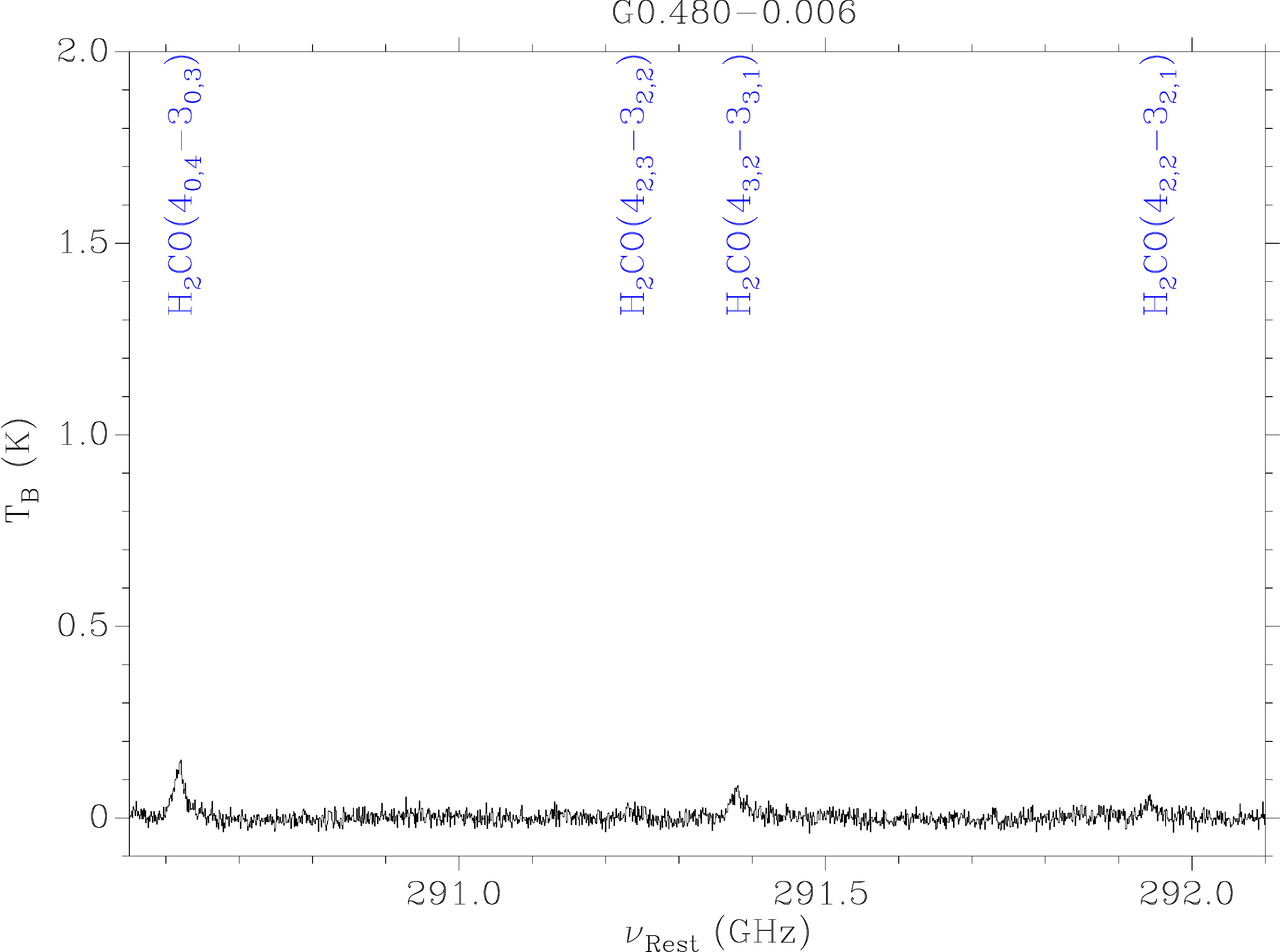}}\\
	\hspace{0.7cm} Sgr C \hspace{7cm} Sgr D\\
	\subfloat{\includegraphics[width=8cm]{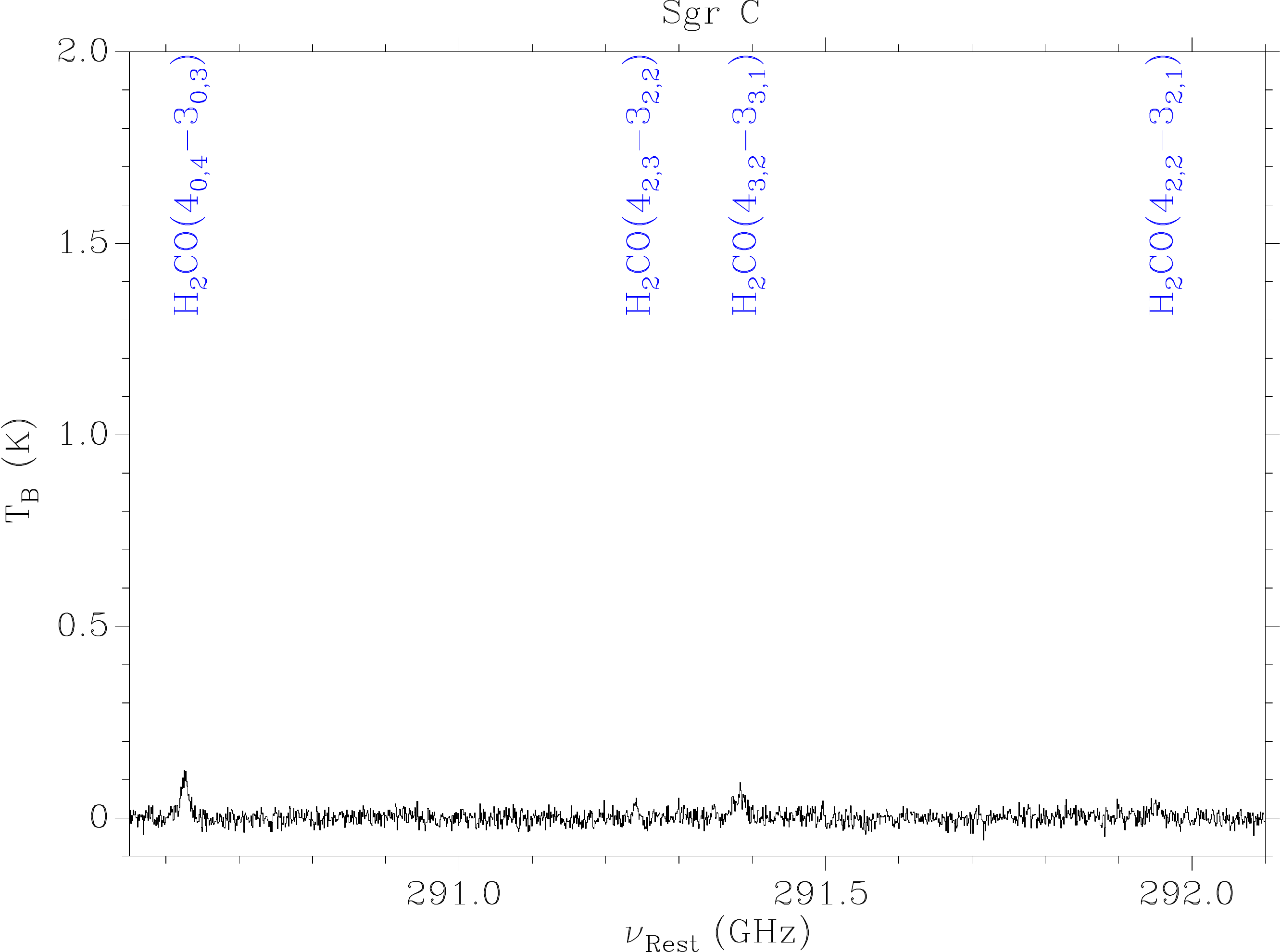}}
	\subfloat{\includegraphics[width=8cm]{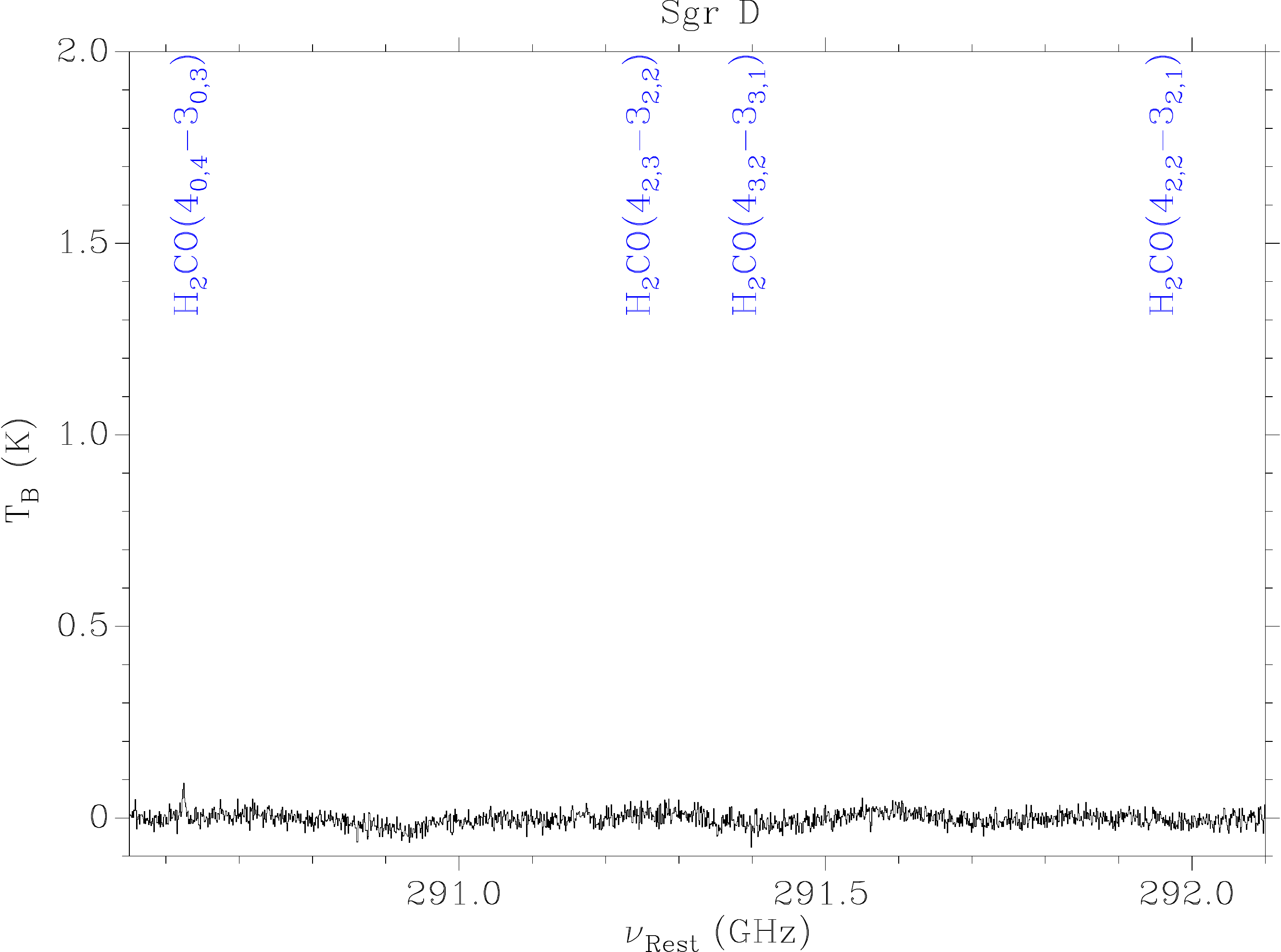}}
\end{figure*}

 \begin{figure*}
	\caption{The spectra of the H$_{2}$CO(3$_{0,3}-$2$_{0,2}$) and H$_{2}$CO(4$_{0,4}-$3$_{0,3}$) transitions for the 20 km/s cloud (left) 
	and the 50 km/s cloud (right) plotted on top of each other. The spectra of H$_{2}$CO(4$_{0,4}-$3$_{0,3}$) were multiplied by a factor of 3.8 and 2.9 in the left and right panels, respectively, to bring the two lines on the same scale. The plots show that the two transitions have comparable width in both sources.}
	\centering
        \hspace{0.8cm}	20 km/s cloud   \hspace{6.2cm} 50 km/s cloud\\ 
	\subfloat{\includegraphics[width=8cm]{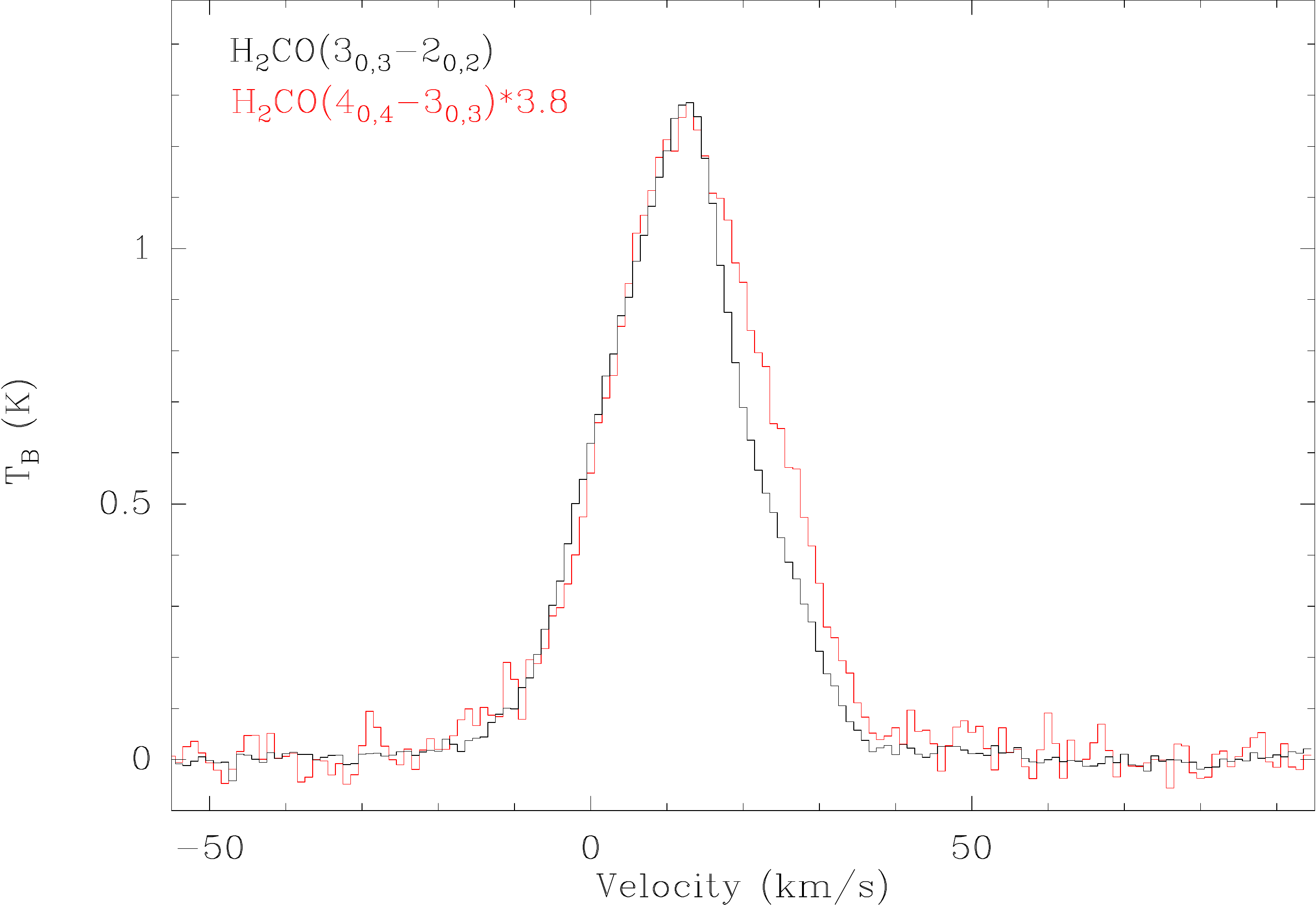}}
\hspace{0.1cm}
	\subfloat{\includegraphics[width=8cm]{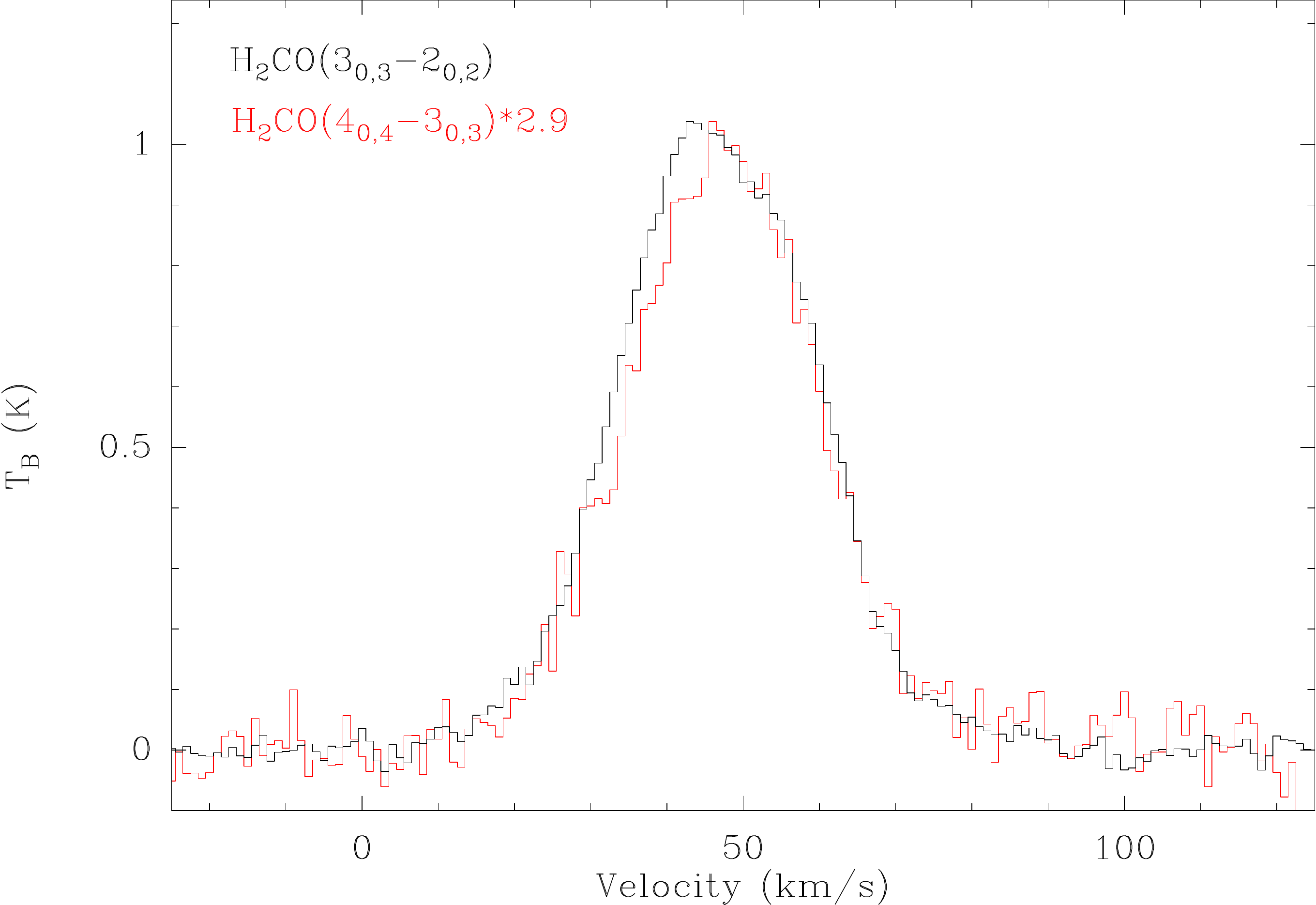}}\\
	\label{Comp218290-Spectra}
\end{figure*}

\clearpage

\section{Methanol contamination}

 \begin{figure*}
	\caption{Comparison of the integrated emission of H$_{2}$CO(3$_{0,3}-$2$_{0,2}$), H$_{2}$CO(3$_{2,1}-$2$_{2,0}$) and 
	H$_{2}$CO(3$_{2,2}-$2$_{2,1}$) of the 20 km/s cloud over the velocity range 27$-$33 km s$^{-1}$, showing the contamination of the 
	H$_{2}$CO(3$_{2,2}-$2$_{2,1})$ line with methanol emission at the south of the cloud. The contours show the moment 0 map of the 
	H$_{2}$CO(3$_{0,3}-$2$_{0,2}$) transition, produced over the whole velocity range of the source (levels: 30\%$-$90\% of the 
	maximum in steps of 10\%) The circle in the lower left corner shows the 33$\arcsec$ beam.}
	\centering
	\subfloat[H$_{2}$CO(3$_{0,3}-$2$_{0,2}$)]{\includegraphics[bb = 0 00 650 560, clip,height=6cm]{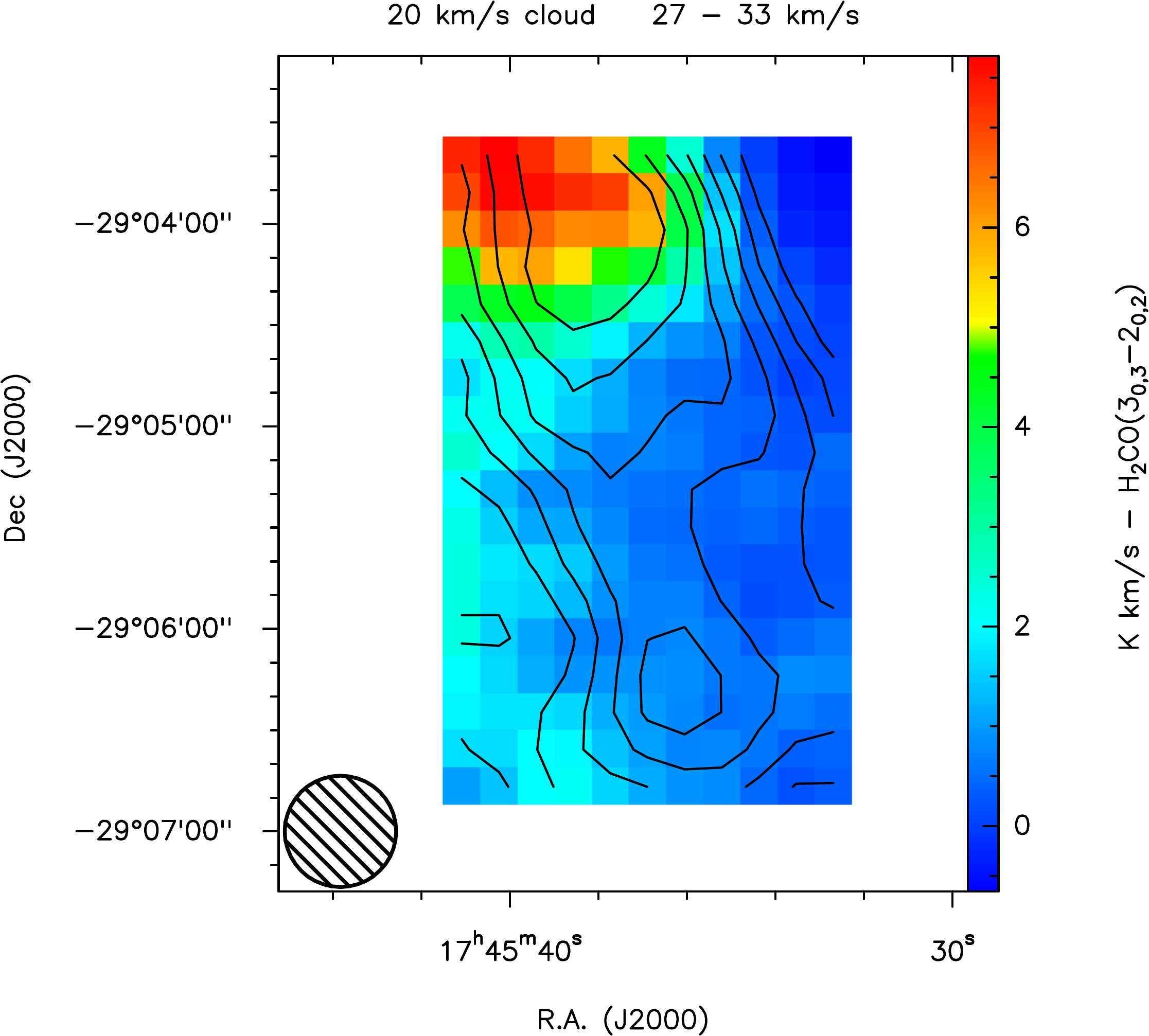}}
	\hspace{0.1cm}
	\subfloat[H$_{2}$CO(3$_{2,1}-$2$_{2,0}$)]{\includegraphics[bb = 140 00 650 560, clip,height=6cm]{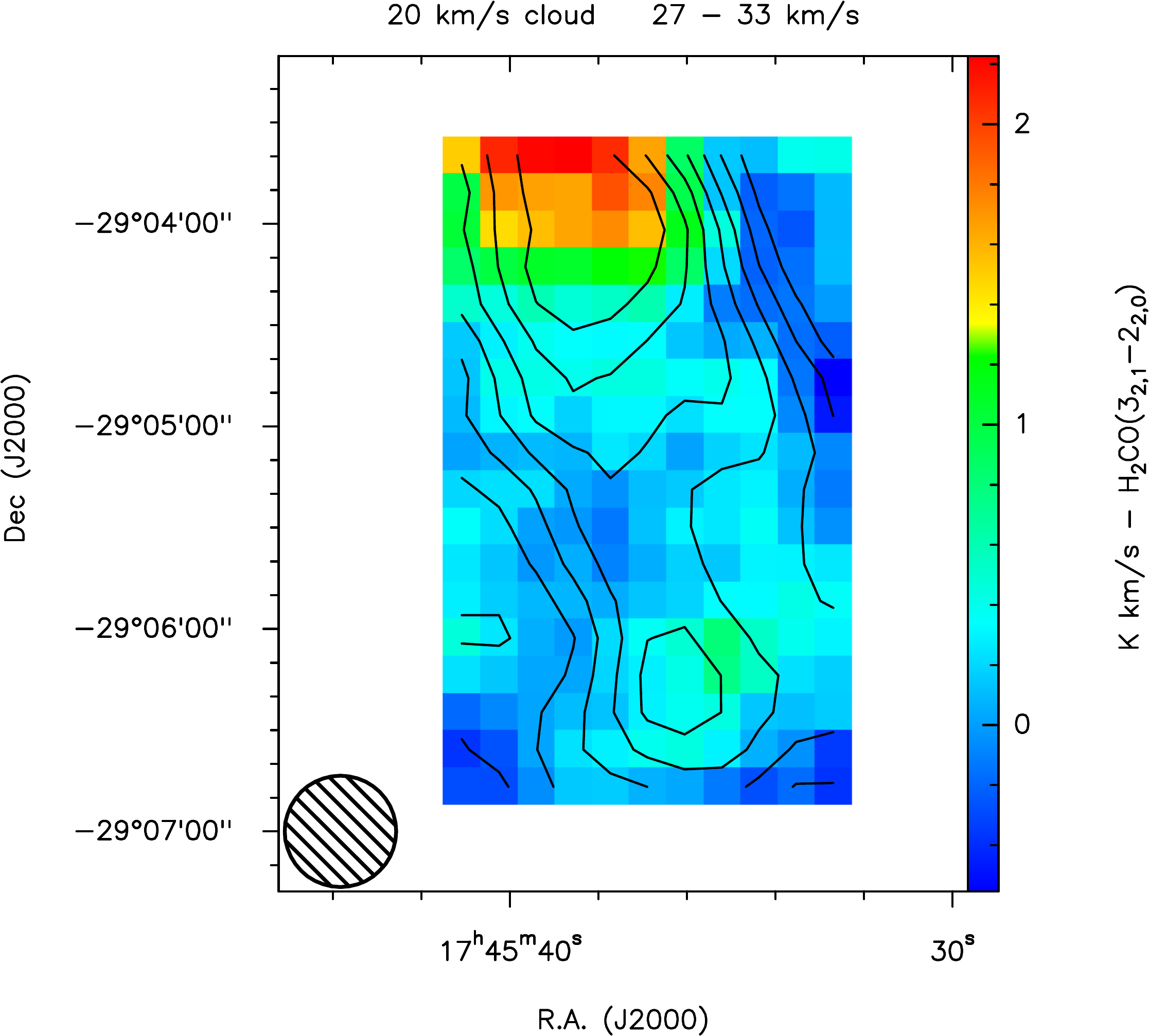}}
	\hspace{0.1cm}
	\subfloat[H$_{2}$CO(3$_{2,2}-$2$_{2,1}$)]{\includegraphics[bb = 140 00 650 560, clip,height=6cm]{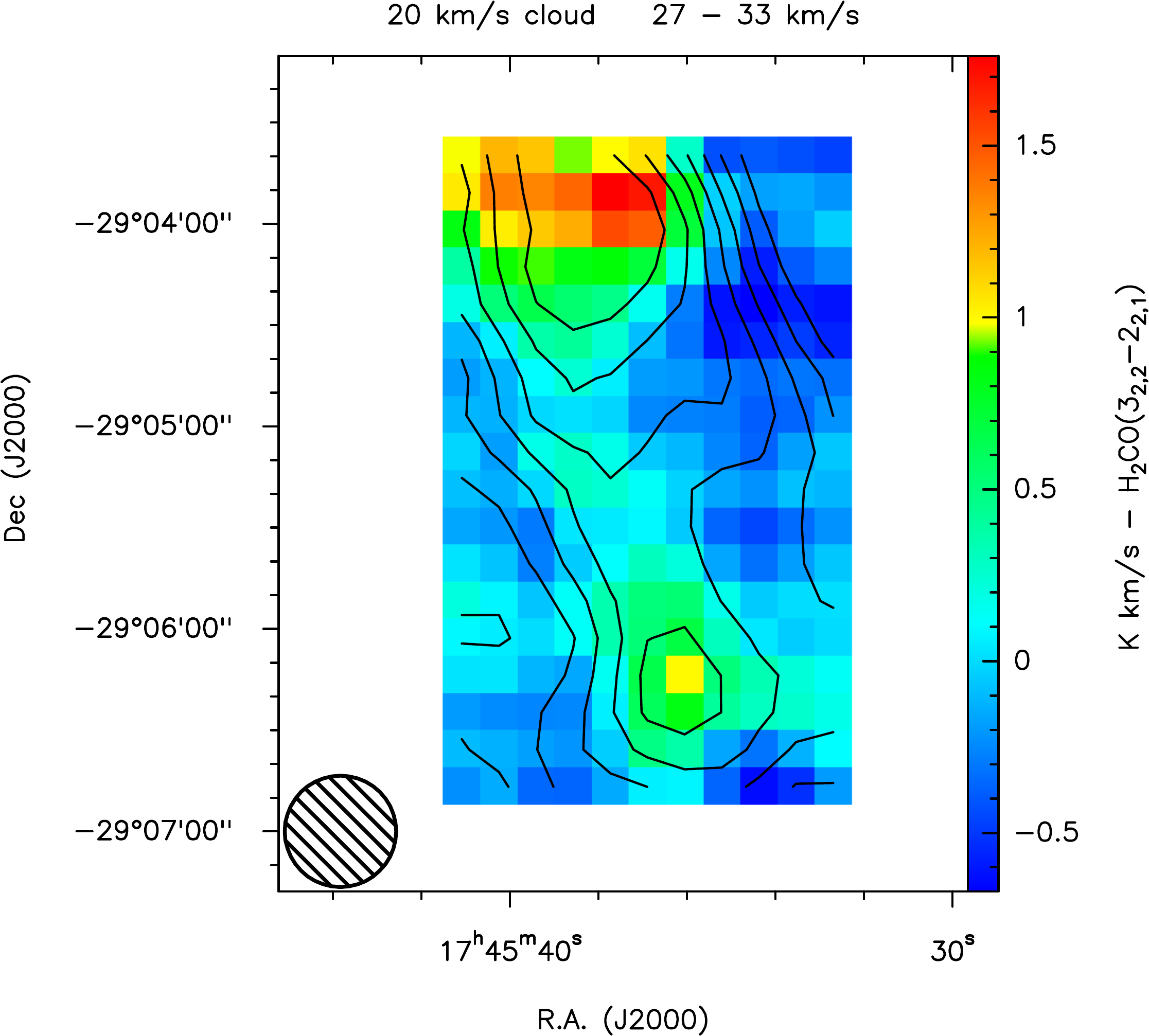}}
	\label{Methanol-Contamination}
\end{figure*}

\clearpage

\section{H$_{2}$CO(3$_{0,3}-$2$_{0,2}$) spectra}

 \begin{figure*}
	\caption{H$_{2}$CO(3$_{0,3}-$2$_{0,2}$) spectra, averaged over the whole OTF map of each source. The blue lines mark the velocity range of the whole source where the line emission is above 3$\sigma$. The red line shows the 
	fitted baseline (rms value in upper right corner).}
	\centering
	\subfloat{\includegraphics[width=7.5cm]{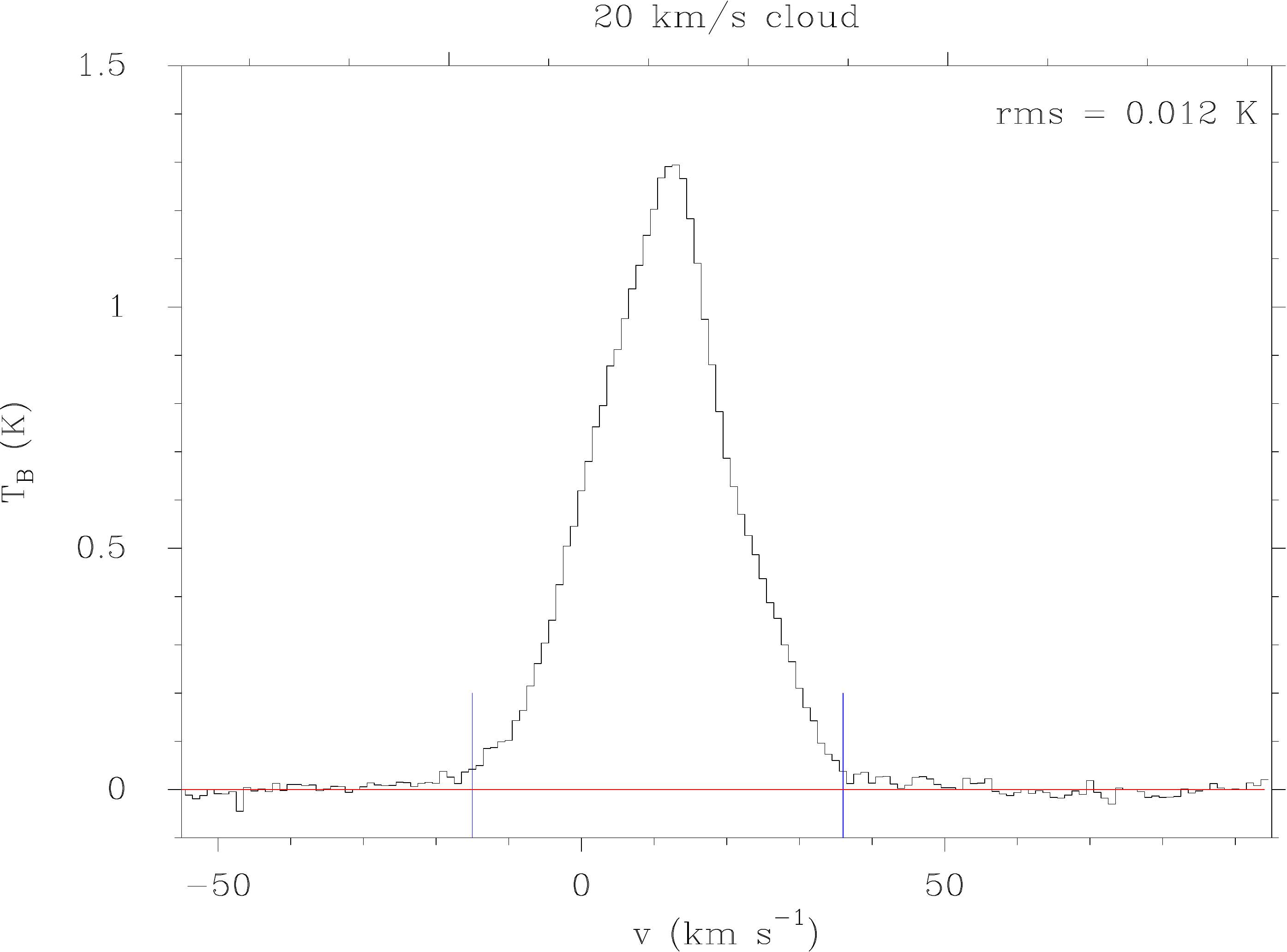}}
\hspace{0.1cm}
	\subfloat{\includegraphics[width=7.5cm]{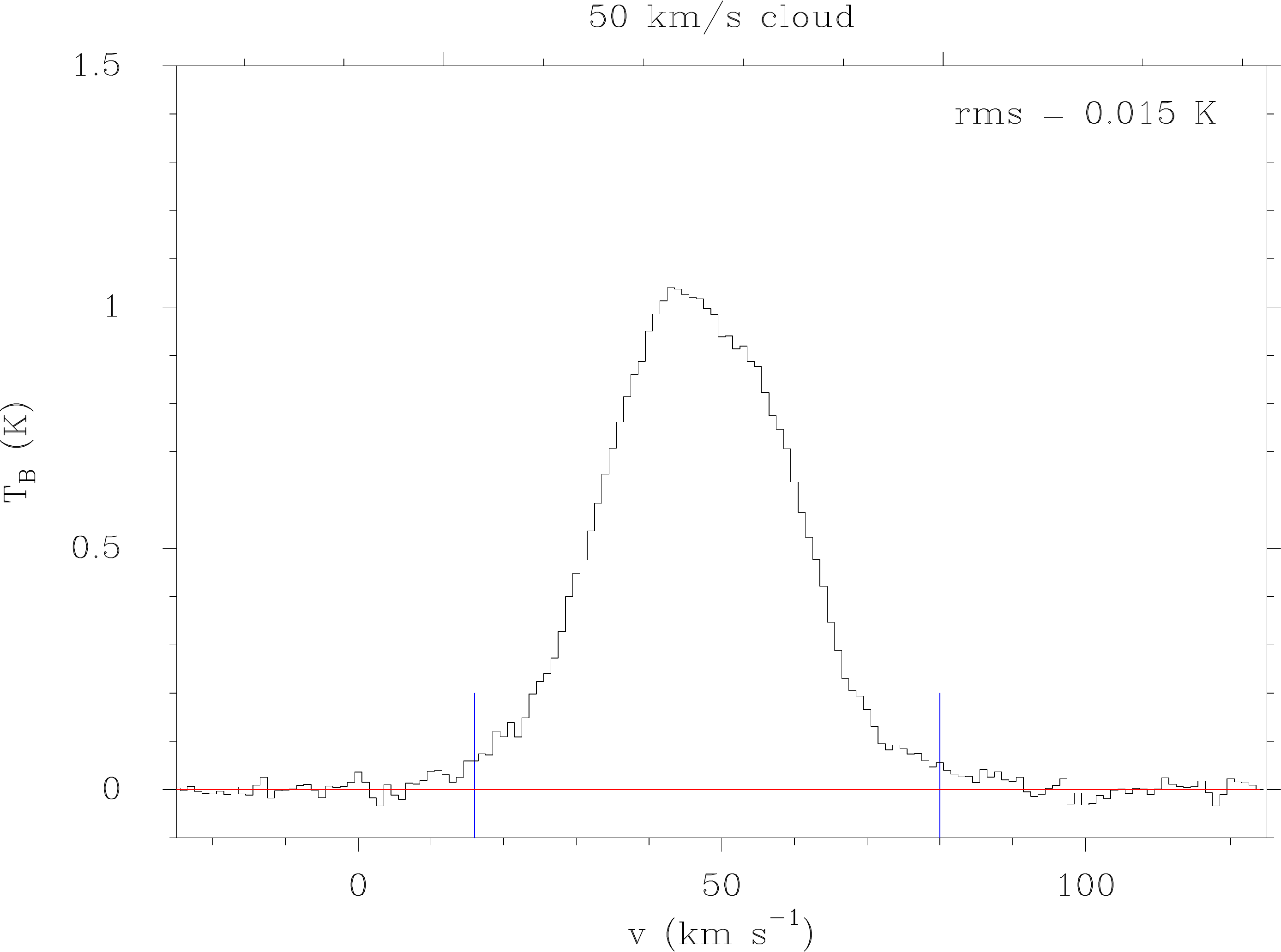}}\\
\hspace{0.1cm}
	\subfloat{\includegraphics[width=7.5cm]{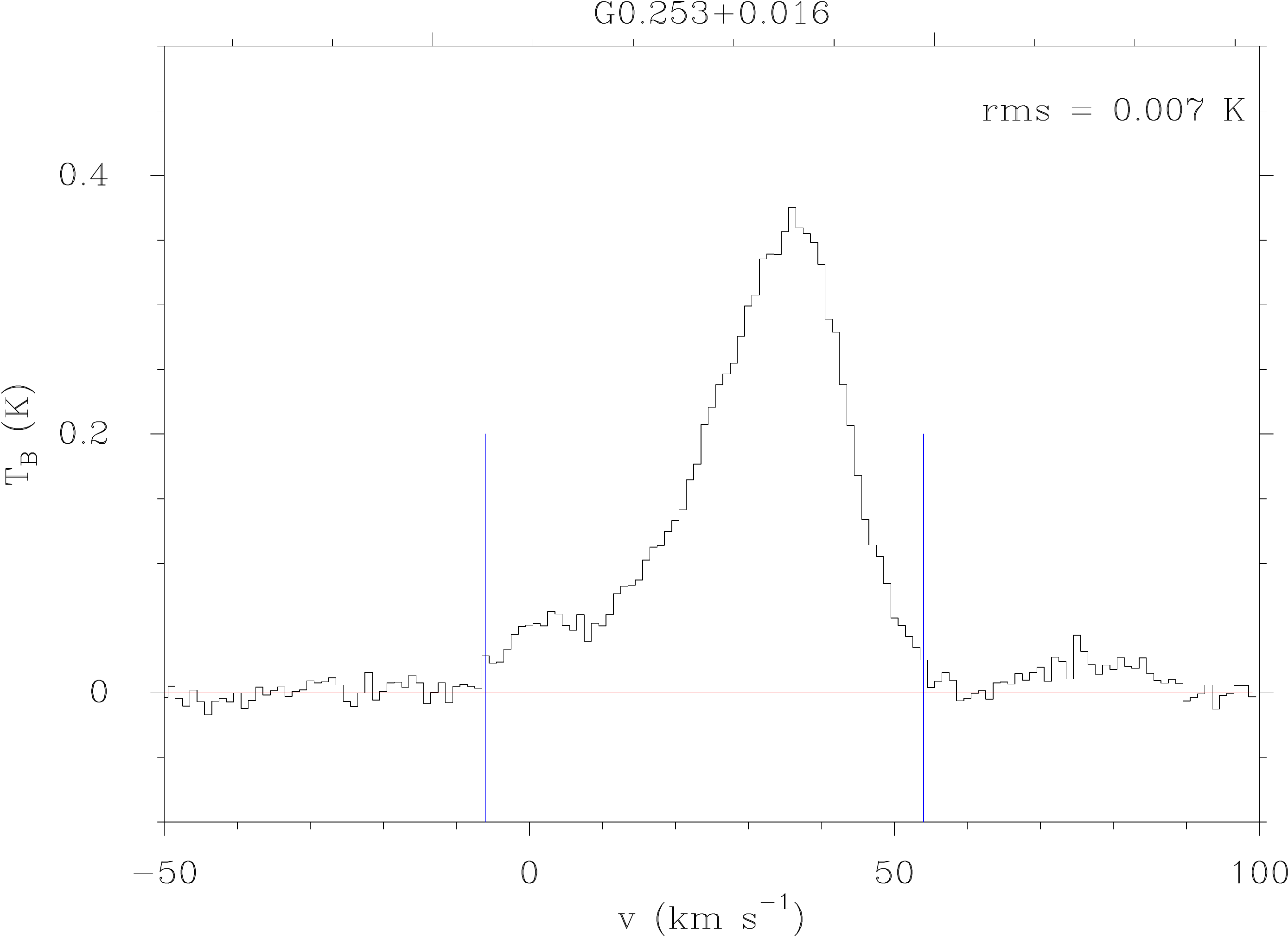}}
	\subfloat{\includegraphics[width=7.5cm]{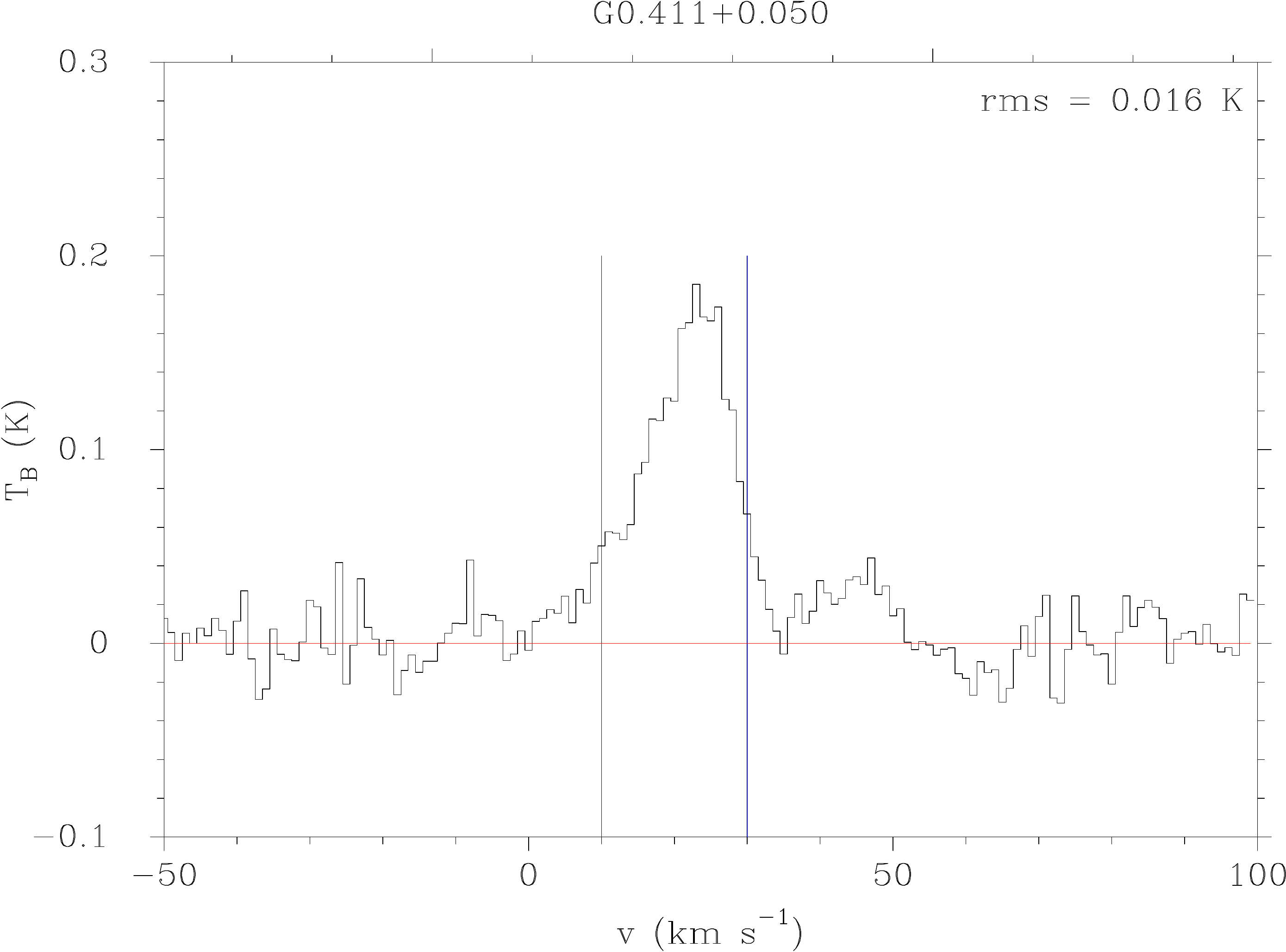}}\\
	\subfloat{\includegraphics[width=7.5cm]{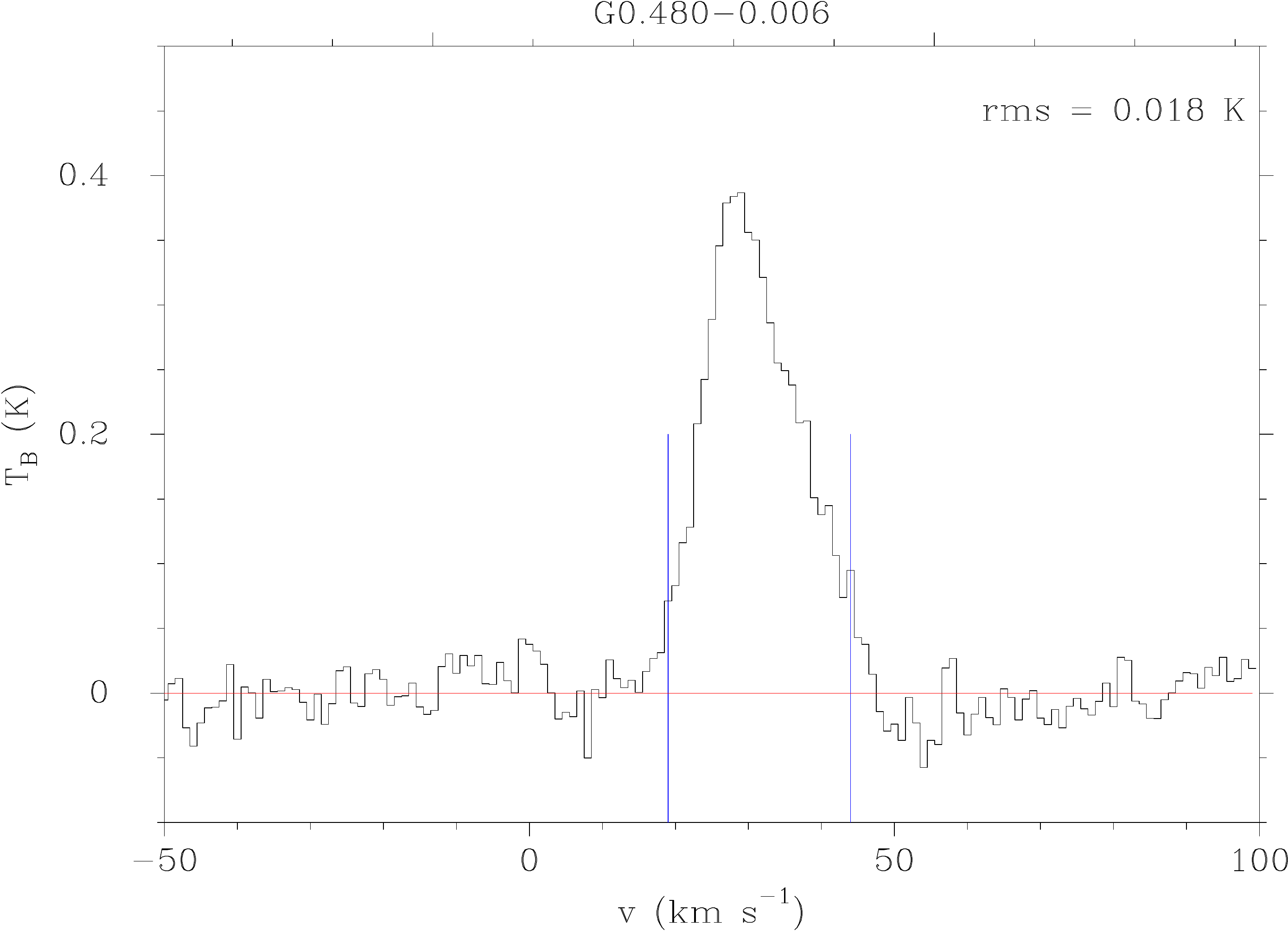}}
	\subfloat{\includegraphics[width=7.5cm]{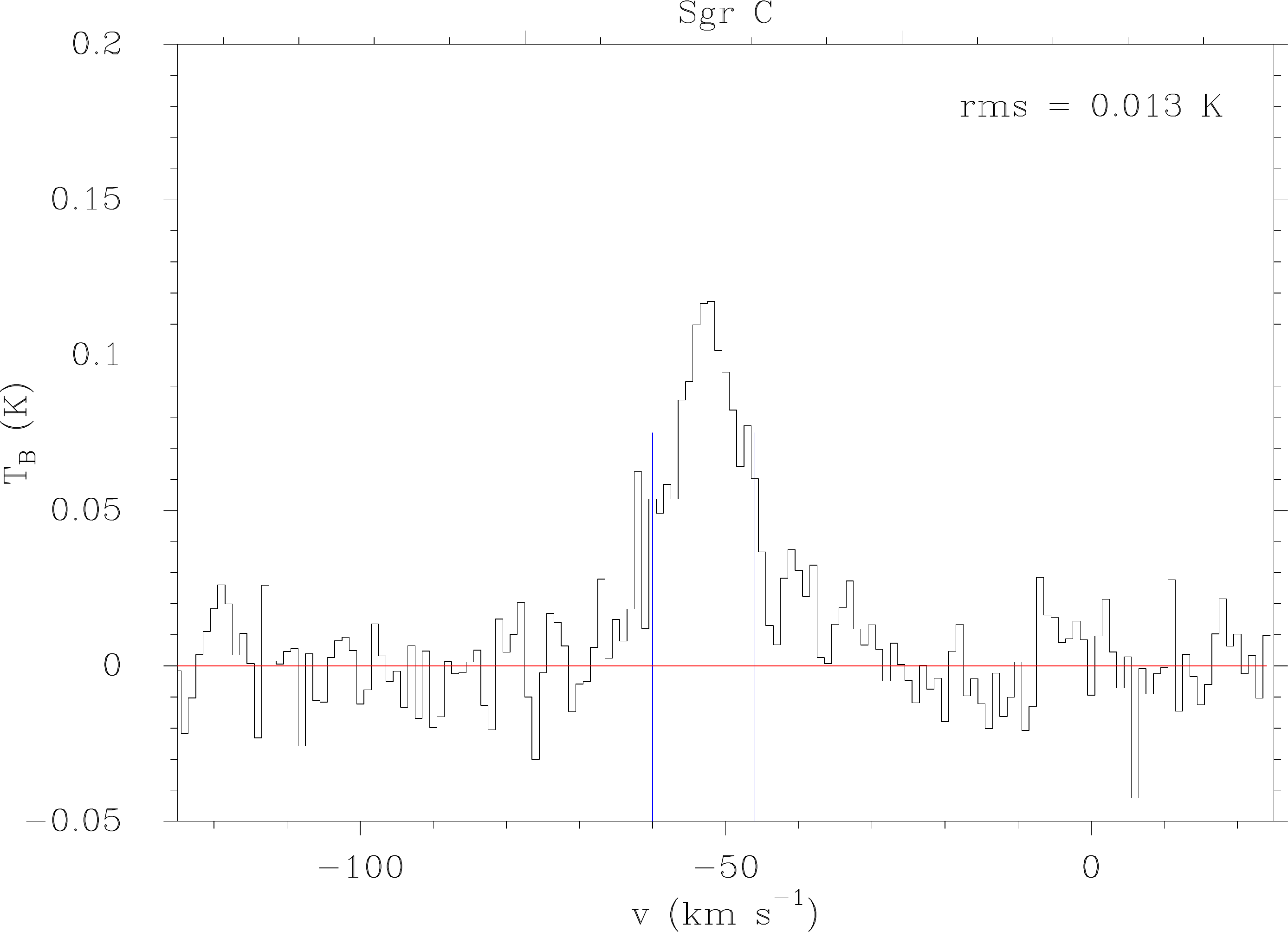}}\\
	\subfloat{\includegraphics[width=7.5cm]{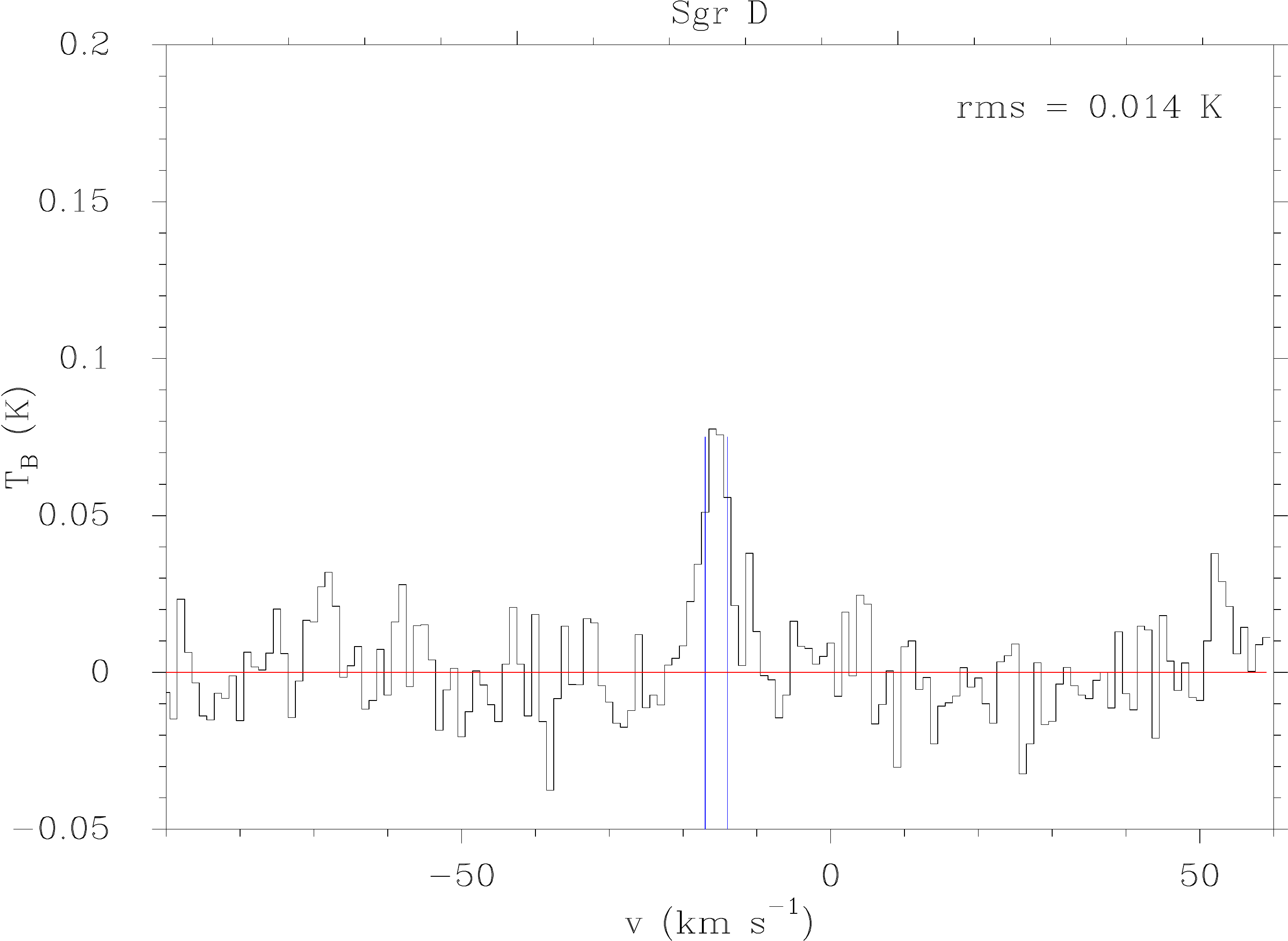}}
	\label{SpectraWholeVelRange}
\end{figure*}

\clearpage

\section{Integrated Intensity Maps}

\begin{figure*}
	\caption{Integrated intensities of the para-H$_{2}$CO lines at 218 and 291 GHz in the 20 km/s cloud. The contours show the moment 
	0 map of the H$_{2}$CO(3$_{0,3}-$2$_{0,2}$) transition, produced over the whole velocity range of the source 
	(levels: 30\%$-$90\% of the maximum in steps of 10\%). The circle in the lower left corner shows the 33$\arcsec$ beam.}
	\centering
	H$_{2}$CO(3$_{0,3}-$2$_{0,2}$)\\
	\subfloat{\includegraphics[bb = 0 0 600 580, clip, height=5cm]{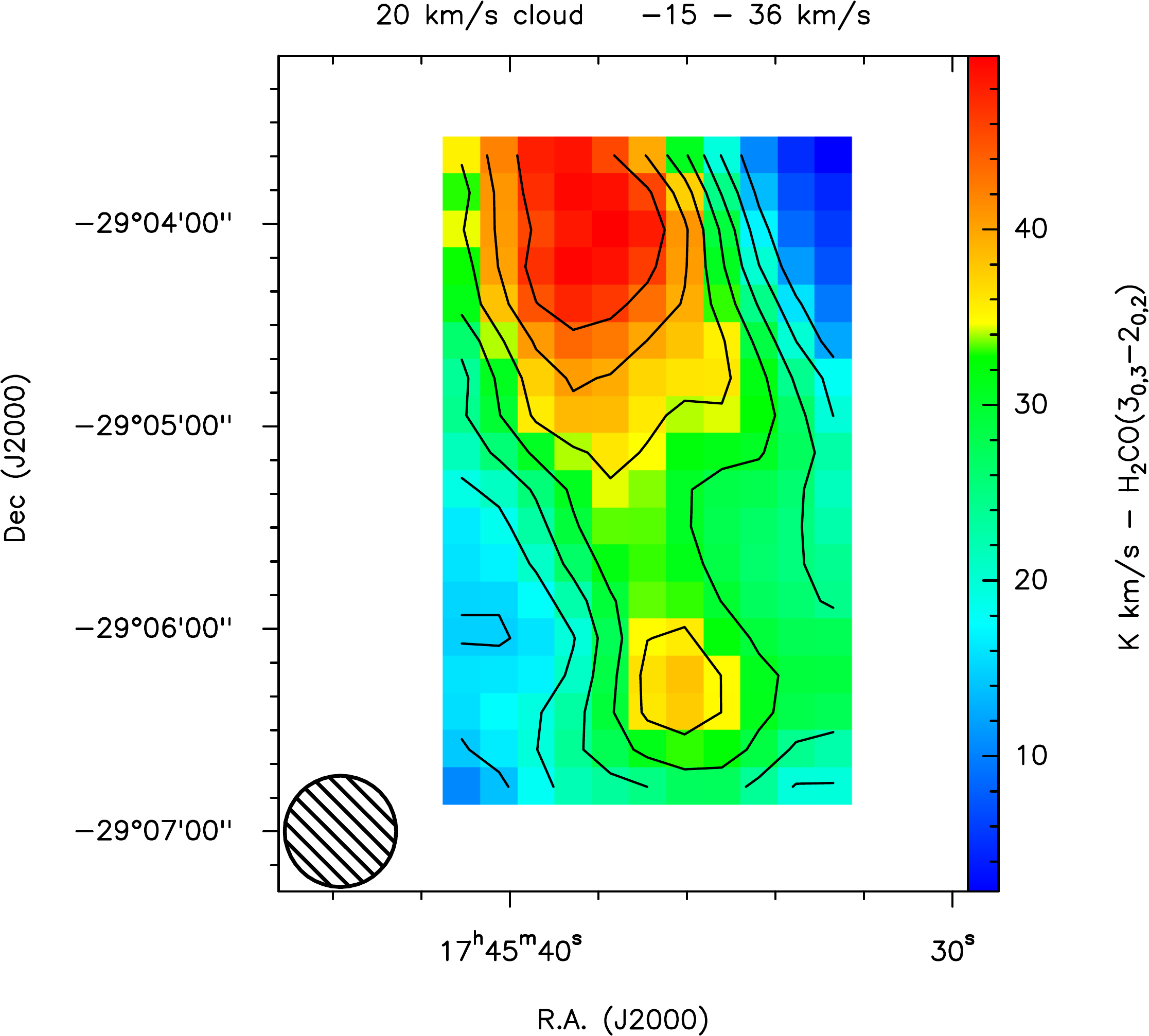}}
	\subfloat{\includegraphics[bb = 140 0 600 580, clip, height=5cm]{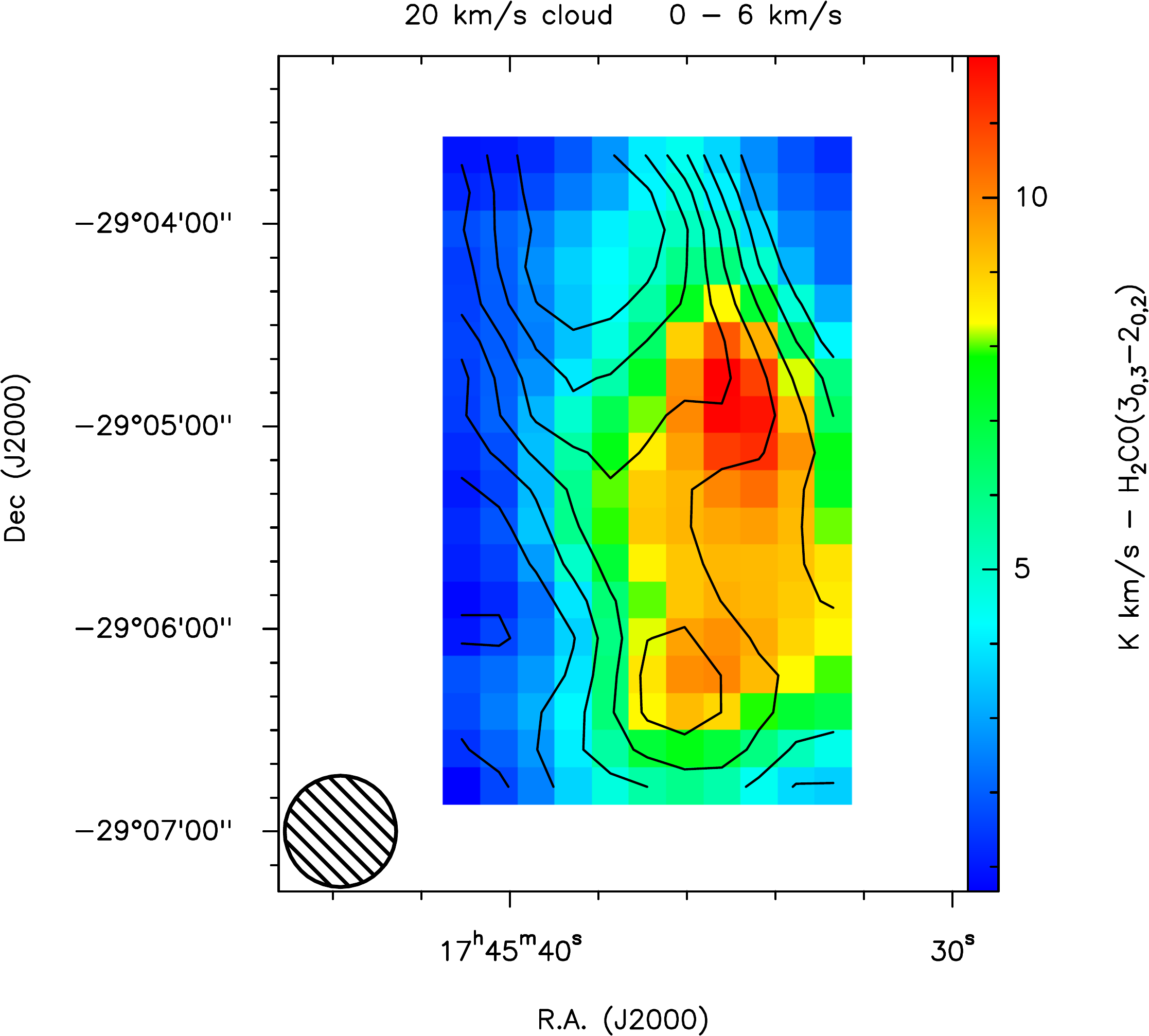}}
	\subfloat{\includegraphics[bb = 140 0 600 580, clip, height=5cm]{20kms-H2CO_303-Int_8-14kms.pdf}}
	\subfloat{\includegraphics[bb = 140 0 650 580, clip, height=5cm]{20kms-H2CO_303-Int_27-33kms.pdf}}\\
	H$_{2}$CO(3$_{2,1}-$2$_{2,0}$)\\
	\subfloat{\includegraphics[bb = 0 0 600 580, clip, height=5cm]{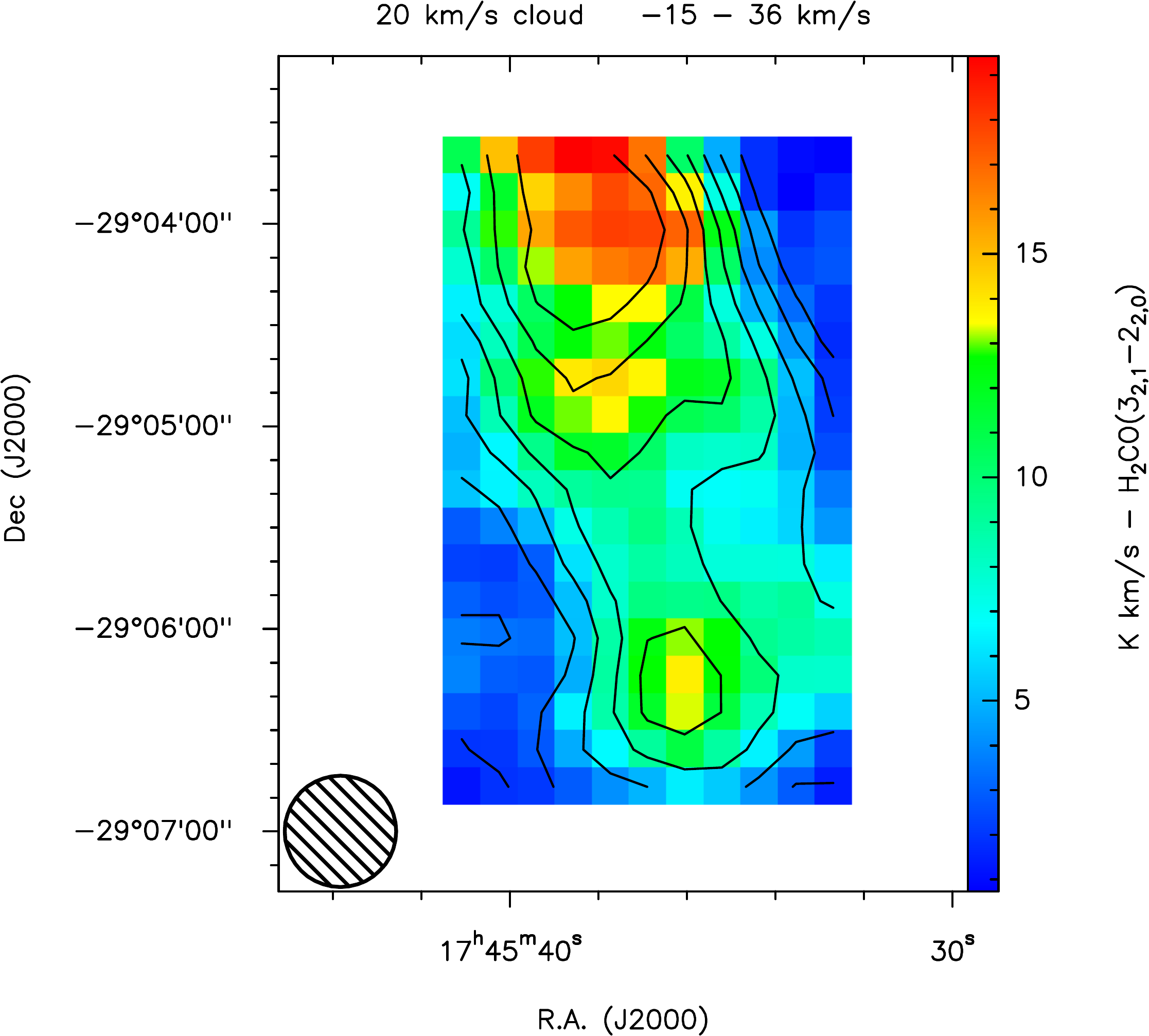}}
	\subfloat{\includegraphics[bb = 140 0 600 580, clip, height=5cm]{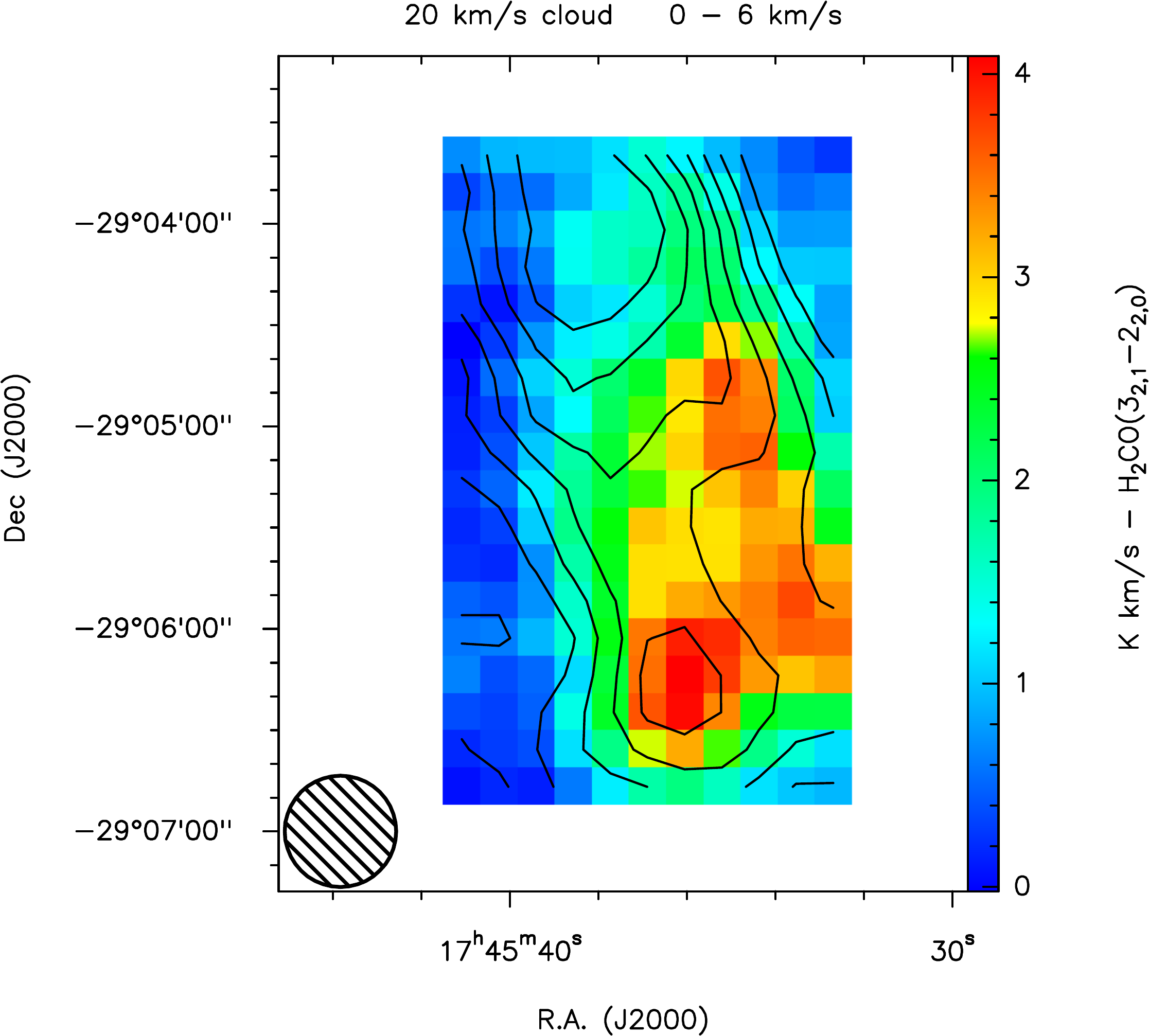}}
	\subfloat{\includegraphics[bb = 140 0 600 580, clip, height=5cm]{20kms-H2CO_321-Int_8-14kms.pdf}}
	\subfloat{\includegraphics[bb = 140 0 650 580, clip, height=5cm]{20kms-H2CO_321-Int_27-33kms.pdf}}\\
	H$_{2}$CO(4$_{0,3}-$3$_{0,3}$)\\
	\subfloat{\includegraphics[bb = 0 0 600 580, clip, height=5cm]{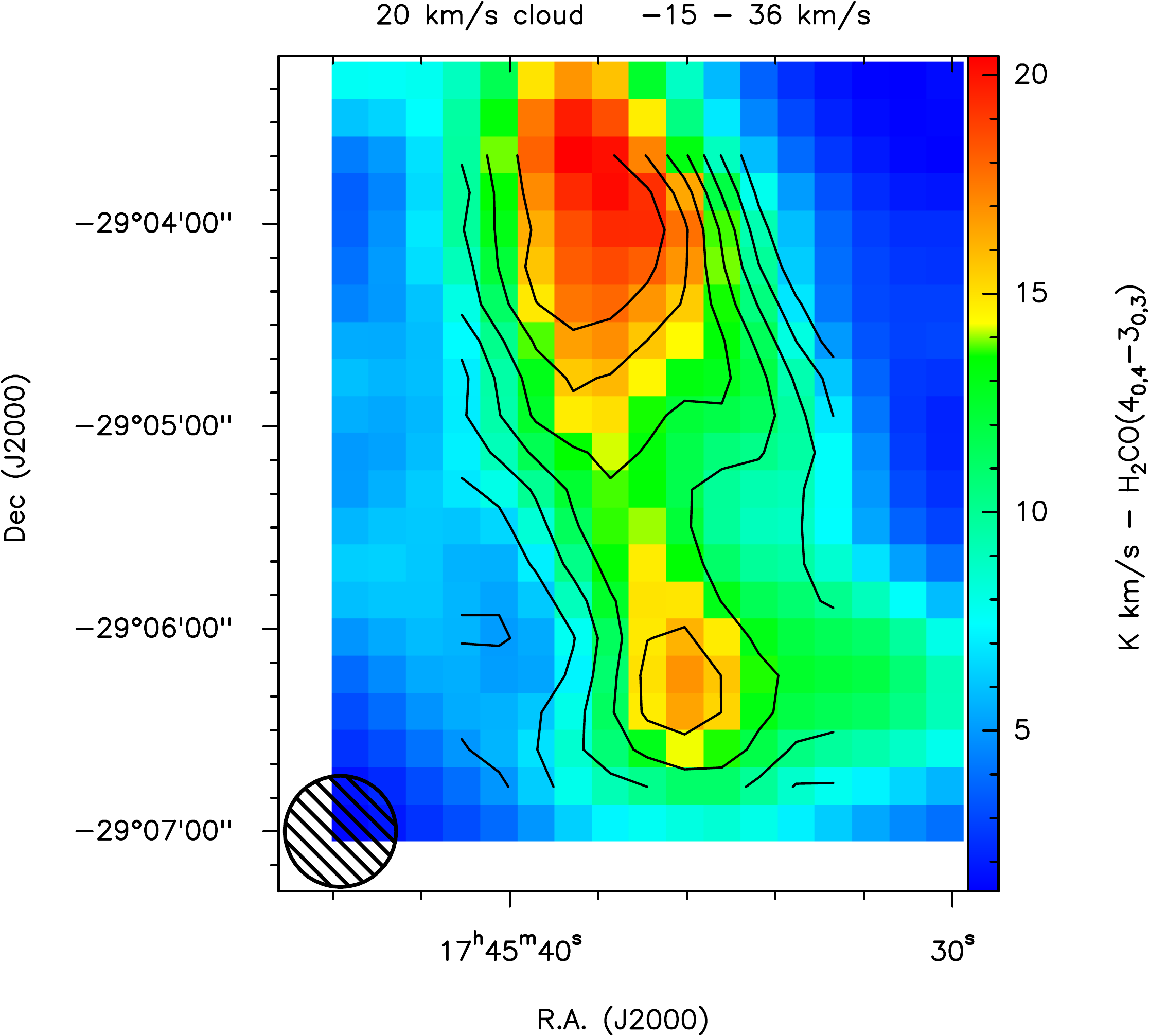}}
	\subfloat{\includegraphics[bb = 140 0 600 580, clip, height=5cm]{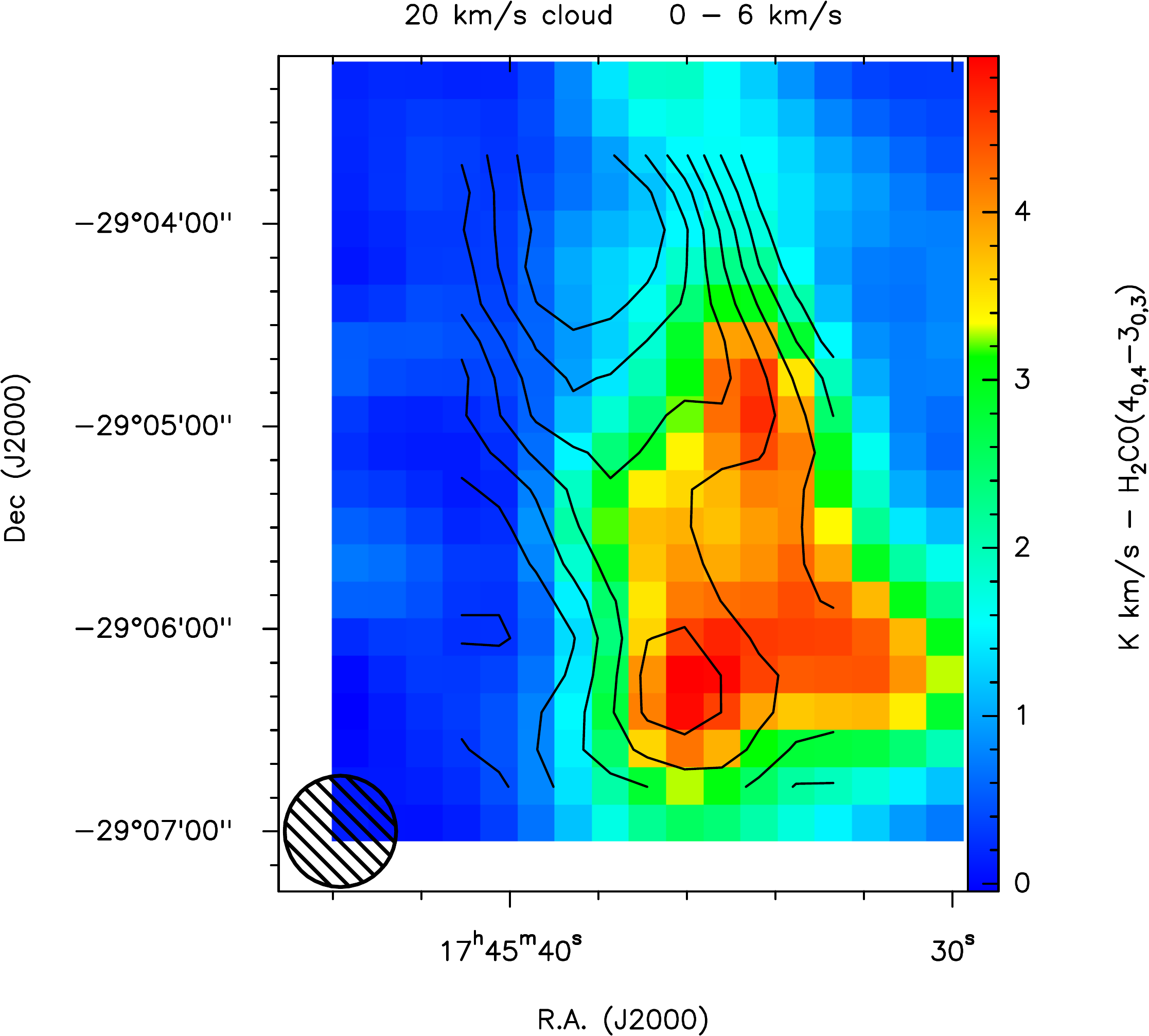}}
	\subfloat{\includegraphics[bb = 140 0 600 580, clip, height=5cm]{20kms-H2CO_404-Int_8-14kms.pdf}}
	\subfloat{\includegraphics[bb = 140 0 650 580, clip, height=5cm]{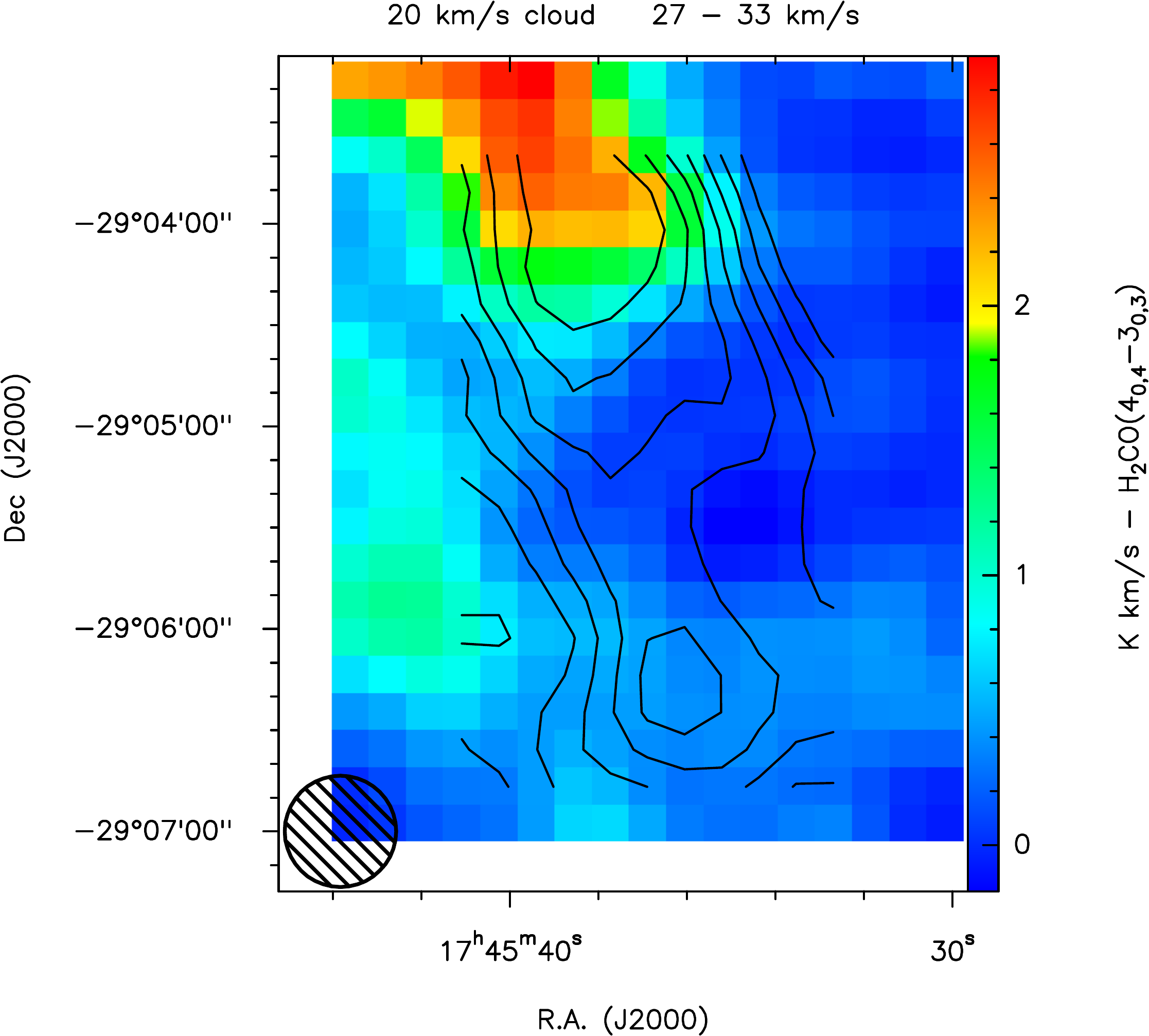}}\\
	H$_{2}$CO(4$_{2,2}-$3$_{2,1}$)\\
	\subfloat{\includegraphics[bb = 0 0 600 580, clip, height=5cm]{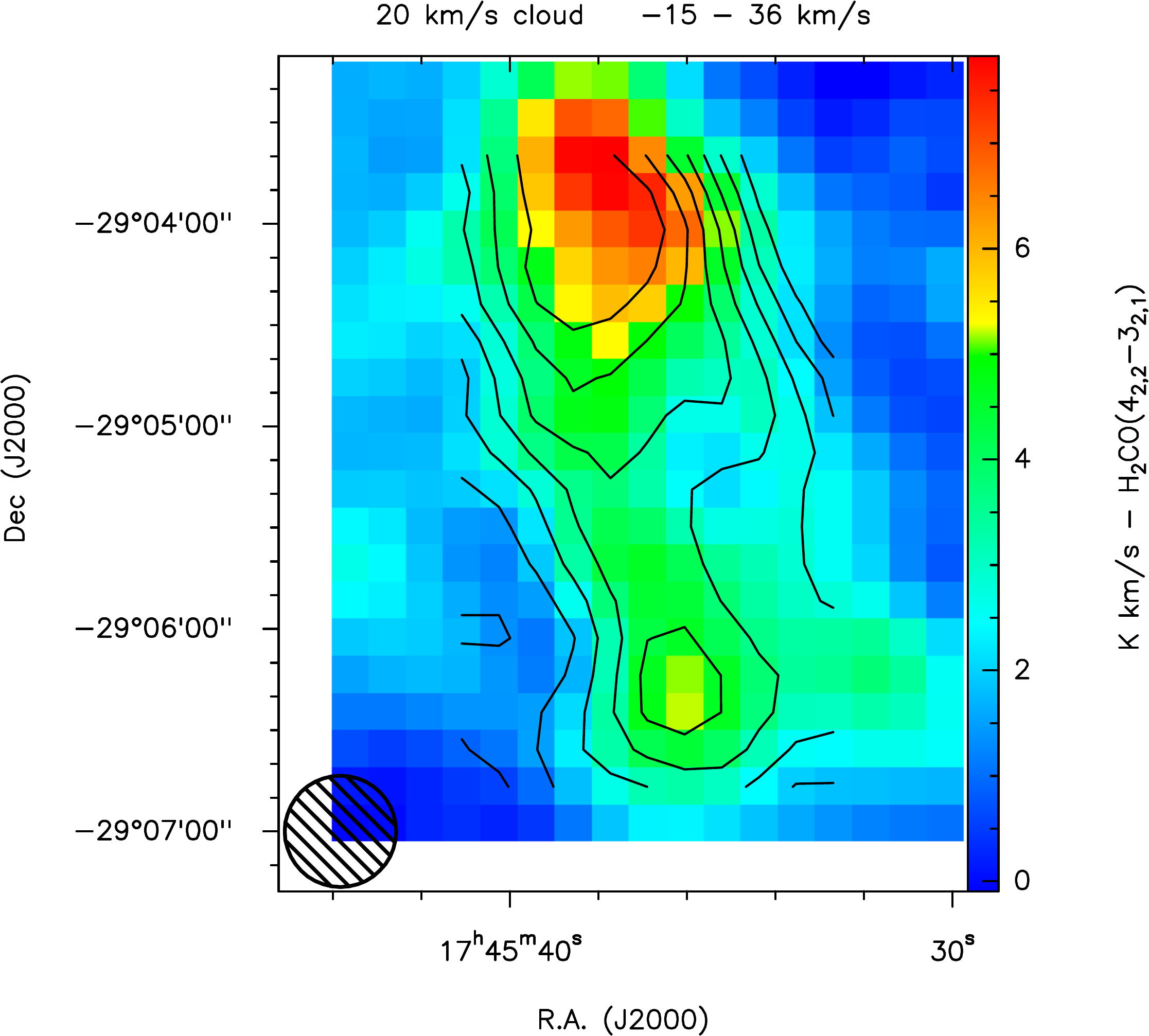}}
	\subfloat{\includegraphics[bb = 140 0 600 580, clip, height=5cm]{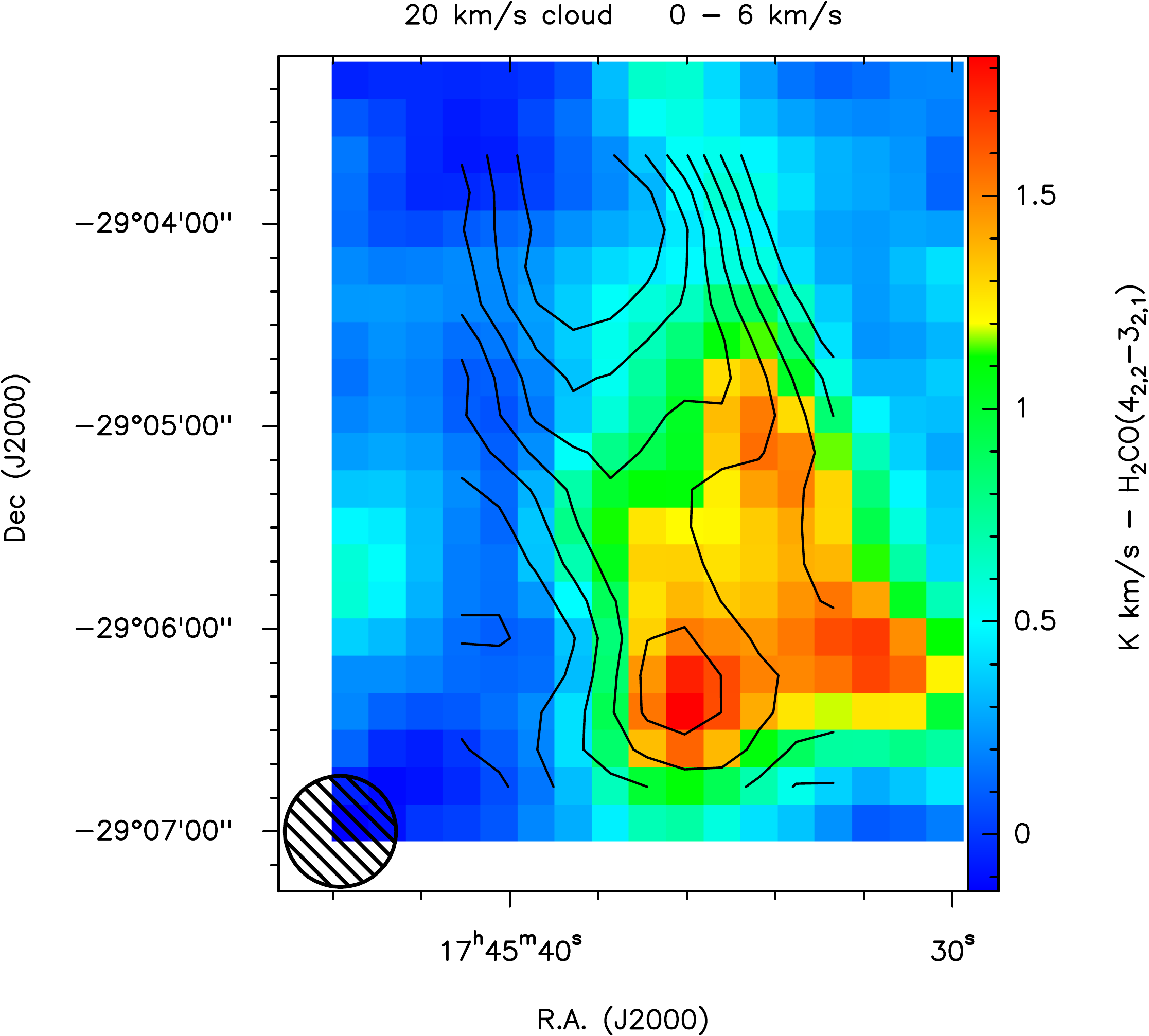}}
	\subfloat{\includegraphics[bb = 140 0 600 580, clip, height=5cm]{20kms-H2CO_422-Int_8-14kms.pdf}}
	\subfloat{\includegraphics[bb = 140 0 650 580, clip, height=5cm]{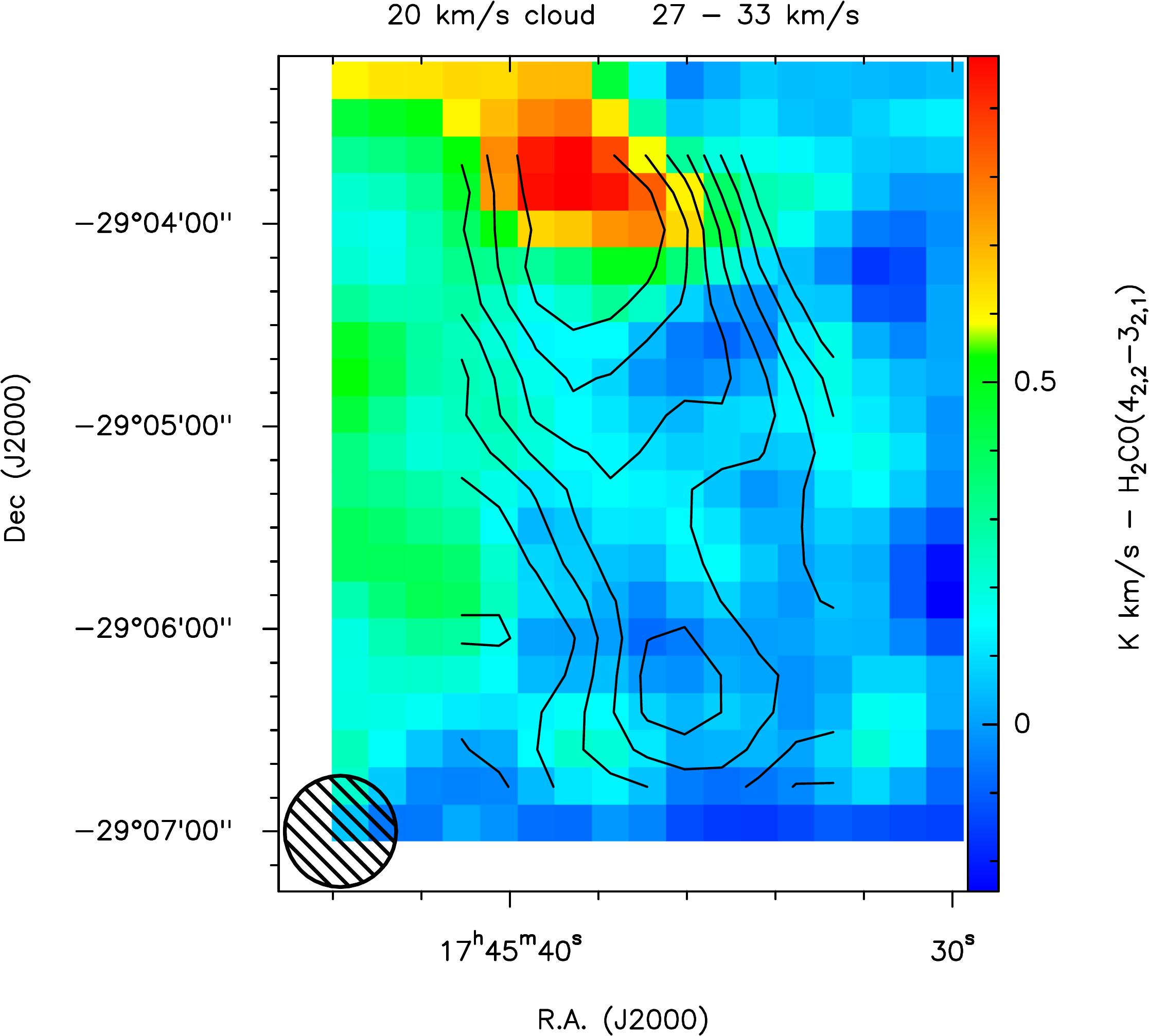}}\\
	\label{20kms-Int-H2CO}
\end{figure*}

\begin{figure*}
	\caption{As Fig. \ref{20kms-Int-H2CO} for the 50 km/s cloud.}
	\centering
	H$_{2}$CO(3$_{0,3}-$2$_{0,2}$)\\
	\subfloat{\includegraphics[bb = 0 0 690 580, clip, height=5cm]{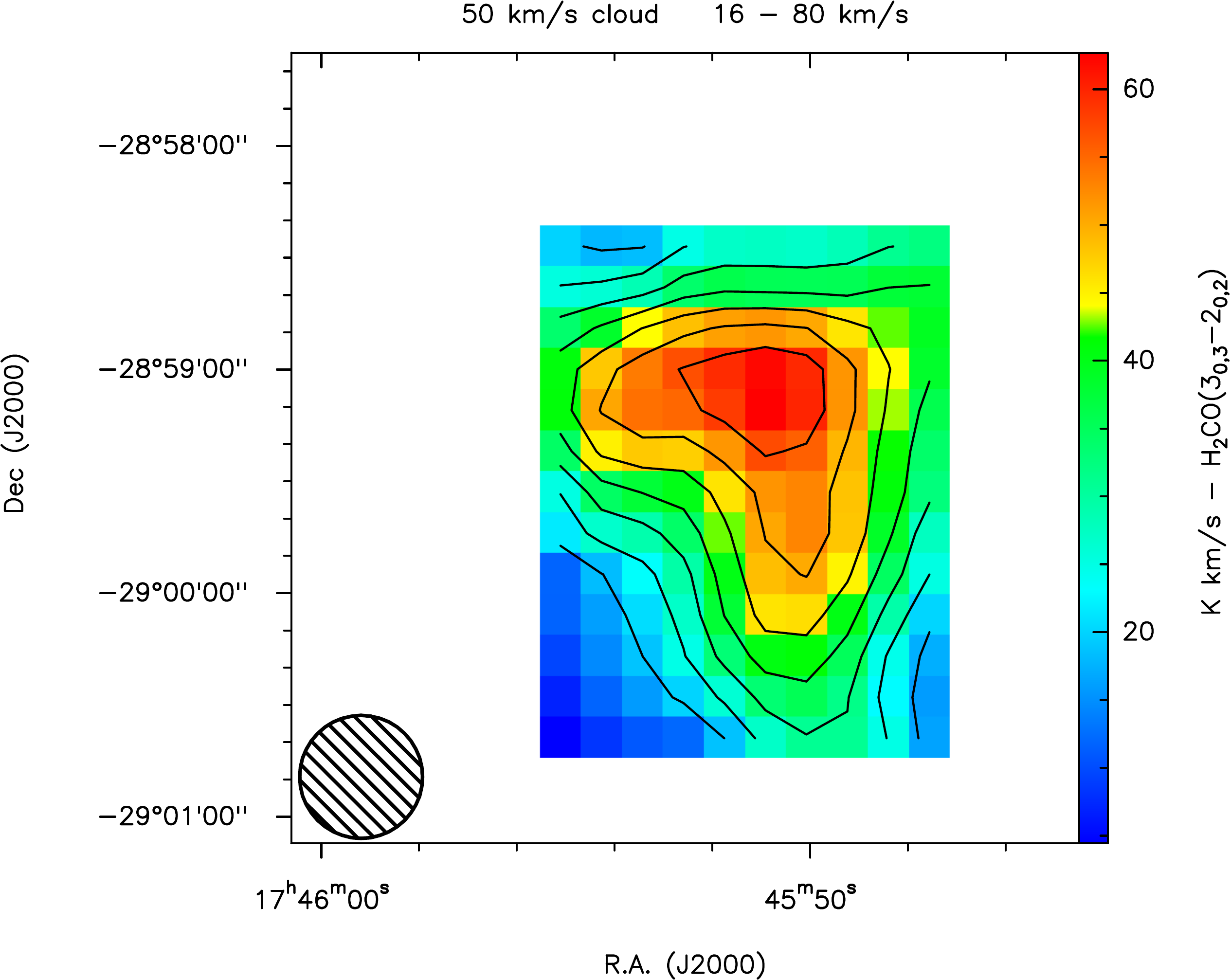}}
	\subfloat{\includegraphics[bb = 150 0 690 580, clip, height=5cm]{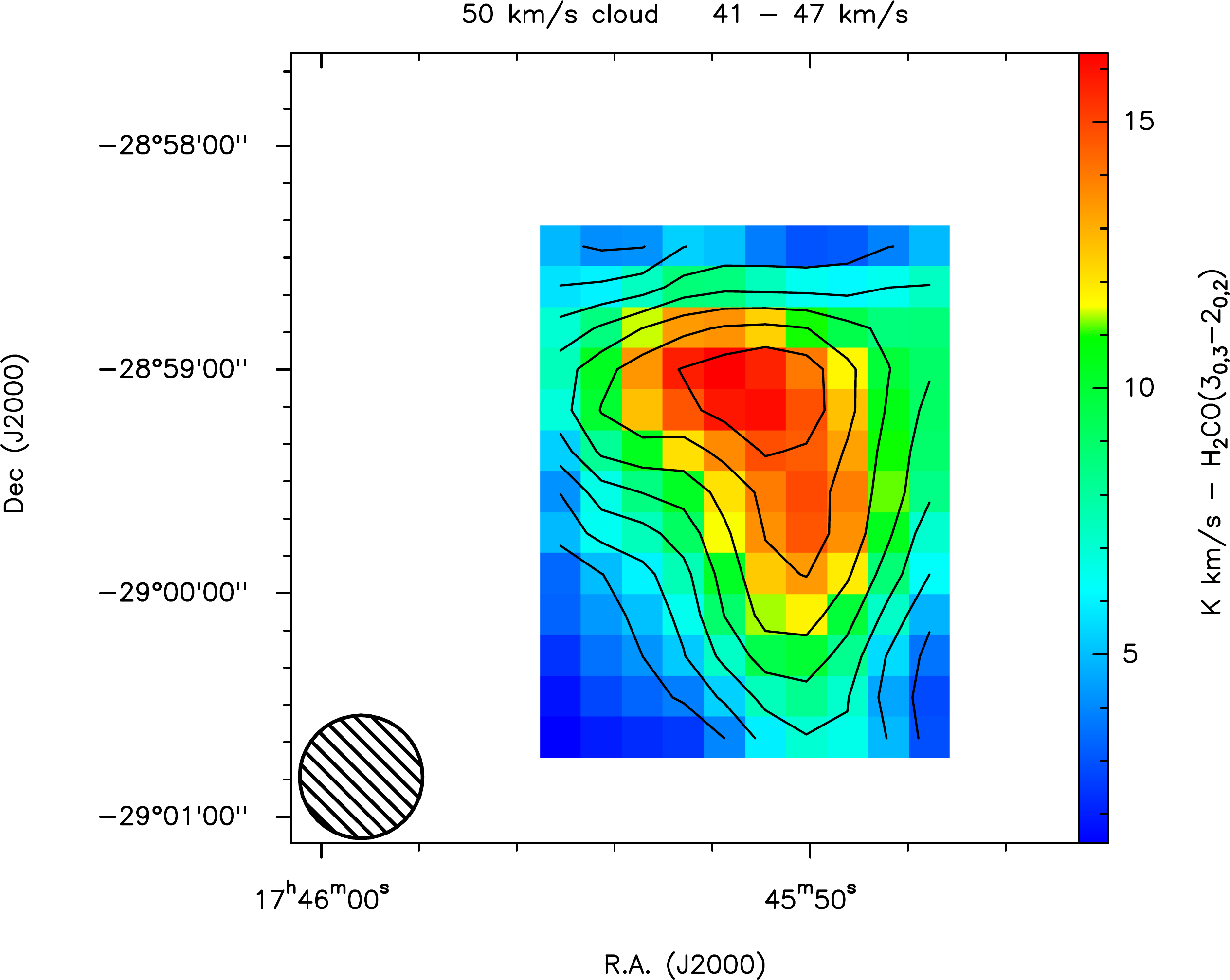}}
	\subfloat{\includegraphics[bb = 150 0 730 580, clip, height=5cm]{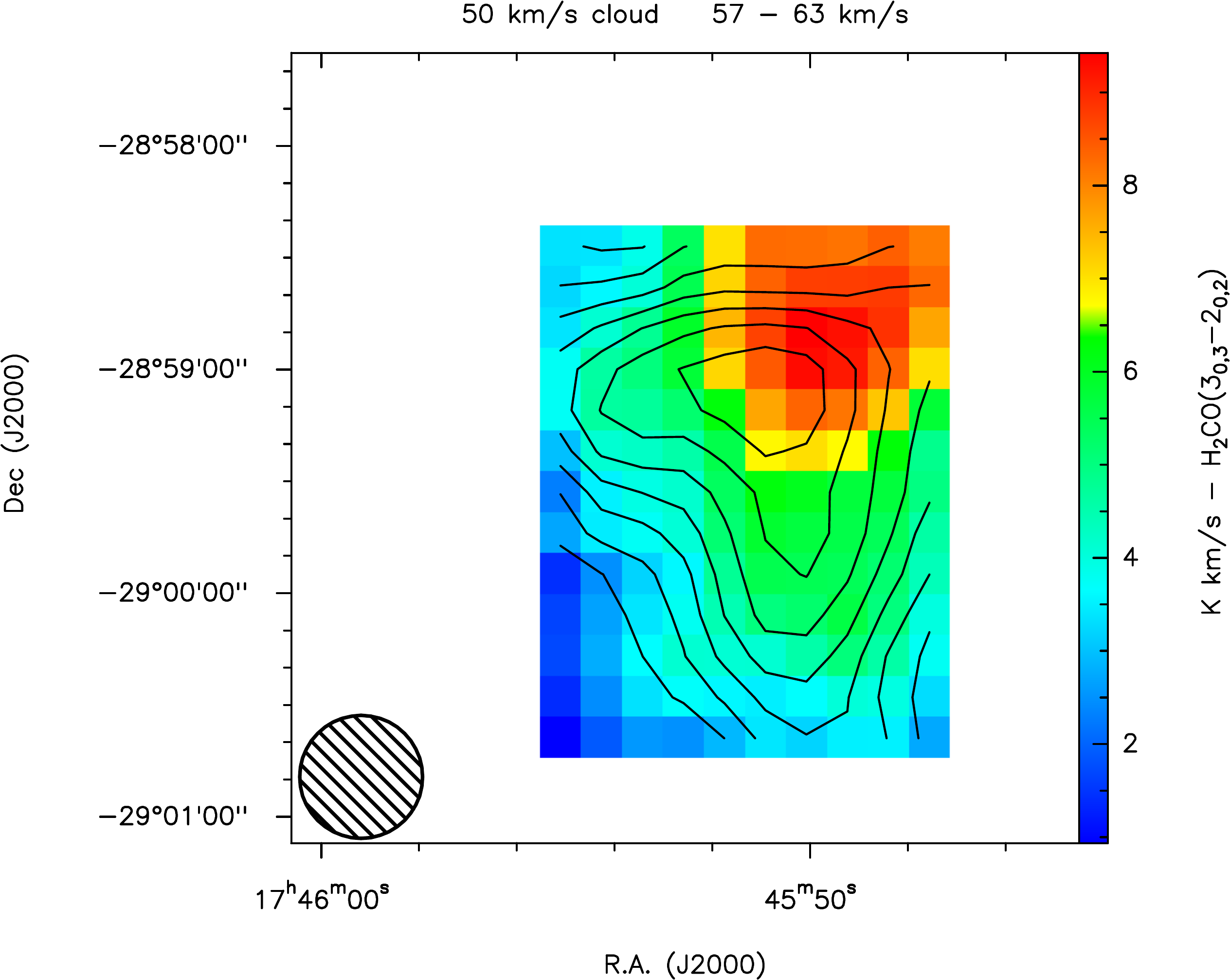}}\\
	H$_{2}$CO(3$_{2,1}-$2$_{2,0}$)\\
	\subfloat{\includegraphics[bb = 0 0 690 580, clip, height=5cm]{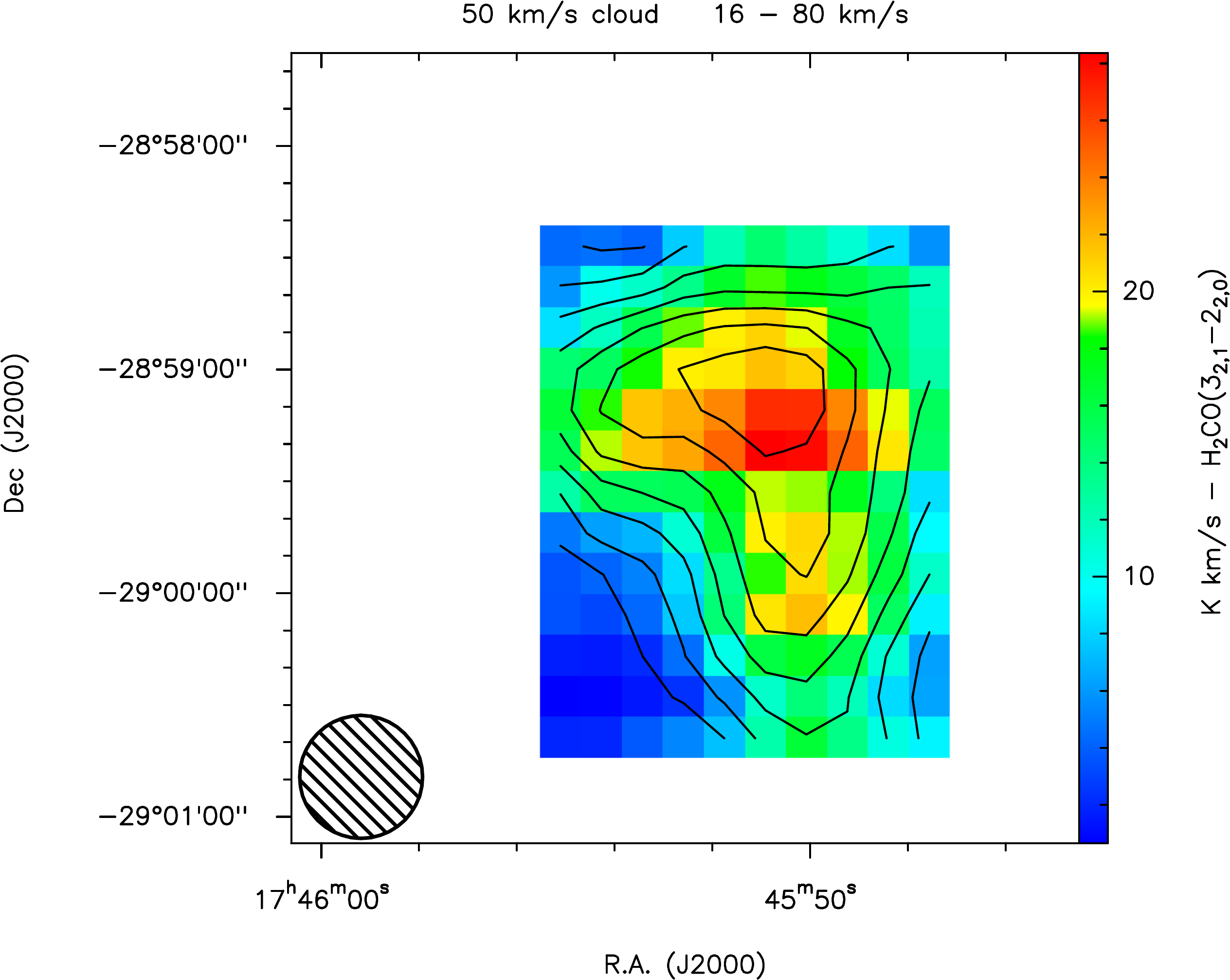}}
	\subfloat{\includegraphics[bb = 150 0 690 580, clip, height=5cm]{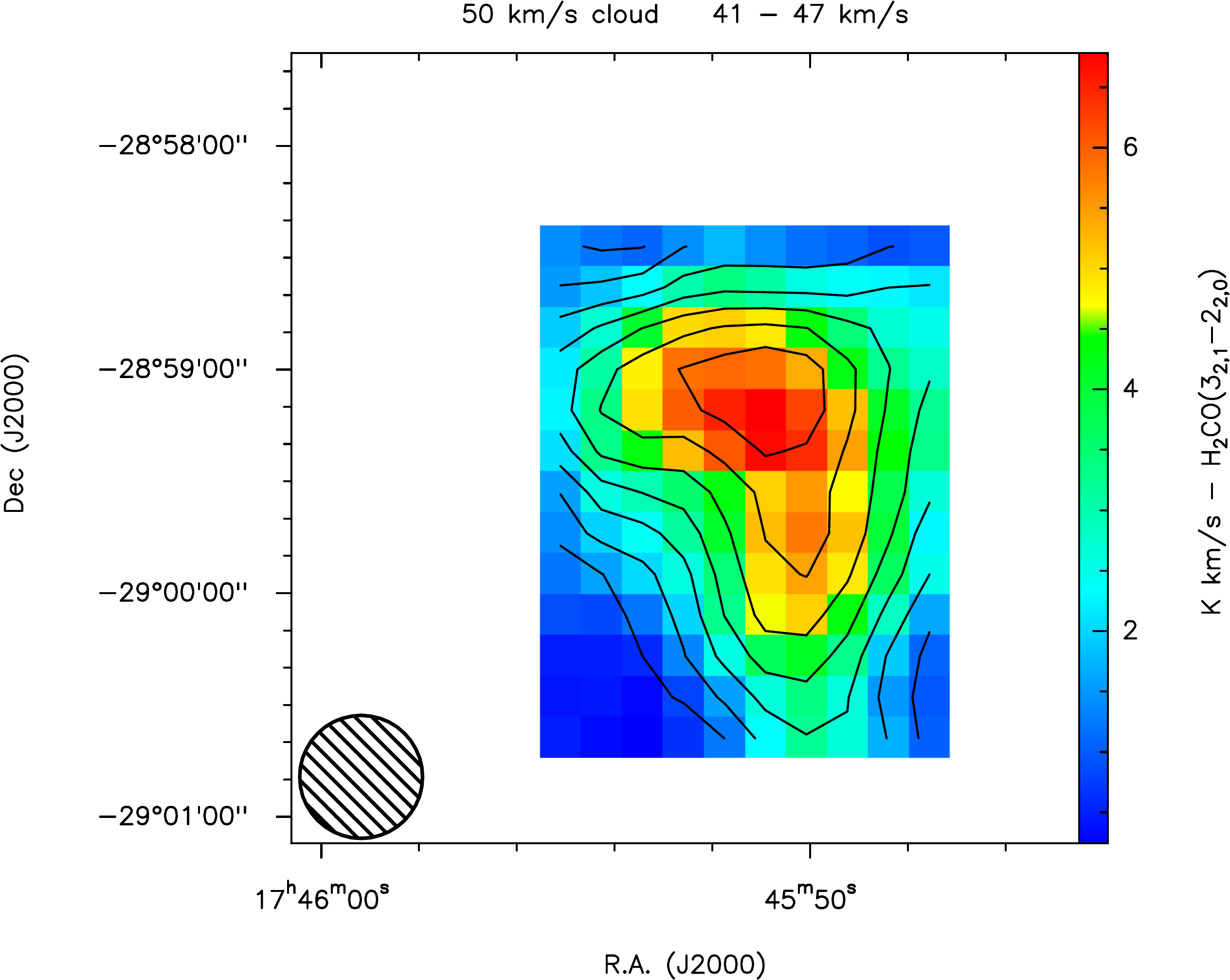}}
	\subfloat{\includegraphics[bb = 150 0 730 580, clip, height=5cm]{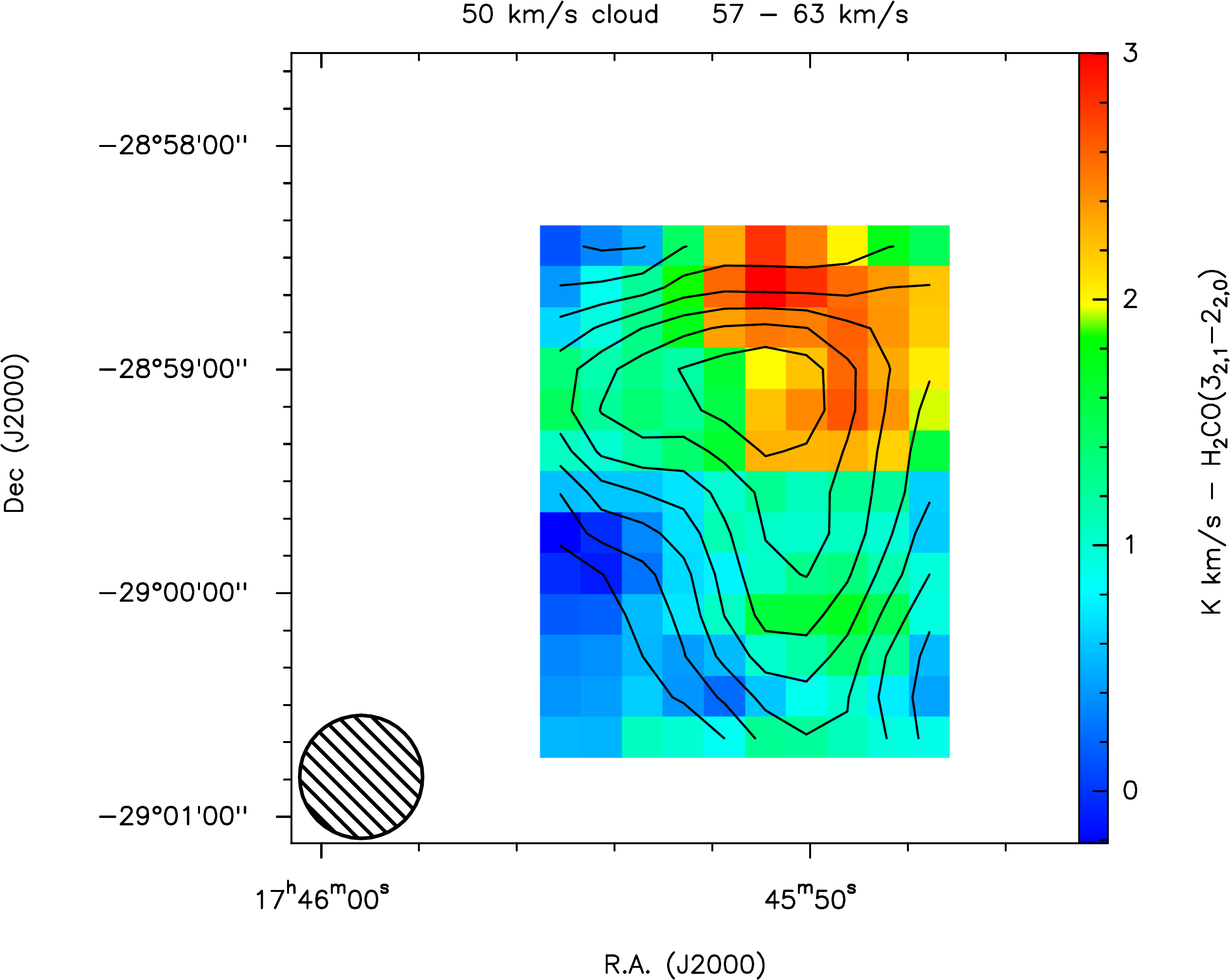}}\\
	H$_{2}$CO(4$_{0,3}-$3$_{0,3}$)\\
	\subfloat{\includegraphics[bb = 0 0 690 580, clip, height=5cm]{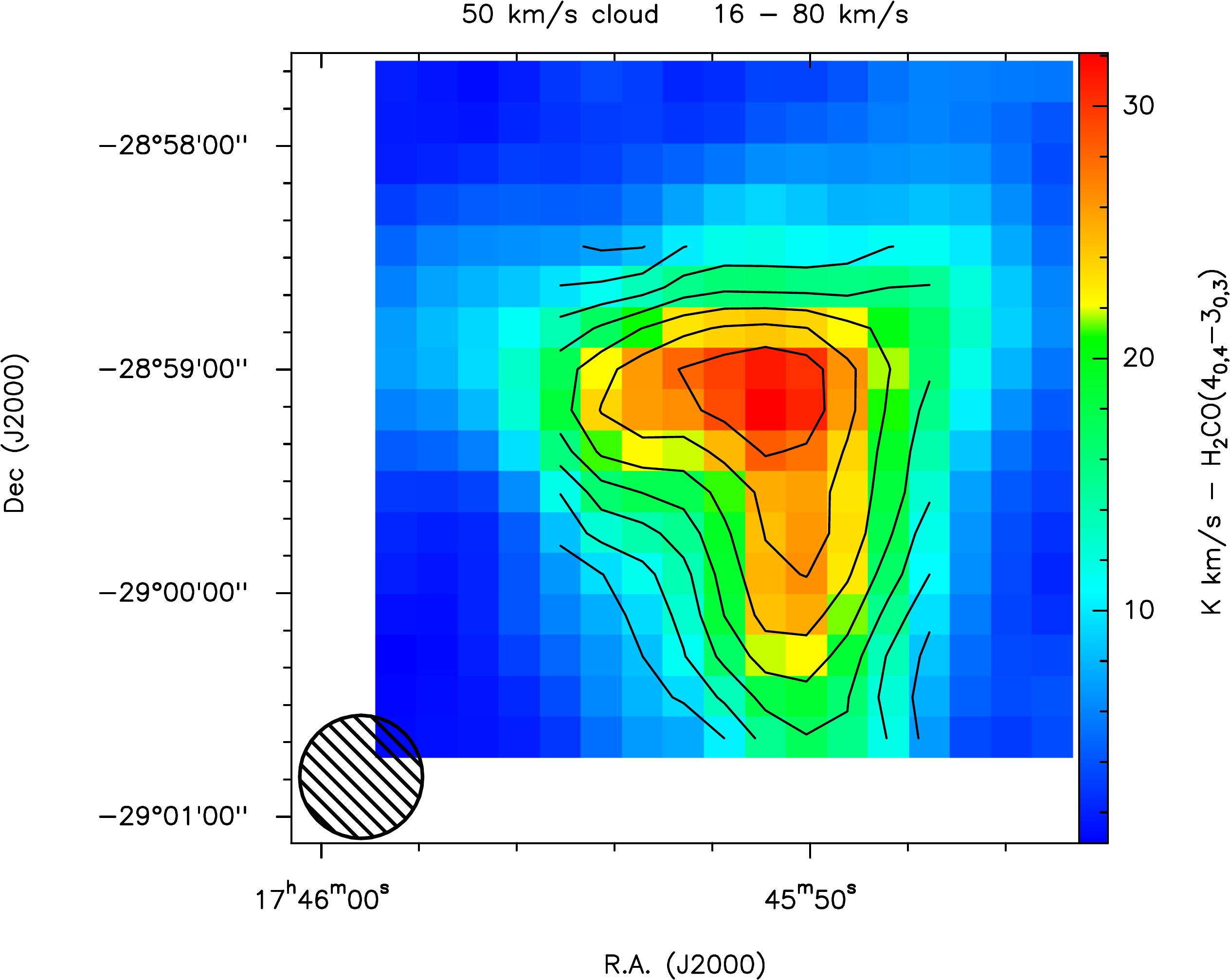}}
	\subfloat{\includegraphics[bb = 150 0 690 580, clip, height=5cm]{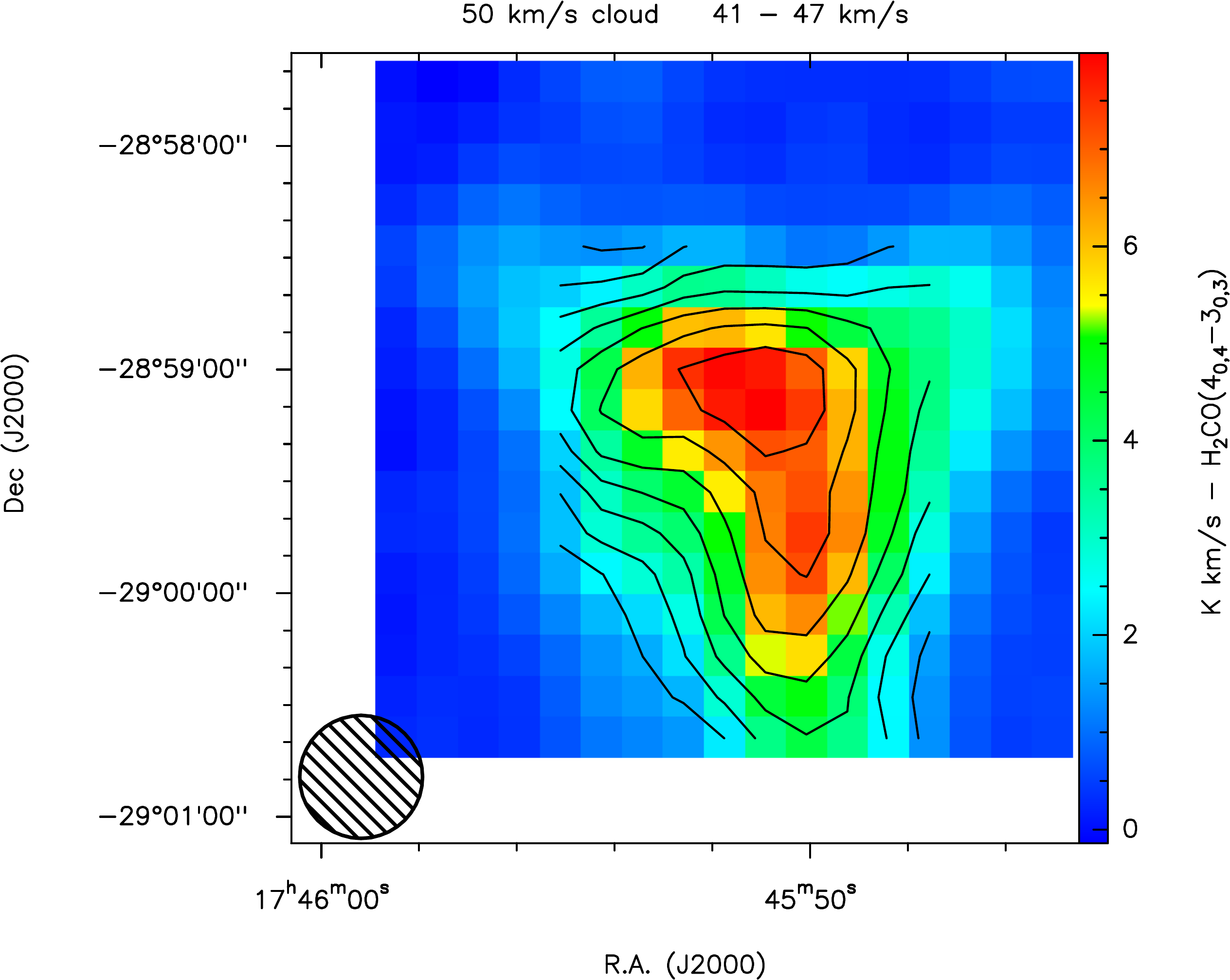}}
	\subfloat{\includegraphics[bb = 150 0 730 580, clip, height=5cm]{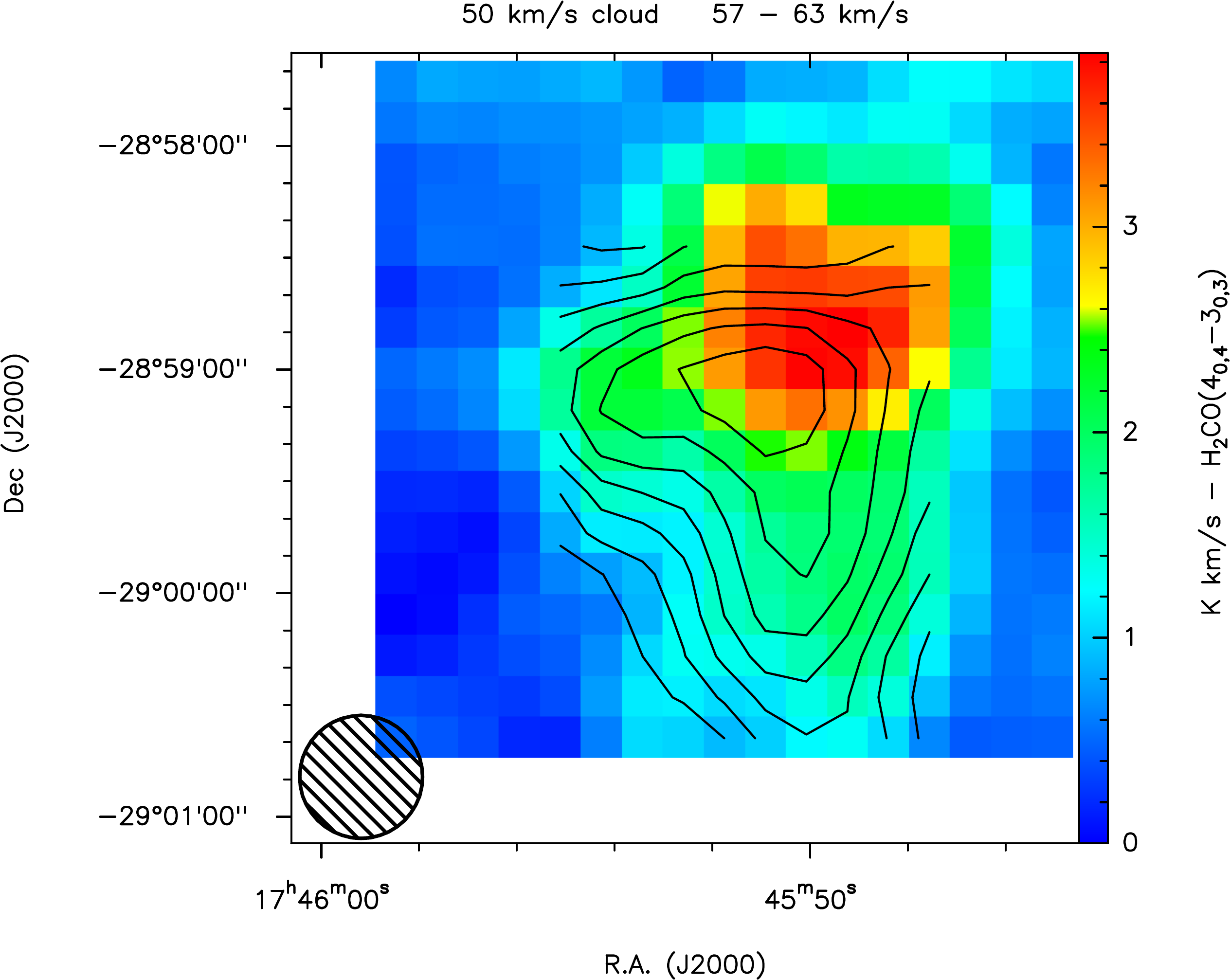}}\\
	H$_{2}$CO(4$_{2,2}-$3$_{2,1}$)\\
	\subfloat{\includegraphics[bb = 0 0 690 580, clip, height=5cm]{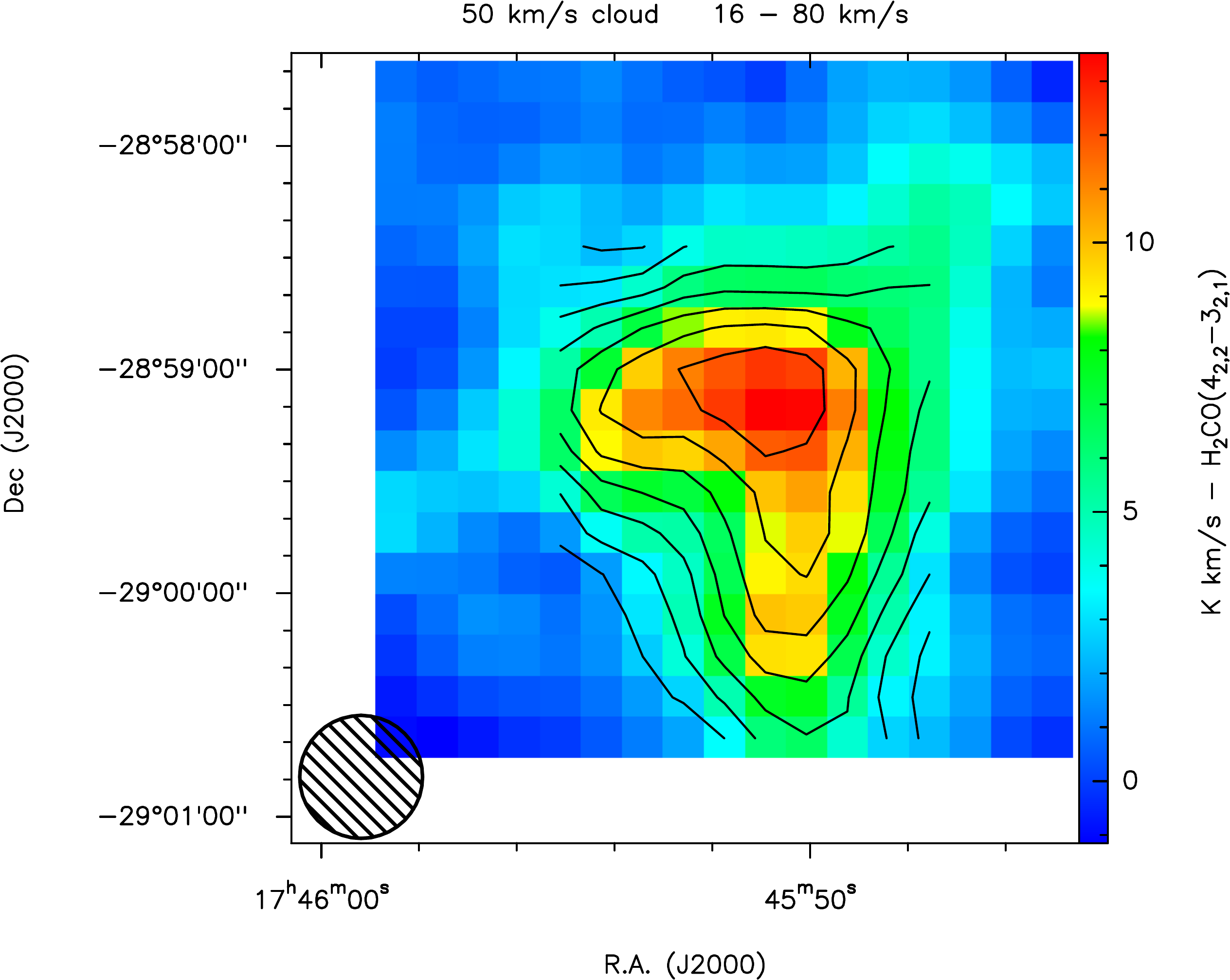}}
	\subfloat{\includegraphics[bb = 150 0 690 580, clip, height=5cm]{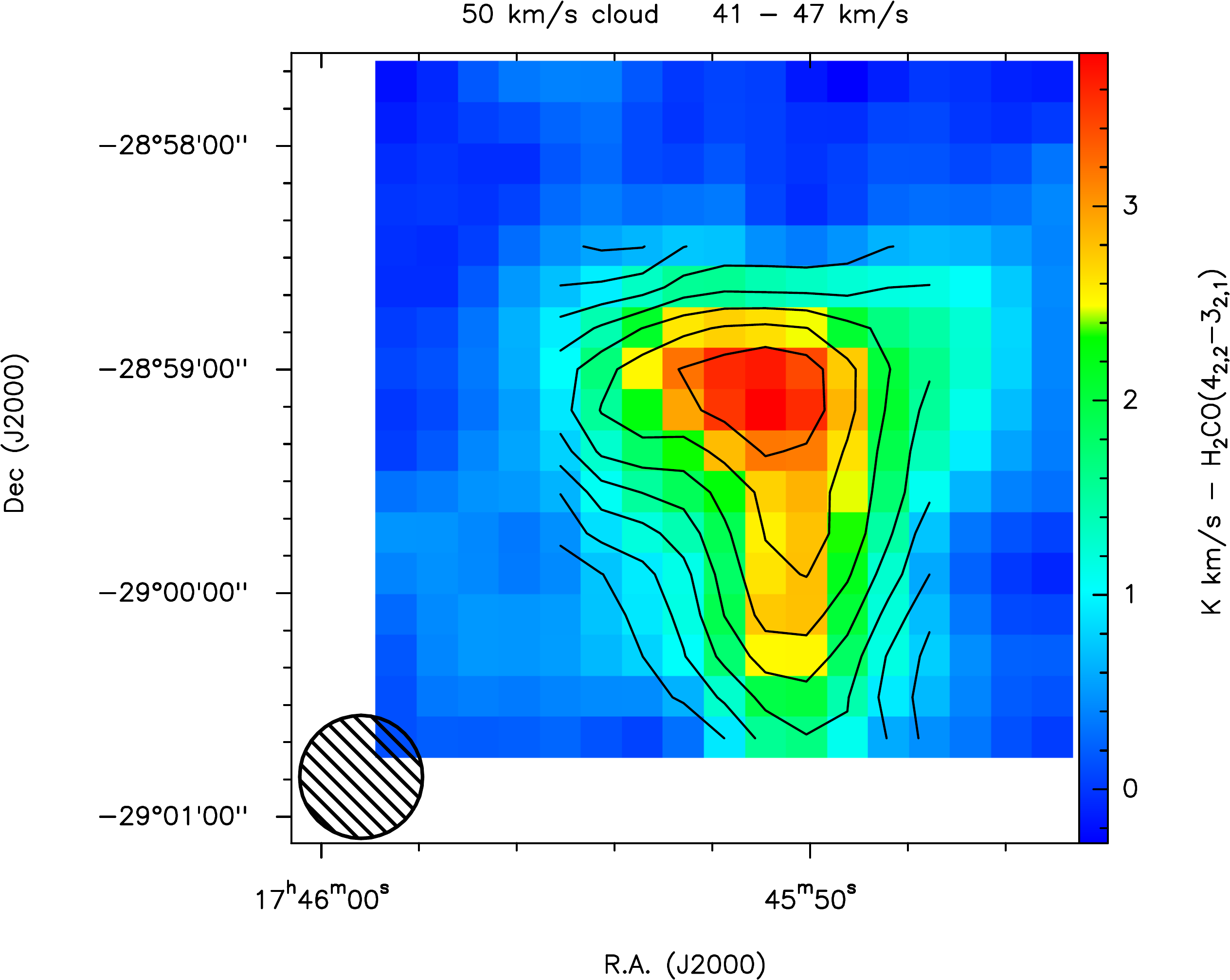}}
	\subfloat{\includegraphics[bb = 150 0 730 580, clip, height=5cm]{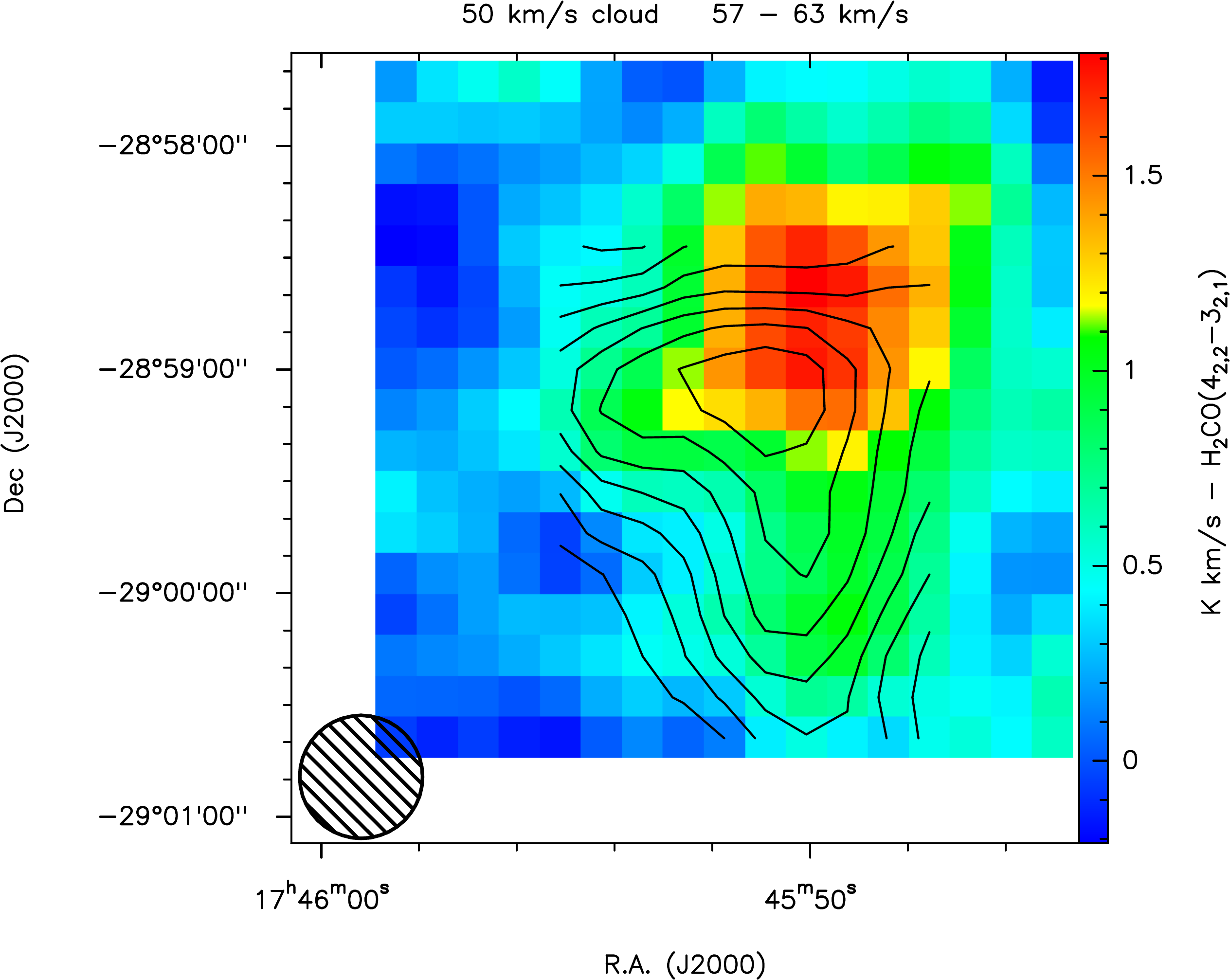}}\\
	\label{50kms-Int-H2CO}
\end{figure*}

\begin{figure*}
	\caption{As Fig. \ref{20kms-Int-H2CO} for G0.253+0.016}
	\centering
	H$_{2}$CO(3$_{0,3}-$2$_{0,2}$)\\
	\subfloat{\includegraphics[bb = 0 0 540 580, clip, height=4.5cm]{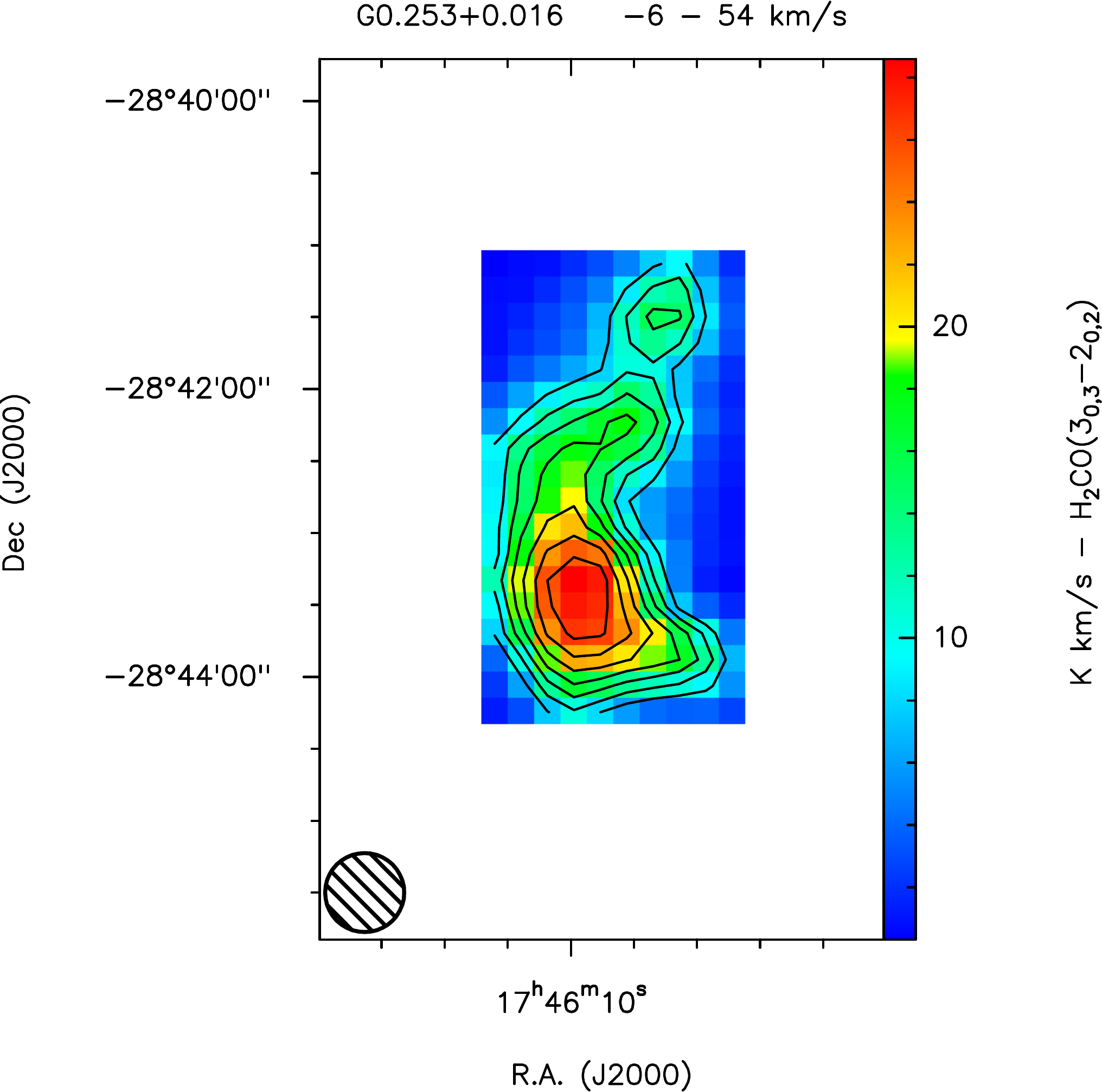}}
	\subfloat{\includegraphics[bb = 150 0 540 580, clip, height=4.5cm]{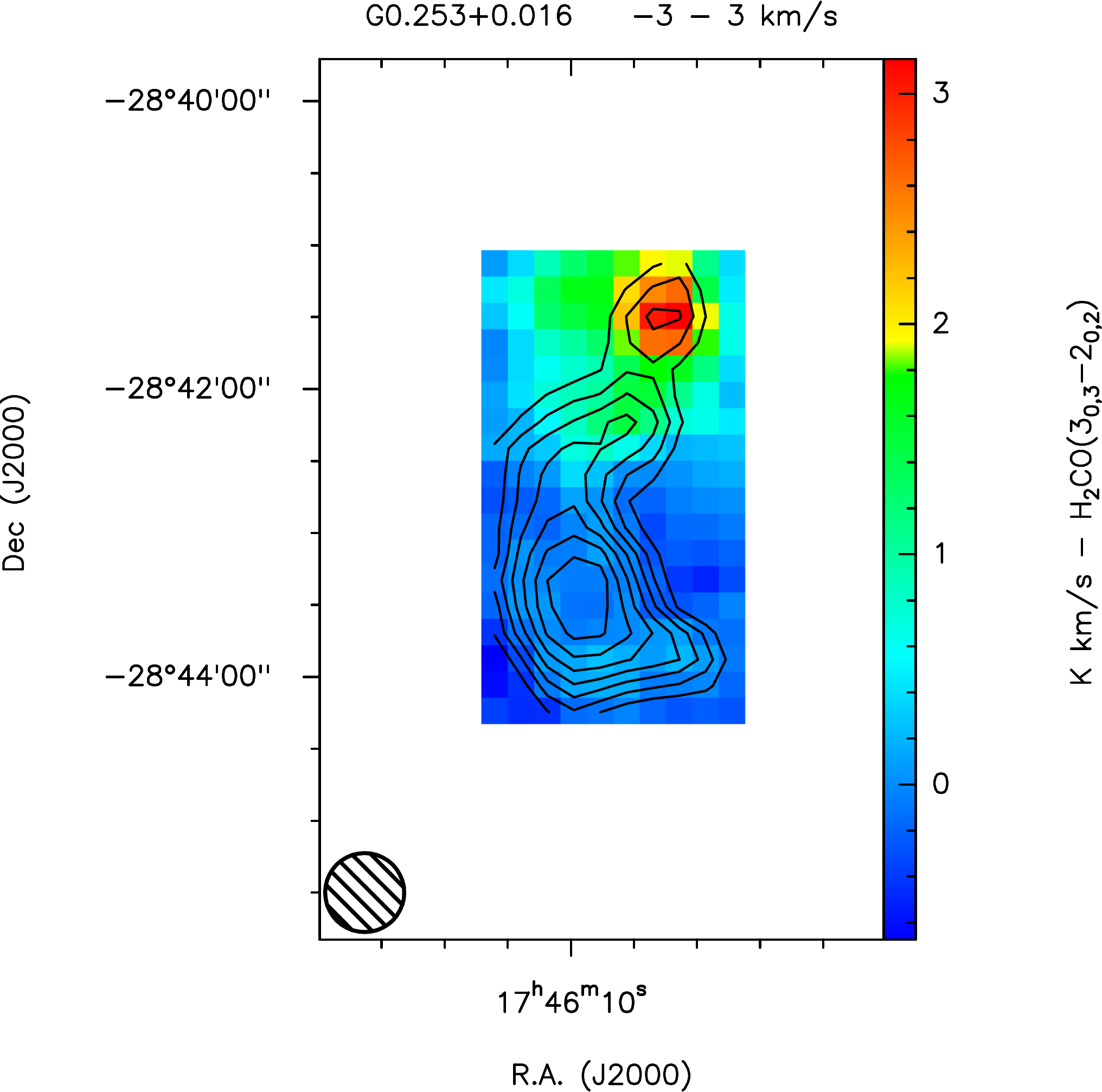}}
	\subfloat{\includegraphics[bb = 150 0 540 580, clip, height=4.5cm]{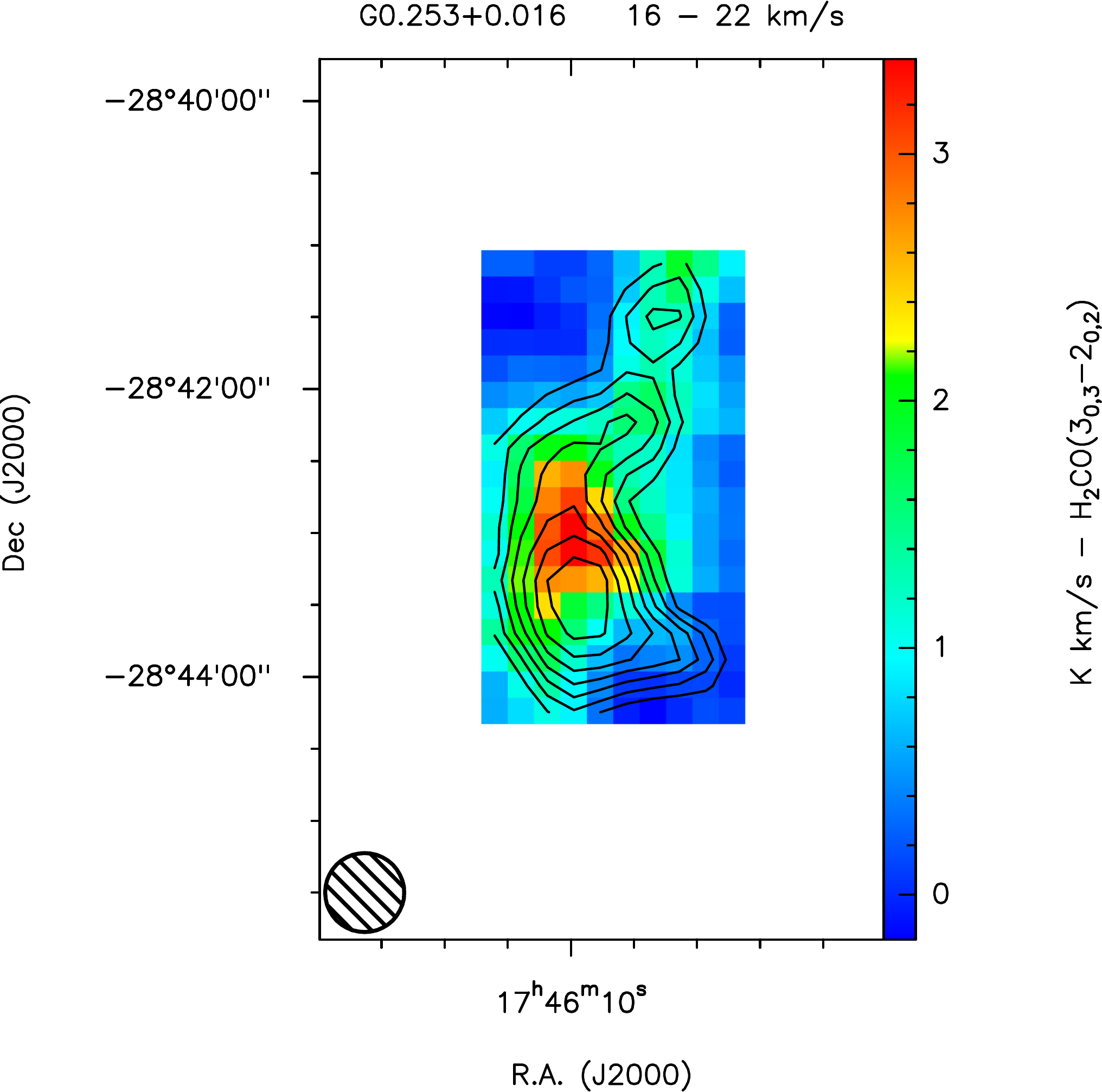}}
	\subfloat{\includegraphics[bb = 150 0 540 580, clip, height=4.5cm]{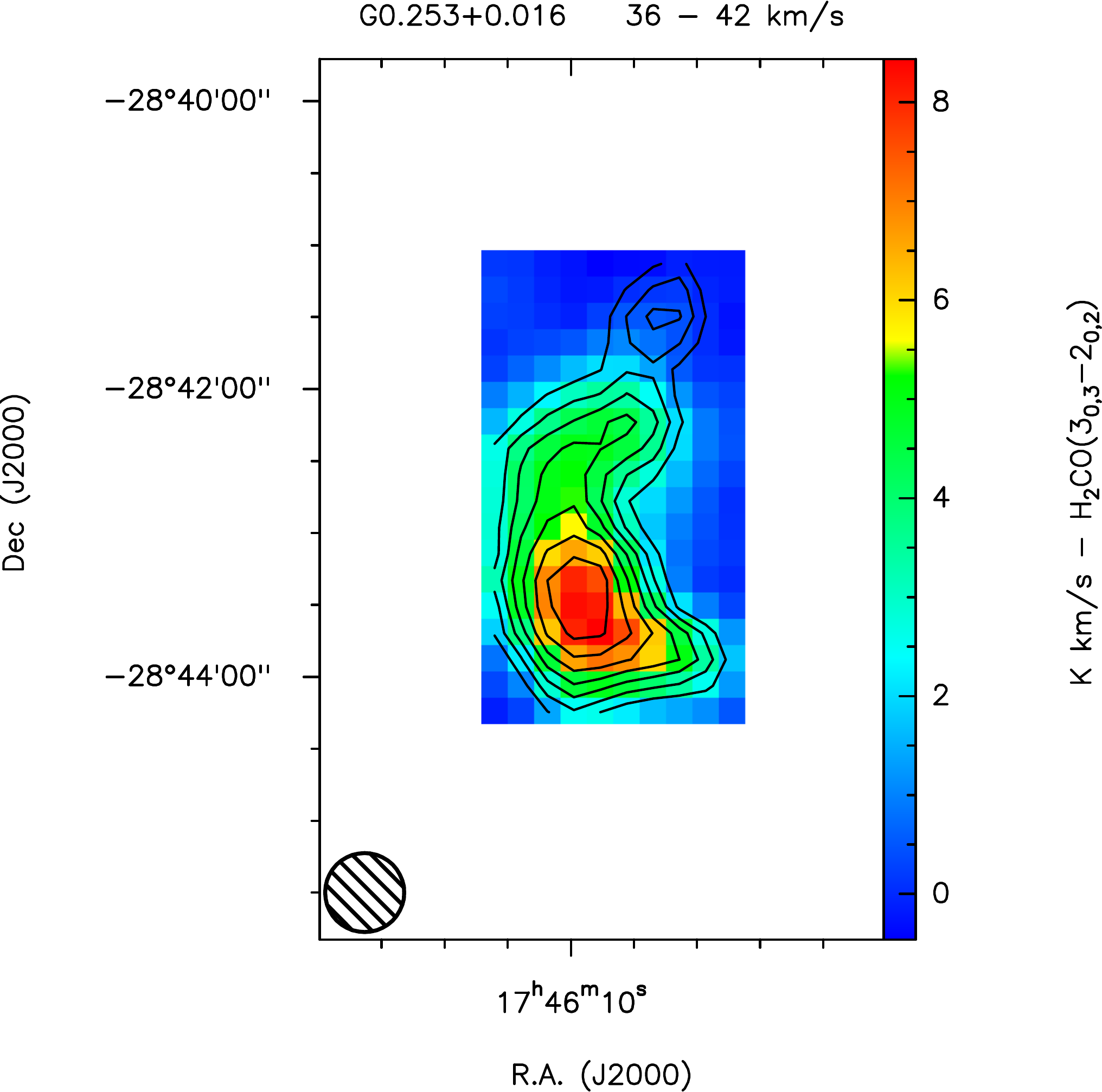}}
	\subfloat{\includegraphics[bb = 150 0 600 580, clip, height=4.5cm]{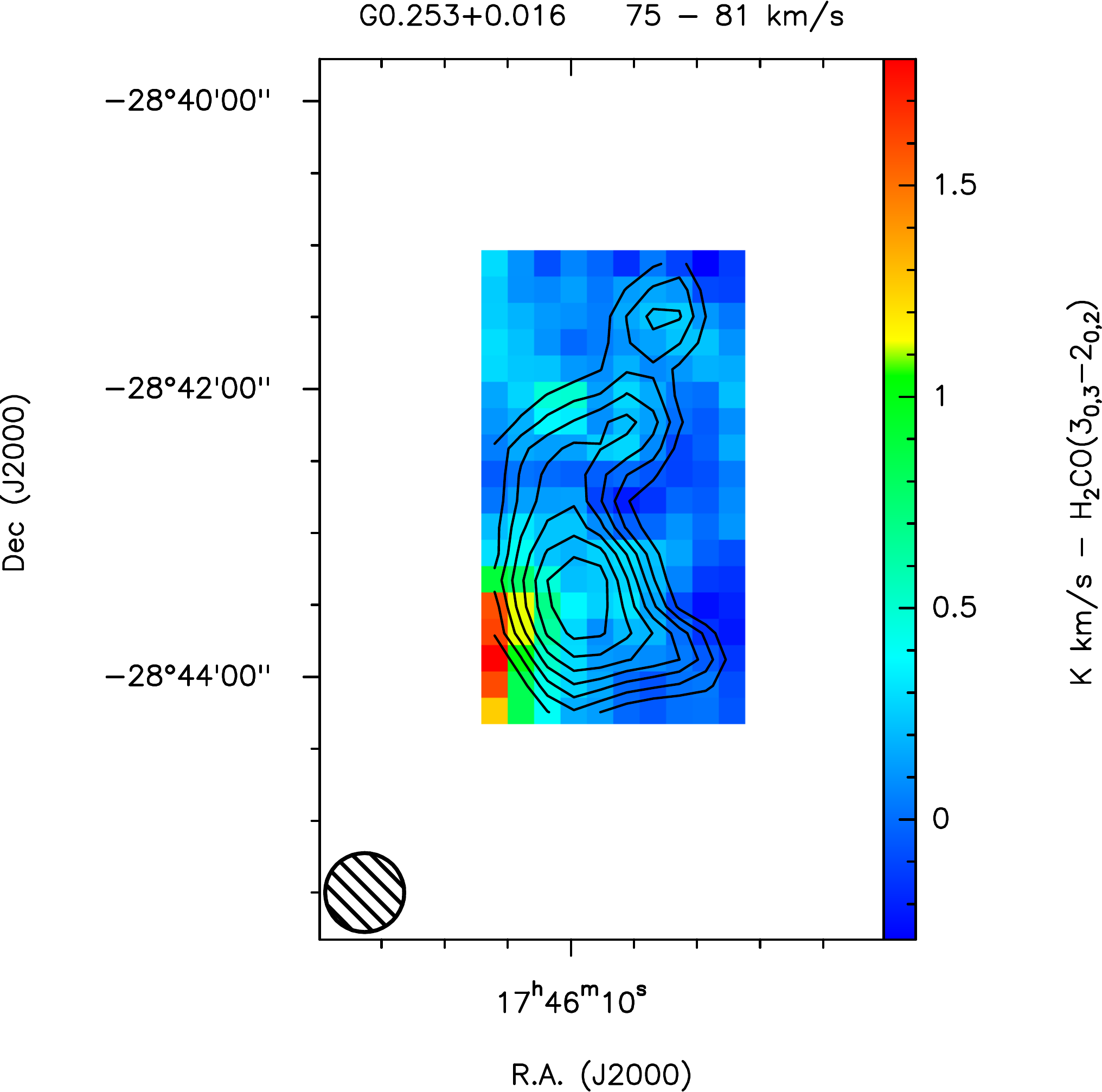}}\\
	H$_{2}$CO(3$_{2,1}-$2$_{2,0}$)\\
	\subfloat{\includegraphics[bb = 0 0 540 580, clip, height=4.5cm]{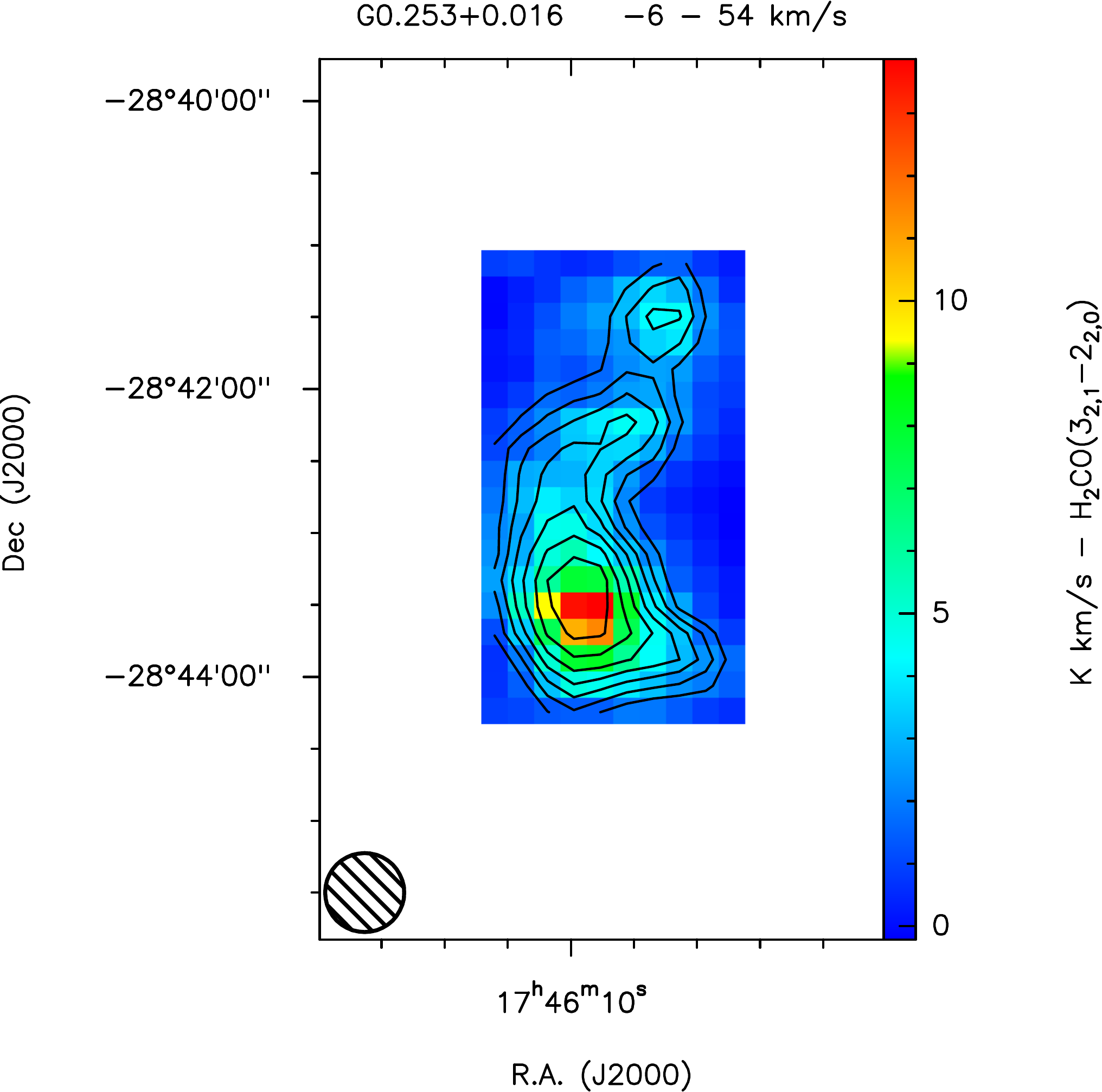}}
	\subfloat{\includegraphics[bb = 150 0 540 580, clip, height=4.5cm]{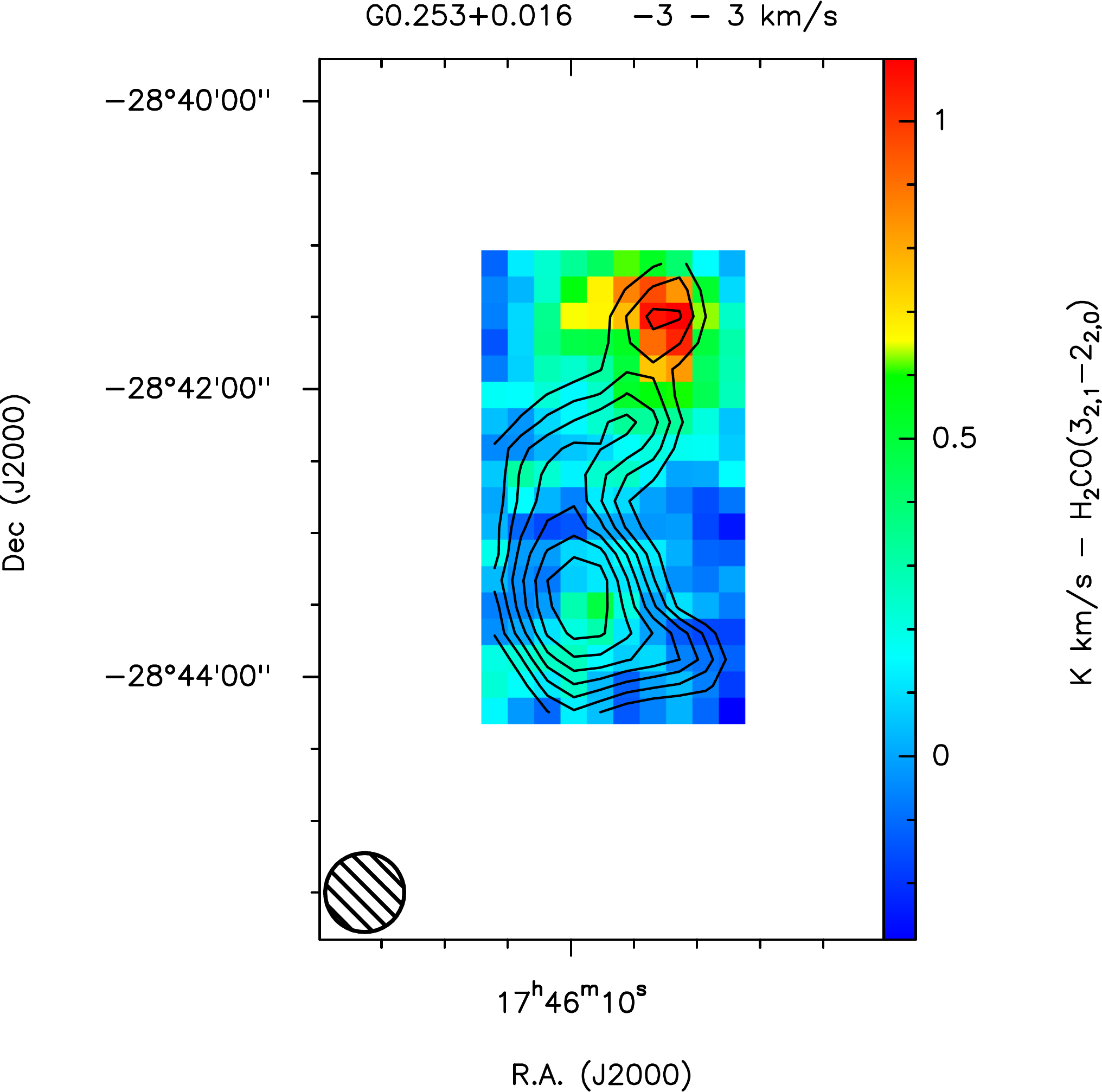}}
	\subfloat{\includegraphics[bb = 150 0 540 580, clip, height=4.5cm]{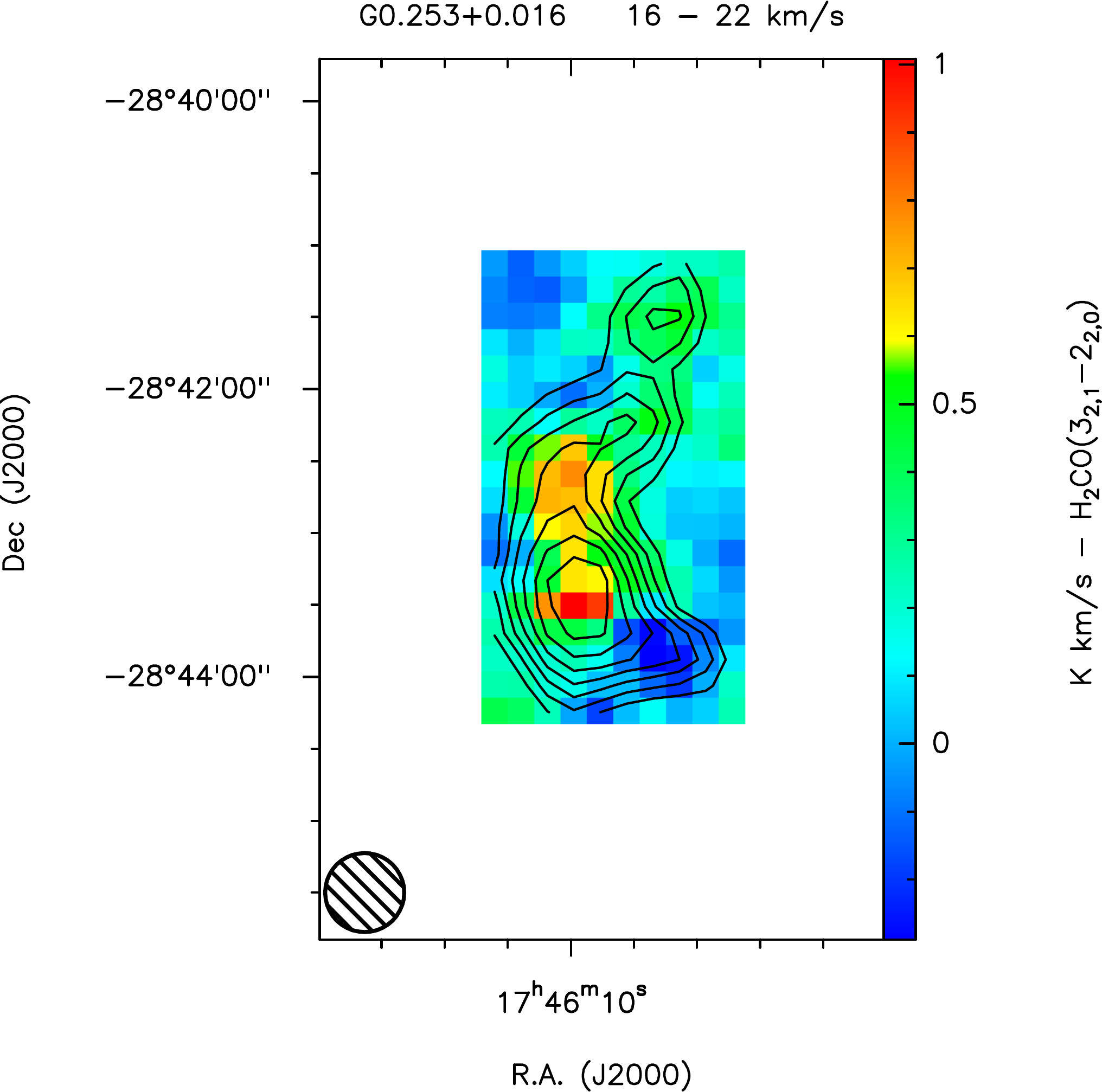}}
	\subfloat{\includegraphics[bb = 150 0 540 580, clip, height=4.5cm]{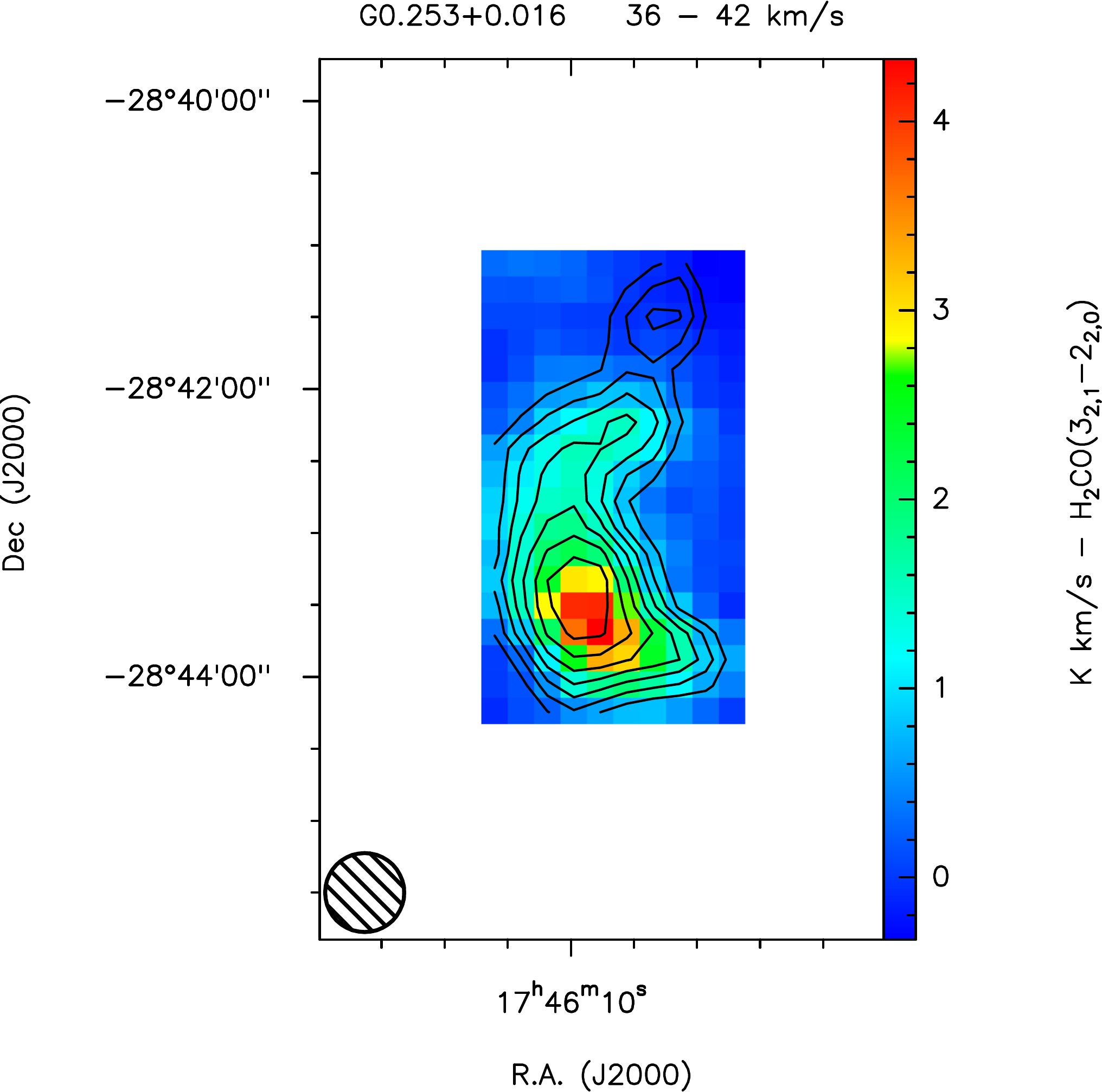}}
	\subfloat{\includegraphics[bb = 150 0 600 580, clip, height=4.5cm]{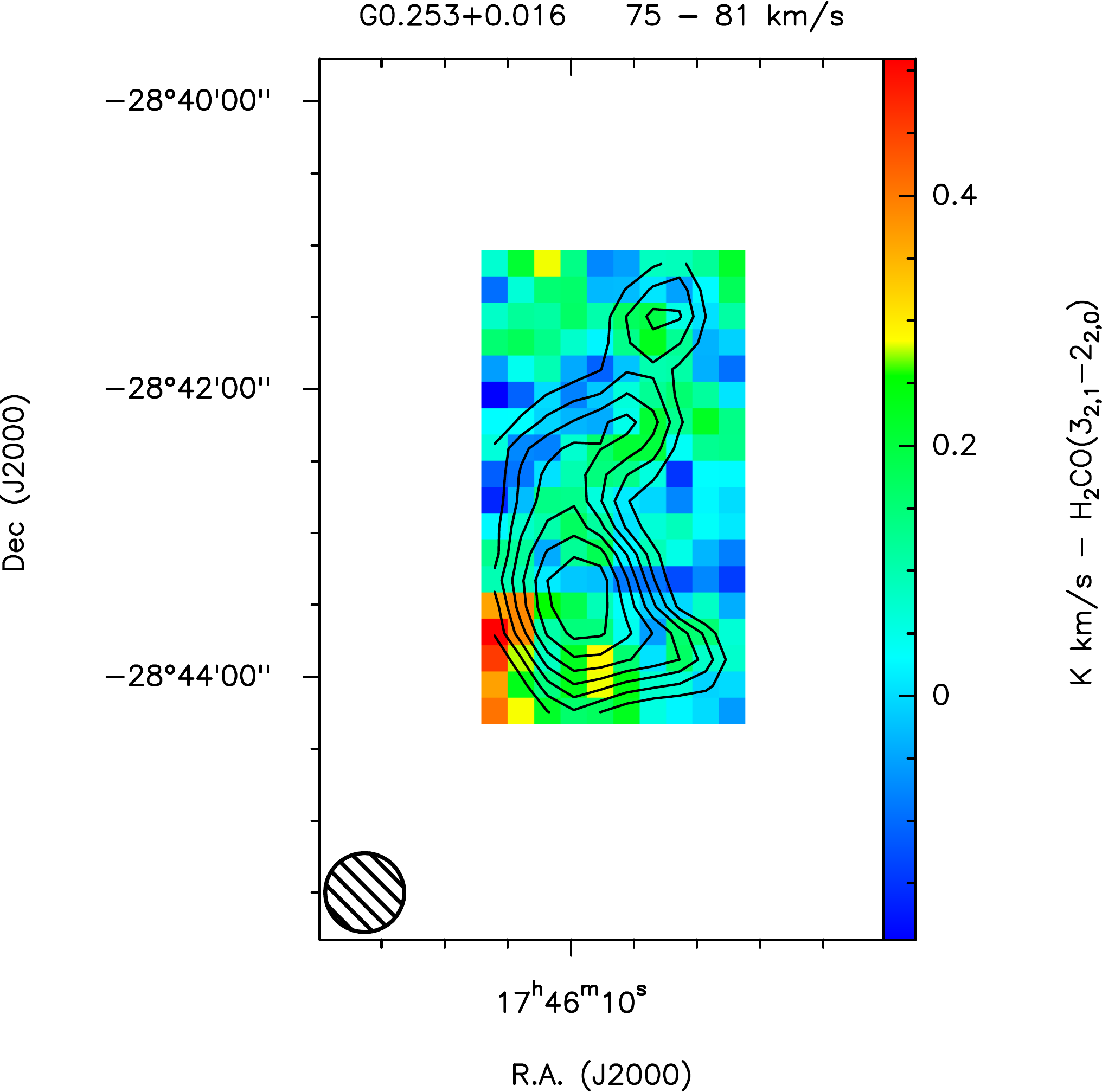}}\\
	H$_{2}$CO(4$_{0,3}-$3$_{0,3}$)\\
	\subfloat{\includegraphics[bb = 0 0 540 580, clip, height=4.5cm]{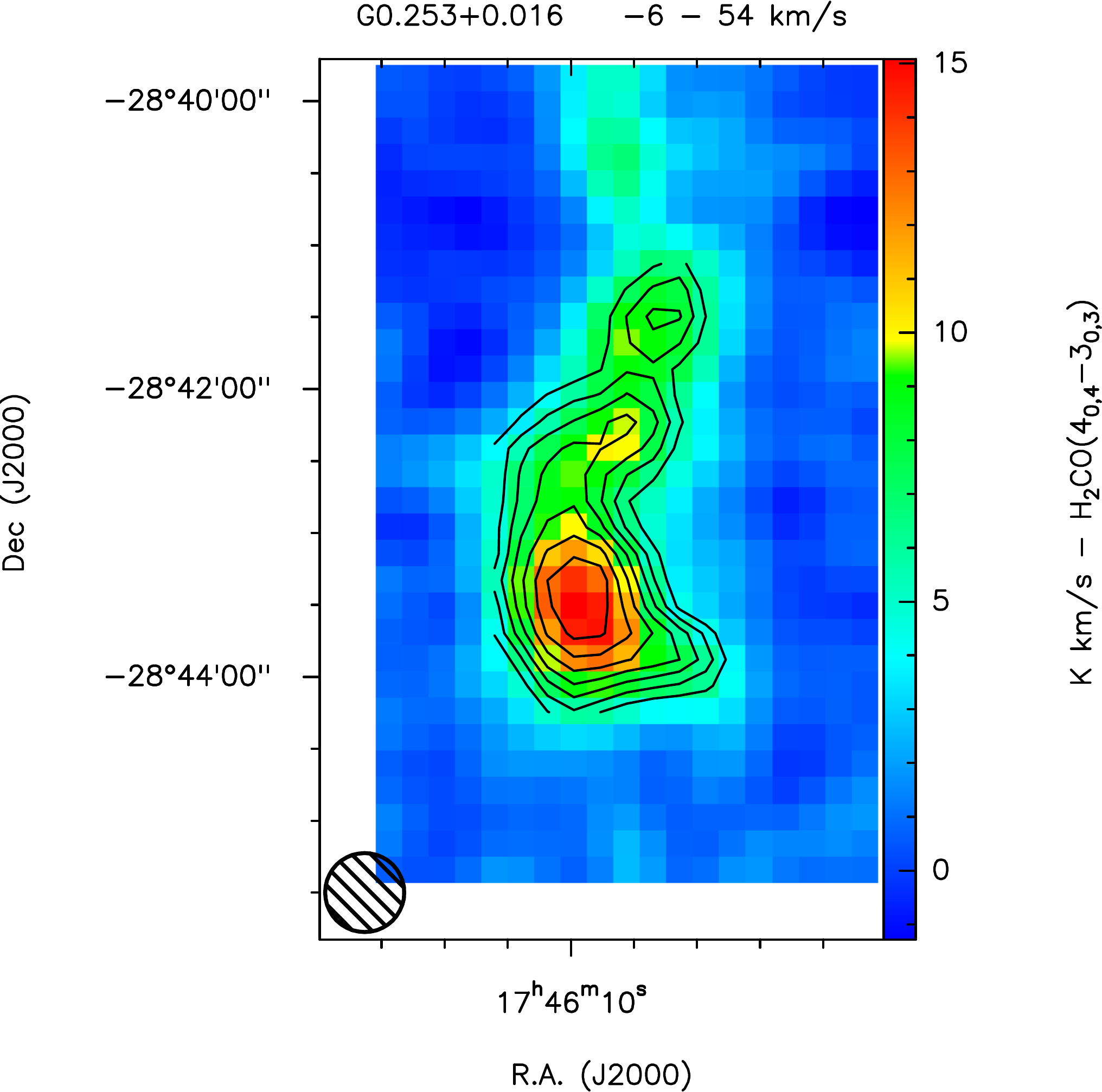}}
	\subfloat{\includegraphics[bb = 150 0 540 580, clip, height=4.5cm]{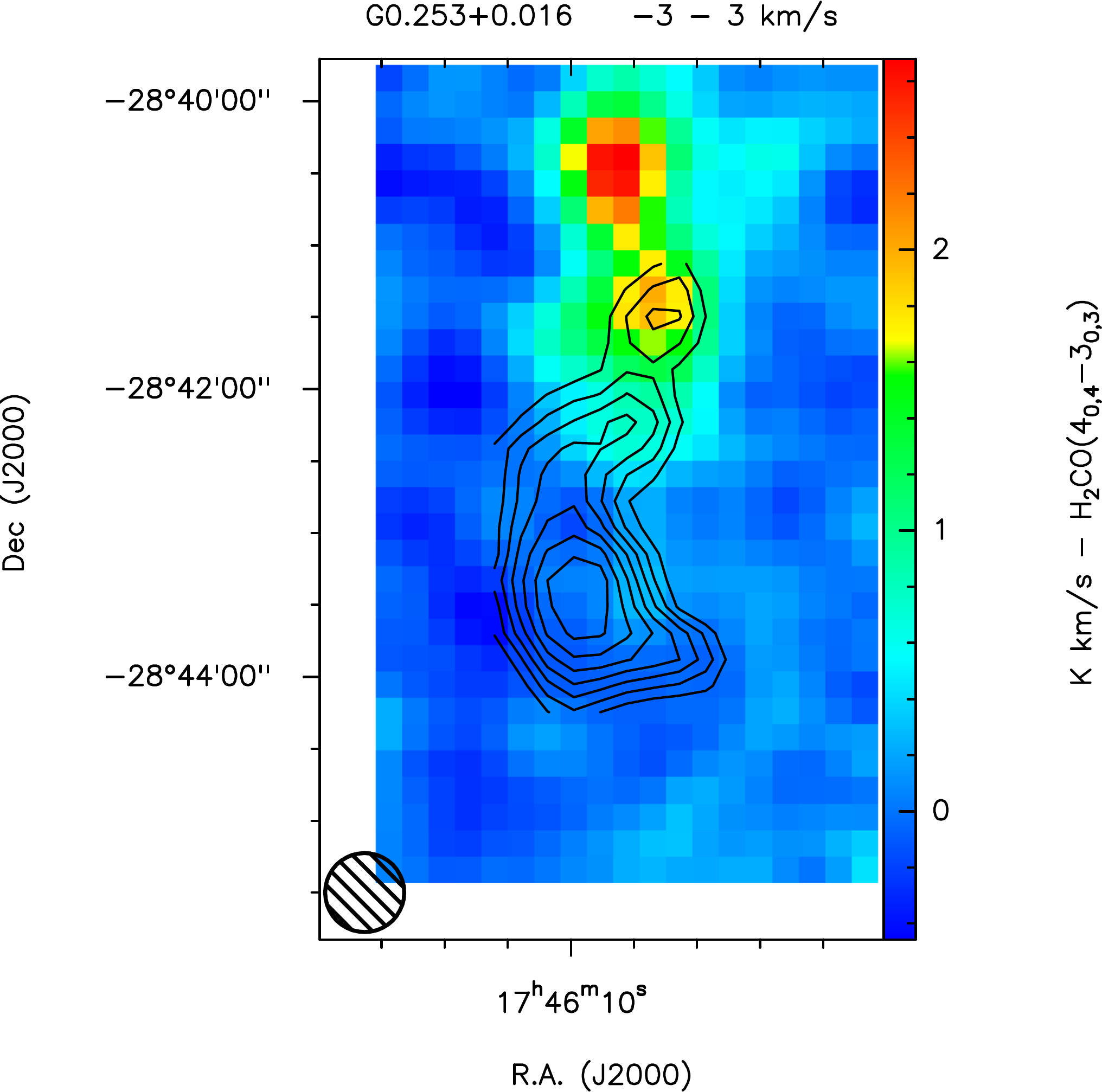}}
	\subfloat{\includegraphics[bb = 150 0 540 580, clip, height=4.5cm]{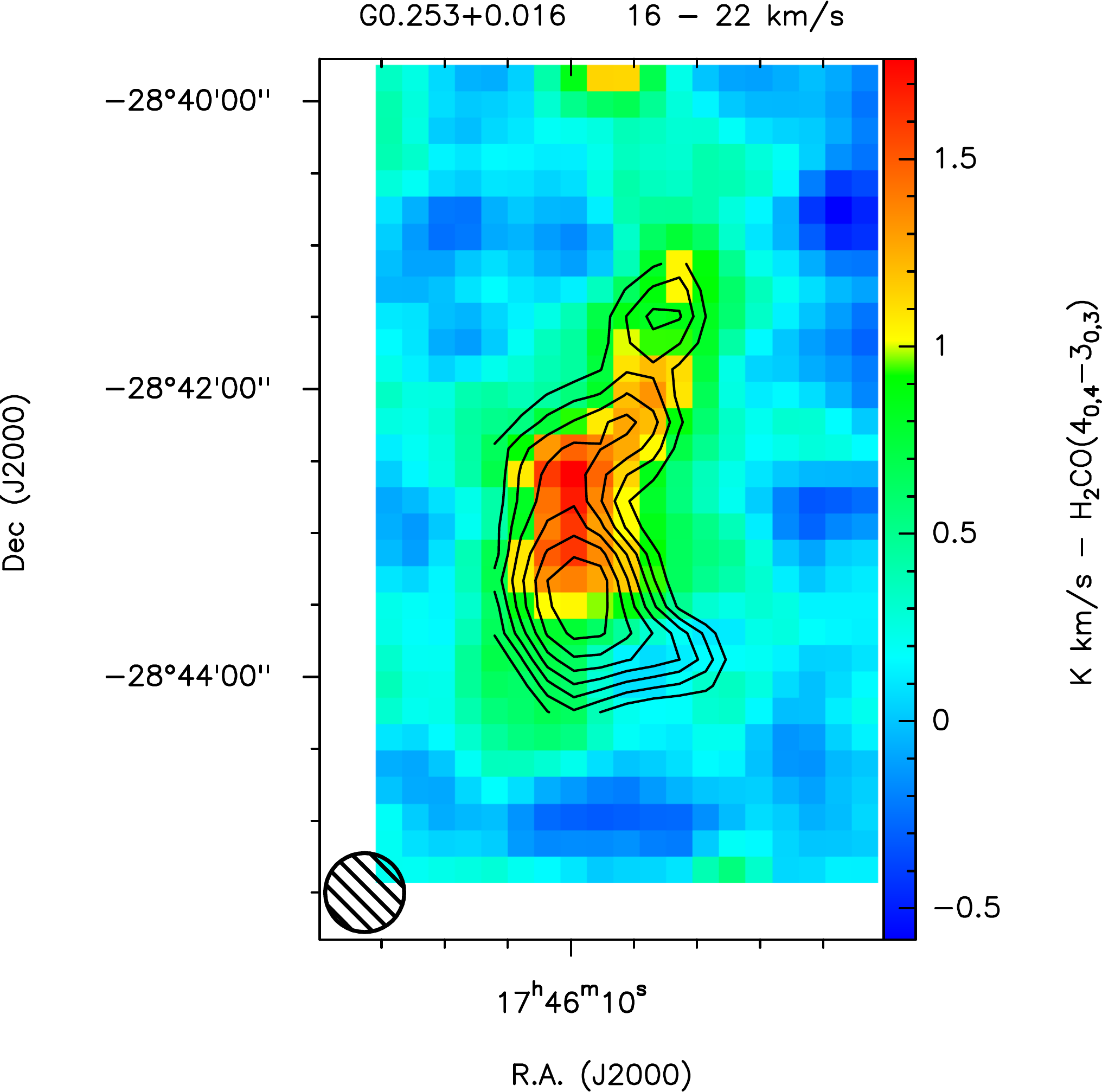}}
	\subfloat{\includegraphics[bb = 150 0 540 580, clip, height=4.5cm]{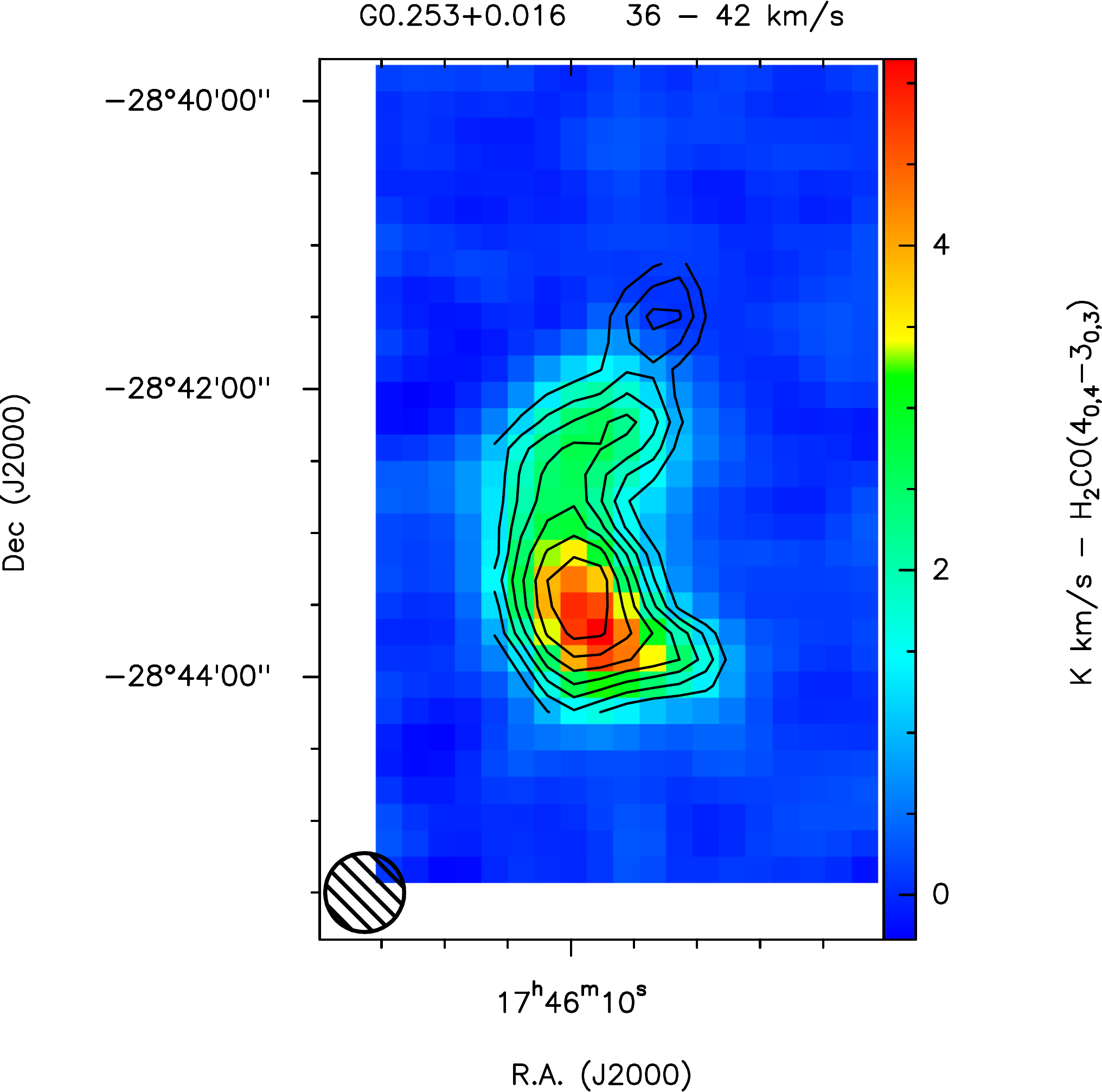}}
	\subfloat{\includegraphics[bb = 150 0 600 580, clip, height=4.5cm]{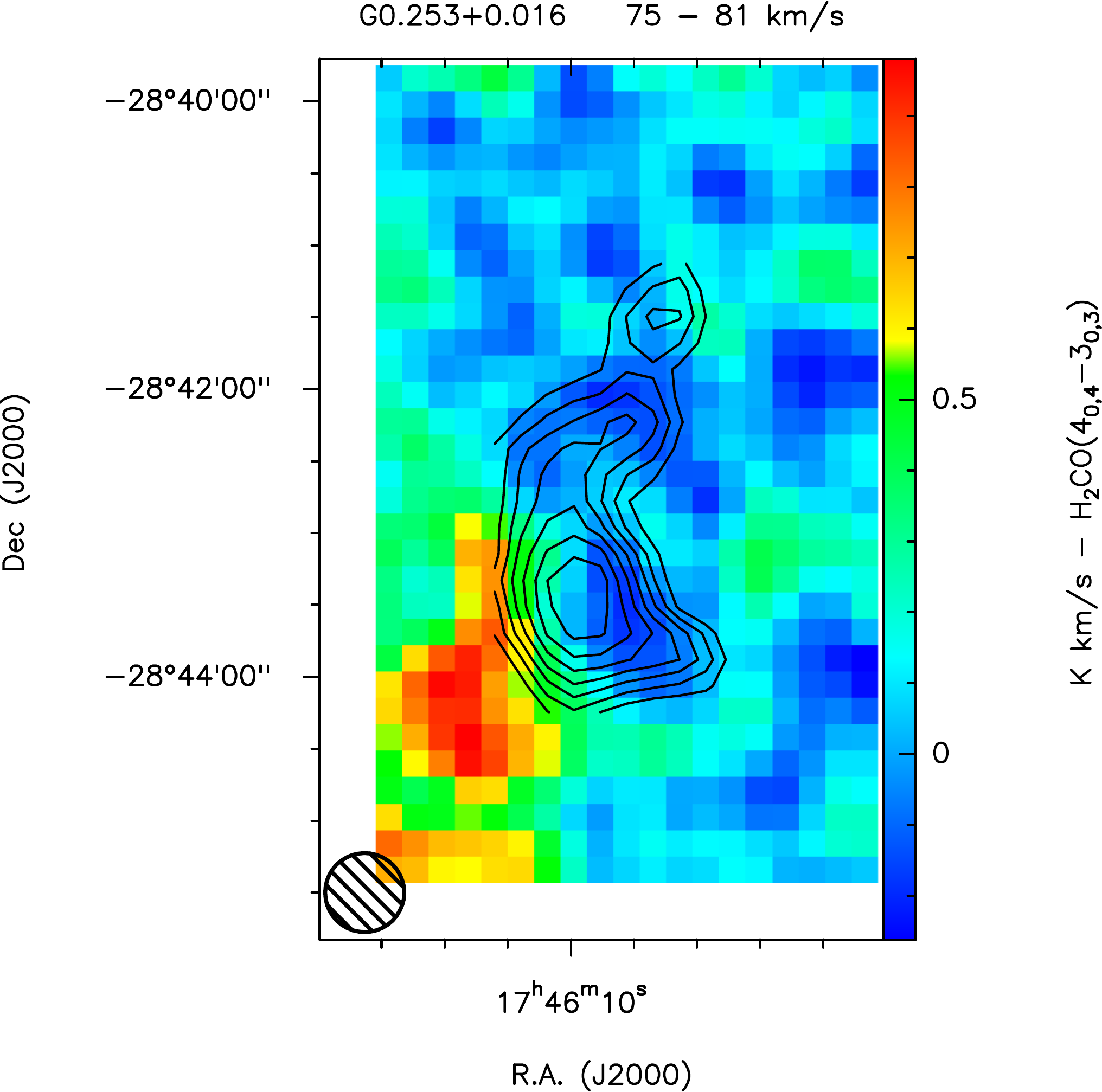}}\\
	H$_{2}$CO(4$_{2,2}-$3$_{2,1}$)\\
	\subfloat{\includegraphics[bb = 0 0 540 580, clip, height=4.5cm]{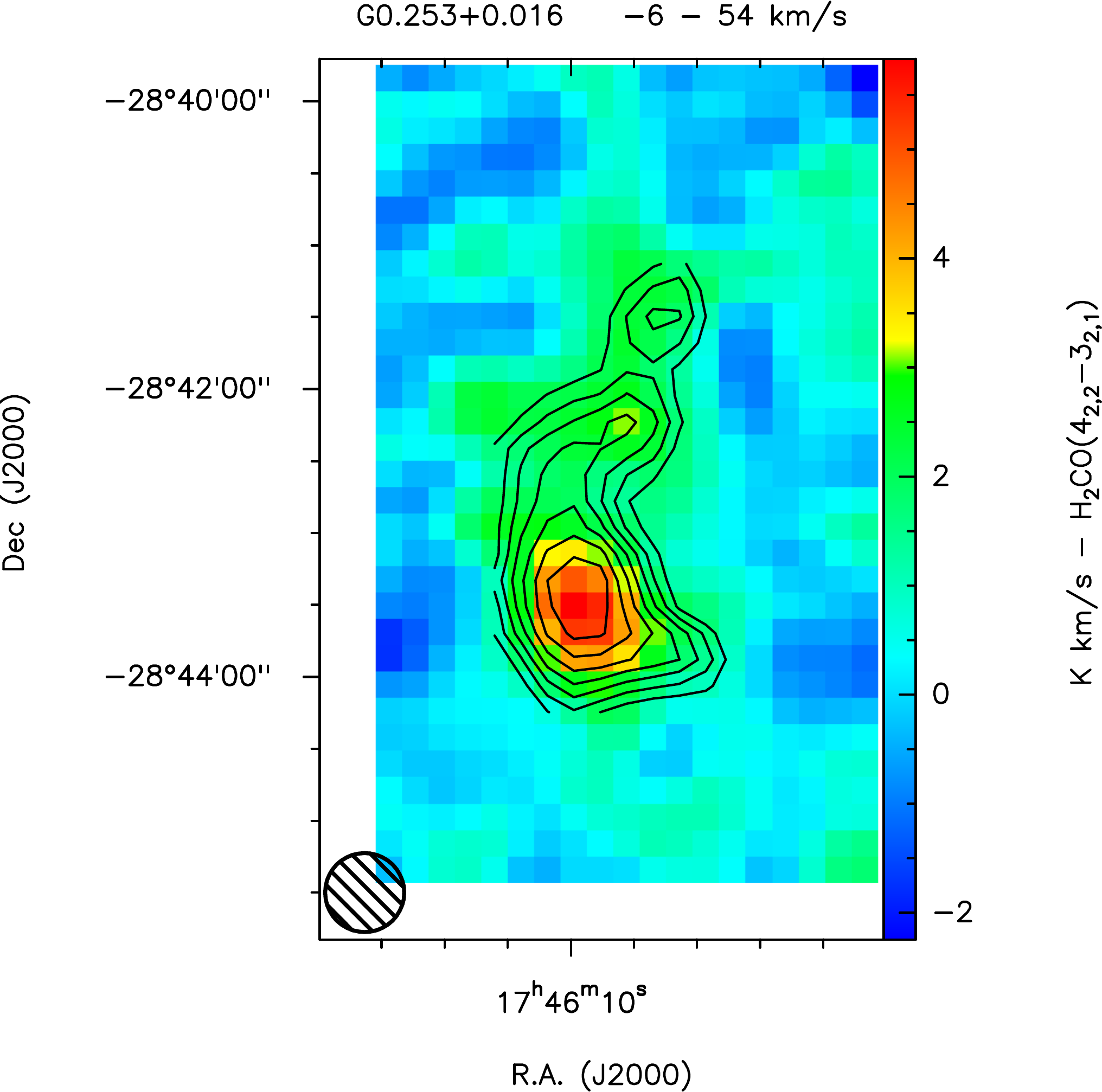}}
	\subfloat{\includegraphics[bb = 150 0 540 580, clip, height=4.5cm]{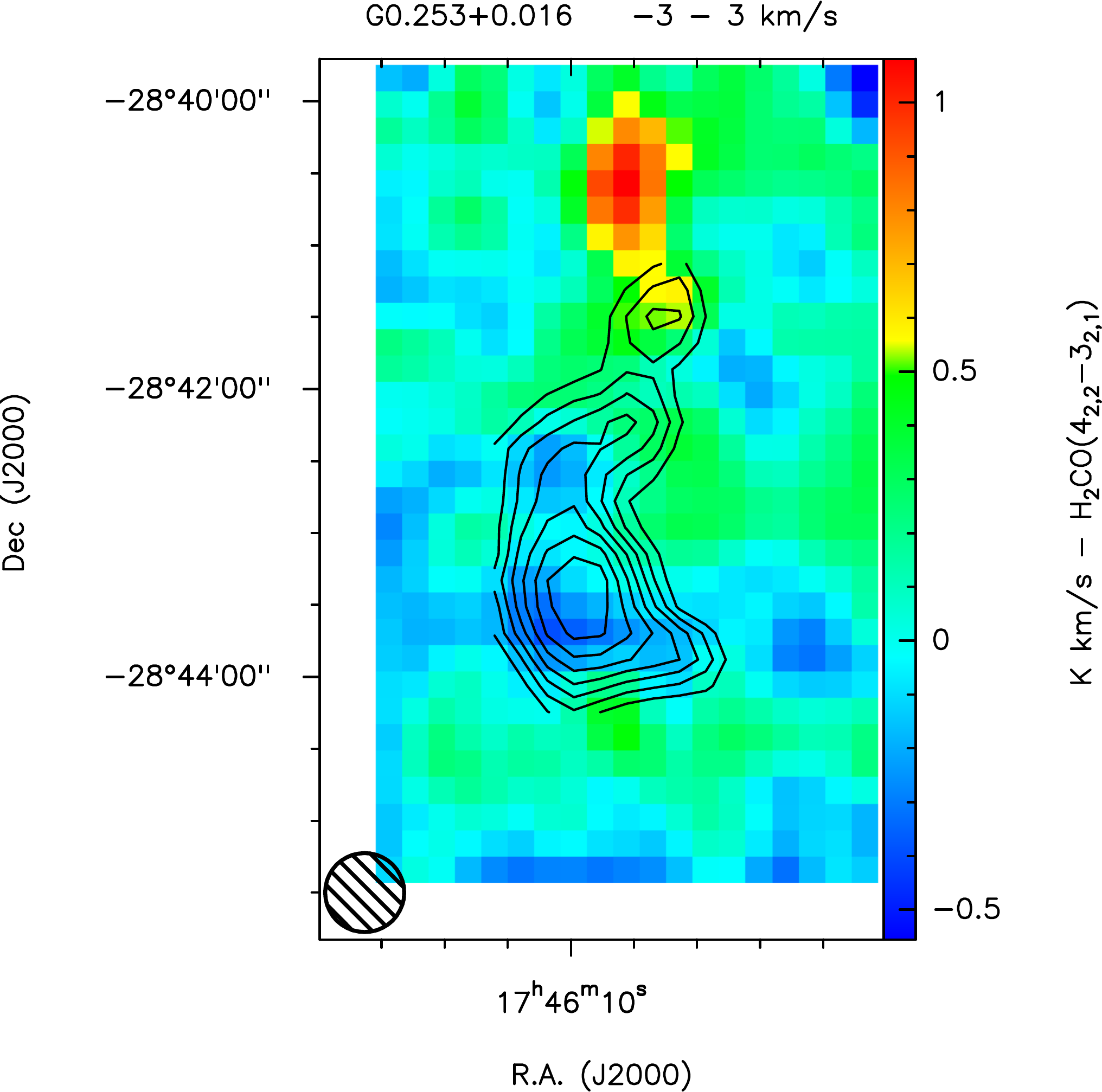}}
	\subfloat{\includegraphics[bb = 150 0 540 580, clip, height=4.5cm]{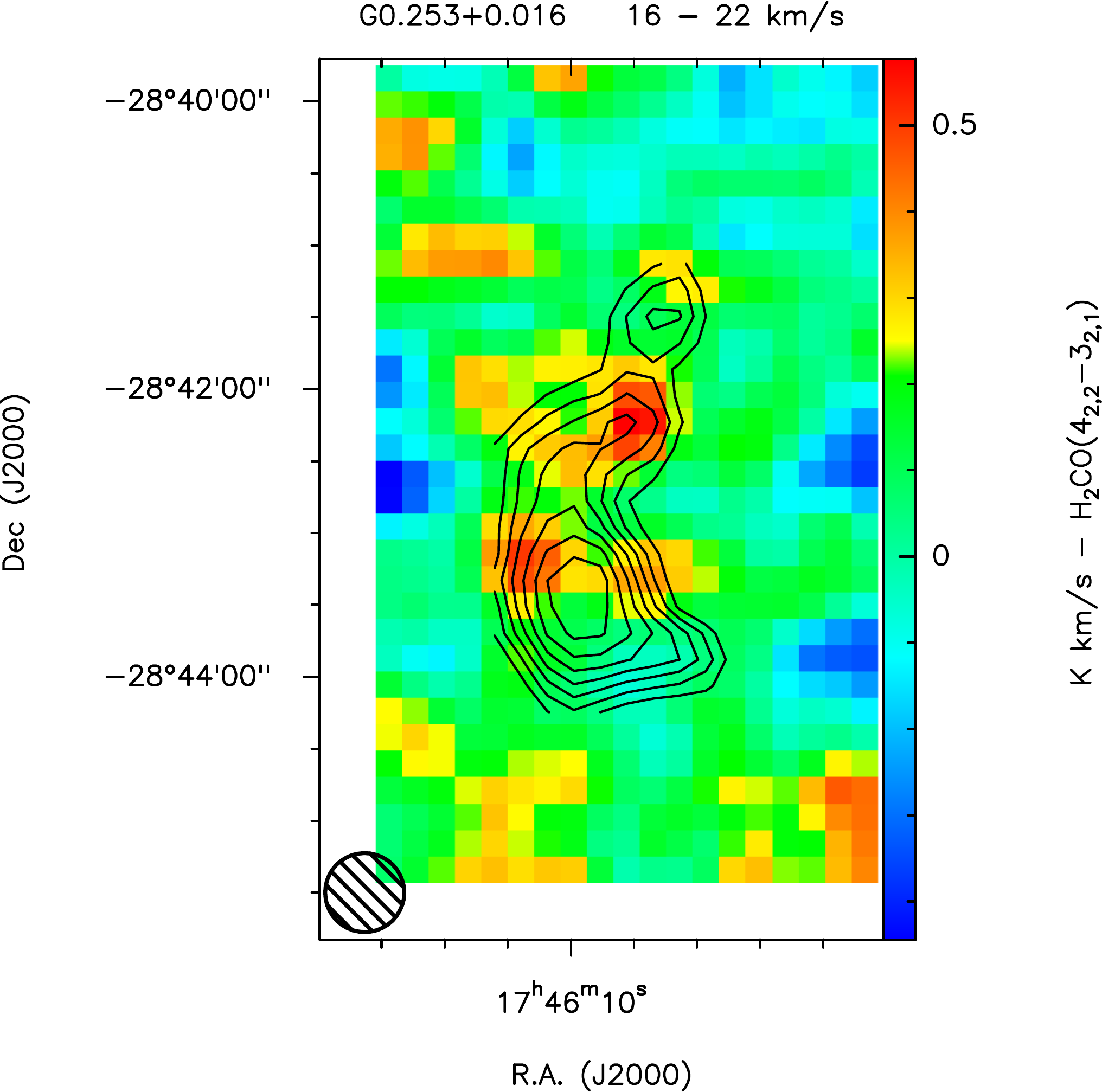}}
	\subfloat{\includegraphics[bb = 150 0 540 580, clip, height=4.5cm]{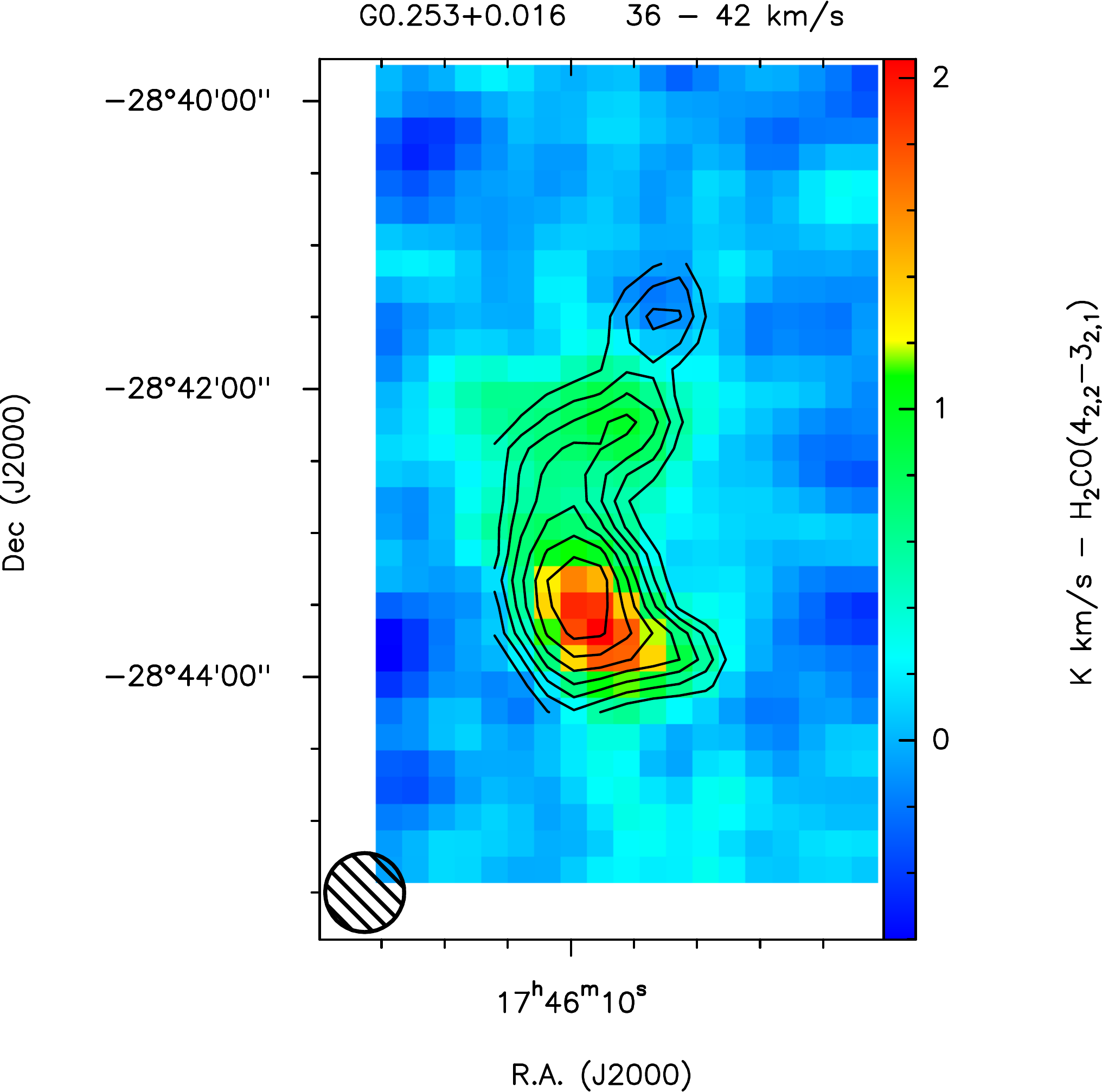}}
	\subfloat{\includegraphics[bb = 150 0 600 580, clip, height=4.5cm]{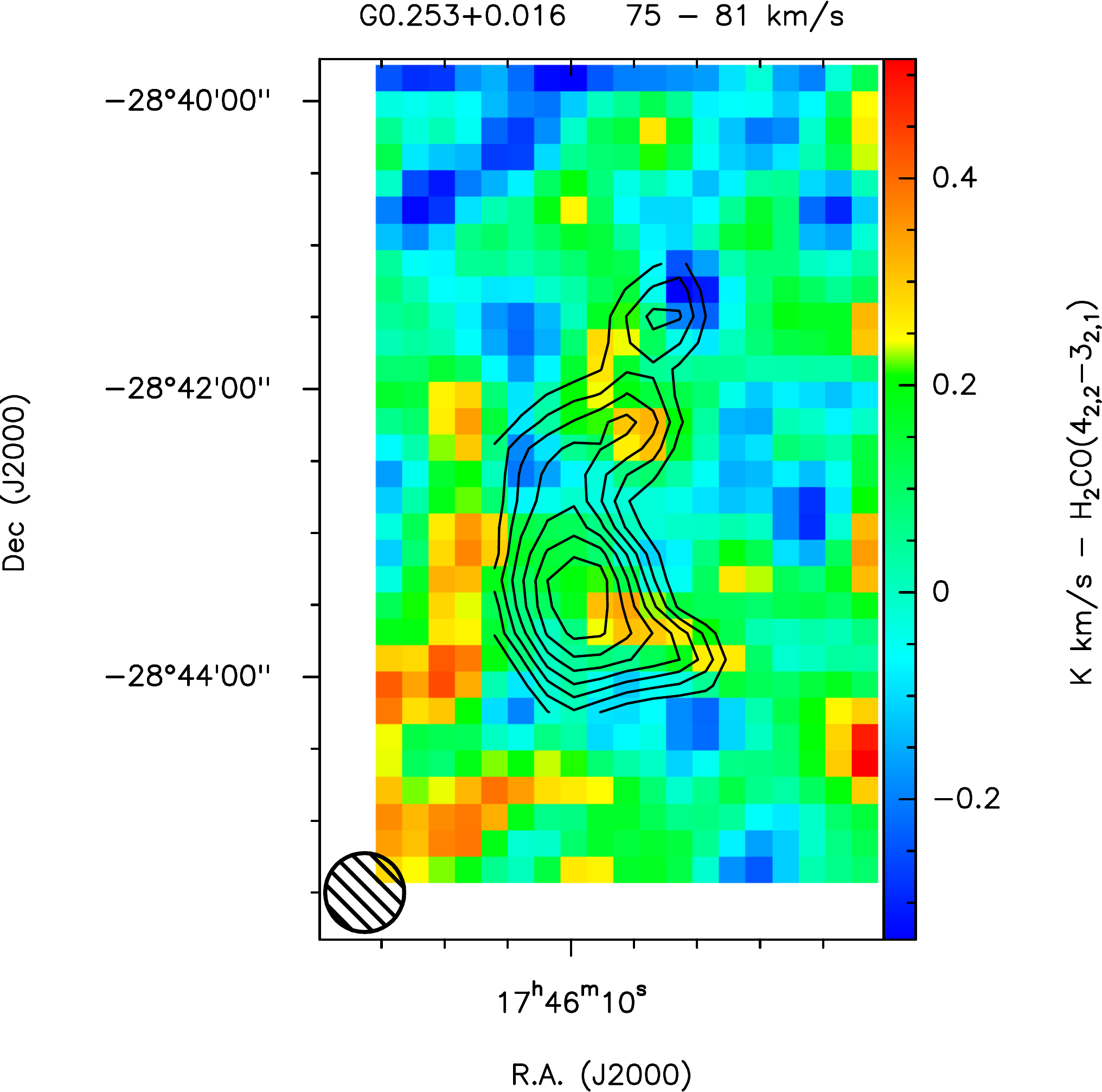}}\\
	\label{G0253-Int-H2CO}
\end{figure*}

\begin{figure*}
	\caption{As Fig. \ref{20kms-Int-H2CO} for G0.411+0.050.}
	\centering
	H$_{2}$CO(3$_{0,3}-$2$_{0,2}$)\\
	\subfloat{\includegraphics[bb = 0 0 610 580, clip, height=5.5cm]{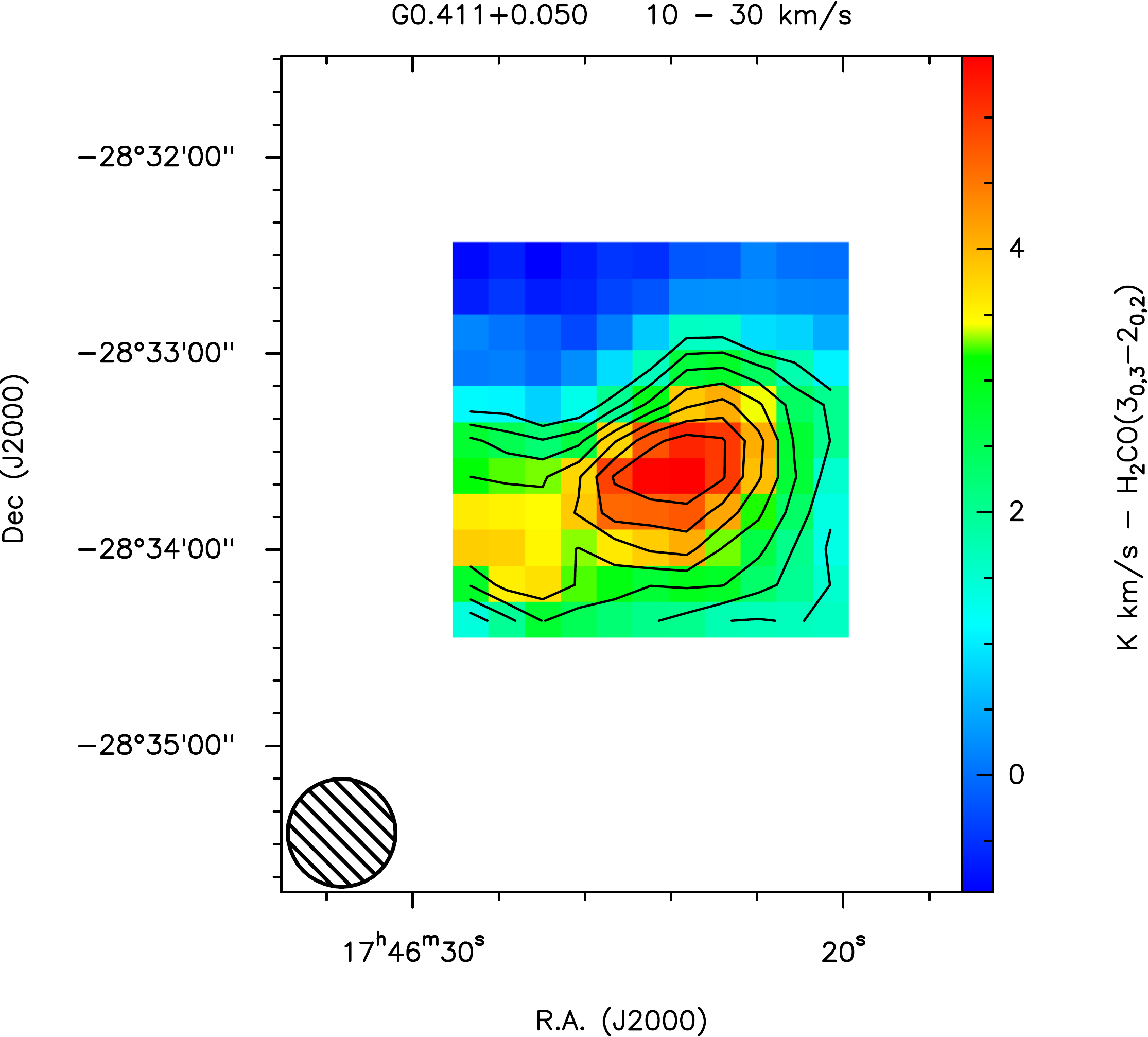}}
	\subfloat{\includegraphics[bb = 150 0 640 580, clip, height=5.5cm]{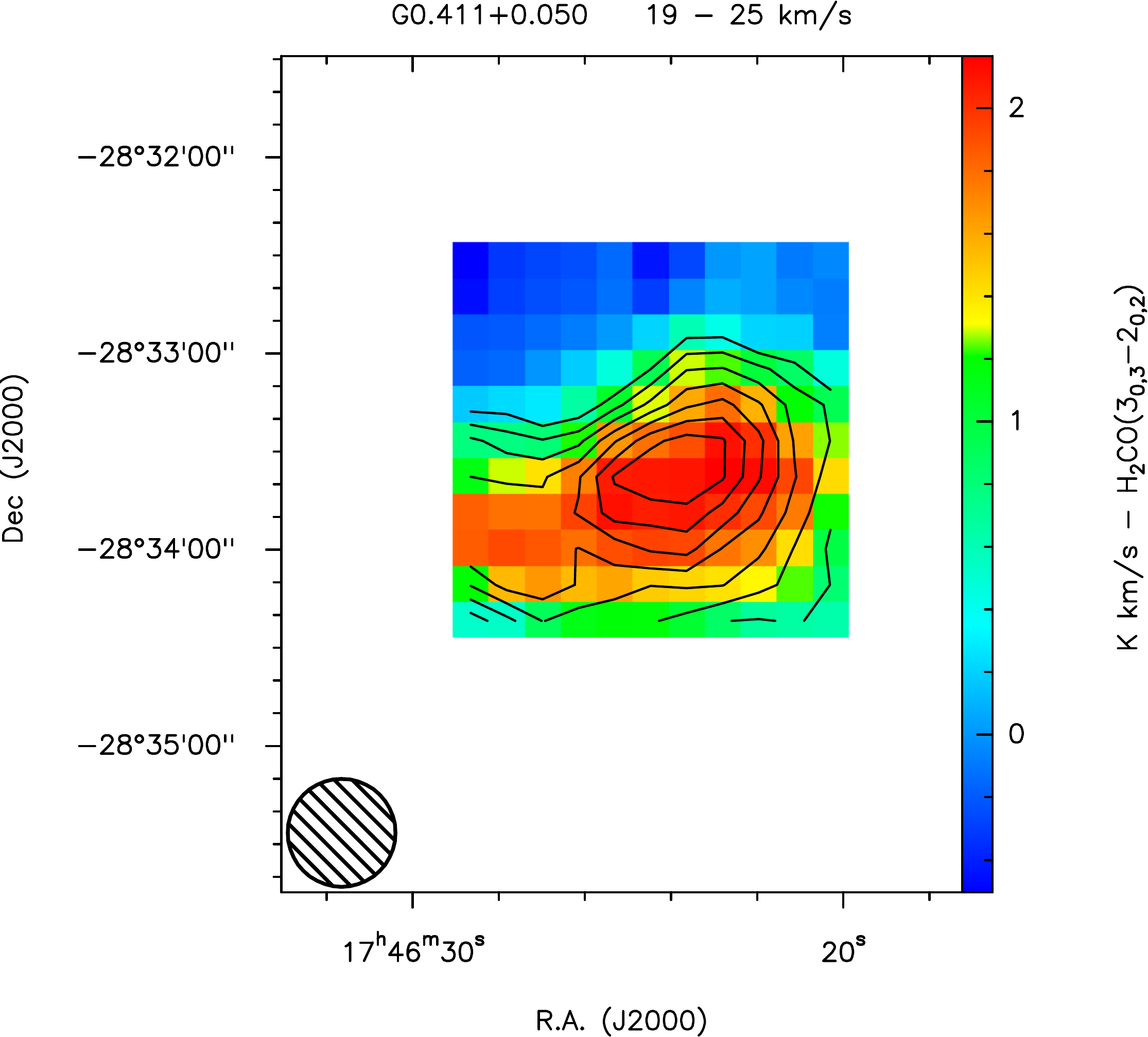}}\\
	H$_{2}$CO(3$_{2,1}-$2$_{2,0}$)\\
	\subfloat{\includegraphics[bb = 0 0 610 580, clip, height=5.5cm]{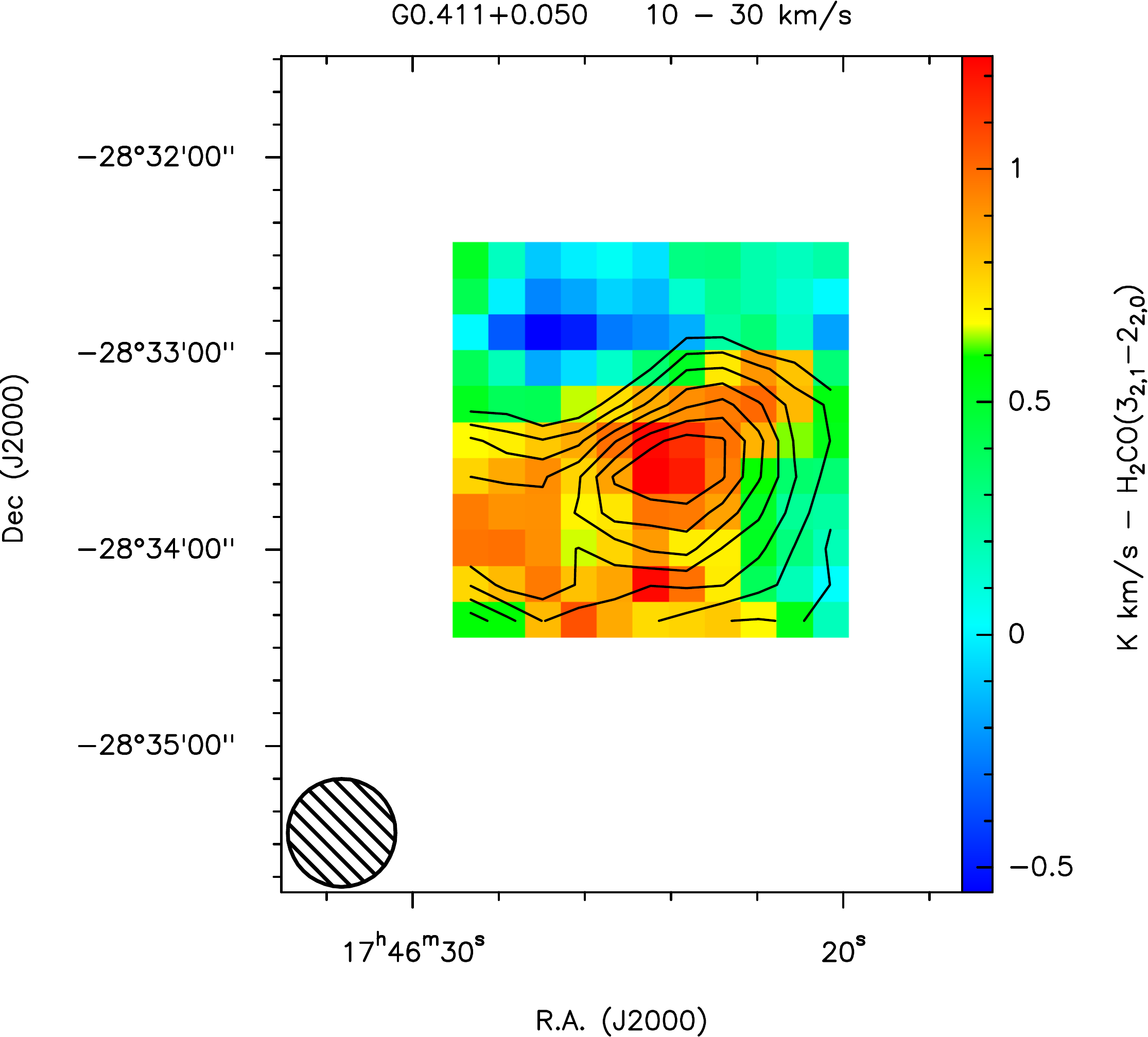}}
	\subfloat{\includegraphics[bb = 150 0 640 580, clip, height=5.5cm]{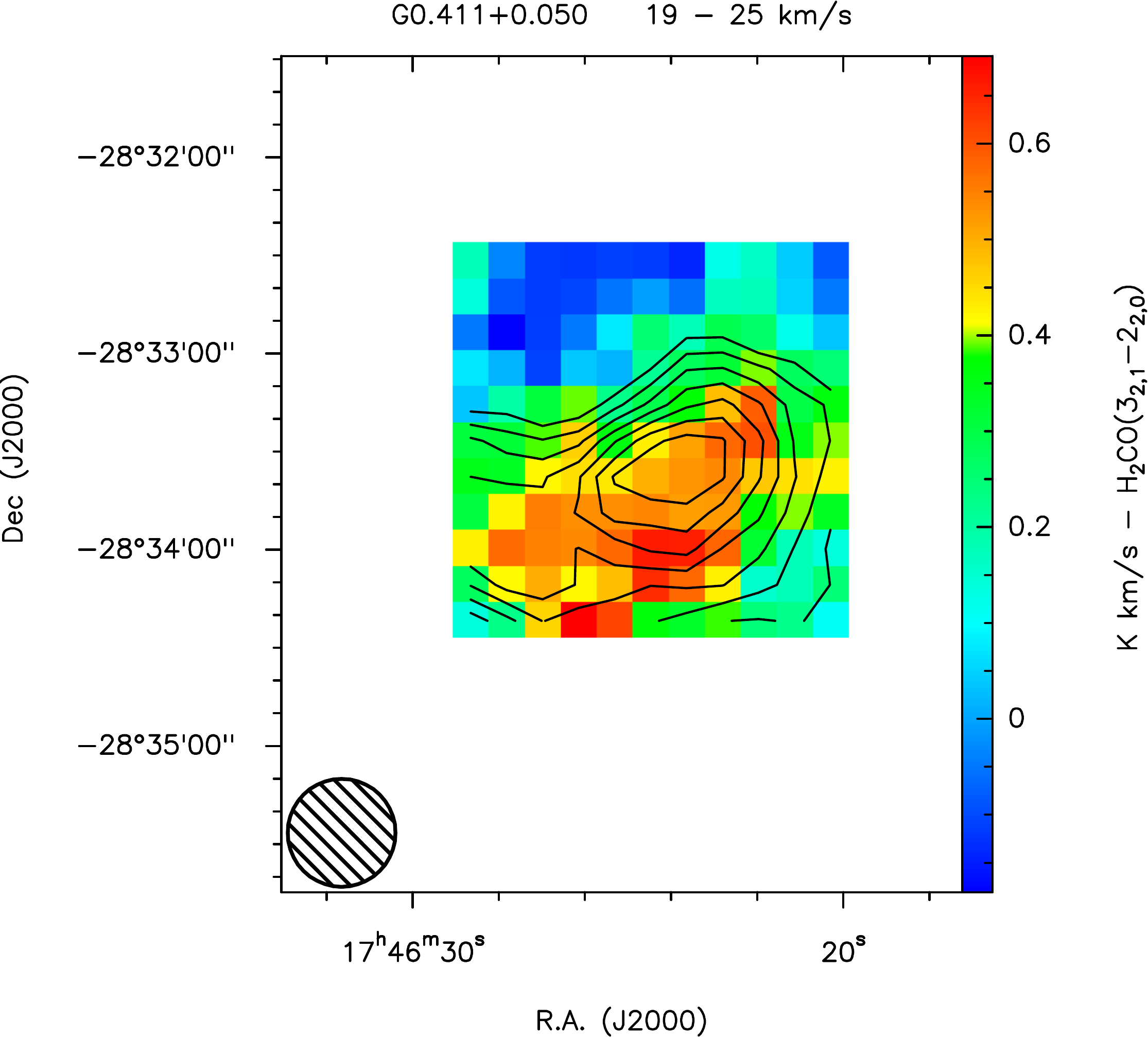}}\\
	H$_{2}$CO(4$_{0,3}-$3$_{0,3}$)\\
	\subfloat{\includegraphics[bb = 0 0 610 580, clip, height=5.5cm]{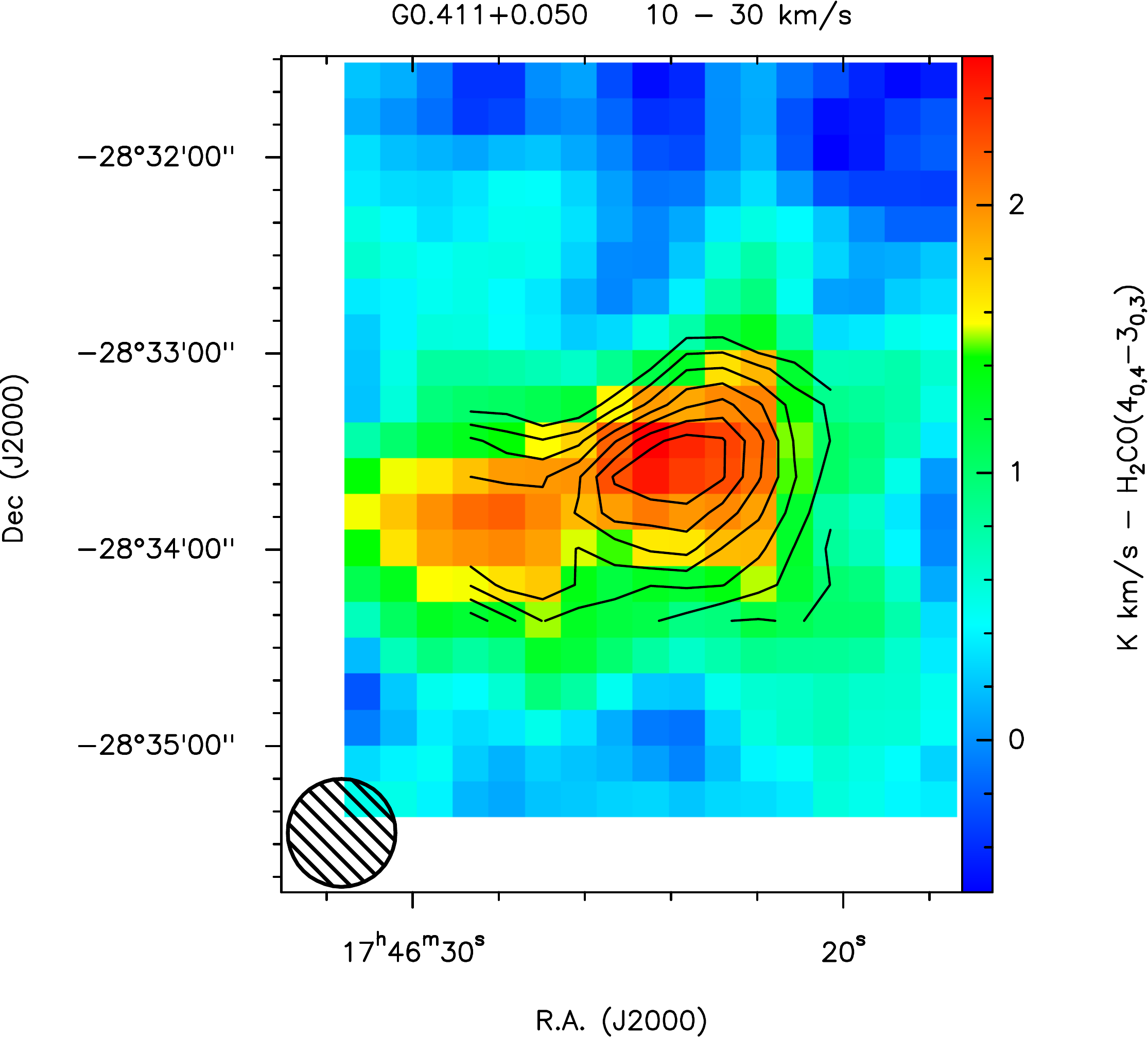}}
	\subfloat{\includegraphics[bb = 150 0 640 580, clip, height=5.5cm]{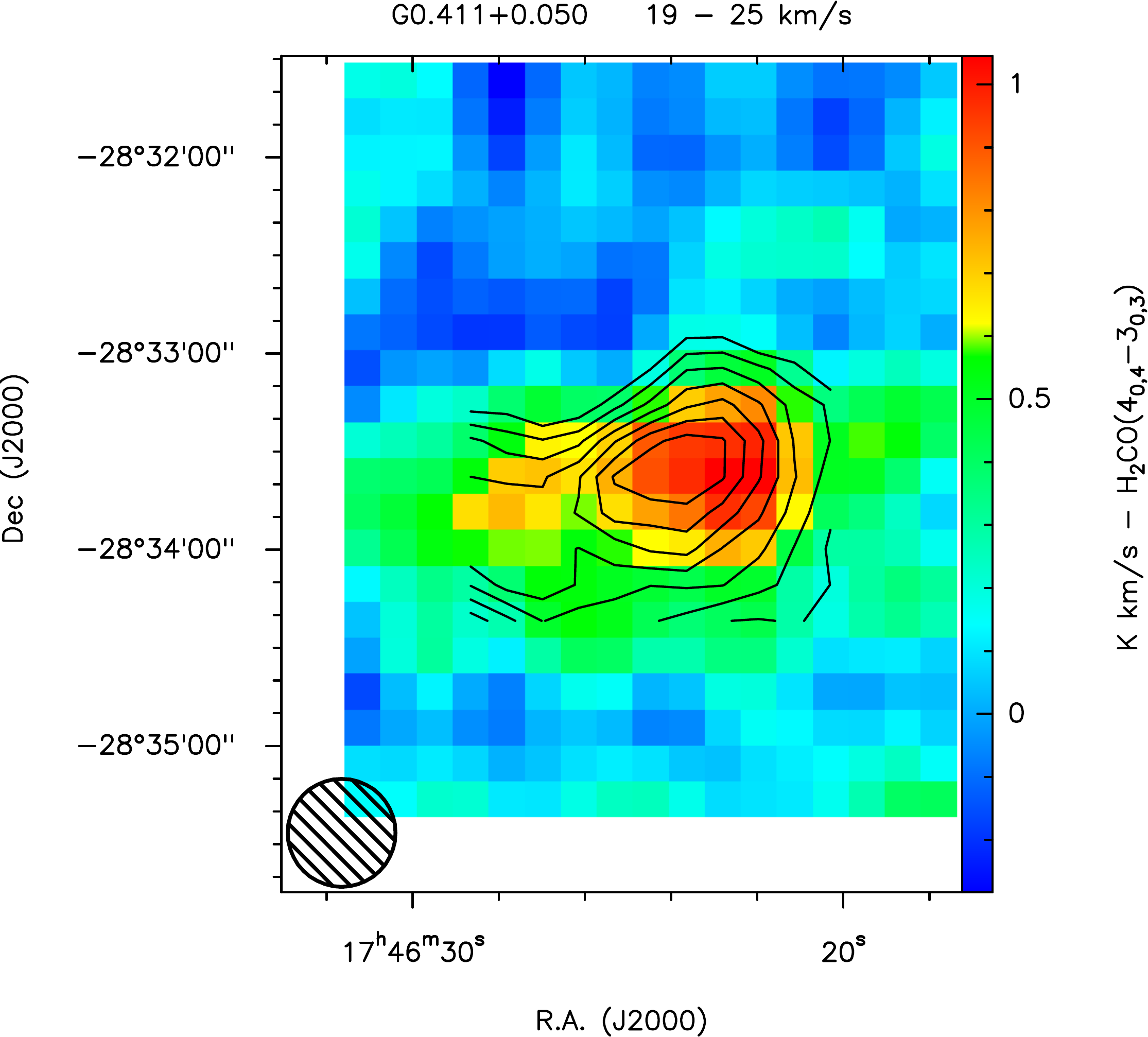}}\\
	H$_{2}$CO(4$_{2,2}-$3$_{2,1}$)\\
	\subfloat{\includegraphics[bb = 0 0 610 580, clip, height=5.5cm]{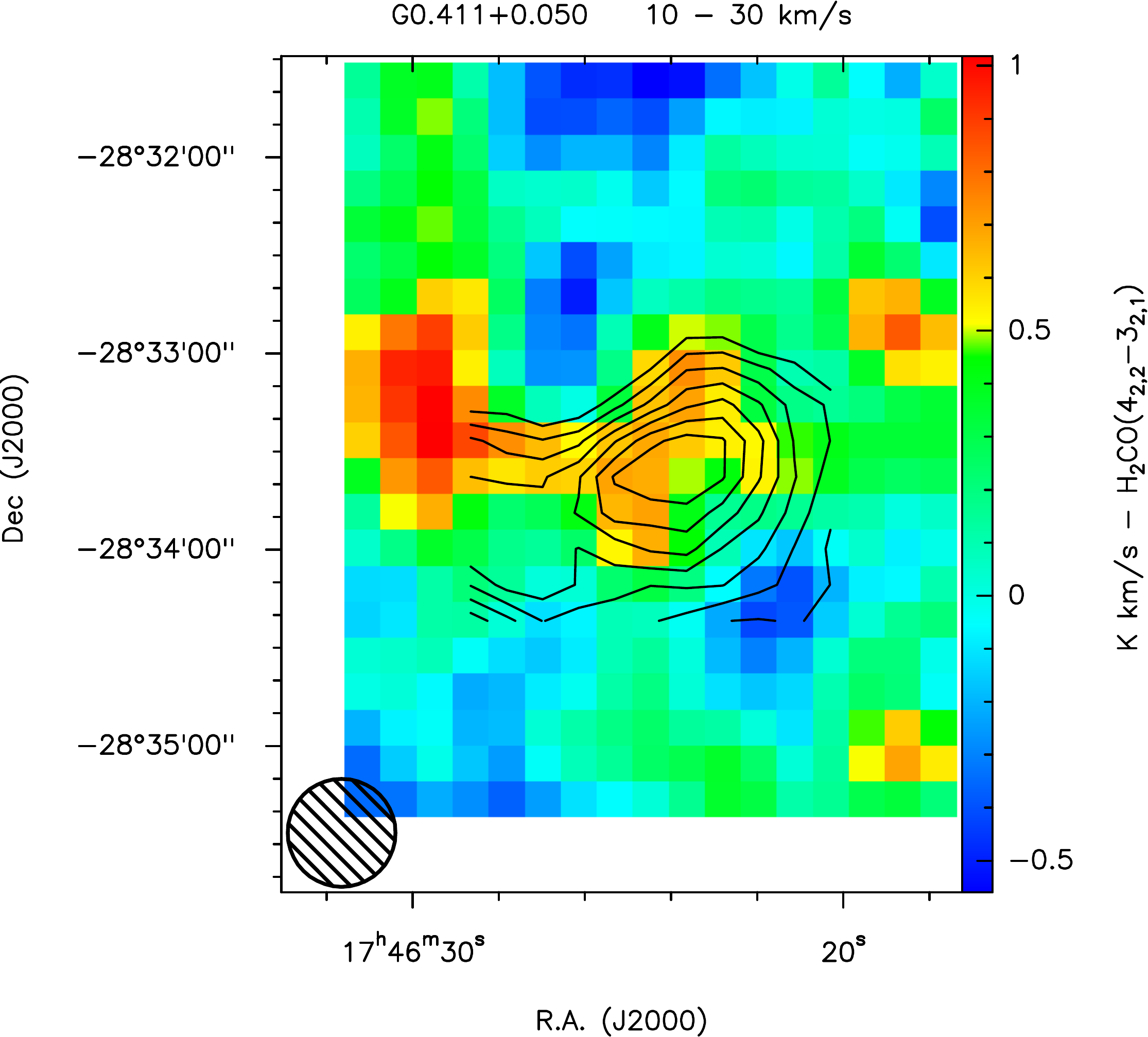}}
	\subfloat{\includegraphics[bb = 150 0 640 580, clip, height=5.5cm]{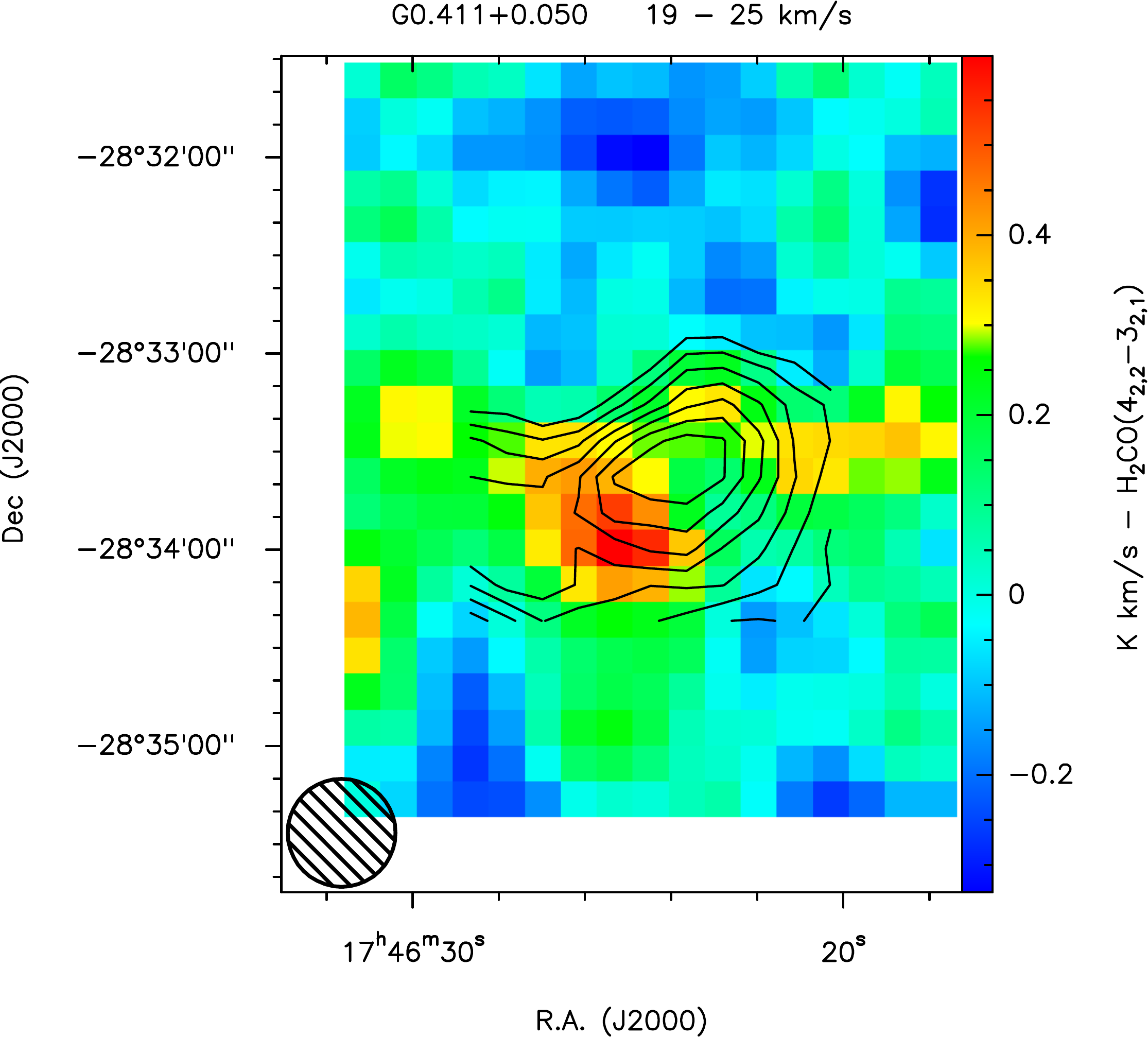}}\\
	\label{G0411-Int-H2CO}
\end{figure*}

\begin{figure*}
	\caption{As Fig. \ref{20kms-Int-H2CO} for G0.480$-$0.006.}
	\centering
	H$_{2}$CO(3$_{0,3}-$2$_{0,2}$)\\
	\subfloat{\includegraphics[bb = 0 0 610 580, clip, height=5.5cm]{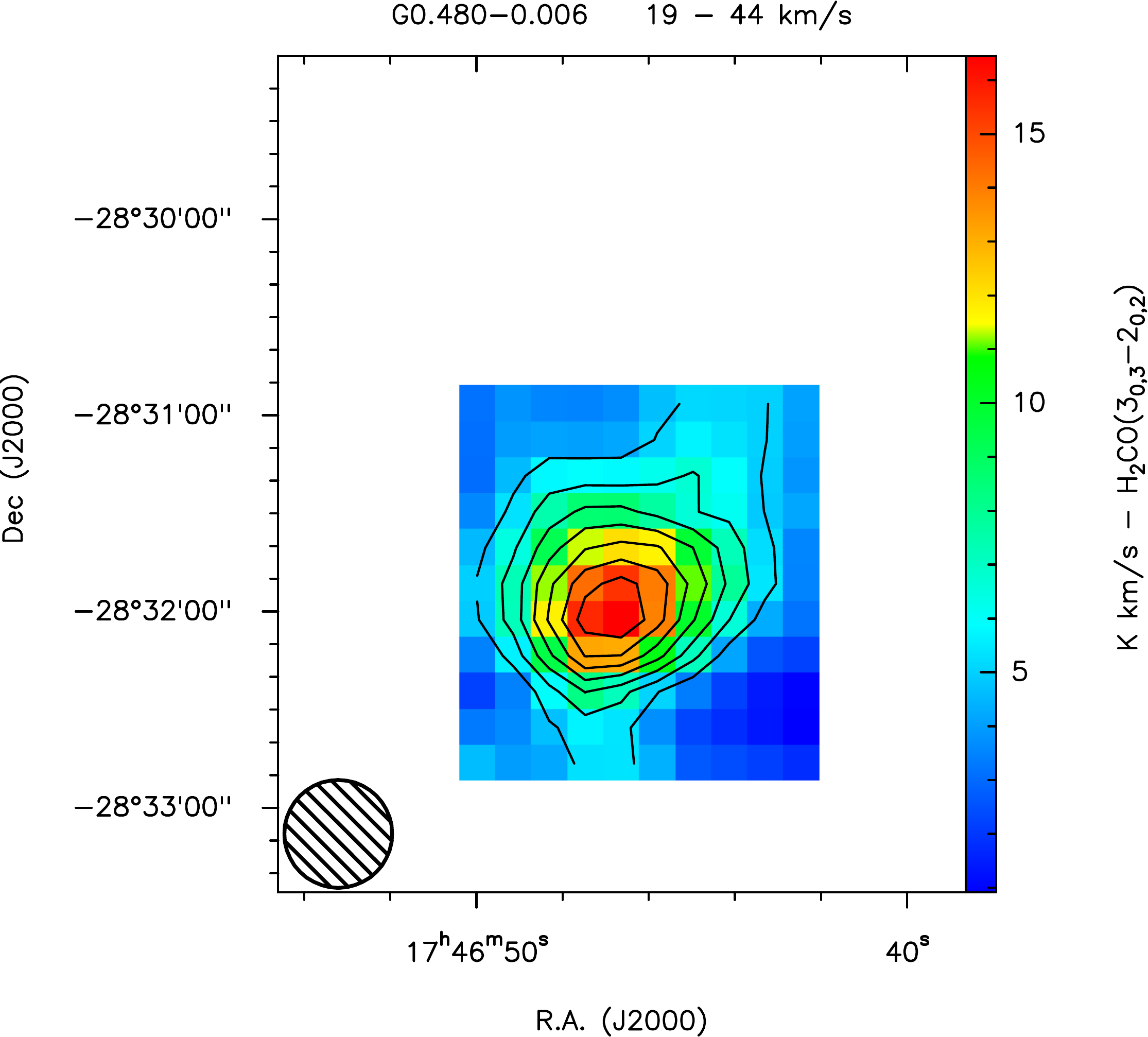}}
	\subfloat{\includegraphics[bb = 150 0 640 580, clip, height=5.5cm]{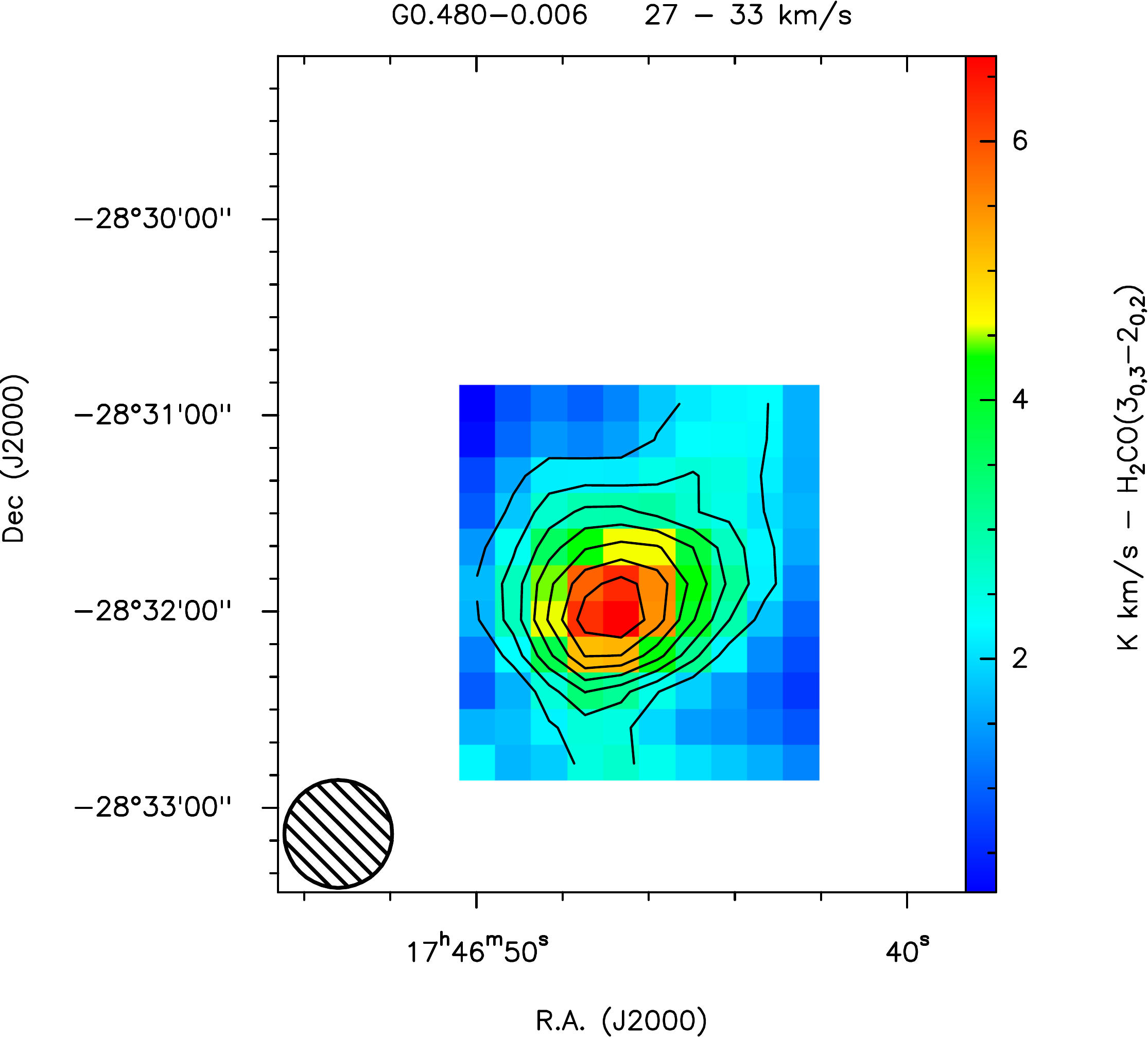}}\\
	H$_{2}$CO(3$_{2,1}-$2$_{2,0}$)\\
	\subfloat{\includegraphics[bb = 0 0 610 580, clip, height=5.5cm]{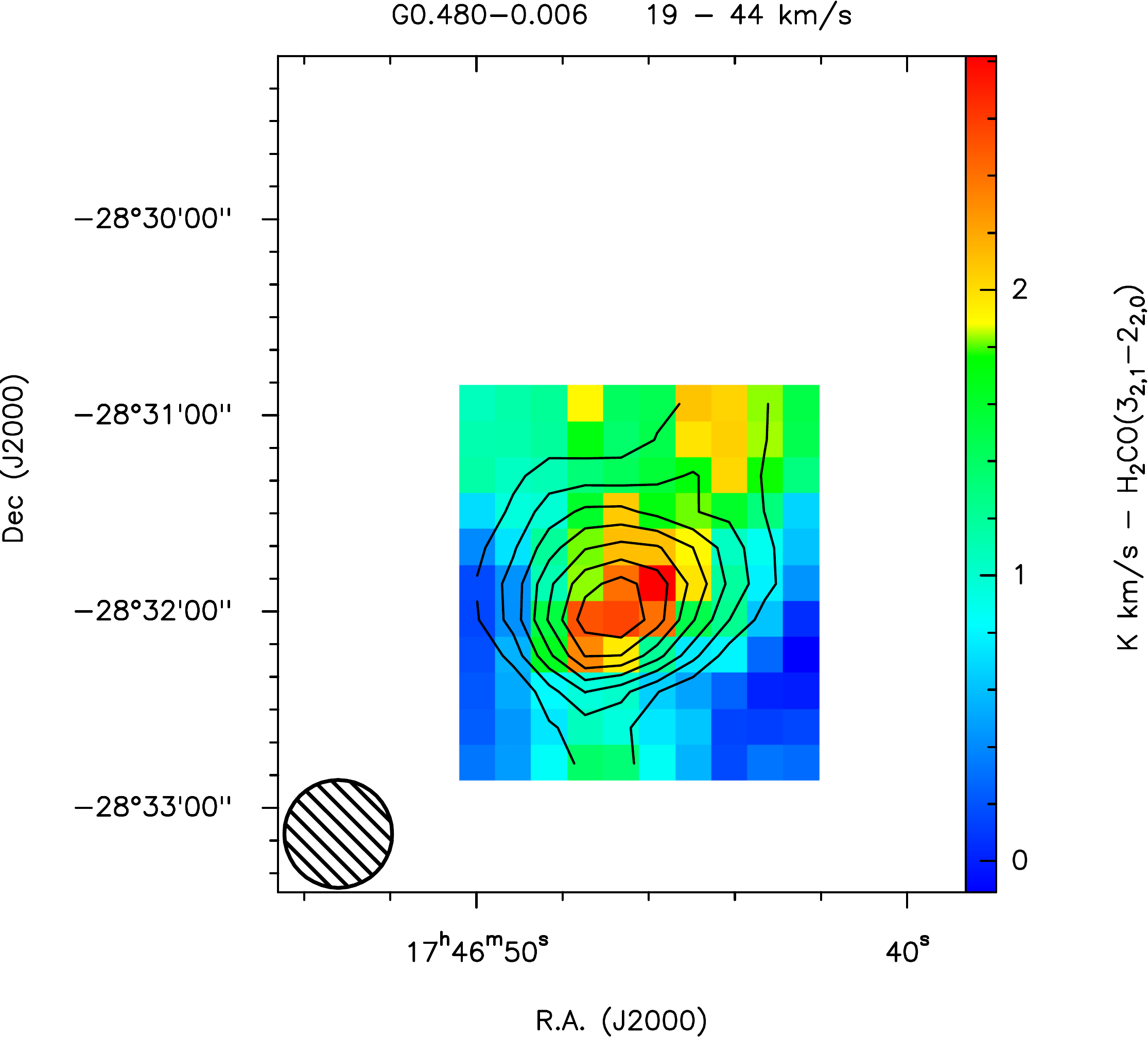}}
	\subfloat{\includegraphics[bb = 150 0 640 580, clip, height=5.5cm]{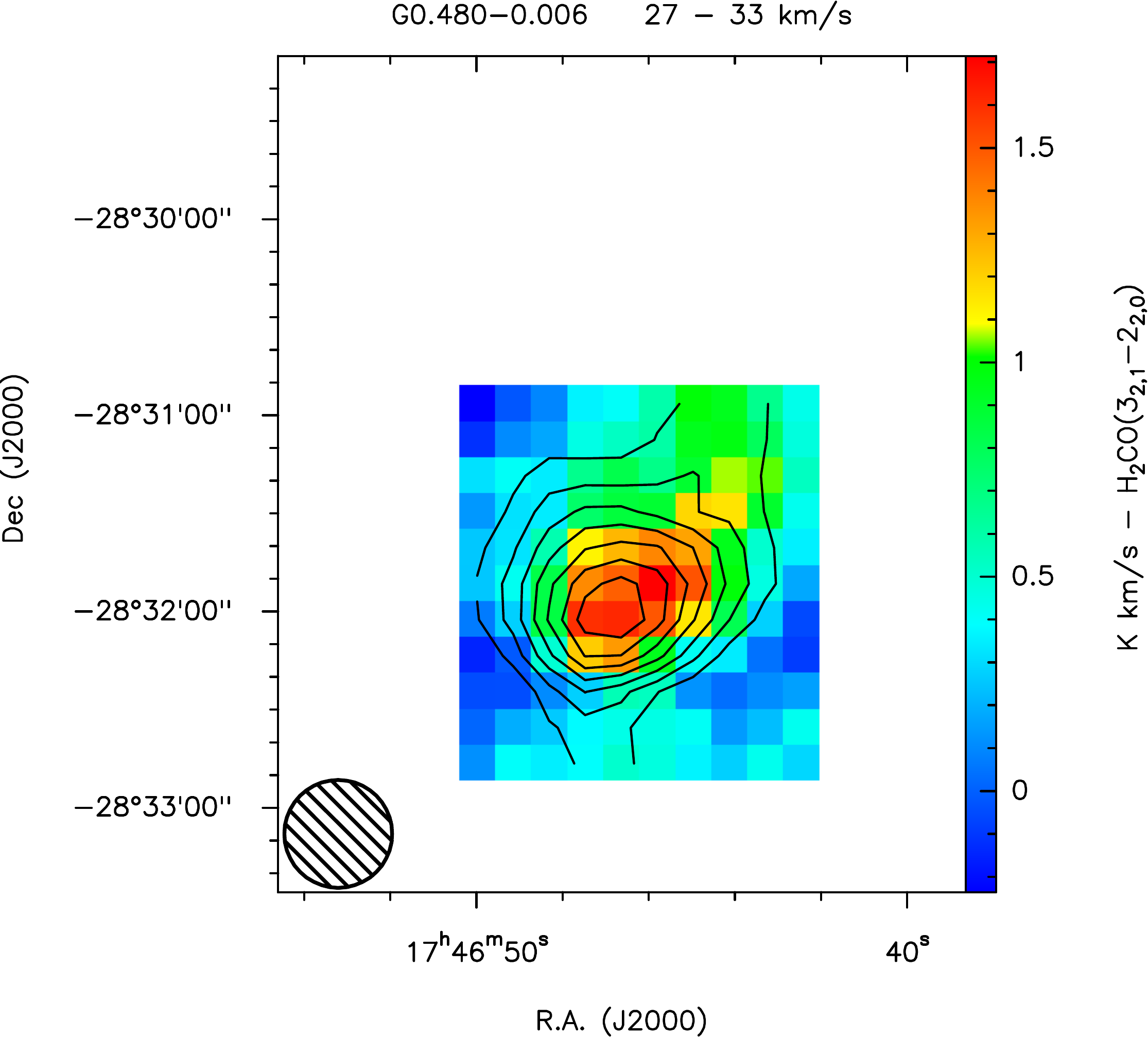}}\\
	H$_{2}$CO(4$_{0,3}-$3$_{0,3}$)\\
	\subfloat{\includegraphics[bb = 0 0 610 580, clip, height=5.5cm]{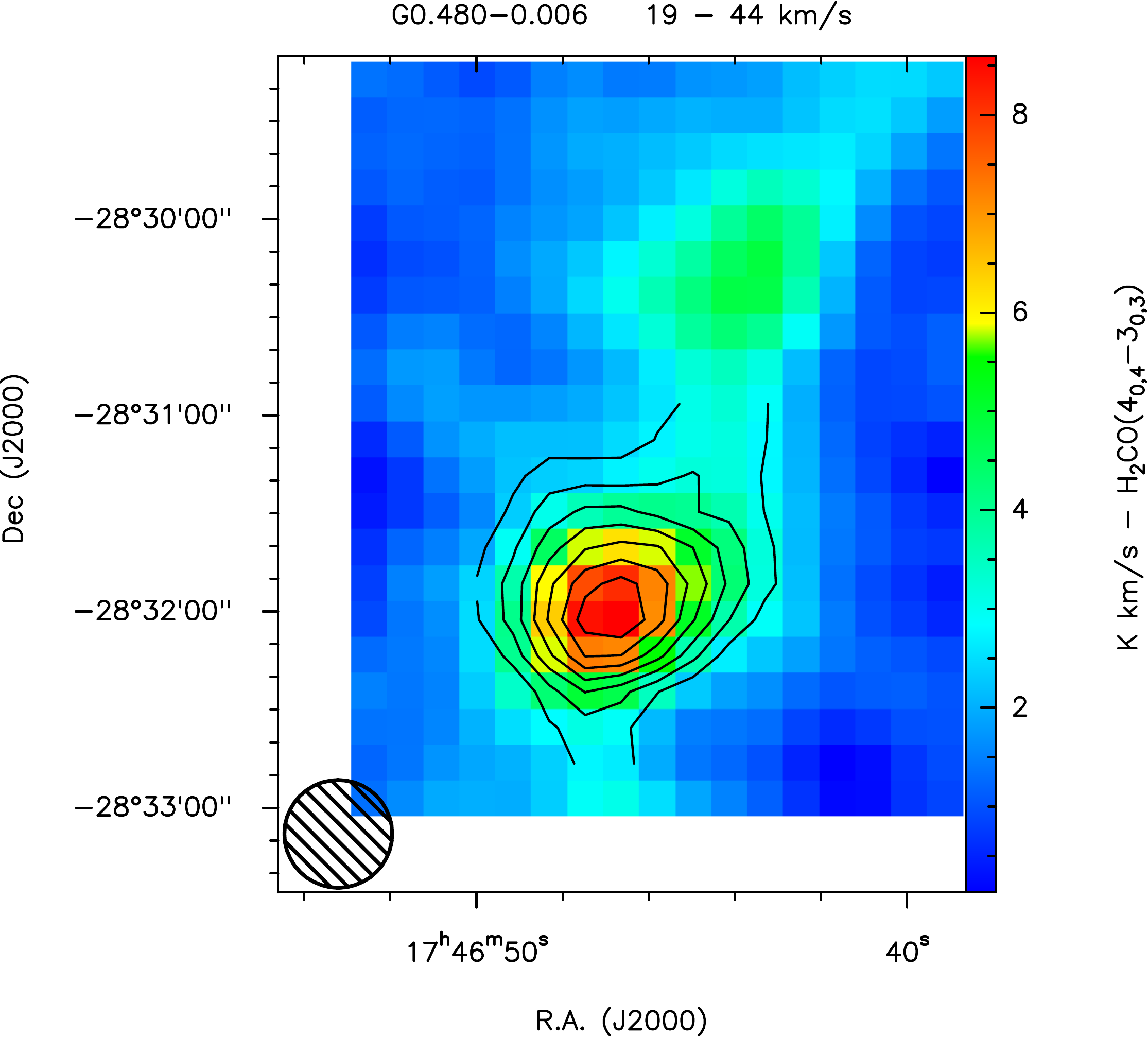}}
	\subfloat{\includegraphics[bb = 150 0 640 580, clip, height=5.5cm]{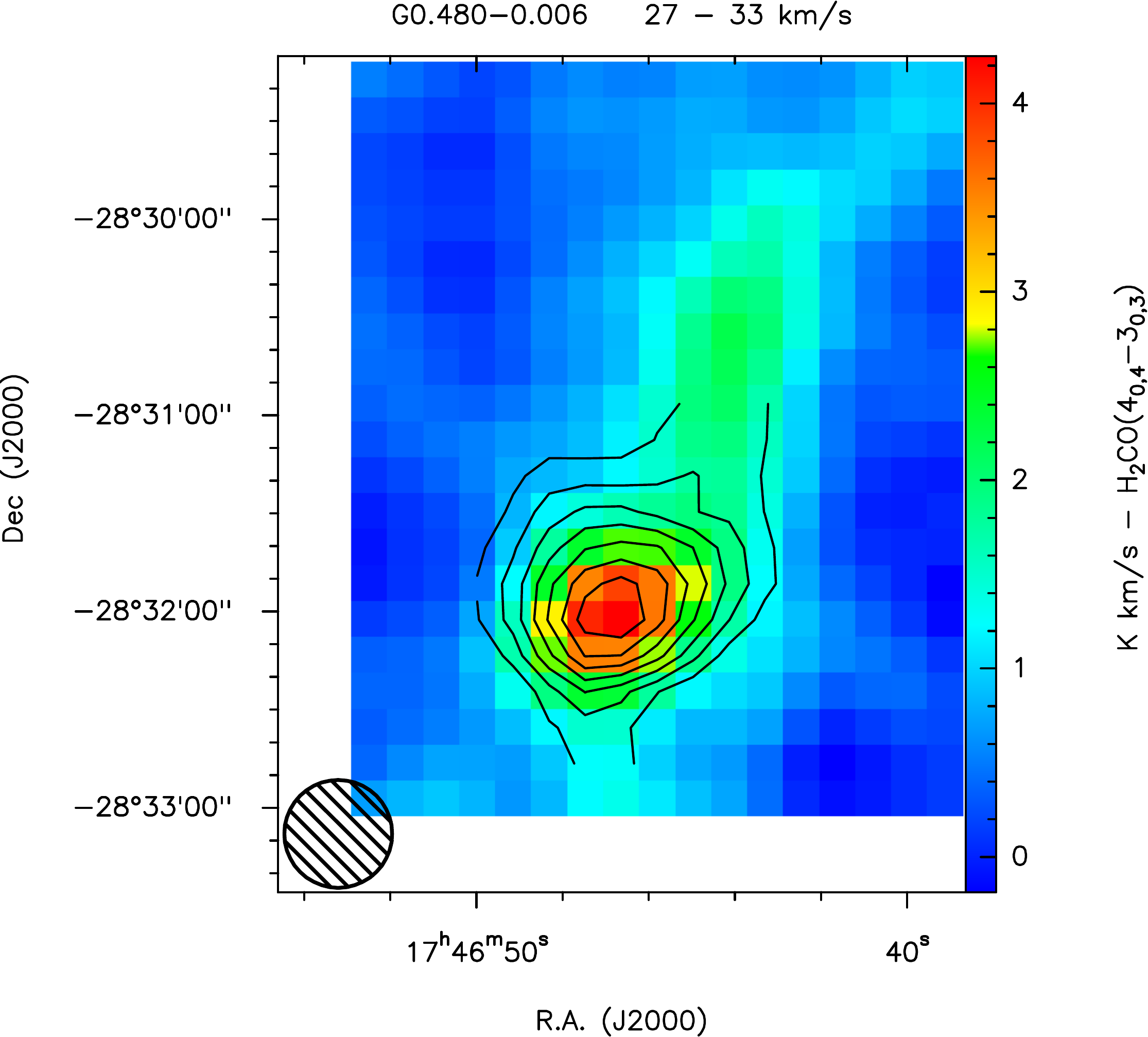}}\\
	H$_{2}$CO(4$_{2,2}-$3$_{2,1}$)\\
	\subfloat{\includegraphics[bb = 0 0 610 580, clip, height=5.5cm]{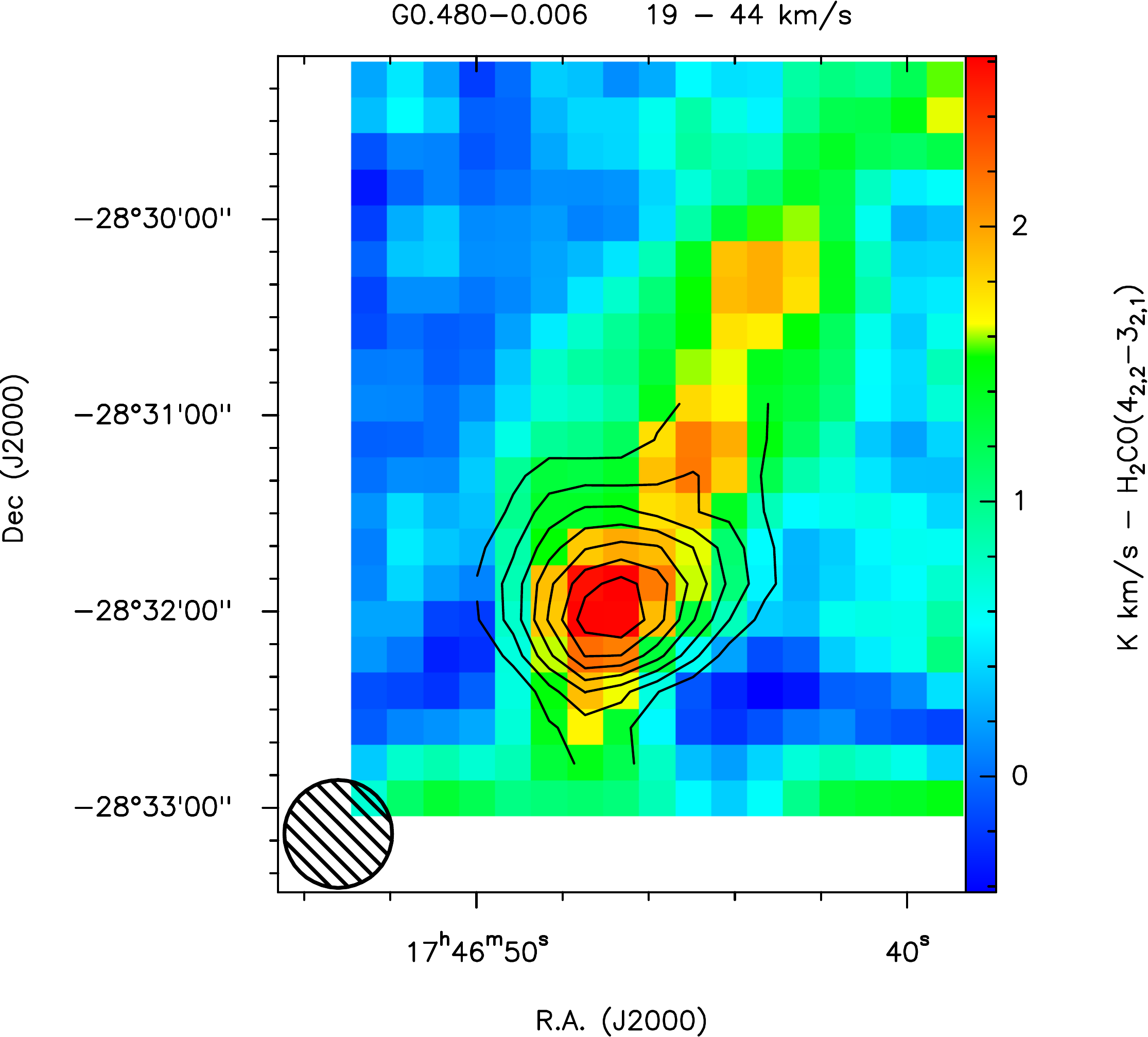}}
	\subfloat{\includegraphics[bb = 150 0 640 580, clip, height=5.5cm]{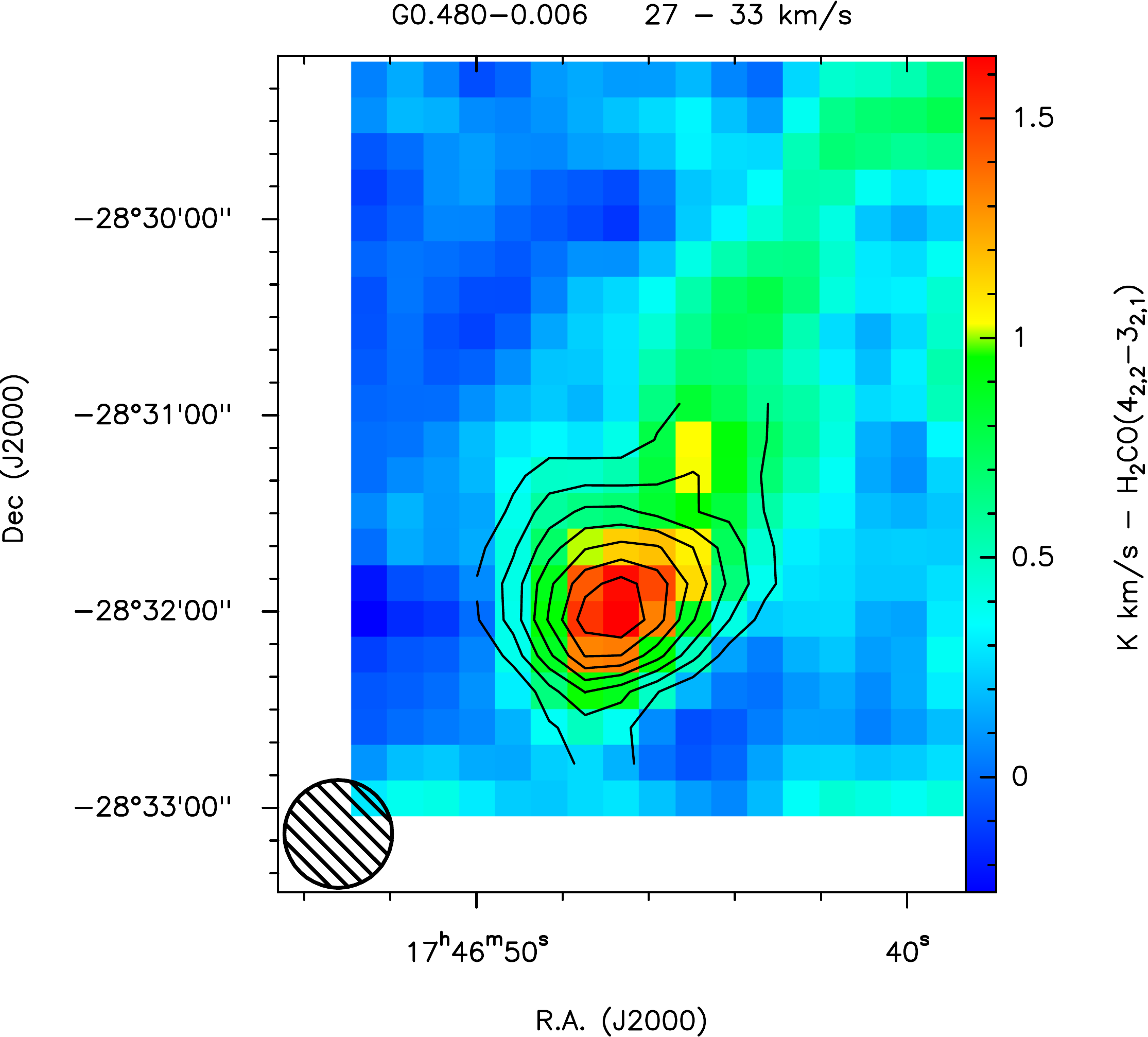}}\\
	\label{G0480-Int-H2CO}
\end{figure*}

\begin{figure*}
	\caption{As Fig. \ref{20kms-Int-H2CO} for Sgr C.}
	\centering
	H$_{2}$CO(4$_{0,3}-$3$_{0,3}$)\\
	\subfloat{\includegraphics[bb = 0 0 720 580, clip, height=5.5cm]{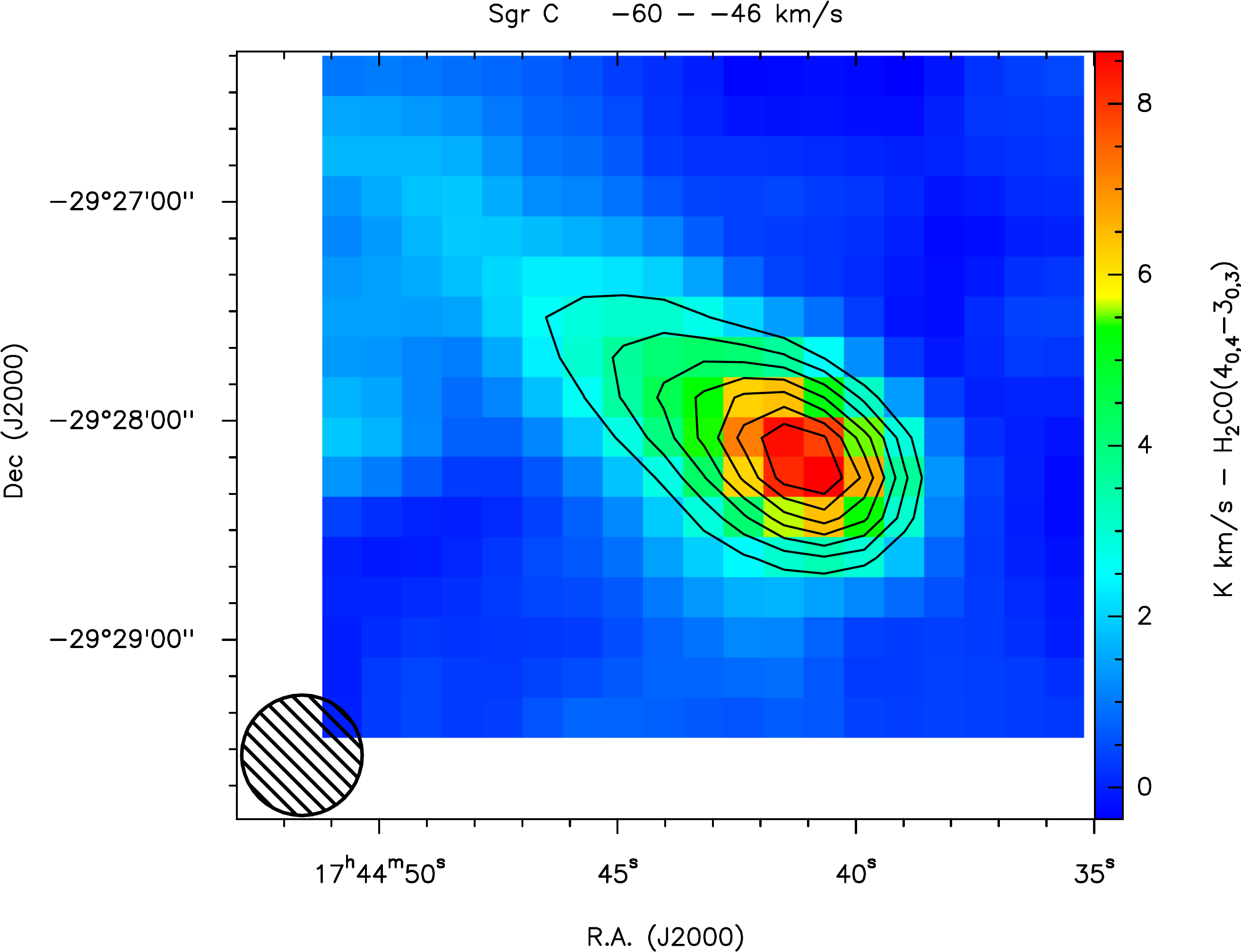}}
	\subfloat{\includegraphics[bb = 130 0 760 580, clip, height=5.5cm]{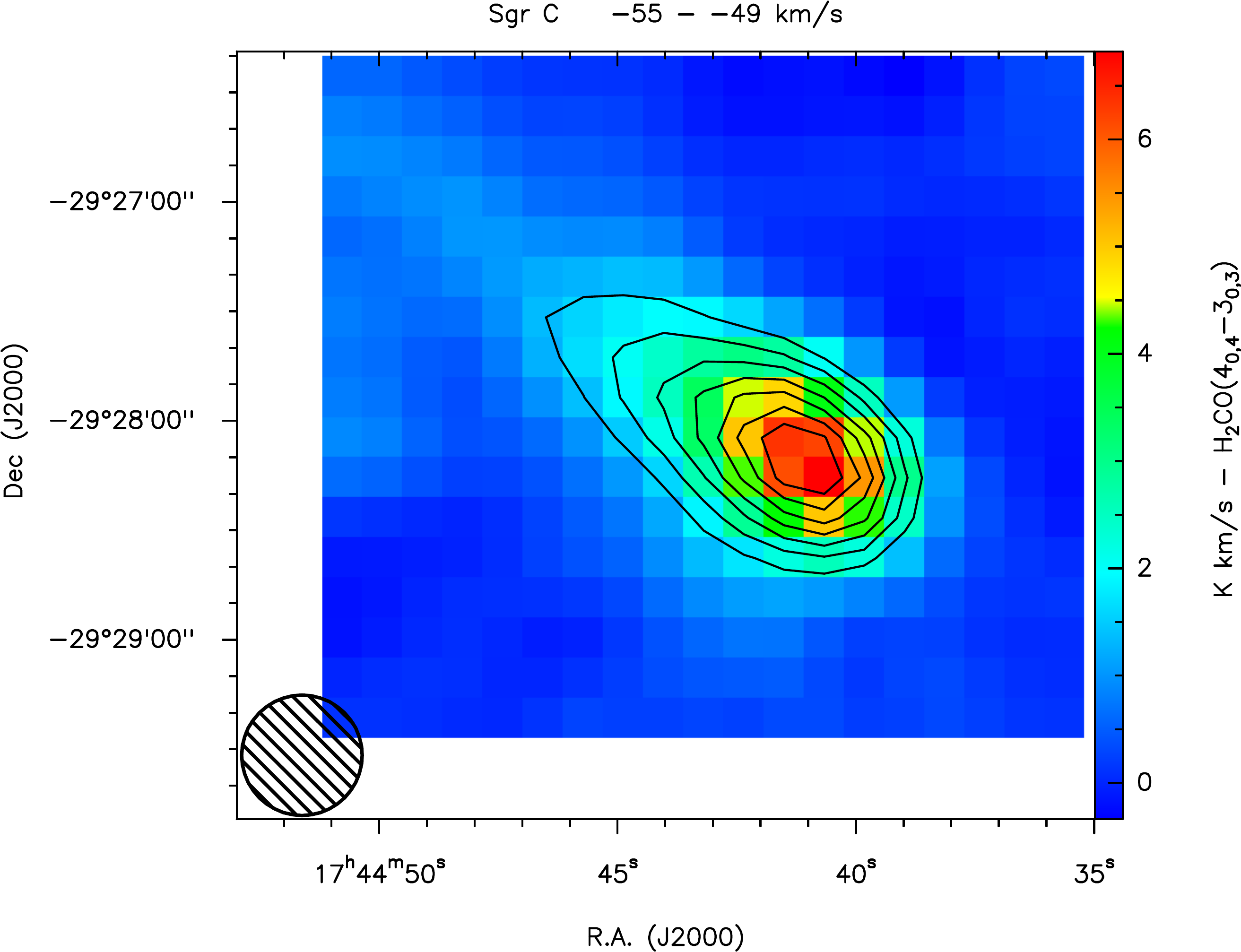}}\\
	H$_{2}$CO(4$_{2,2}-$3$_{2,1}$)\\
	\subfloat{\includegraphics[bb = 0 0 720 580, clip, height=5.5cm]{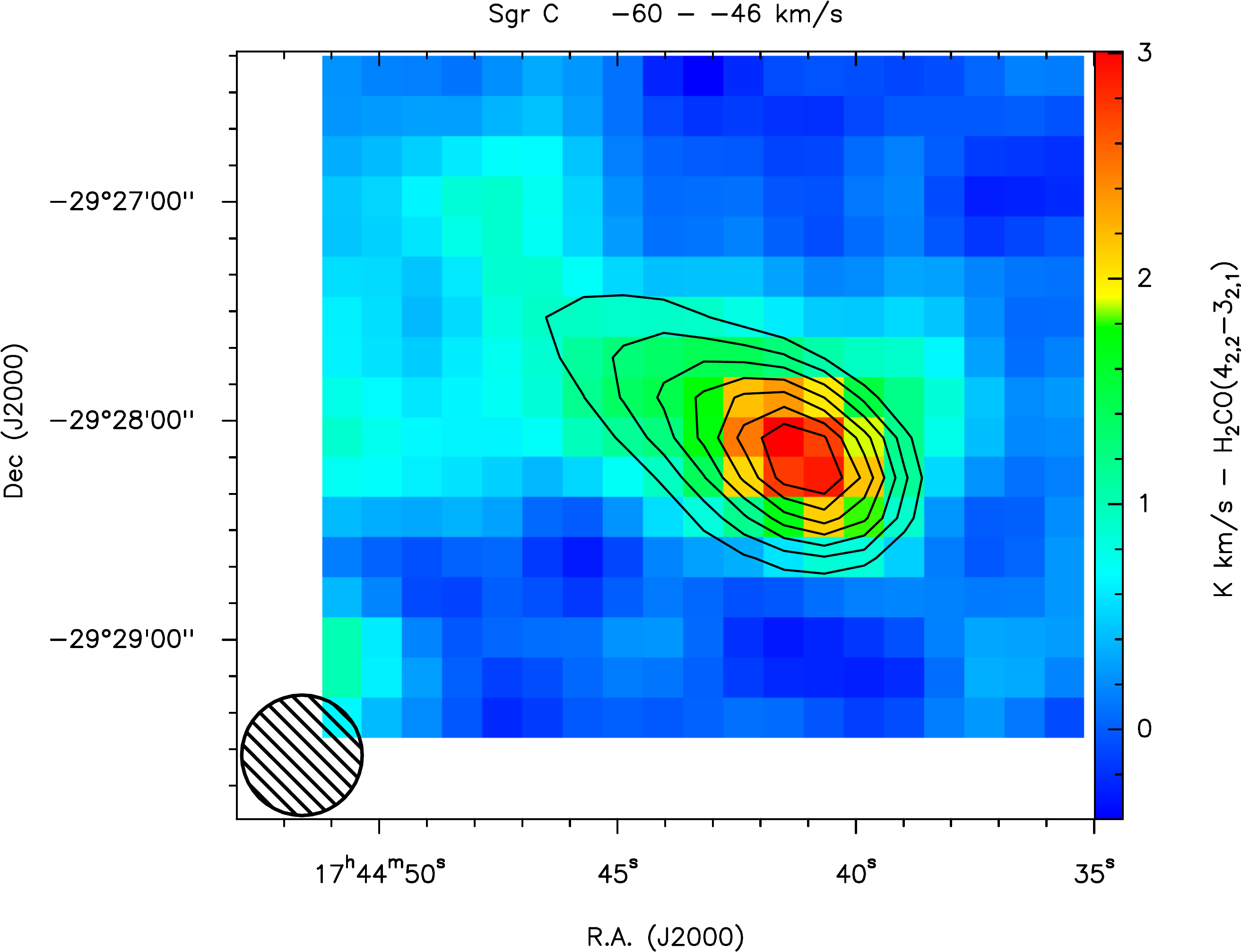}}
	\subfloat{\includegraphics[bb = 130 0 760 580, clip, height=5.5cm]{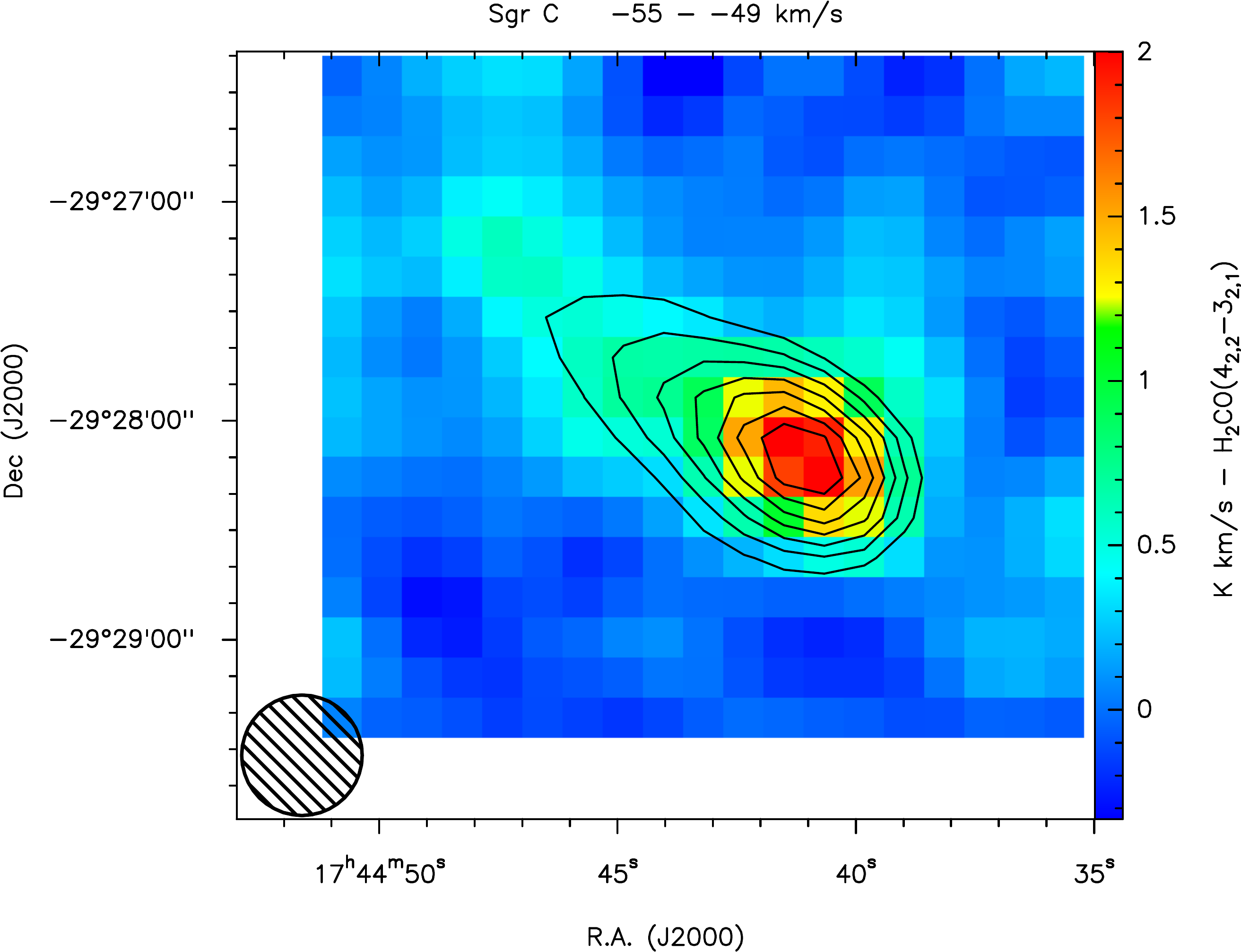}}
	\label{SGRC-Int-H2CO}
\end{figure*}

\begin{figure*}
	\caption{As Fig. \ref{20kms-Int-H2CO} for Sgr D.}
	\centering
	H$_{2}$CO(4$_{0,3}-$3$_{0,3}$)\\
	\subfloat{\includegraphics[bb = 0 0 700 580, clip, height=5.5cm]{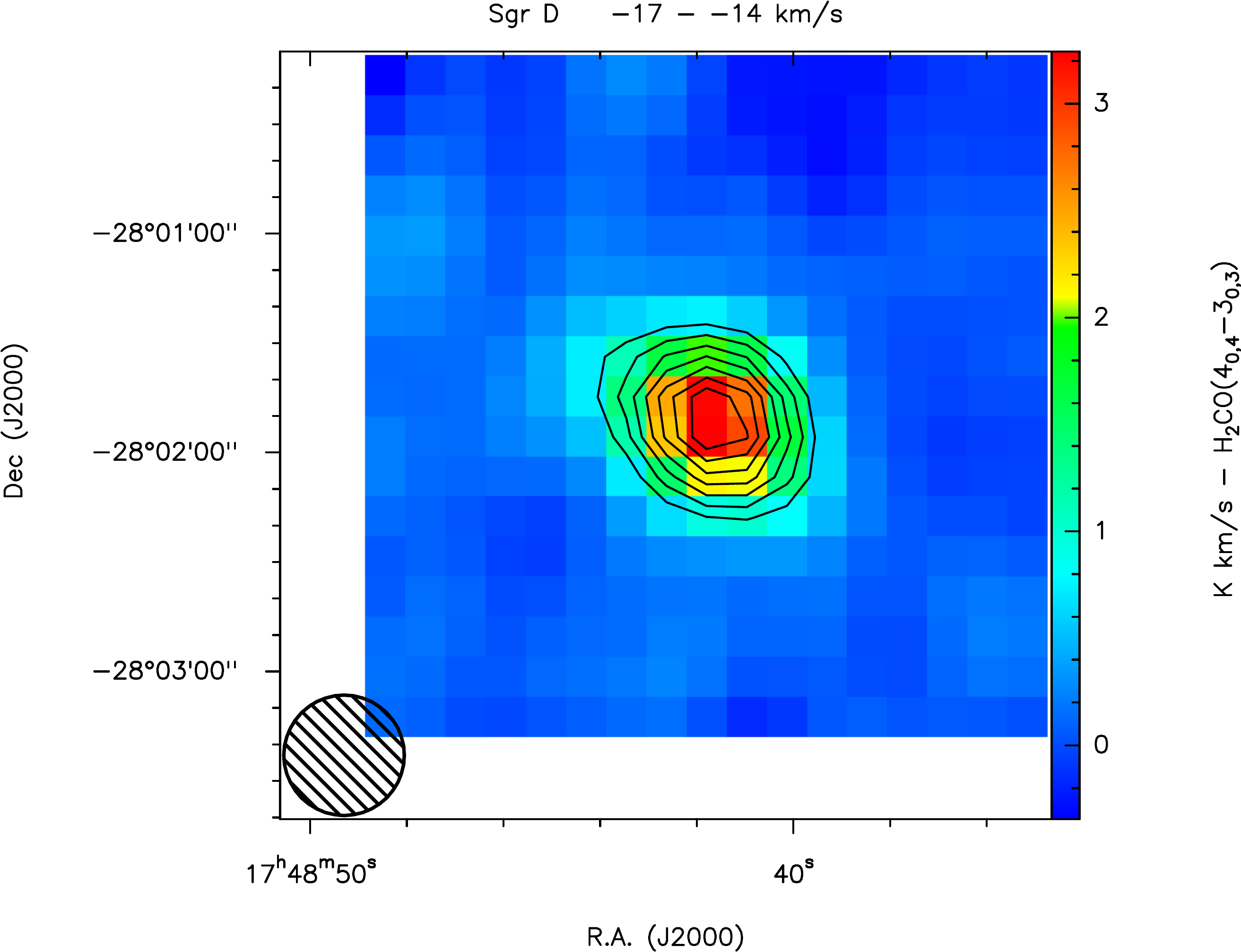}}
	\subfloat{\includegraphics[bb = 150 0 760 580, clip, height=5.5cm]{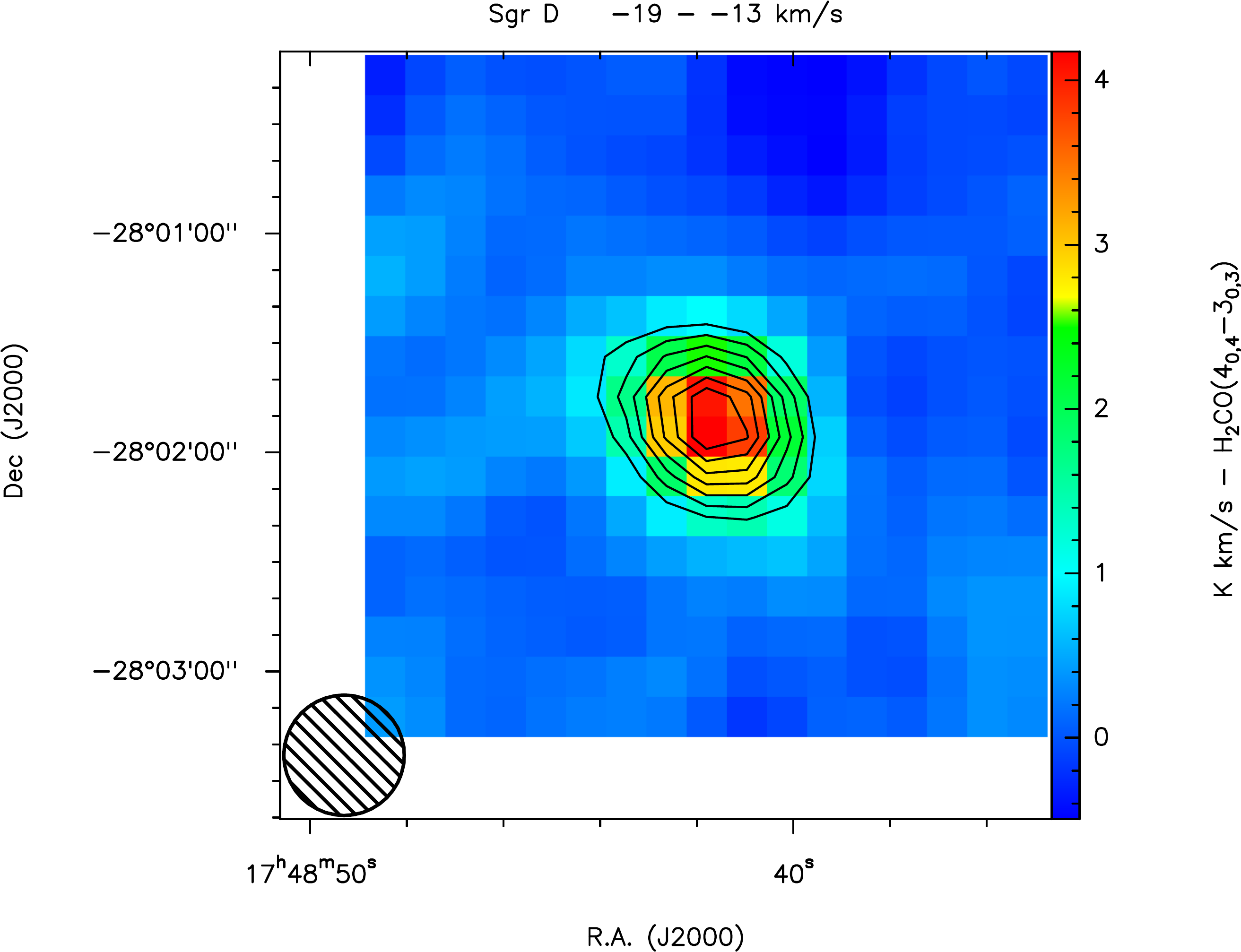}}\\
	H$_{2}$CO(4$_{2,2}-$3$_{2,1}$)\\
	\subfloat{\includegraphics[bb = 0 0 700 580, clip, height=5.5cm]{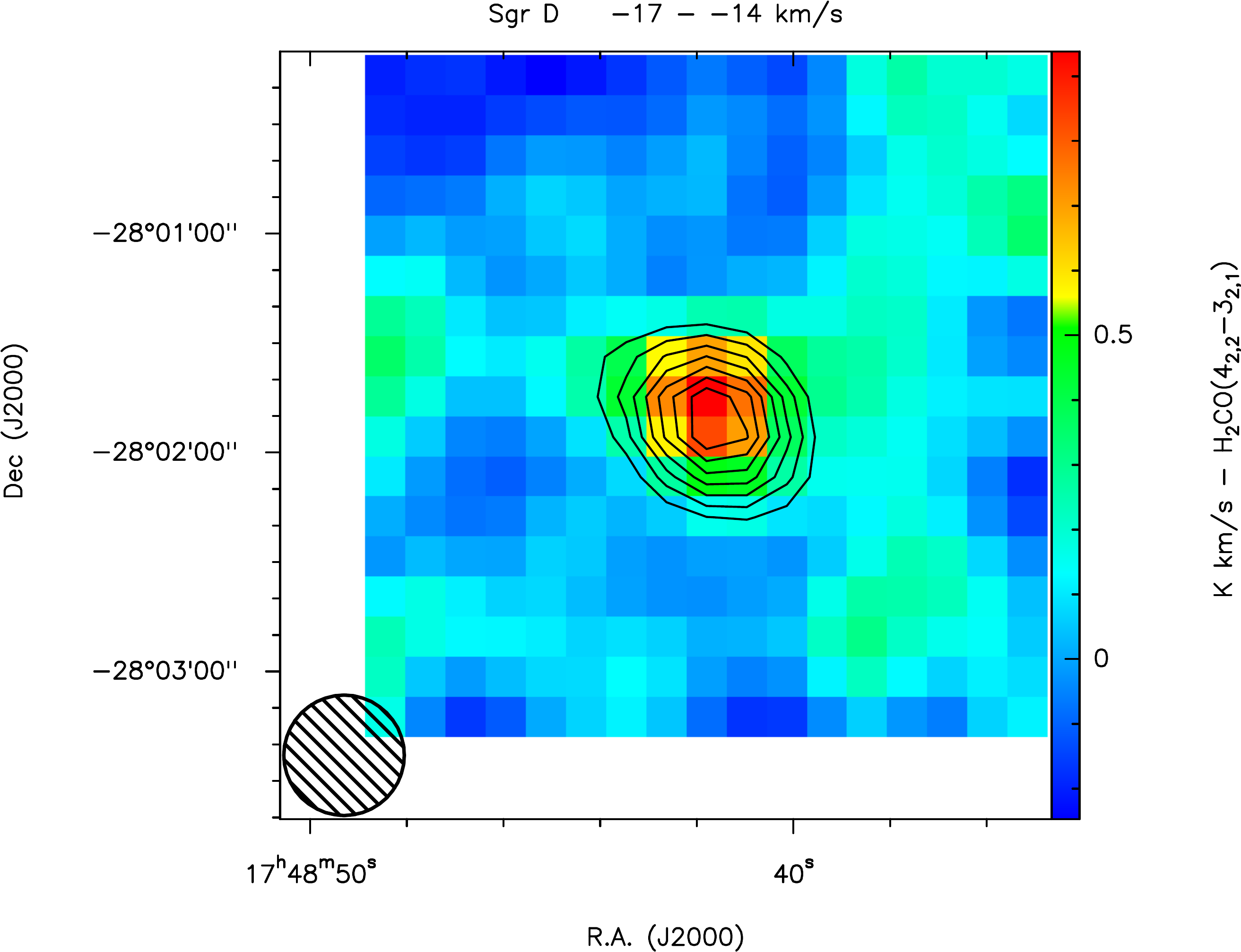}}
	\subfloat{\includegraphics[bb = 150 0 760 580, clip, height=5.5cm]{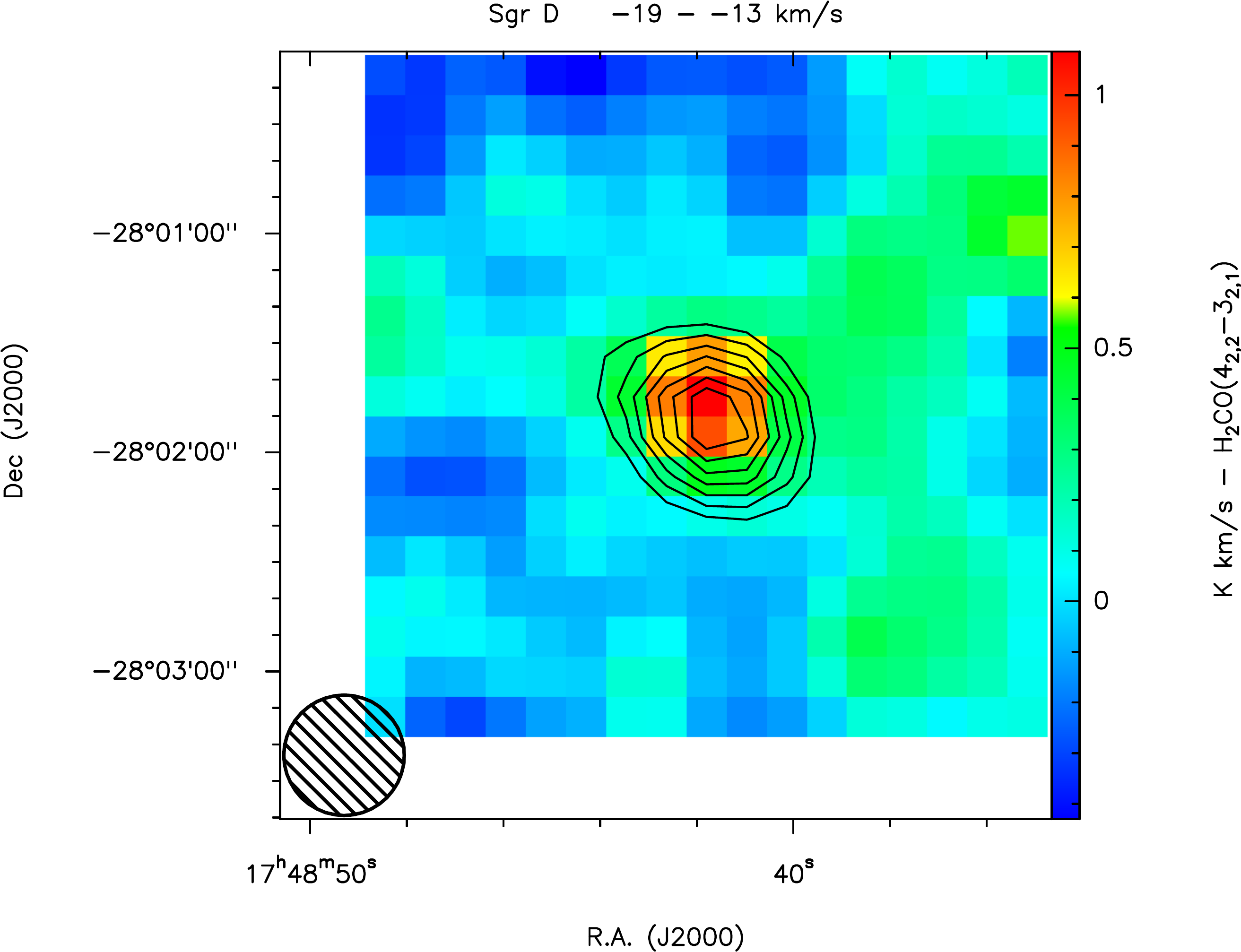}}
	\label{SGRD-Int-H2CO}
\end{figure*}

\clearpage

\section{Integrated Intensity Ratio and Uncertainty Maps}

\begin{figure*}
	\caption{Integrated intensity ratio (upper panels) and uncertainty (lower panels) maps of the 20 km/s cloud (contours as in Fig. 
	\ref{20kms-Int-H2CO}). 
	Upper limits of the ratios are marked with Xs. The corresponding pixels in the uncertainty maps are shown in 
	grey. The circle in the lower left corner shows the 33$\arcsec$ beam.}
	\centering
        R$_{321}$\\
	\subfloat{\includegraphics[bb = 0 60 600 580, clip, height=4.5cm]{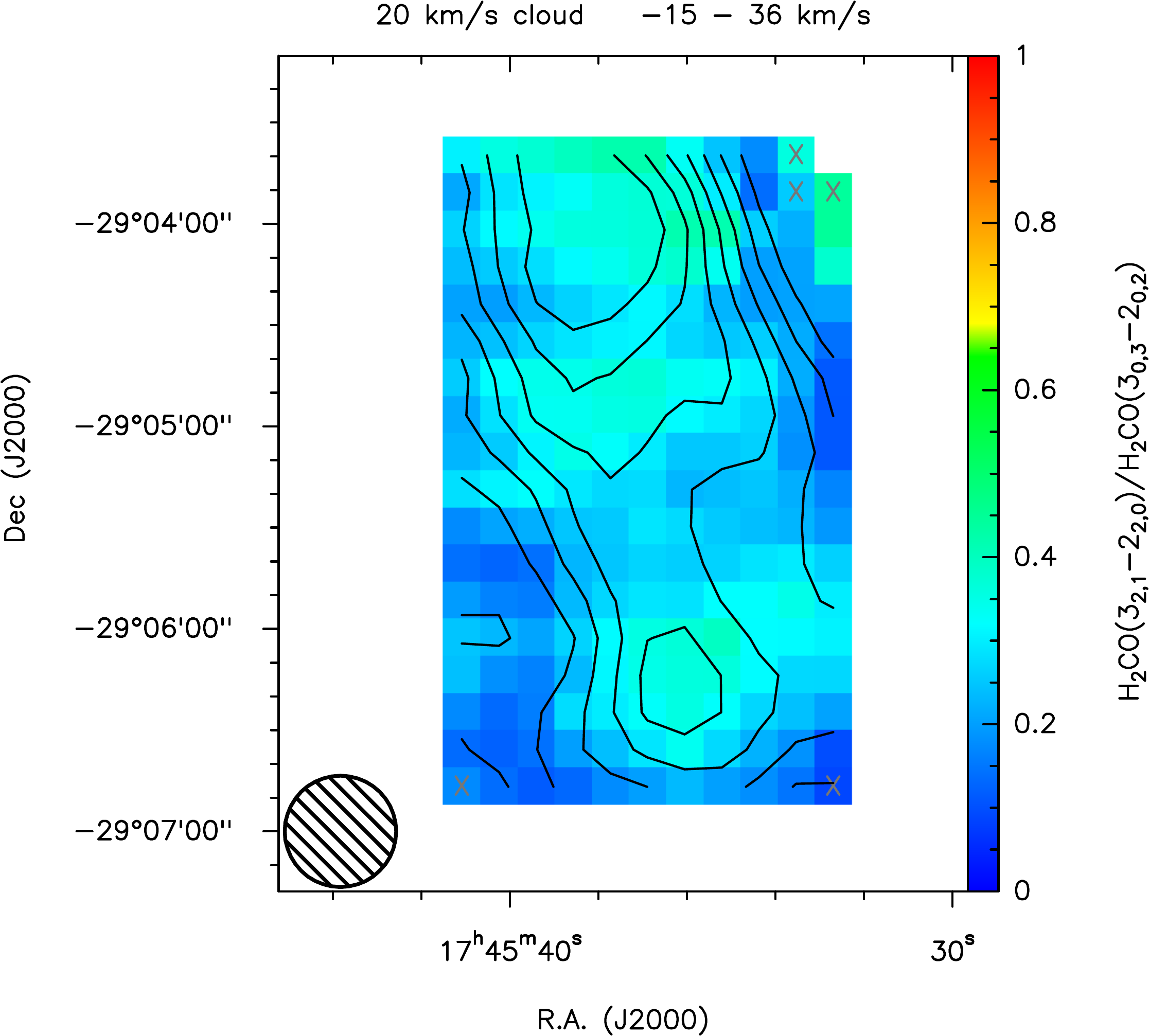}}
	\subfloat{\includegraphics[bb = 140 60 600 580, clip, height=4.5cm]{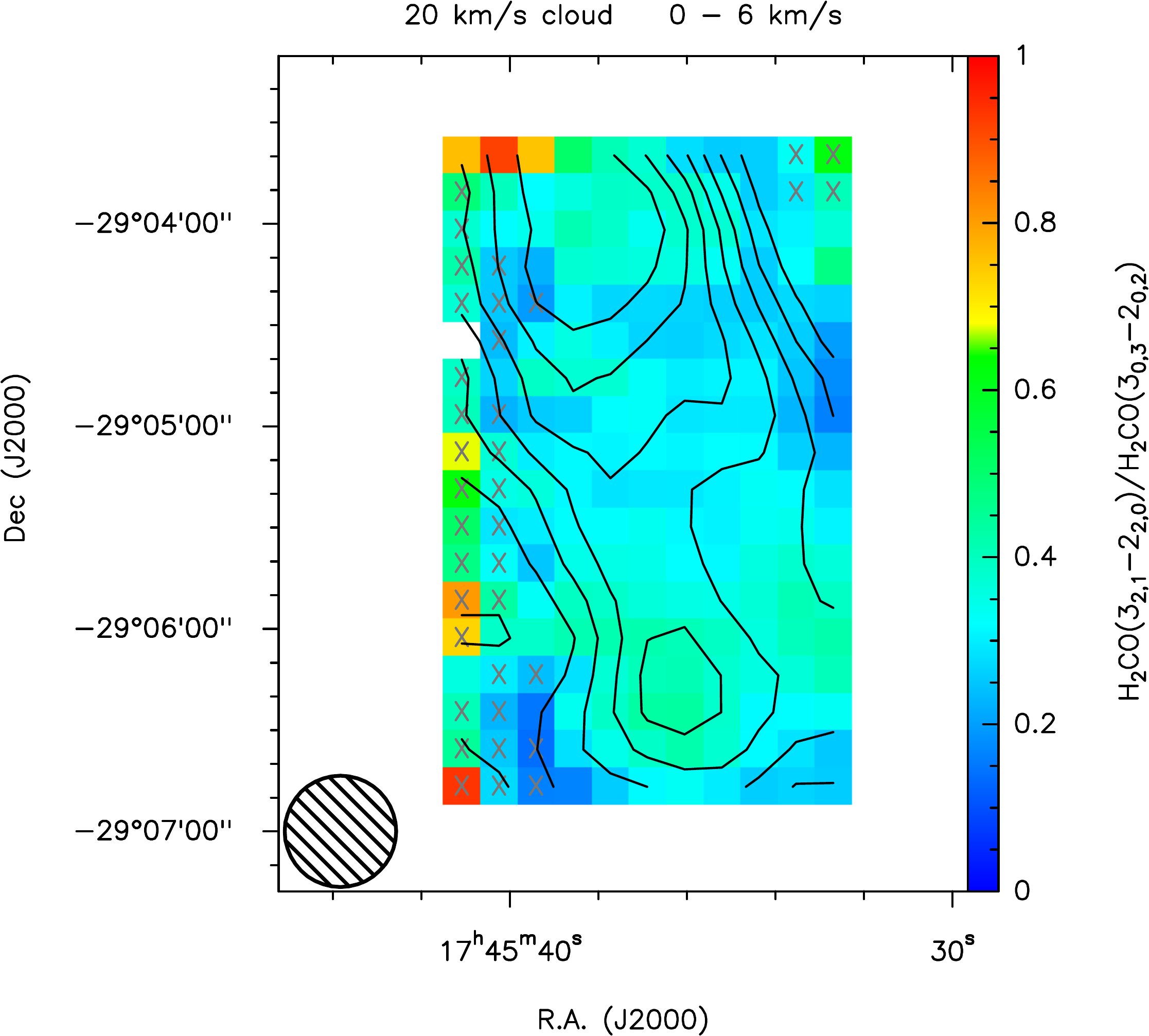}}
	\subfloat{\includegraphics[bb = 140 60 600 580, clip, height=4.5cm]{20kms-H2CO-8-14-Ratio_321_303.pdf}}
	\subfloat{\includegraphics[bb = 140 60 650 580, clip, height=4.5cm]{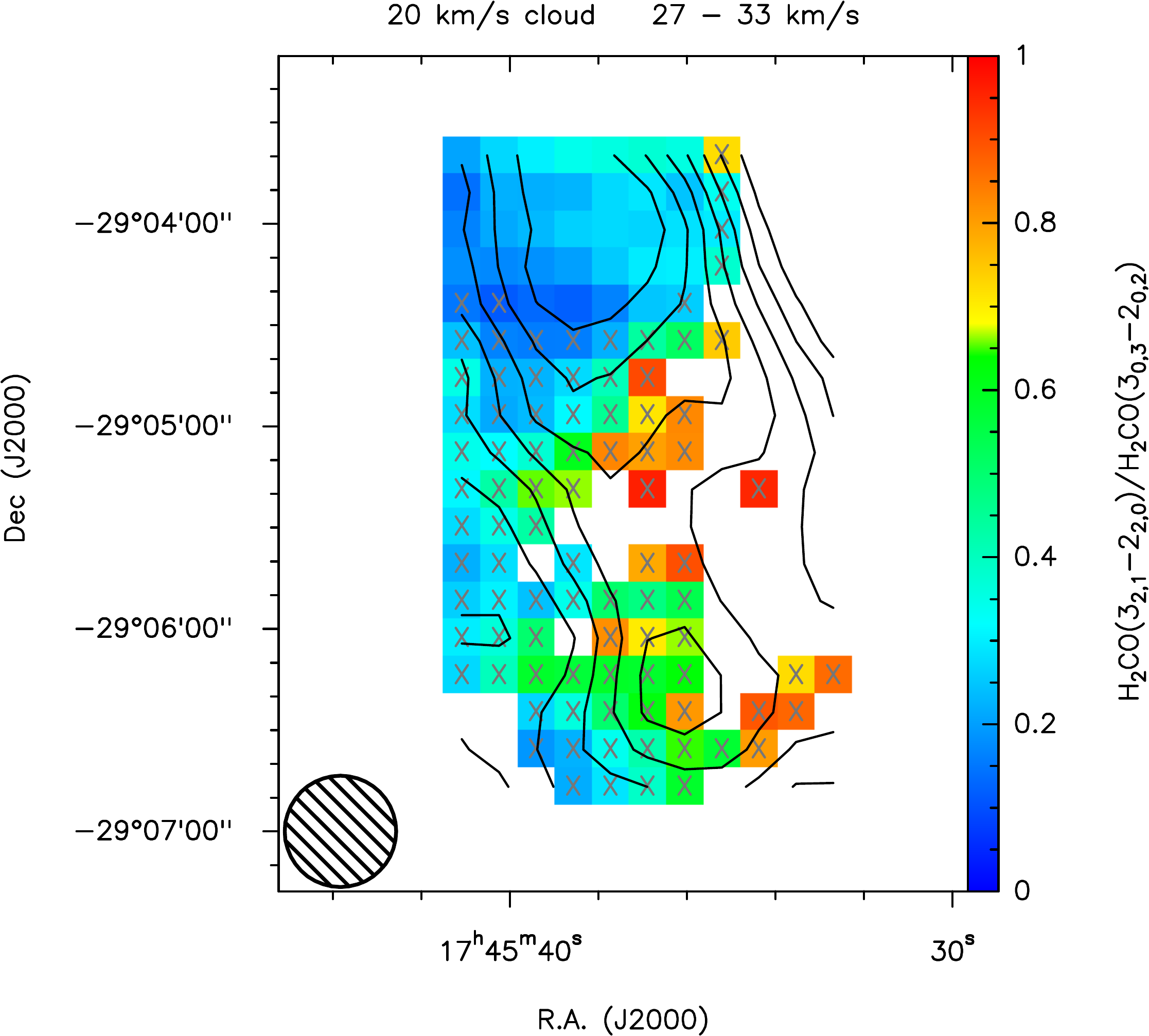}}\\\vspace{-0.5cm}
	\subfloat{\includegraphics[bb = 0 0 600 560, clip, height=4.845cm]{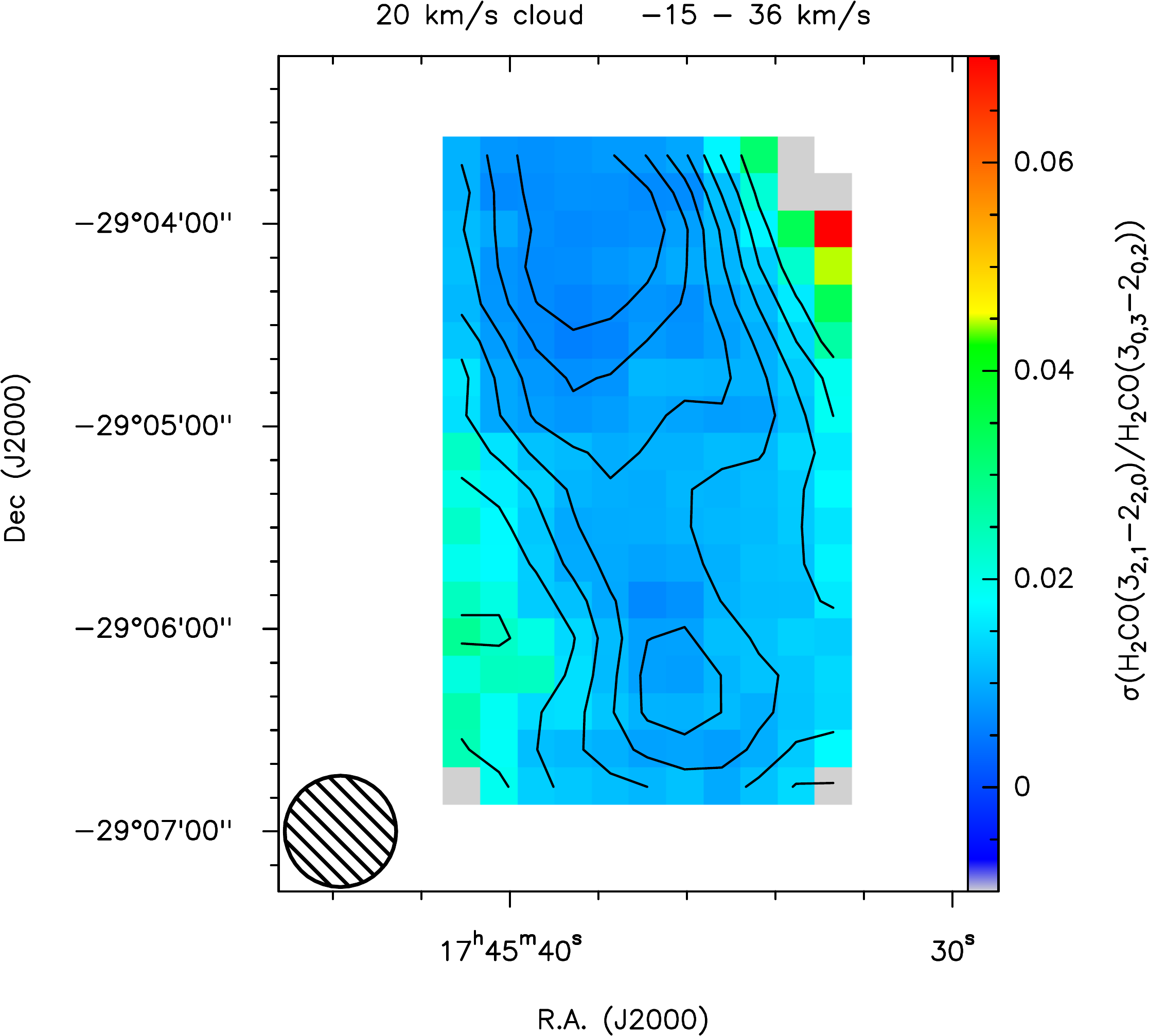}}
	\subfloat{\includegraphics[bb = 140 0 600 560, clip, height=4.845cm]{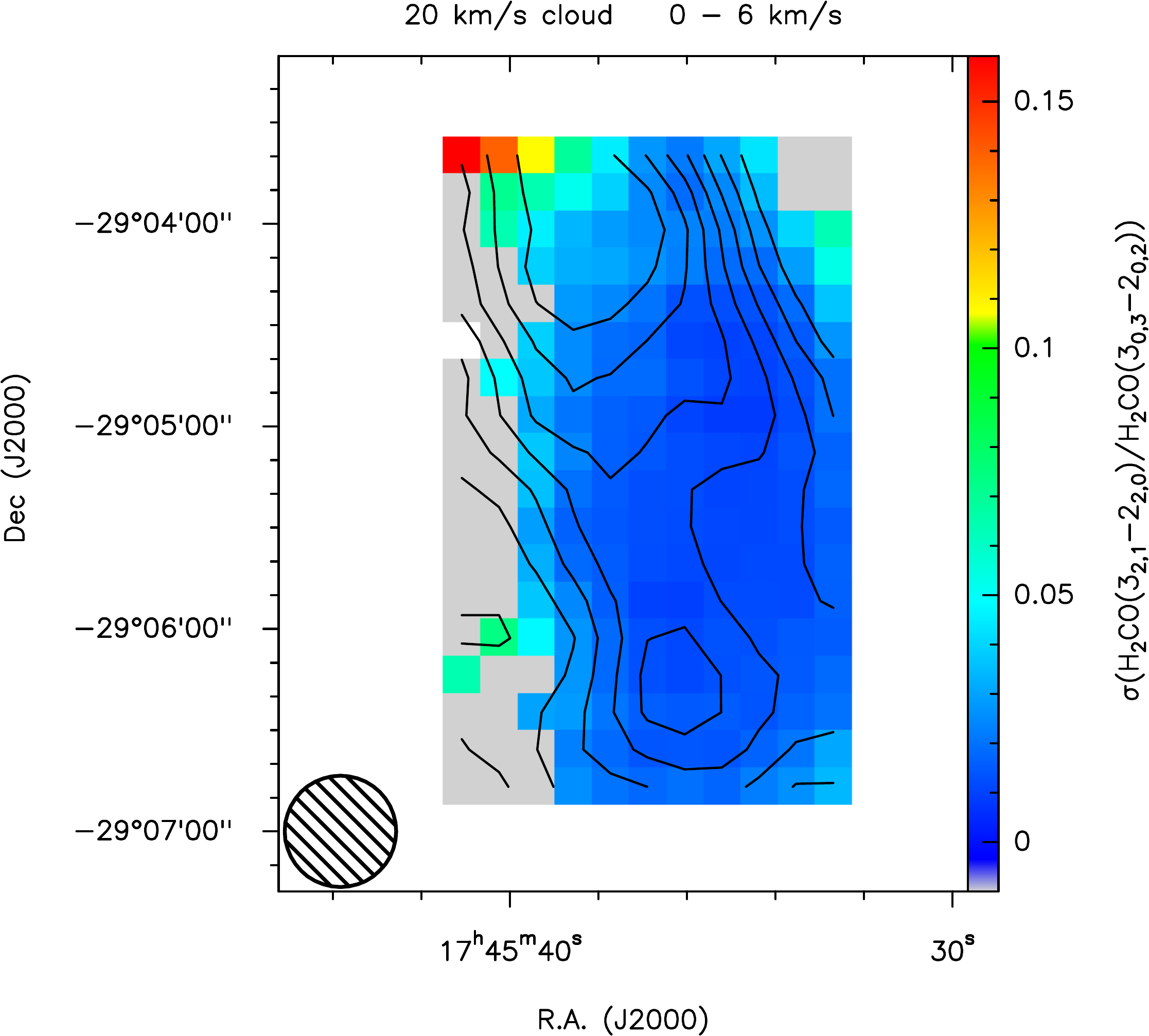}}
	\subfloat{\includegraphics[bb = 140 0 600 560, clip, height=4.845cm]{20kms-H2CO-8-14-Uncertainty-Ratio_321_303.pdf}}
	\subfloat{\includegraphics[bb = 140 0 650 560, clip, height=4.845cm]{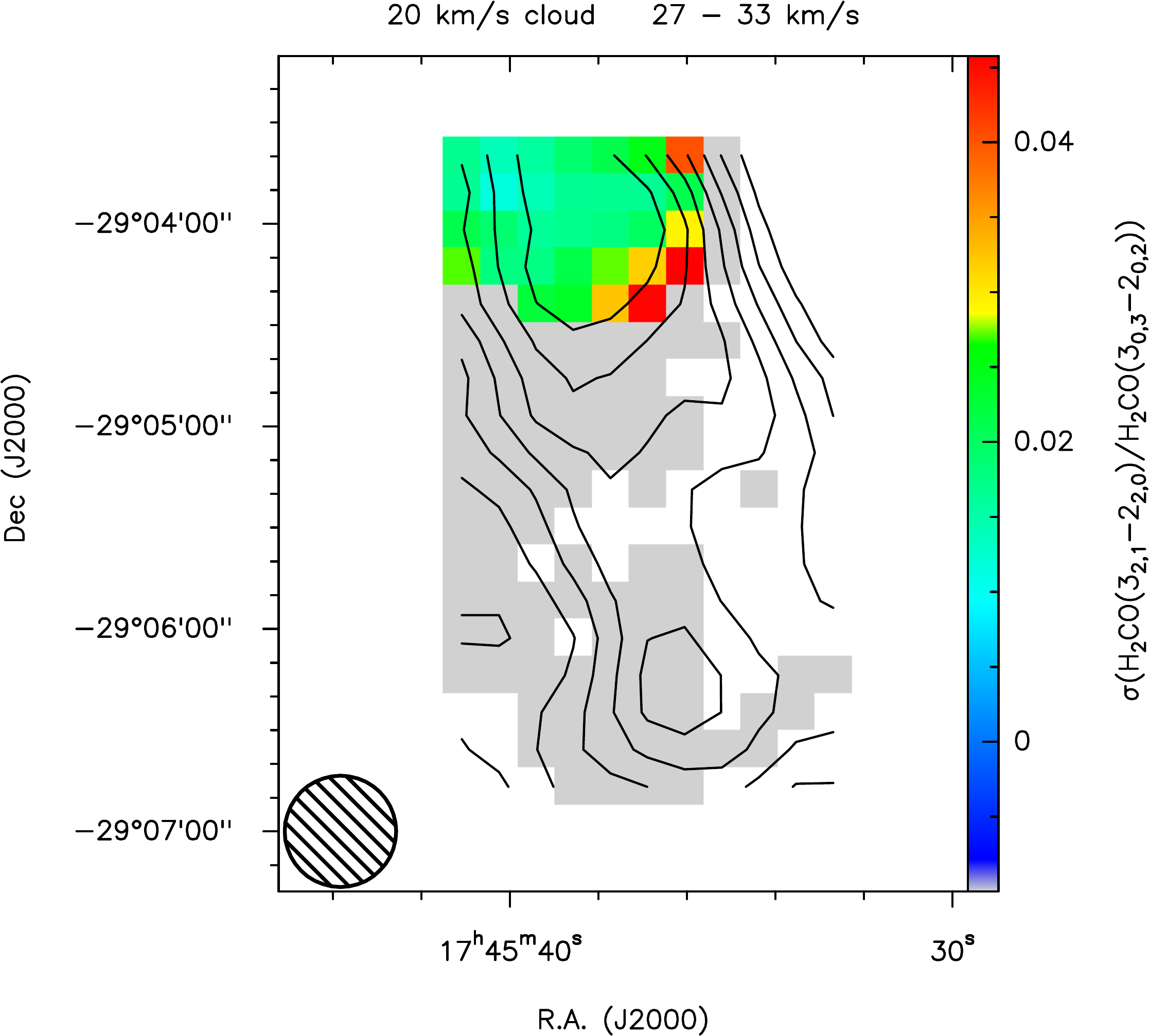}}\\ 
	\vspace{0.1cm}
        R$_{422}$\\
	\subfloat{\includegraphics[bb = 0 60 600 580, clip, height=4.5cm]{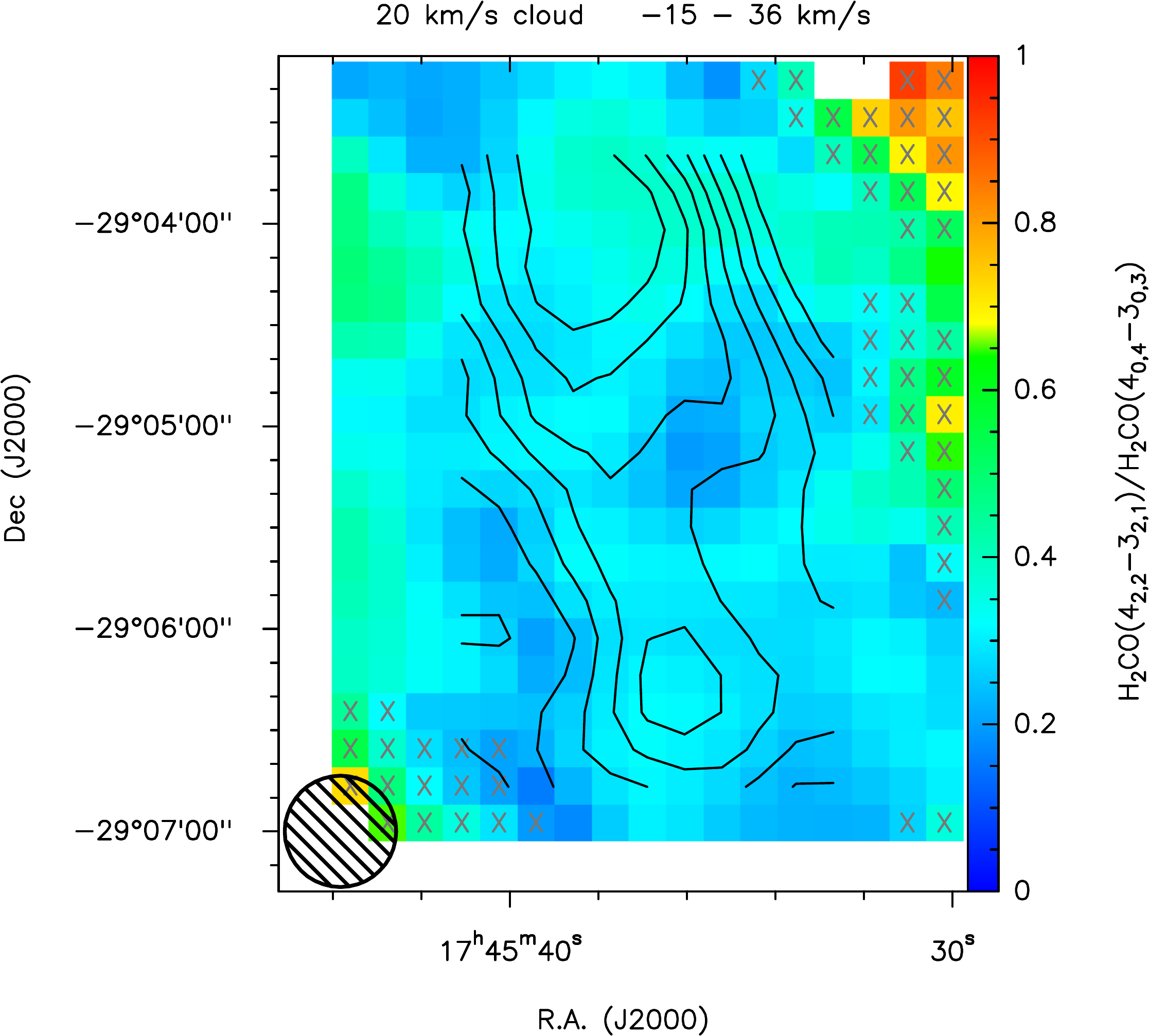}}
	\subfloat{\includegraphics[bb = 140 60 600 580, clip, height=4.5cm]{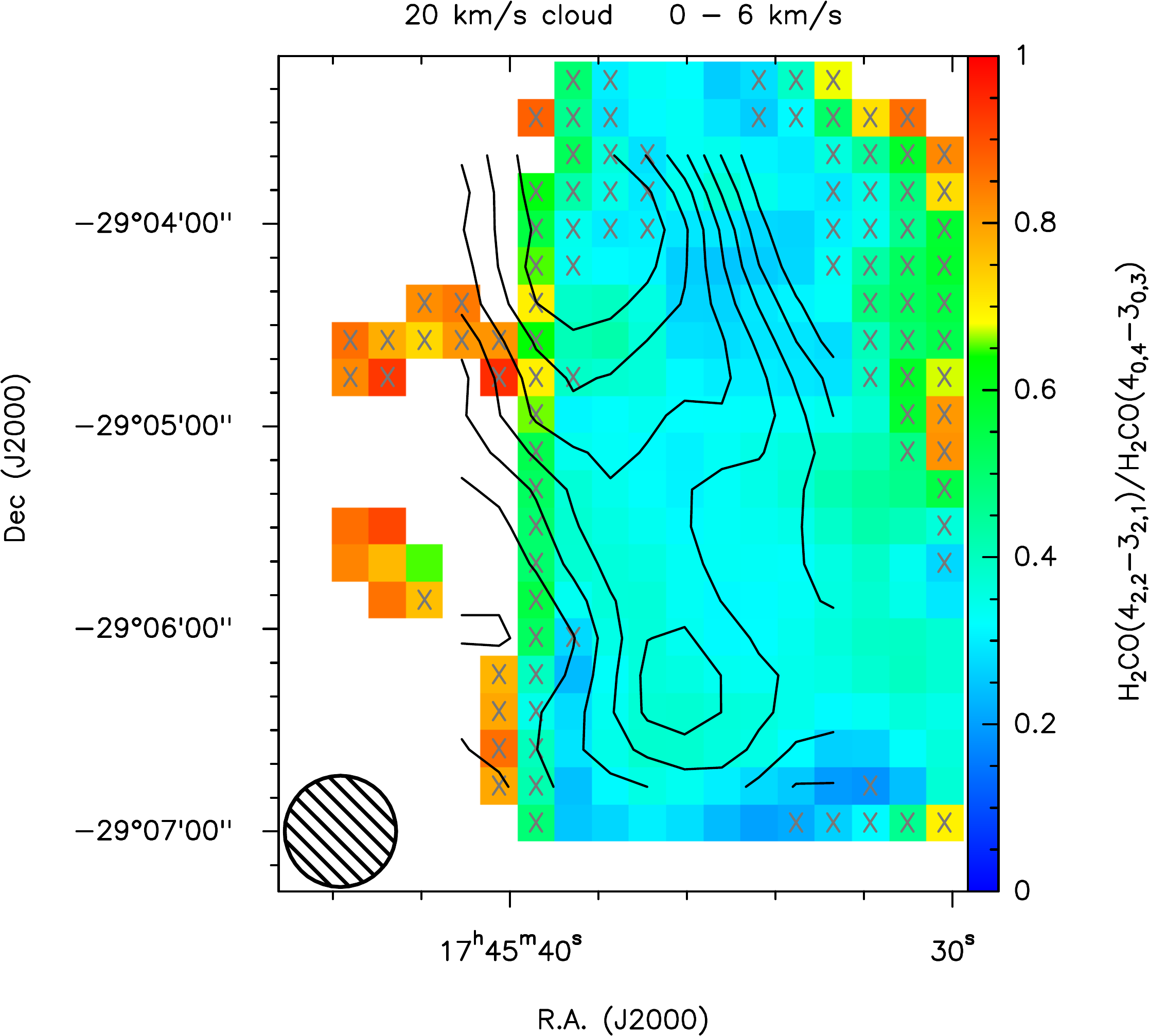}}
	\subfloat{\includegraphics[bb = 140 60 600 580, clip, height=4.5cm]{20kms-H2CO-8-14-Ratio_422_404.pdf}}
	\subfloat{\includegraphics[bb = 140 60 650 580, clip, height=4.5cm]{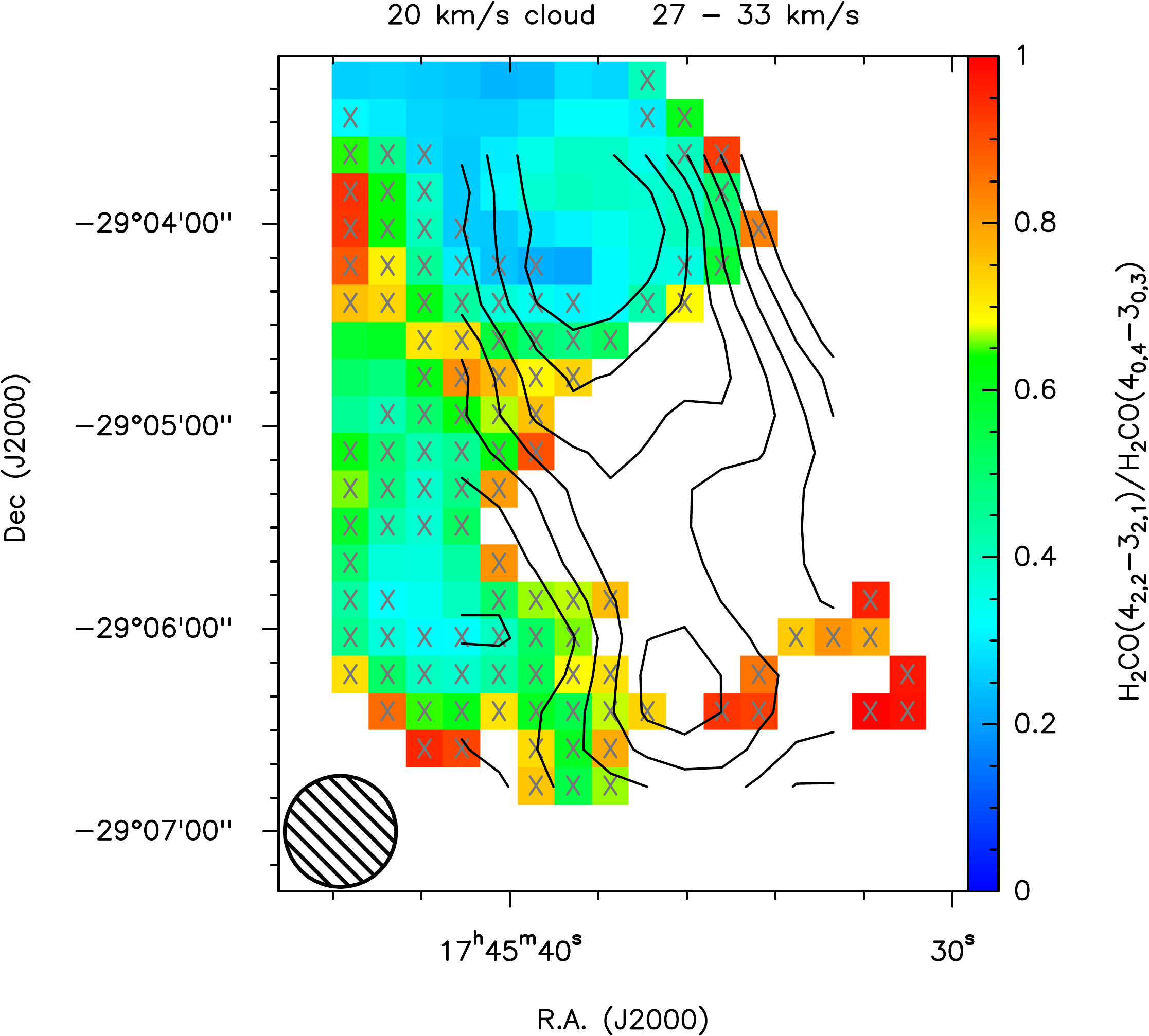}}\\\vspace{-0.5cm}
	\subfloat{\includegraphics[bb = 0 0 600 560, clip, height=4.845cm]{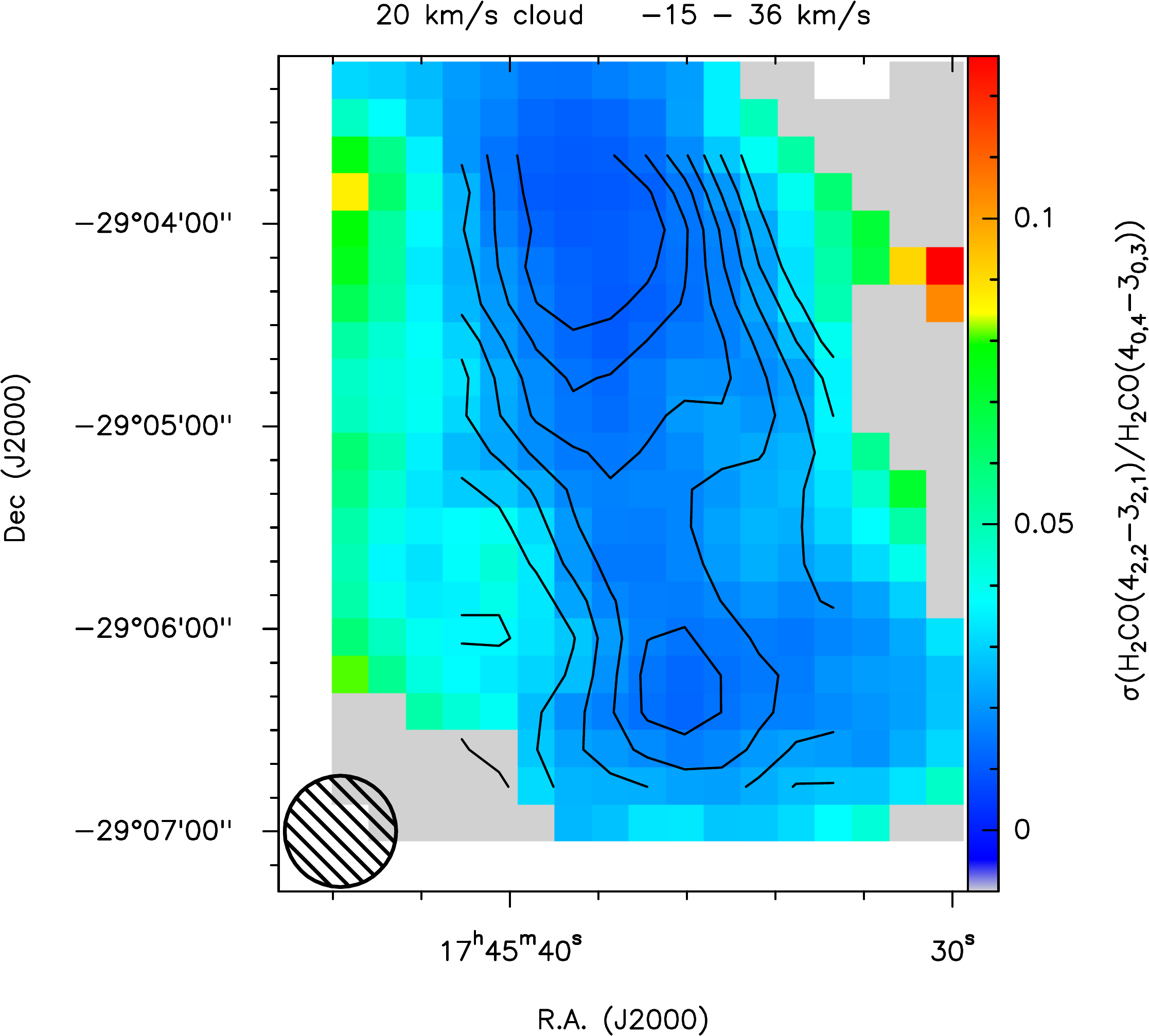}}
	\subfloat{\includegraphics[bb = 140 0 600 560, clip, height=4.845cm]{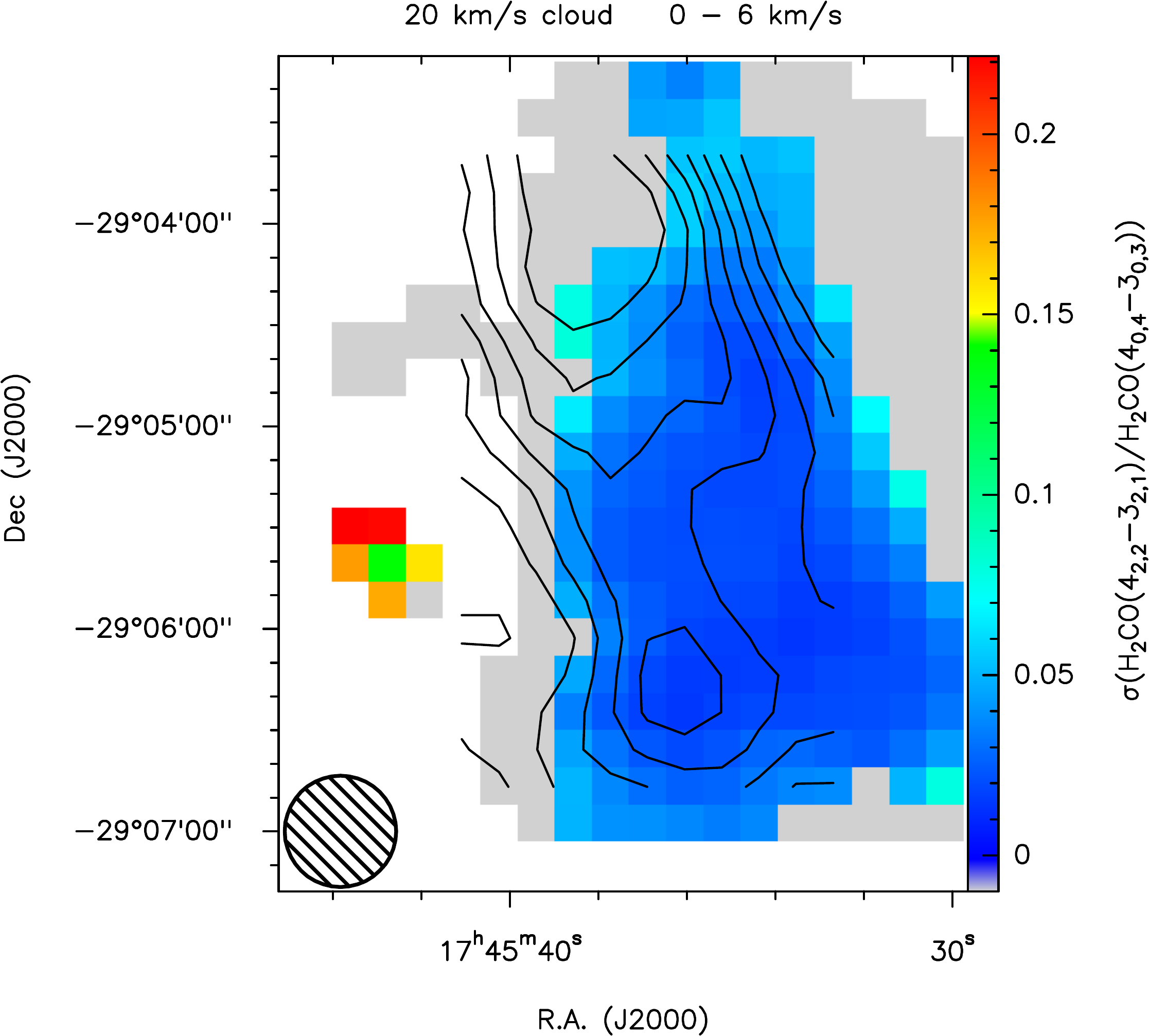}}
	\subfloat{\includegraphics[bb = 140 0 600 560, clip, height=4.845cm]{20kms-H2CO-8-14-Uncertainty-Ratio_422_404.pdf}}
	\subfloat{\includegraphics[bb = 140 0 650 560, clip, height=4.845cm]{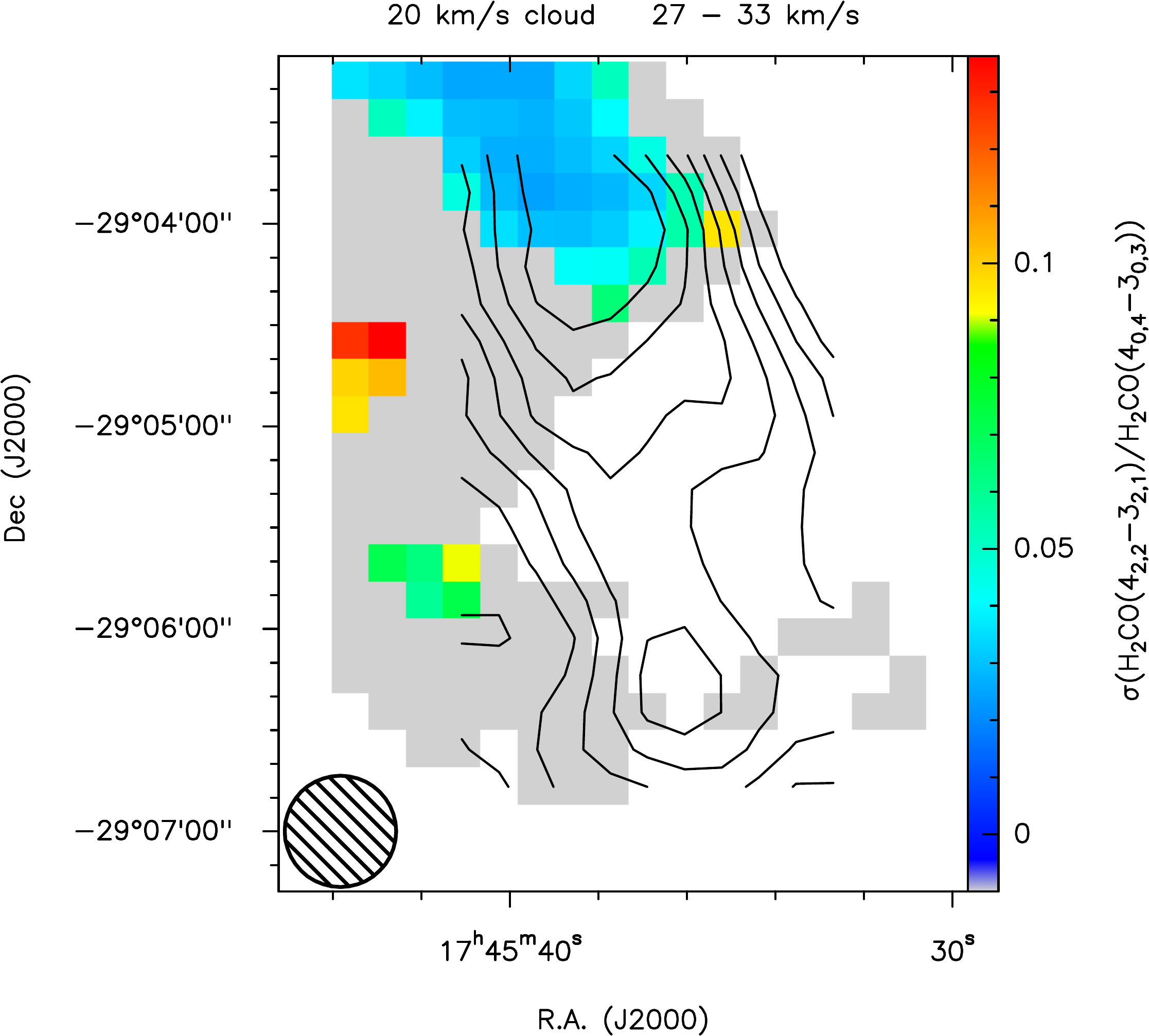}}\\
	\label{20kms-All-Ratio-H2CO}
\end{figure*}

\begin{figure*}
	\centering
	R$_{404}$ \\
	\subfloat{\includegraphics[bb = 0 60 600 580, clip, height=4.5cm]{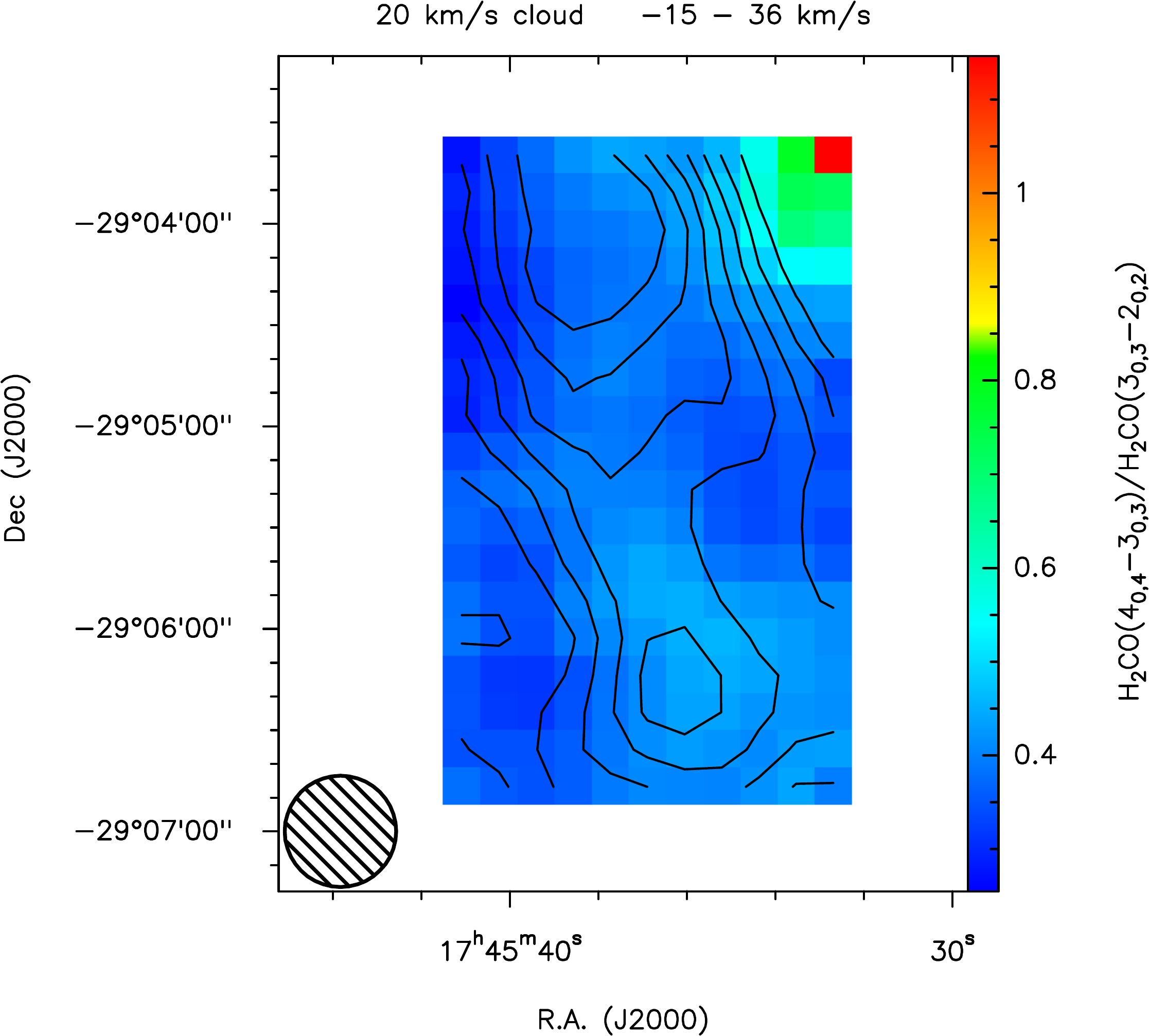}}
	\subfloat{\includegraphics[bb = 140 60 600 580, clip, height=4.5cm]{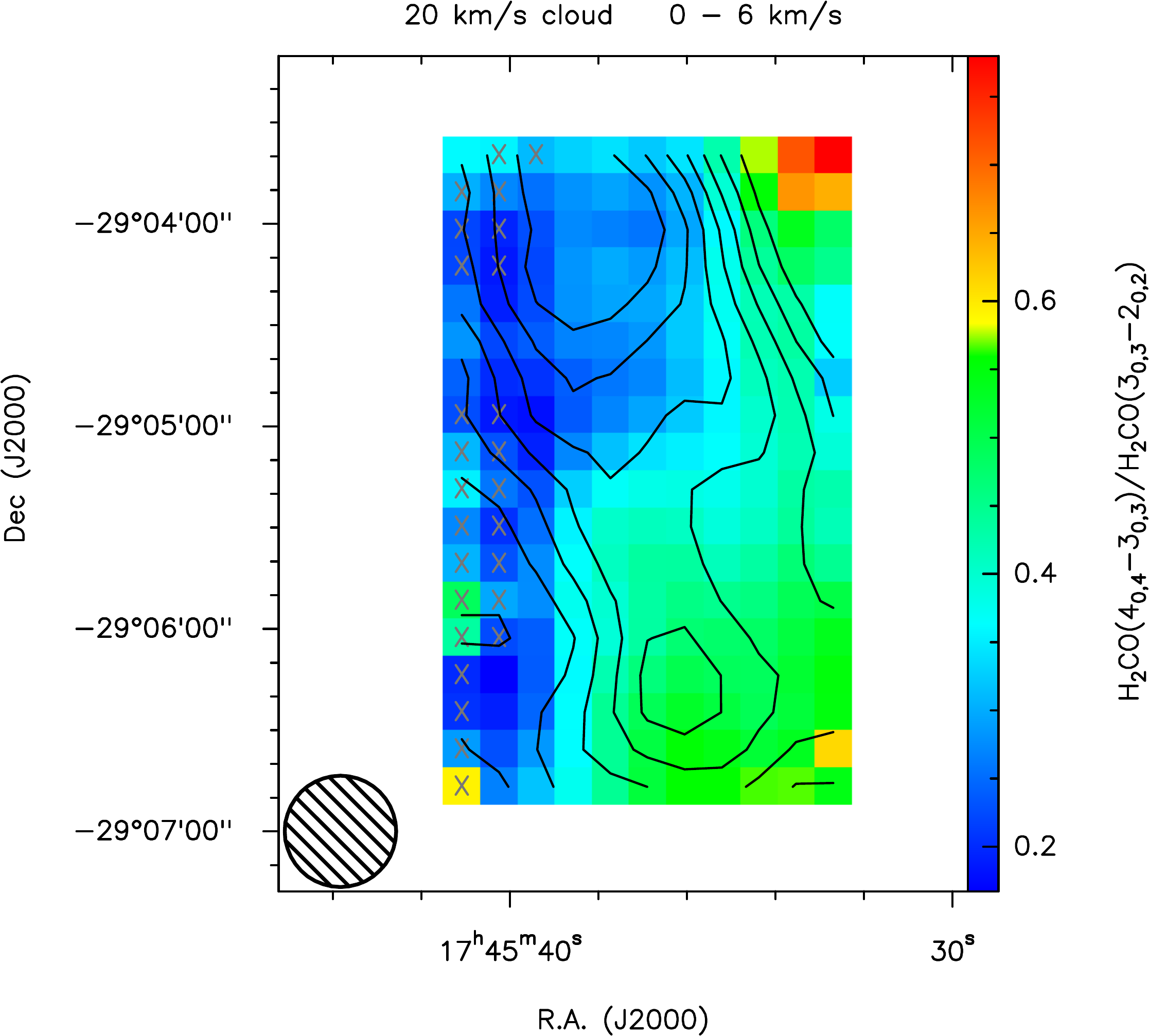}}
	\subfloat{\includegraphics[bb = 140 60 600 580, clip, height=4.5cm]{20kms-H2CO-8-14-Ratio_404_303.pdf}}
	\subfloat{\includegraphics[bb = 140 60 650 580, clip, height=4.5cm]{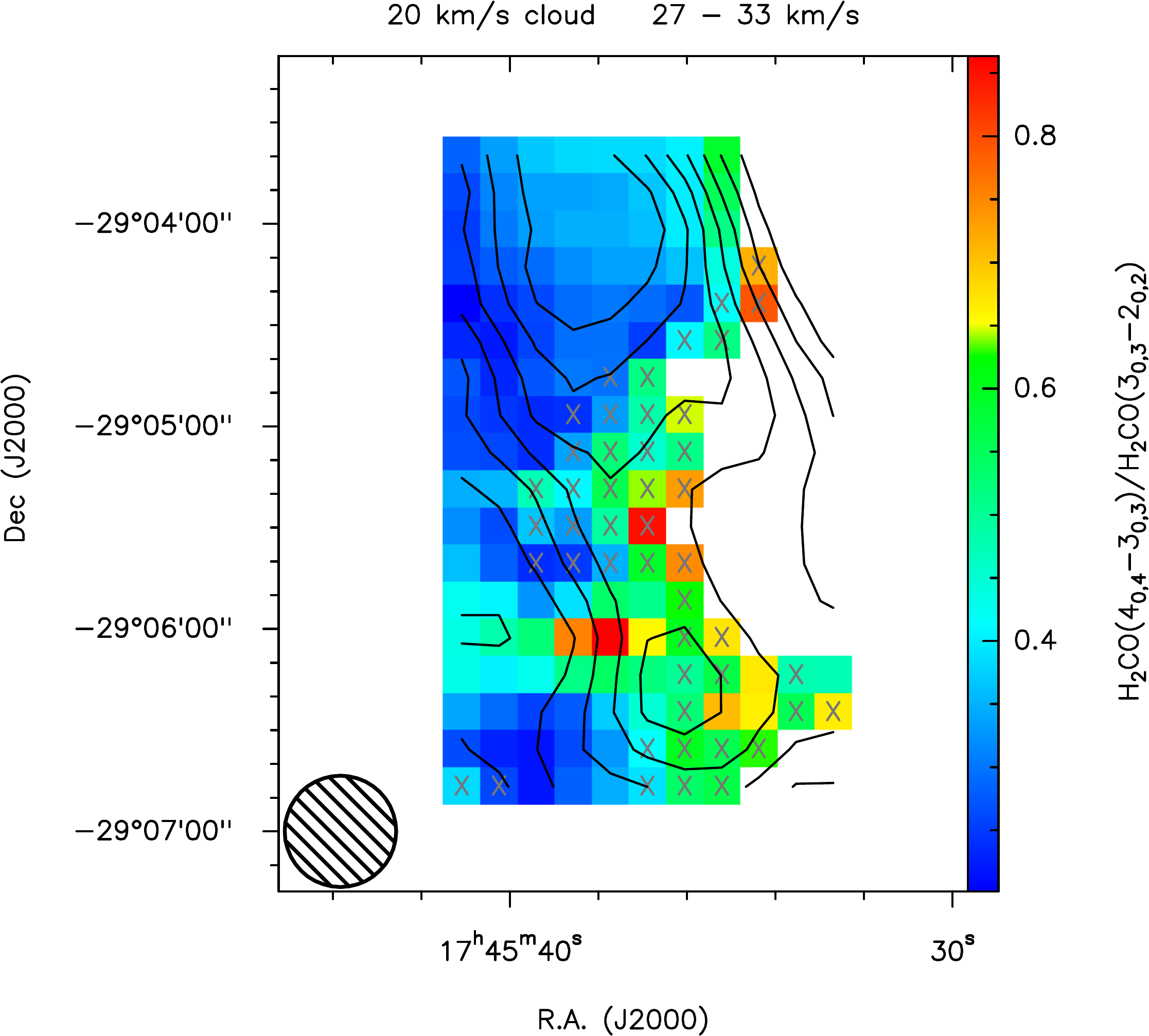}}\\\vspace{-0.5cm}
	\subfloat{\includegraphics[bb = 0 0 600 560, clip, height=4.845cm]{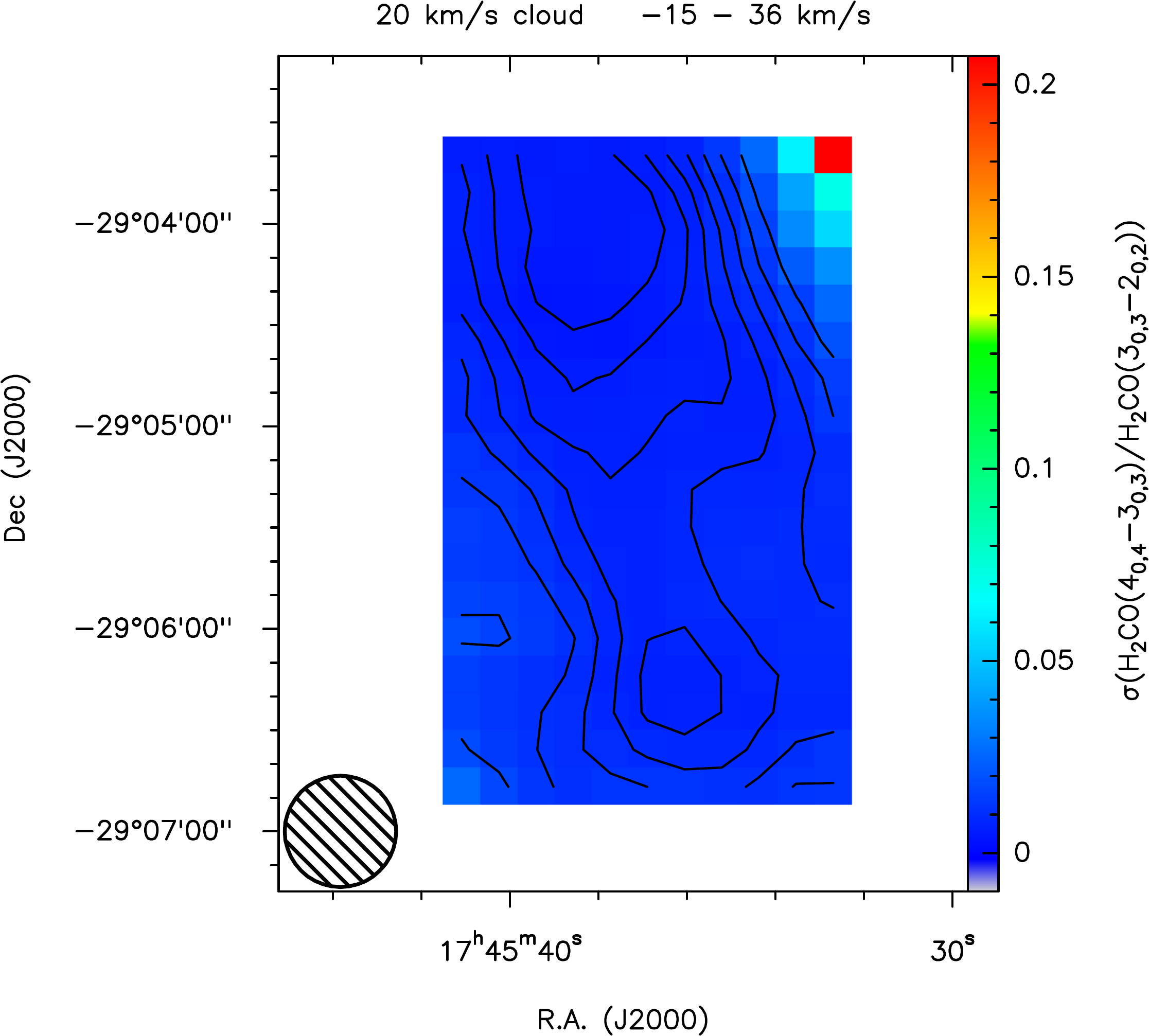}}
	\subfloat{\includegraphics[bb = 140 0 600 560, clip, height=4.845cm]{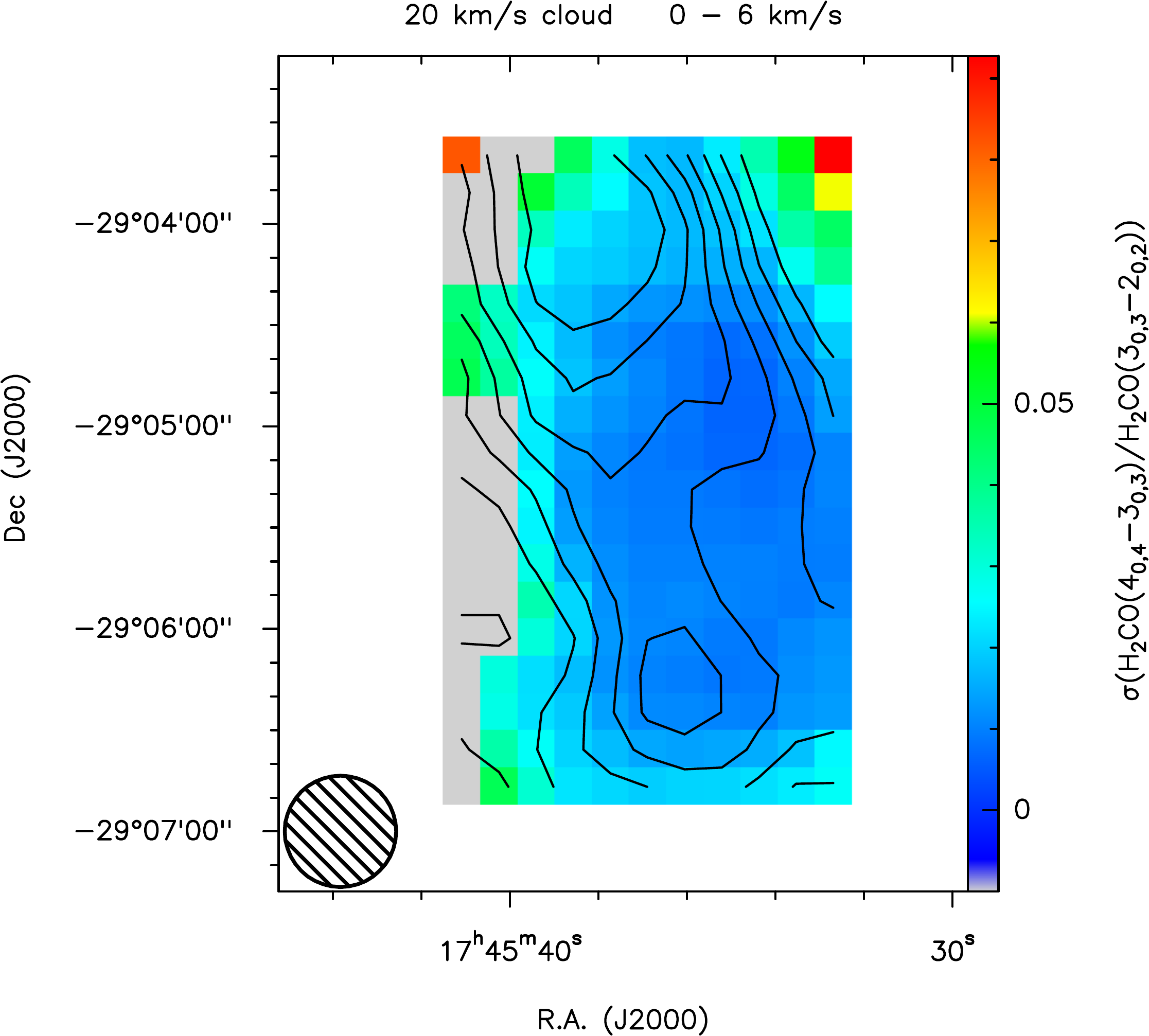}}
	\subfloat{\includegraphics[bb = 140 0 600 560, clip, height=4.845cm]{20kms-H2CO-8-14-Uncertainty-Ratio_404_303.pdf}}
	\subfloat{\includegraphics[bb = 140 0 650 560, clip, height=4.845cm]{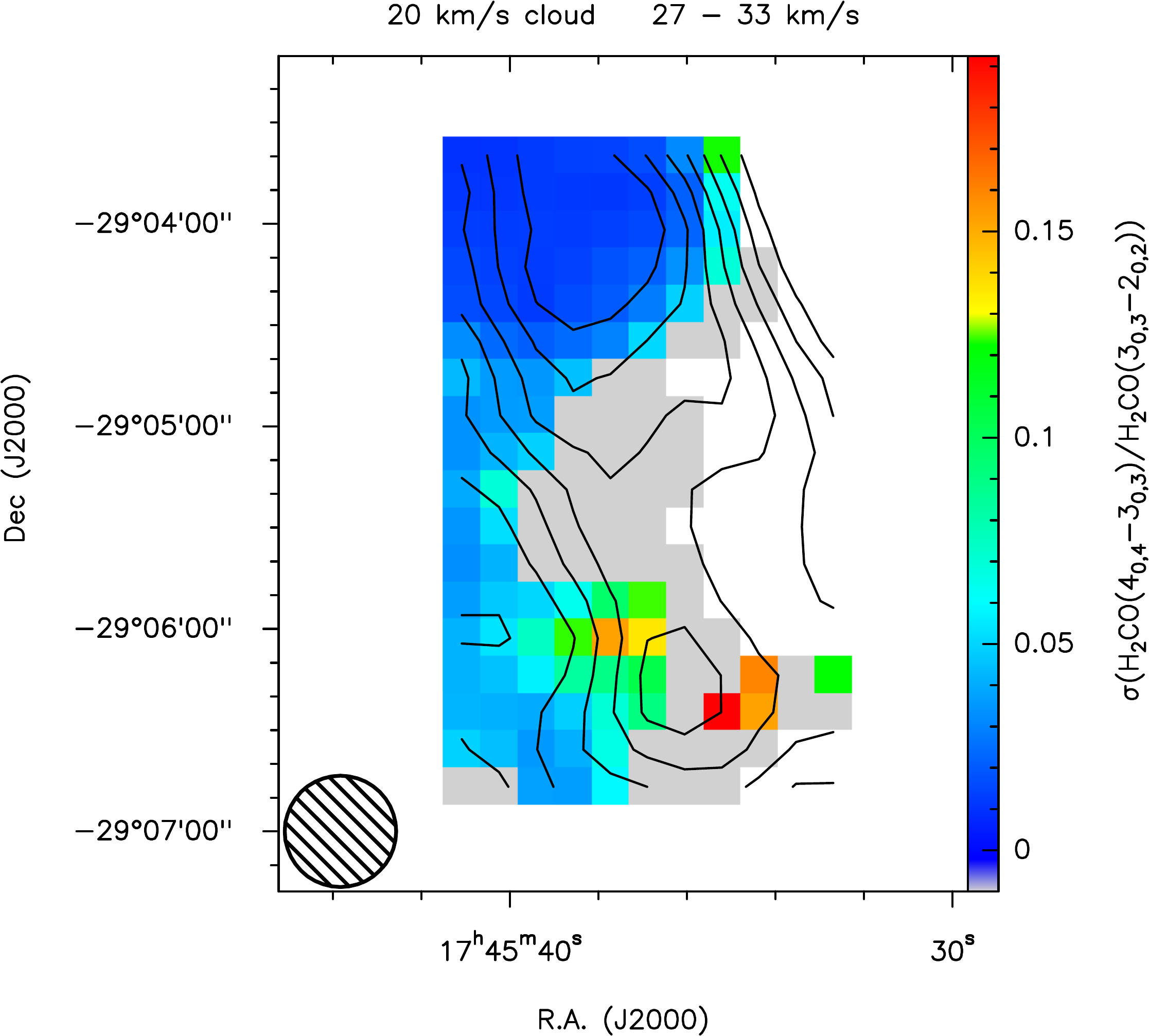}}
\end{figure*}

\begin{figure*}
	\caption{As Fig. \ref{20kms-All-Ratio-H2CO} for the 50 km/s cloud.}
	\centering
        R$_{321}$\\
	\subfloat{\includegraphics[bb = 0 60 700 580, clip, height=5cm]{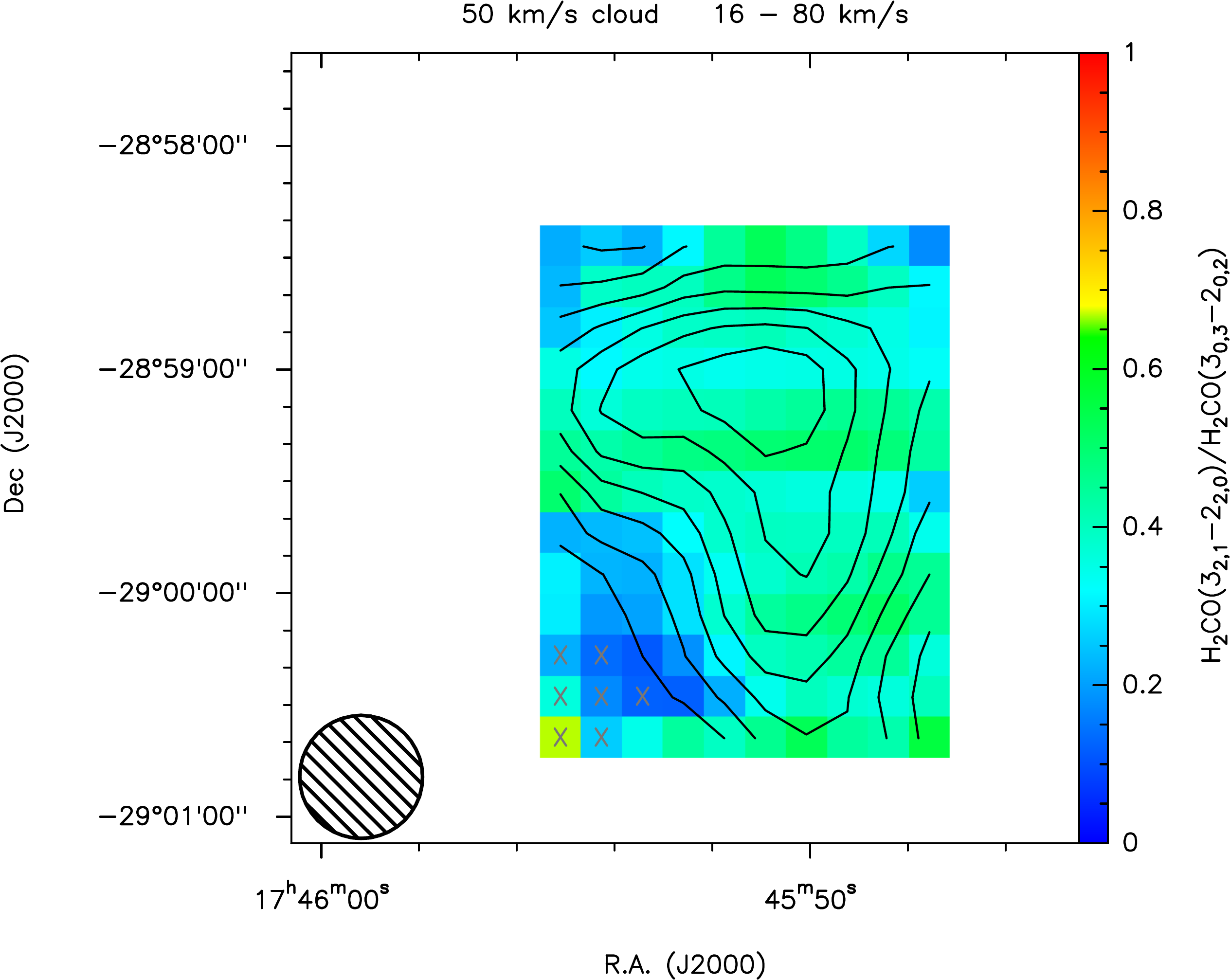}}
	\subfloat{\includegraphics[bb = 150 60 700 580, clip, height=5cm]{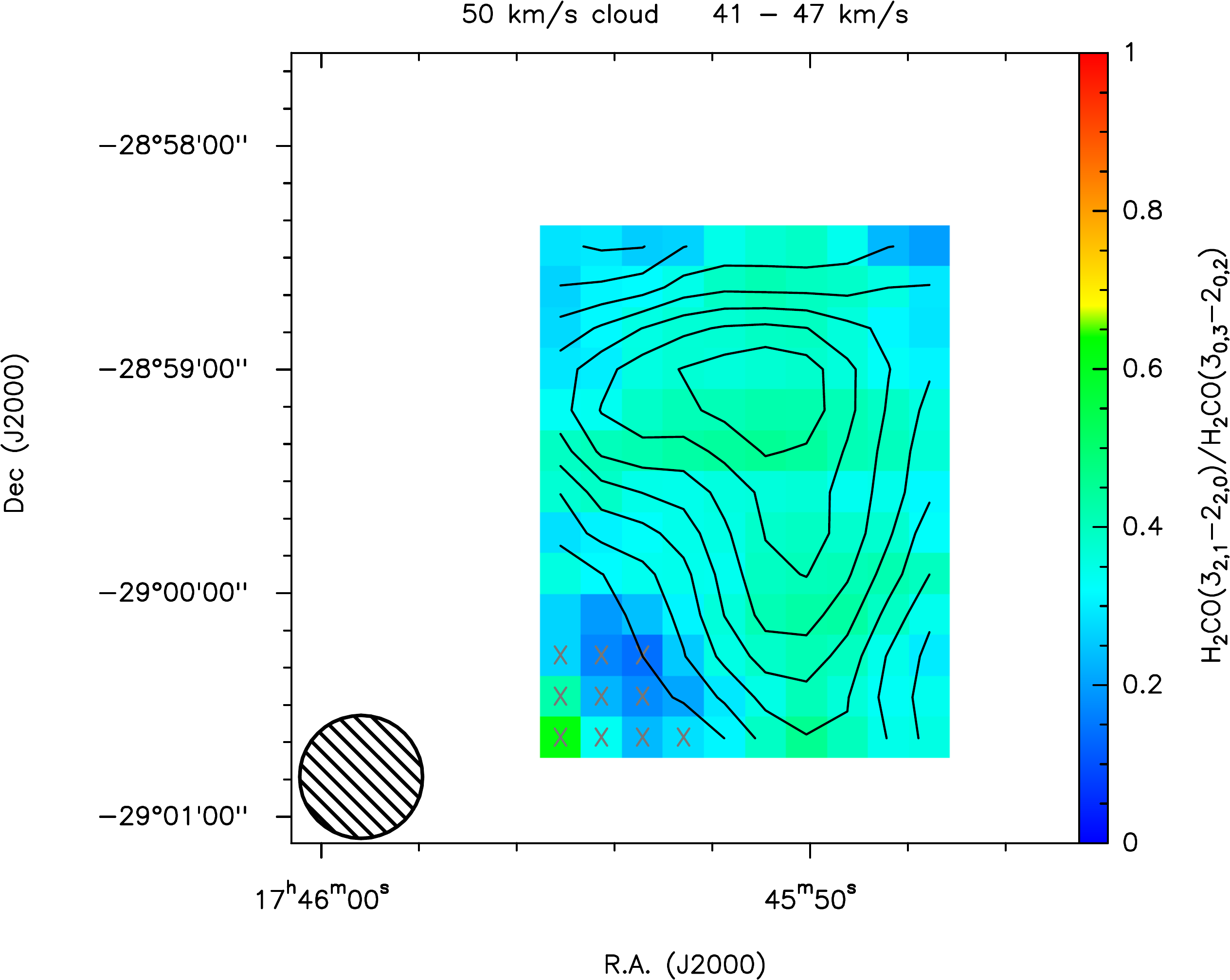}}
	\subfloat{\includegraphics[bb = 150 60 730 580, clip, height=5cm]{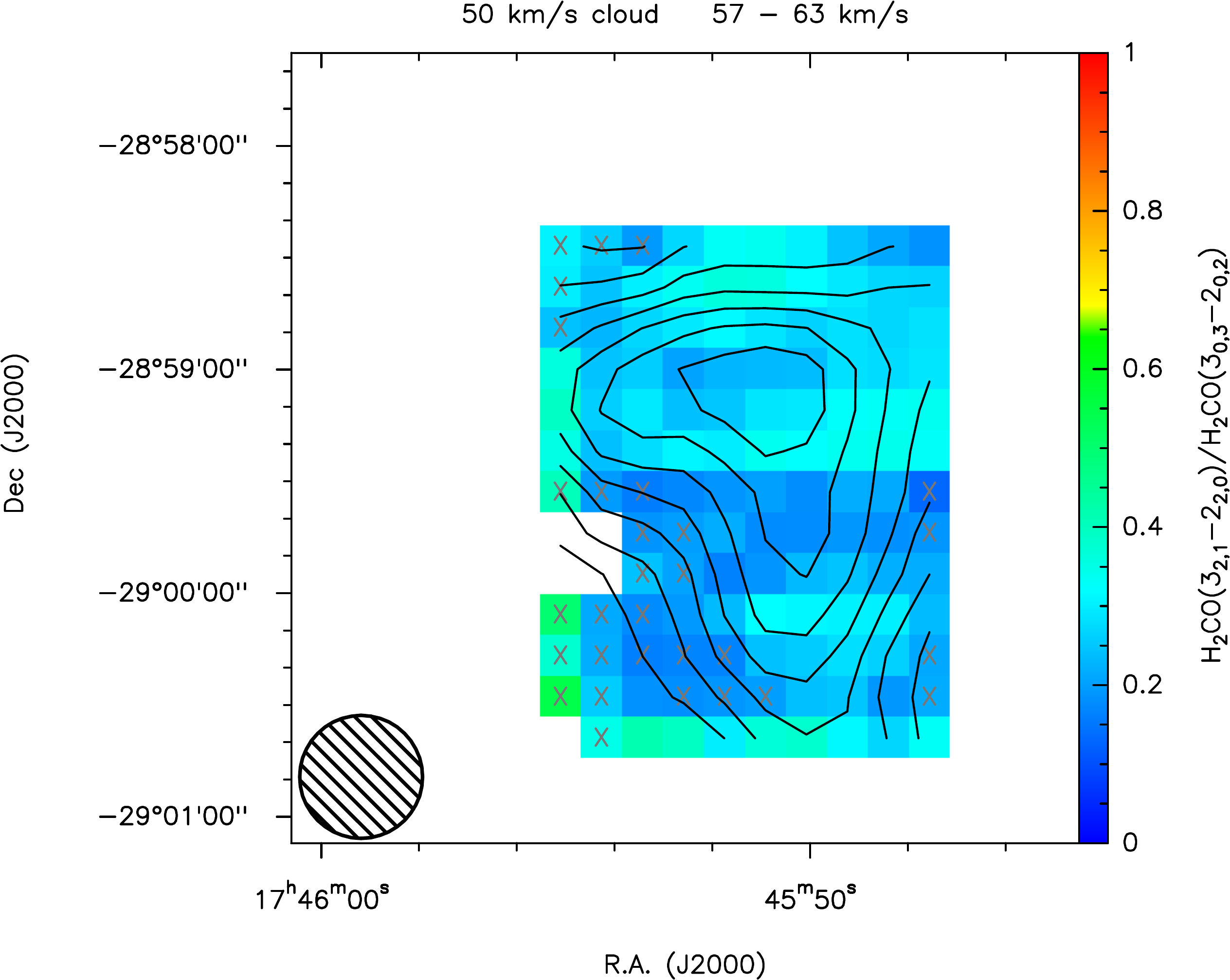}}\\\vspace{-0.5cm}
	\subfloat{\includegraphics[bb = 0 0 700 560, clip, height=5.385cm]{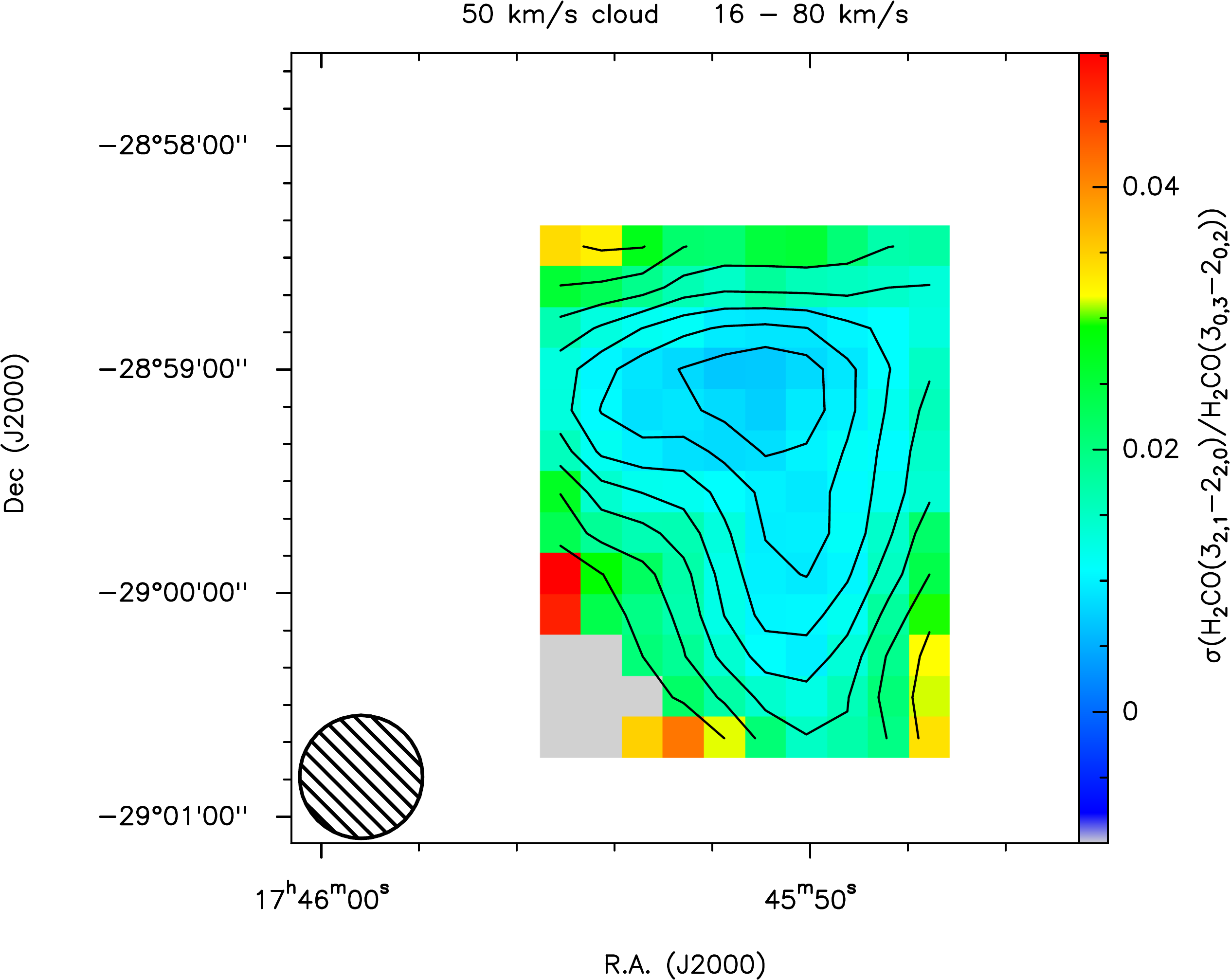}}
	\subfloat{\includegraphics[bb = 150 0 700 560, clip, height=5.385cm]{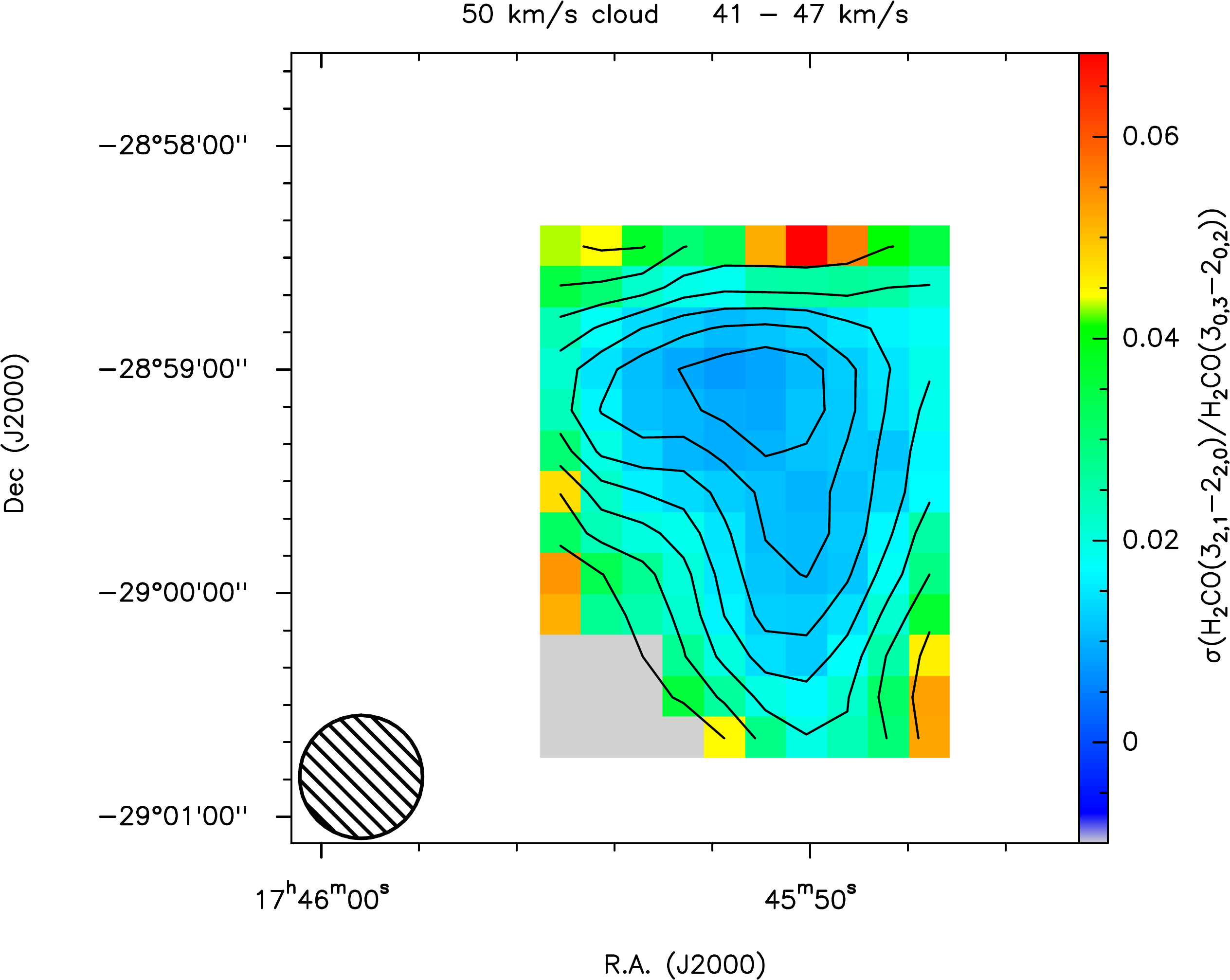}}
	\subfloat{\includegraphics[bb = 150 0 730 560, clip, height=5.385cm]{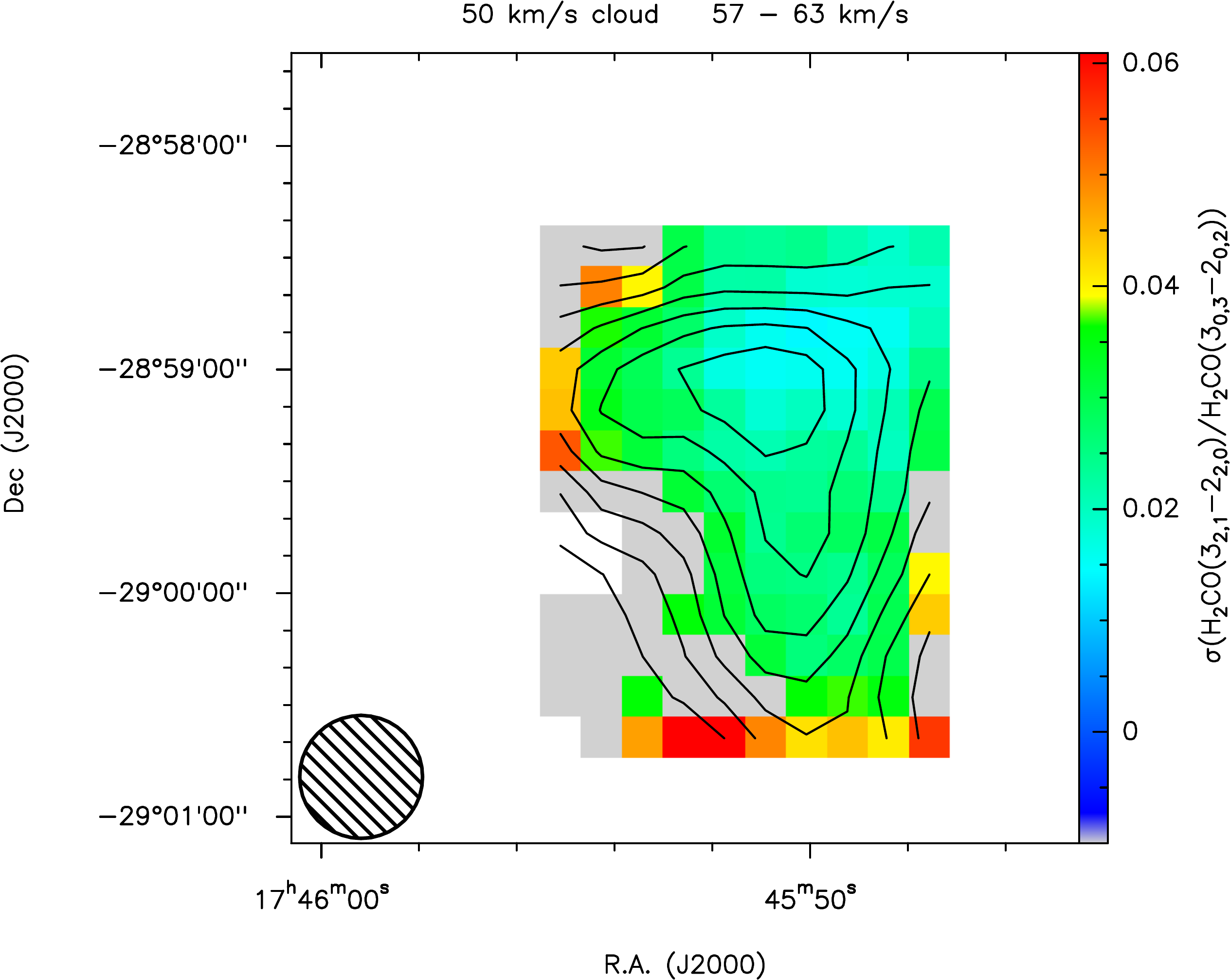}}\\
	\vspace{0.1cm}
        R$_{422}$\\
	\subfloat{\includegraphics[bb = 0 60 700 580, clip, height=5cm]{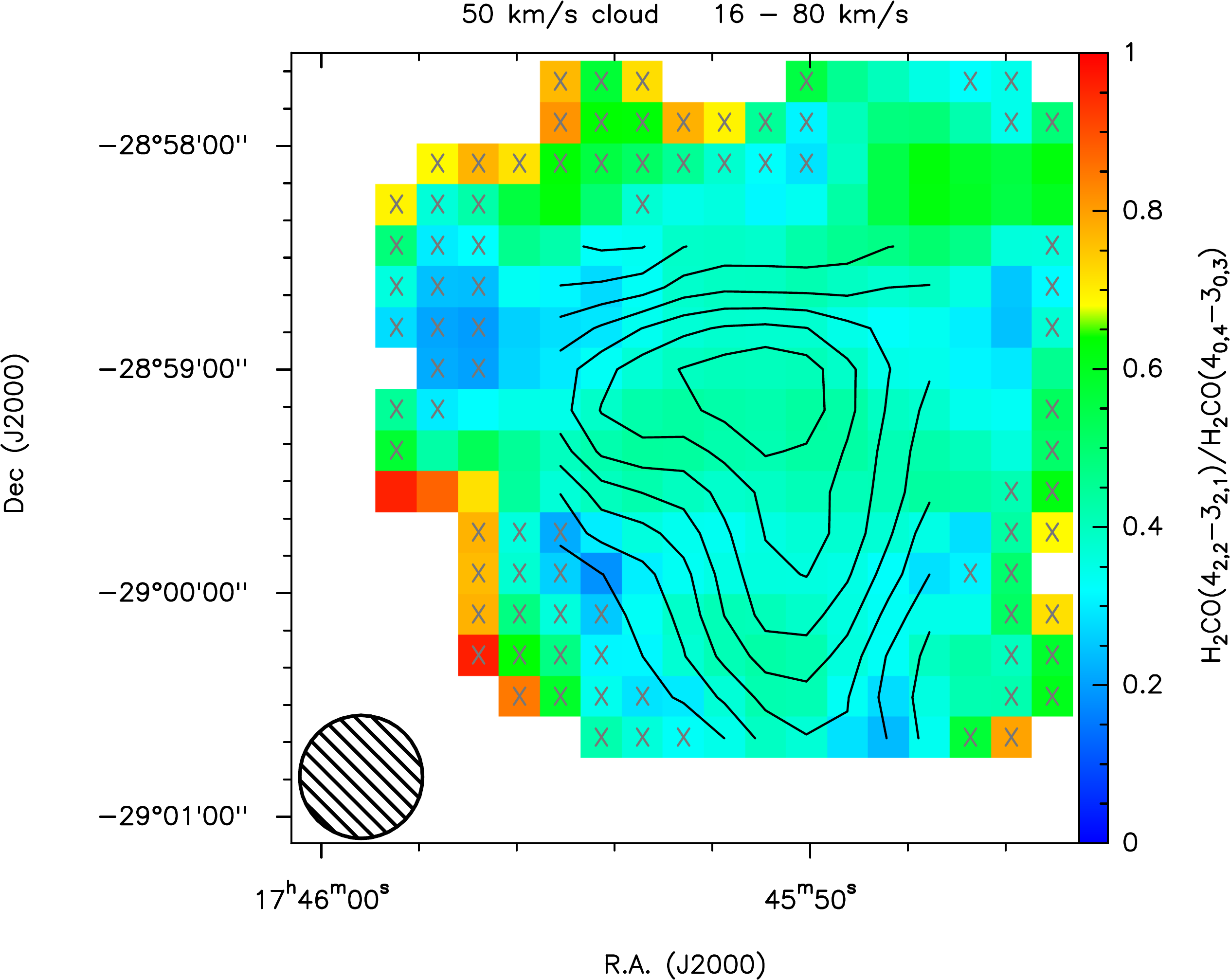}}
	\subfloat{\includegraphics[bb = 150 60 700 580, clip, height=5cm]{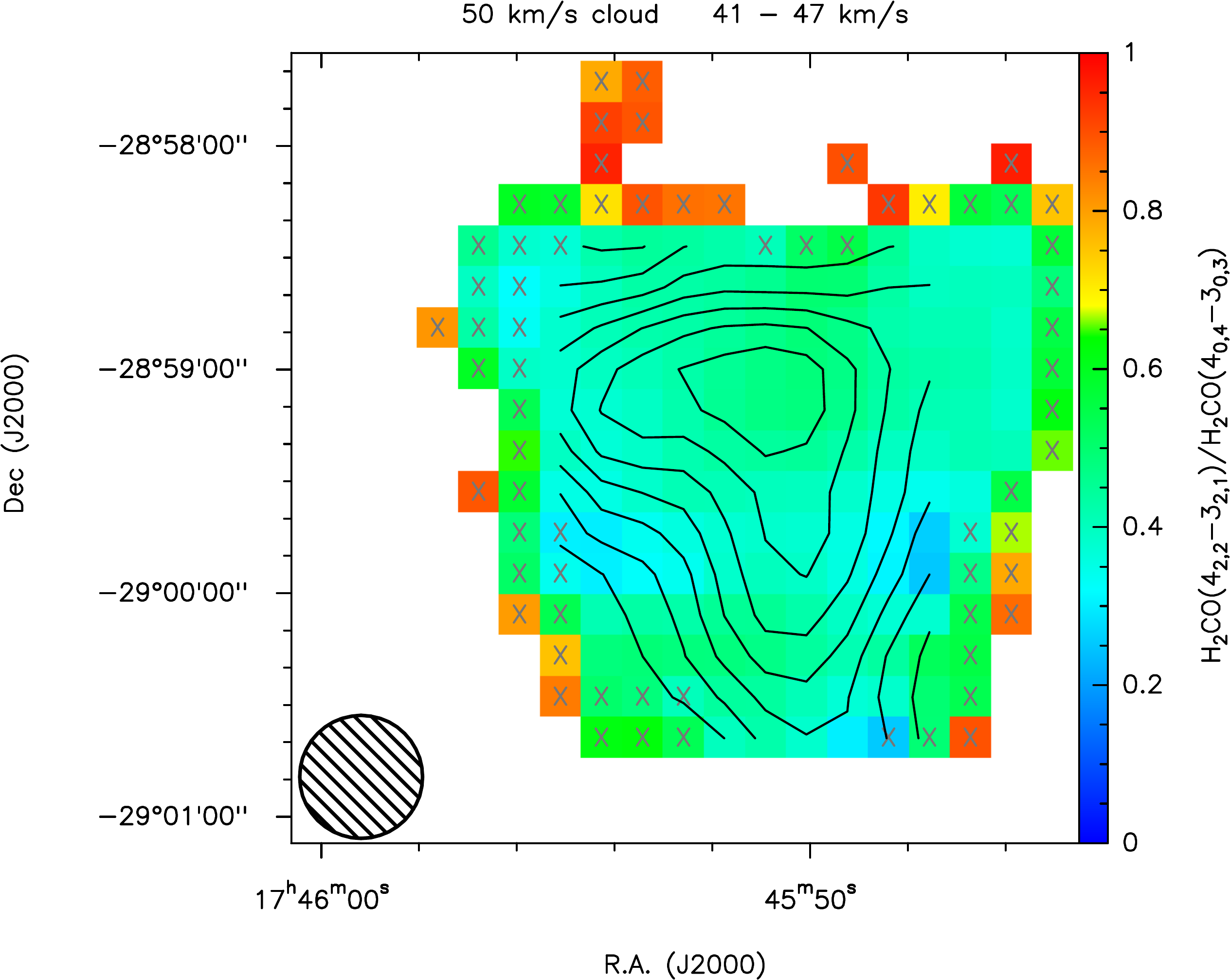}}
	\subfloat{\includegraphics[bb = 150 60 730 580, clip, height=5cm]{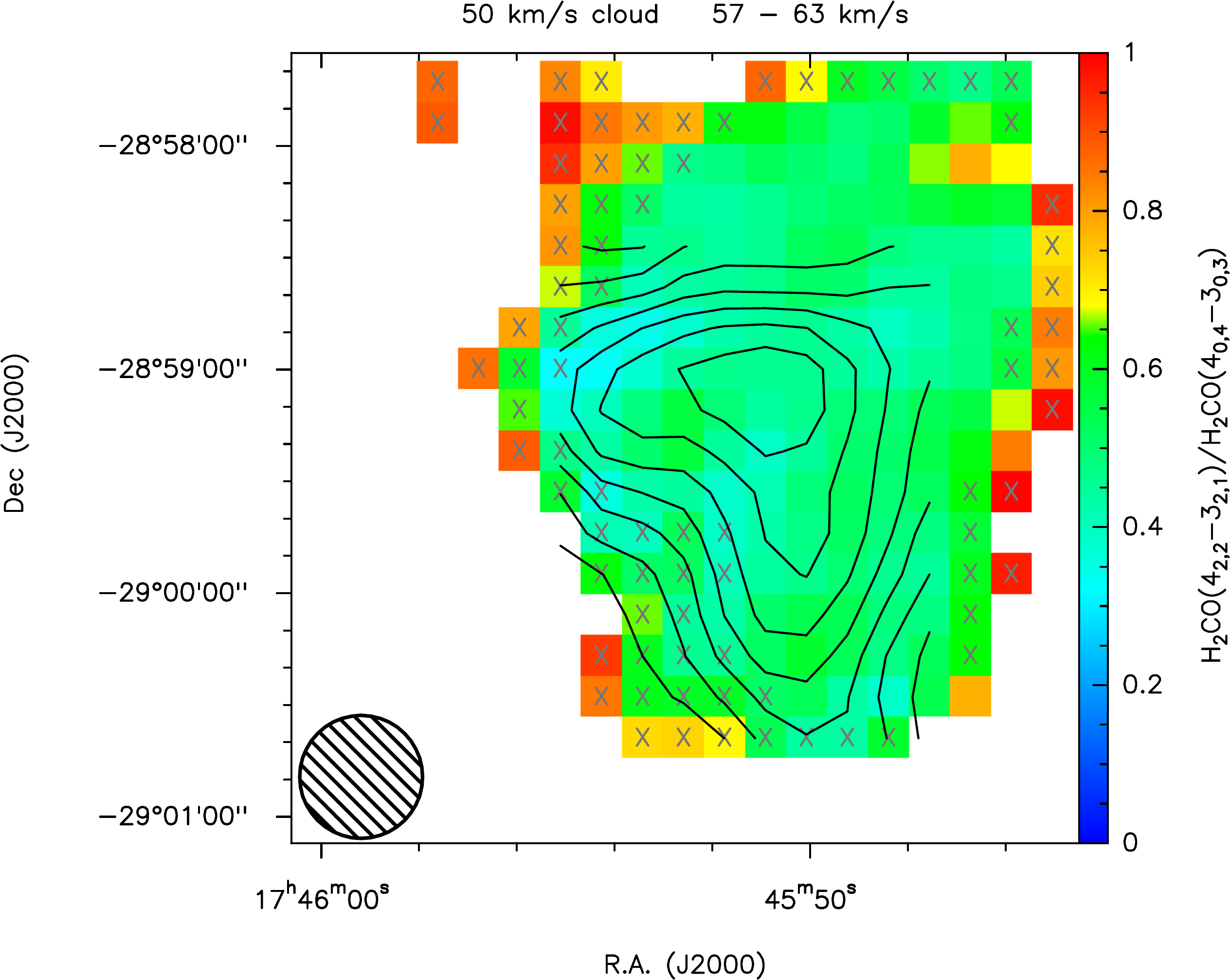}}\\\vspace{-0.5cm}
	\subfloat{\includegraphics[bb = 0 0 700 560, clip, height=5.385cm]{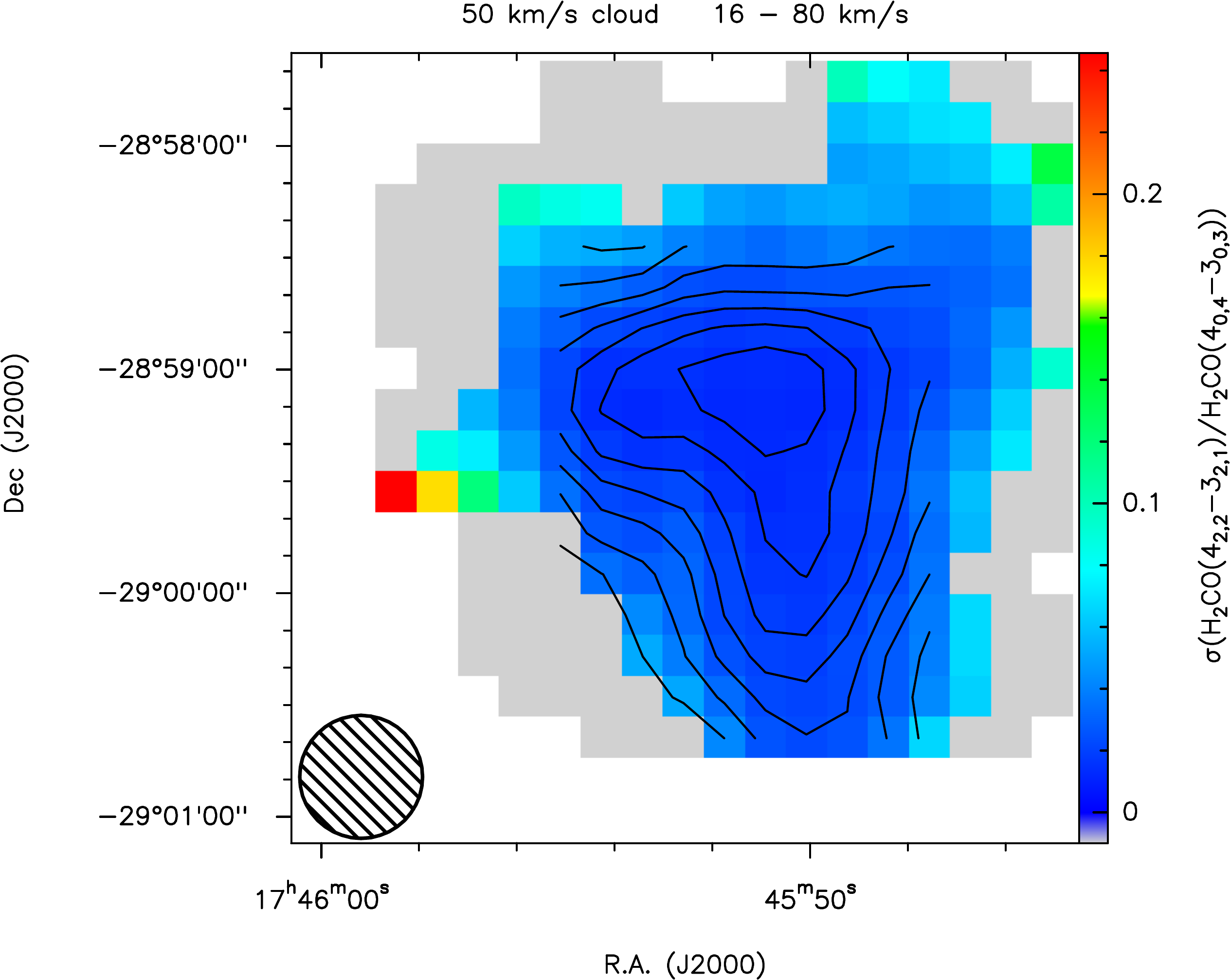}}
	\subfloat{\includegraphics[bb = 150 0 700 560, clip, height=5.385cm]{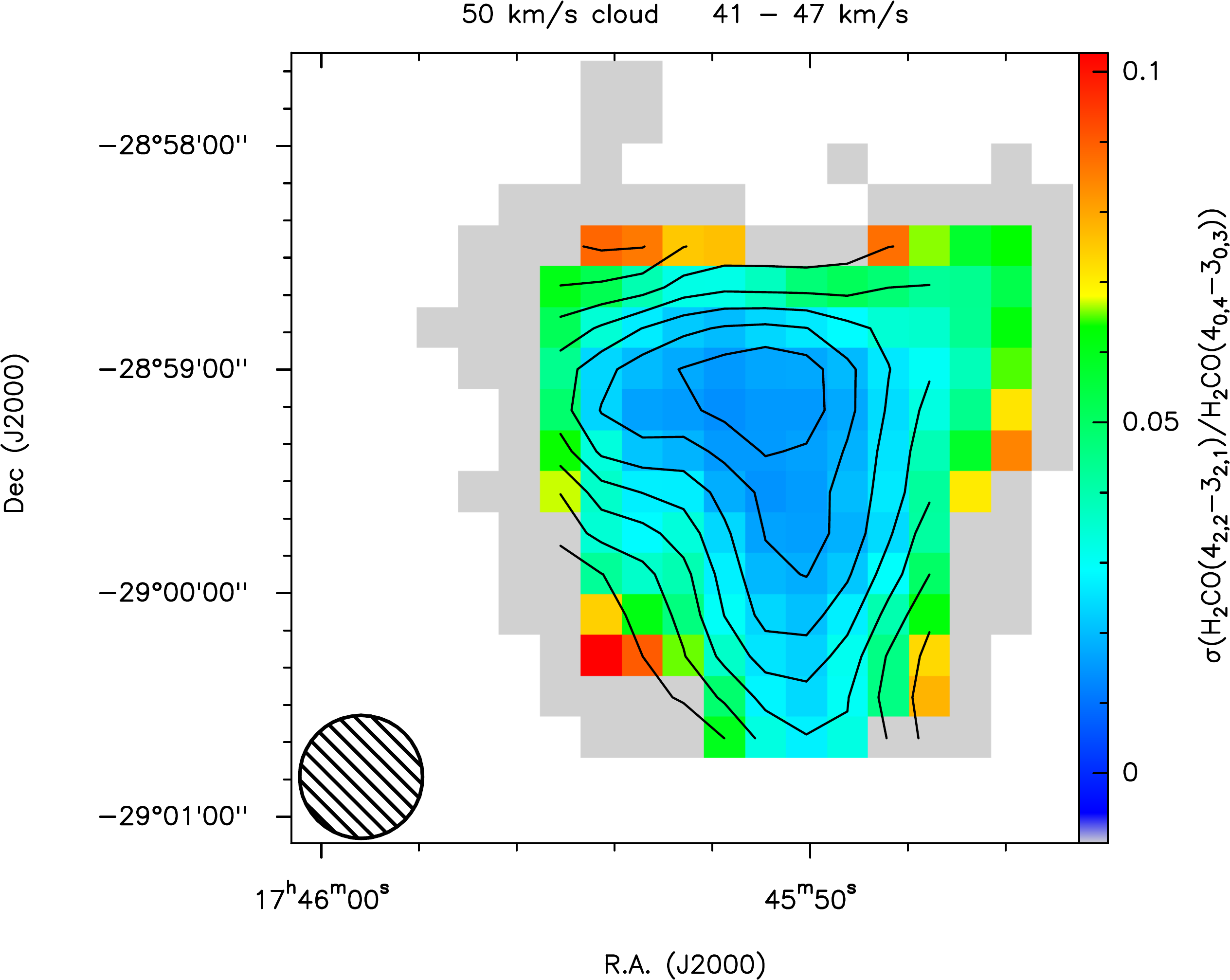}}
	\subfloat{\includegraphics[bb = 150 0 730 560, clip, height=5.385cm]{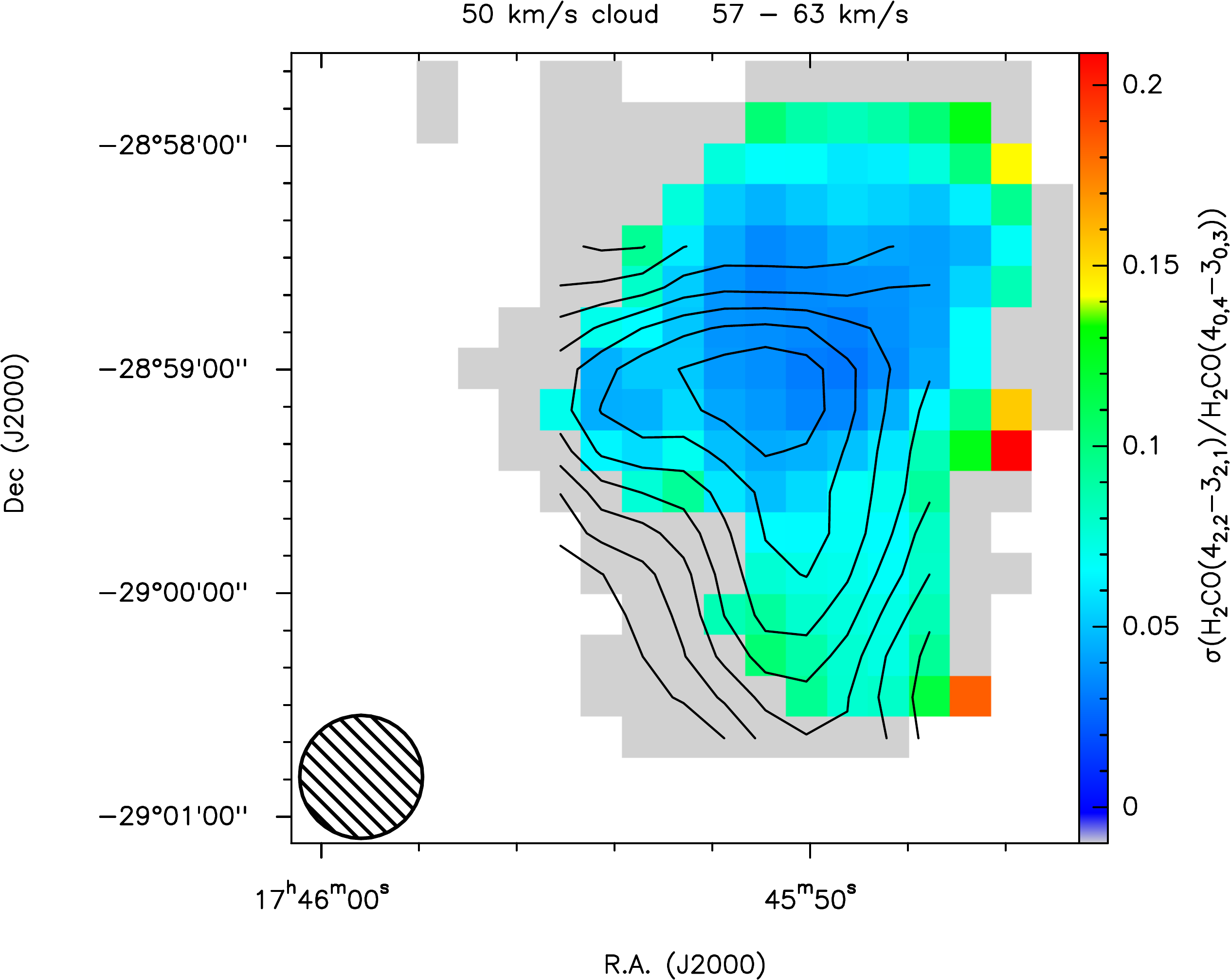}}\\
	\label{50kms-All-Ratio-H2CO}
\end{figure*}

\begin{figure*}
	\centering
	R$_{404}$ \\
	\subfloat{\includegraphics[bb = 0 60 700 580, clip, height=5cm]{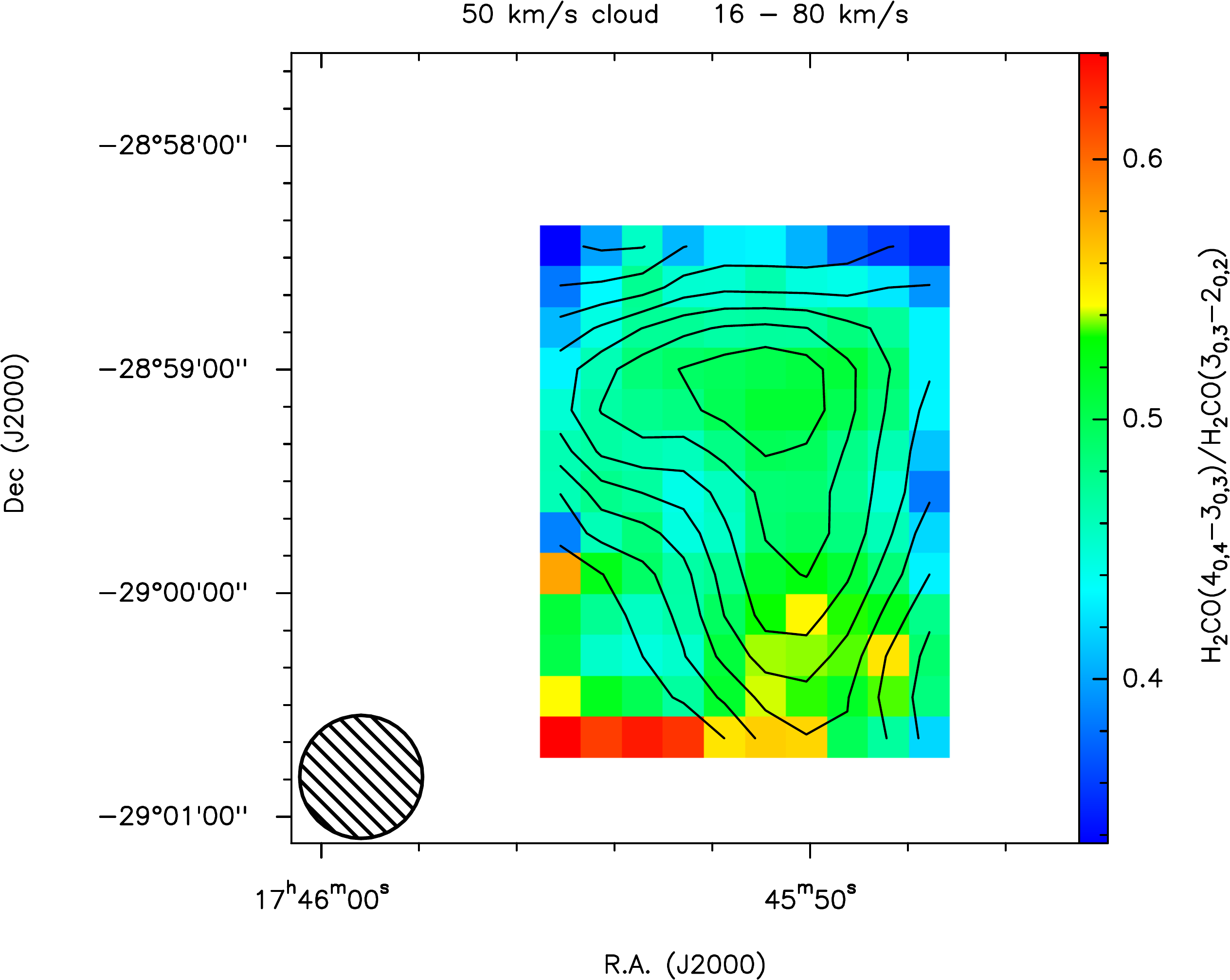}}
	\subfloat{\includegraphics[bb = 150 60 700 580, clip, height=5cm]{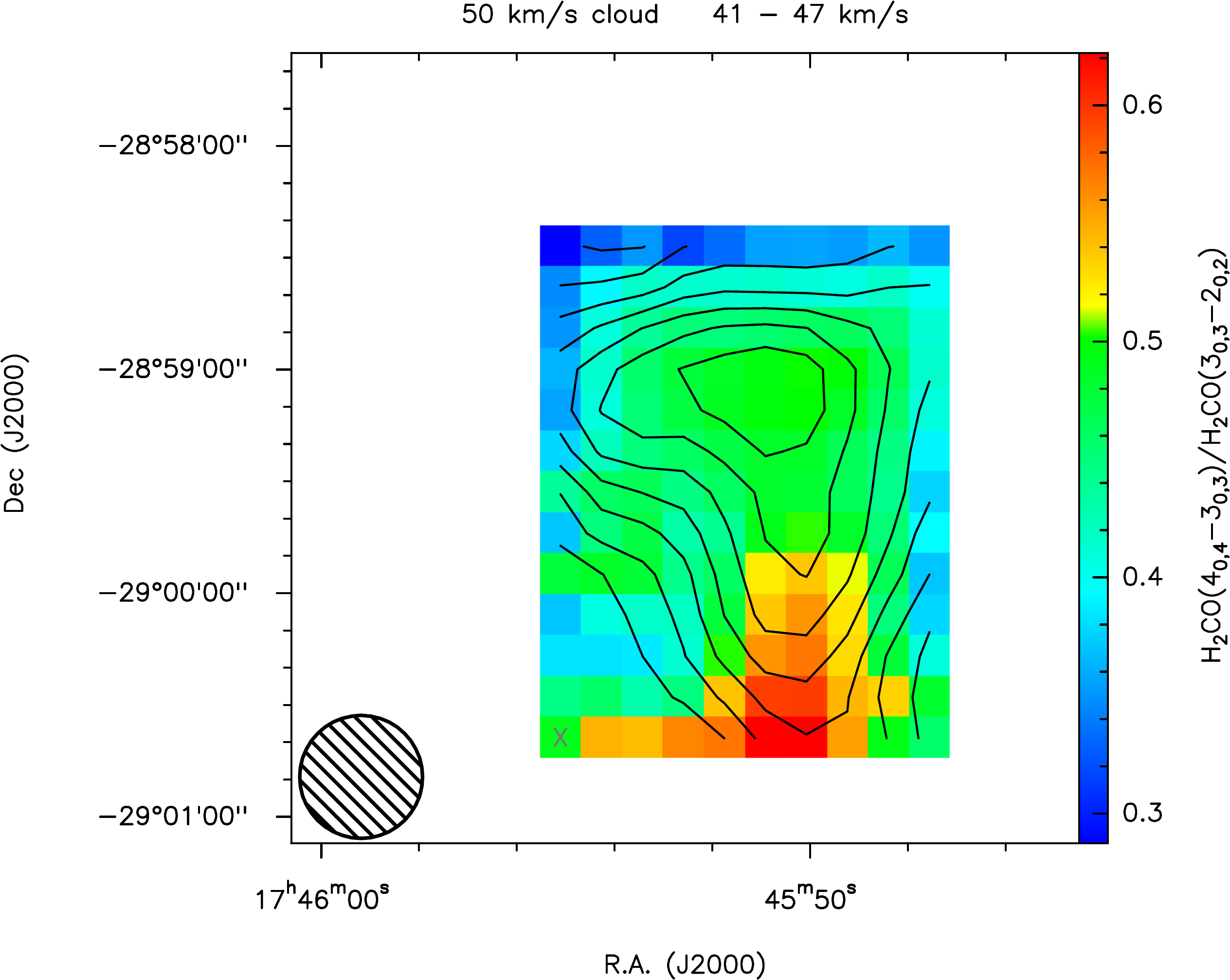}}
	\subfloat{\includegraphics[bb = 150 60 730 580, clip, height=5cm]{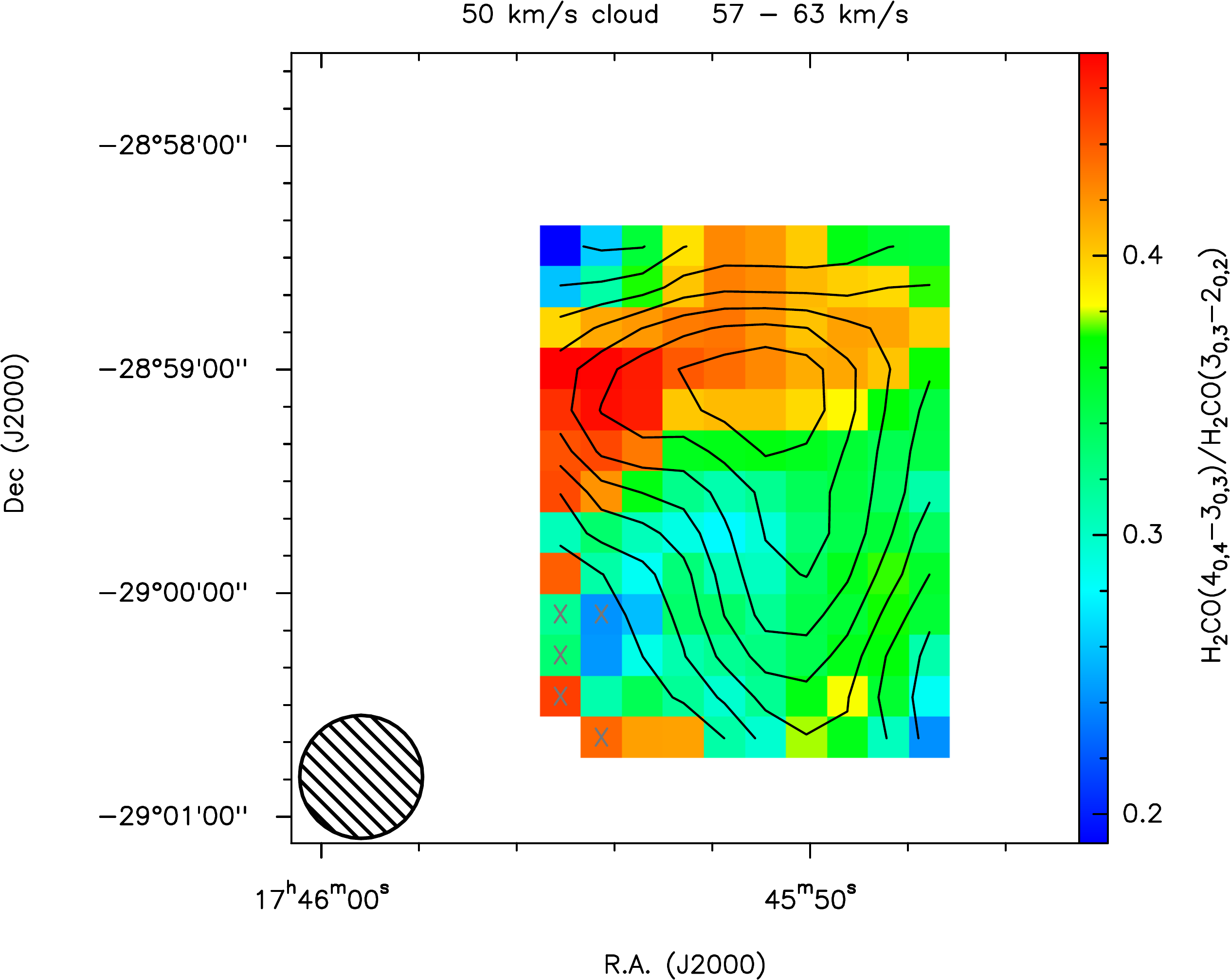}}\\\vspace{-0.5cm}
	\subfloat{\includegraphics[bb = 0 0 700 560, clip, height=5.385cm]{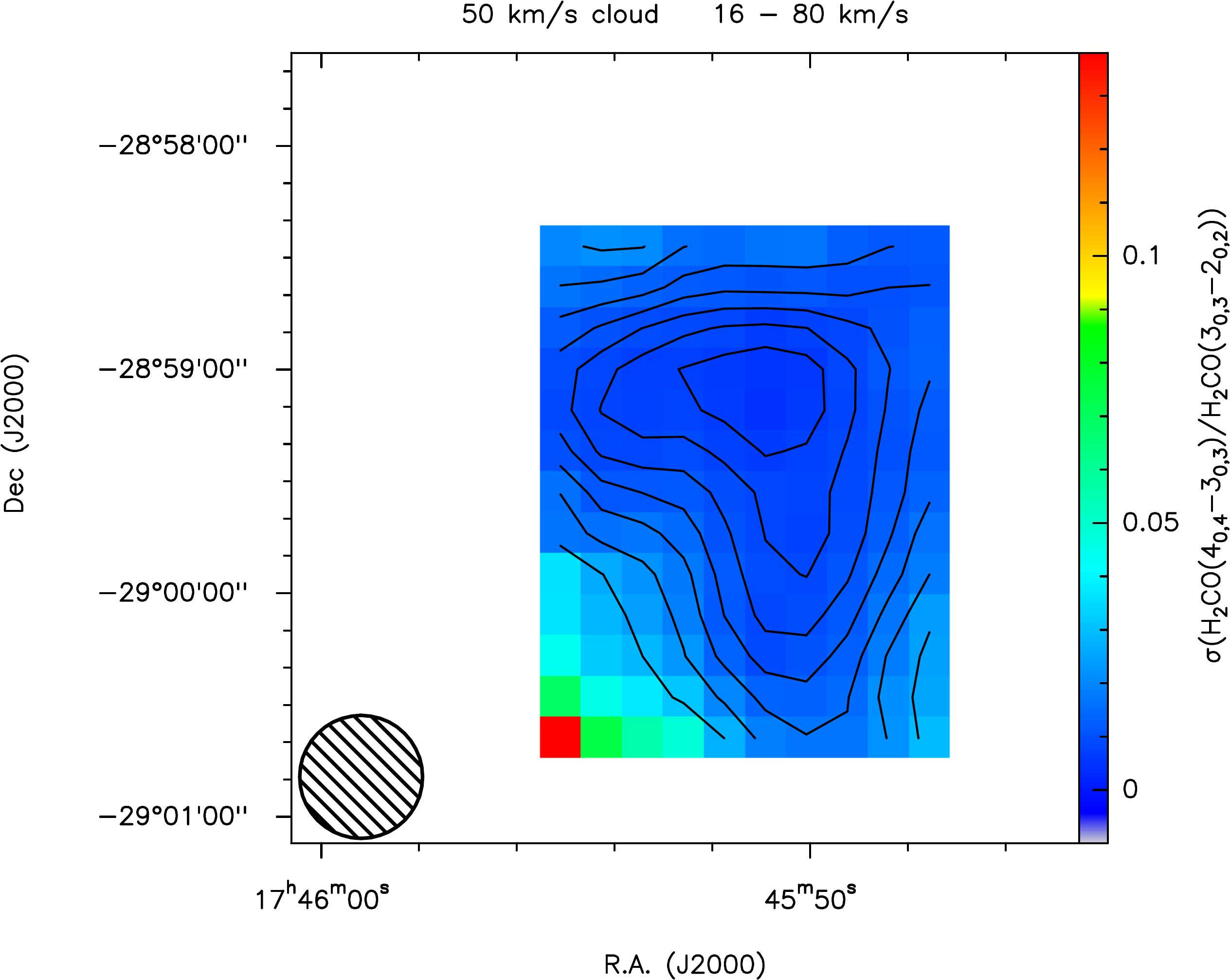}}
	\subfloat{\includegraphics[bb = 150 0 700 560, clip, height=5.385cm]{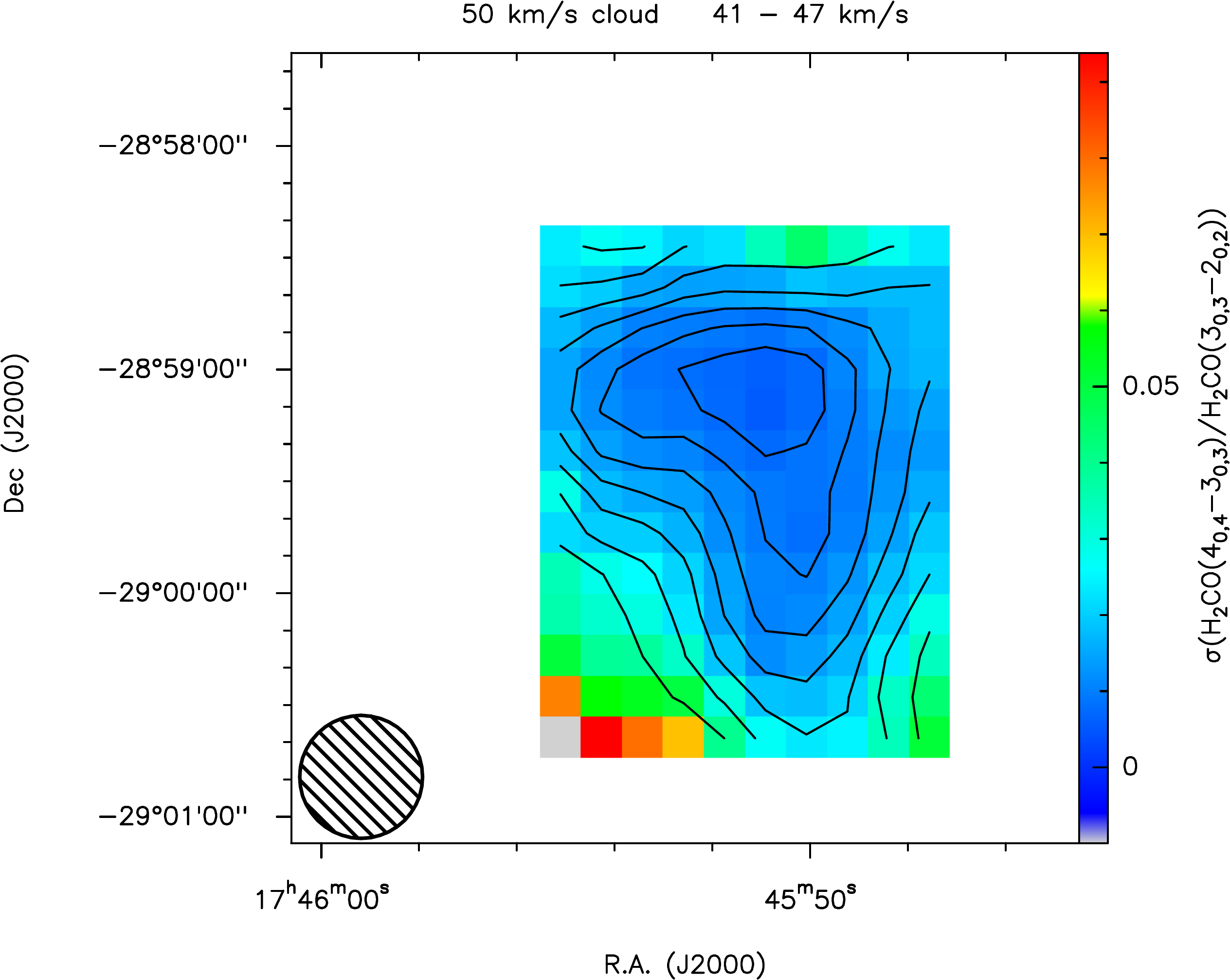}}
	\subfloat{\includegraphics[bb = 150 0 730 560, clip, height=5.385cm]{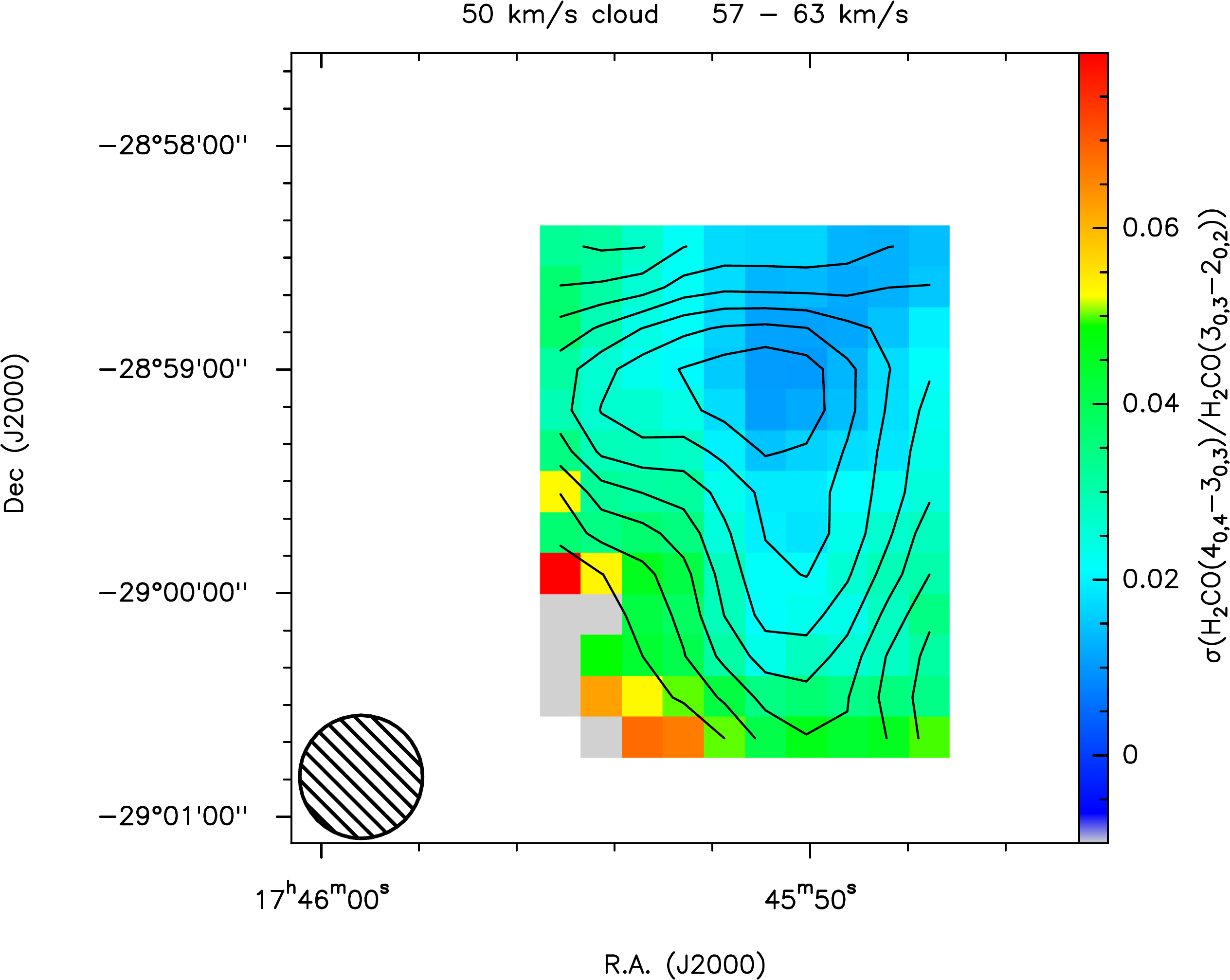}}
\end{figure*}

\begin{figure*}
	\caption{As Fig. \ref{20kms-All-Ratio-H2CO} for G0.253+0.016.}
	\centering
        R$_{321}$\\
	\subfloat{\includegraphics[bb = 0 60 540 580, clip, height=4.4cm]{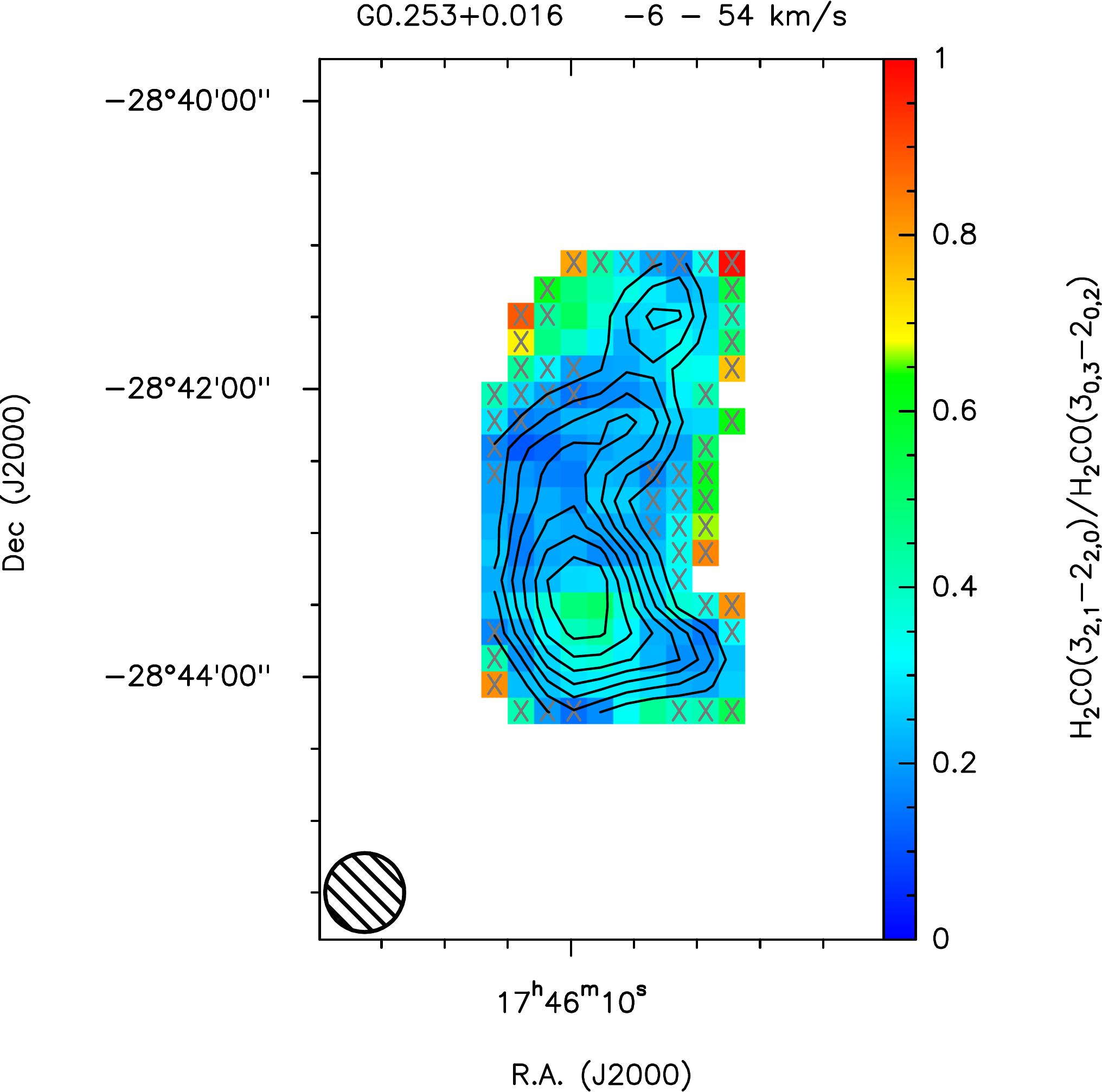}}
	\subfloat{\includegraphics[bb = 150 60 540 580, clip, height=4.4cm]{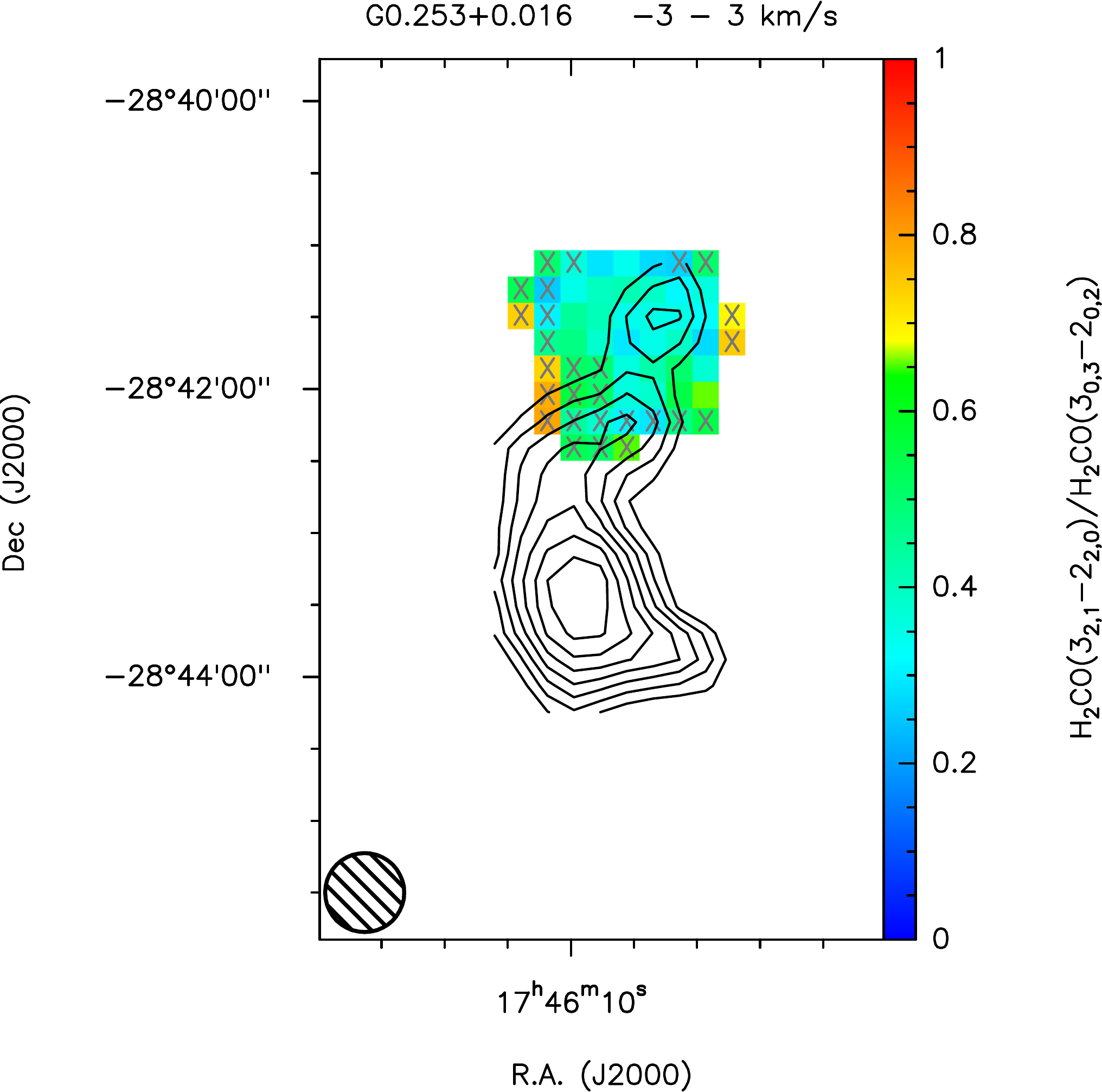}}
	\subfloat{\includegraphics[bb = 150 60 540 580, clip, height=4.4cm]{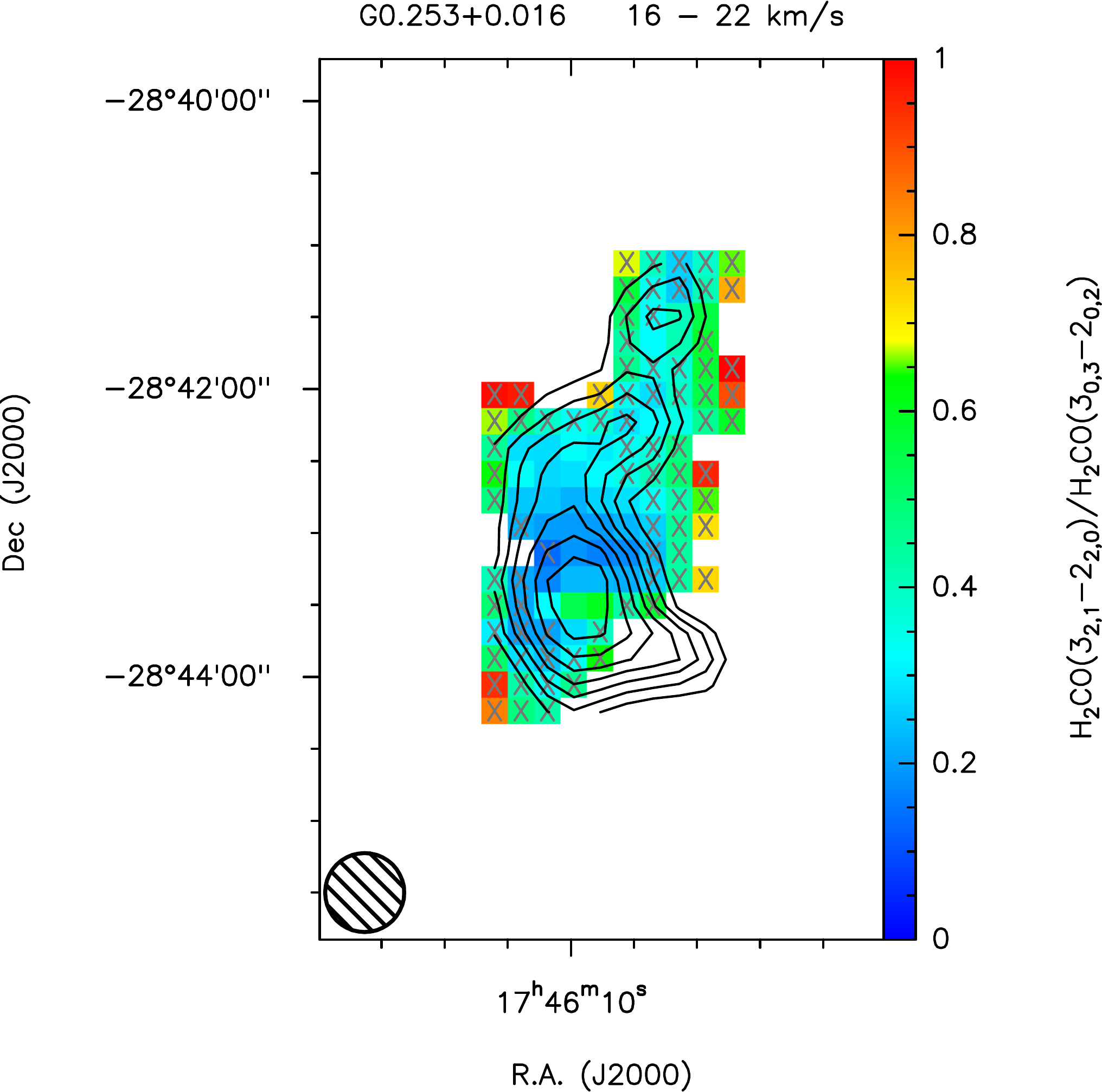}}
	\subfloat{\includegraphics[bb = 150 60 540 580, clip, height=4.4cm]{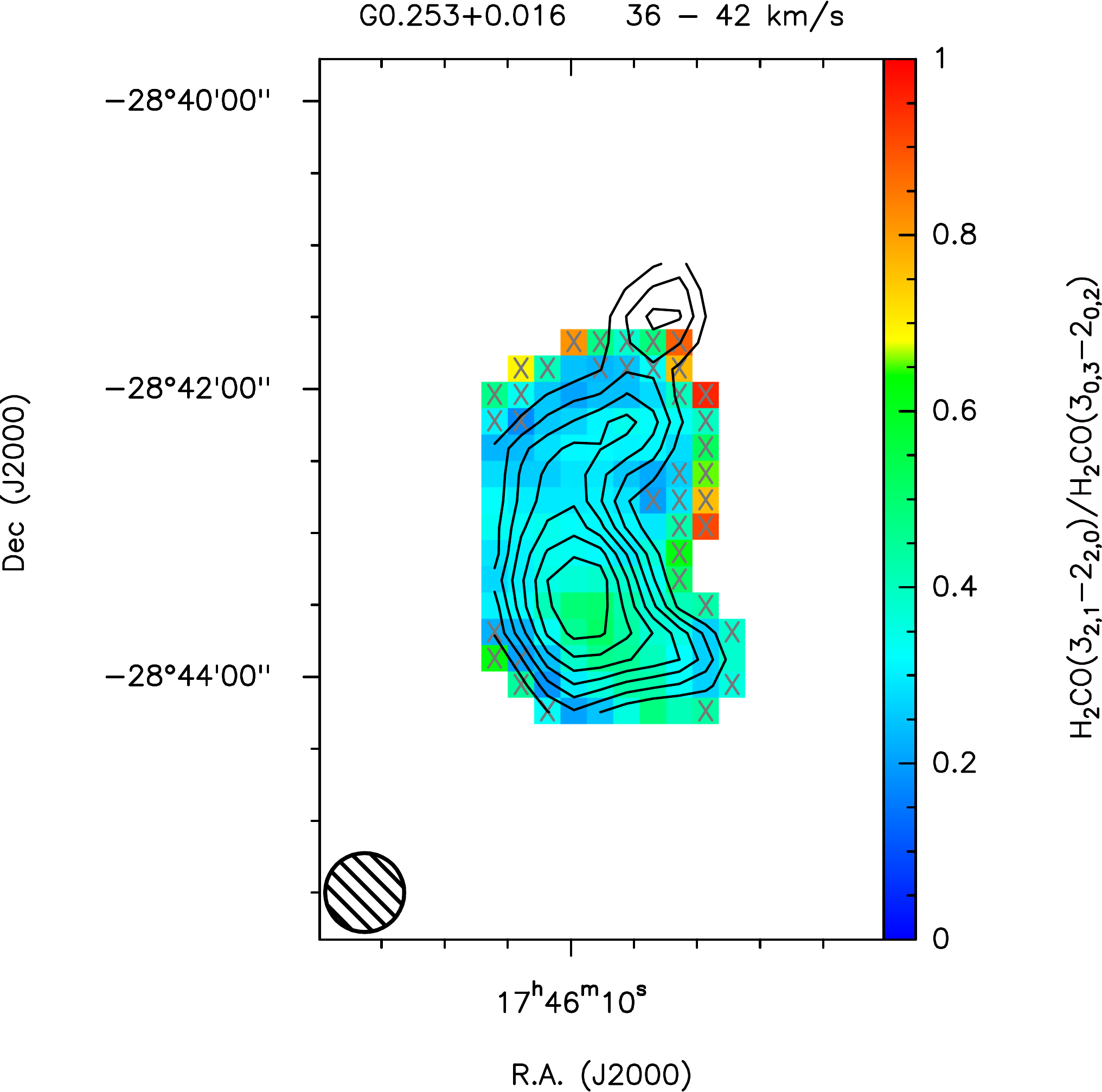}}
	\subfloat{\includegraphics[bb = 150 60 600 580, clip, height=4.4cm]{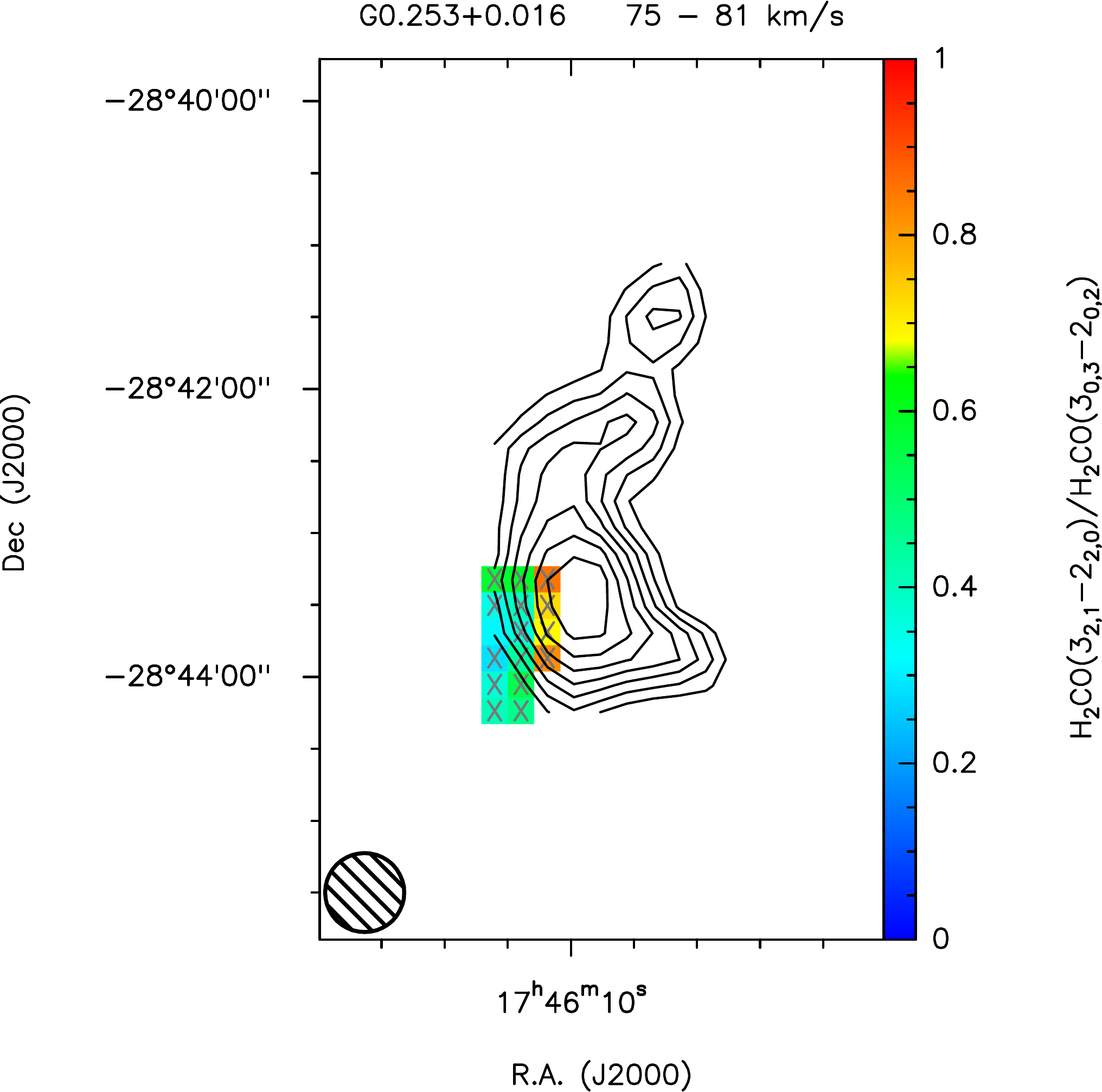}}\\\vspace{-0.5cm}
	\subfloat{\includegraphics[bb = 0 0 540 560, clip, height=4.737cm]{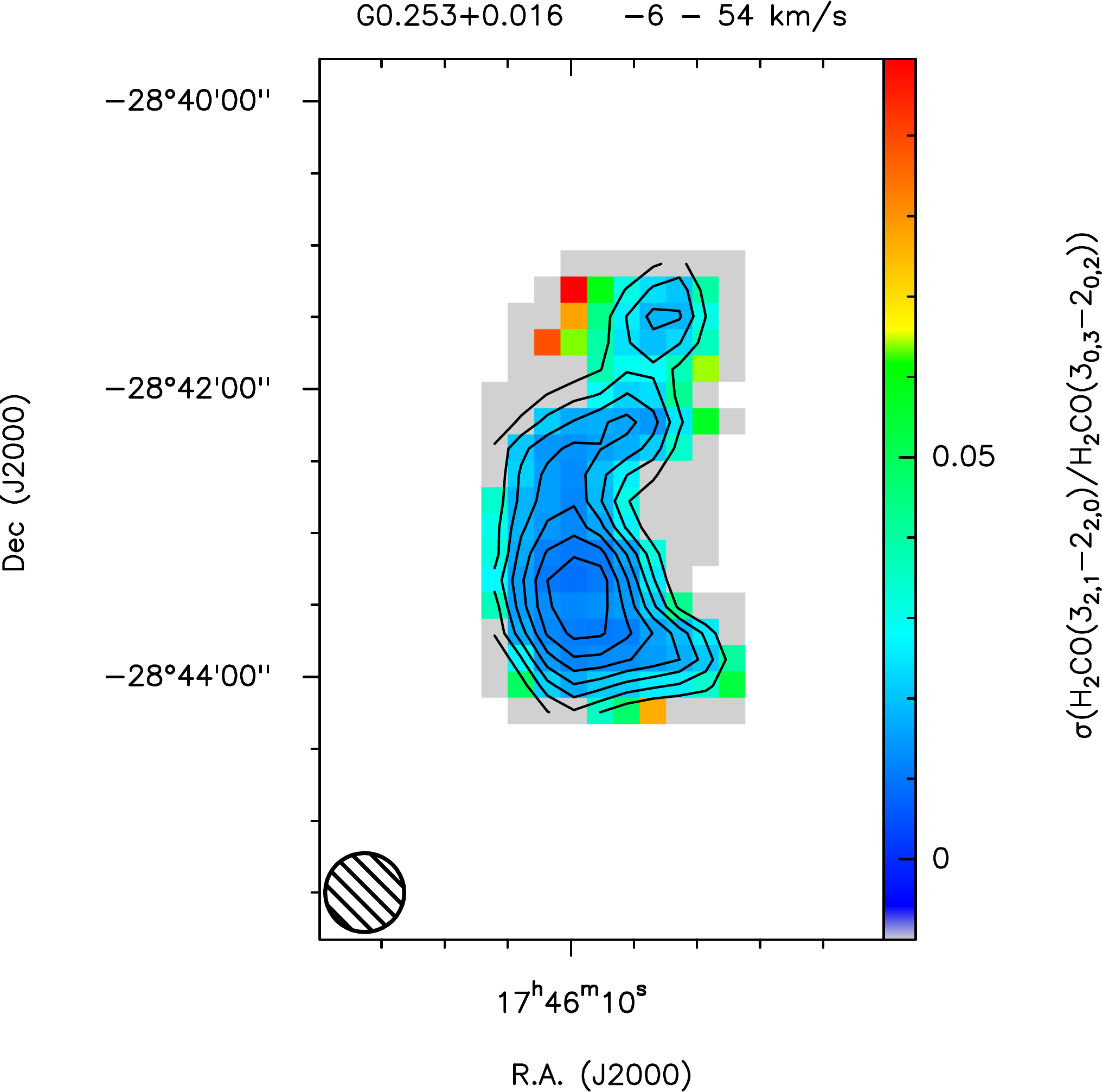}}
	\subfloat{\includegraphics[bb = 150 0 540 560, clip, height=4.737cm]{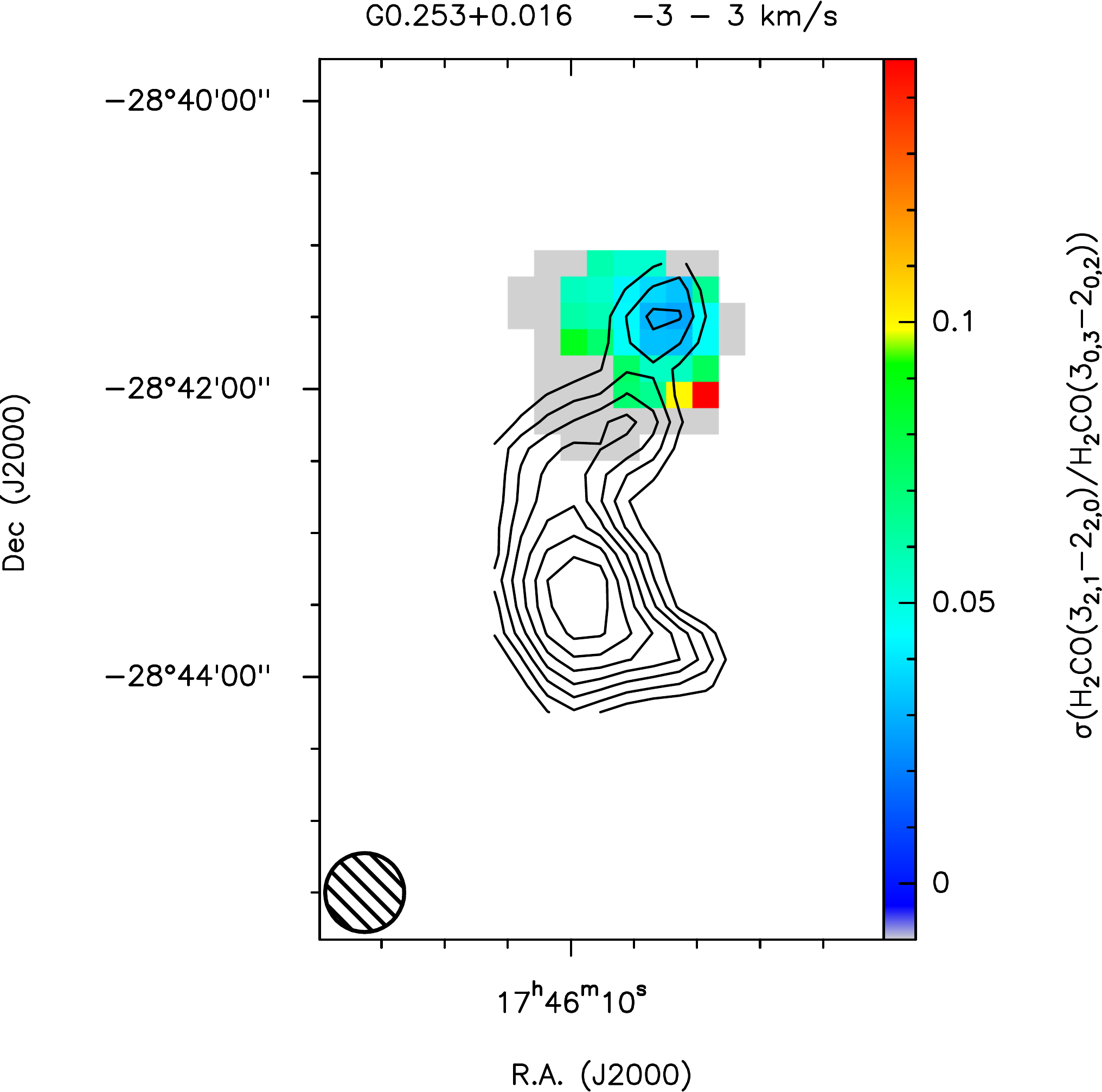}}
	\subfloat{\includegraphics[bb = 150 0 540 560, clip, height=4.737cm]{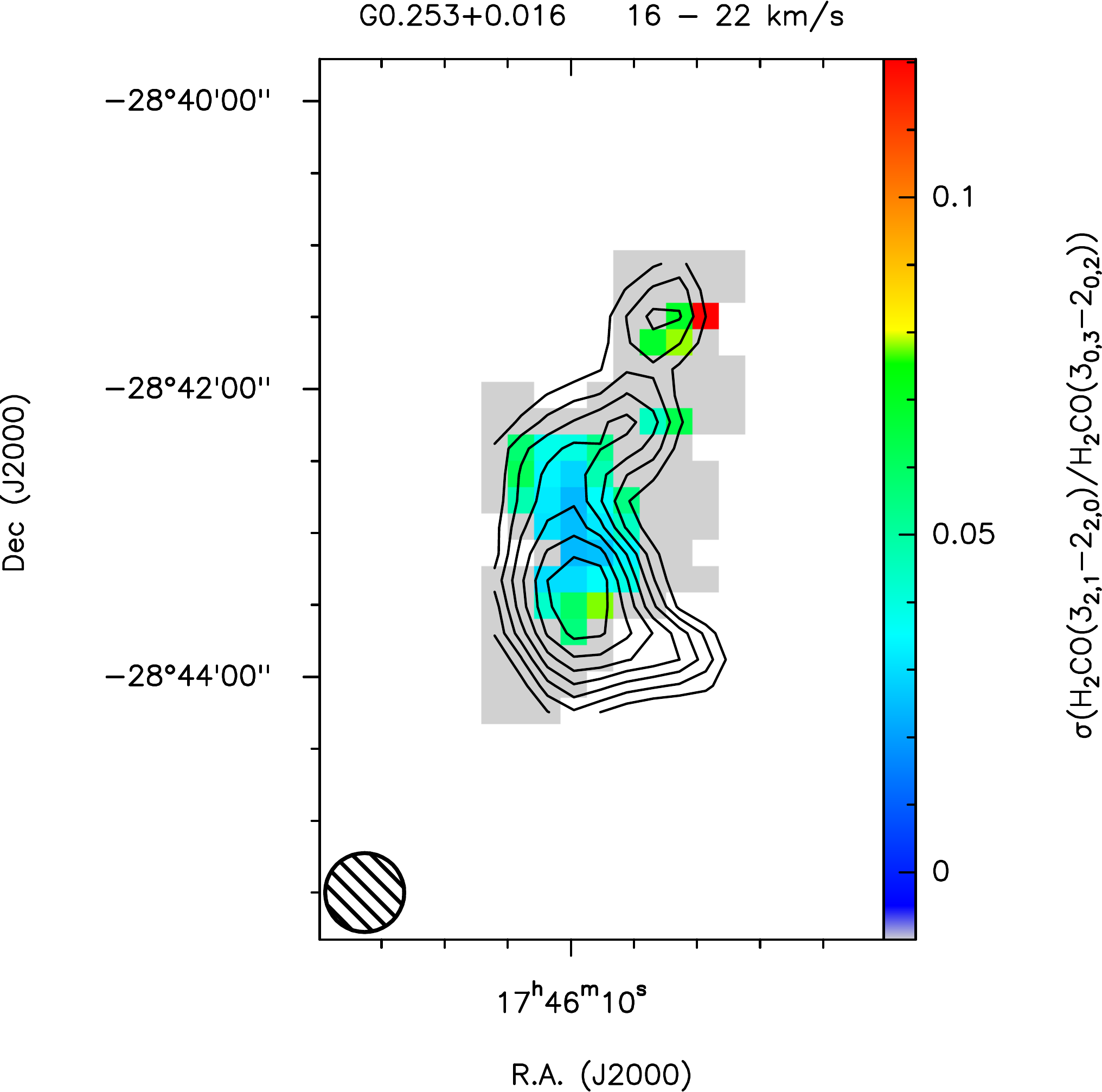}}
	\subfloat{\includegraphics[bb = 150 0 540 560, clip, height=4.737cm]{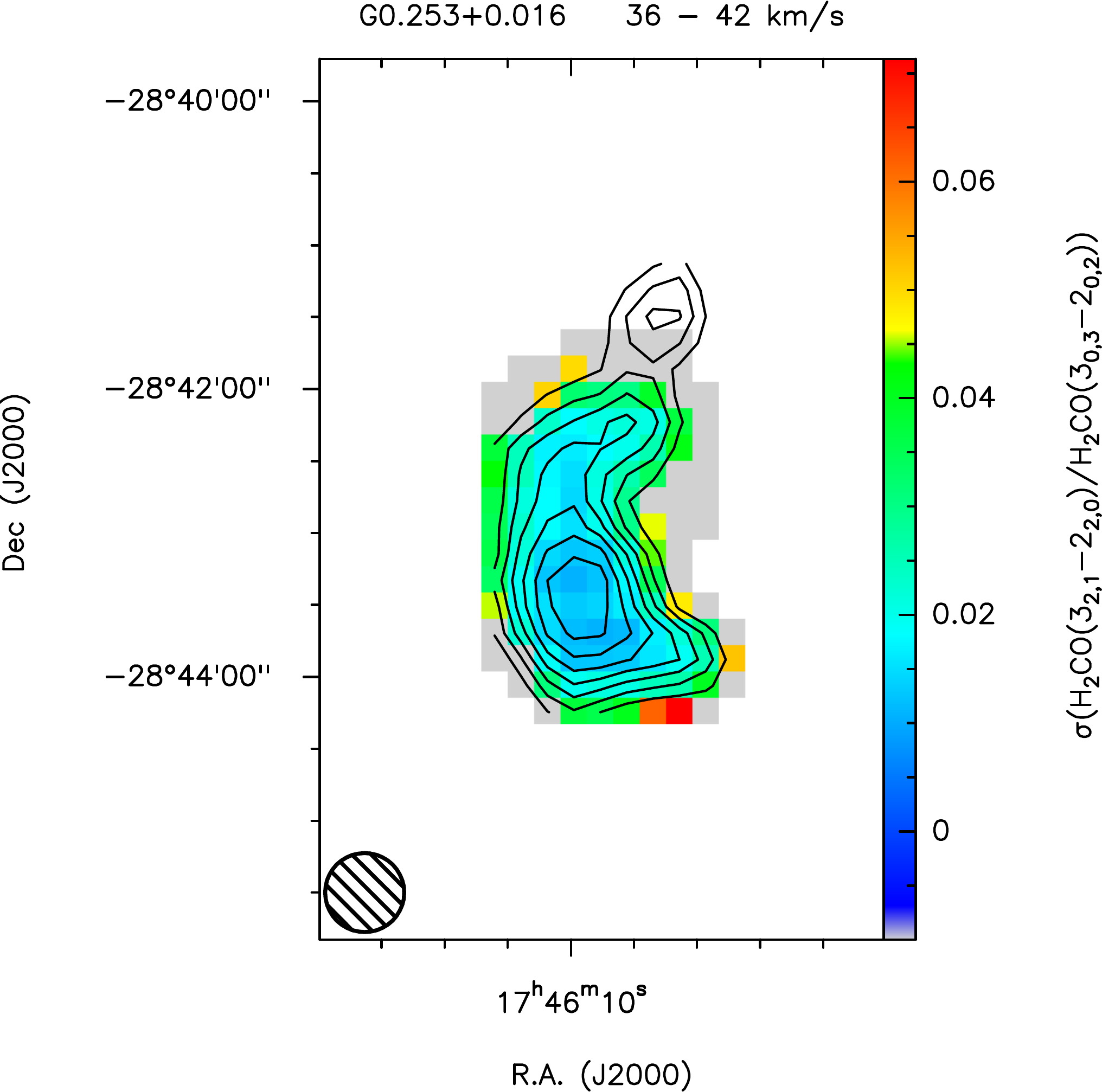}}
	\subfloat{\includegraphics[bb = 150 0 600 560, clip, height=4.737cm]{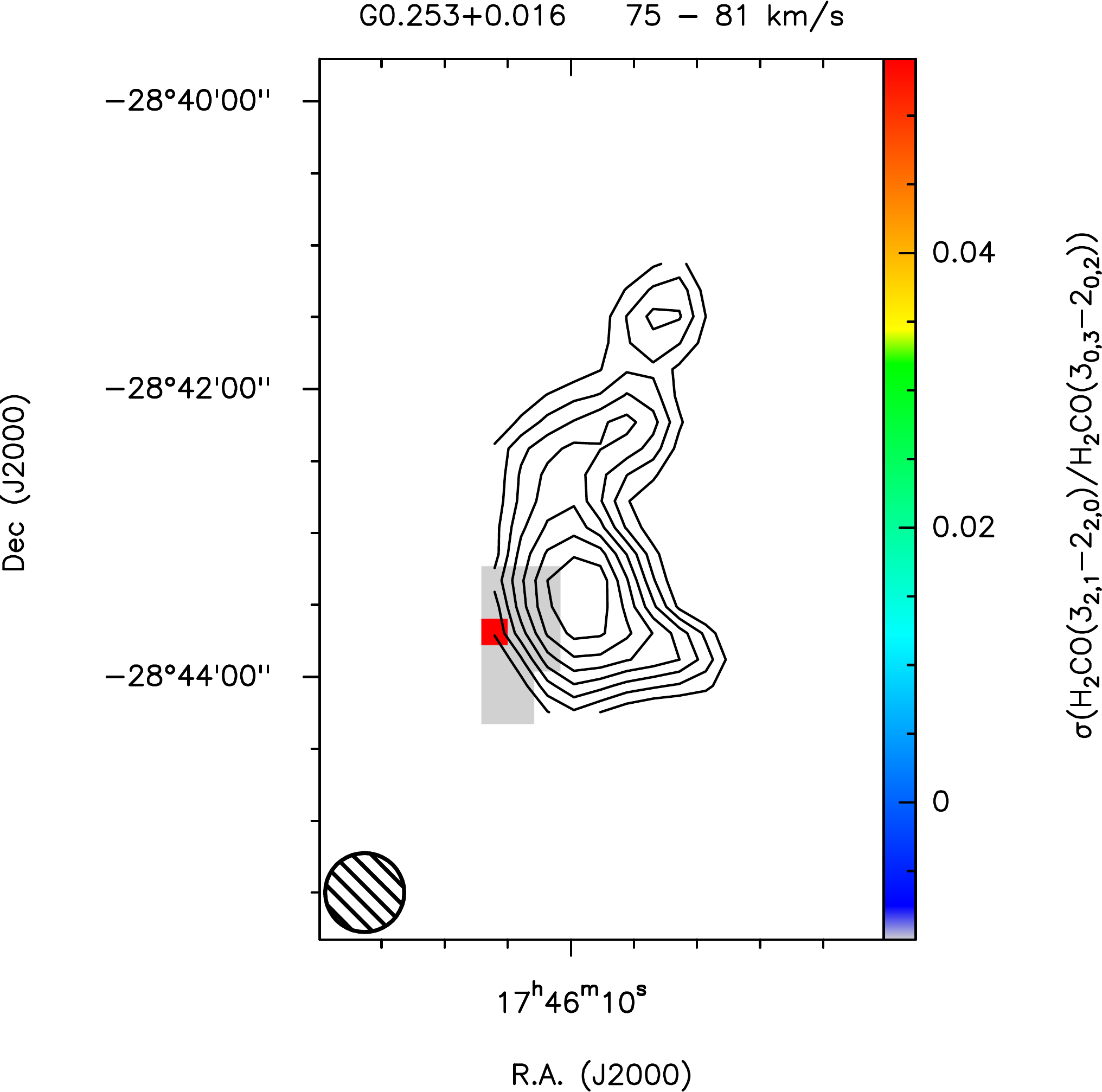}}\\
	\vspace{0.1cm}
        R$_{422}$\\
	\subfloat{\includegraphics[bb = 0 60 540 580, clip, height=4.4cm]{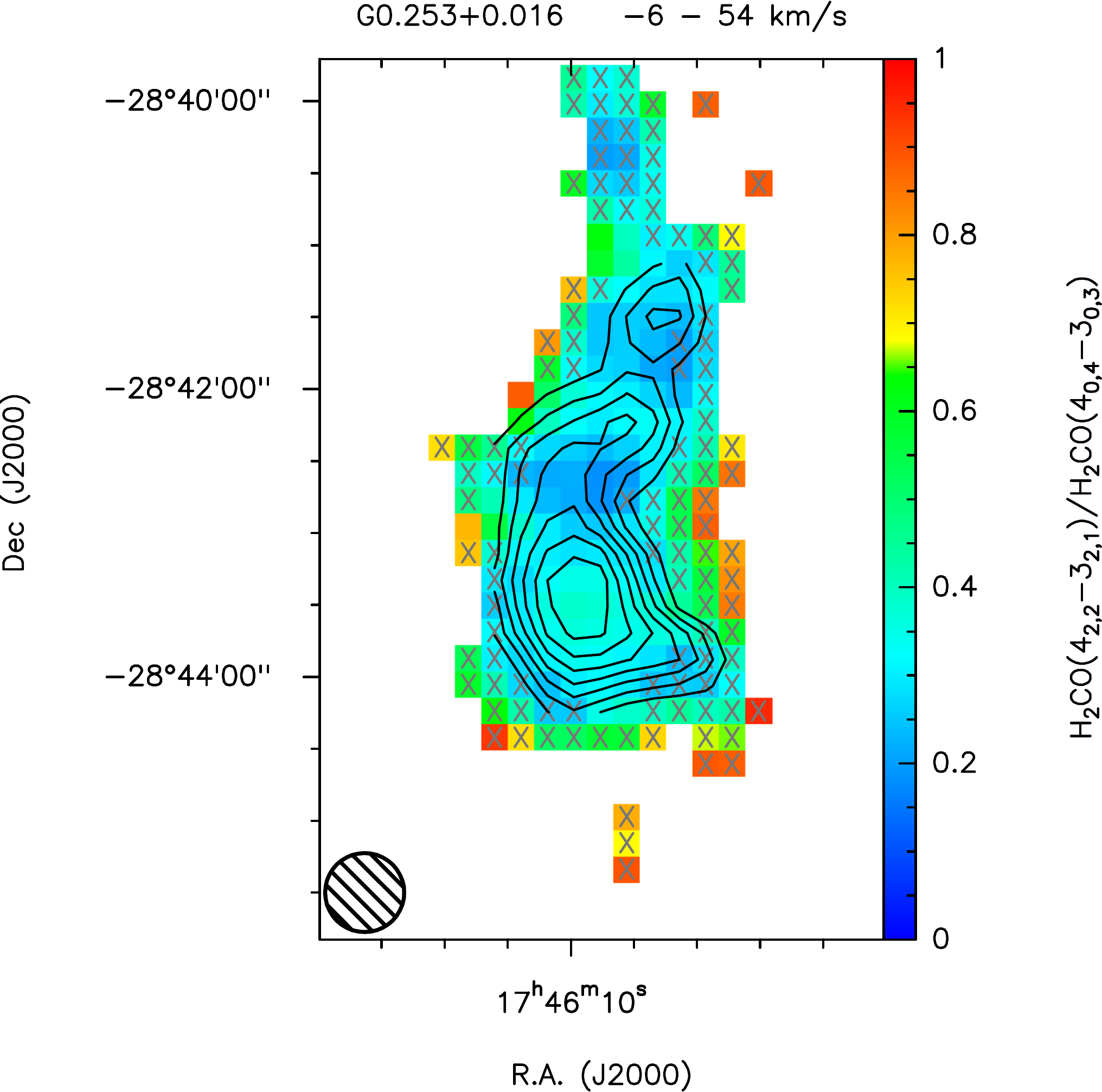}}
	\subfloat{\includegraphics[bb = 150 60 540 580, clip, height=4.4cm]{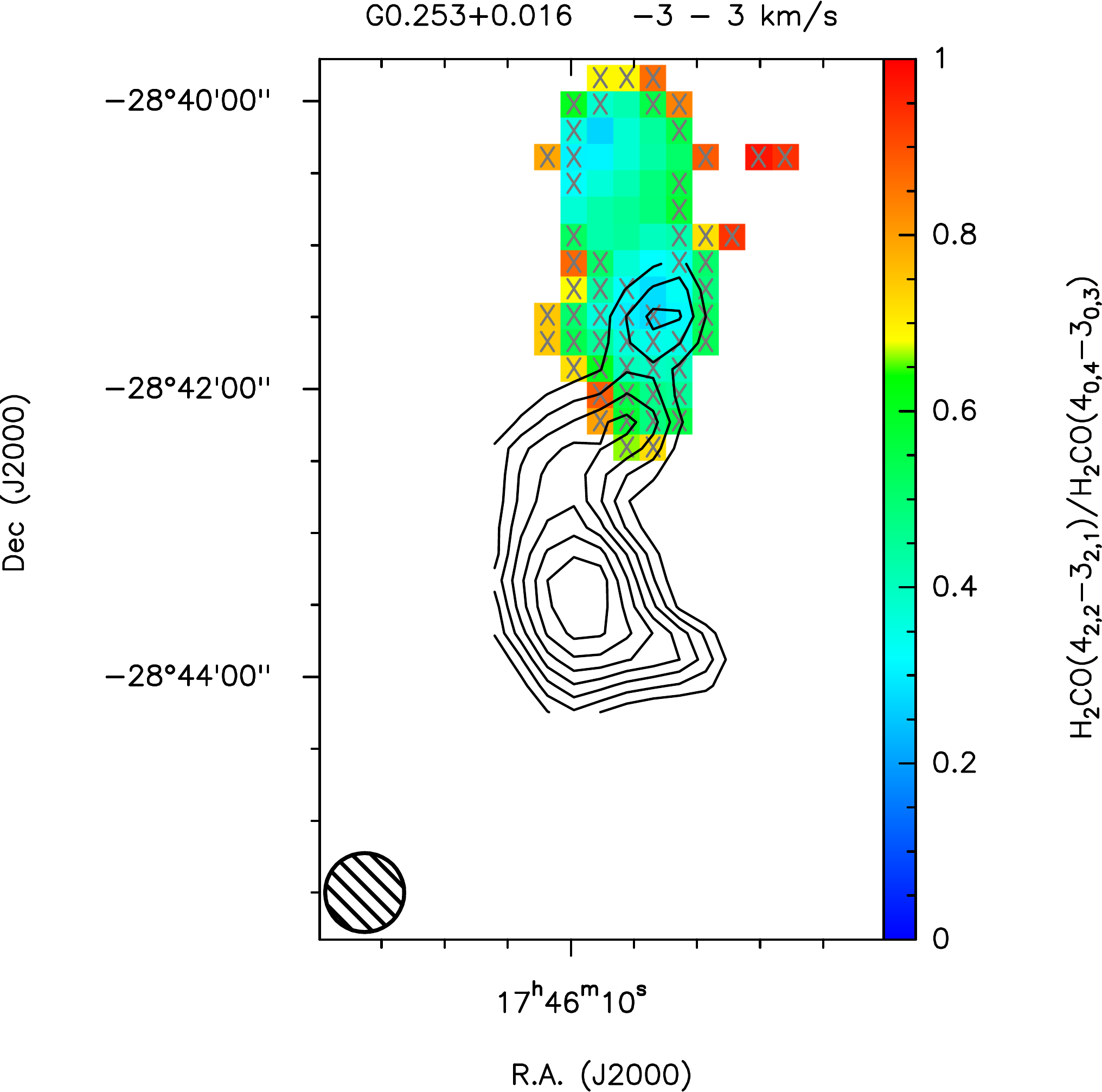}}
	\subfloat{\includegraphics[bb = 150 60 540 580, clip, height=4.4cm]{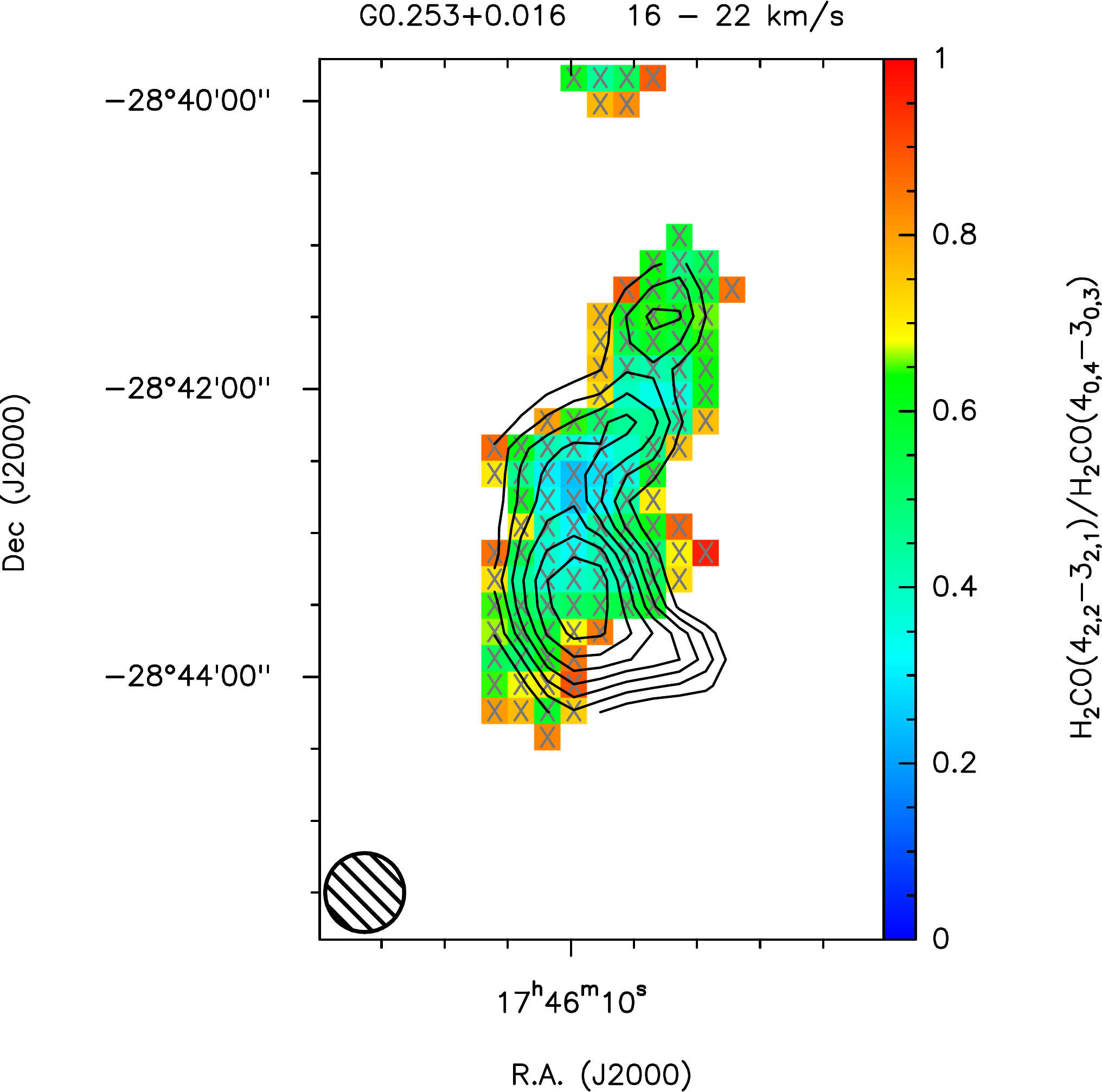}}
	\subfloat{\includegraphics[bb = 150 60 540 580, clip, height=4.4cm]{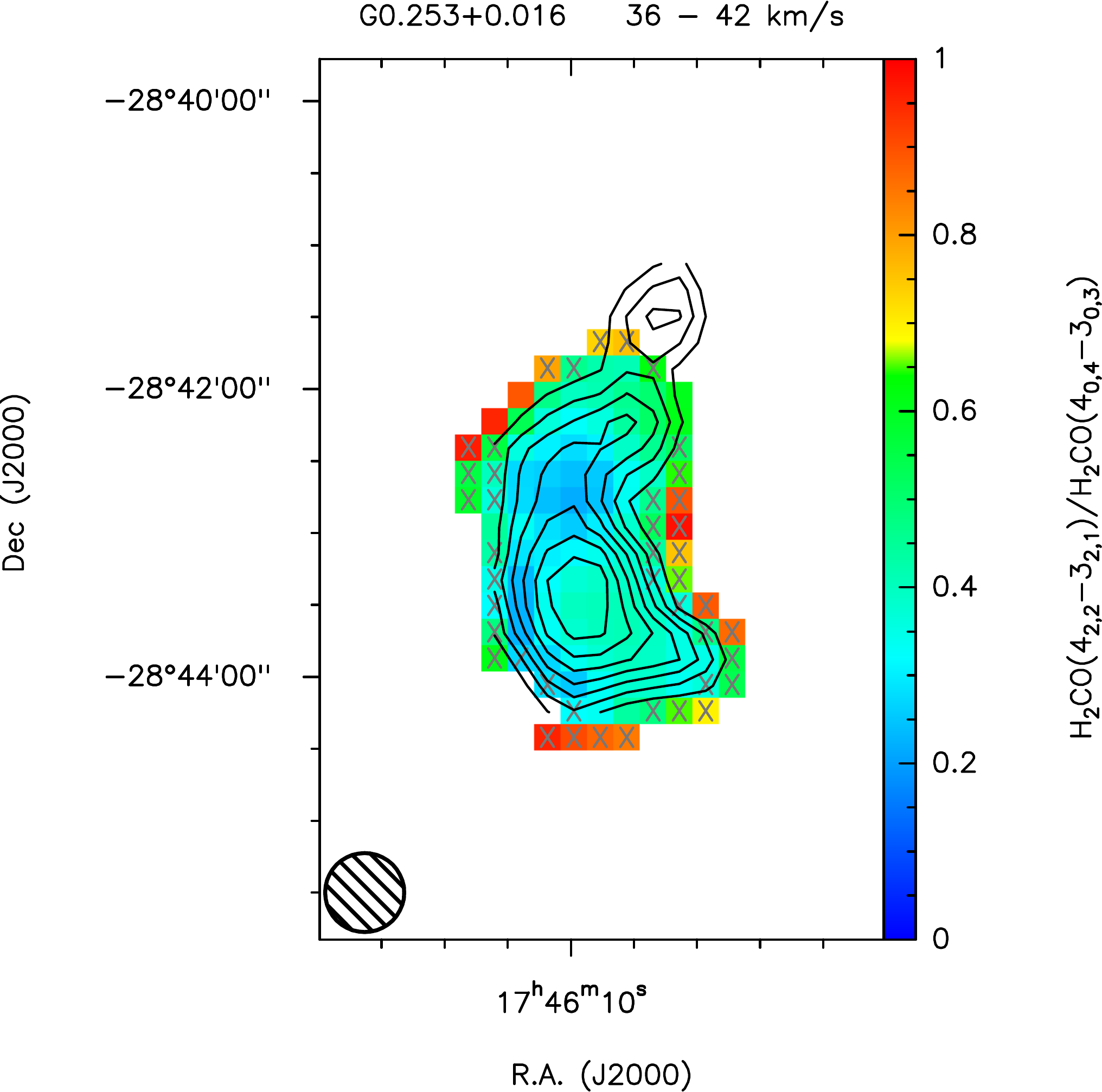}}
	\subfloat{\includegraphics[bb = 150 60 600 580, clip, height=4.4cm]{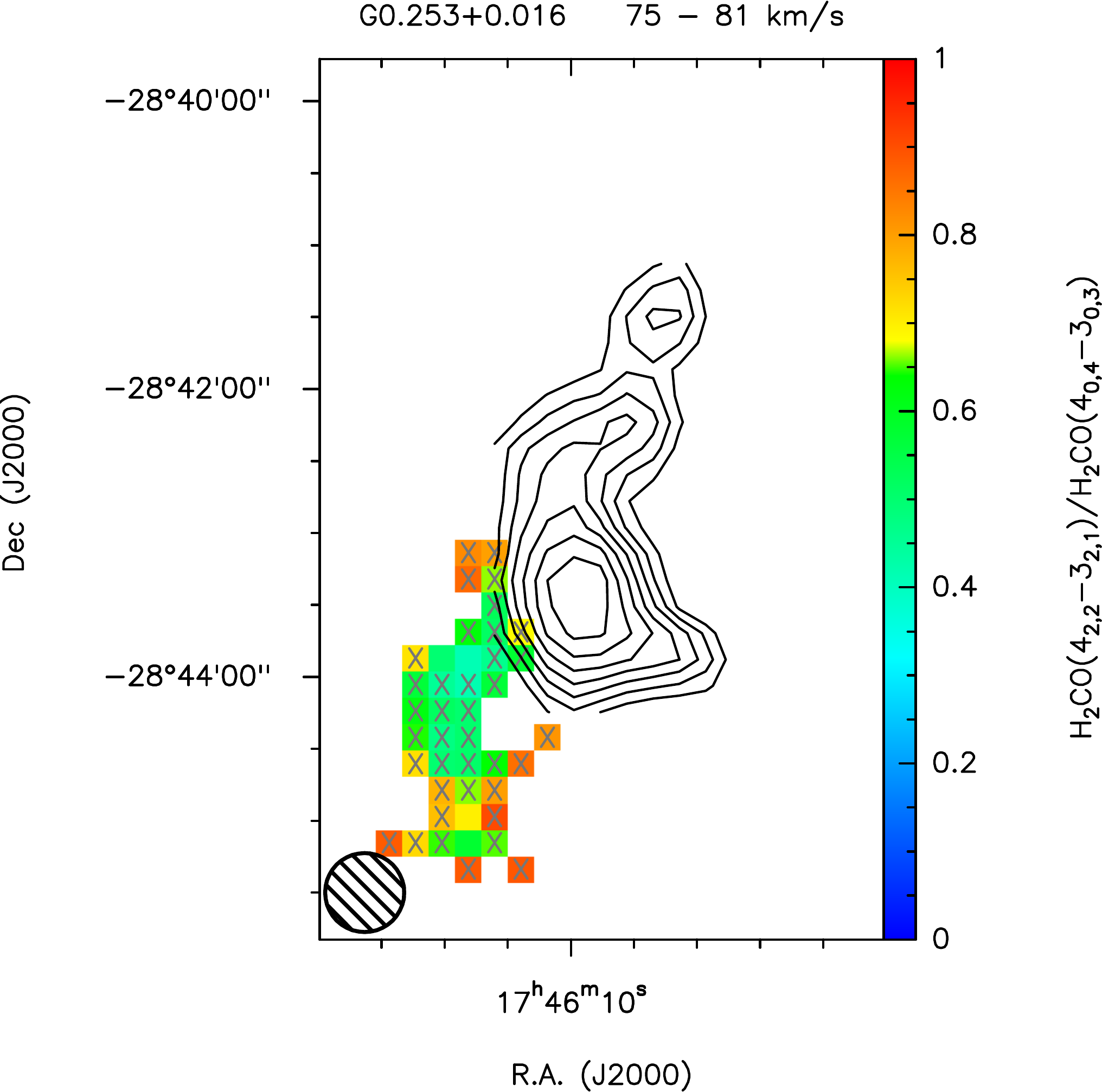}}\\\vspace{-0.5cm}
	\subfloat{\includegraphics[bb = 0 0 540 560, clip, height=4.737cm]{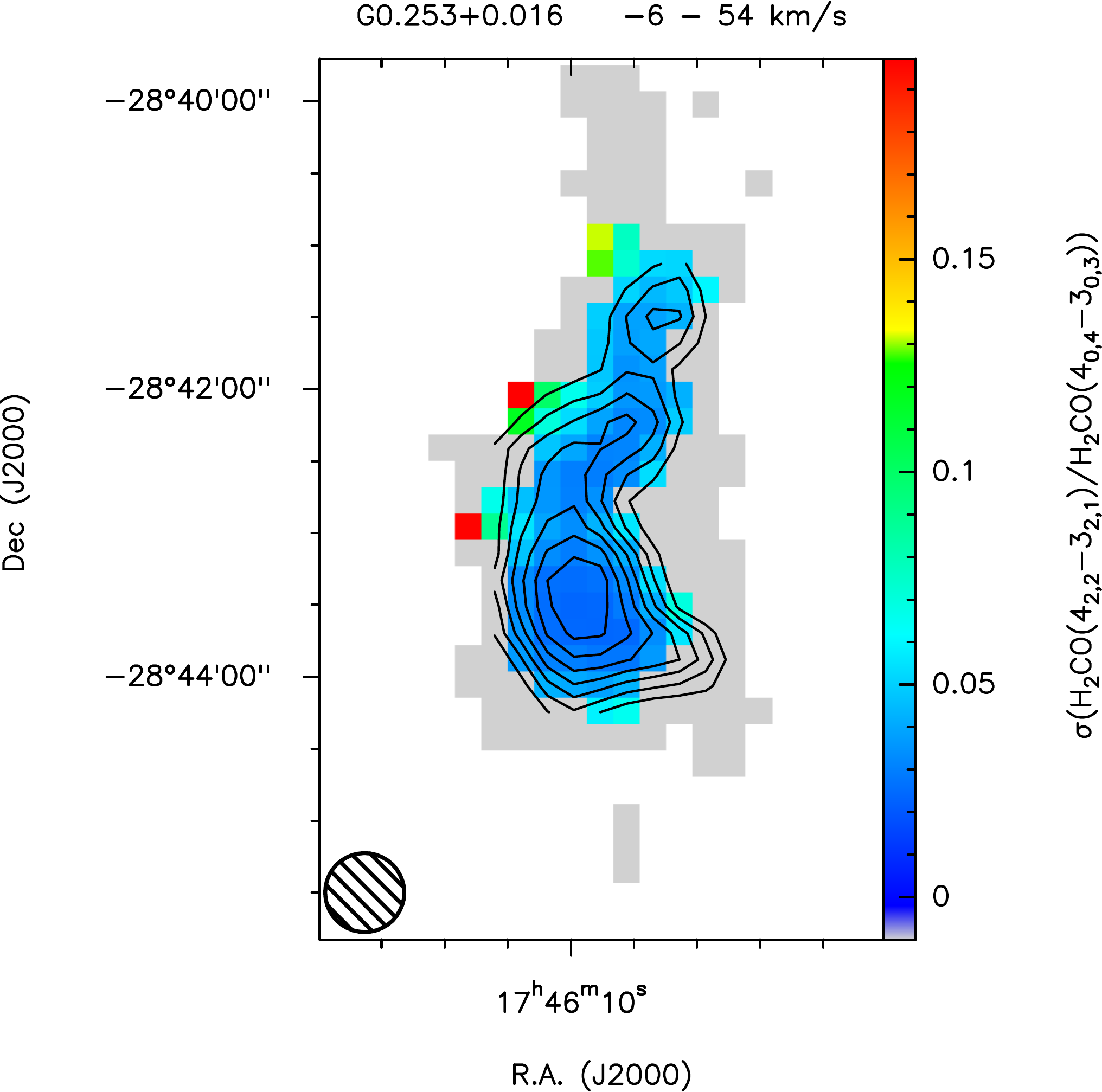}}
	\subfloat{\includegraphics[bb = 150 0 540 560, clip, height=4.737cm]{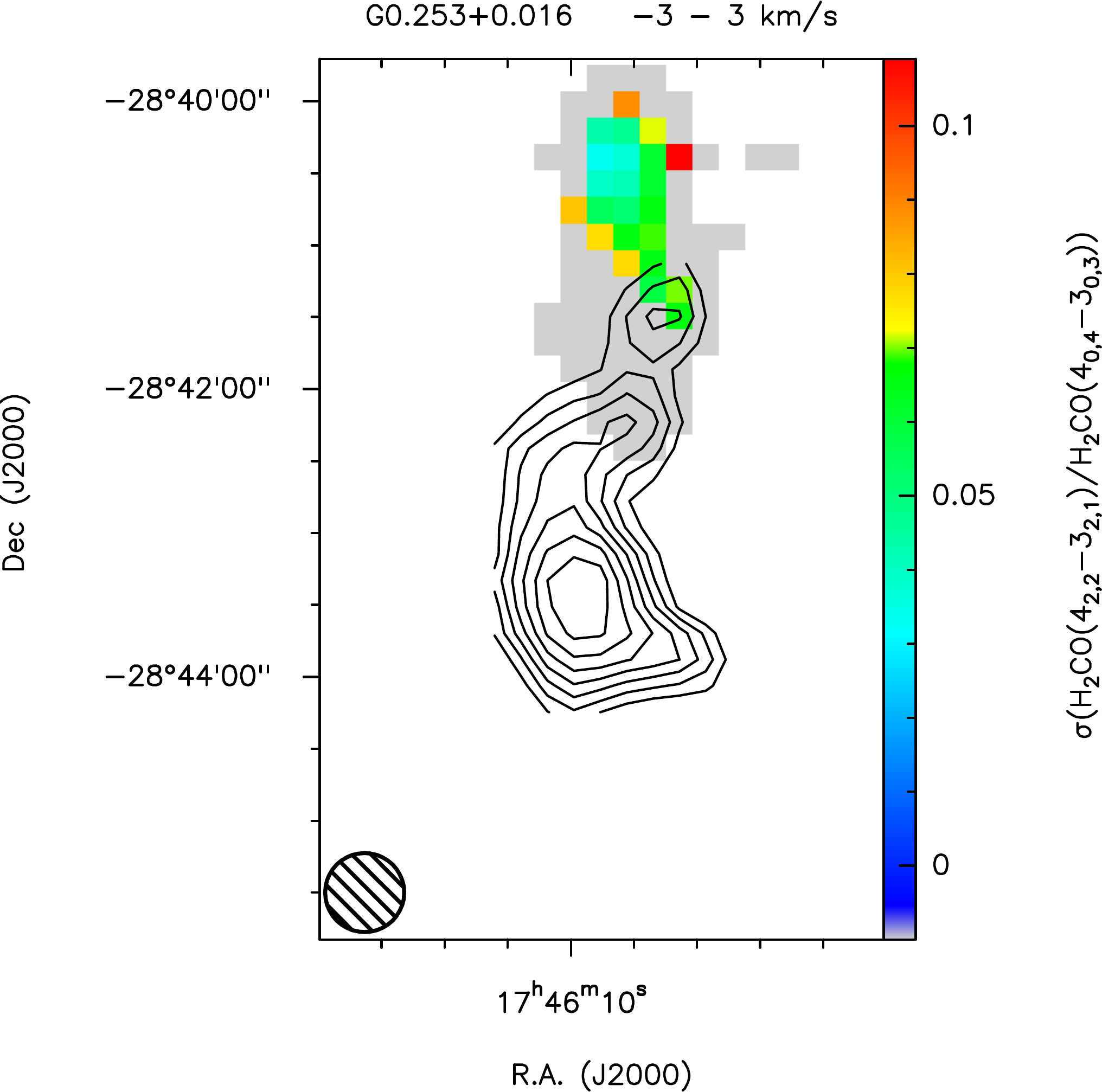}}
	\subfloat{\includegraphics[bb = 150 0 540 560, clip, height=4.737cm]{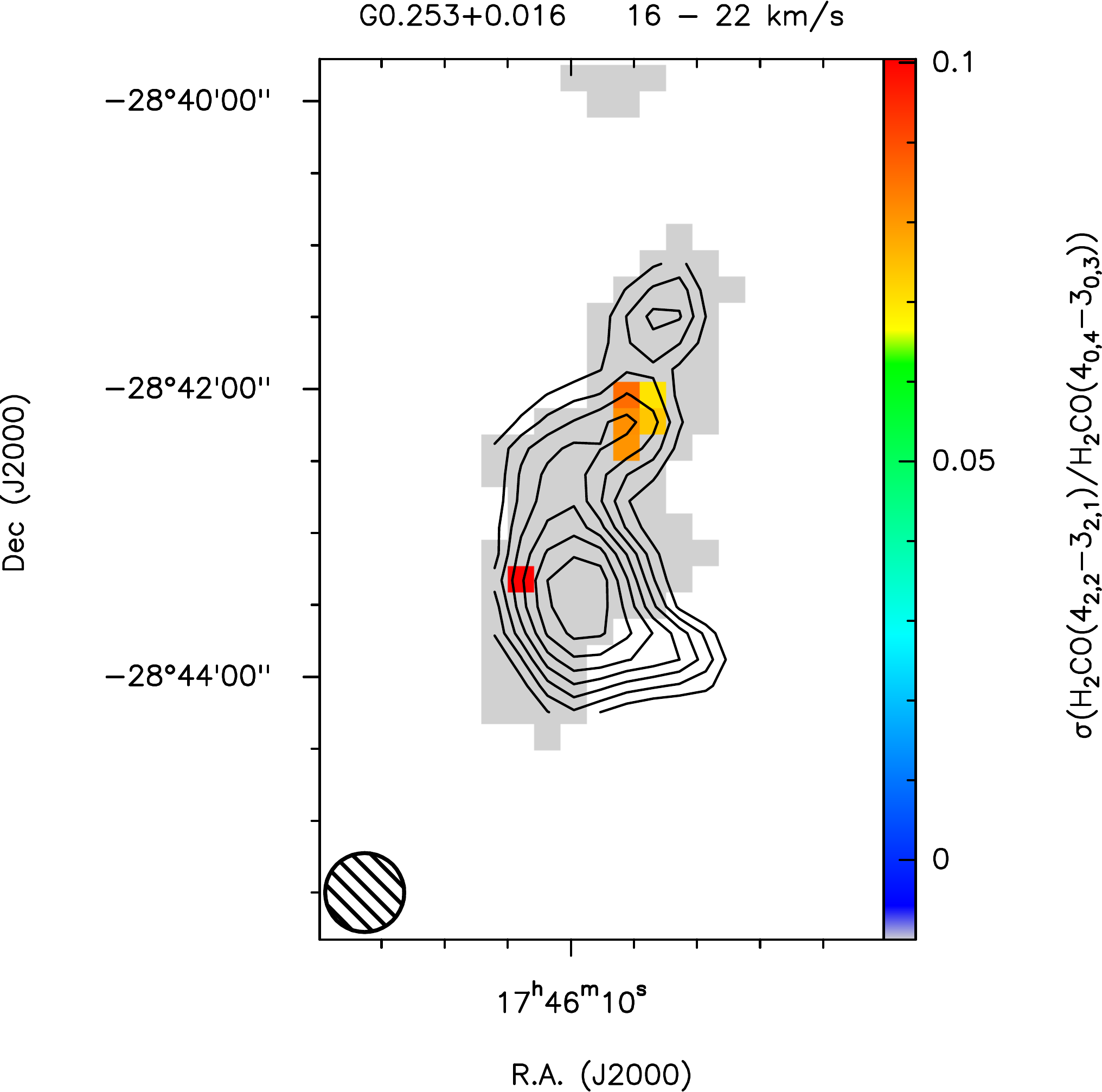}}
	\subfloat{\includegraphics[bb = 150 0 540 560, clip, height=4.737cm]{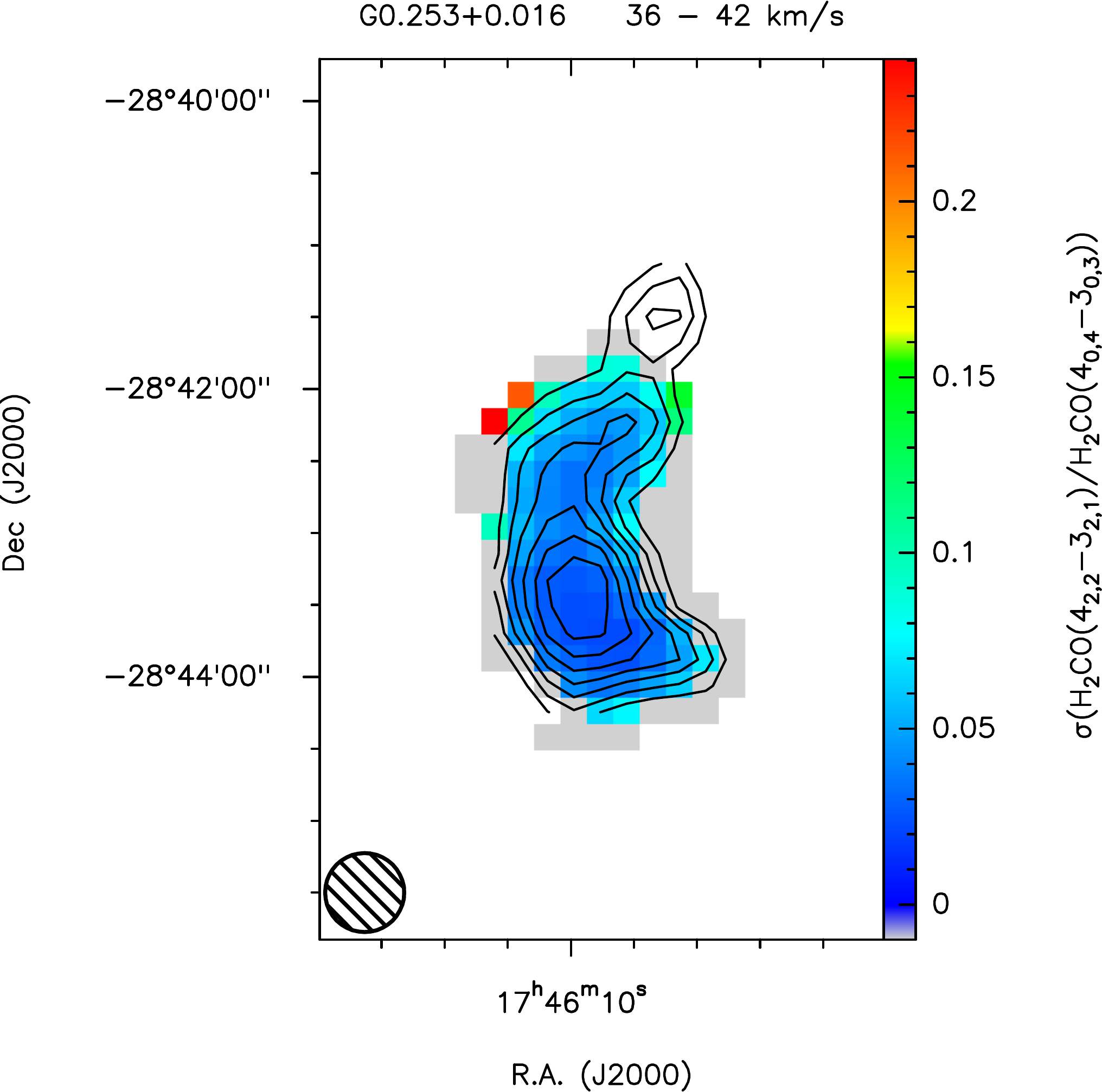}}
	\subfloat{\includegraphics[bb = 150 0 600 560, clip, height=4.737cm]{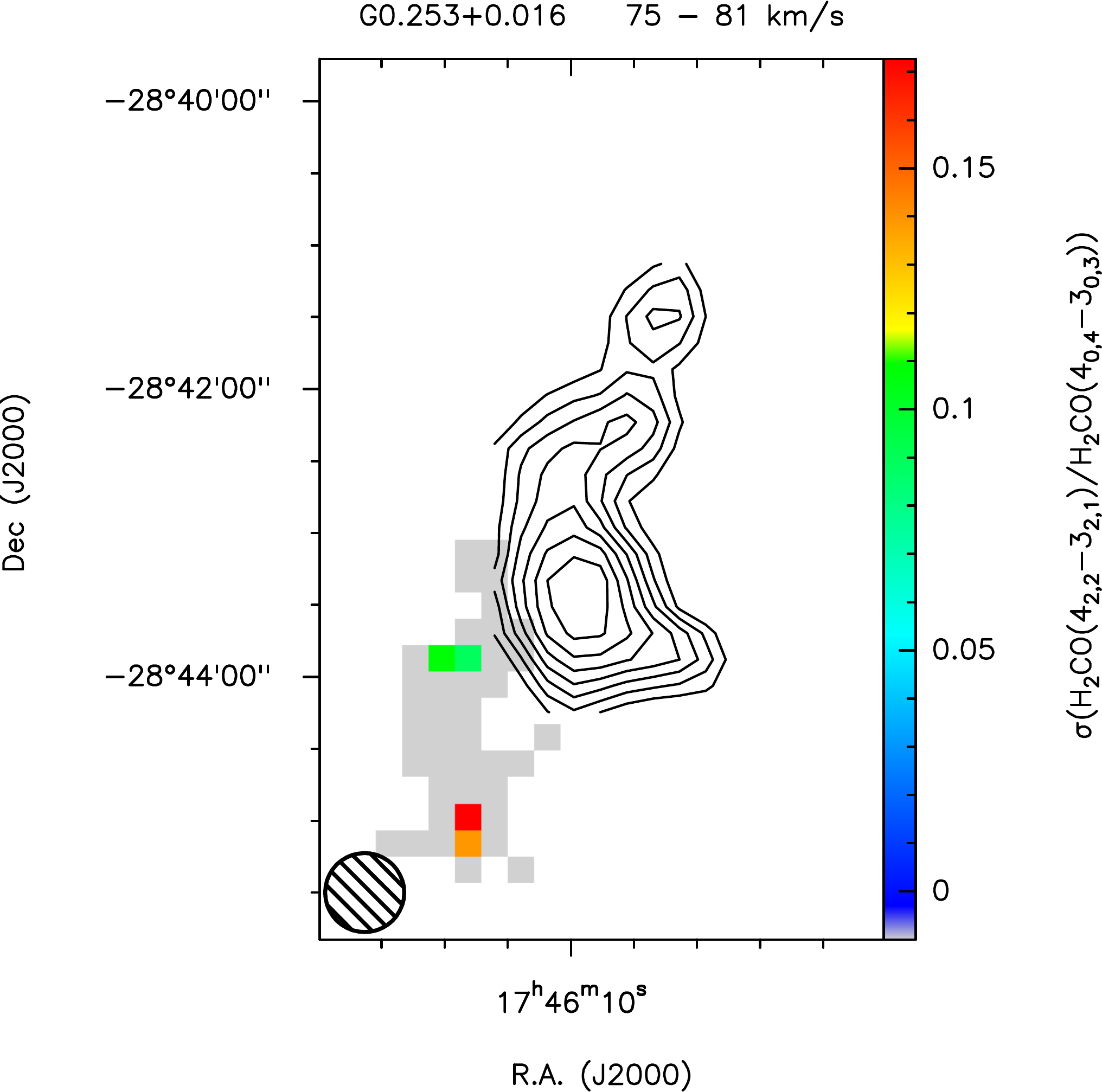}}\\
	\label{G0253-All-Ratio-H2CO}
\end{figure*}

\begin{figure*}
	\centering
	R$_{404}$ \\
	\subfloat{\includegraphics[bb = 0 60 540 580, clip, height=4.4cm]{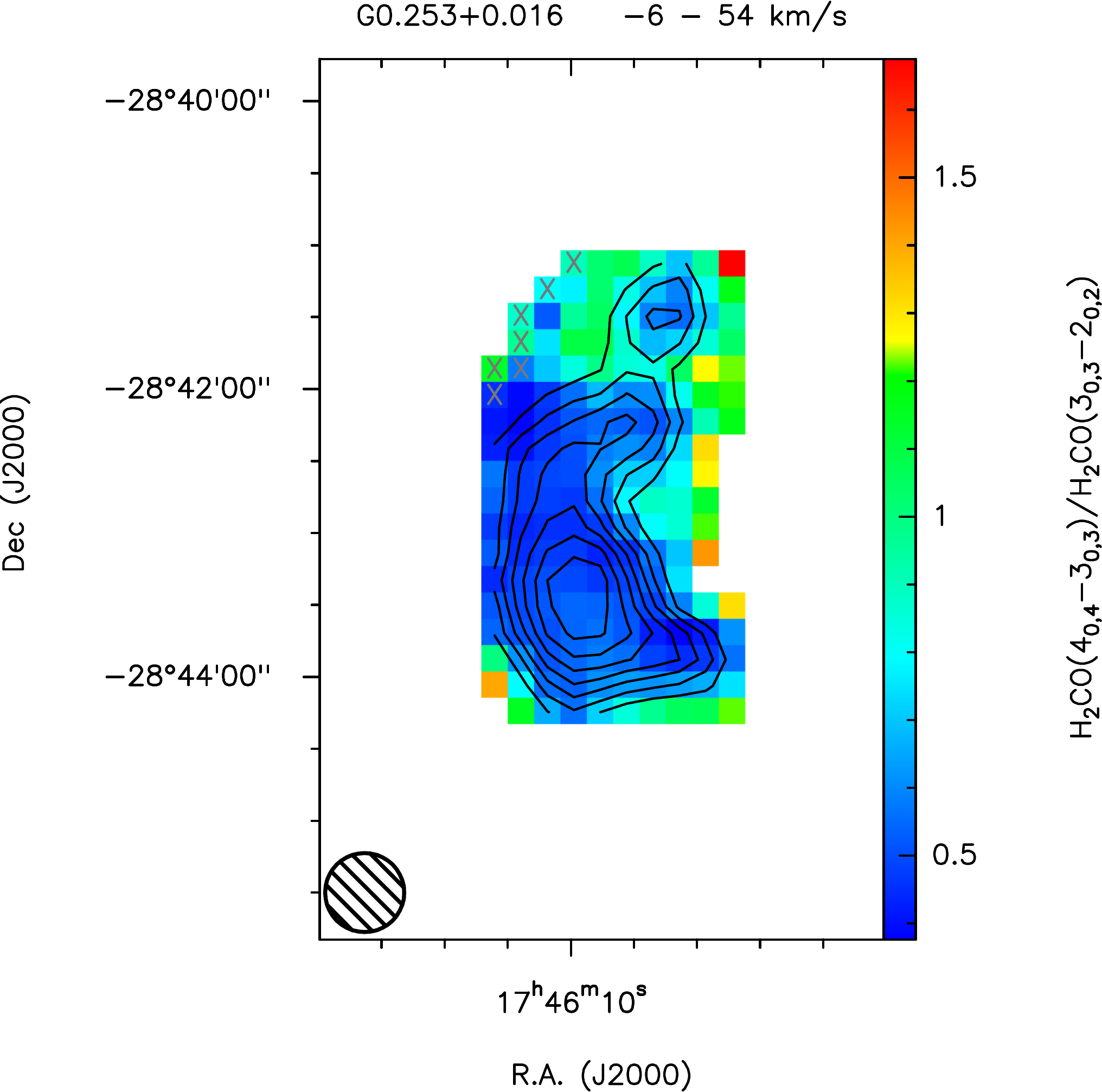}}
	\subfloat{\includegraphics[bb = 150 60 540 580, clip, height=4.4cm]{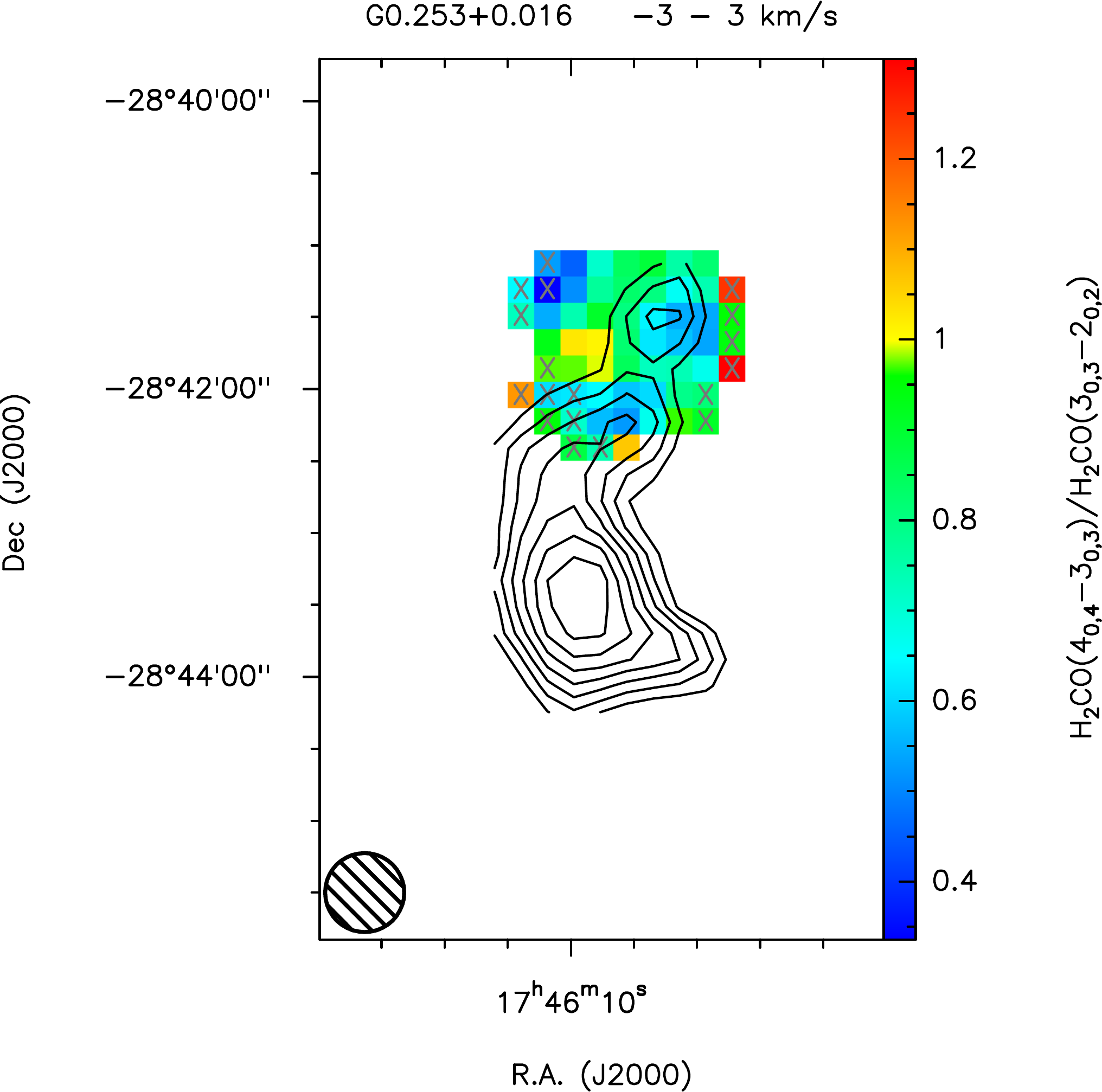}}
	\subfloat{\includegraphics[bb = 150 60 540 580, clip, height=4.4cm]{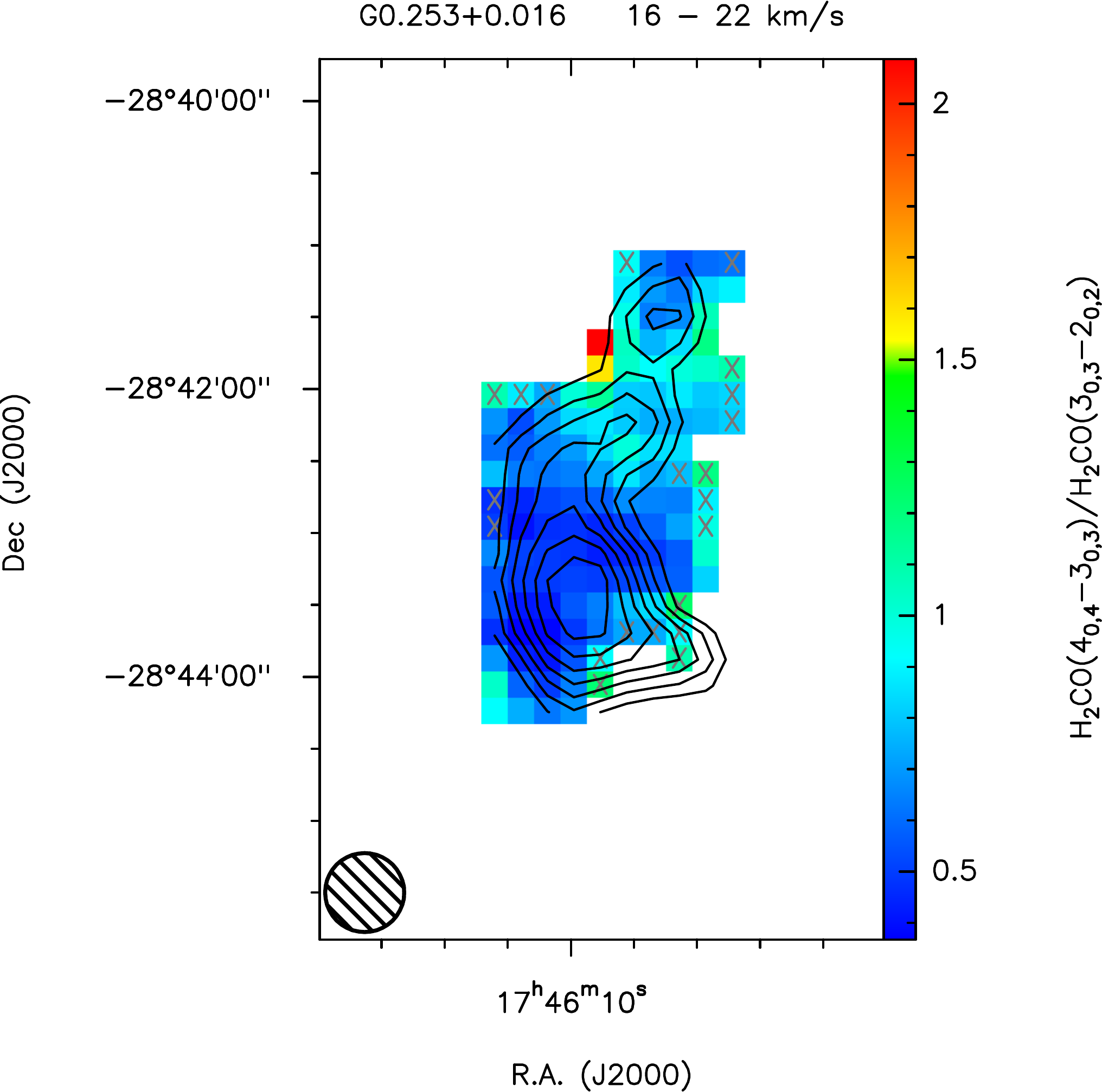}}
	\subfloat{\includegraphics[bb = 150 60 540 580, clip, height=4.4cm]{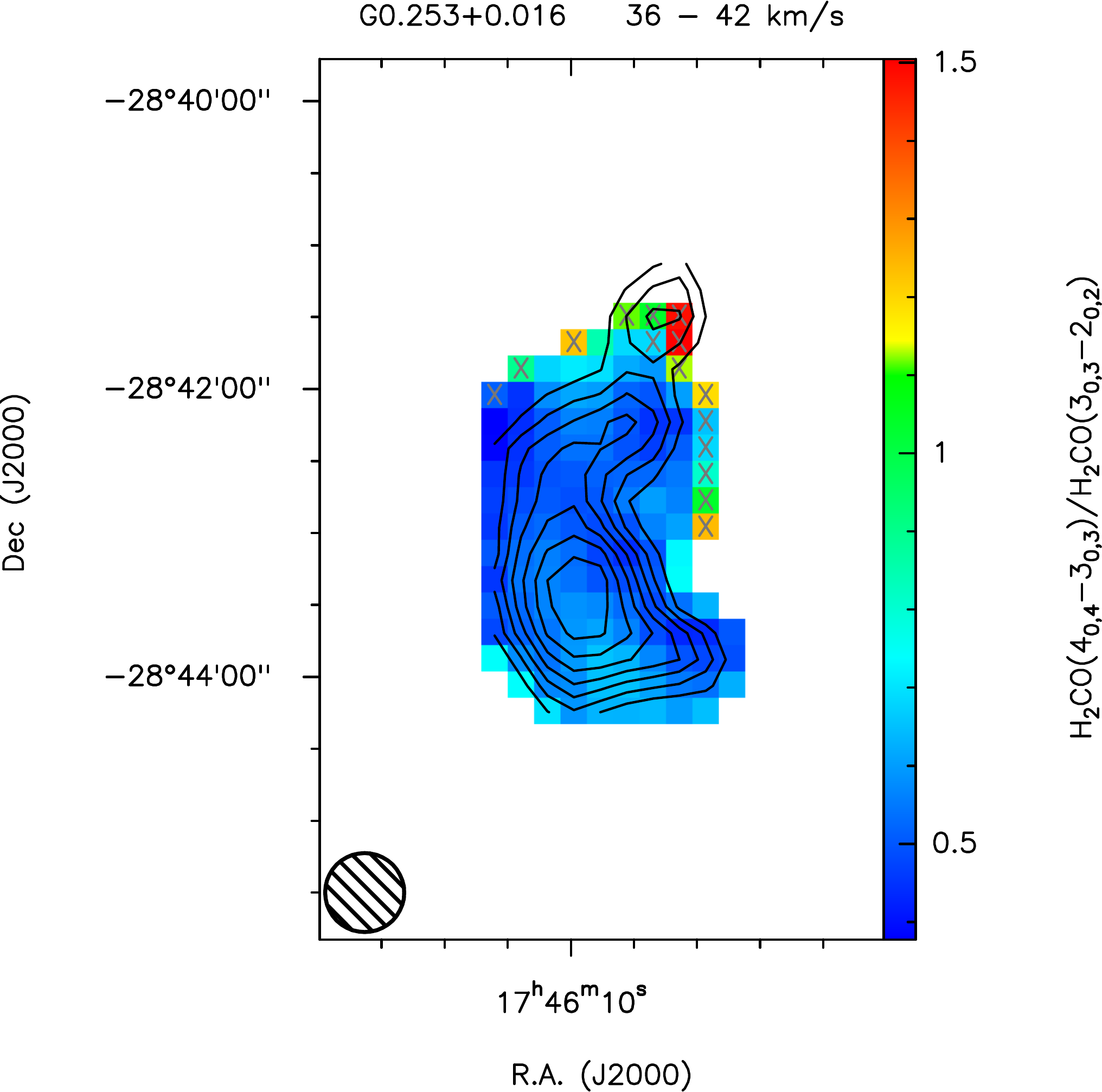}}
	\subfloat{\includegraphics[bb = 150 60 600 580, clip, height=4.4cm]{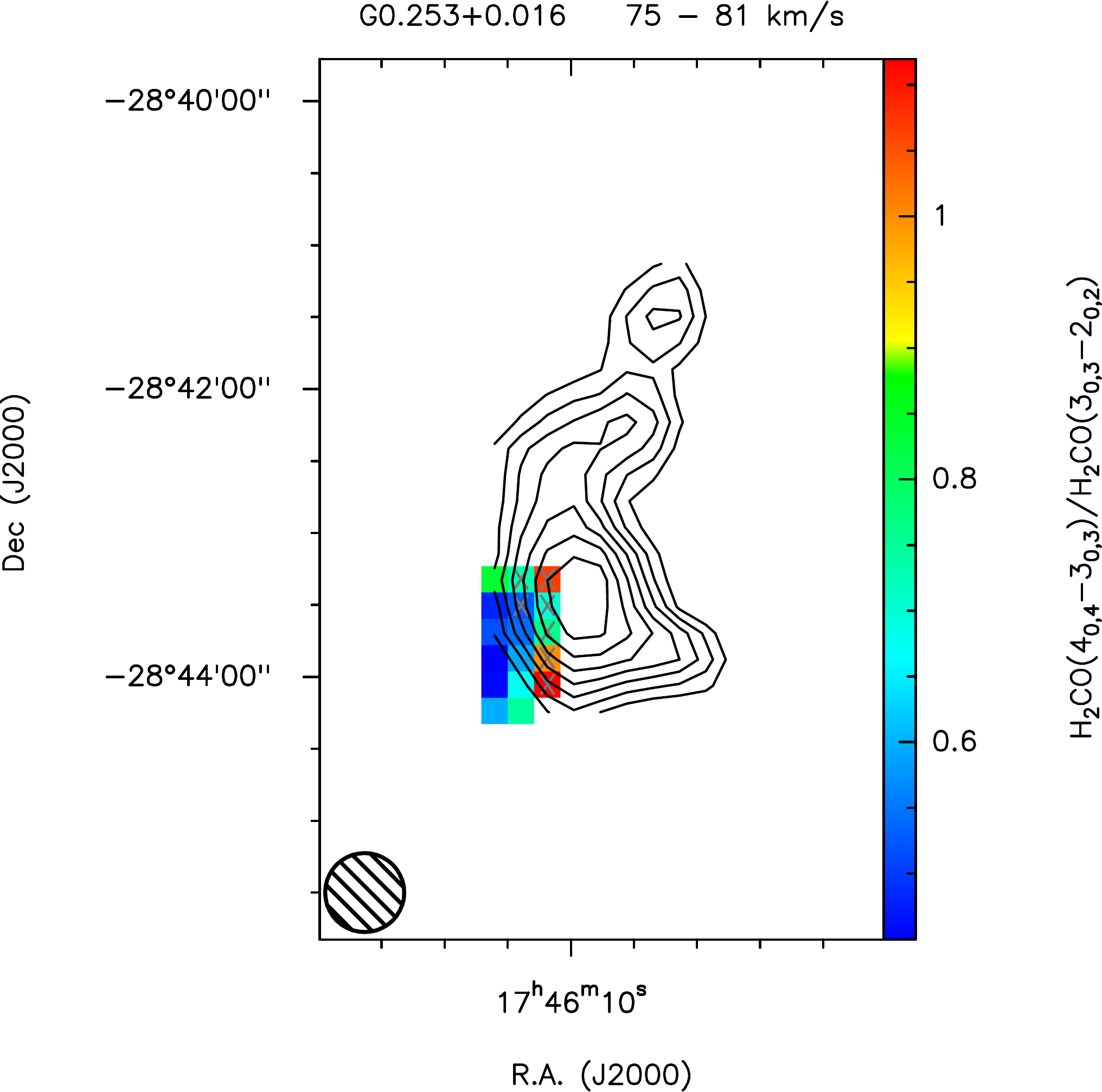}}\\\vspace{-0.5cm}
	\subfloat{\includegraphics[bb = 0 0 540 560, clip, height=4.737cm]{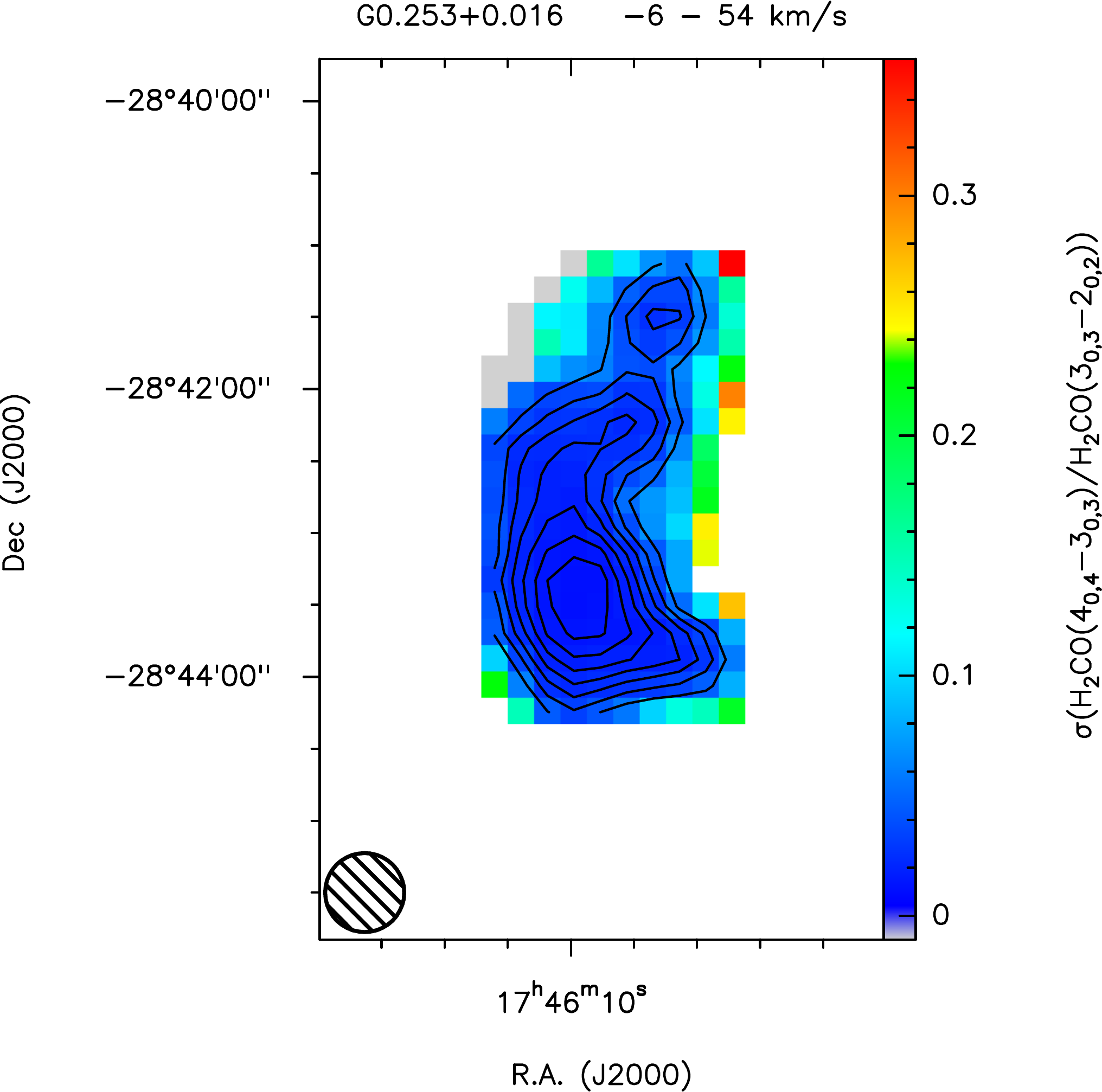}}
	\subfloat{\includegraphics[bb = 150 0 540 560, clip, height=4.737cm]{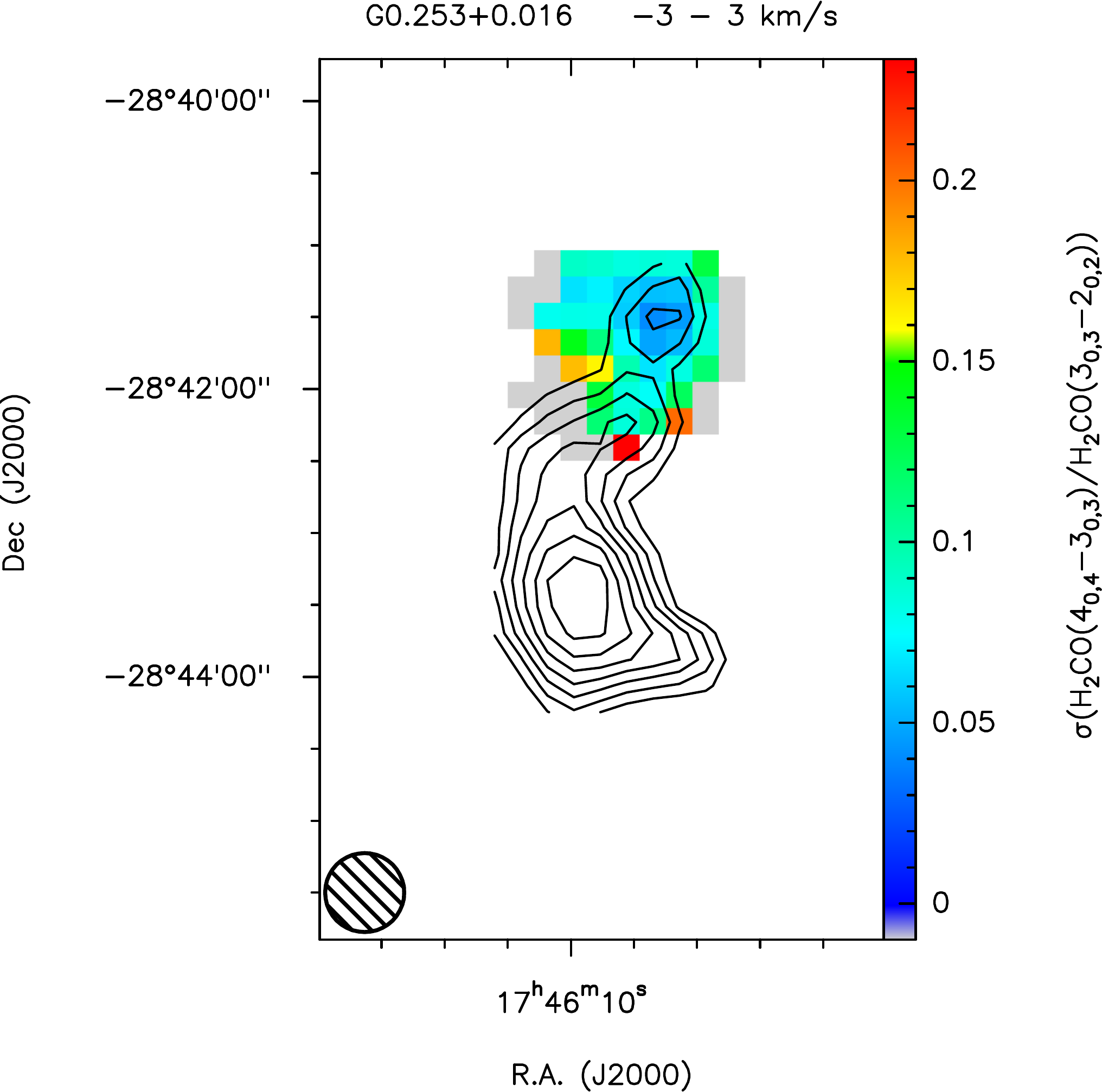}}
	\subfloat{\includegraphics[bb = 150 0 540 560, clip, height=4.737cm]{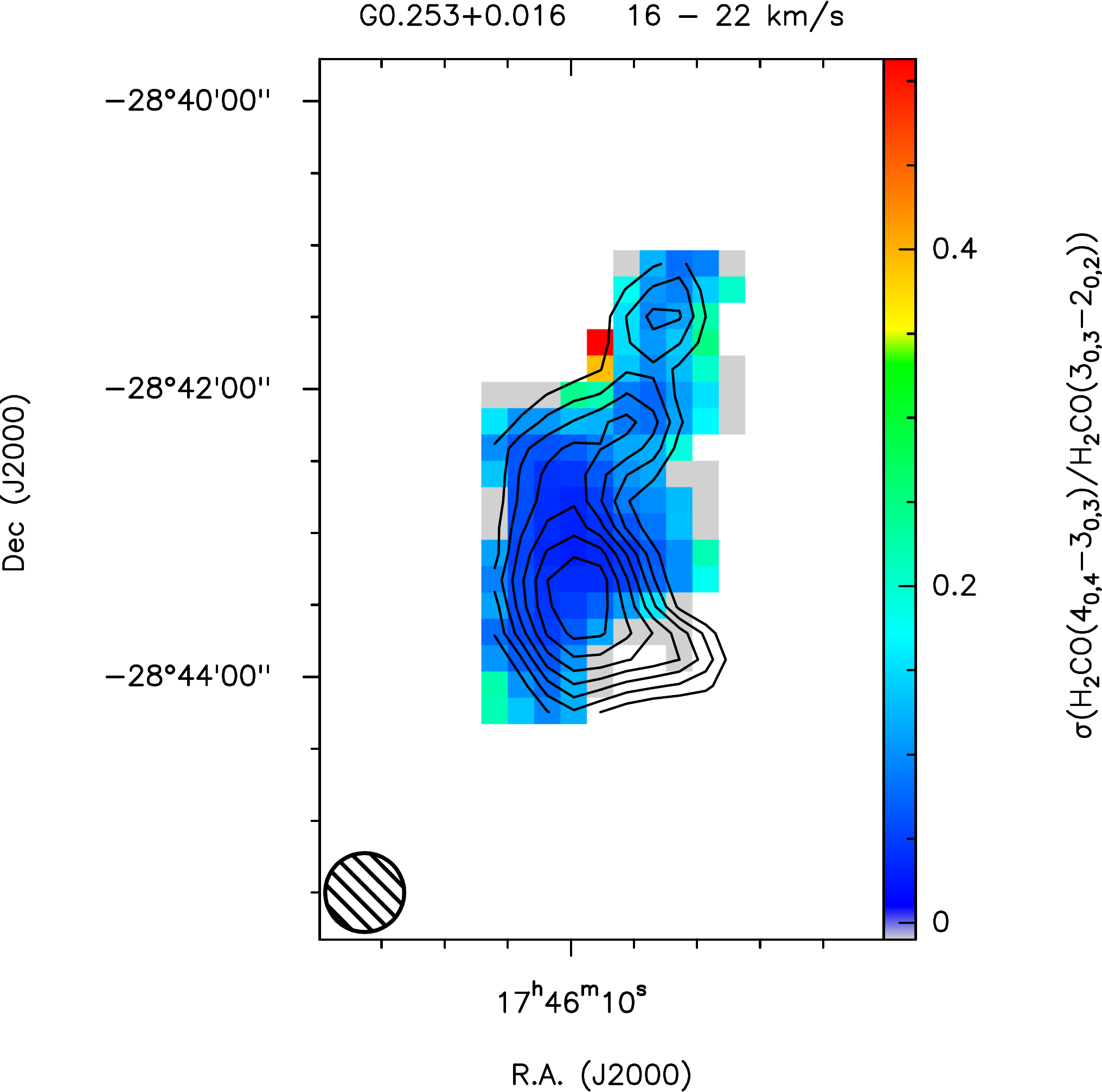}}
	\subfloat{\includegraphics[bb = 150 0 540 560, clip, height=4.737cm]{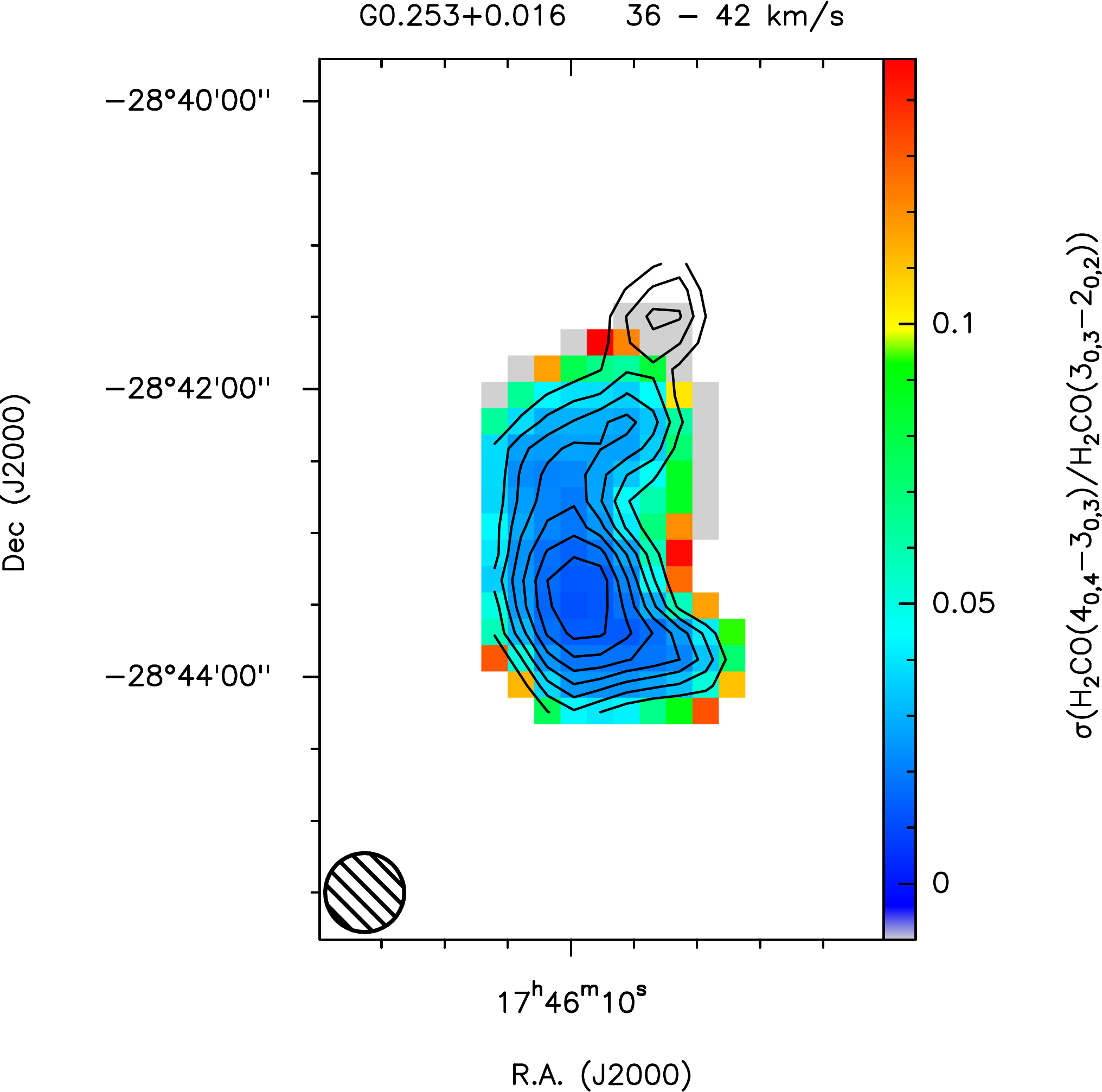}}
	\subfloat{\includegraphics[bb = 150 0 600 560, clip, height=4.737cm]{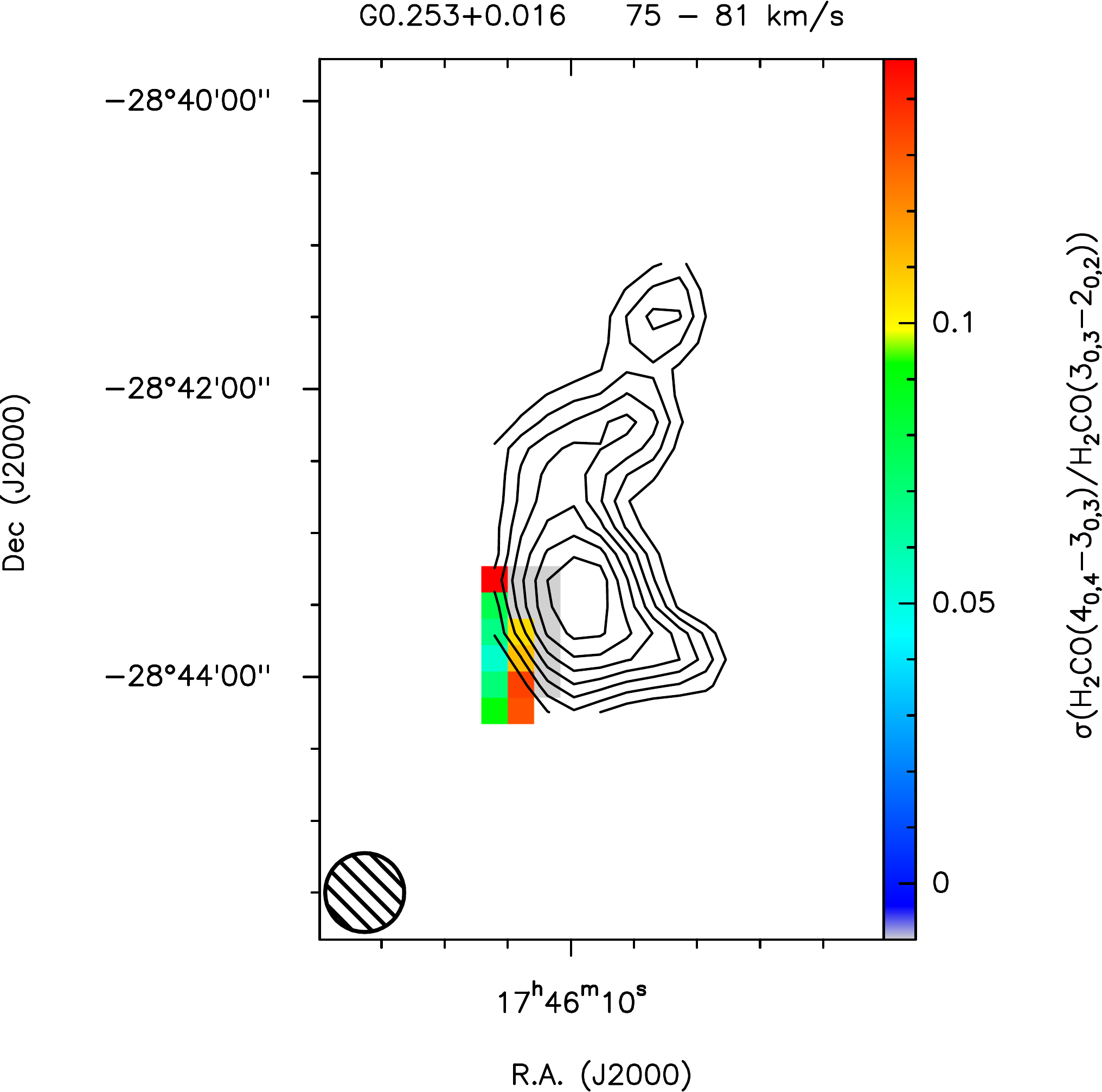}}
\end{figure*}

\begin{figure*}
	\caption{As Fig. \ref{20kms-All-Ratio-H2CO} for G0.411+0.050.}
	\centering
        R$_{321}$\\
	\subfloat{\includegraphics[bb = 0 60 610 580, clip, height=5cm]{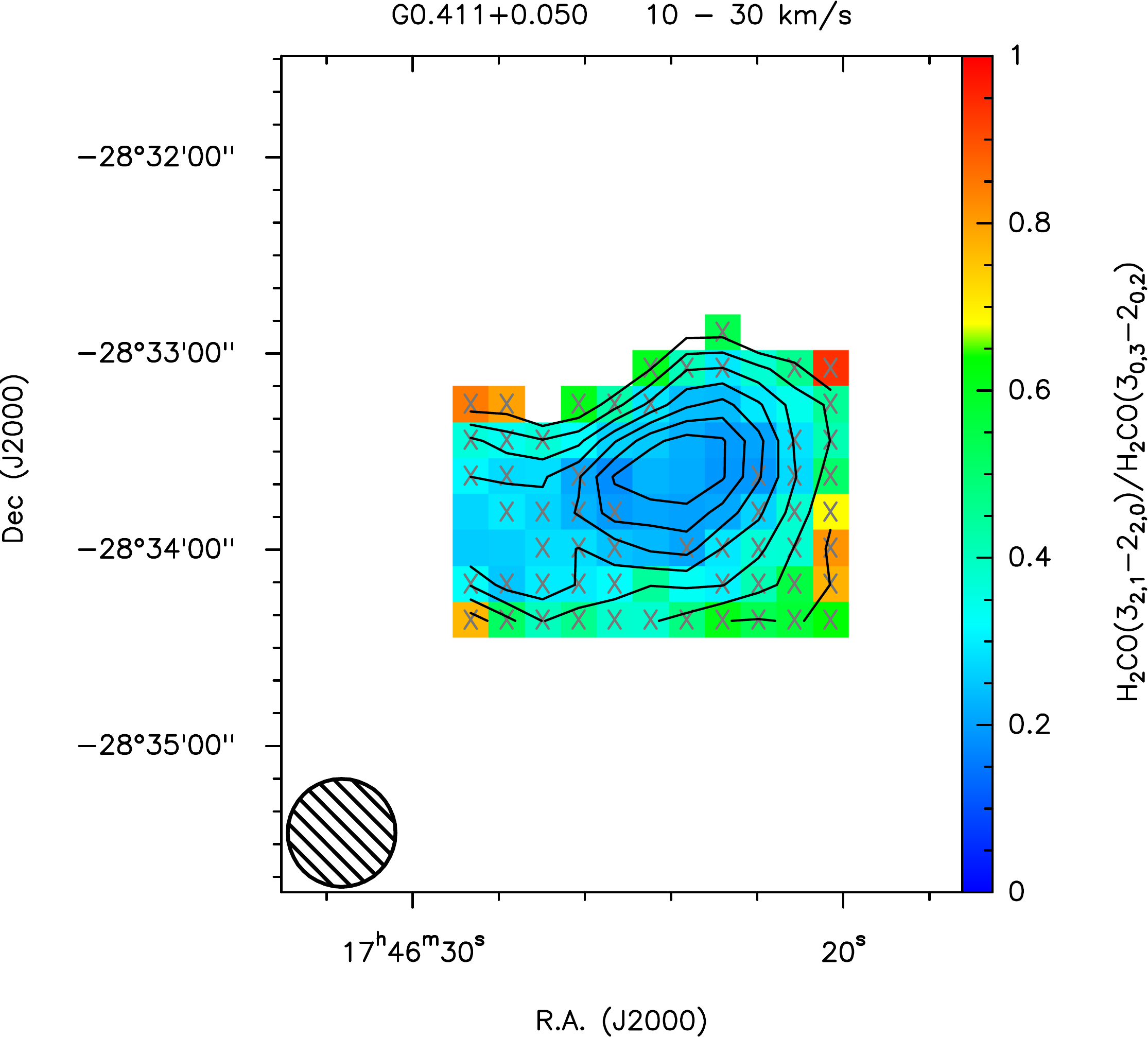}}
	\subfloat{\includegraphics[bb = 150 60 640 580, clip, height=5cm]{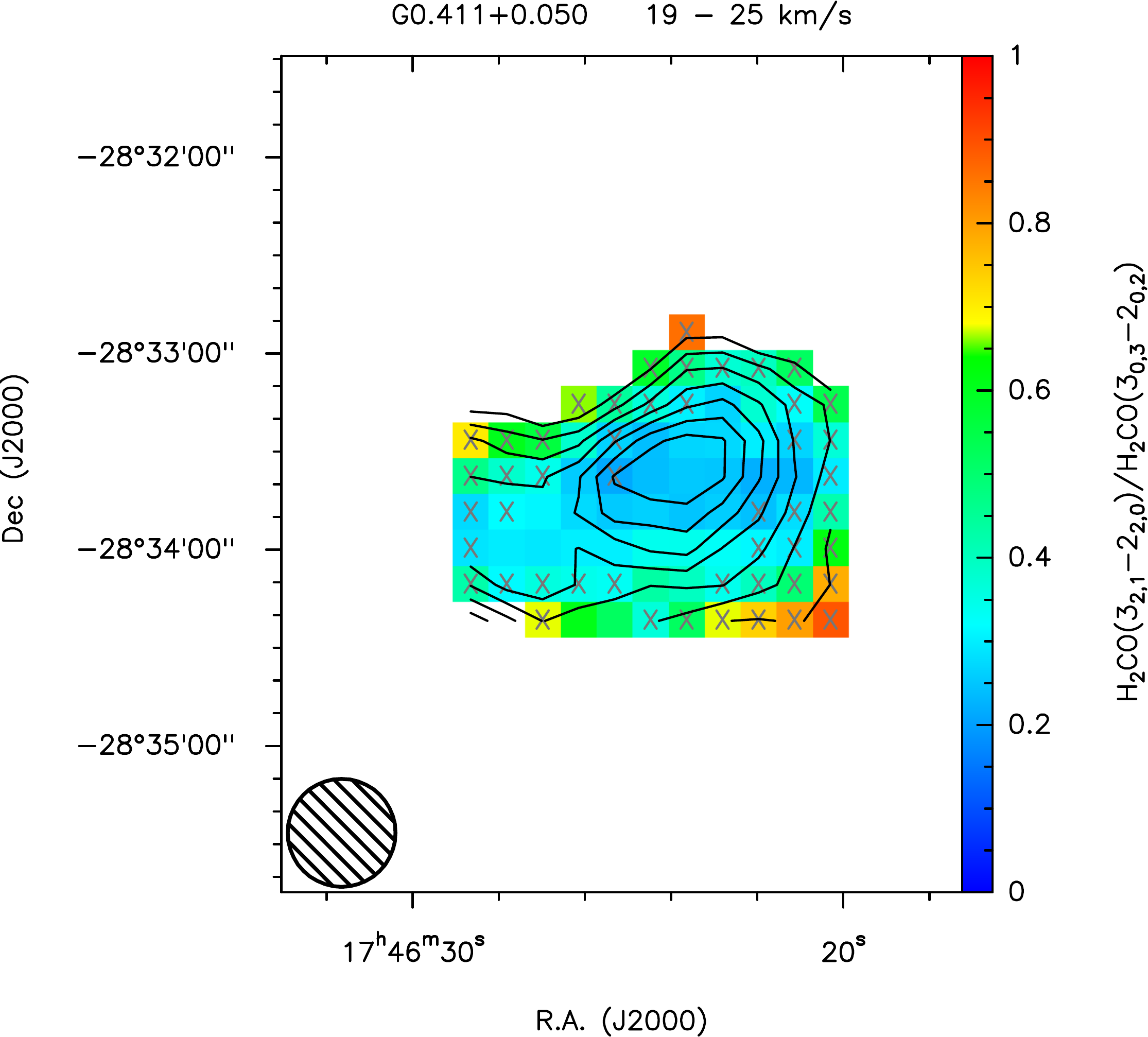}}\\
	\vspace{-0.5cm}
	\subfloat{\includegraphics[bb = 0 0 610 560, clip, height=5.39cm]{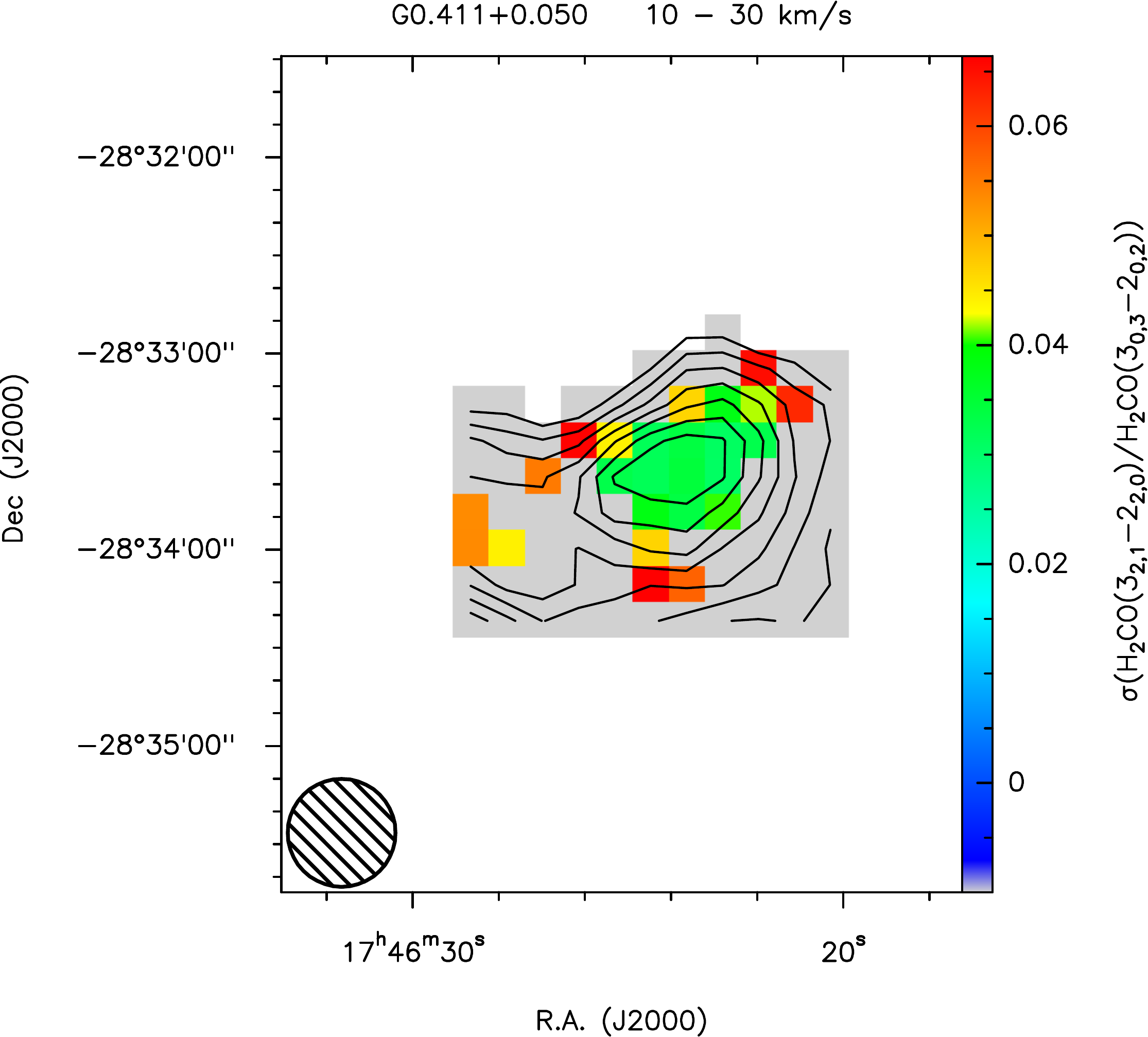}}
	\subfloat{\includegraphics[bb = 150 0 640 560, clip, height=5.39cm]{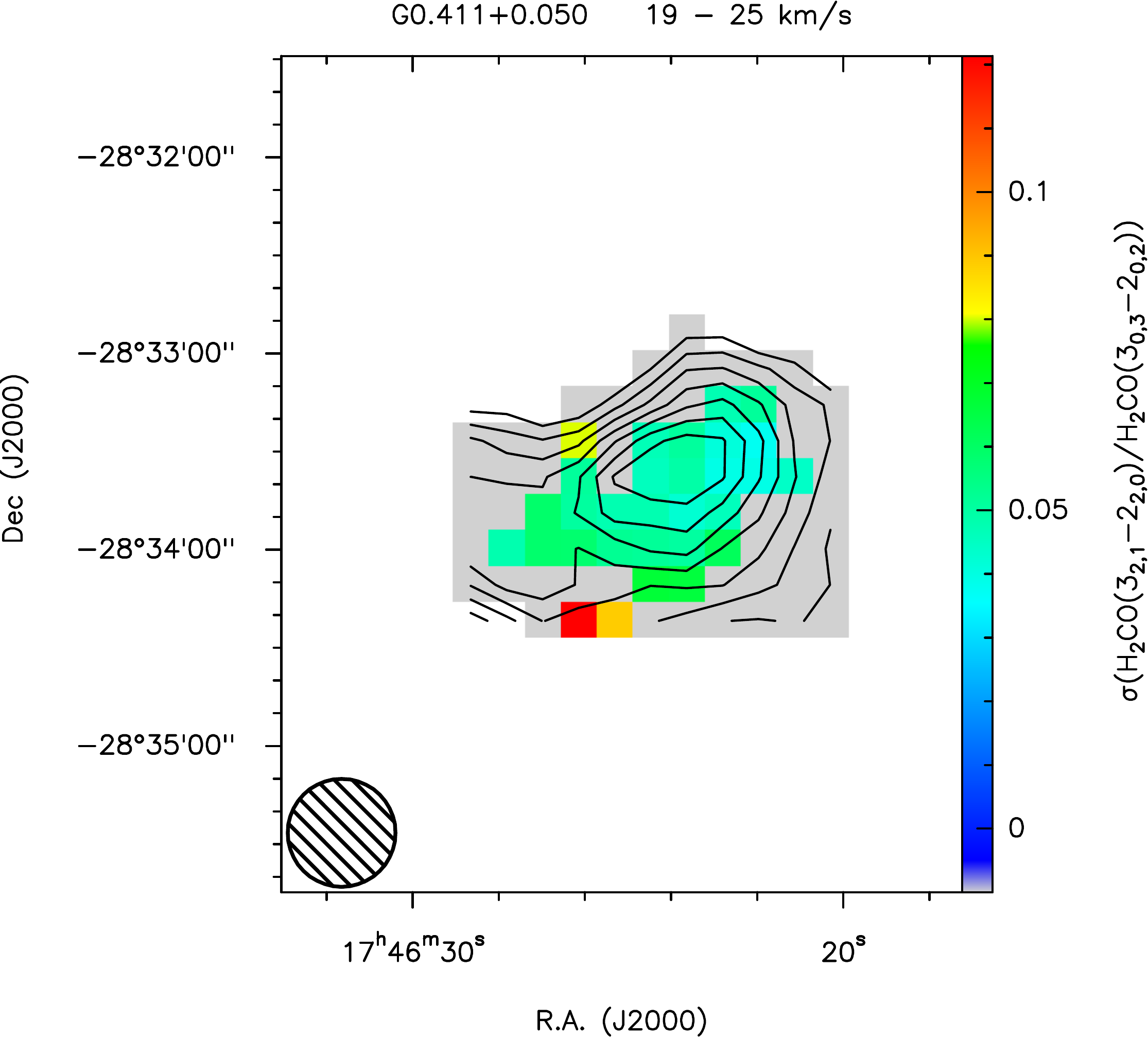}}\\
	\vspace{0.1cm}
        R$_{422}$\\
	\subfloat{\includegraphics[bb = 0 60 610 580, clip, height=5cm]{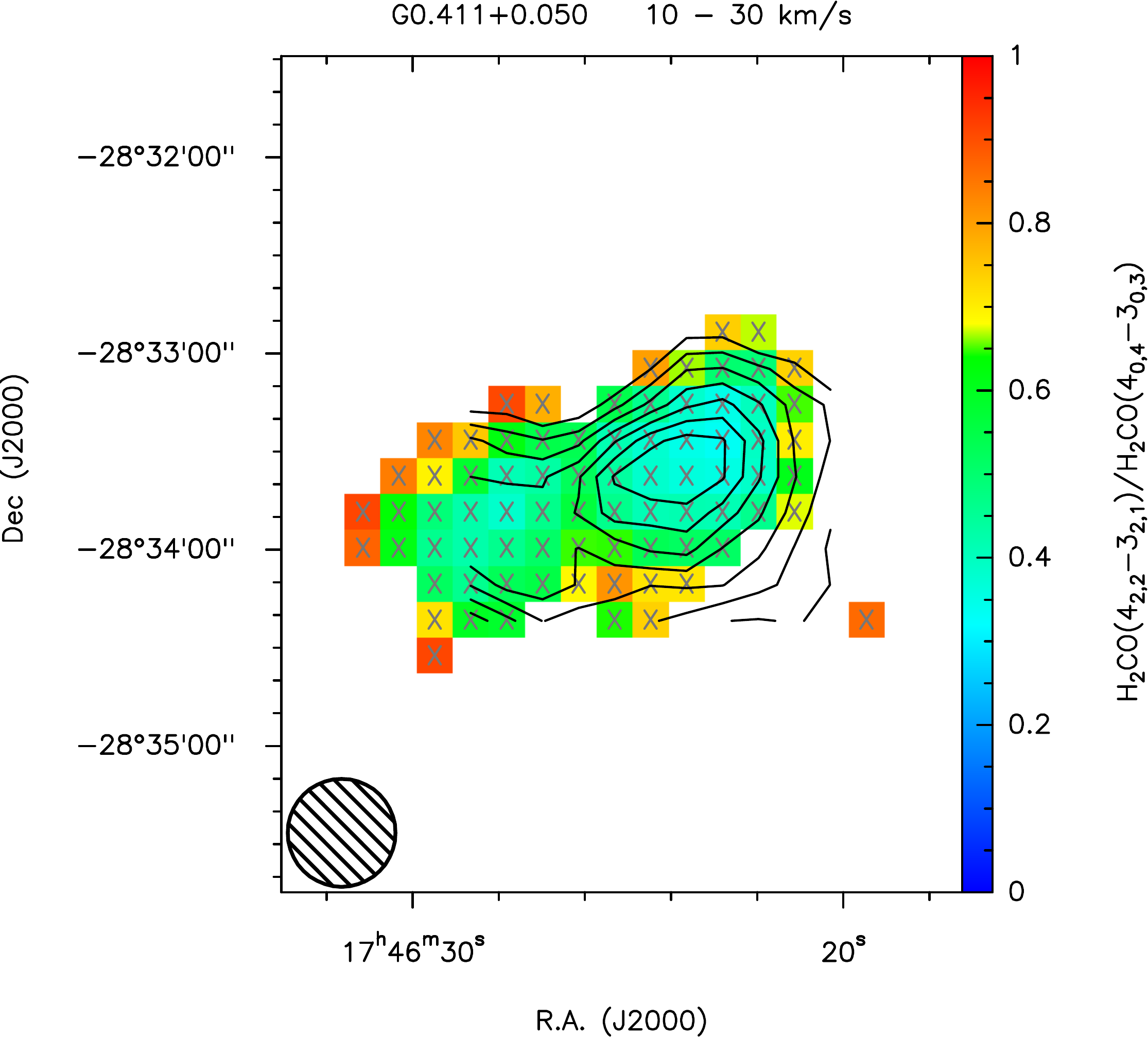}}
	\subfloat{\includegraphics[bb = 150 60 640 580, clip, height=5cm]{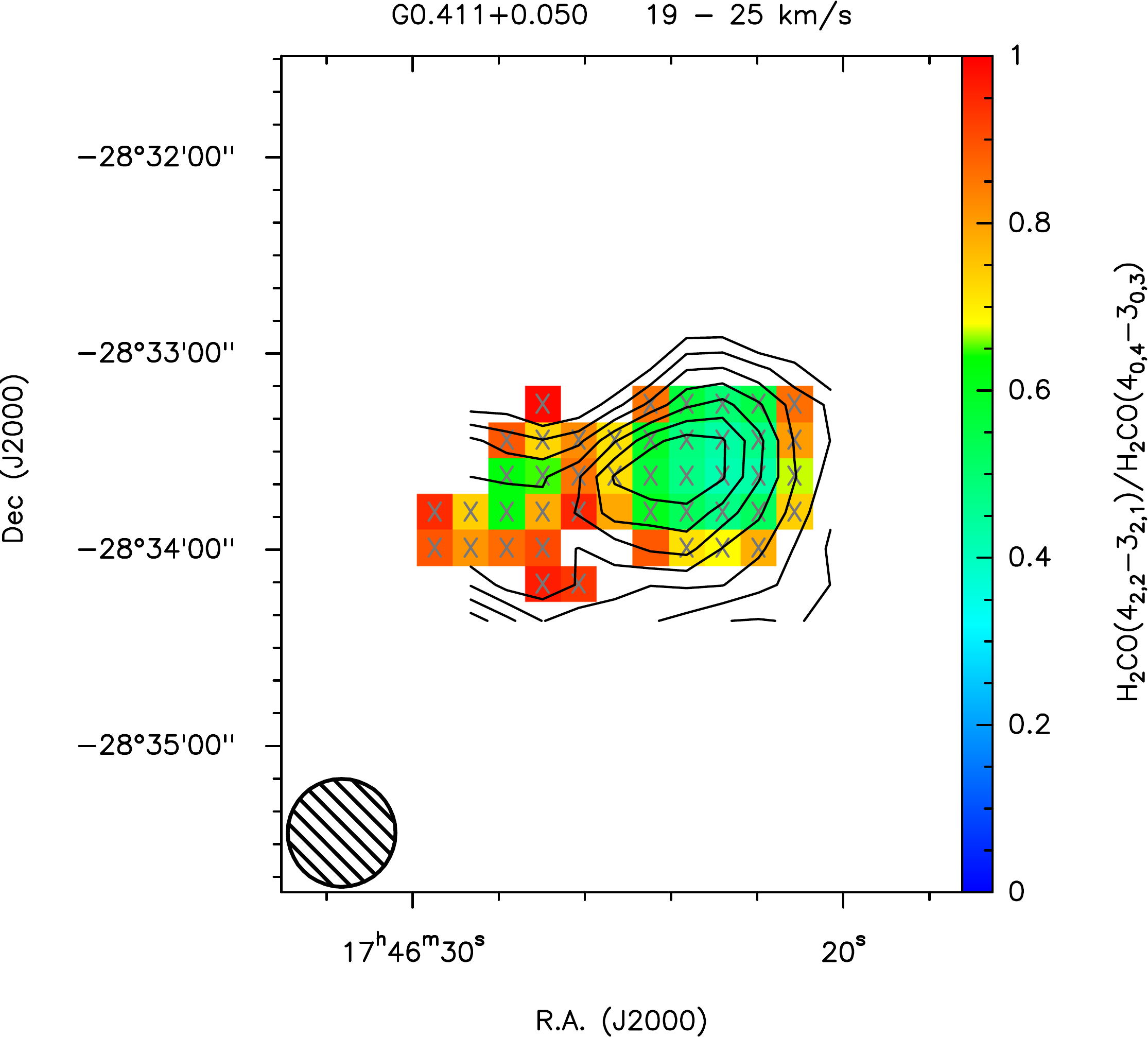}}\\
	\vspace{-0.5cm}
	\subfloat{\includegraphics[bb = 0 0 610 560, clip, height=5.39cm]{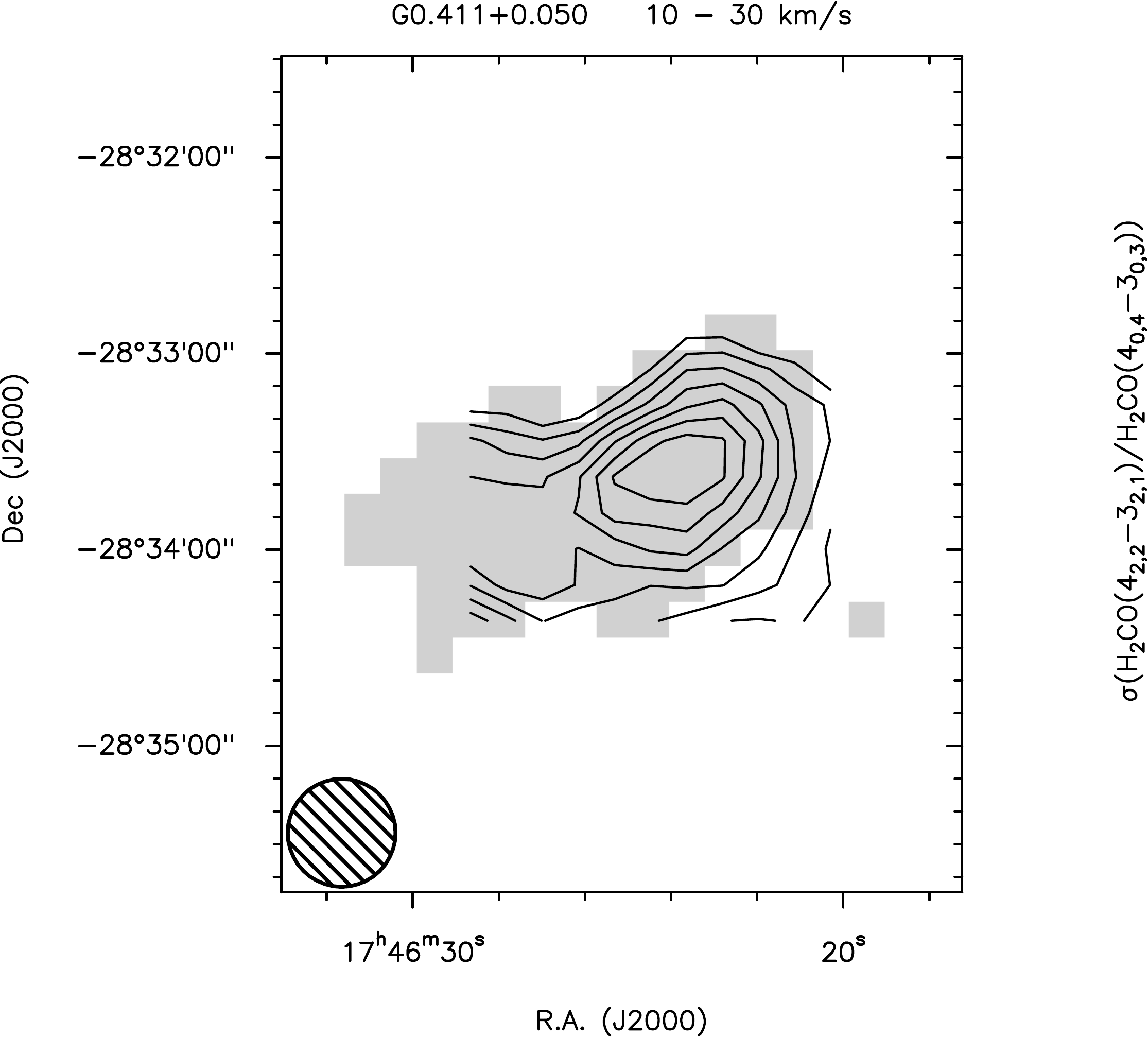}}
	\subfloat{\includegraphics[bb = 150 0 640 560, clip, height=5.39cm]{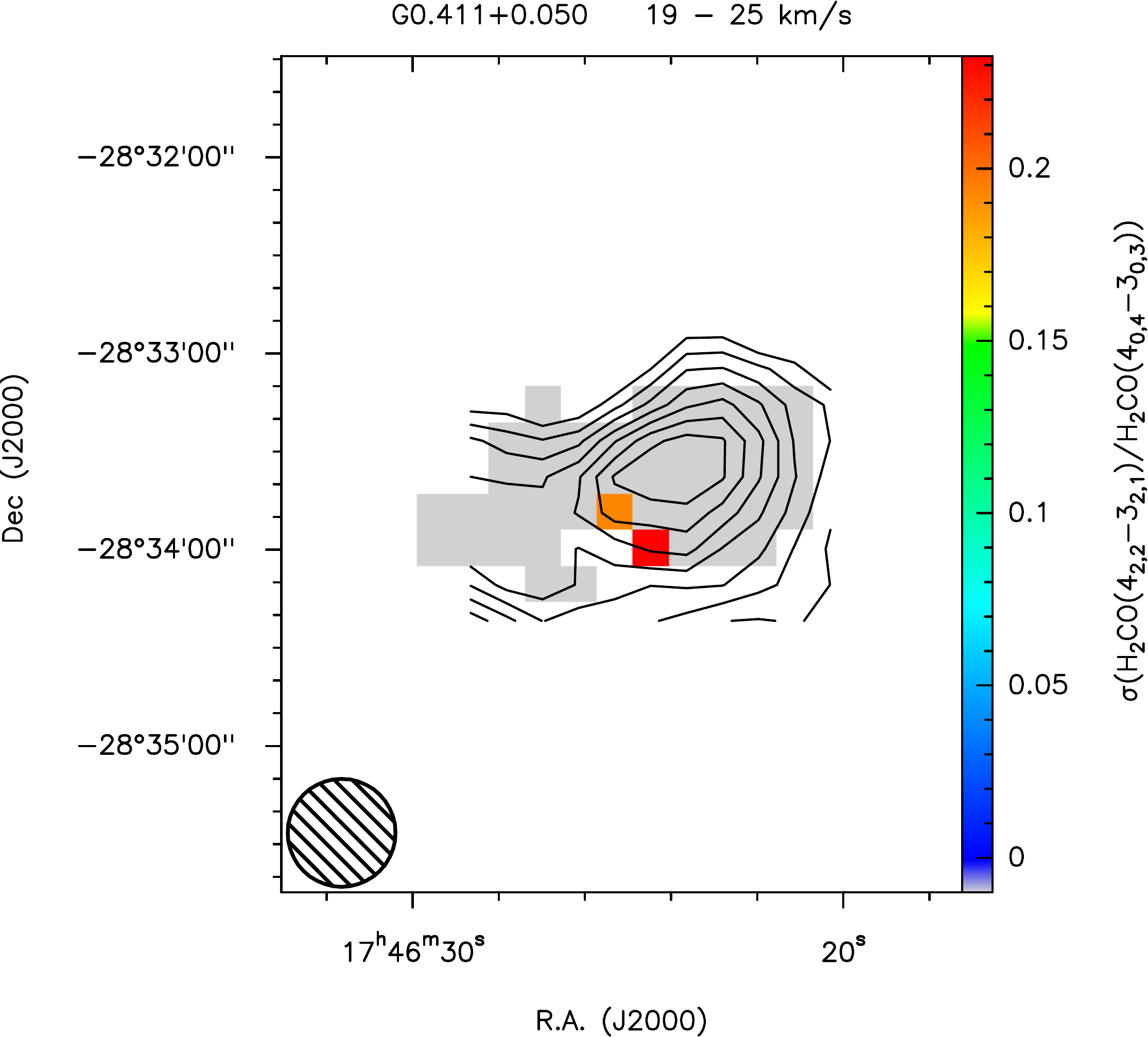}}
	\label{G0411-All-Ratio-H2CO}
\end{figure*}

\begin{figure*}
	\centering
	R$_{404}$ \\
	\subfloat{\includegraphics[bb = 0 60 610 580, clip, height=5cm]{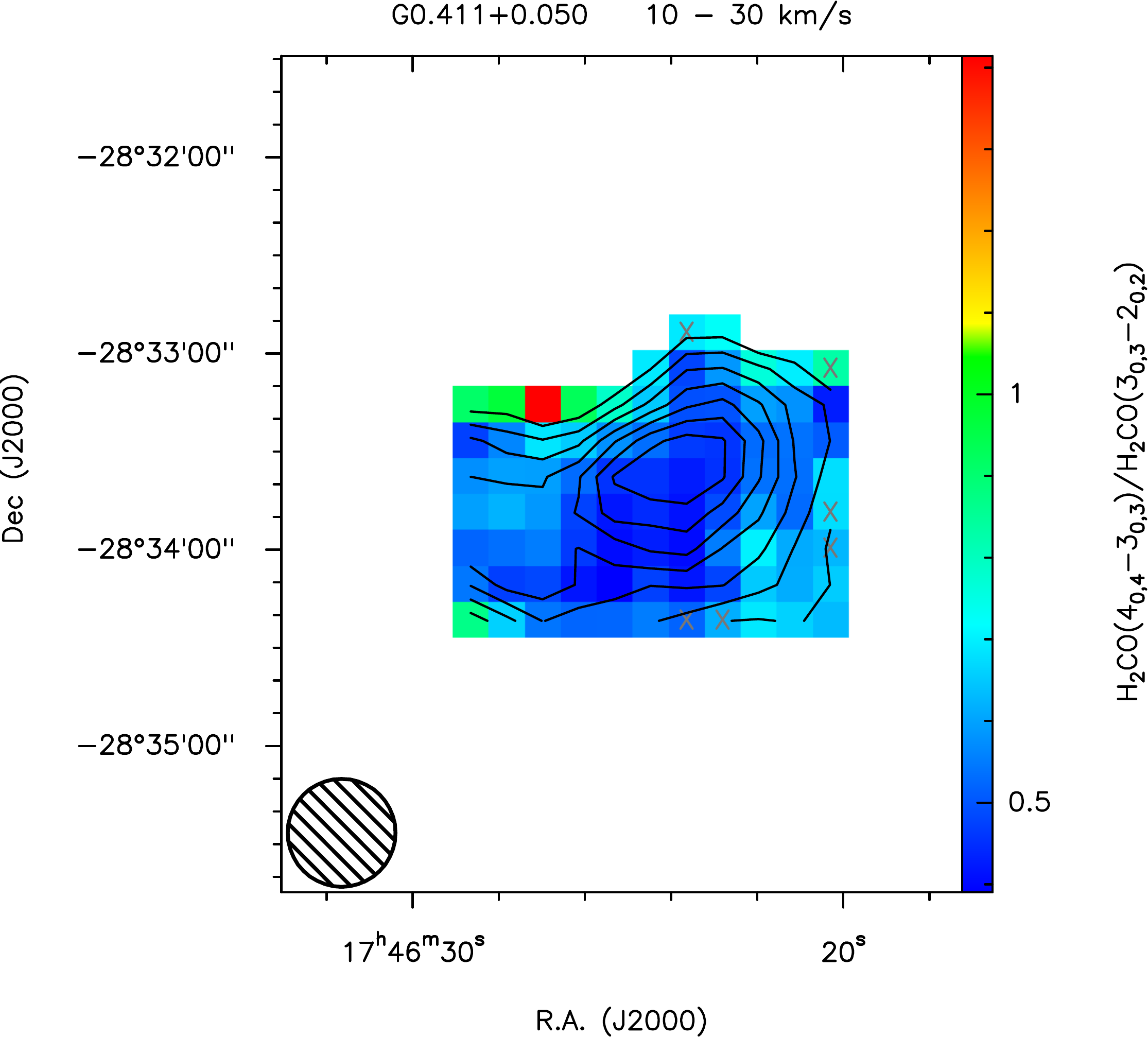}}
	\subfloat{\includegraphics[bb = 150 60 640 580, clip, height=5cm]{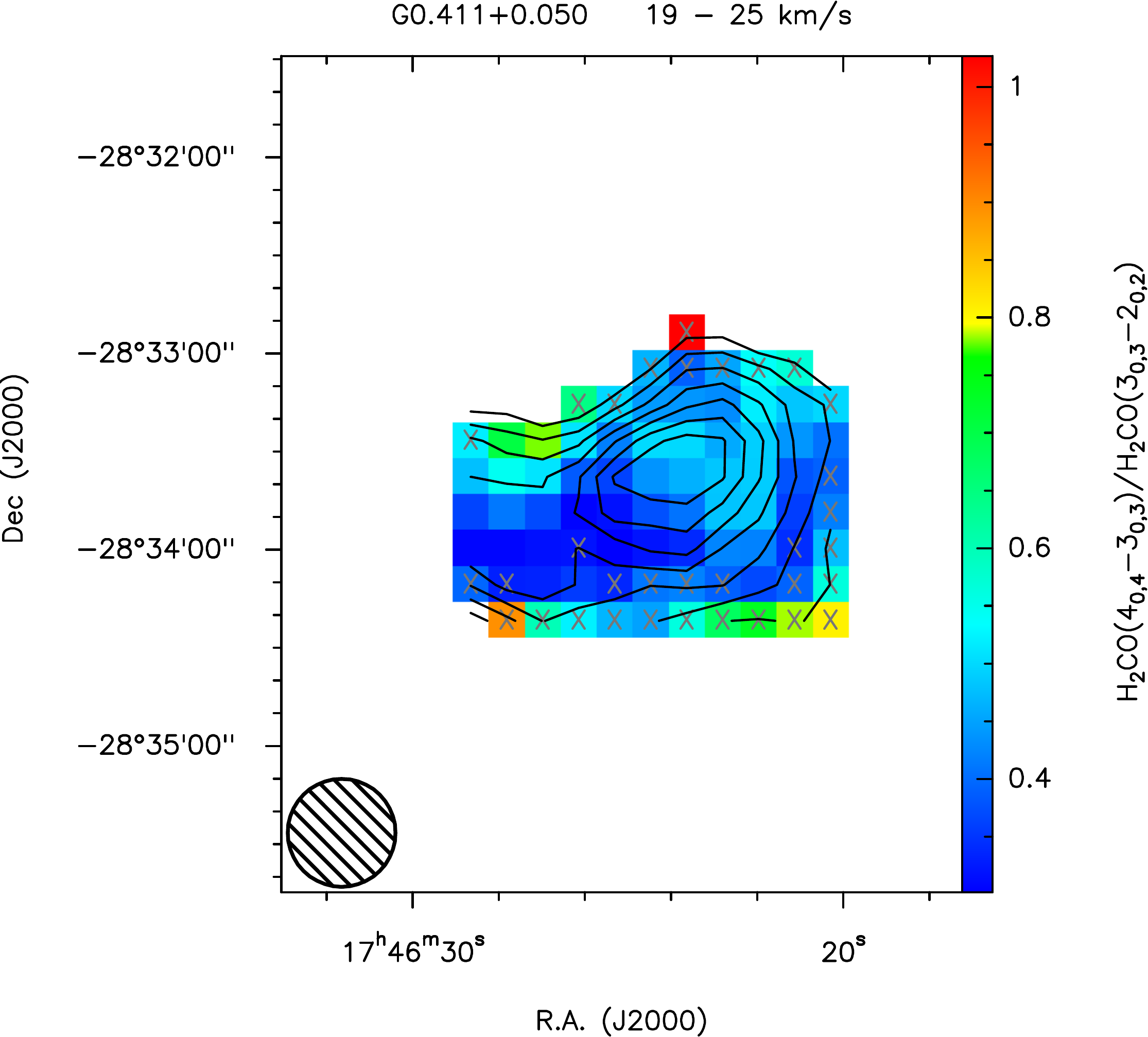}}\\
	\vspace{-0.5cm}
	\subfloat{\includegraphics[bb = 0 0 610 560, clip, height=5.39cm]{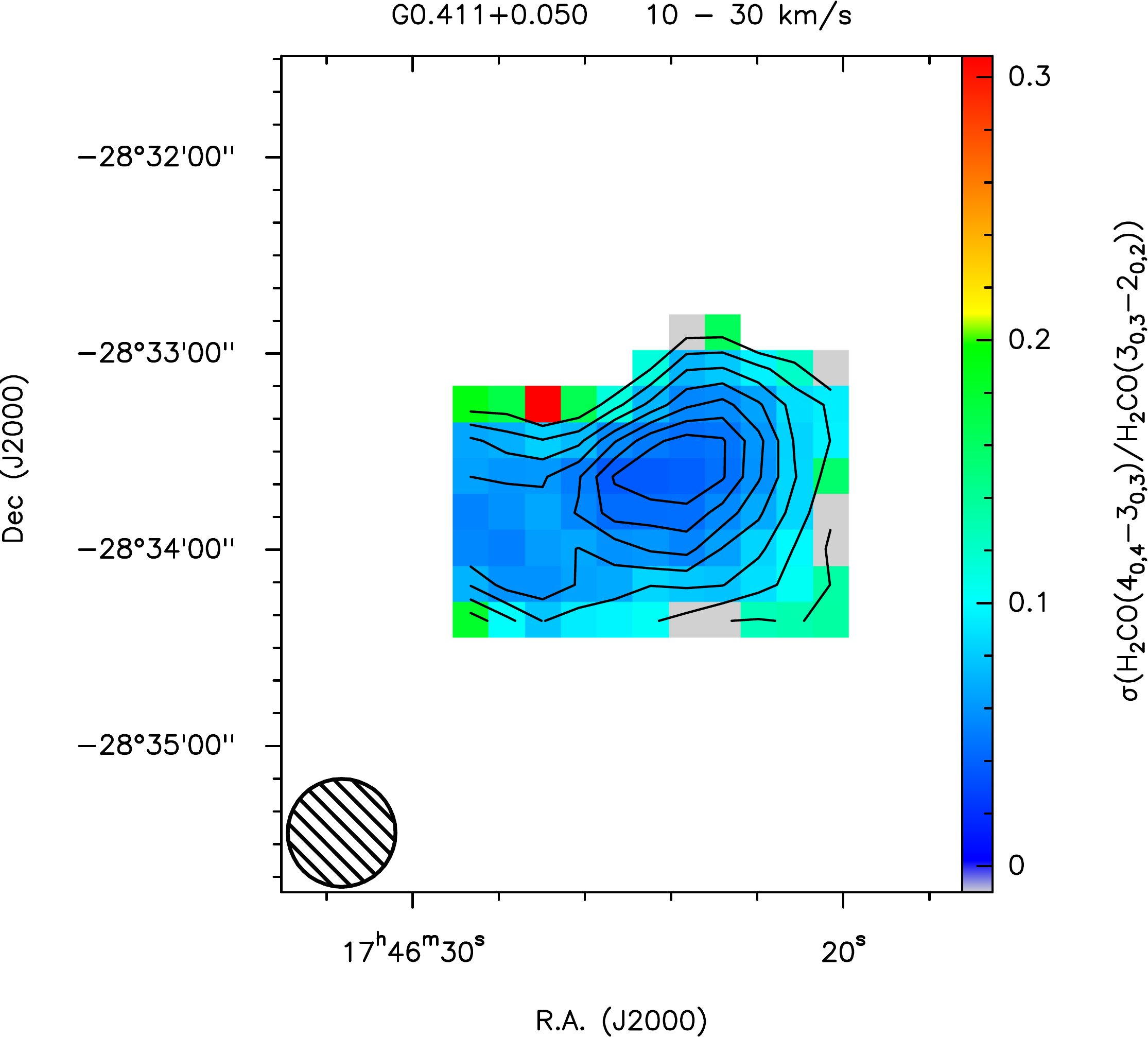}}
	\subfloat{\includegraphics[bb = 150 0 640 560, clip, height=5.39cm]{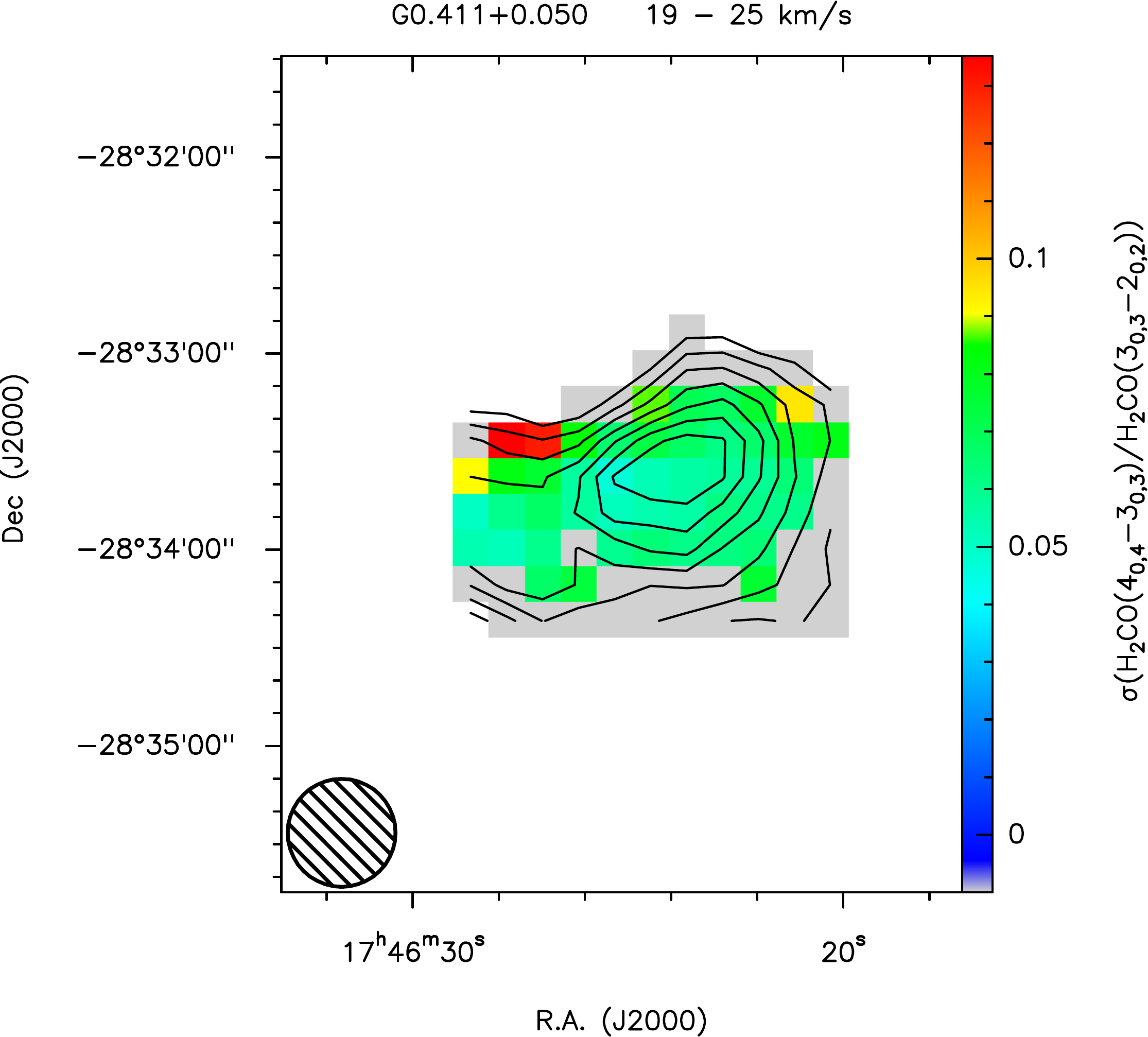}}\\
\end{figure*}

\begin{figure*}
	\caption{As Fig. \ref{20kms-All-Ratio-H2CO} for G0.480$-$0.006.}
	\centering
        R$_{321}$\\
	\subfloat{\includegraphics[bb = 0 60 610 580, clip, height=5cm]{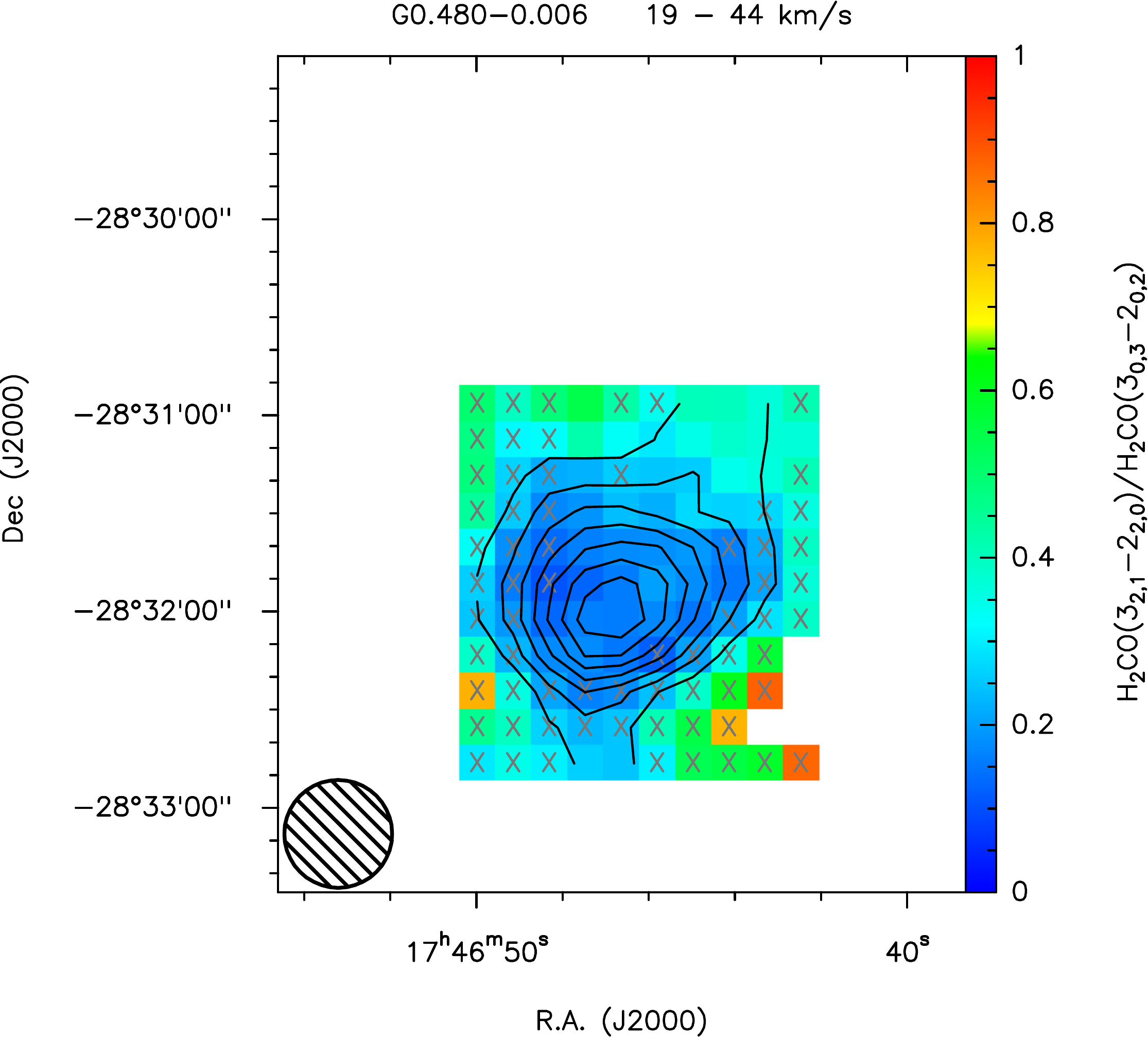}}
	\subfloat{\includegraphics[bb = 150 60 640 580, clip, height=5cm]{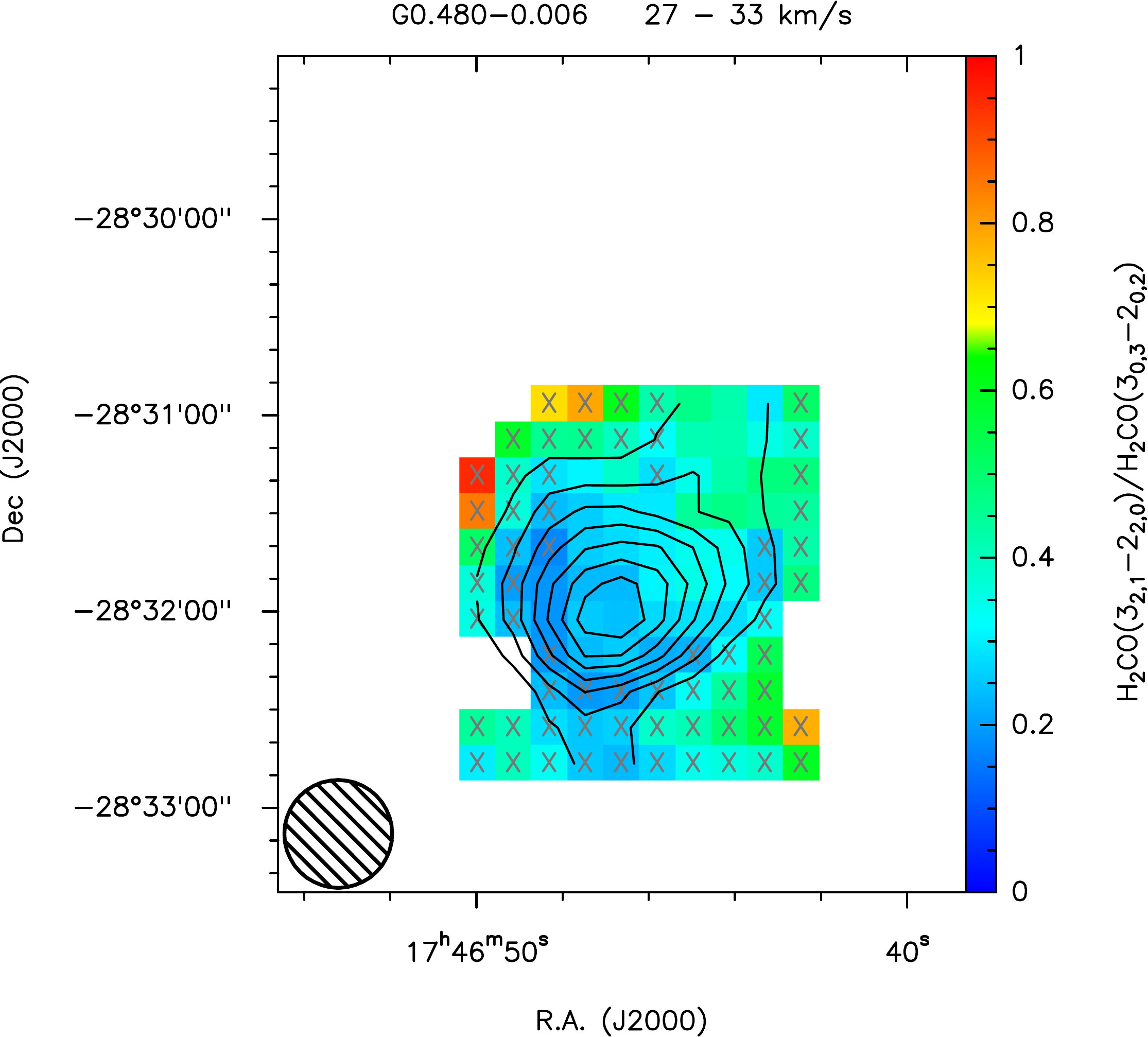}}\\
	\vspace{-0.5cm}
	\subfloat{\includegraphics[bb = 0 0 610 560, clip, height=5.39cm]{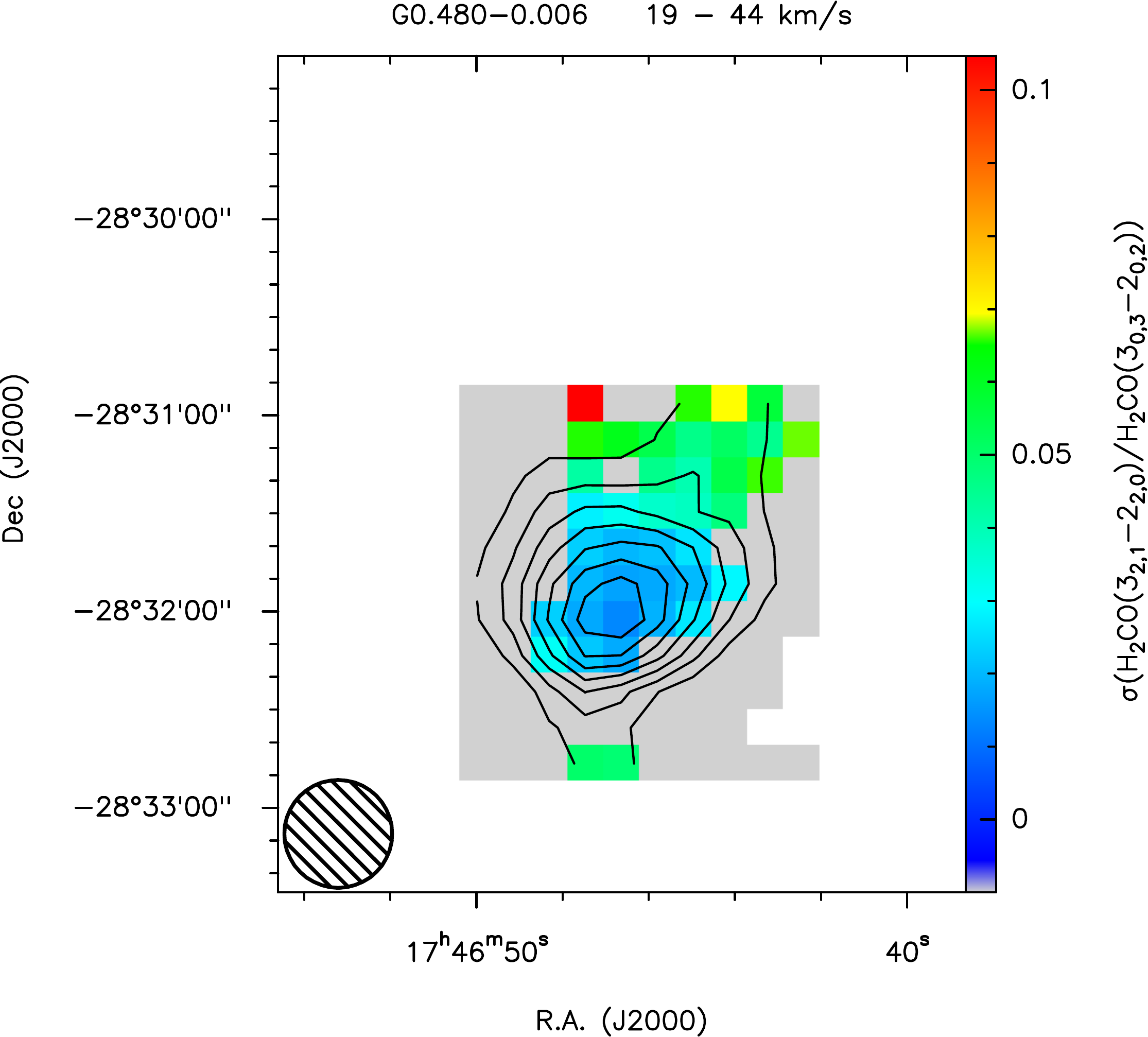}}
	\subfloat{\includegraphics[bb = 150 0 640 560, clip, height=5.39cm]{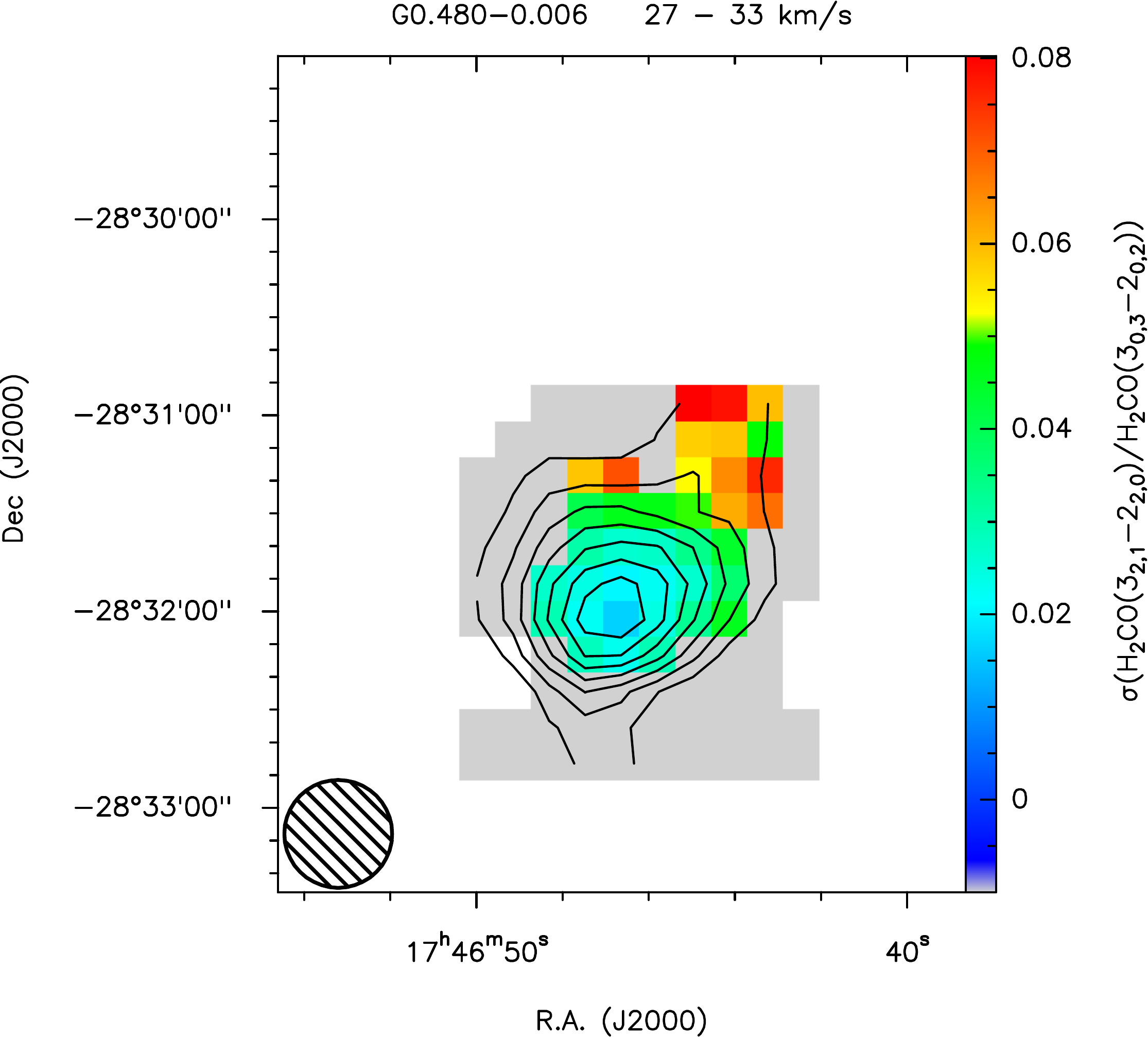}}\\
	\vspace{0.1cm}
        R$_{422}$\\
	\subfloat{\includegraphics[bb = 0 60 610 580, clip, height=5cm]{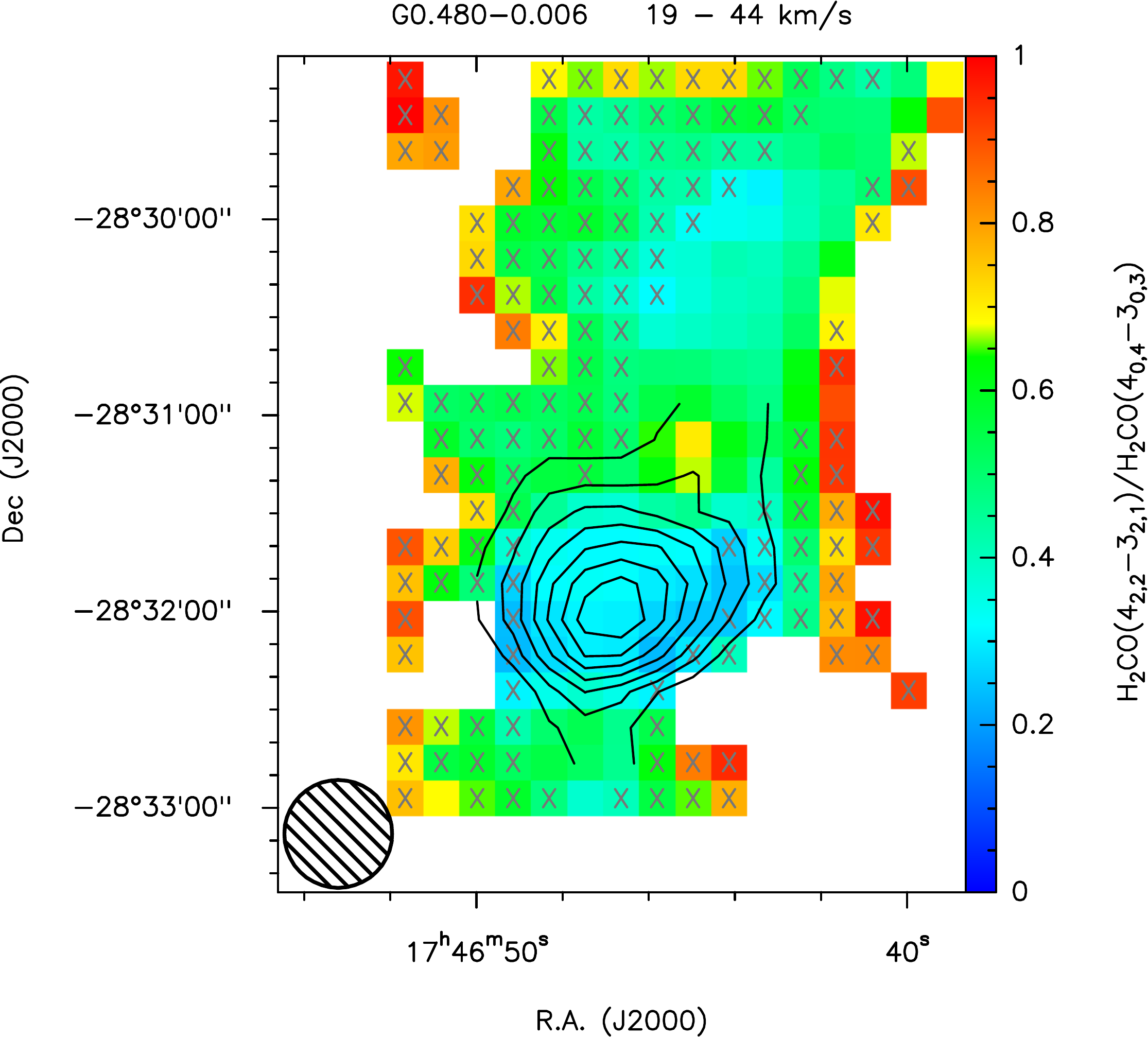}}
	\subfloat{\includegraphics[bb = 150 60 640 580, clip, height=5cm]{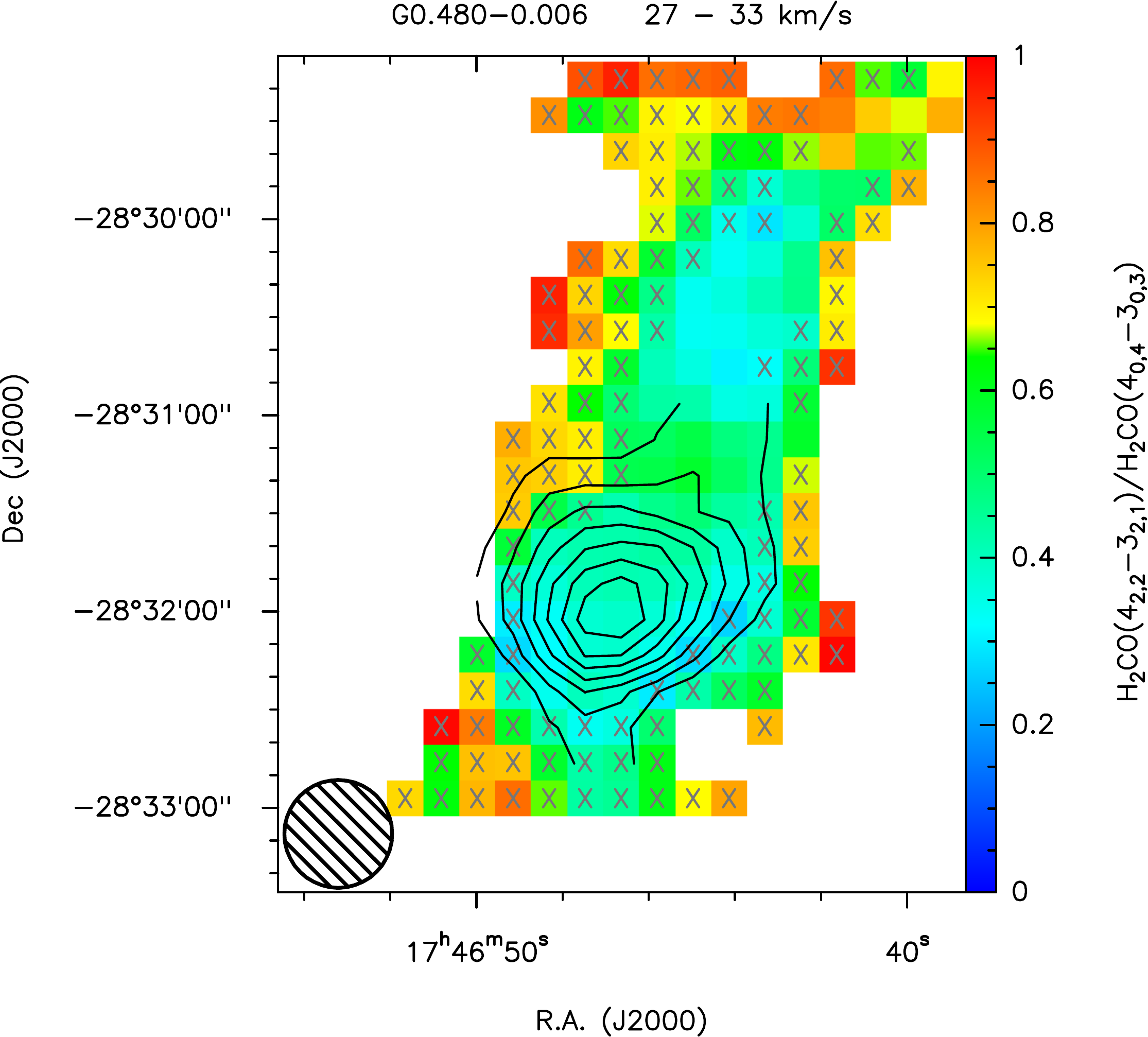}}\\
	\vspace{-0.5cm}
	\subfloat{\includegraphics[bb = 0 0 610 560, clip, height=5.39cm]{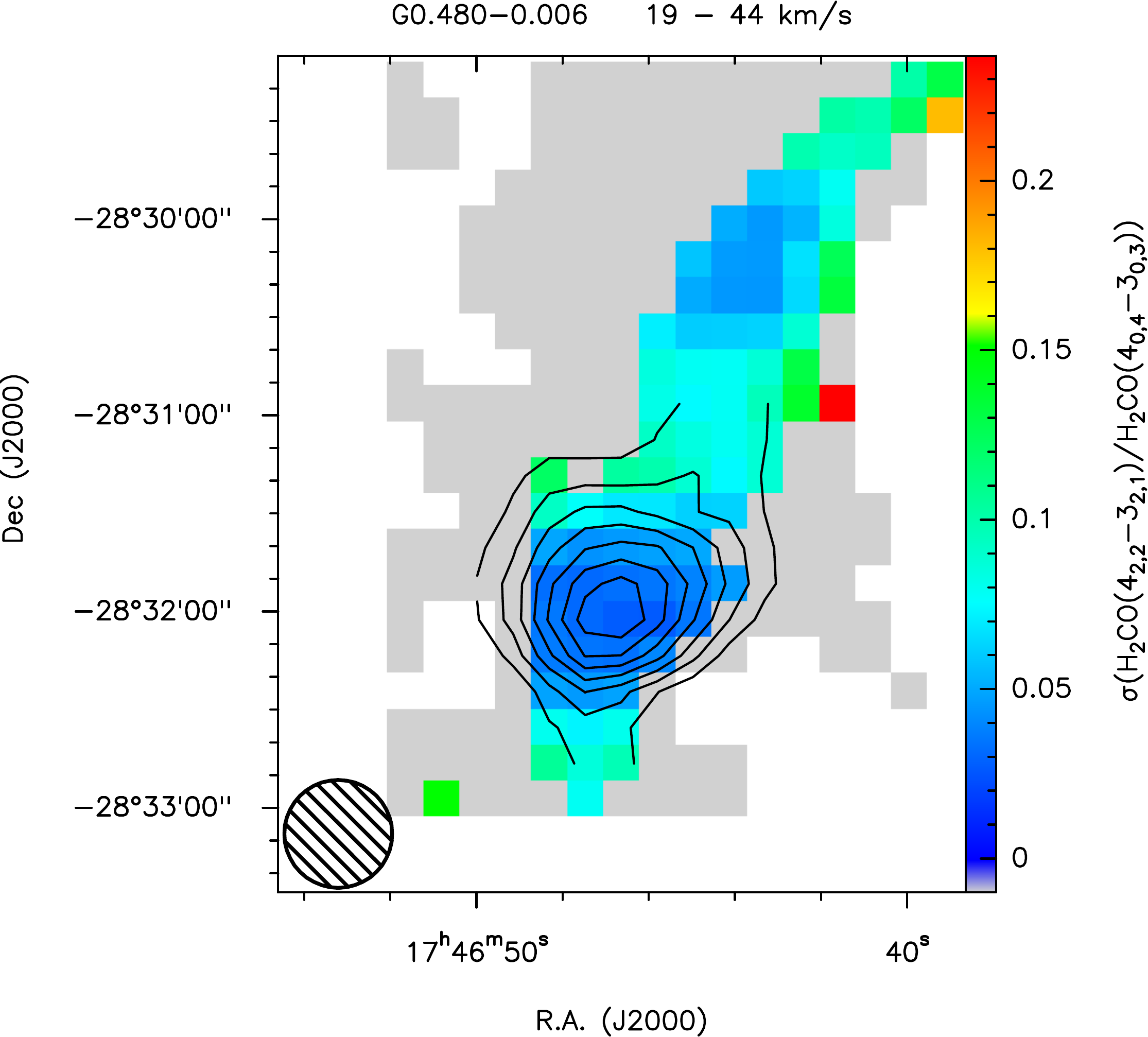}}
	\subfloat{\includegraphics[bb = 150 0 640 560, clip, height=5.39cm]{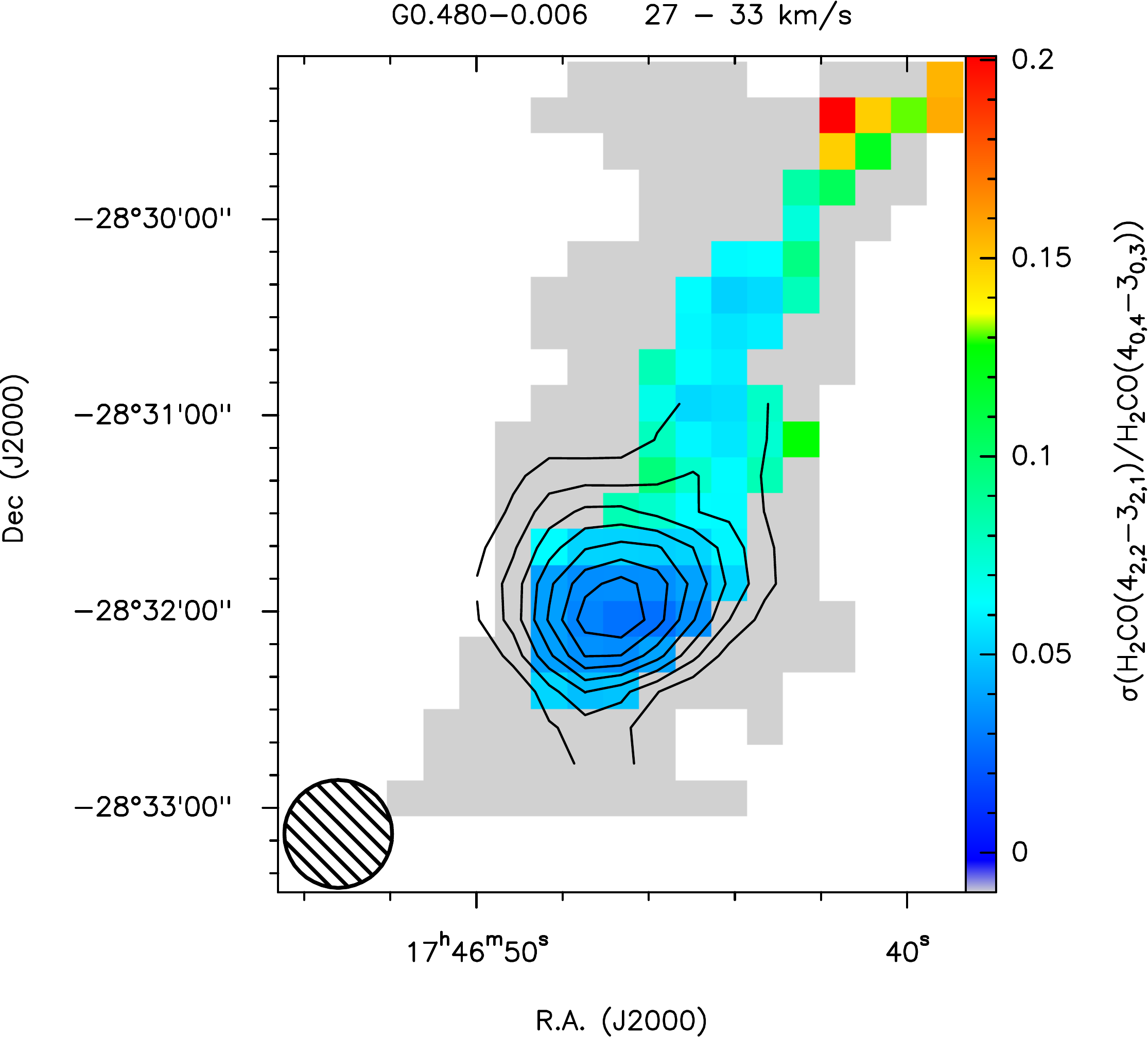}}
	\label{G0480-All-Ratio-H2CO}
\end{figure*}

\begin{figure*}
	\centering
	R$_{404}$ \\
	\subfloat{\includegraphics[bb = 0 60 610 580, clip, height=5cm]{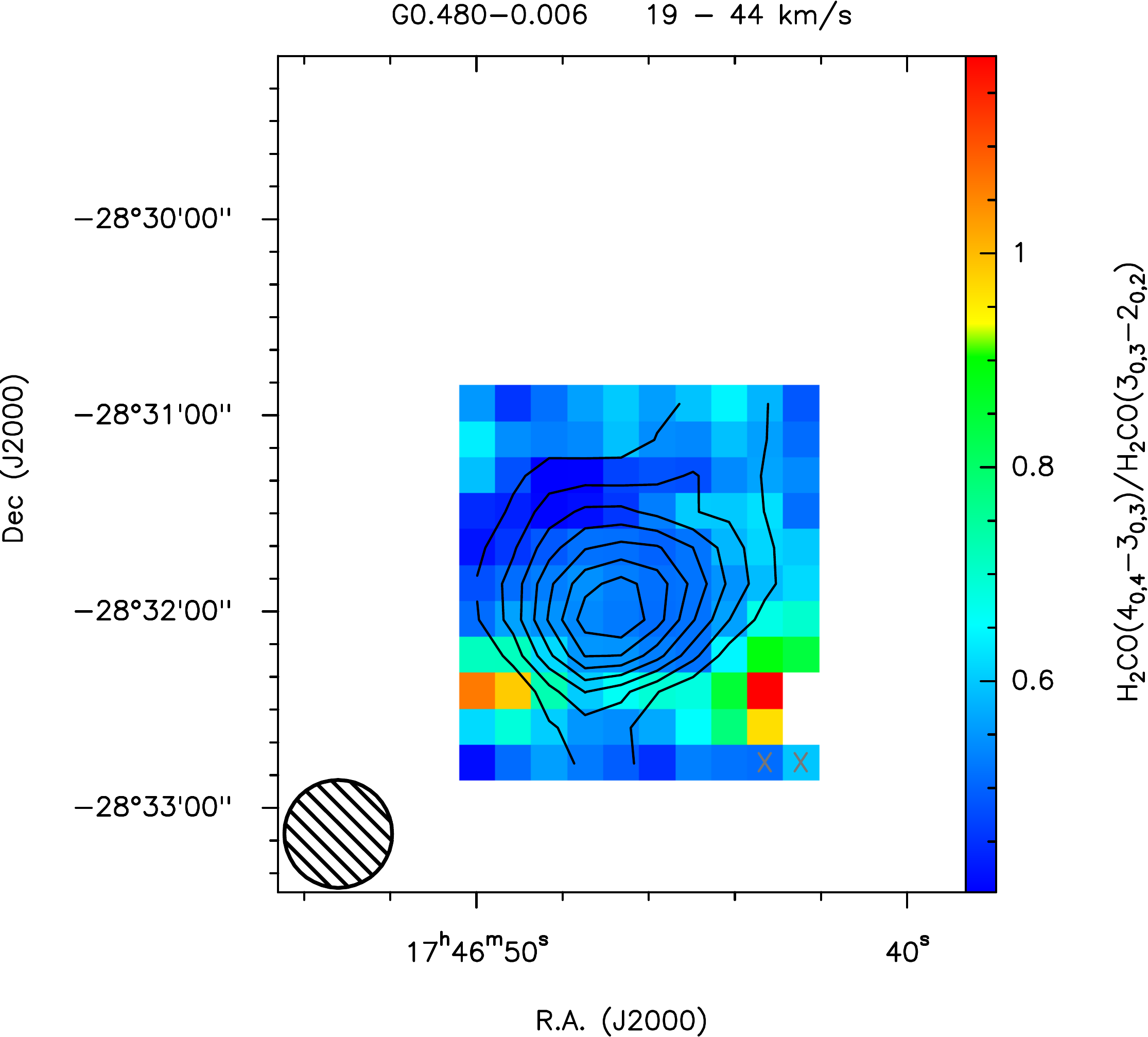}}
	\subfloat{\includegraphics[bb = 150 60 640 580, clip, height=5cm]{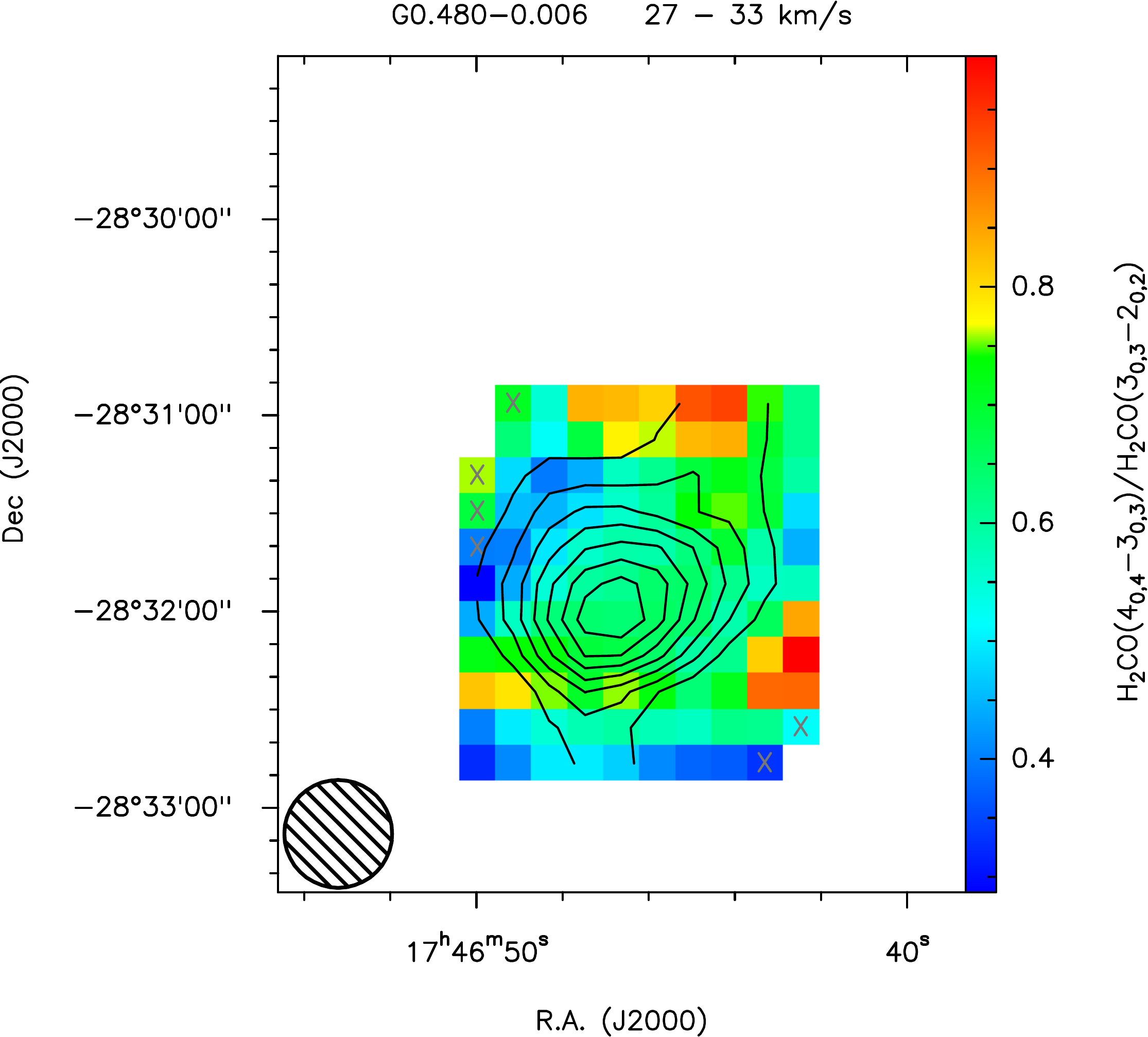}}\\
	\vspace{-0.5cm}
	\subfloat{\includegraphics[bb = 0 0 610 560, clip, height=5.39cm]{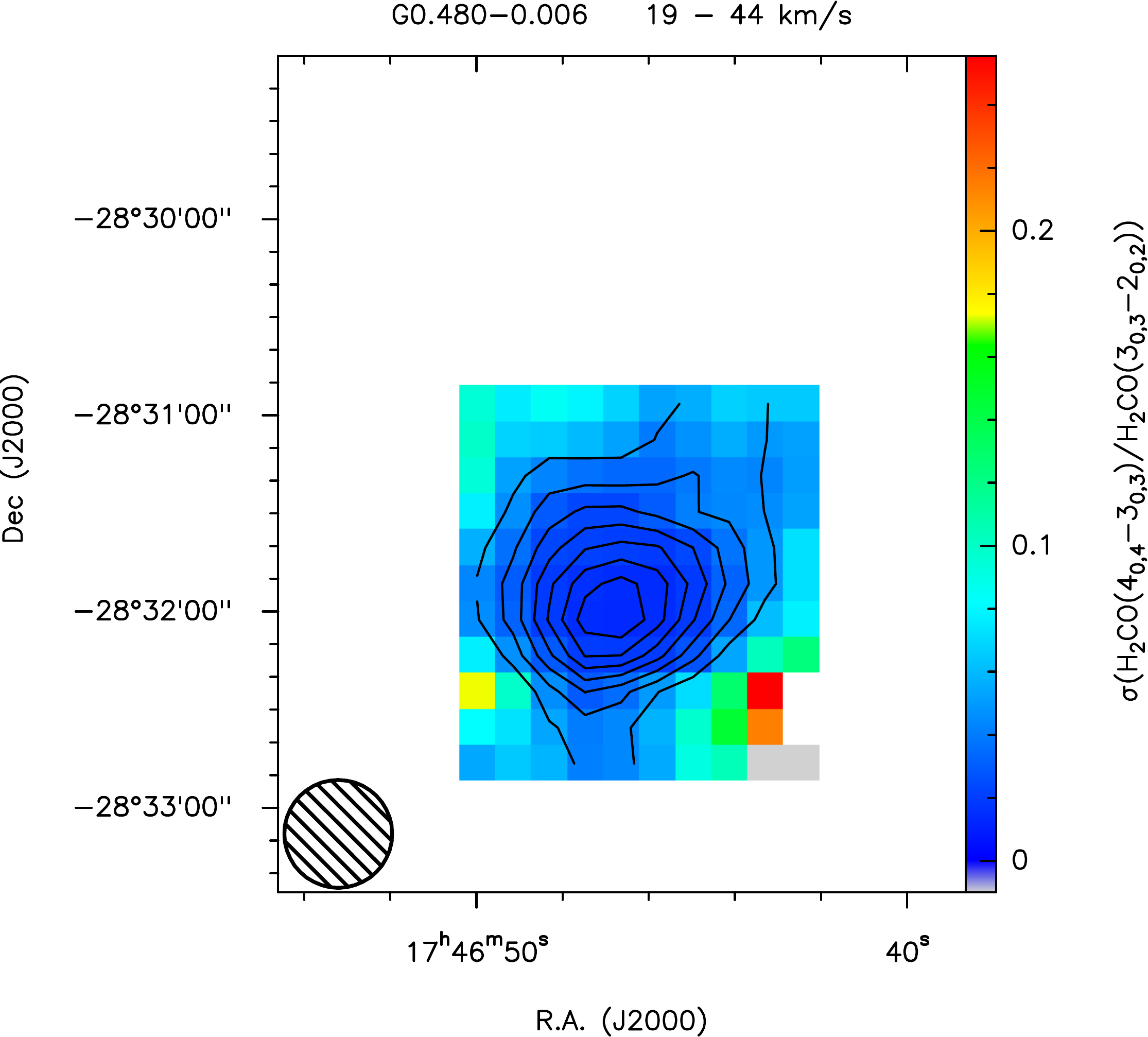}}
	\subfloat{\includegraphics[bb = 150 0 640 560, clip, height=5.39cm]{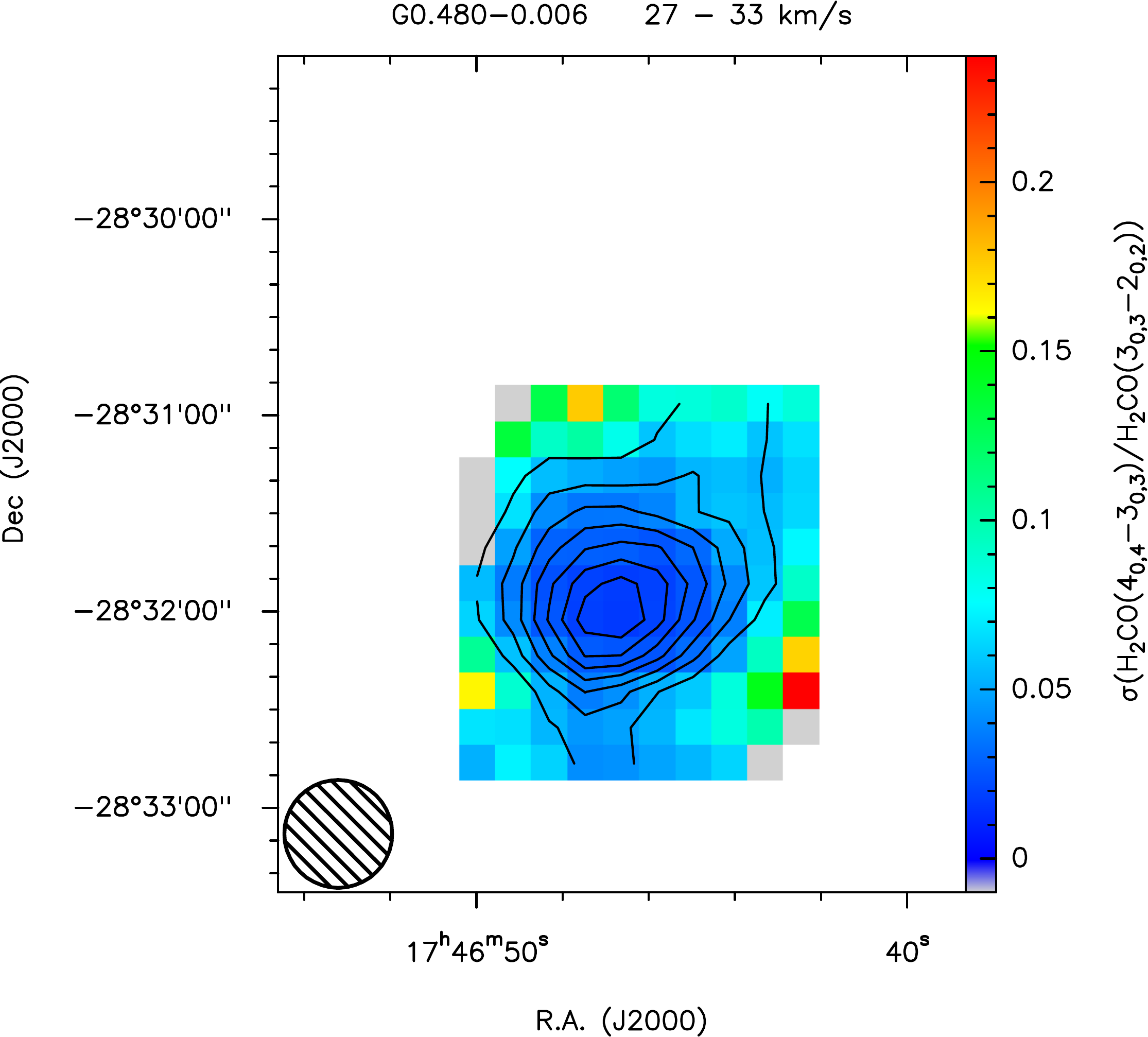}}\\
\end{figure*}

\begin{figure*}
	\caption{As Fig. \ref{20kms-All-Ratio-H2CO} for Sgr C.}
	\centering
        R$_{422}$\\
	\subfloat{\includegraphics[bb = 0 60 730 580, clip, height=5cm]{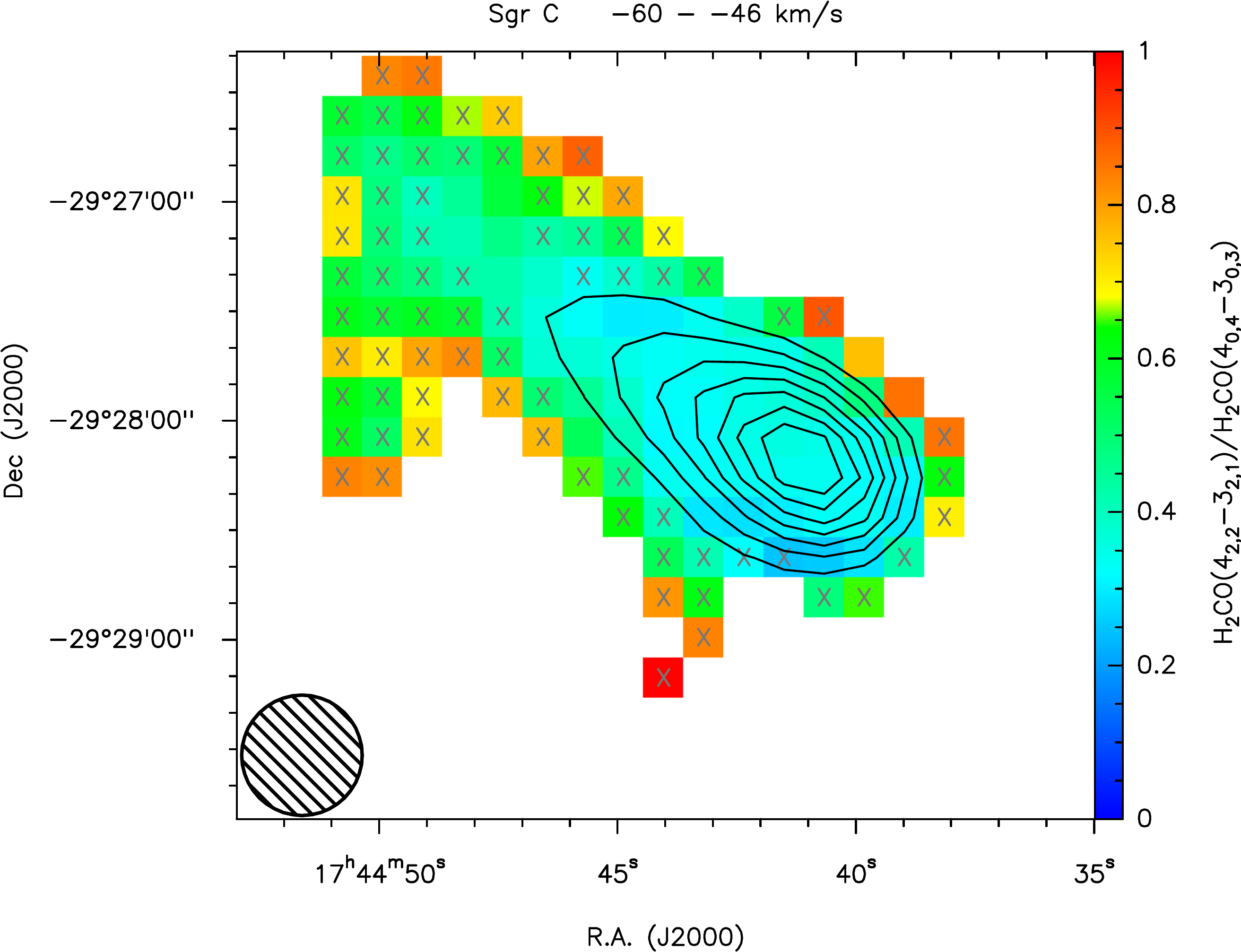}}
	\subfloat{\includegraphics[bb = 130 60 760 580, clip, height=5cm]{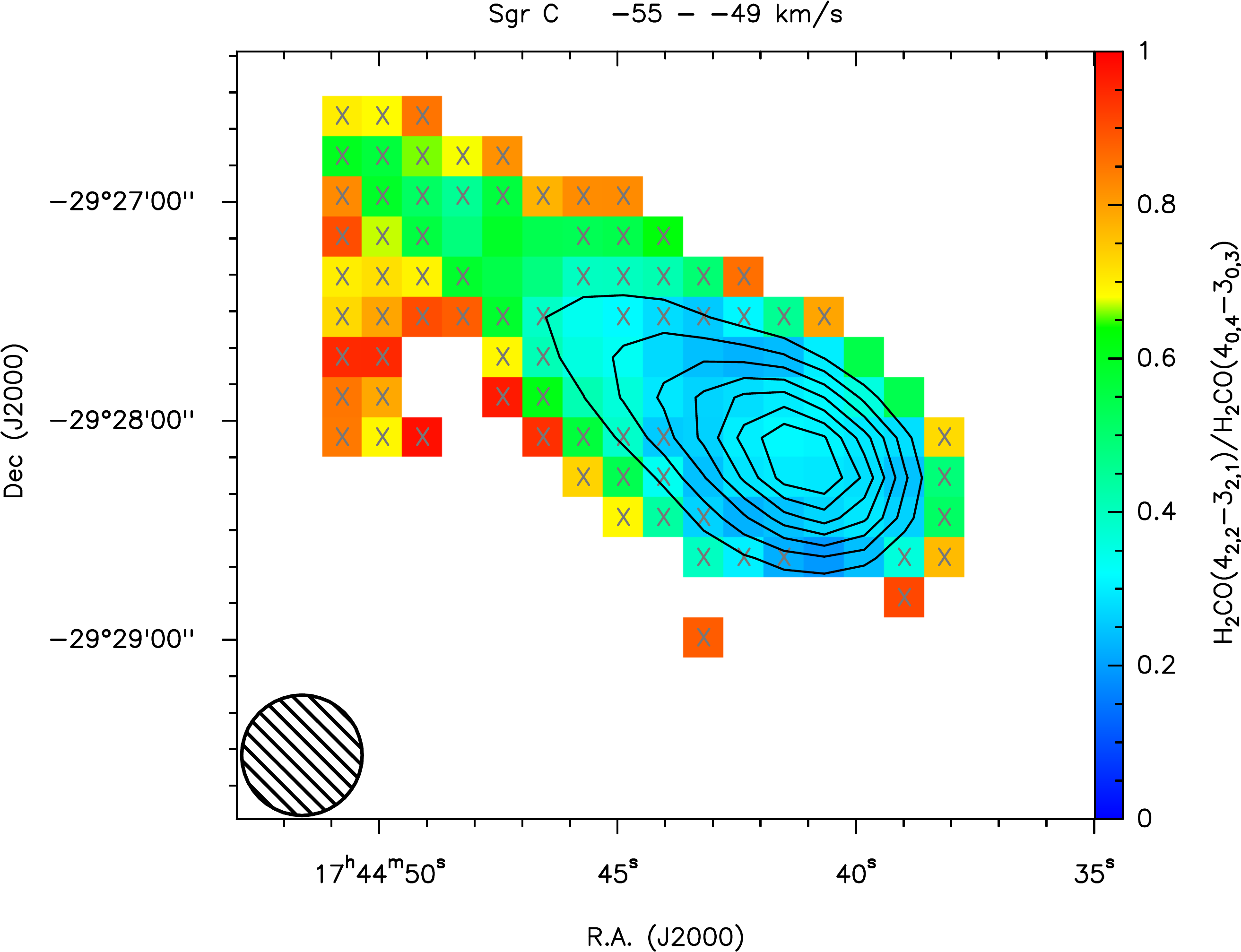}}\\
	\vspace{-0.5cm}
	\subfloat{\includegraphics[bb = 0 0 730 560, clip, height=5.39cm]{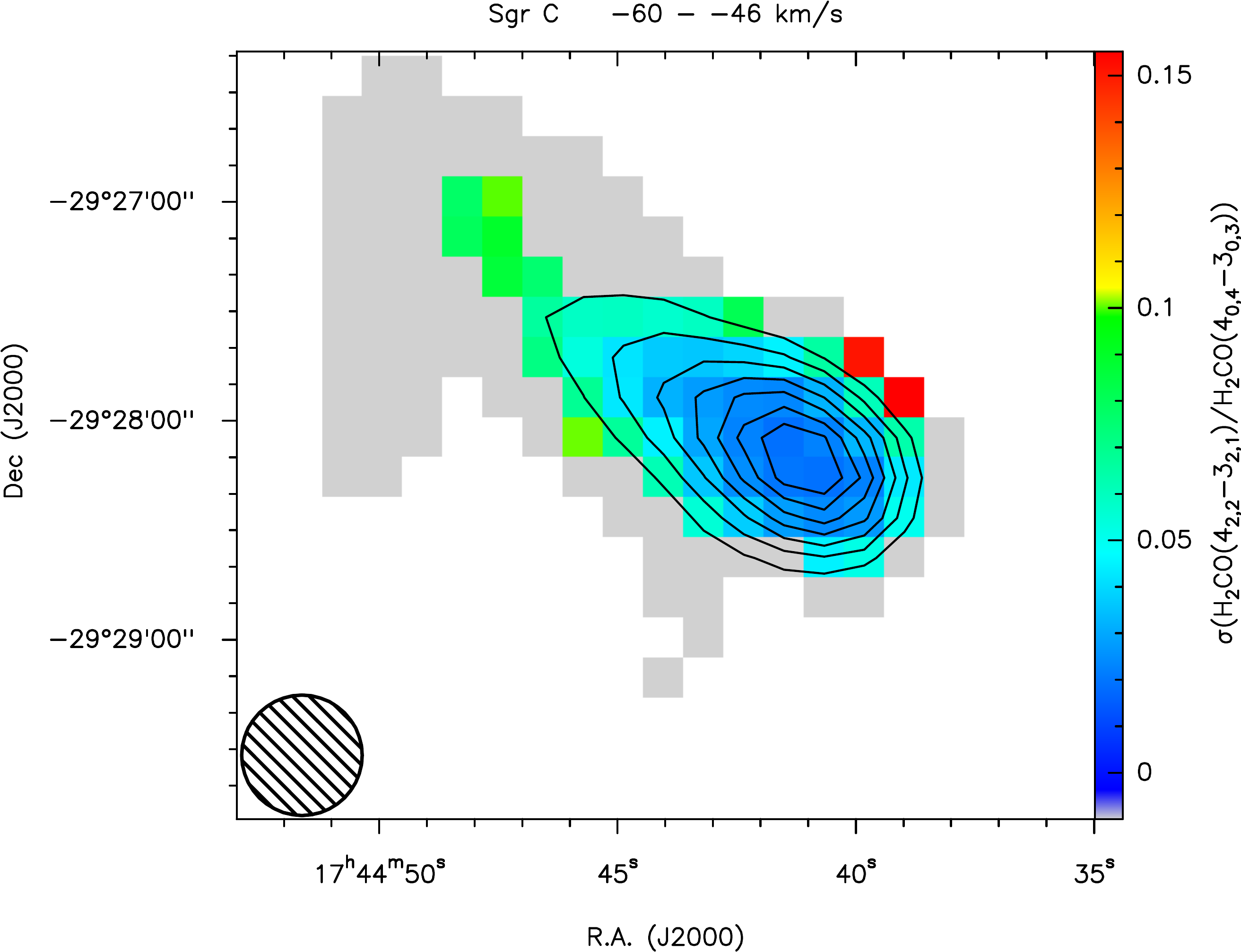}}
	\subfloat{\includegraphics[bb = 130 0 760 560, clip, height=5.39cm]{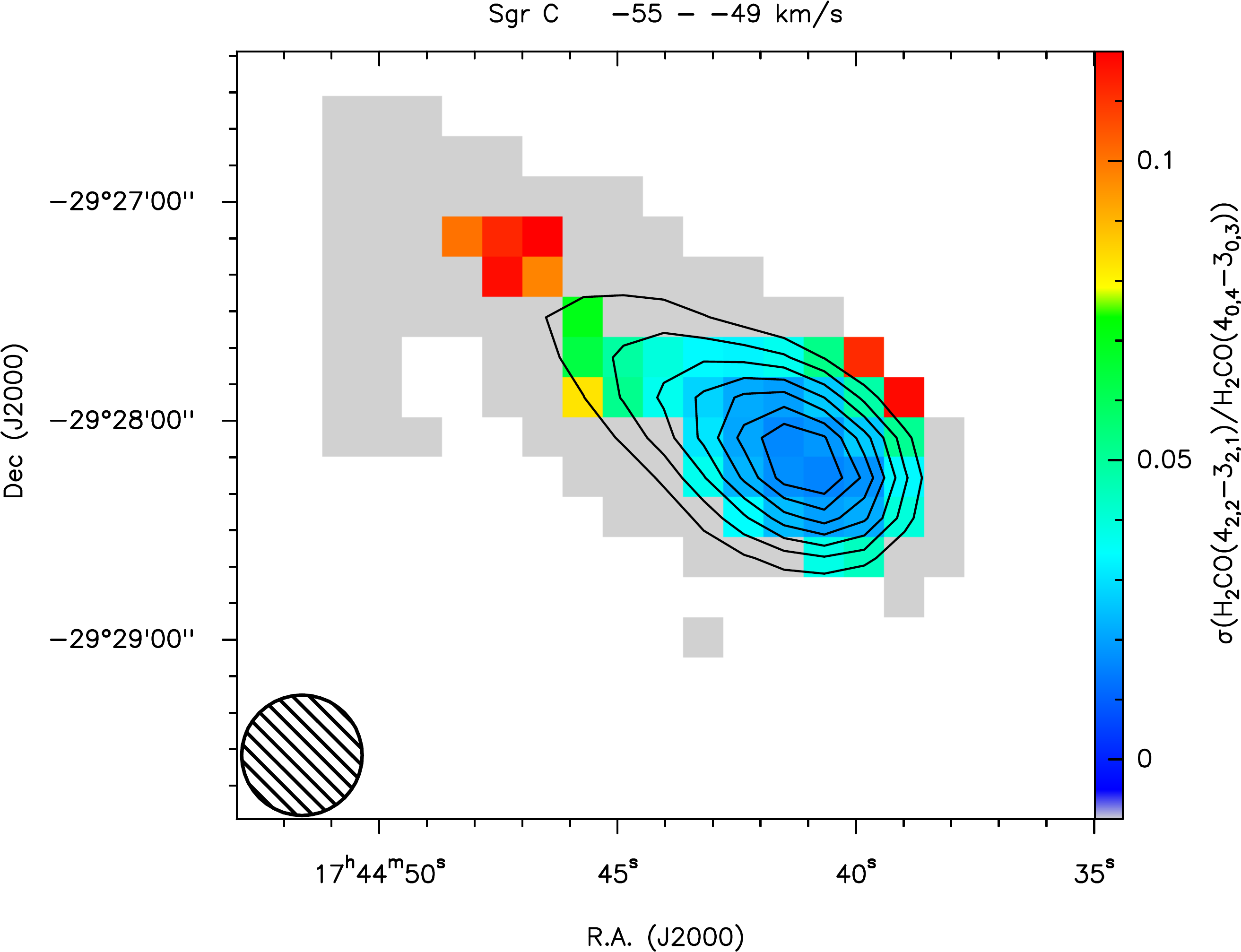}}
	\label{SGRC-All-Ratio-H2CO}
\end{figure*}

\begin{figure*}
	\caption{As Fig. \ref{20kms-All-Ratio-H2CO} for Sgr D.}
	\centering
        R$_{422}$\\
	\subfloat{\includegraphics[bb = 0 60 710 580, clip, height=5cm]{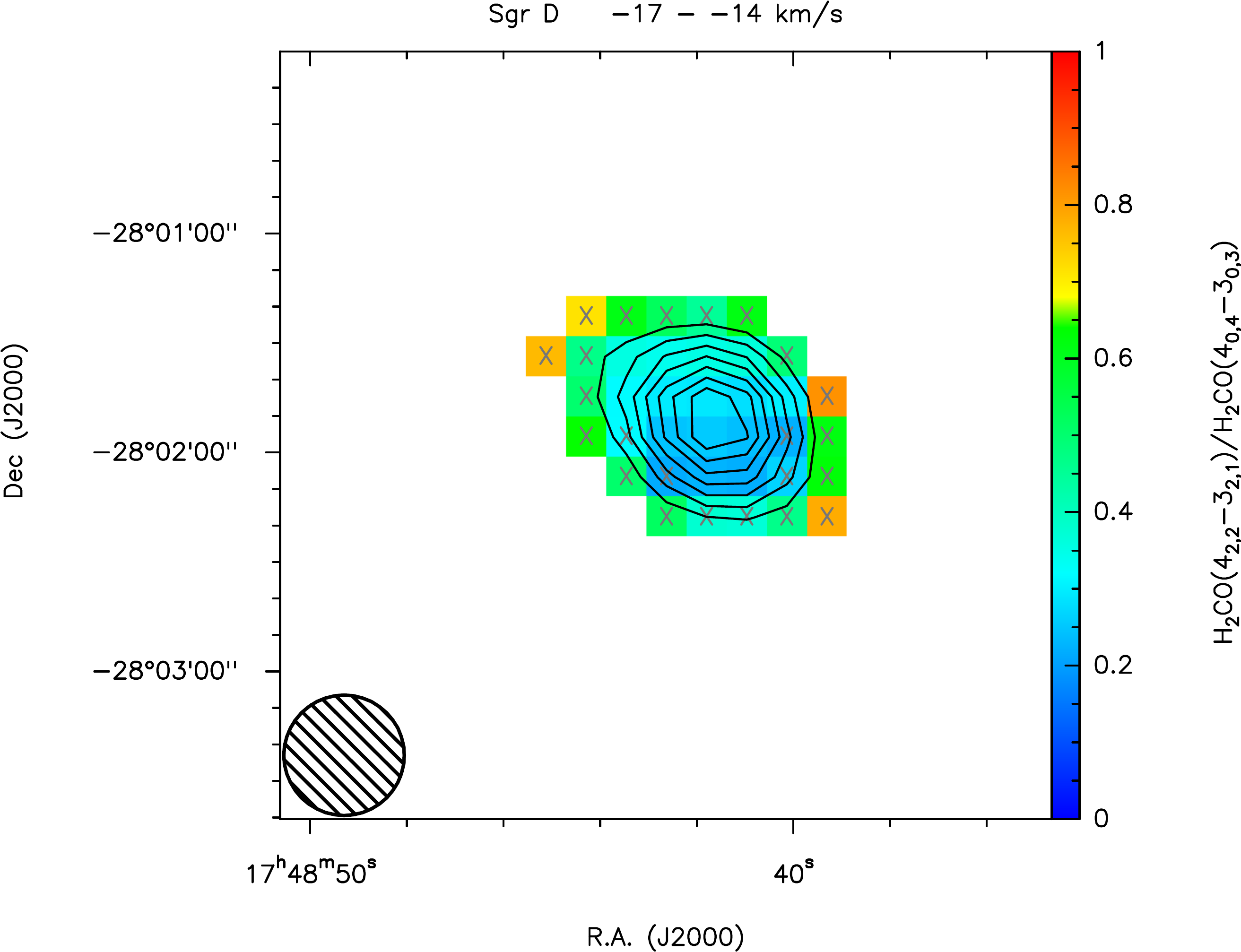}}
	\subfloat{\includegraphics[bb = 150 60 760 580, clip, height=5cm]{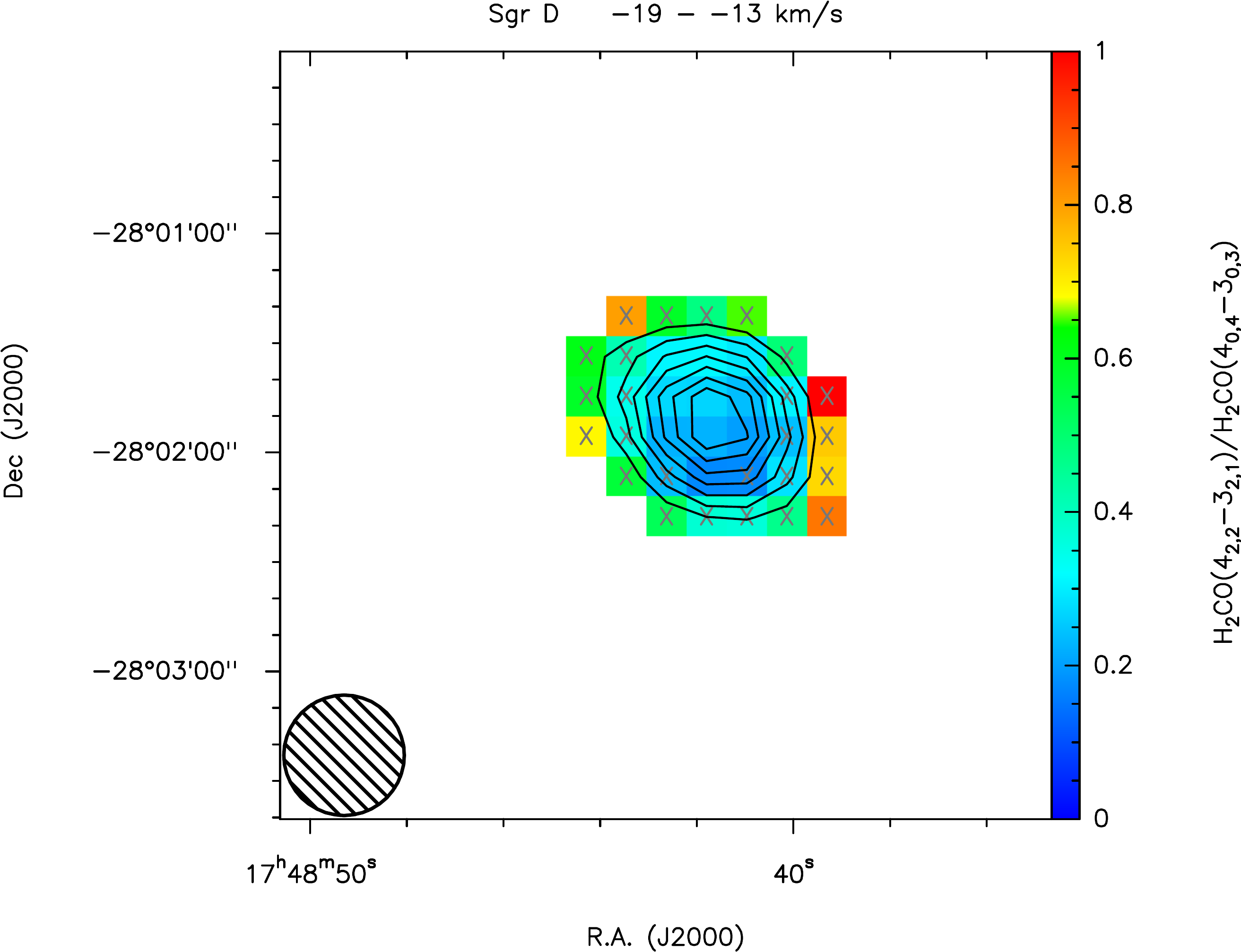}}\\
	\vspace{-0.5cm}
	\subfloat{\includegraphics[bb = 0 0 710 560, clip, height=5.39cm]{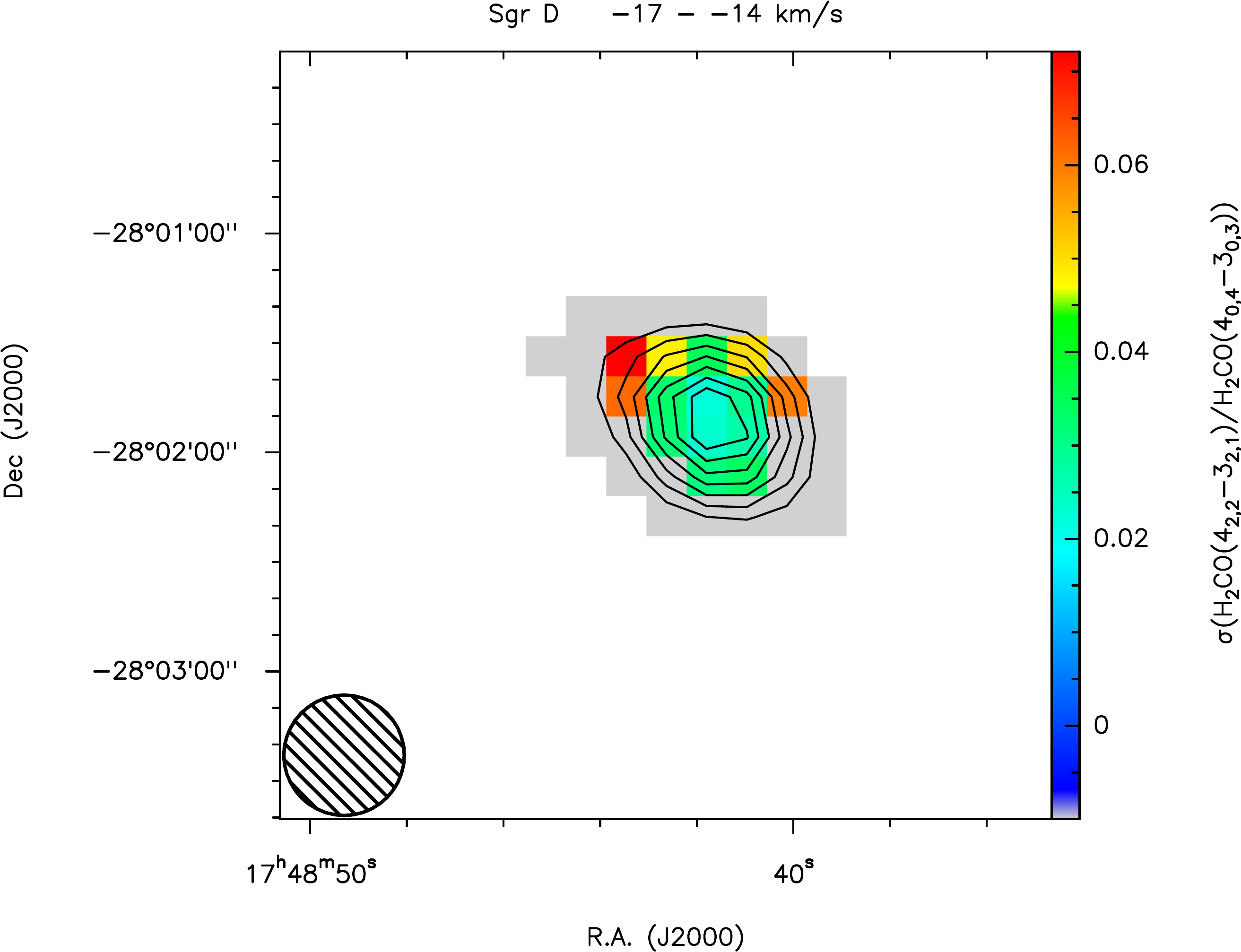}}
	\subfloat{\includegraphics[bb = 150 0 760 560, clip, height=5.39cm]{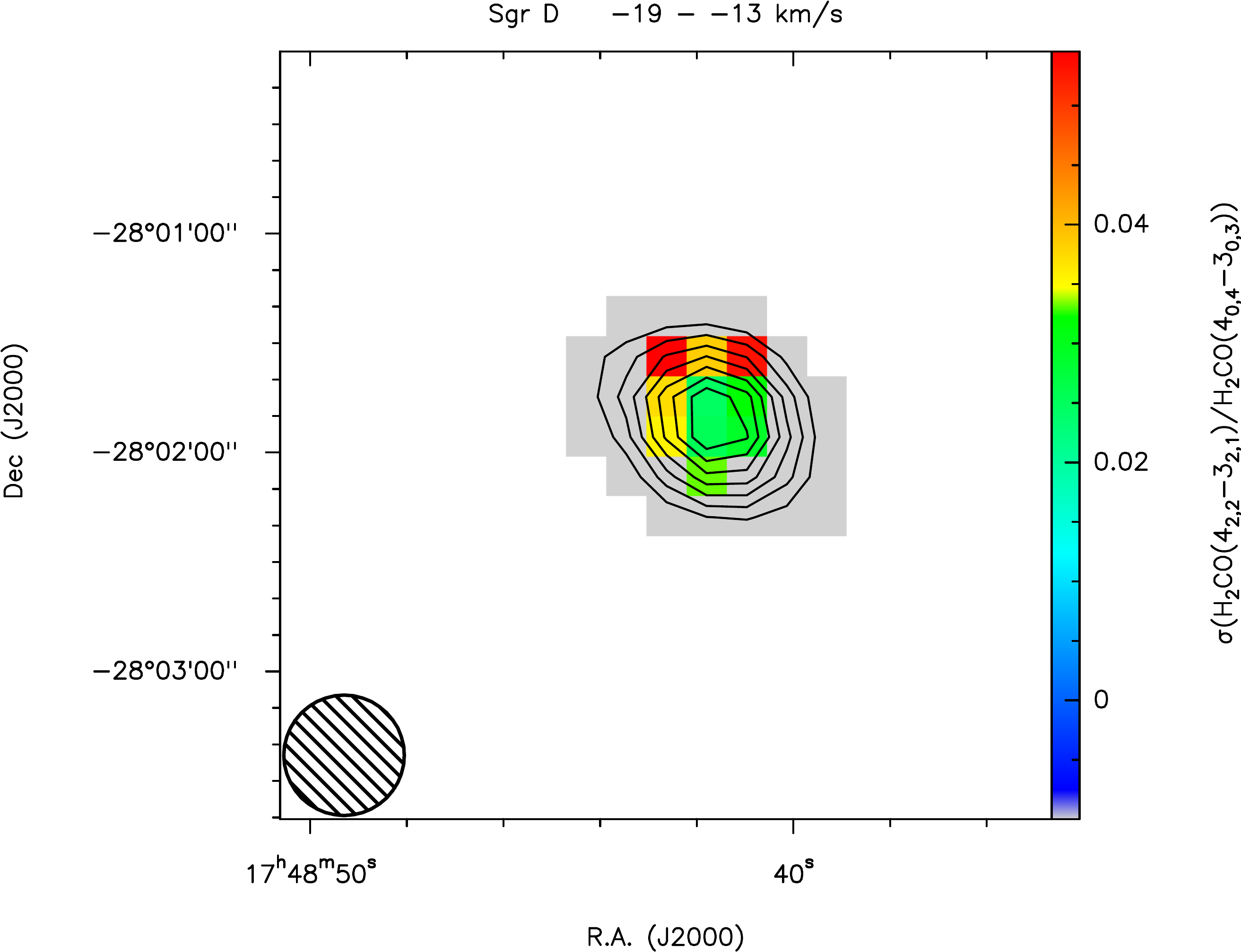}}
	\label{SGRD-All-Ratio-H2CO}
\end{figure*}

\clearpage

\section{Modeling}

\begin{figure*}
	\caption{Impact of different assumptions in column density, density, line width, and geometry on the estimated temperatures. 
	The solid blue line is alway the fiducial case used to obtain the temperatures presented in this paper.}
	\centering
	\subfloat{\includegraphics[height=\textheight]{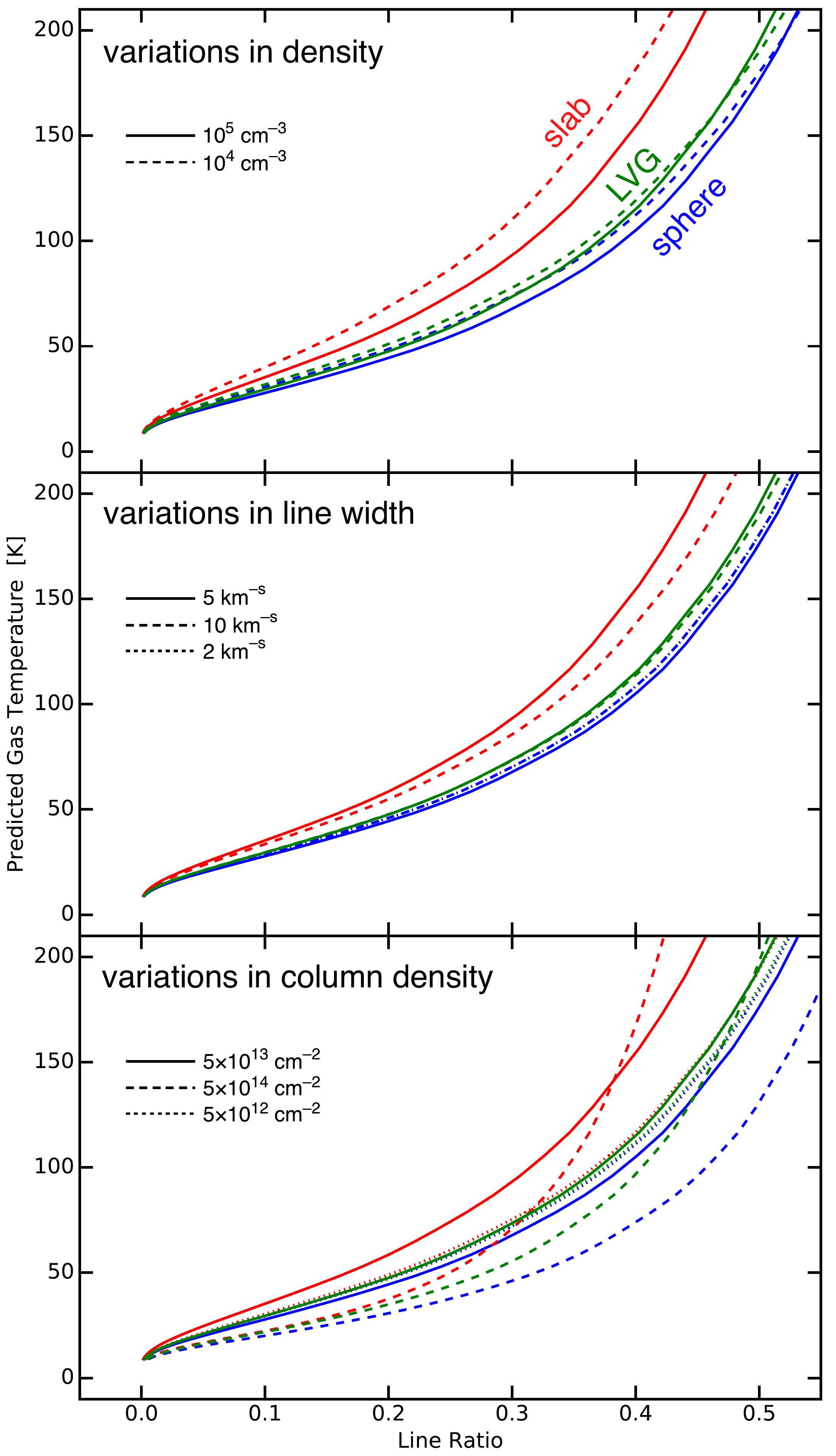}}
	\label{TempUncertainties}
\end{figure*}

\begin{figure*}
	\caption{Impact of line optical depth. The optical depths shown on the left are calculated for the fiducial case (see Section
	 \ref{SystematicUncertainties}), while those on the right are for an increased column density of 5$\cdot$10$^{14}$ cm$^{-2}$. The red 
	 lines indicate the observed line ratios of 0.25 (top panel) and 0.35 (bottom panel) at 218 and 291 GHz, respectively, in the 8$-$14 
	 km s$^{-1}$ slice of the 20 km/s cloud. 
	The blue lines indicate an intensity of 15 K km s$^{-1}$ at 218 GHz (top panel) and of 4 K km s$^{-1}$ at 291 GHz (bottom panel),
	 respectively, in the same slice. Black lines give optical depth contours. The lines come in pairs since we allow for $\pm$ 20\% variation
	  in drawing. Simultaneous fits to line ratios and intensities are found in locations where the blue and red lines intersect. At 218 GHz 
	  (top panel) such simultaneous matches require either relatively high densities (left panel) or high optical depths (right panel).}
	\centering
	\subfloat{\includegraphics[width=0.95\textwidth]{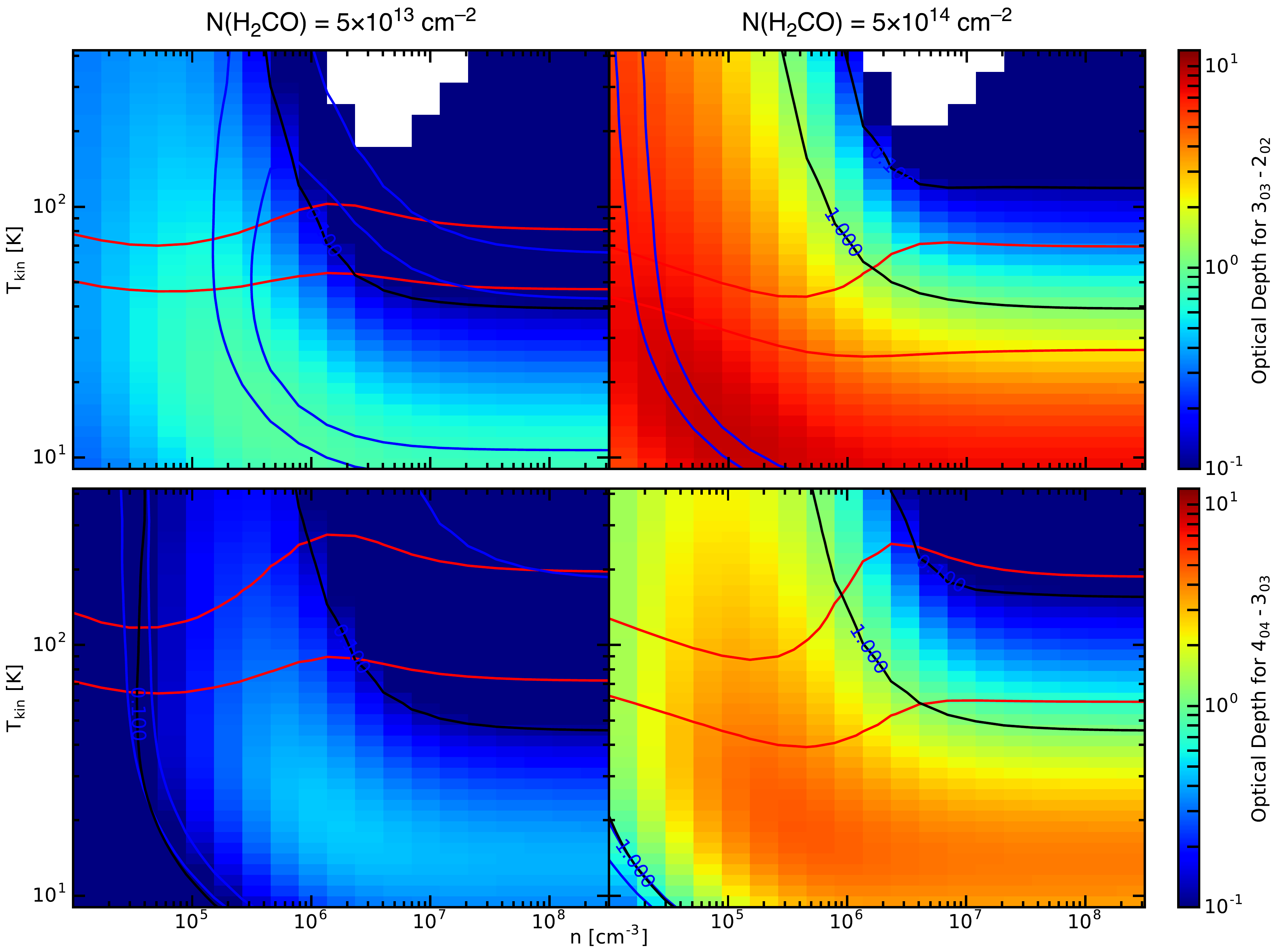}}
	\label{TempOpacity}
\end{figure*}

\clearpage

\section{Temperature Maps}

\begin{figure*}
	\caption{Temperature (upper panels) and uncertainty (lower panels) maps of the 20 km/s cloud (contours as in Fig. \ref{20kms-Int-H2CO}). 
	Upper limits of the temperatures are marked with Xs. The corresponding pixels in the uncertainty maps are blanked. The purple squares 
	present the areas over which the emission of the main H$_{2}$CO line was integrated to determine the line width as well as the 
	average temperature (see Section \ref{SectTempLineWidth}). The circle in the lower left corner shows the 33$\arcsec$ beam.}
	\centering
        218 GHz temperatures\\
	\subfloat{\includegraphics[bb = 0 60 600 580, clip, height=4.5cm]{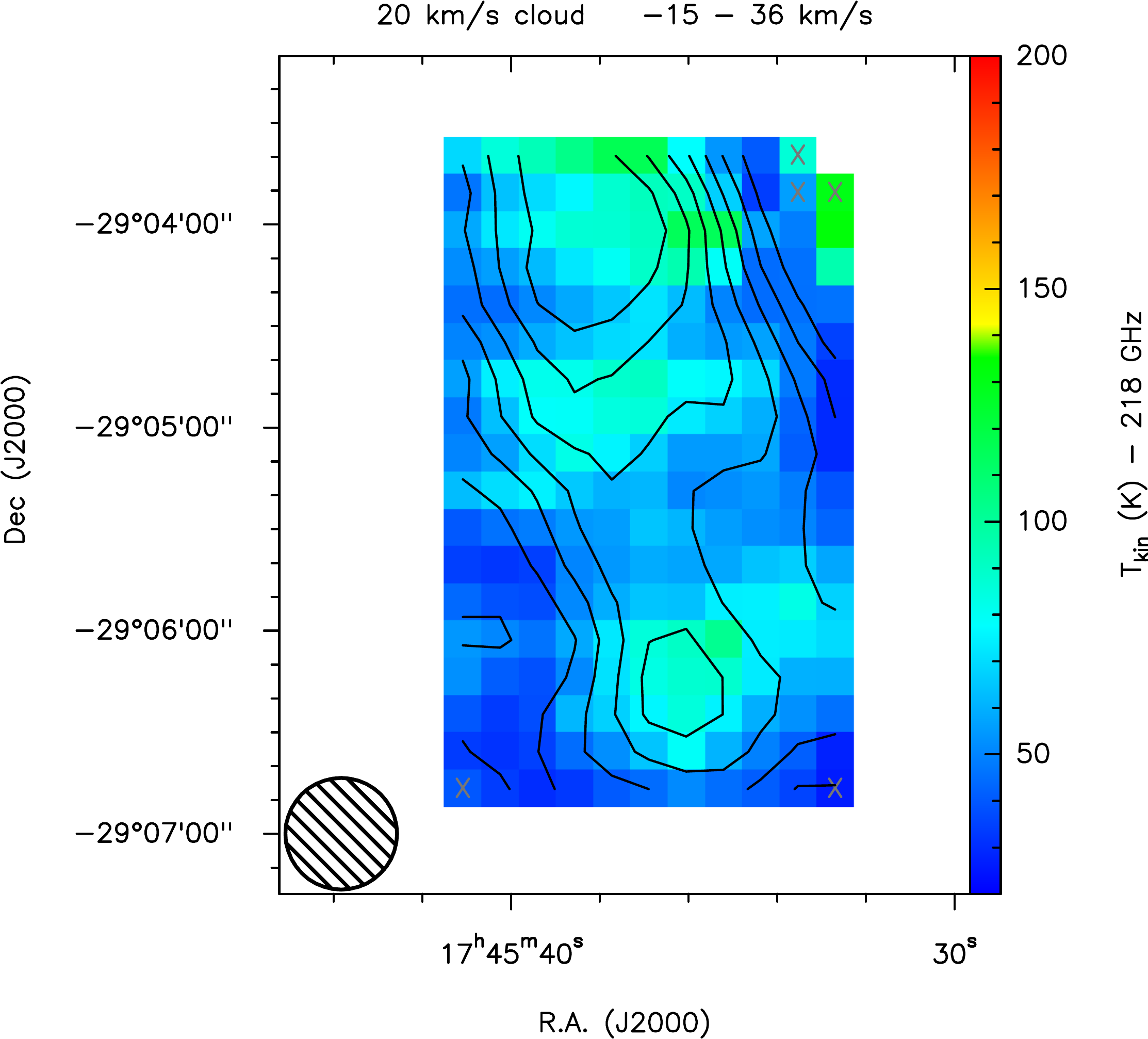}}
	\subfloat{\includegraphics[bb = 140 60 600 580, clip, height=4.5cm]{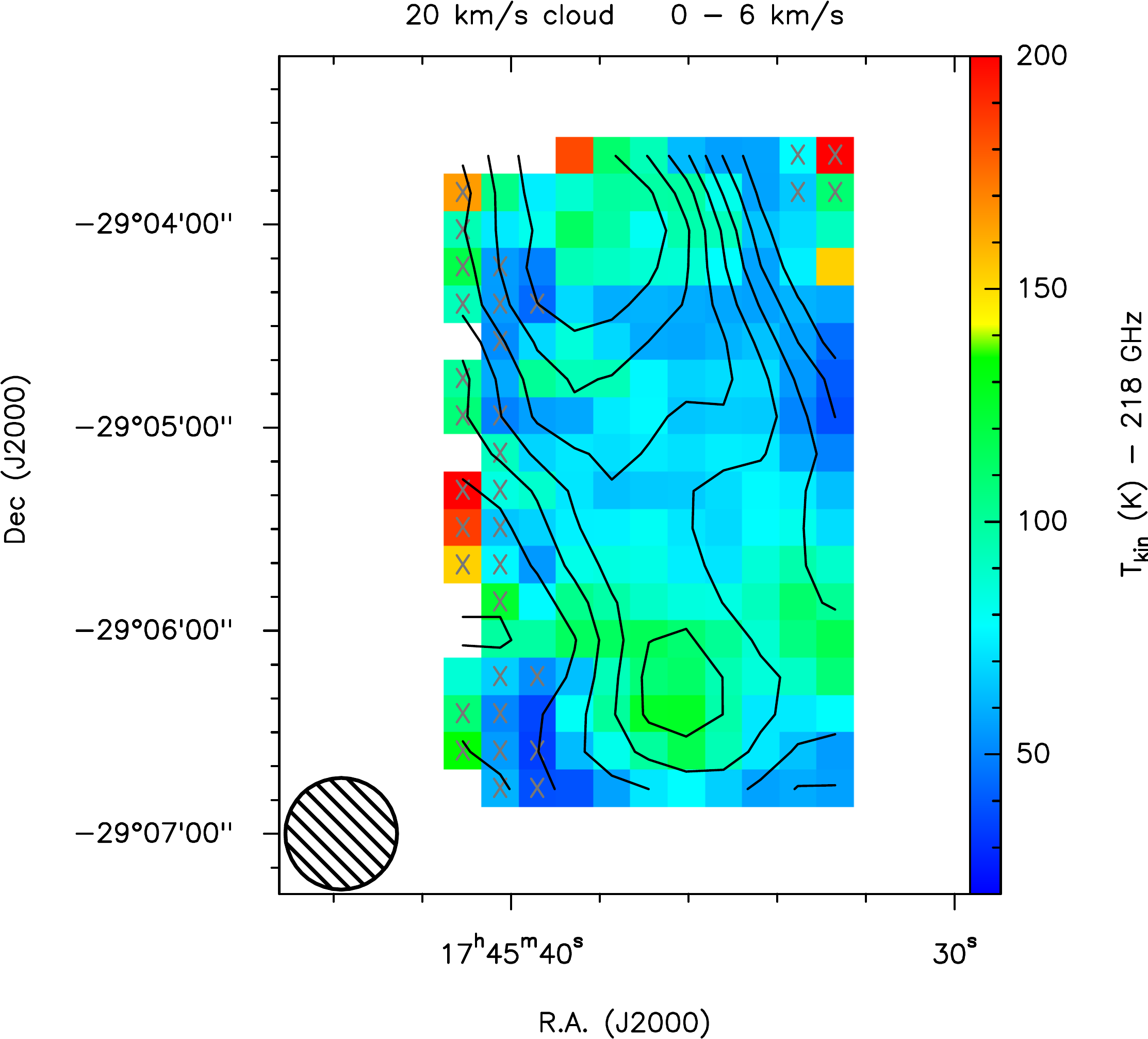}}
	\subfloat{\includegraphics[bb = 140 60 600 580, clip, height=4.5cm]{20kms-H2CO-8-14-Temp-220GHz.pdf}}
	\subfloat{\includegraphics[bb = 140 60 650 580, clip, height=4.5cm]{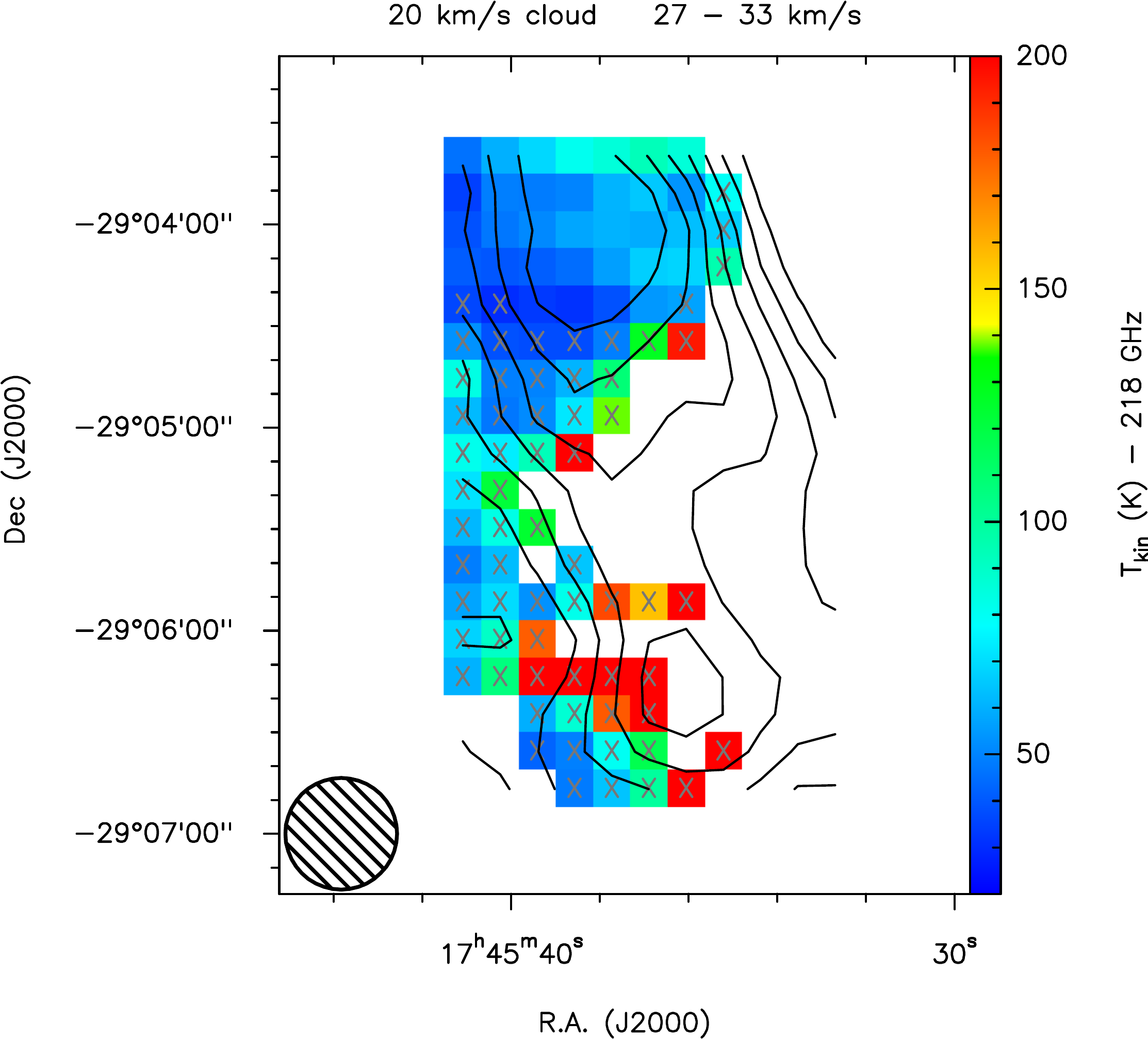}}\\\vspace{-0.5cm}
	\subfloat{\includegraphics[bb = 0 0 600 560, clip, height=4.845cm]{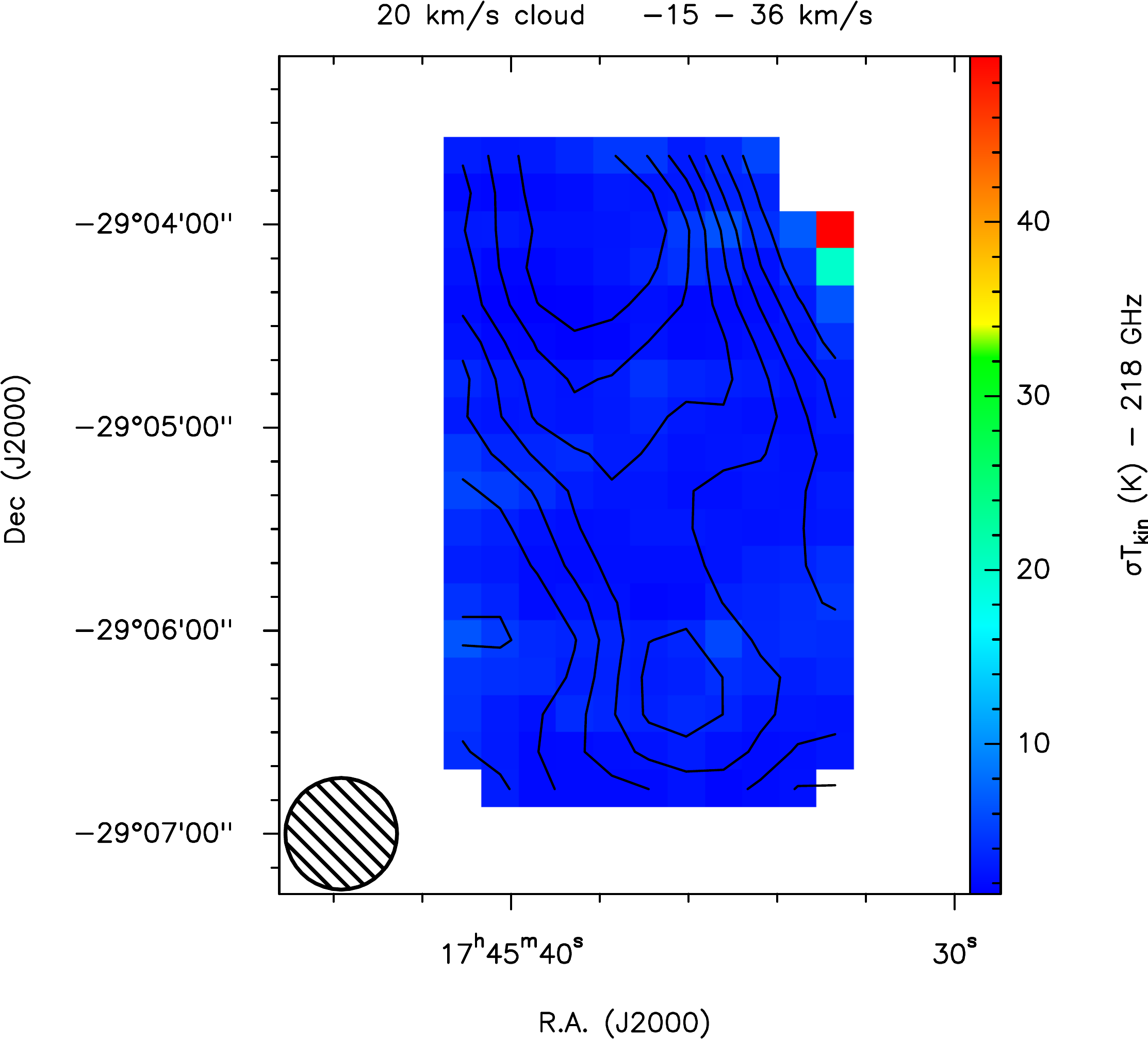}}
	\subfloat{\includegraphics[bb = 140 0 600 560, clip, height=4.845cm]{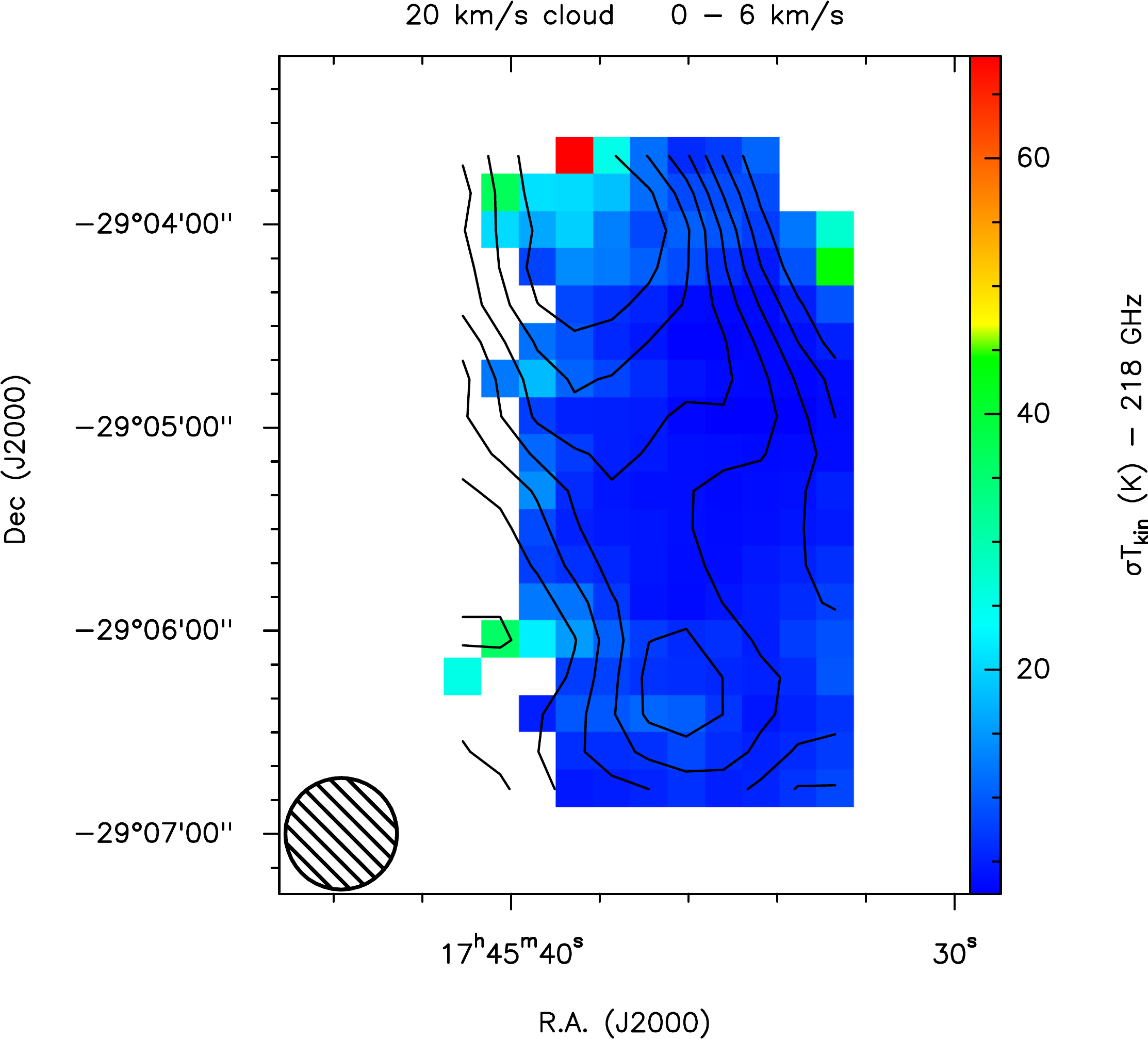}}
	\subfloat{\includegraphics[bb = 140 0 600 560, clip, height=4.845cm]{20kms-H2CO-8-14-Uncertainty-Temp-220GHz.pdf}}
	\subfloat{\includegraphics[bb = 140 0 650 560, clip, height=4.845cm]{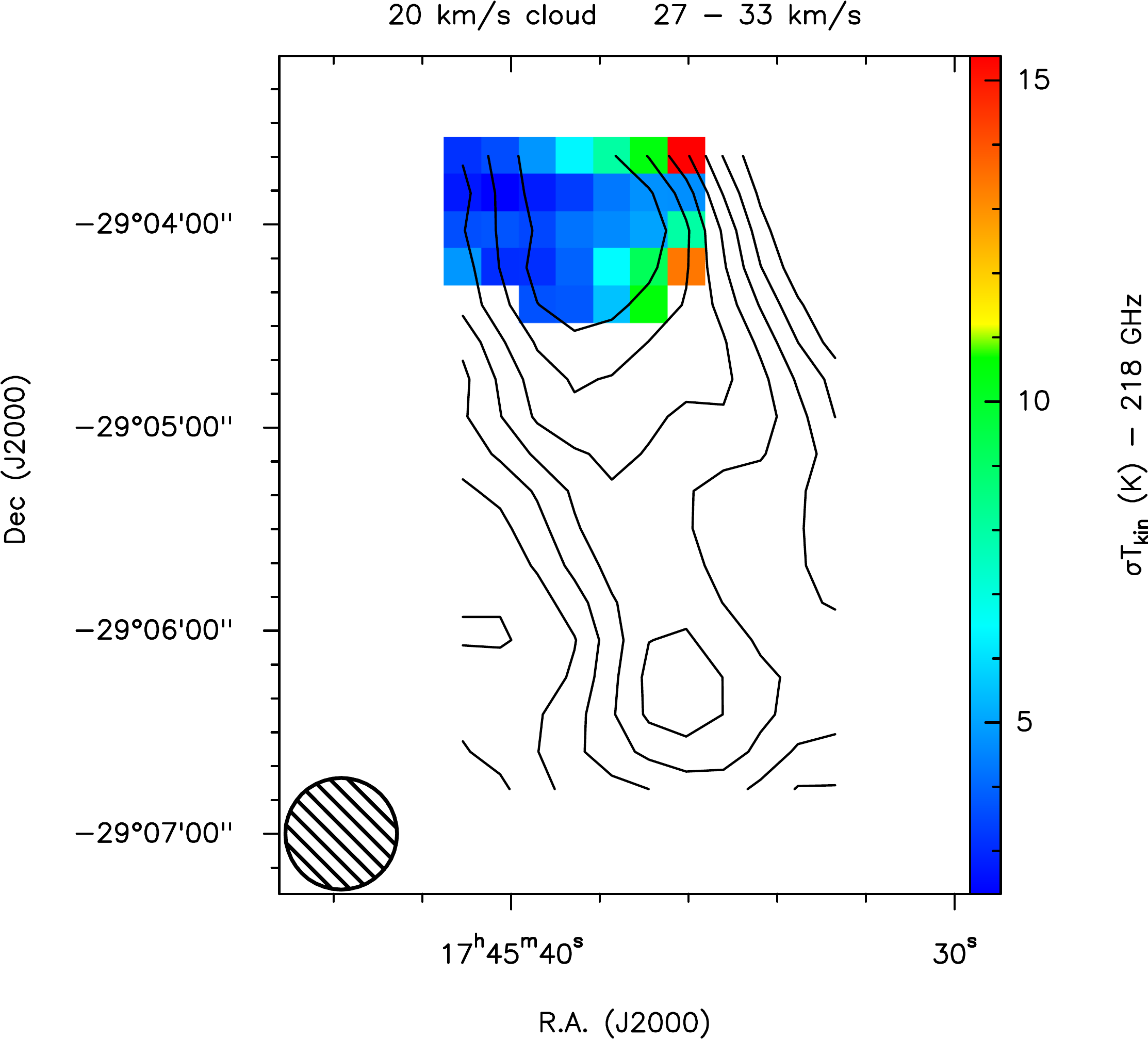}}\\ 
	\vspace{0.1cm}
	291 GHz temperatures \\
	\subfloat{\includegraphics[bb = 0 60 600 580, clip, height=4.5cm]{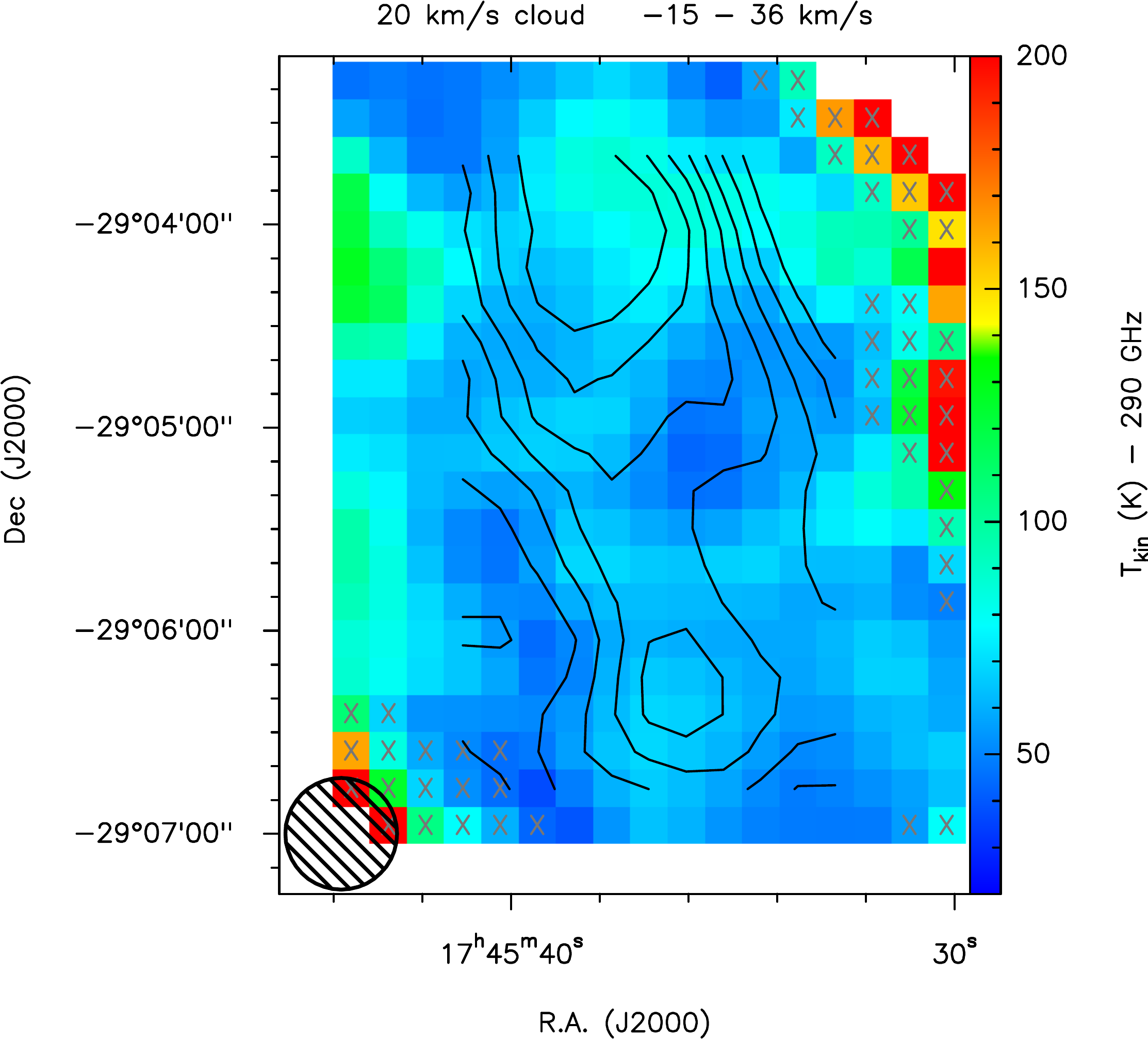}}
	\subfloat{\includegraphics[bb = 140 60 600 580, clip, height=4.5cm]{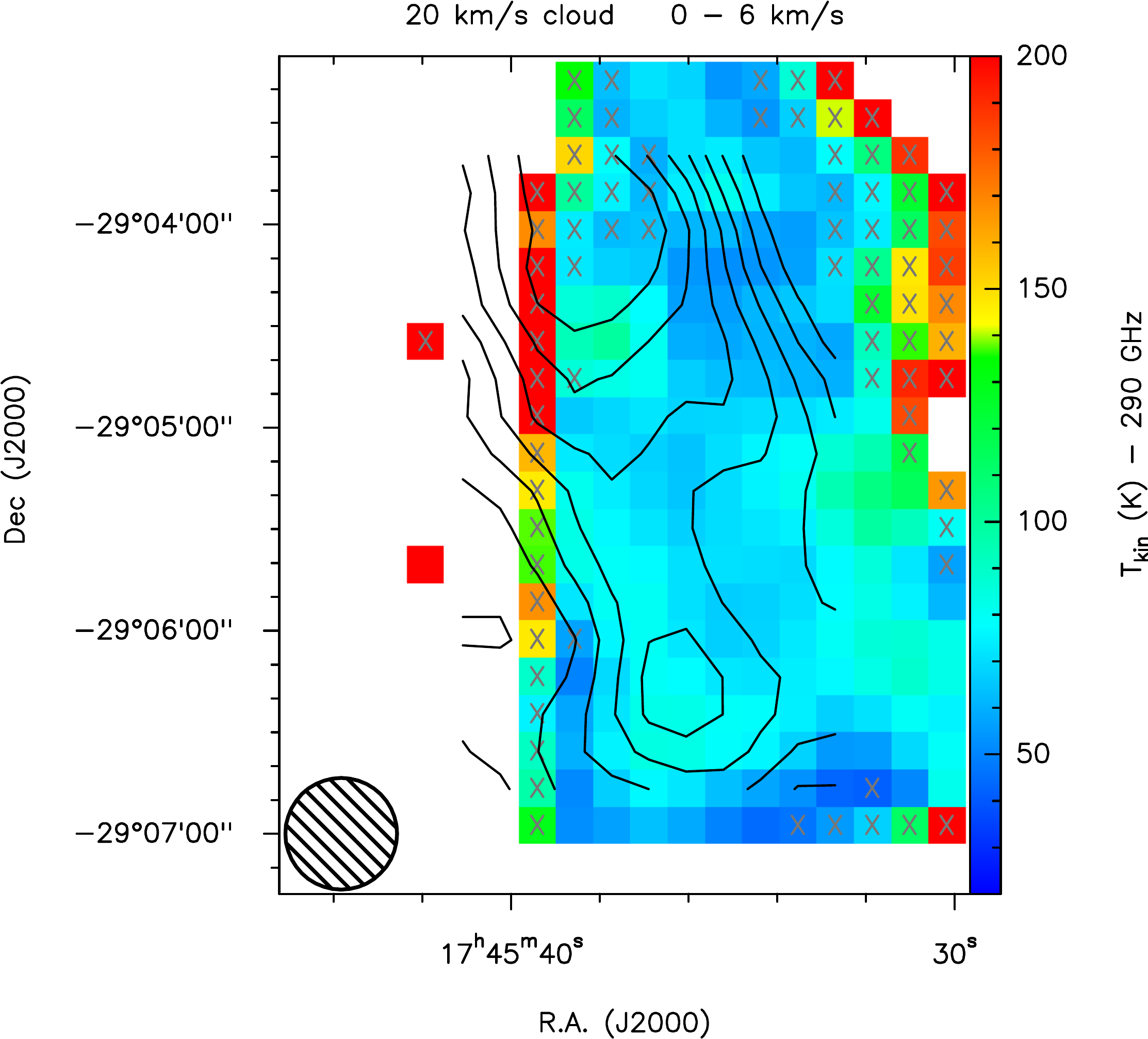}}
	\subfloat{\includegraphics[bb = 140 60 600 580, clip, height=4.5cm]{20kms-H2CO-8-14-Temp-290GHz.pdf}}
	\subfloat{\includegraphics[bb = 140 60 650 580, clip, height=4.5cm]{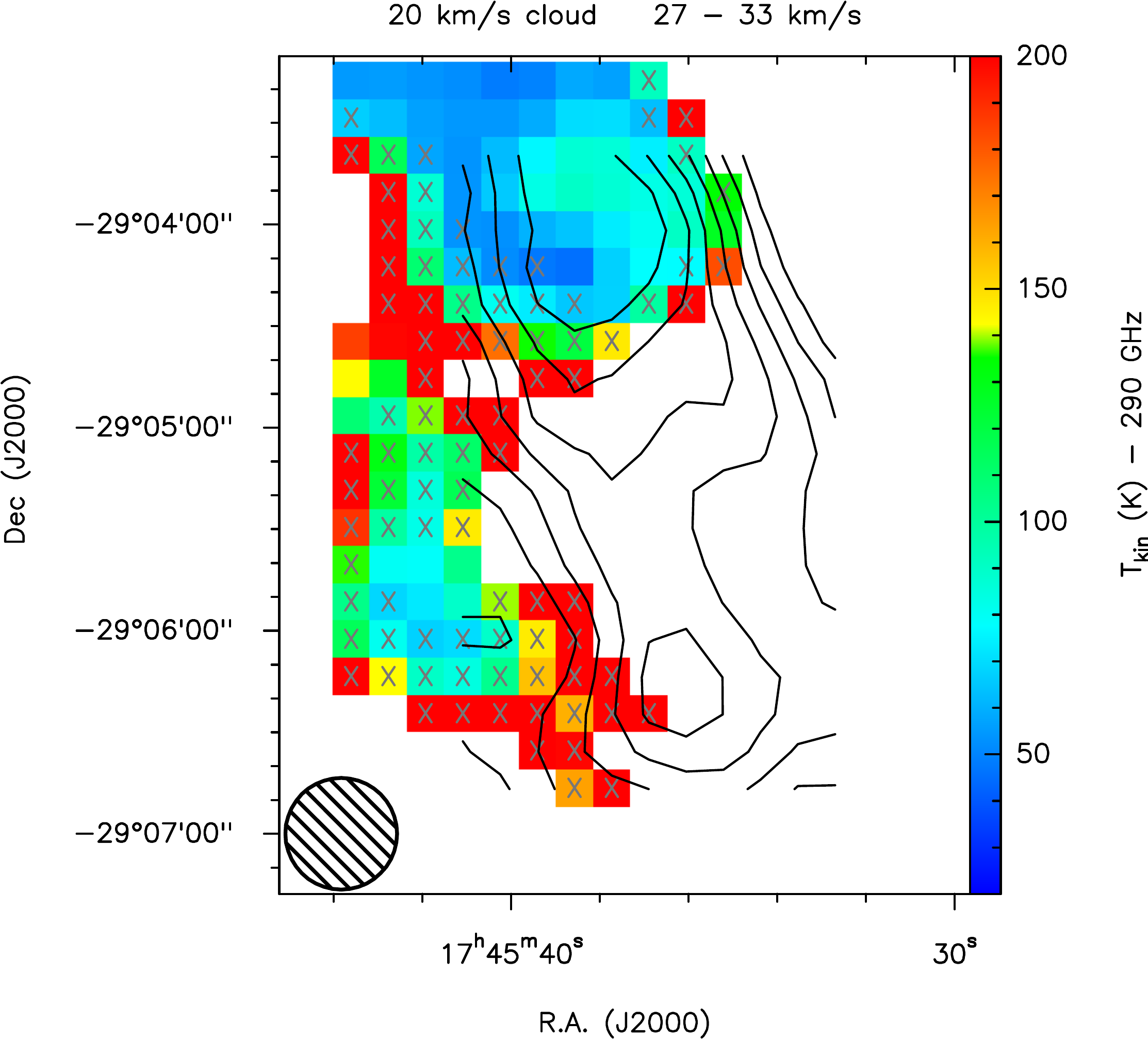}}\\\vspace{-0.5cm}
	\subfloat{\includegraphics[bb = 0 0 600 560, clip, height=4.845cm]{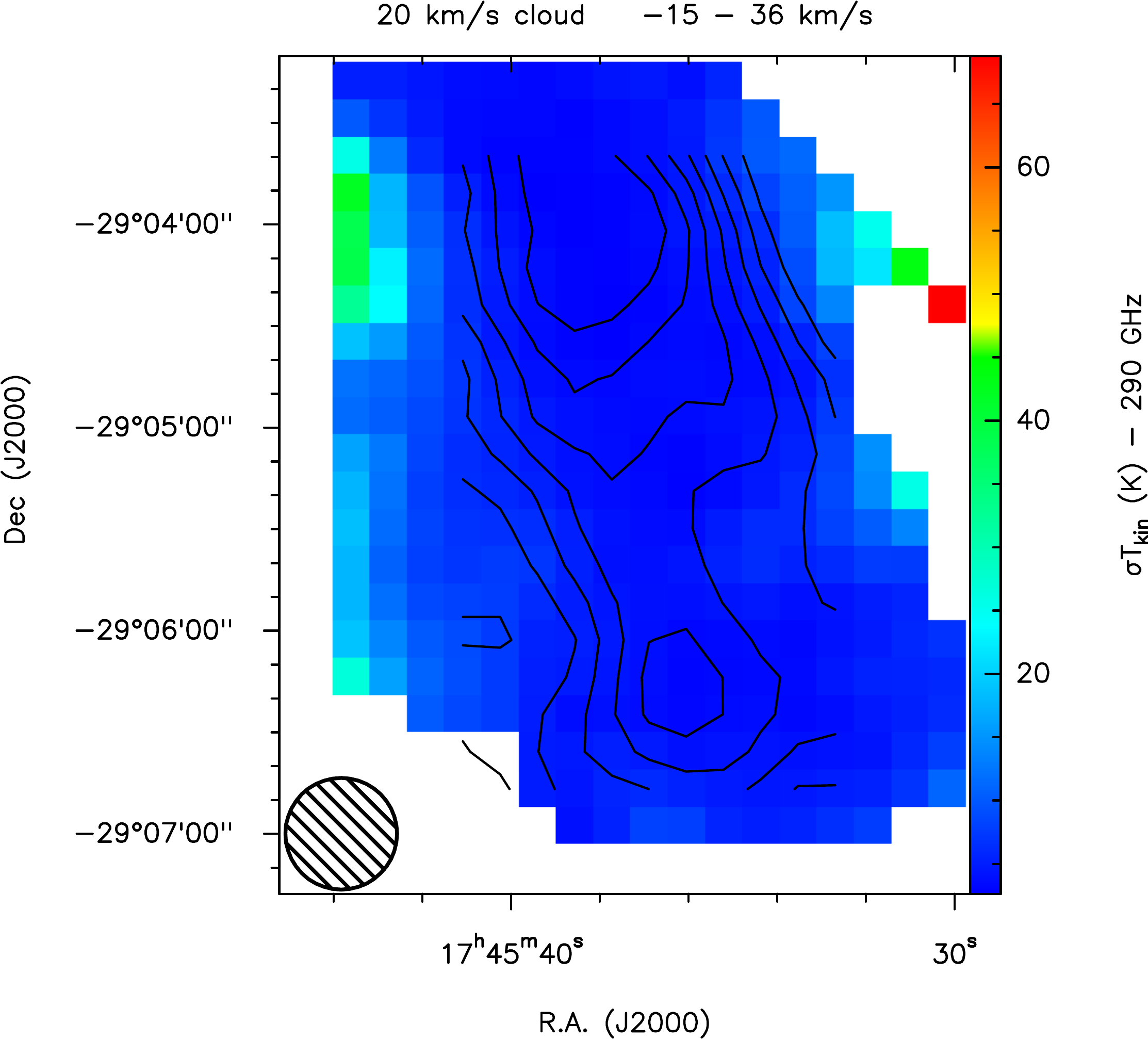}}
	\subfloat{\includegraphics[bb = 140 0 600 560, clip, height=4.845cm]{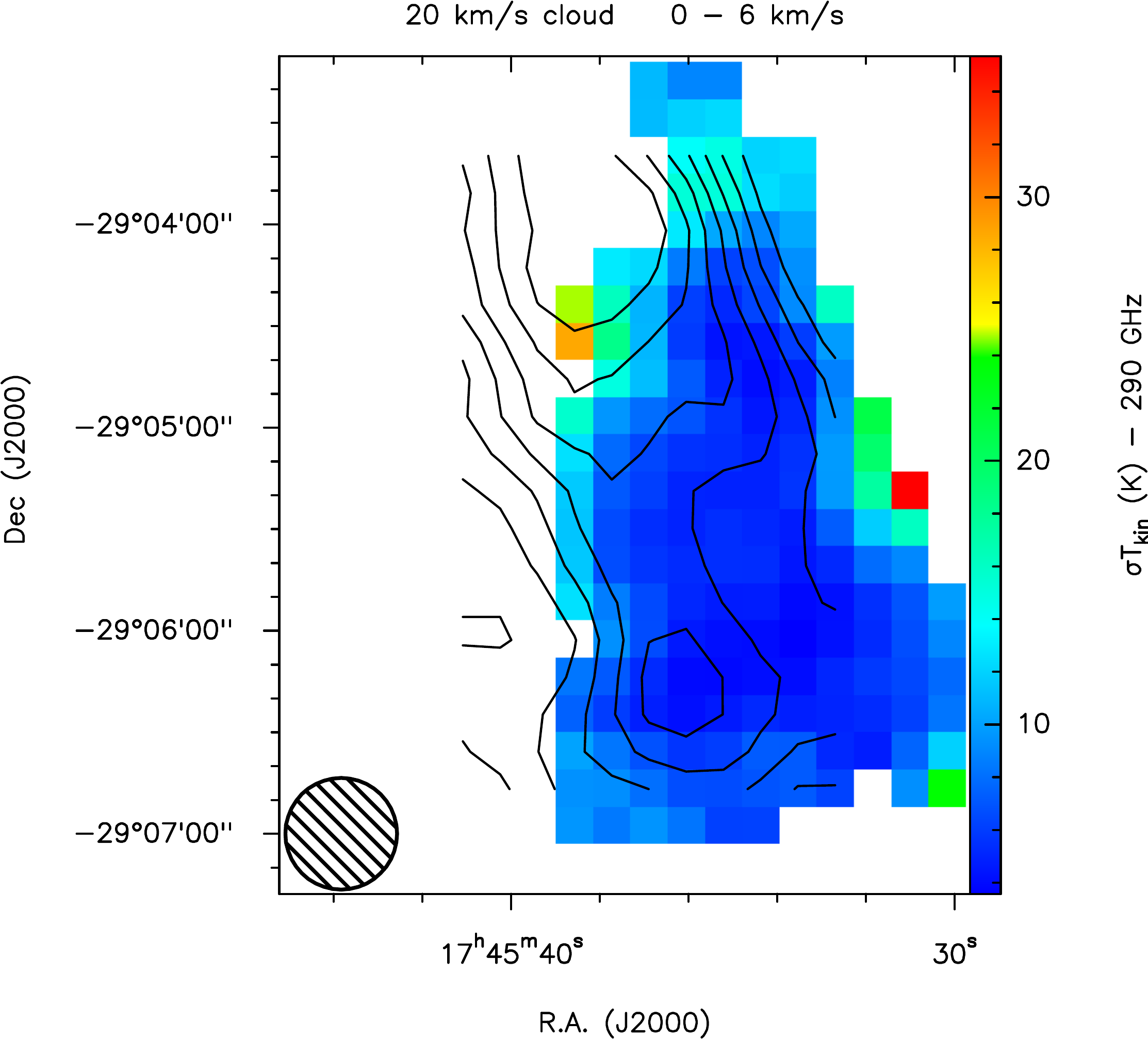}}
	\subfloat{\includegraphics[bb = 140 0 600 560, clip, height=4.845cm]{20kms-H2CO-8-14-Uncertainty-Temp-290GHz.pdf}}
	\subfloat{\includegraphics[bb = 140 0 650 560, clip, height=4.845cm]{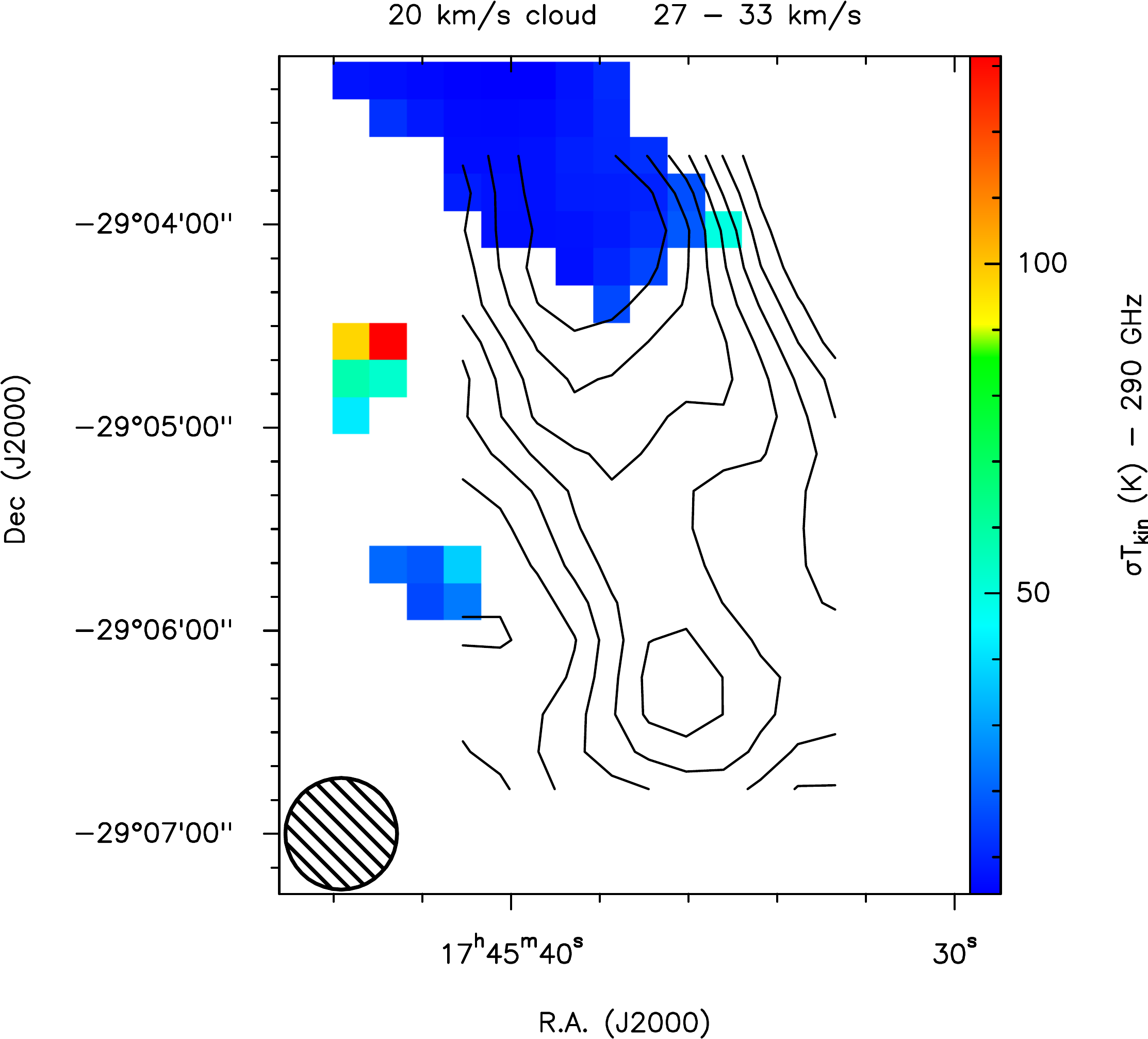}}
	\label{20kms-All-Temp-H2CO}
\end{figure*}

\begin{figure*}
	\caption{As Fig. \ref{20kms-All-Temp-H2CO}, for the 50 km/s cloud.}
	\centering
        218 GHz temperatures\\
	\subfloat{\includegraphics[bb = 0 60 700 580, clip, height=5cm]{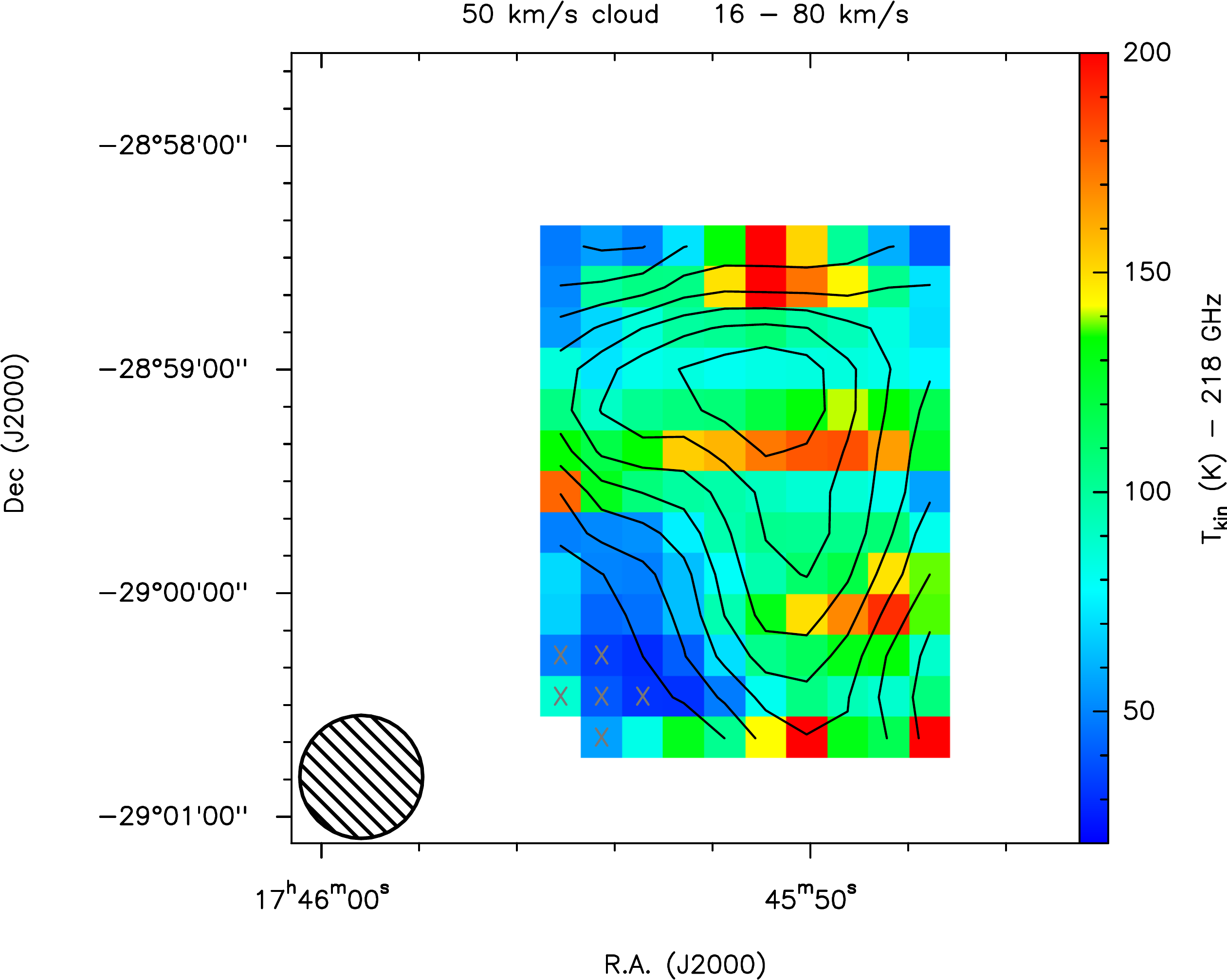}}
	\subfloat{\includegraphics[bb = 150 60 700 580, clip, height=5cm]{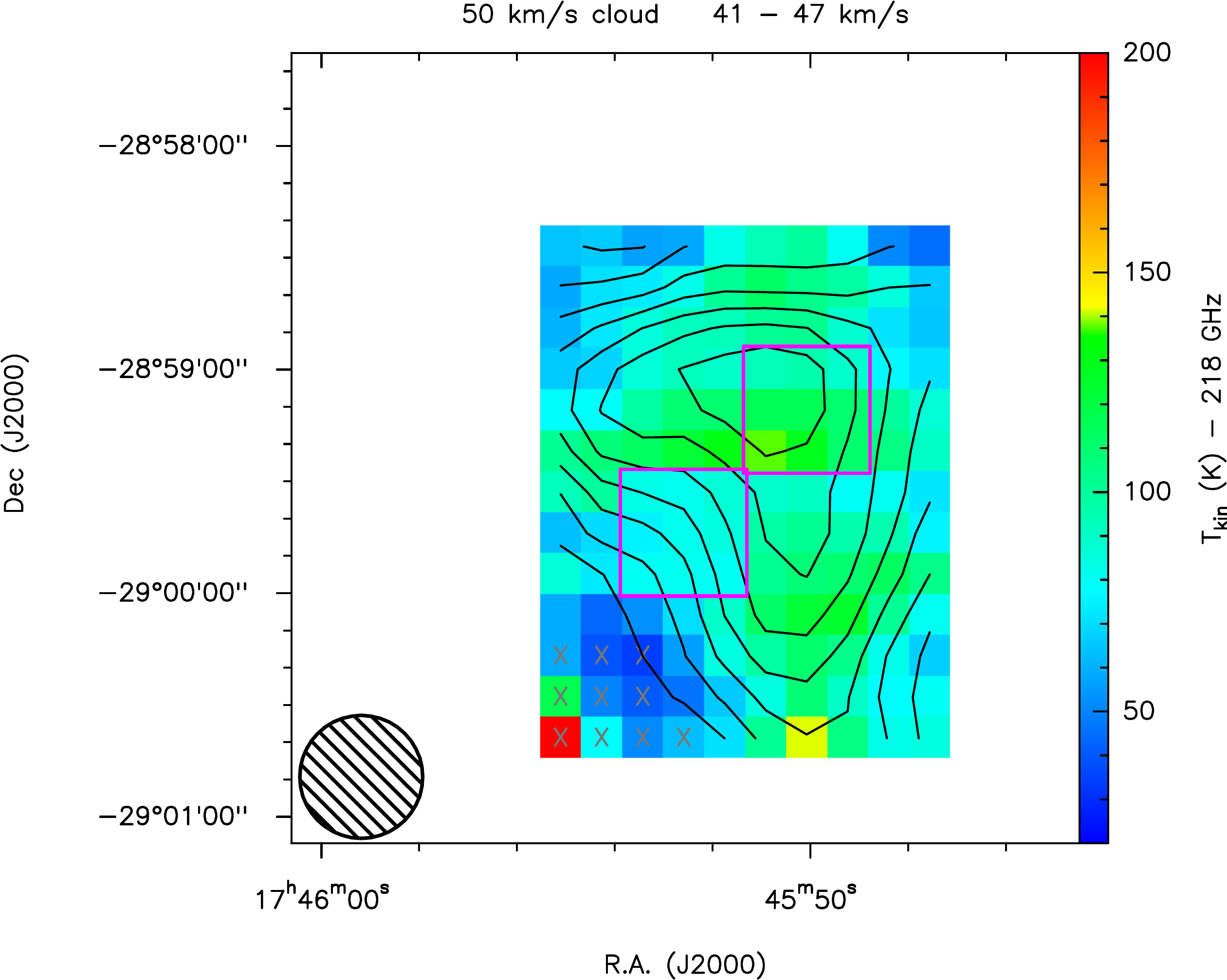}}
	\subfloat{\includegraphics[bb = 150 60 730 580, clip, height=5cm]{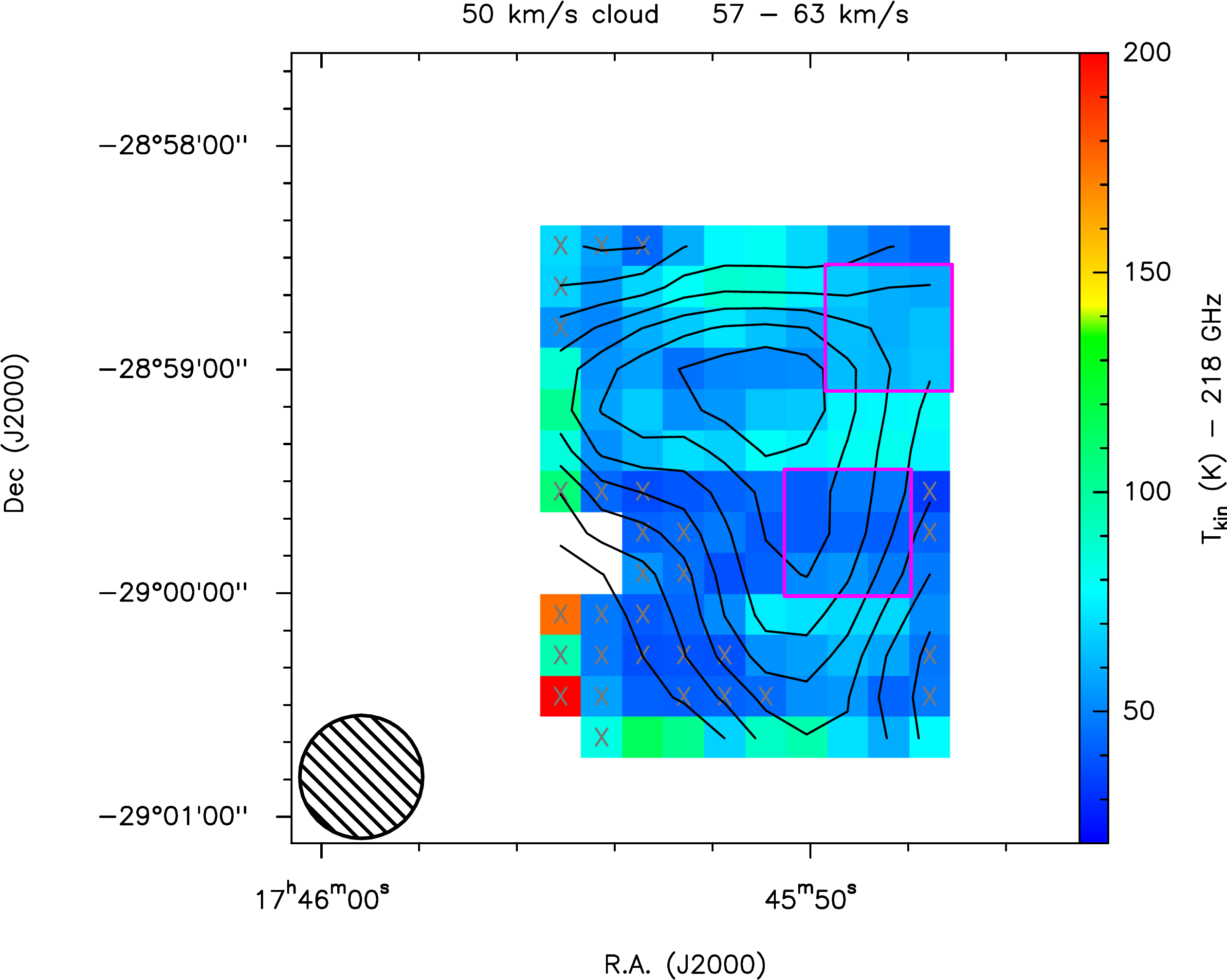}}\\
	\vspace{-0.5cm}
	\subfloat{\includegraphics[bb = 0 0 700 560, clip, height=5.385cm]{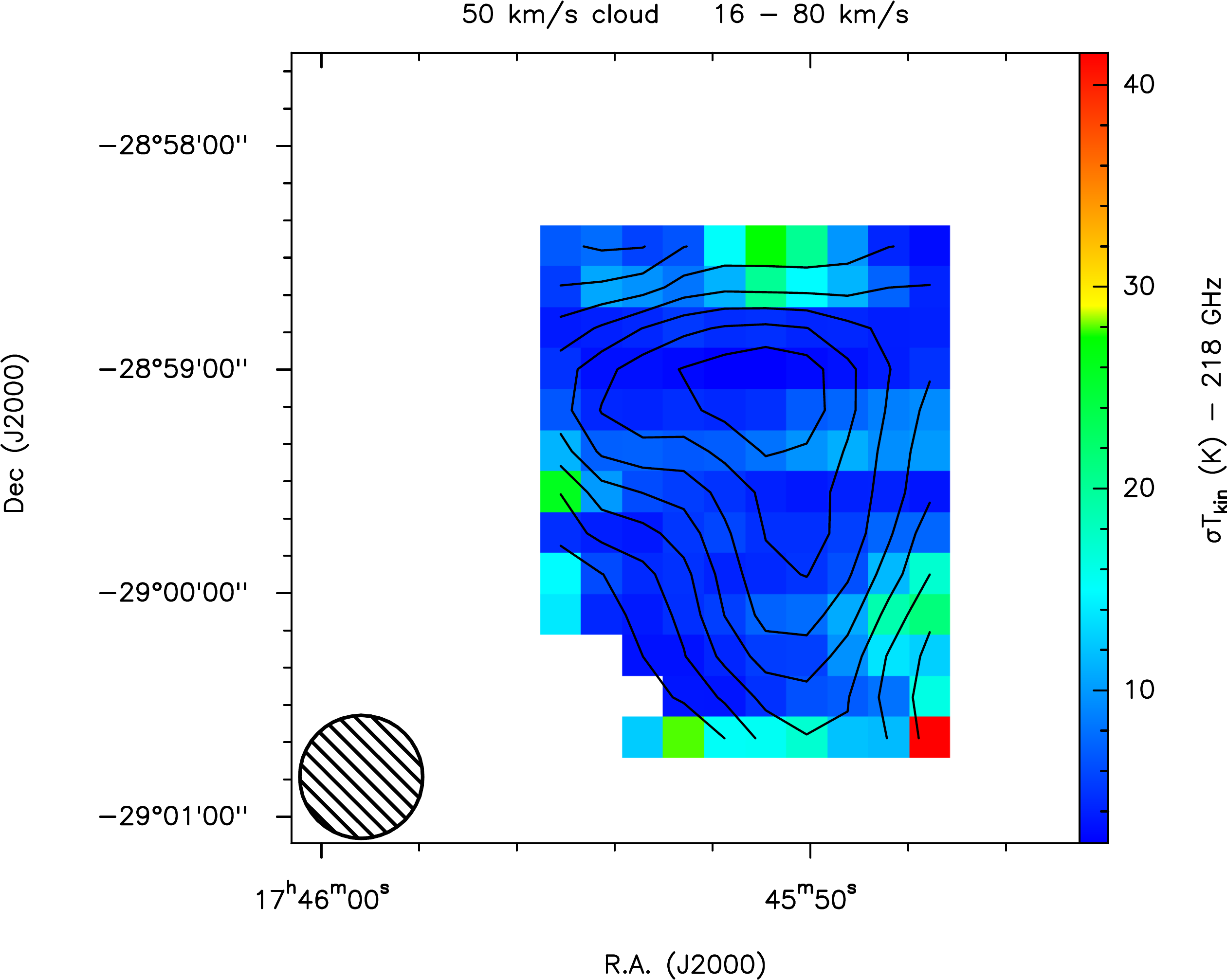}}
	\subfloat{\includegraphics[bb = 150 0 700 560, clip, height=5.385cm]{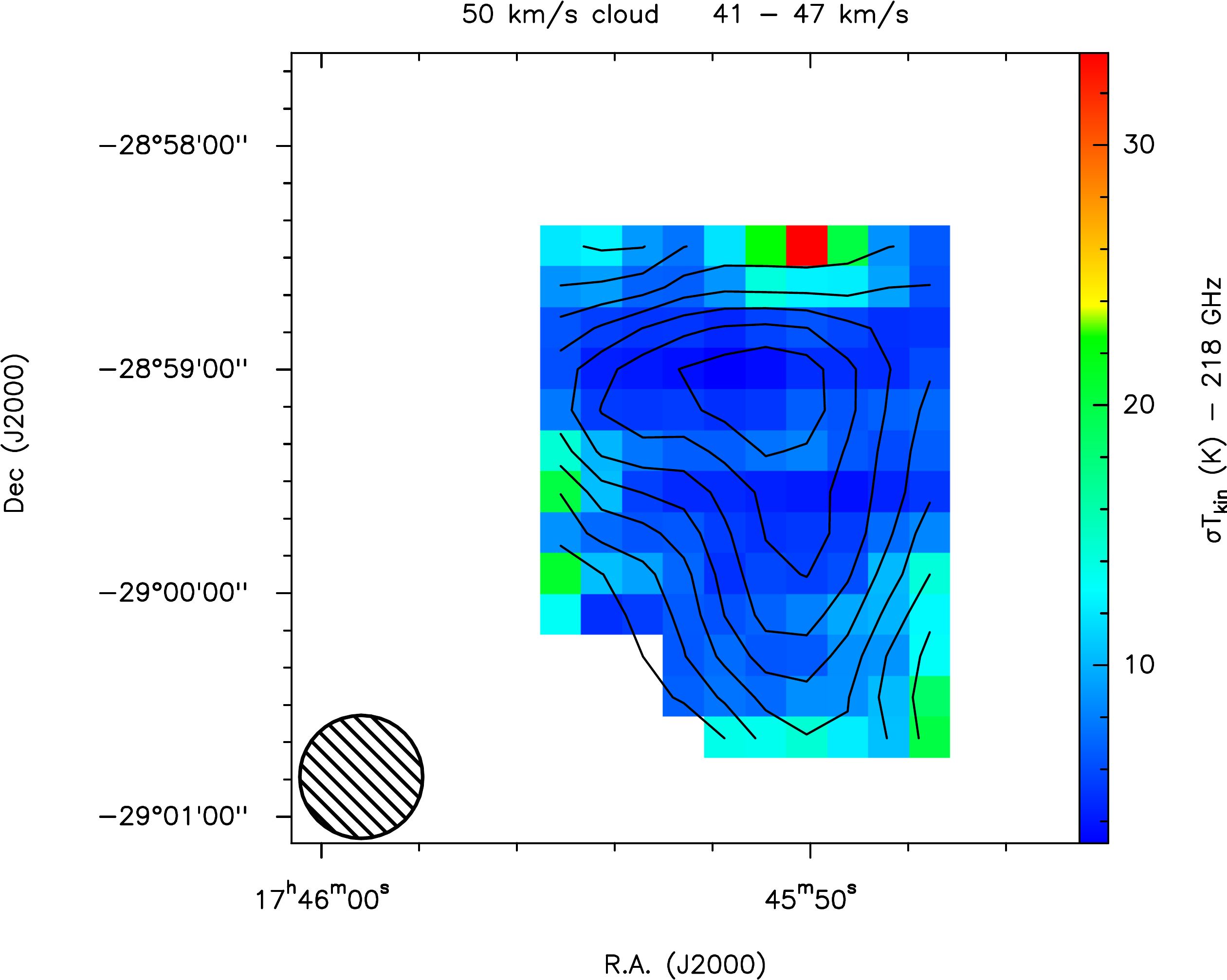}}
	\subfloat{\includegraphics[bb = 150 0 730 560, clip, height=5.385cm]{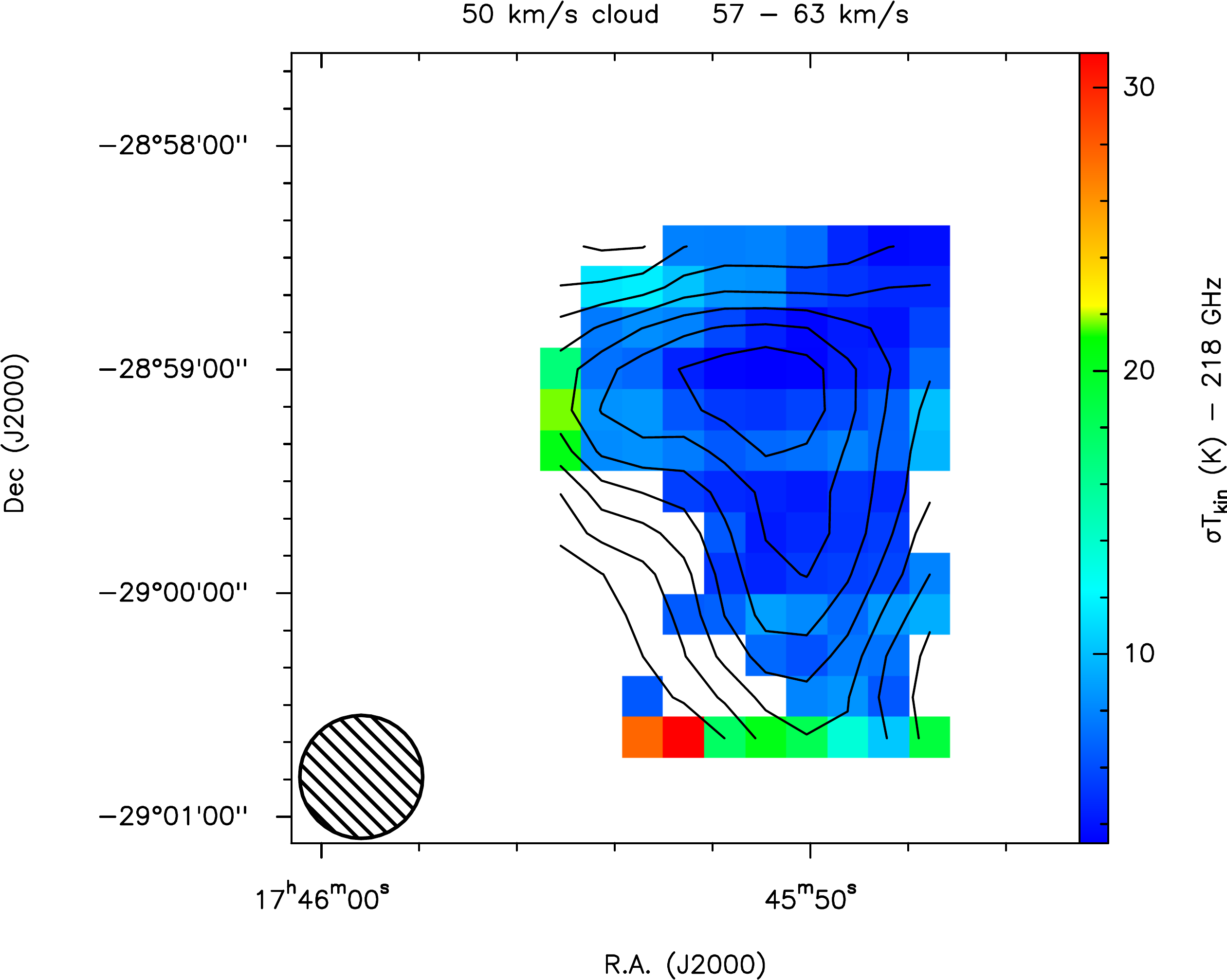}}\\ 
	\vspace{0.1cm}
	291 GHz temperatures \\
	\subfloat{\includegraphics[bb = 0 60 700 580, clip, height=5cm]{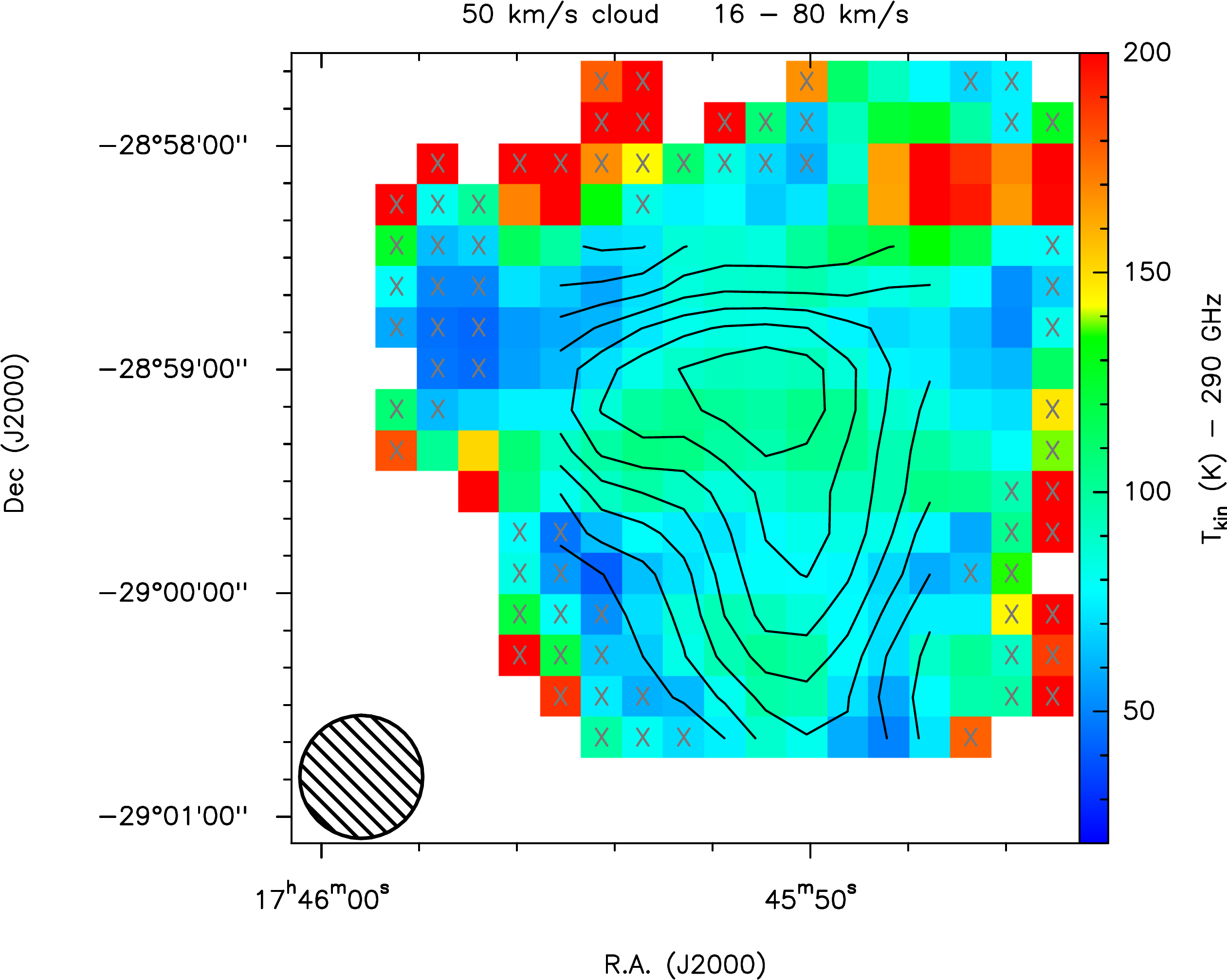}}
	\subfloat{\includegraphics[bb = 150 60 700 580, clip, height=5cm]{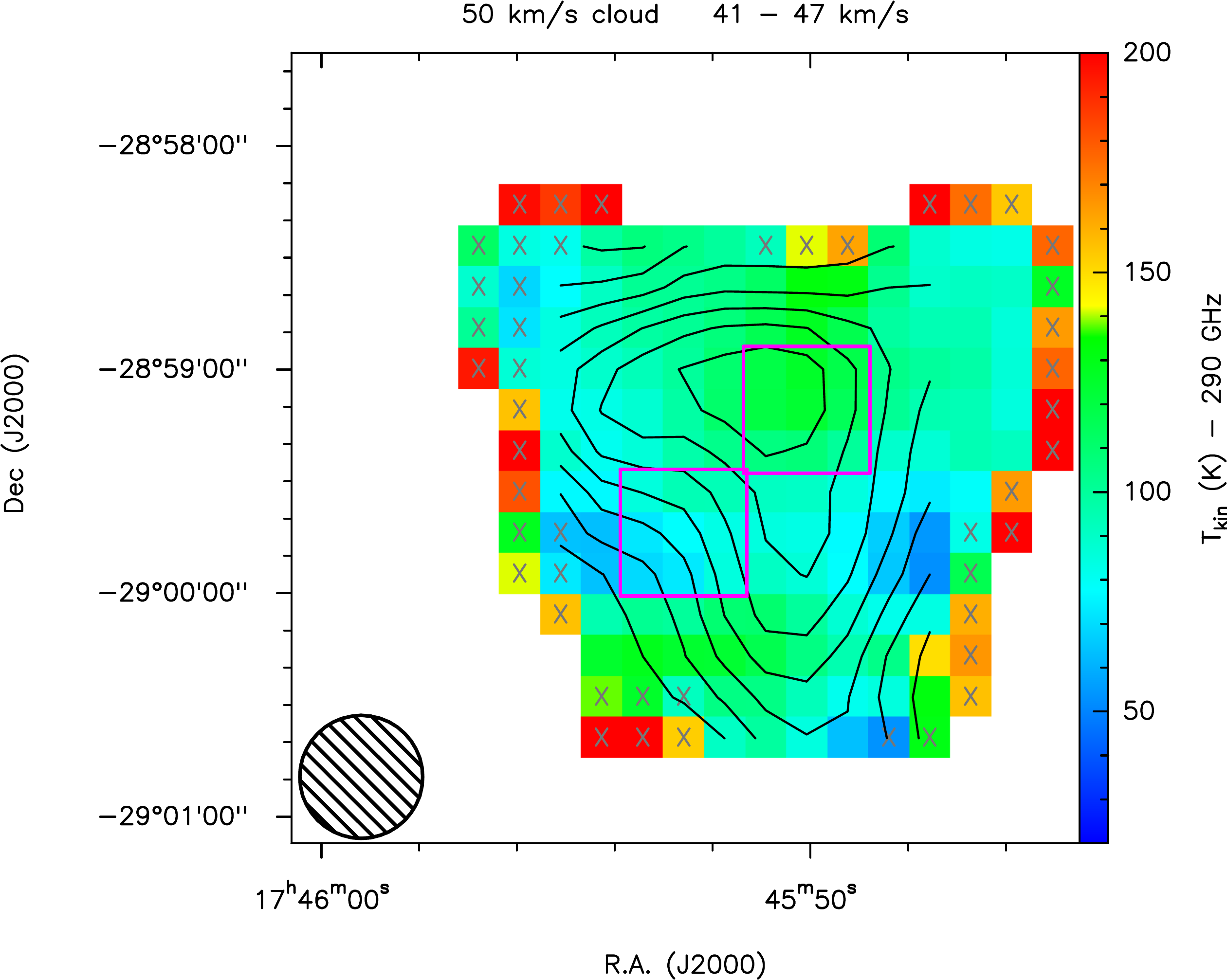}}
	\subfloat{\includegraphics[bb = 150 60 730 580, clip, height=5cm]{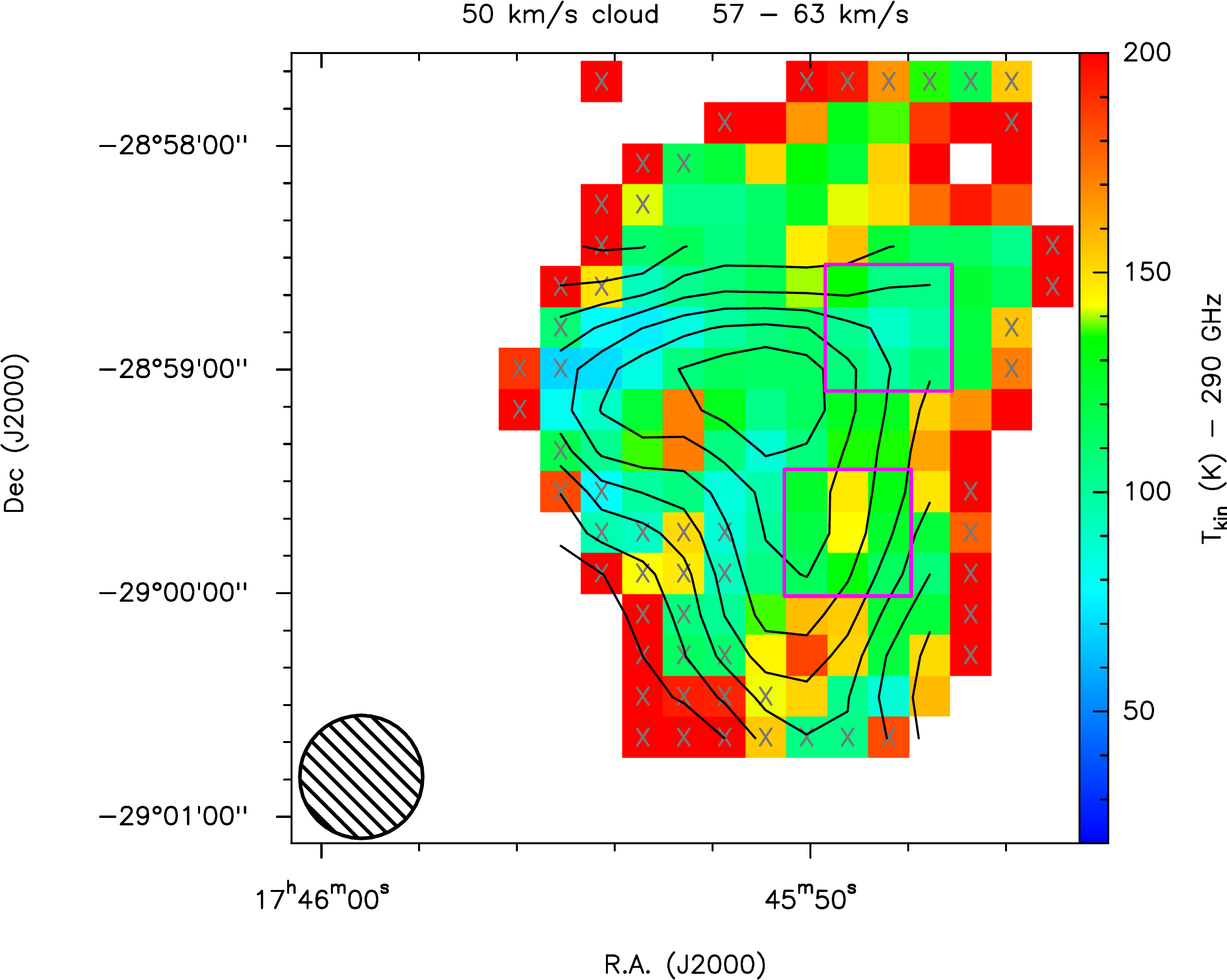}}\\\vspace{-0.5cm}
	\subfloat{\includegraphics[bb = 0 0 700 560, clip, height=5.385cm]{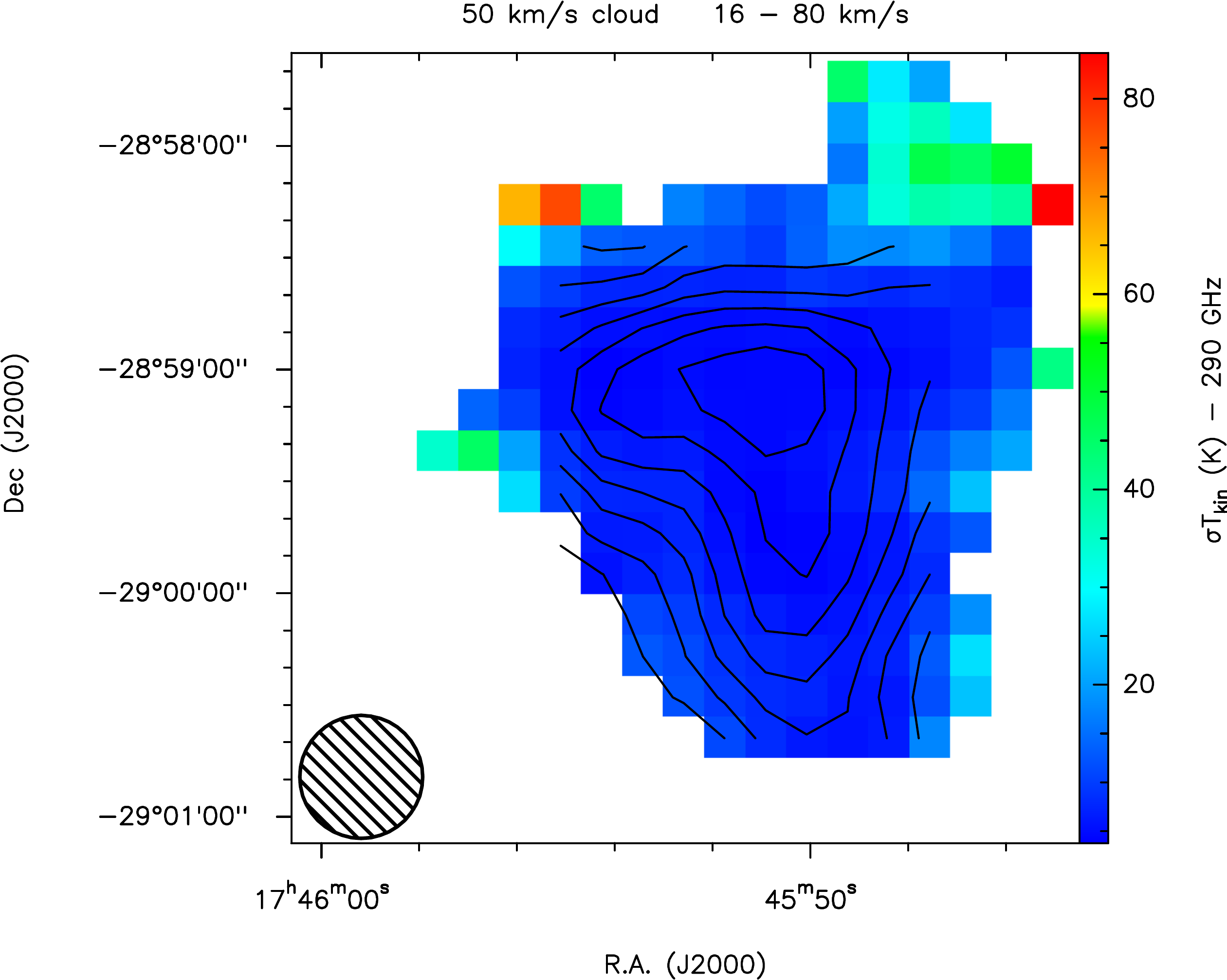}}
	\subfloat{\includegraphics[bb = 150 0 700 560, clip, height=5.385cm]{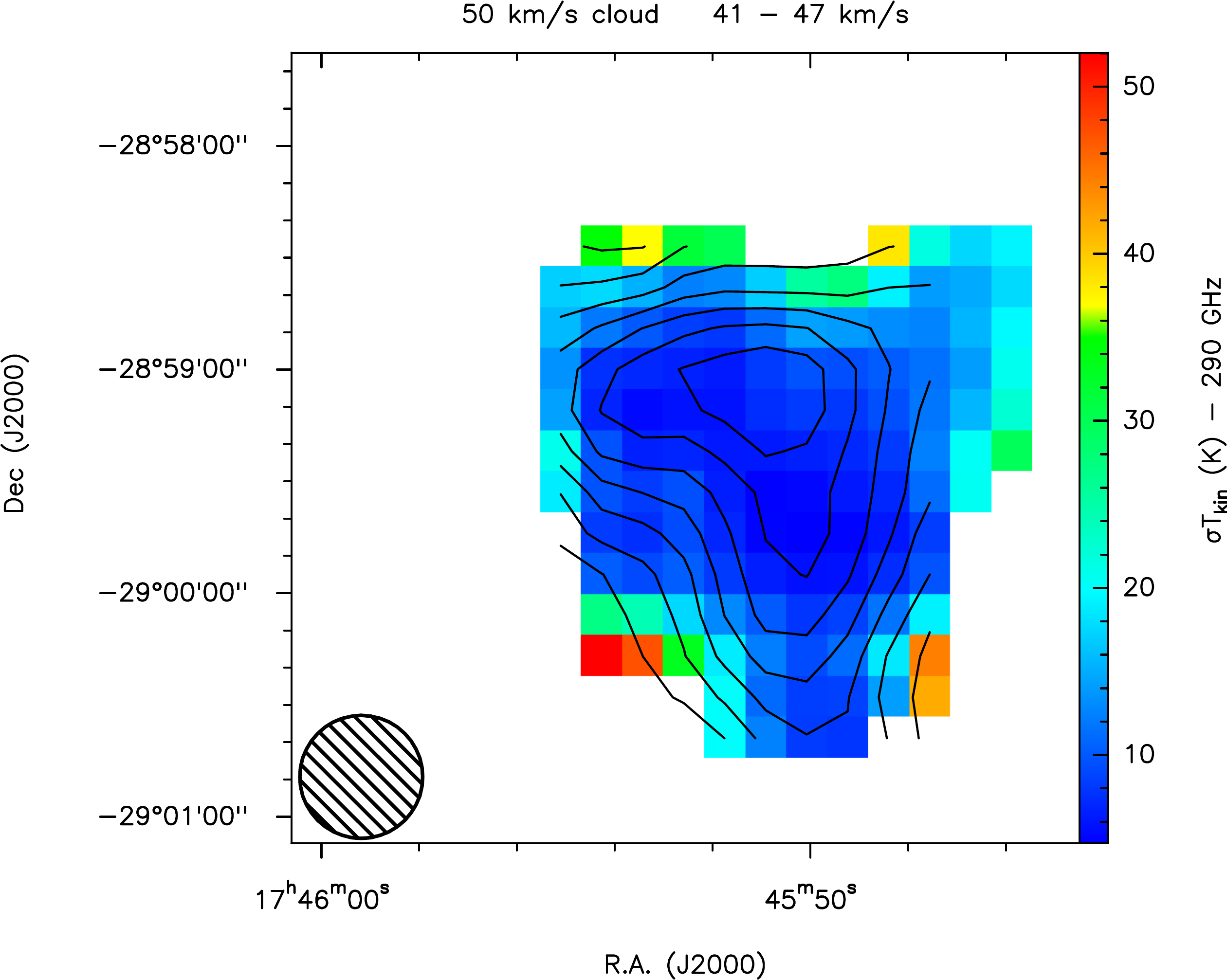}}
	\subfloat{\includegraphics[bb = 150 0 730 560, clip, height=5.385cm]{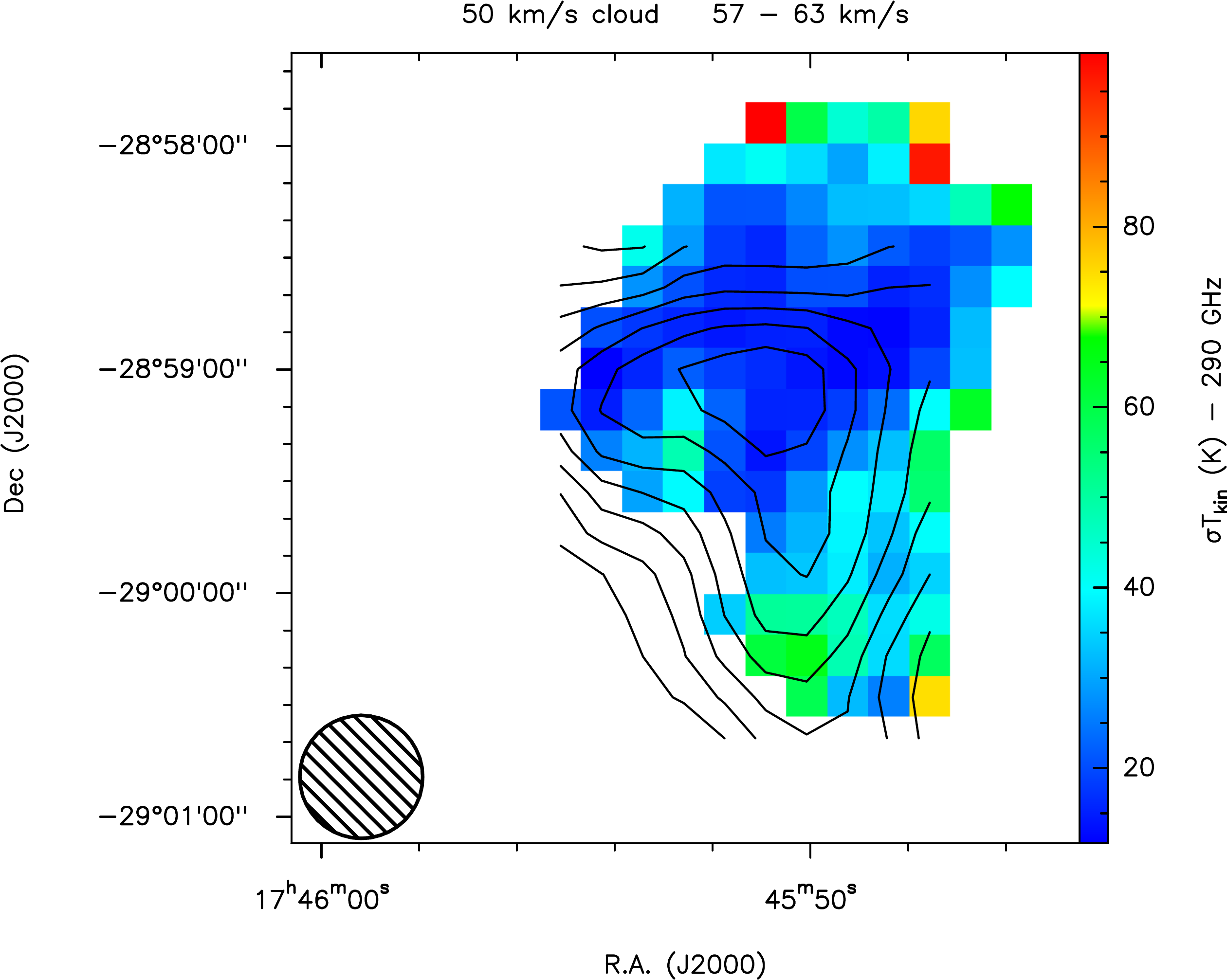}}
	\label{50kms-All-Temp-H2CO}
\end{figure*}

\begin{figure*}
	\caption{As Fig. \ref{20kms-All-Temp-H2CO}, for G0.253+0.016.}
	\centering
        218 GHz temperatures\\
	\subfloat{\includegraphics[bb = 0 60 540 580, clip, height=4.4cm]{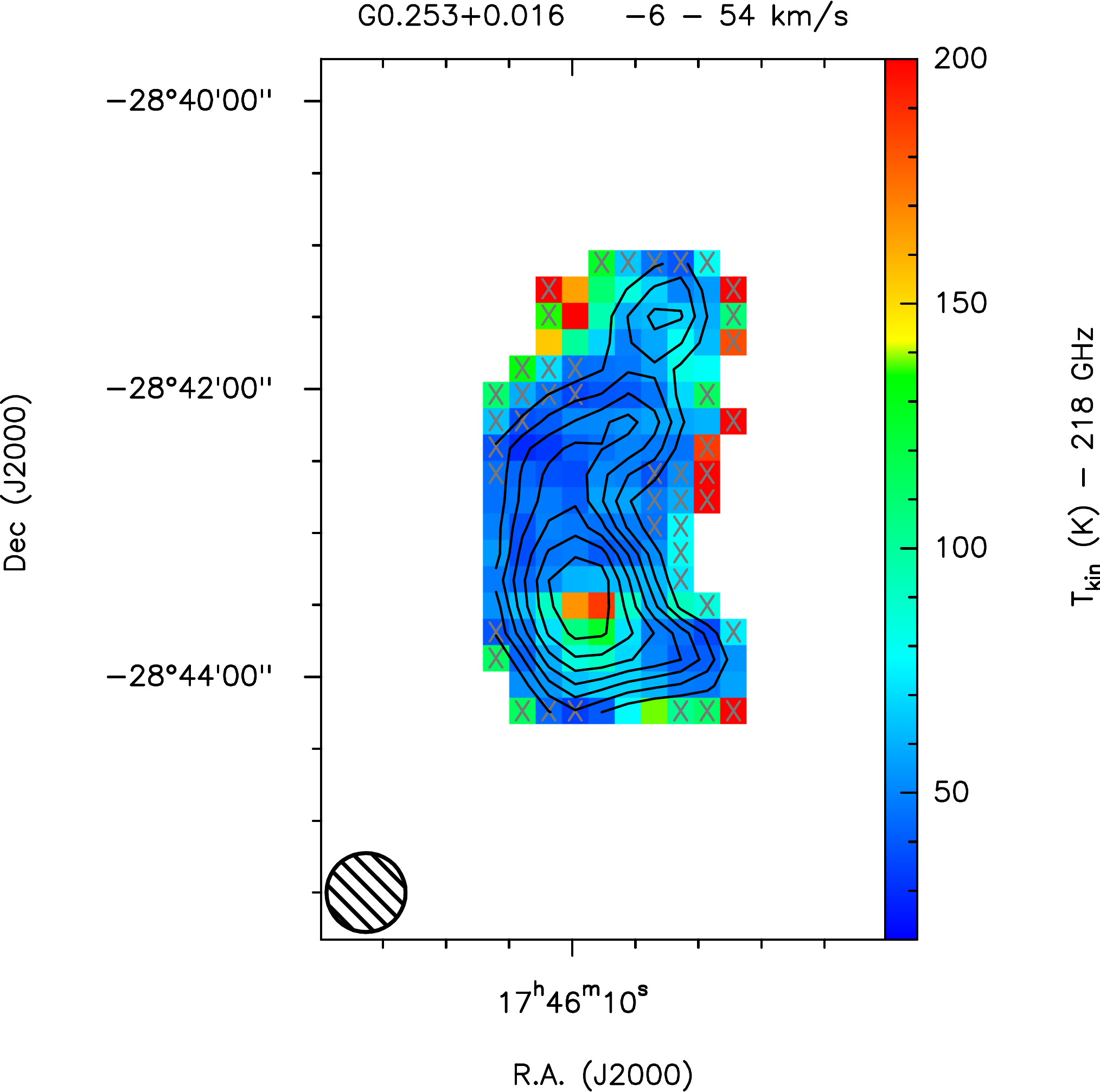}}
	\subfloat{\includegraphics[bb = 150 60 540 580, clip, height=4.4cm]{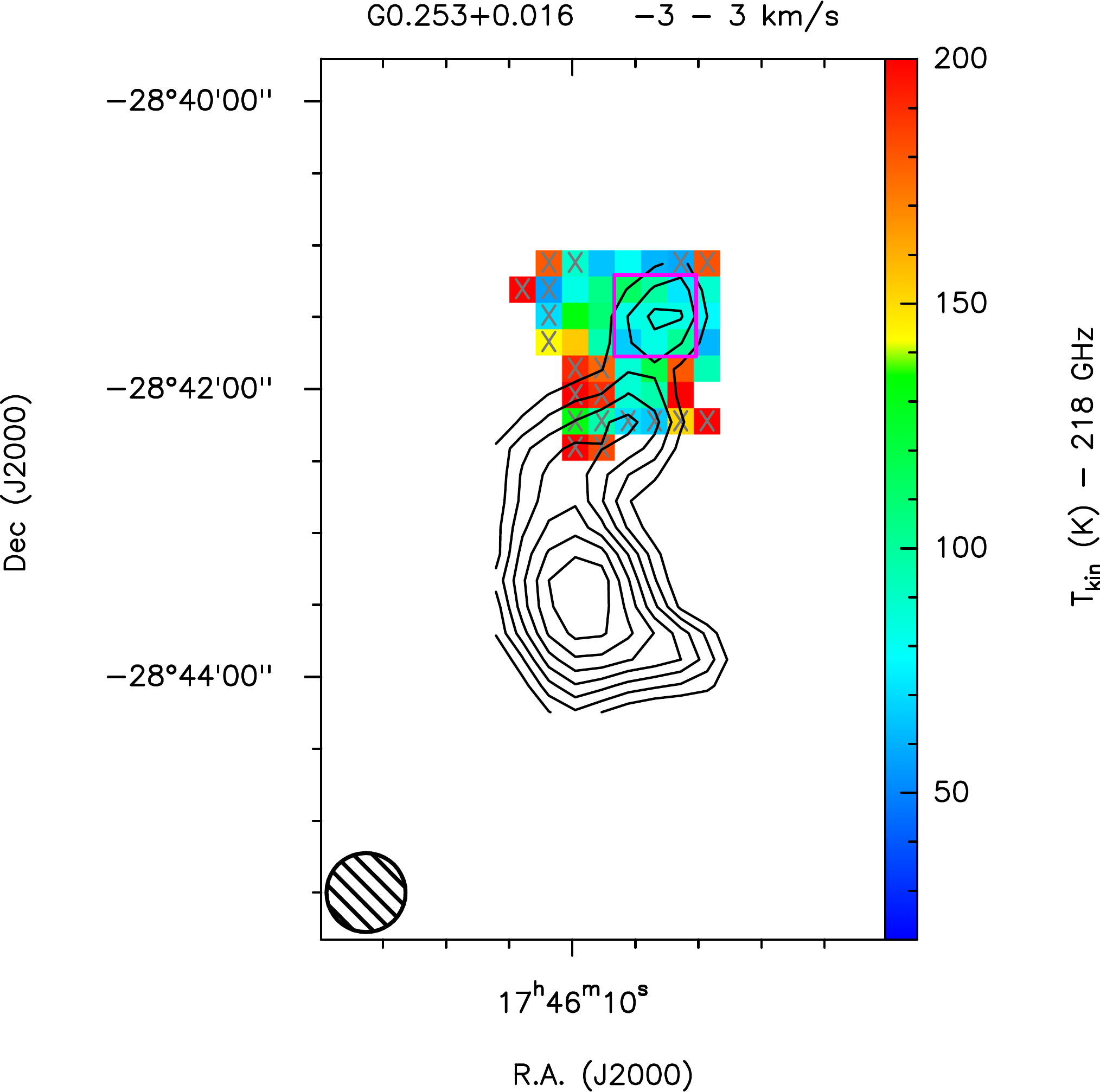}}
	\subfloat{\includegraphics[bb = 150 60 540 580, clip, height=4.4cm]{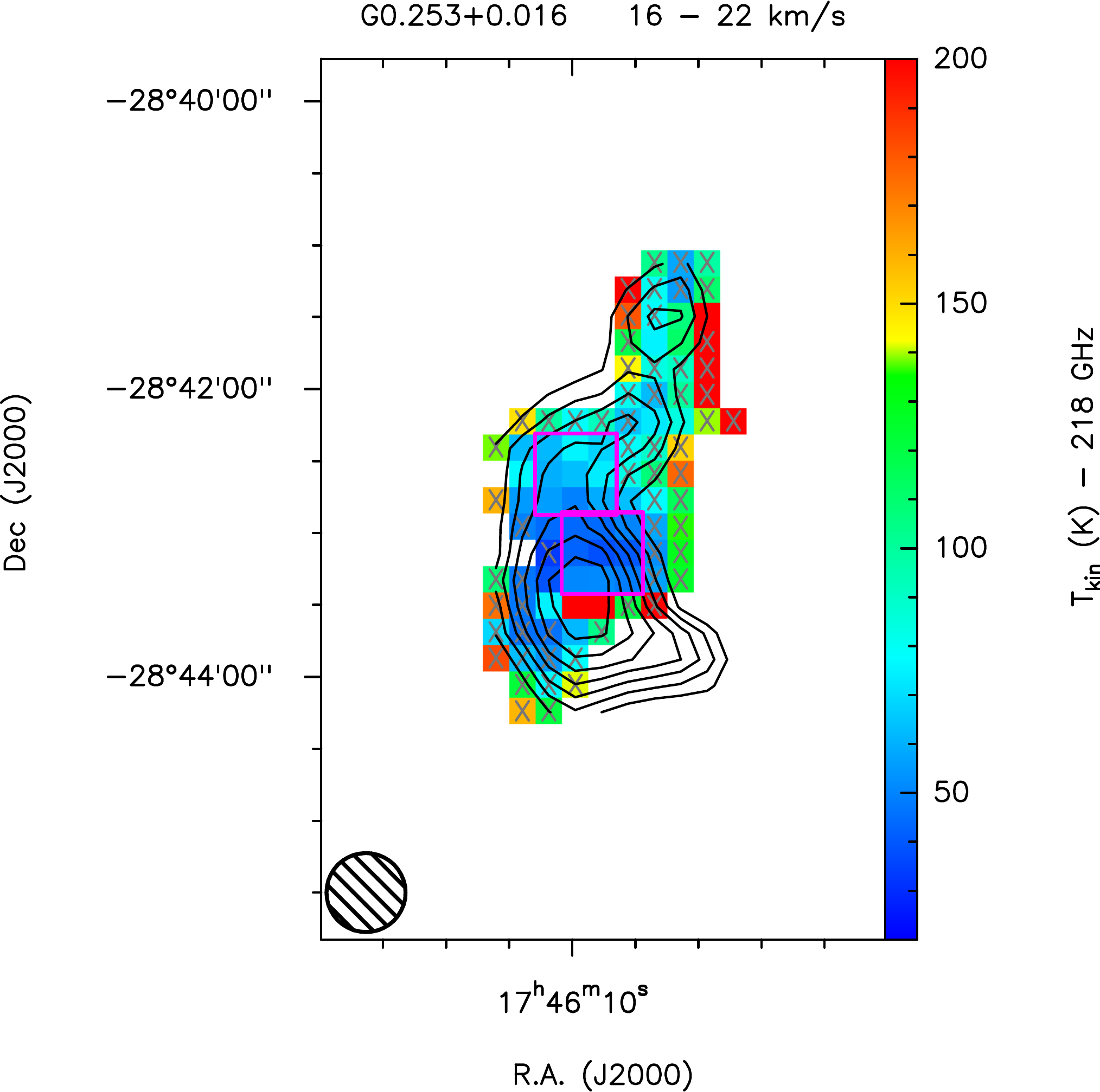}}
	\subfloat{\includegraphics[bb = 150 60 540 580, clip, height=4.4cm]{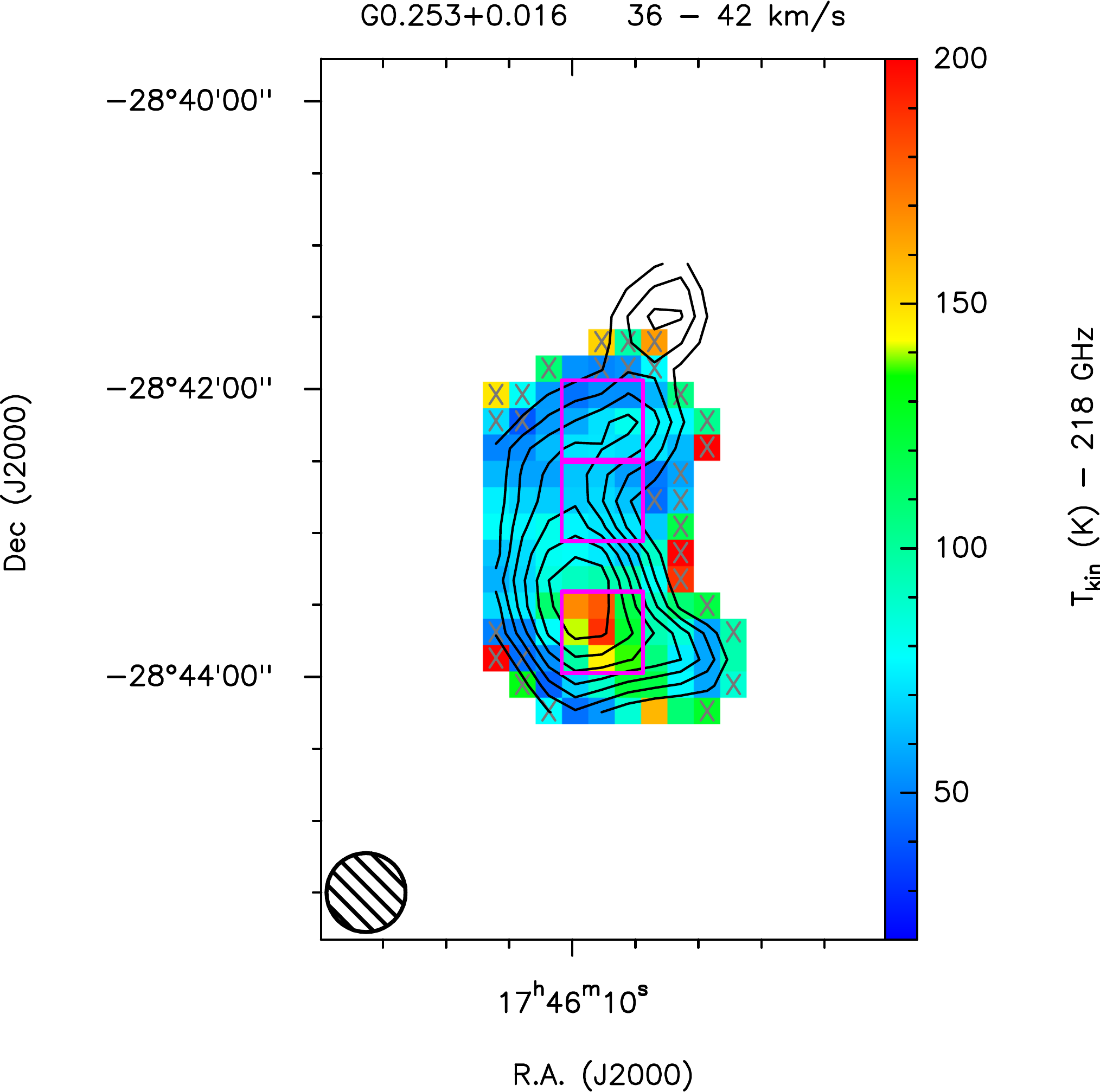}}
	\subfloat{\includegraphics[bb = 150 60 600 580, clip, height=4.4cm]{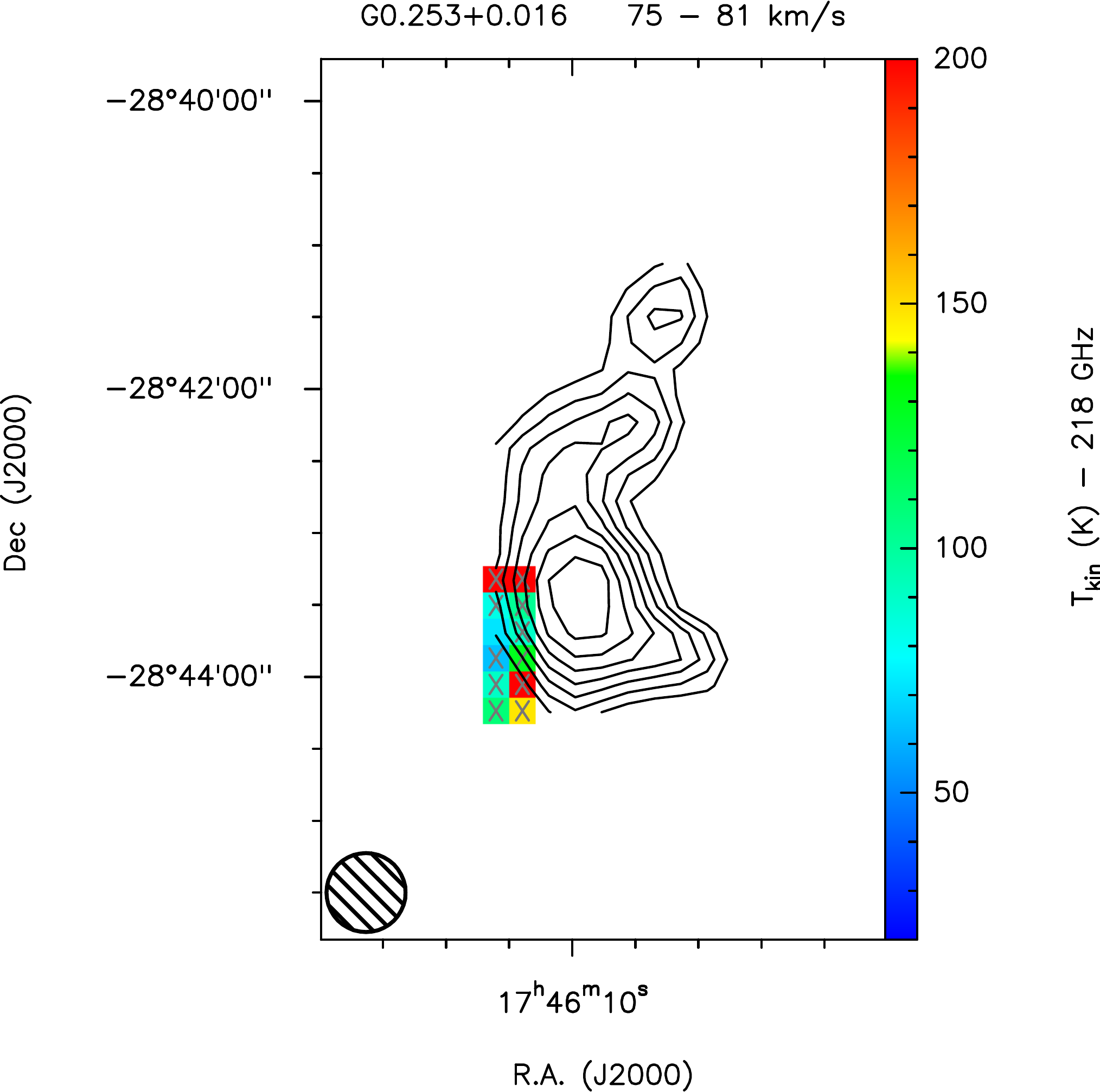}}\\
	\vspace{-0.5cm}
	\subfloat{\includegraphics[bb = 0 0 540 560, clip, height=4.737cm]{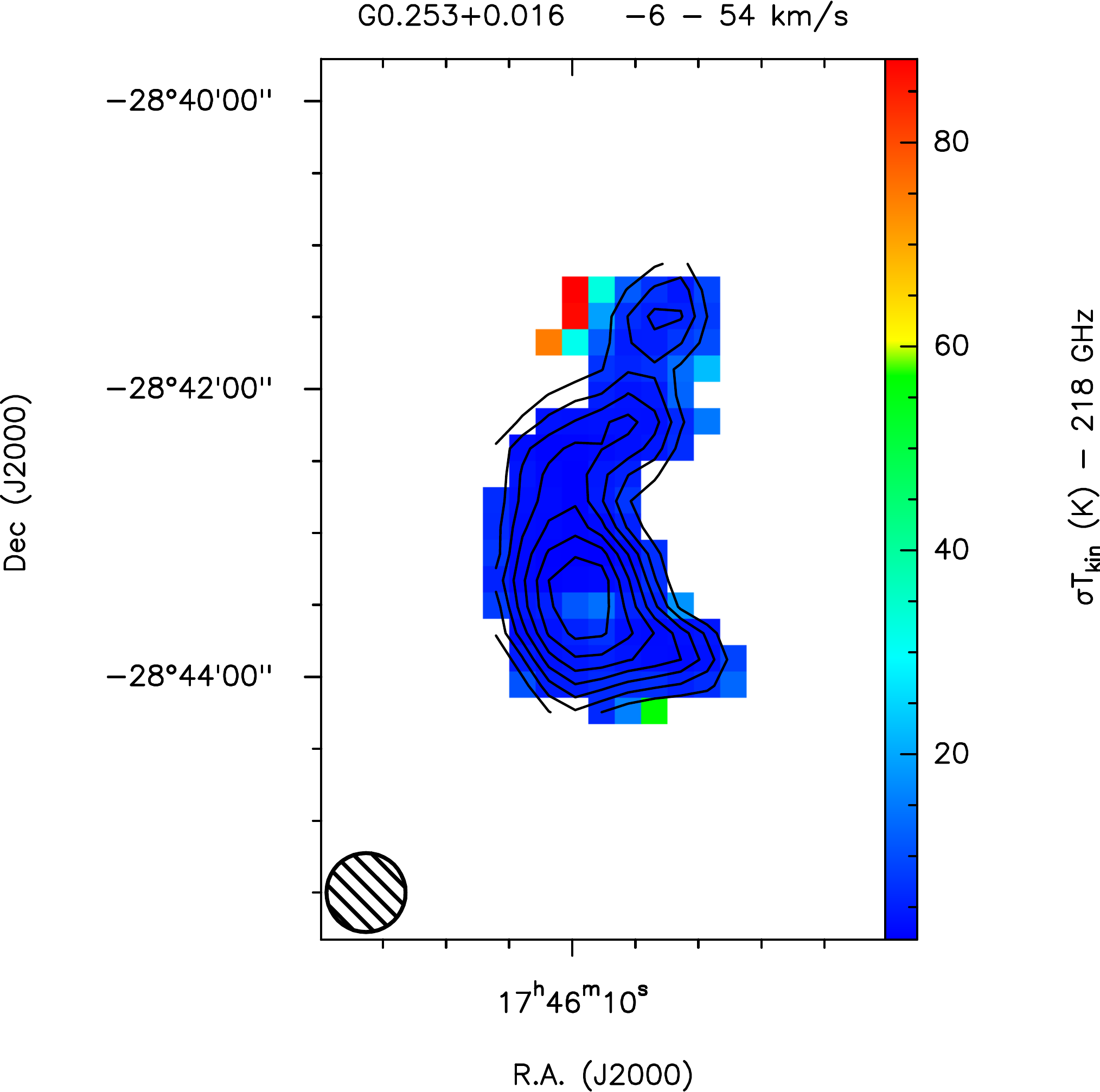}}
	\subfloat{\includegraphics[bb = 150 0 540 560, clip, height=4.737cm]{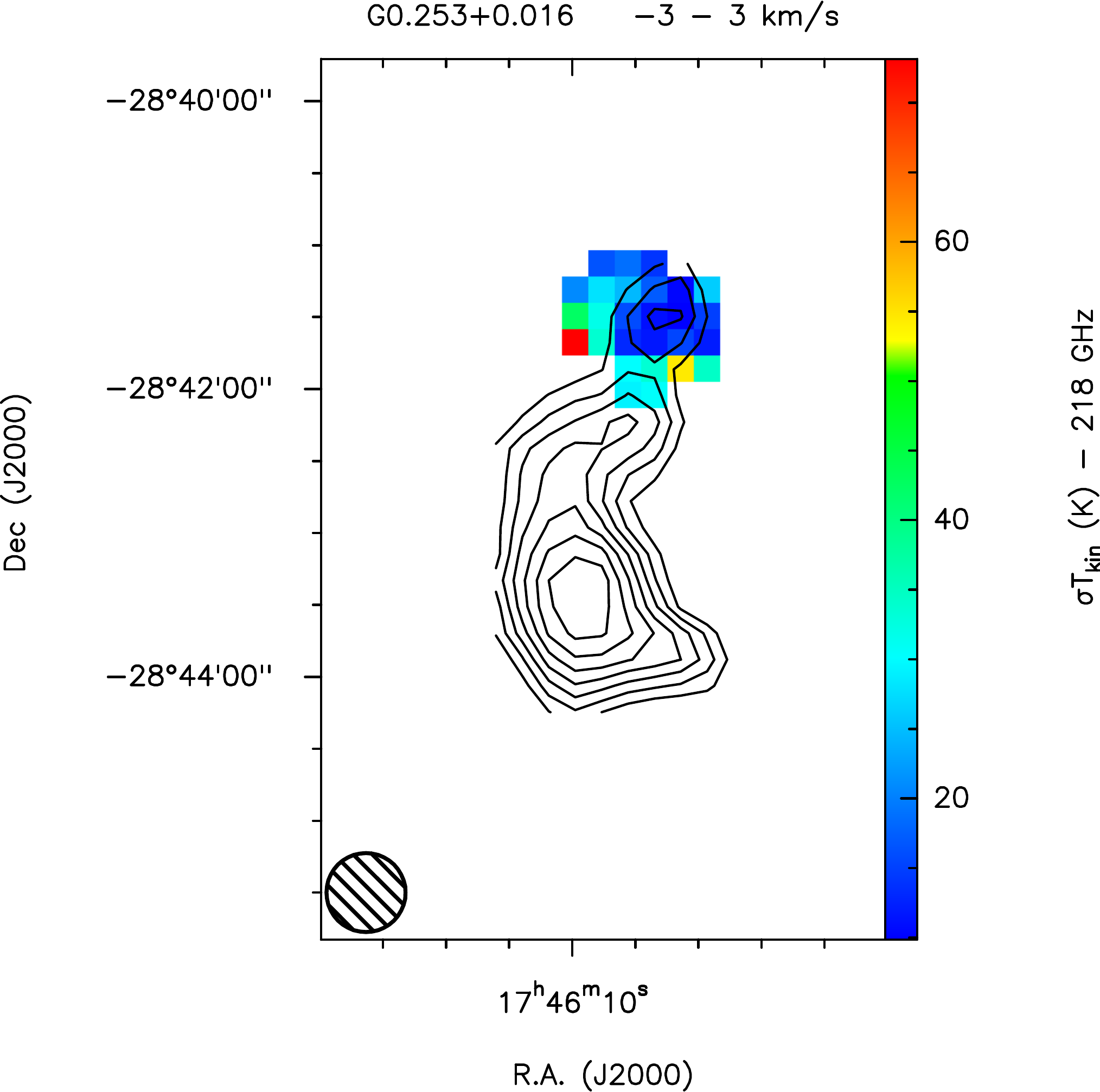}}
	\subfloat{\includegraphics[bb = 150 0 540 560, clip, height=4.737cm]{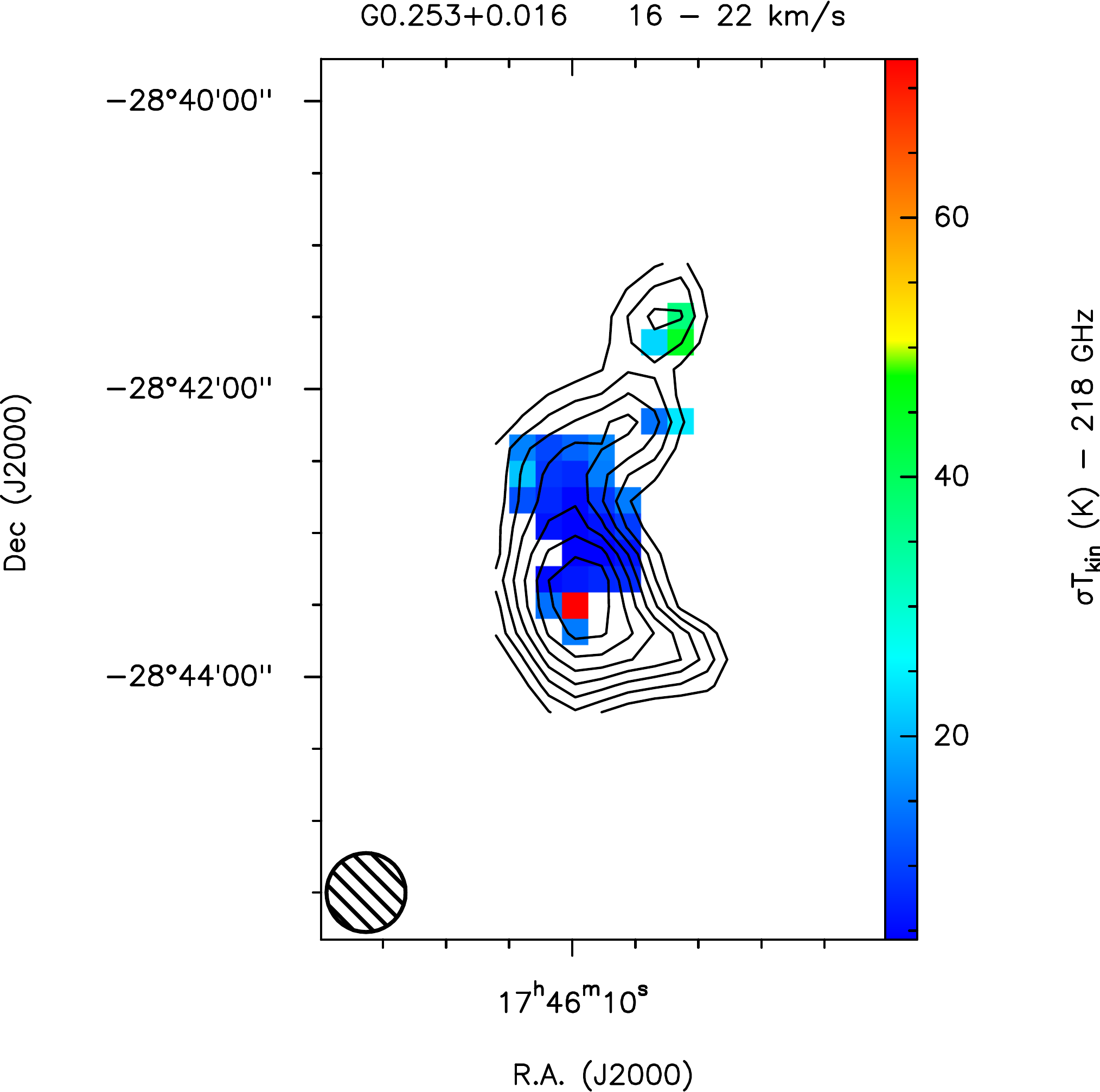}}
	\subfloat{\includegraphics[bb = 150 0 540 560, clip, height=4.737cm]{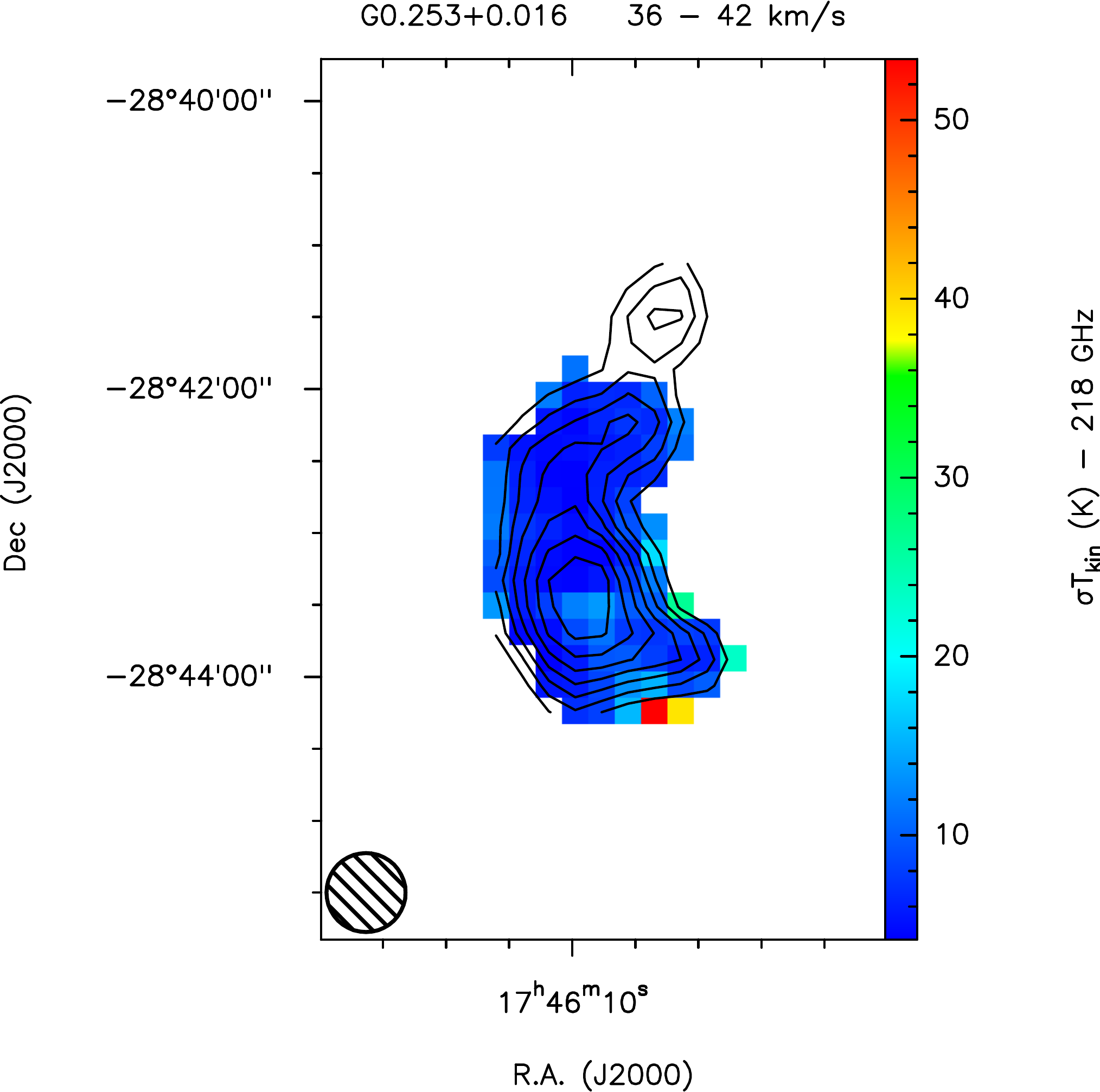}}
	\subfloat{\includegraphics[bb = 150 0 600 560, clip, height=4.737cm]{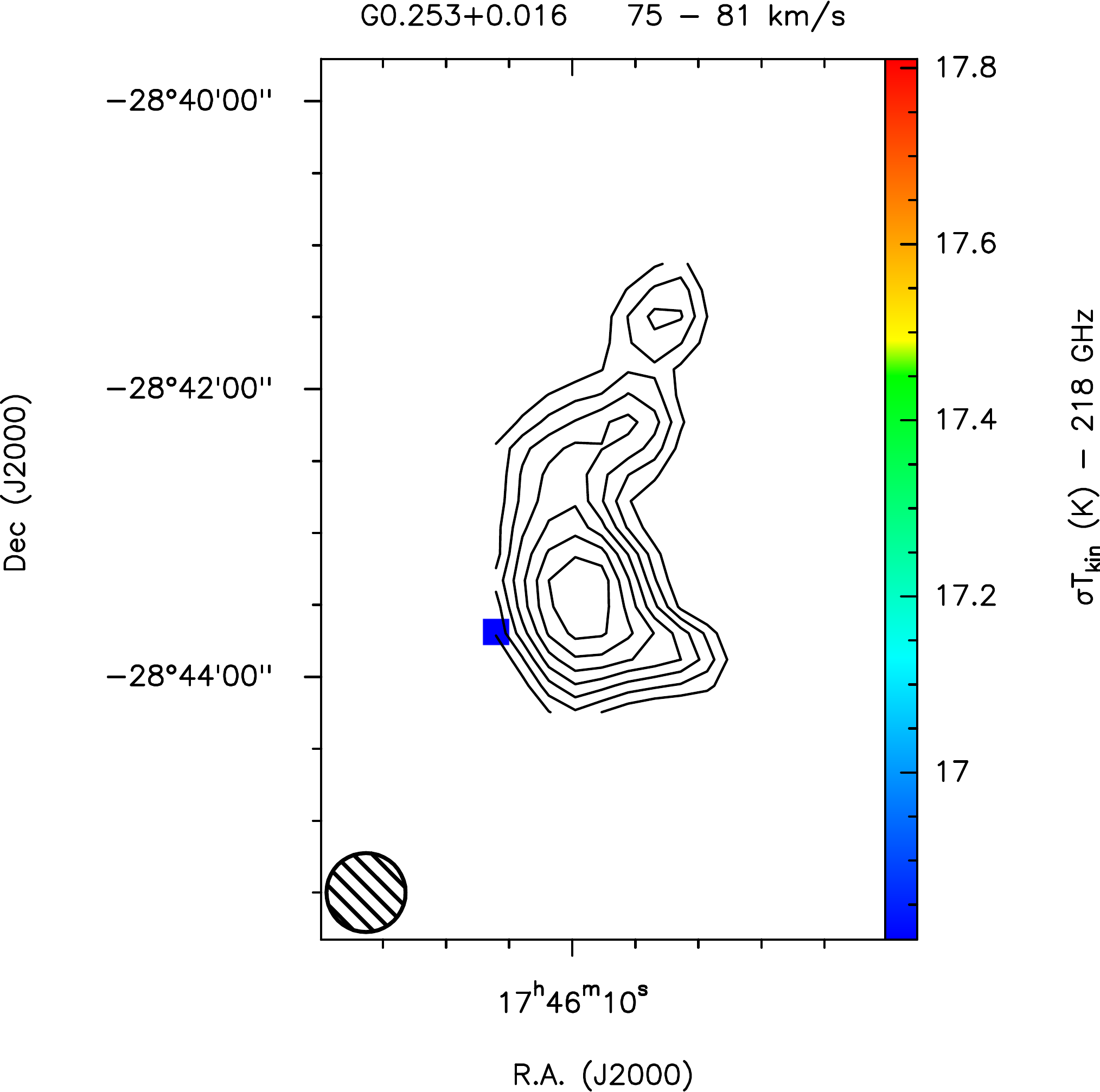}}\\ 
	\vspace{0.1cm}
	291 GHz temperatures \\
	\subfloat{\includegraphics[bb = 0 60 540 580, clip, height=4.4cm]{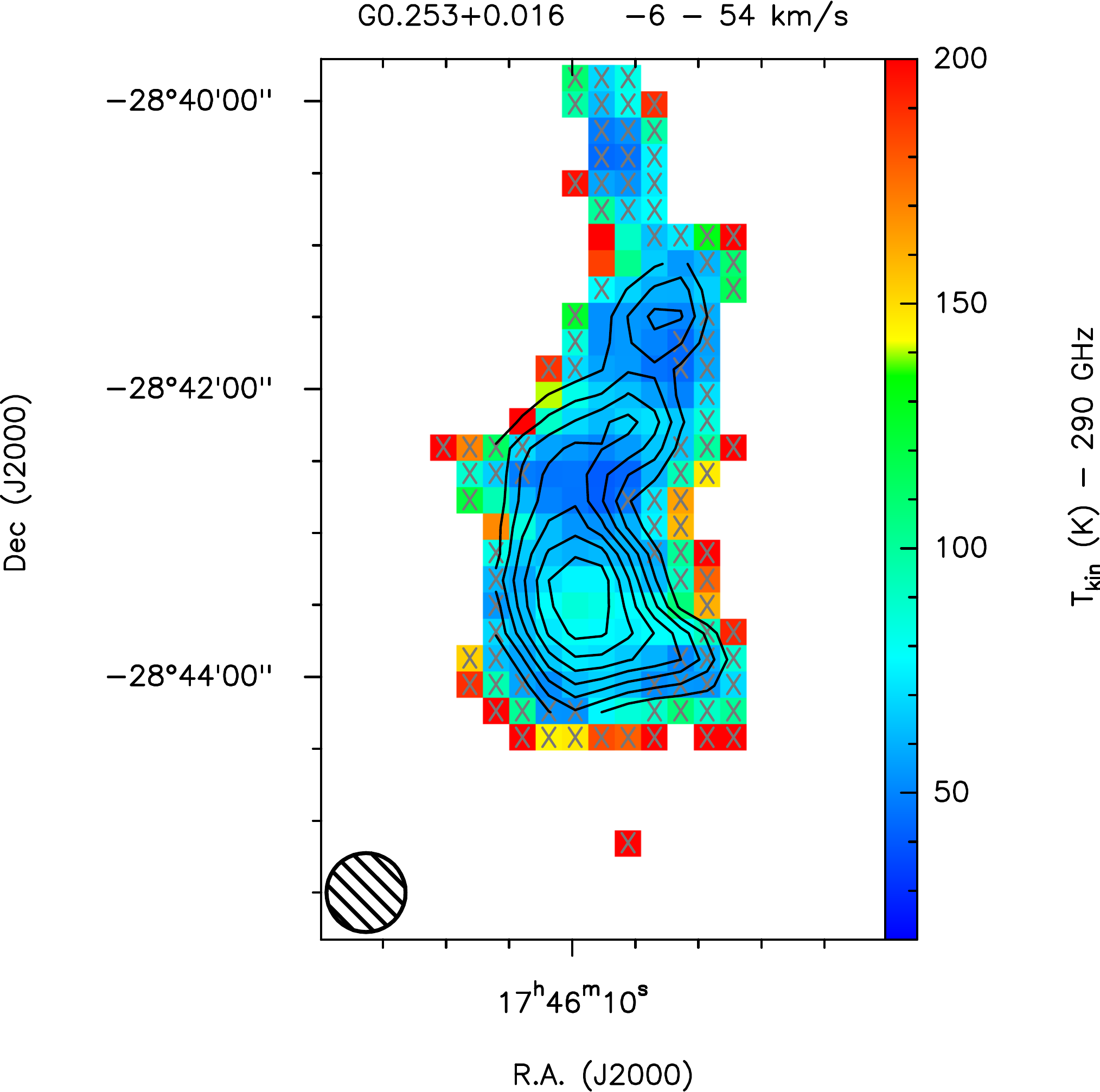}}
	\subfloat{\includegraphics[bb = 150 60 540 580, clip, height=4.4cm]{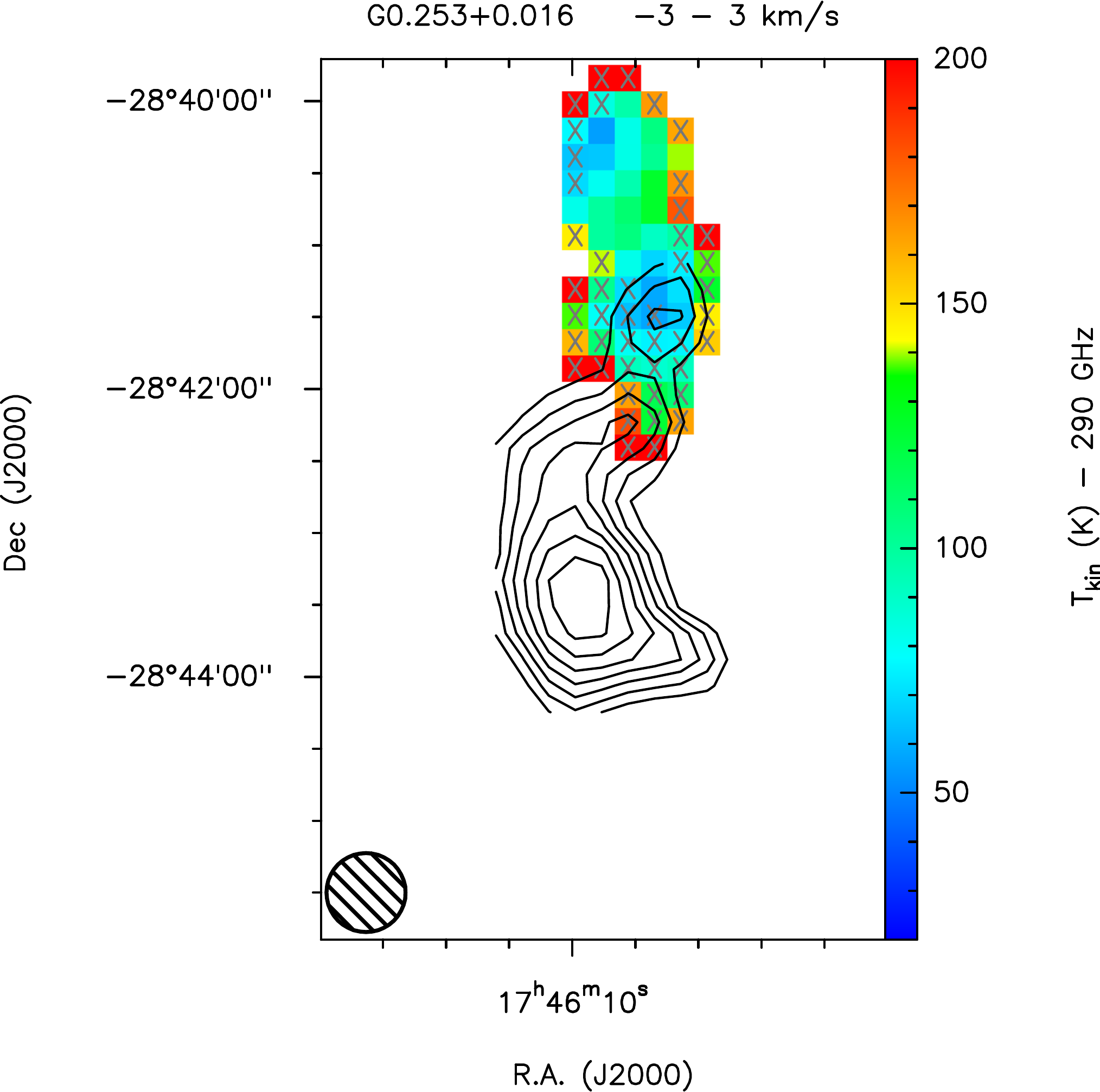}}
	\subfloat{\includegraphics[bb = 150 60 540 580, clip, height=4.4cm]{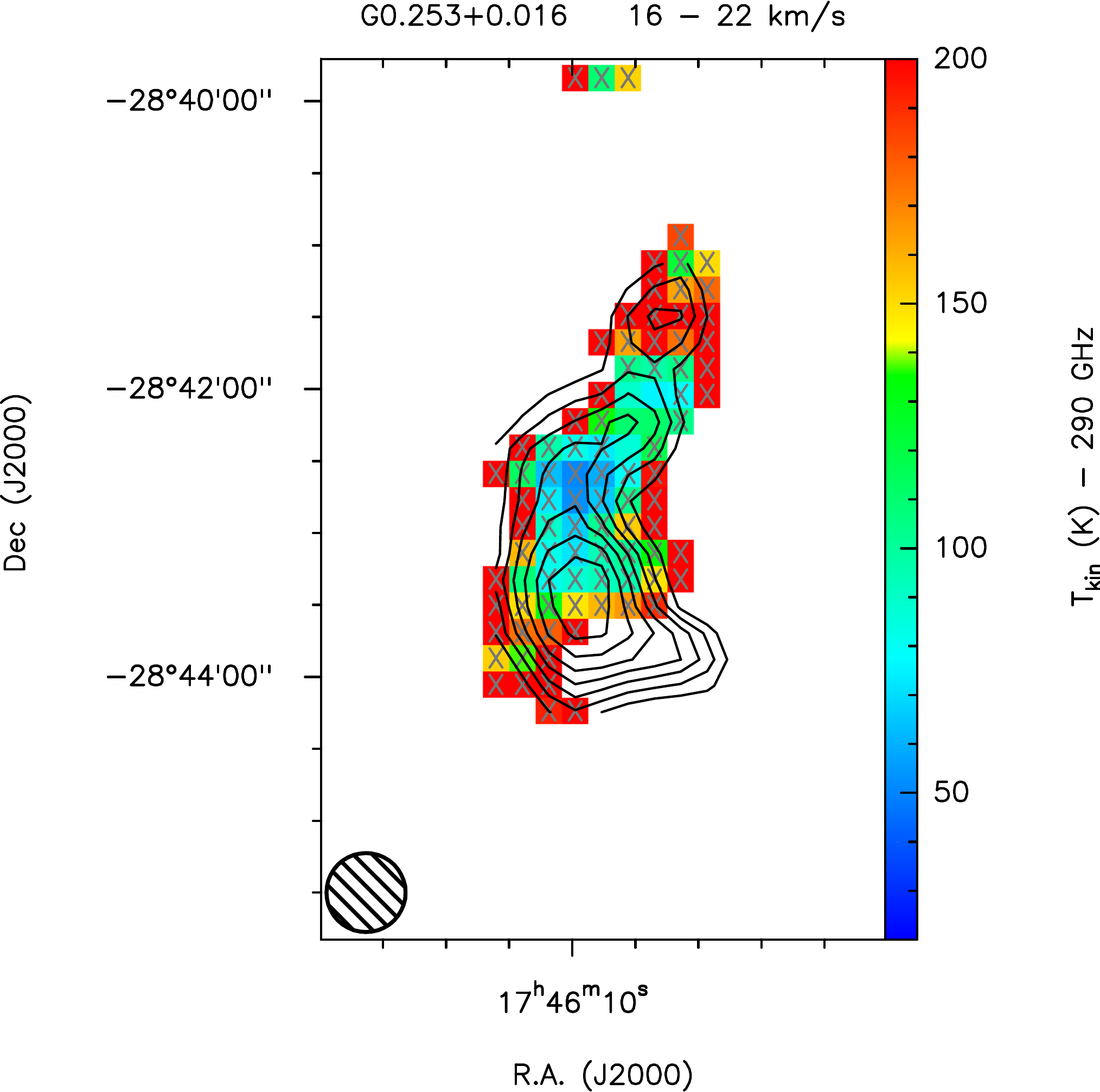}}
	\subfloat{\includegraphics[bb = 150 60 540 580, clip, height=4.4cm]{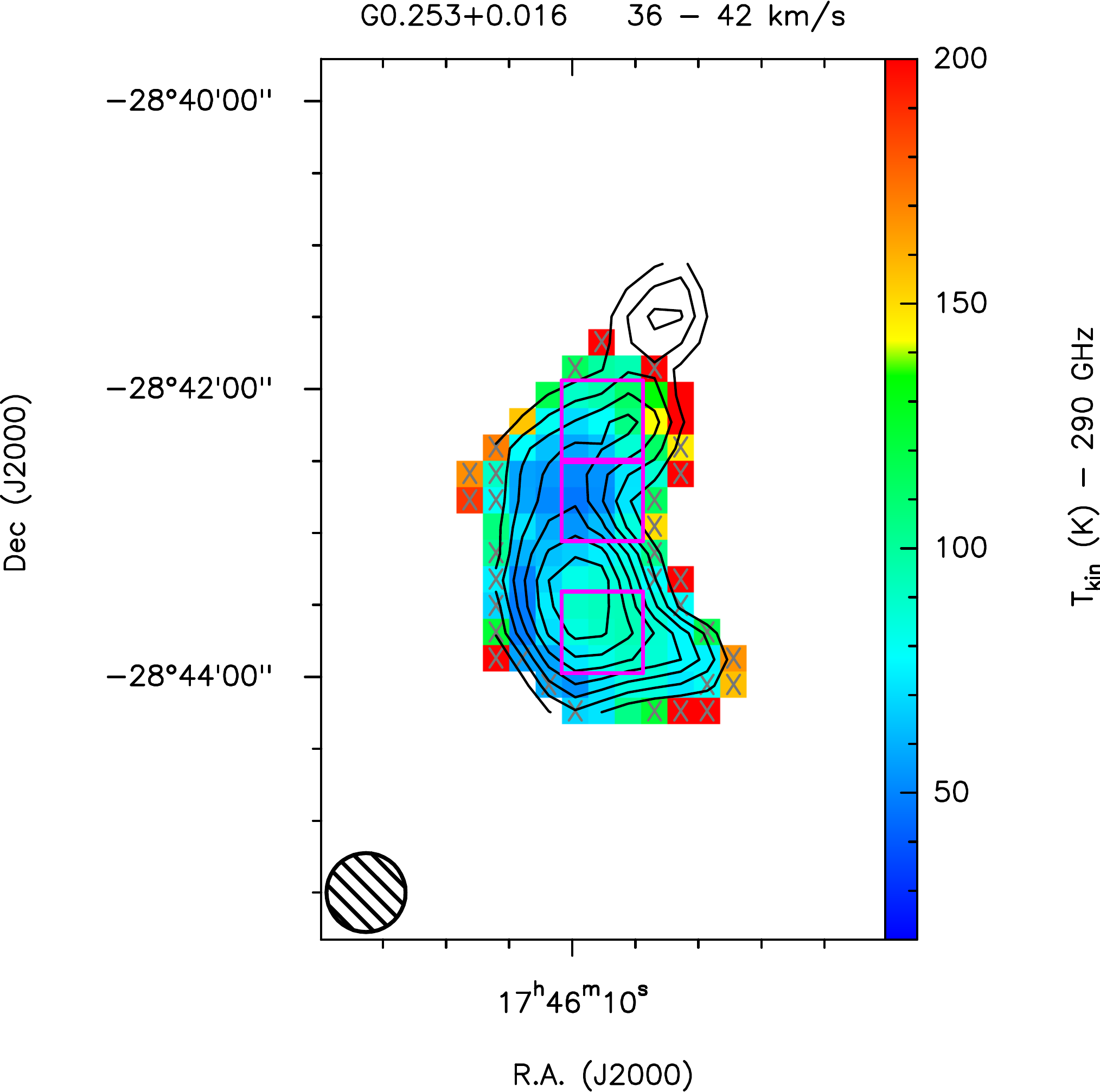}}
	\subfloat{\includegraphics[bb = 150 60 600 580, clip, height=4.4cm]{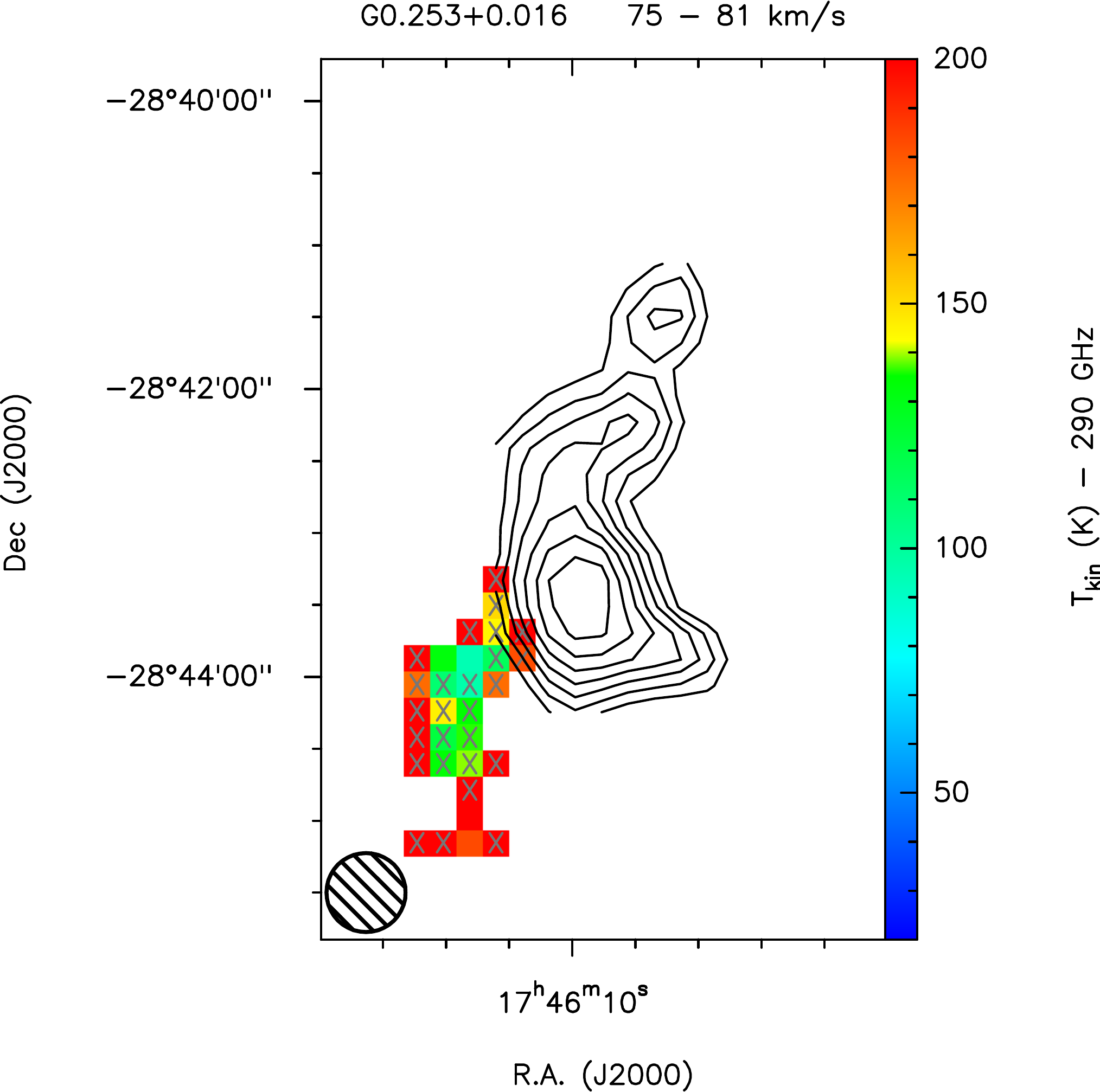}}\\
	\vspace{-0.5cm}
	\subfloat{\includegraphics[bb = 0 0 540 560, clip, height=4.737cm]{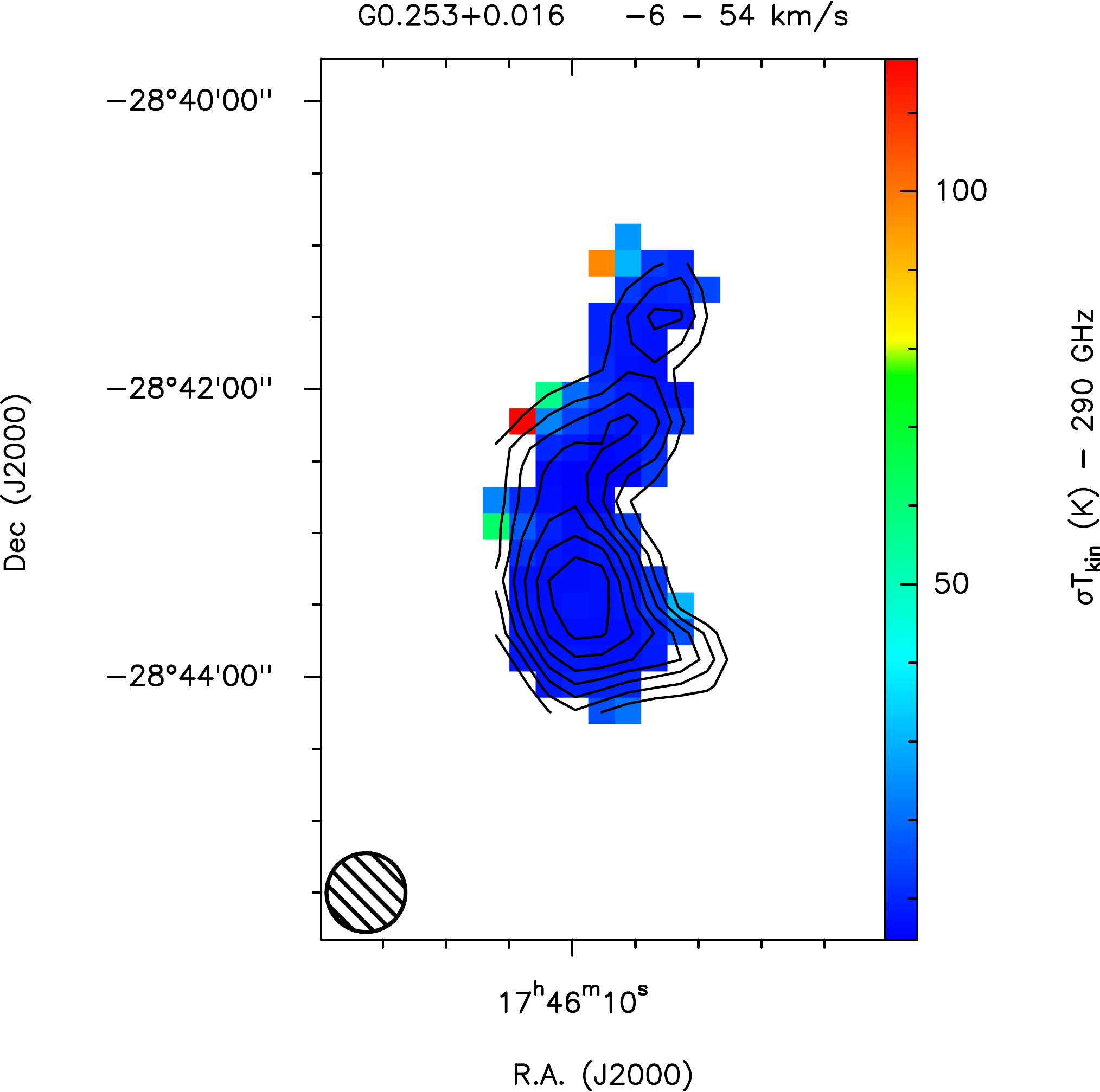}}
	\subfloat{\includegraphics[bb = 150 0 540 560, clip, height=4.737cm]{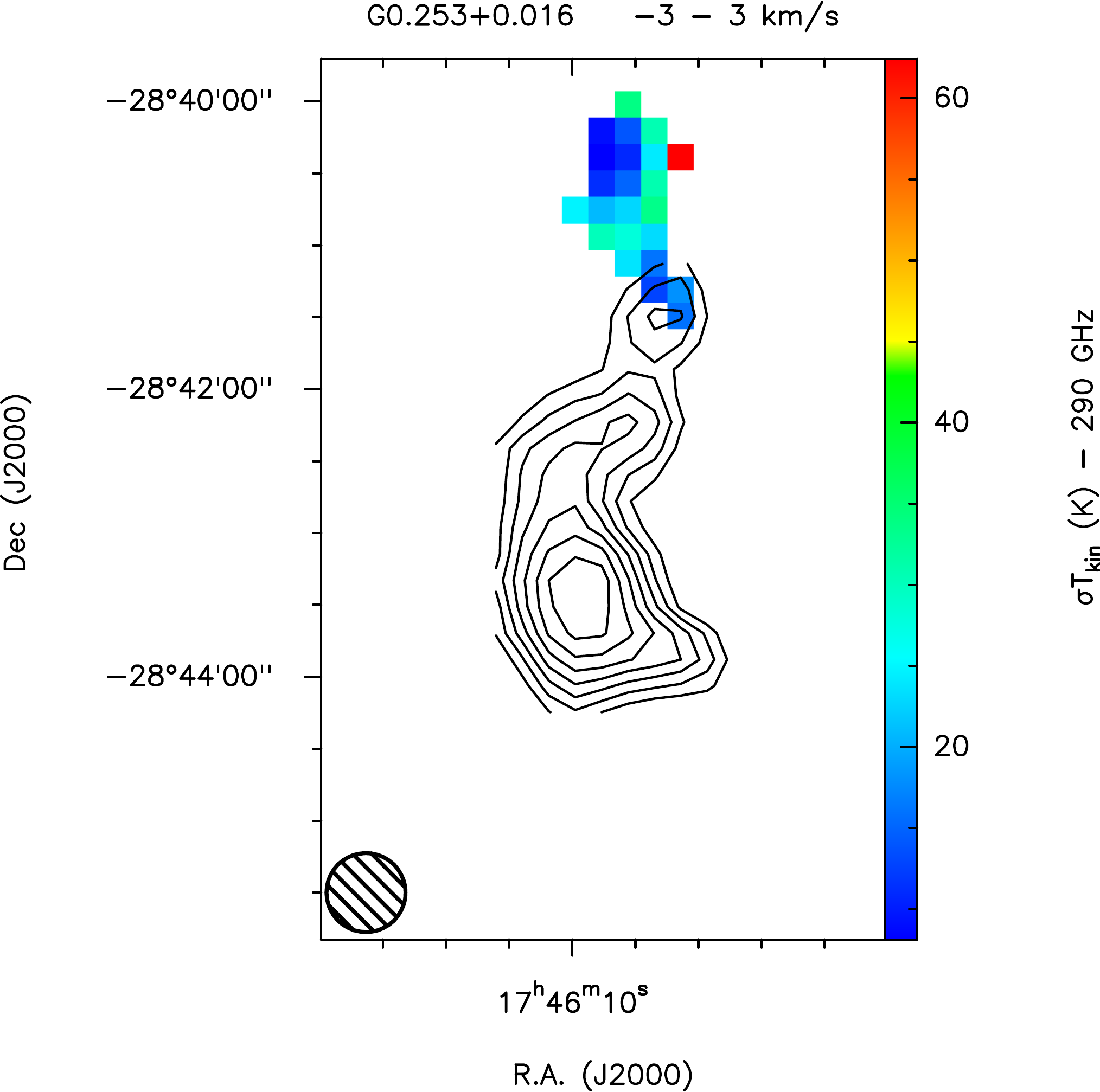}}
	\subfloat{\includegraphics[bb = 150 0 540 560, clip, height=4.737cm]{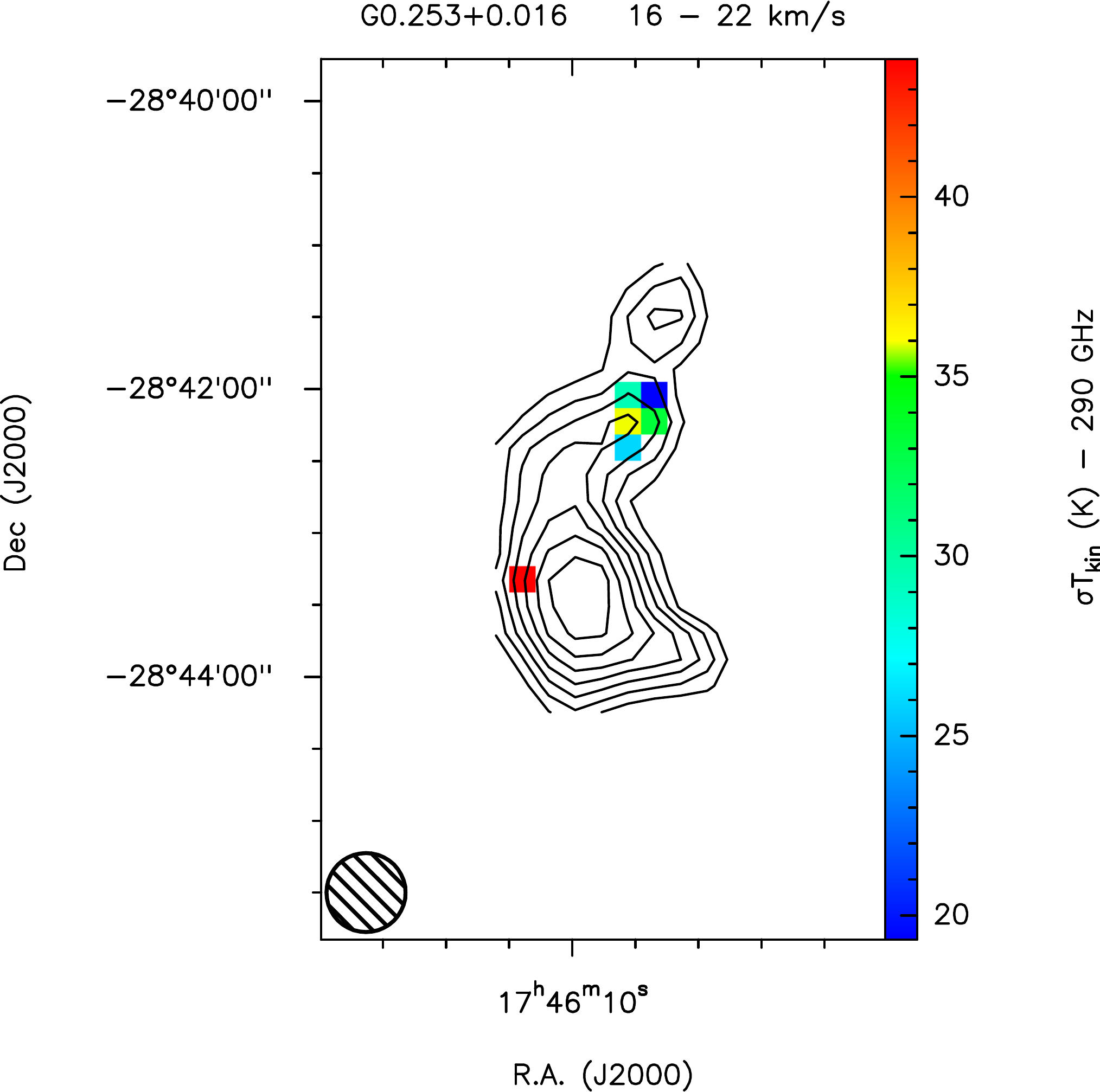}}
	\subfloat{\includegraphics[bb = 150 0 540 560, clip, height=4.737cm]{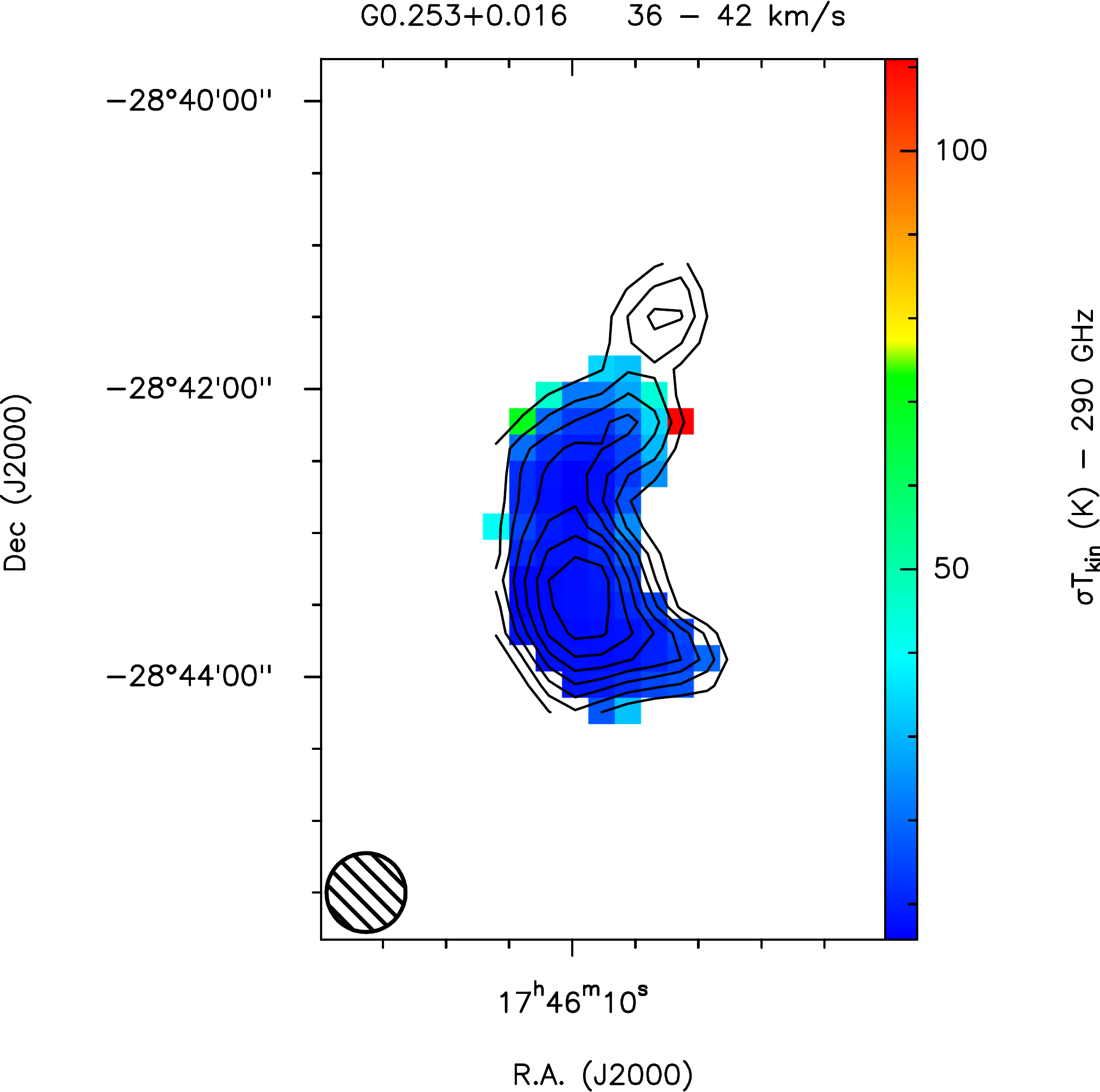}}
	\subfloat{\includegraphics[bb = 150 0 600 560, clip, height=4.737cm]{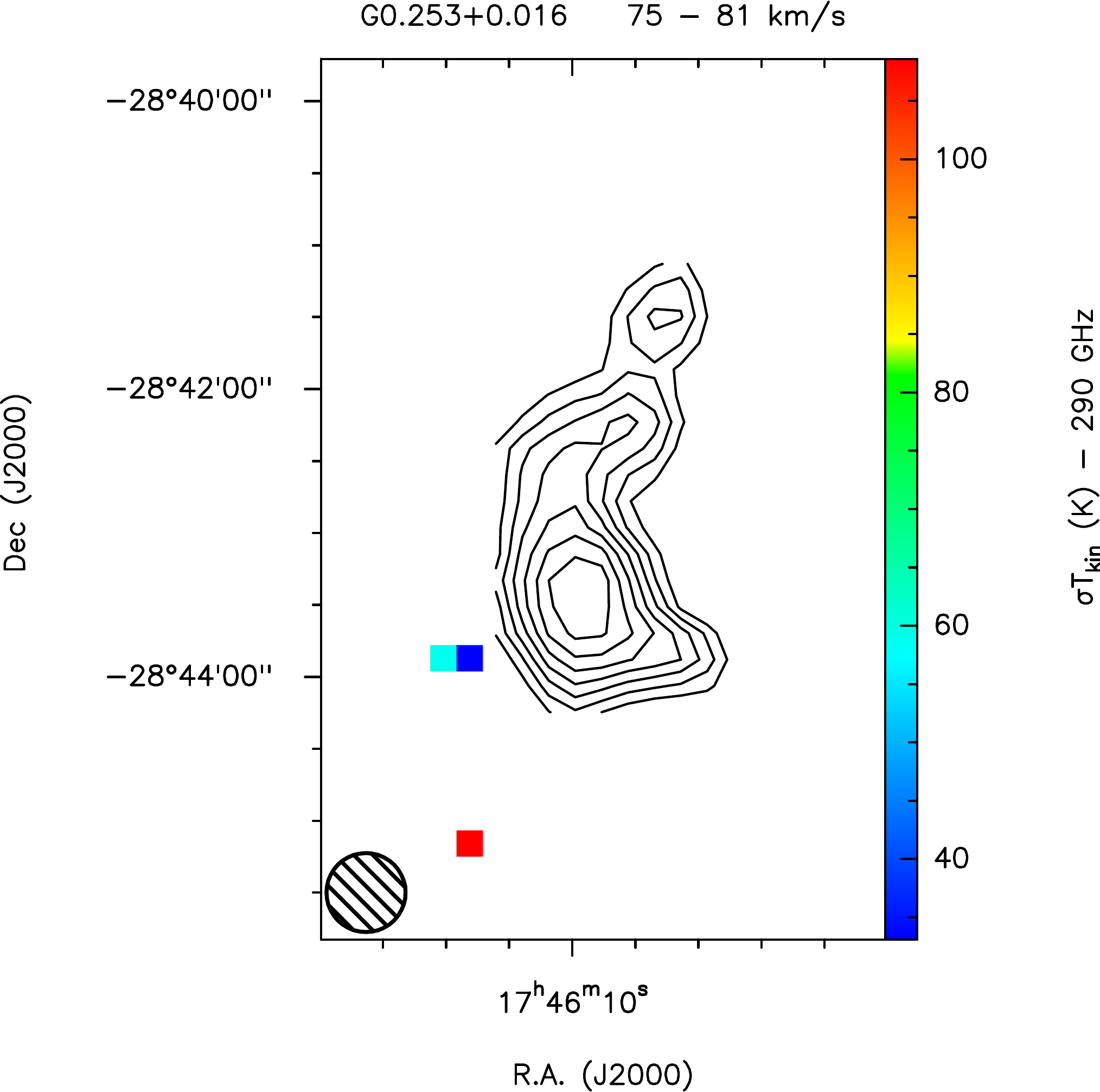}}\\ 
	\label{G0253-All-Temp-H2CO}
\end{figure*}

\begin{figure*}
	\caption{As Fig. \ref{20kms-All-Temp-H2CO}, for G0.411+0.050.}
	\centering
        218 GHz temperatures\\
	\subfloat{\includegraphics[bb = 0 60 600 580, clip, height=5cm]{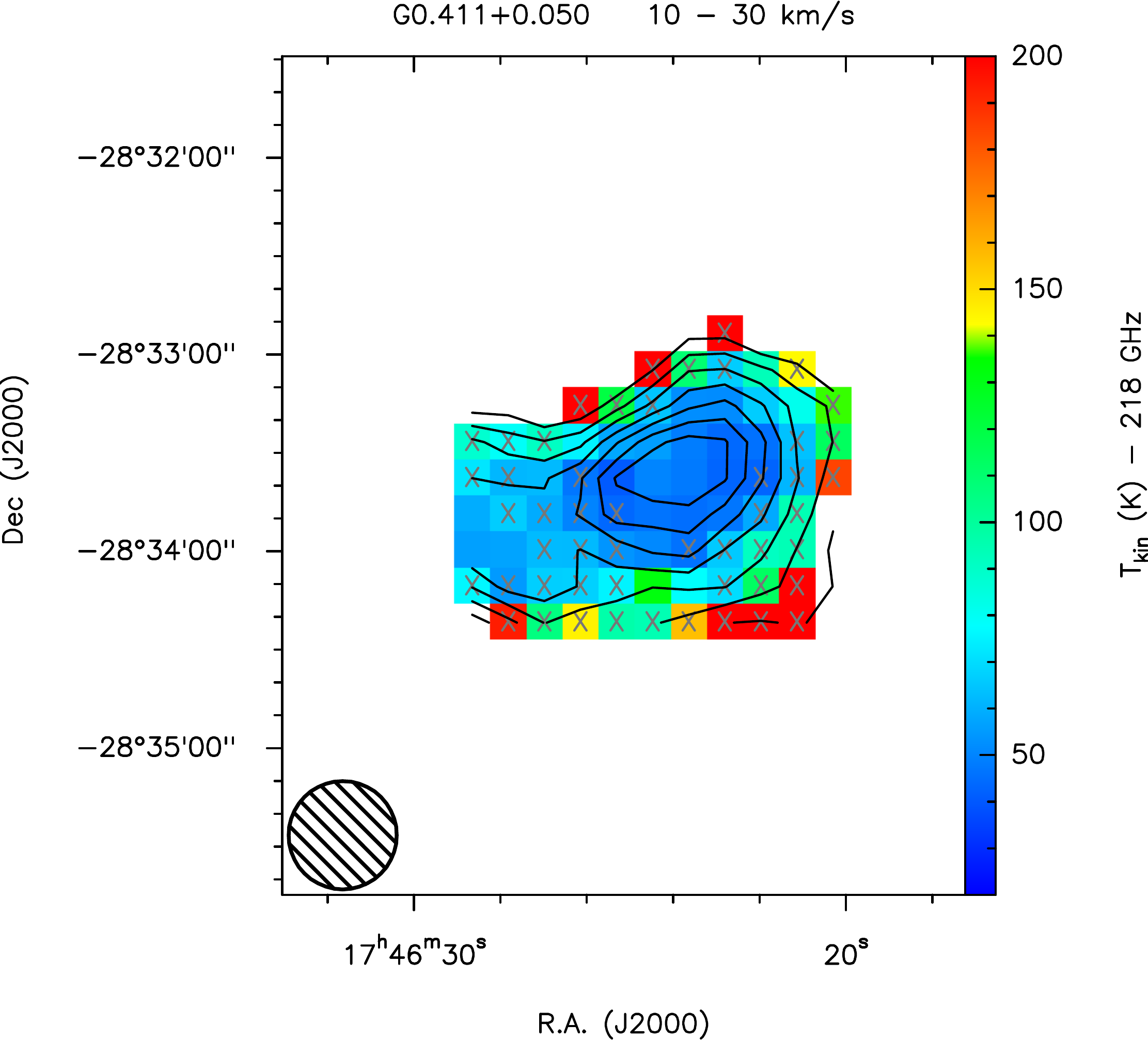}}
	\subfloat{\includegraphics[bb = 150 60 640 580, clip, height=5cm]{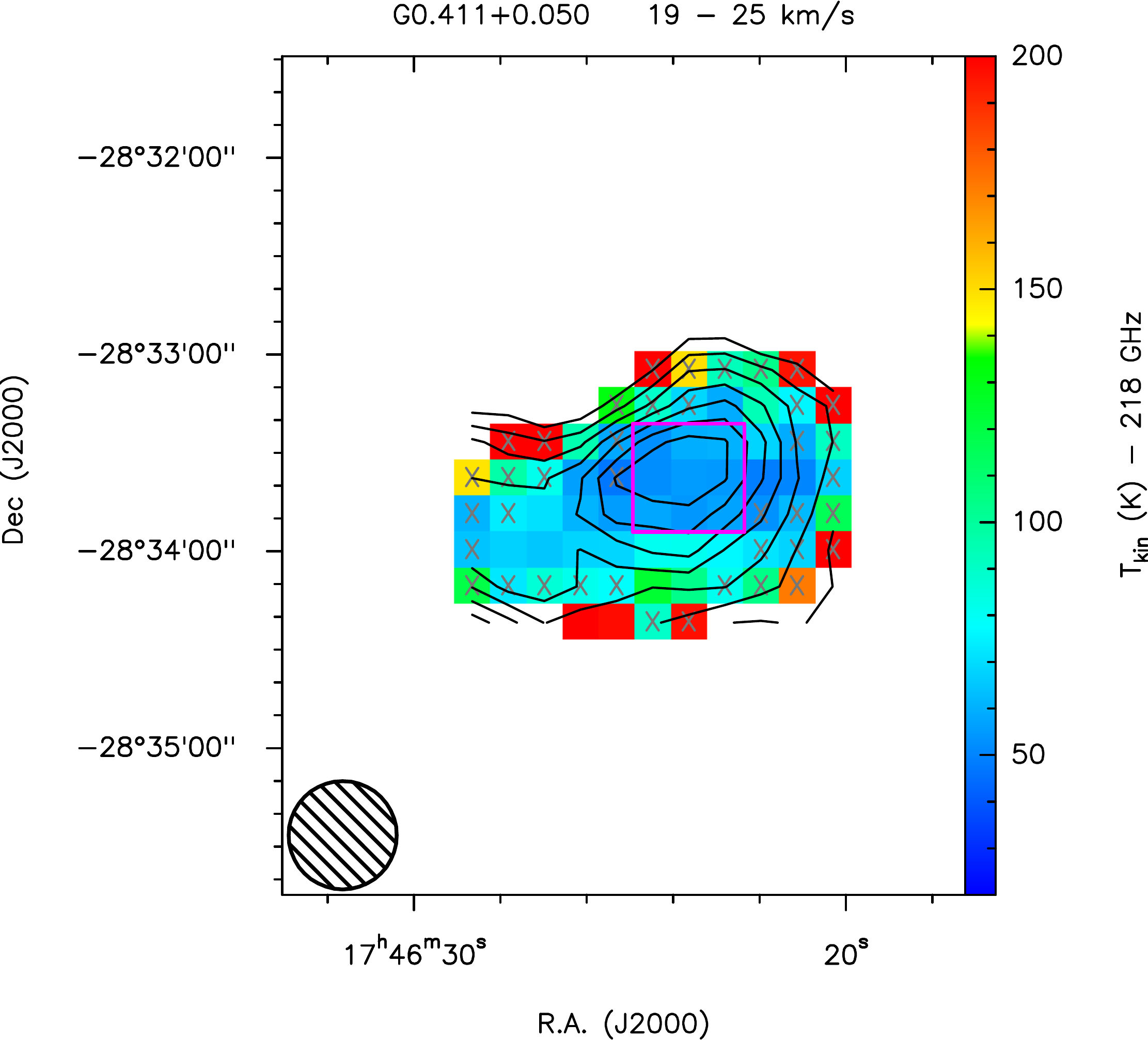}}\\
	\vspace{-0.5cm}
	\subfloat{\includegraphics[bb = 0 0 600 560, clip, height=5.39cm]{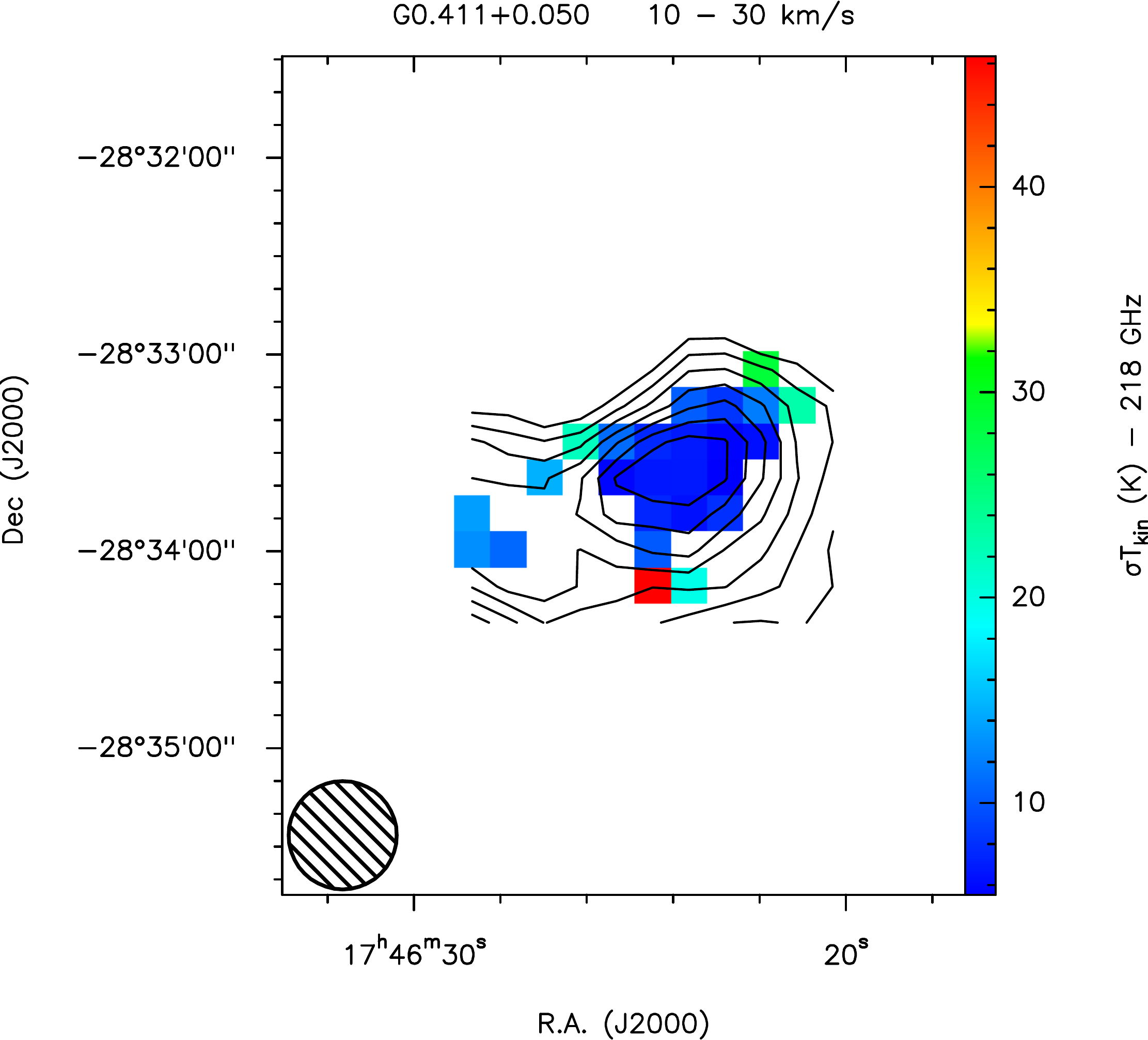}}
	\subfloat{\includegraphics[bb = 150 0 640 560, clip, height=5.39cm]{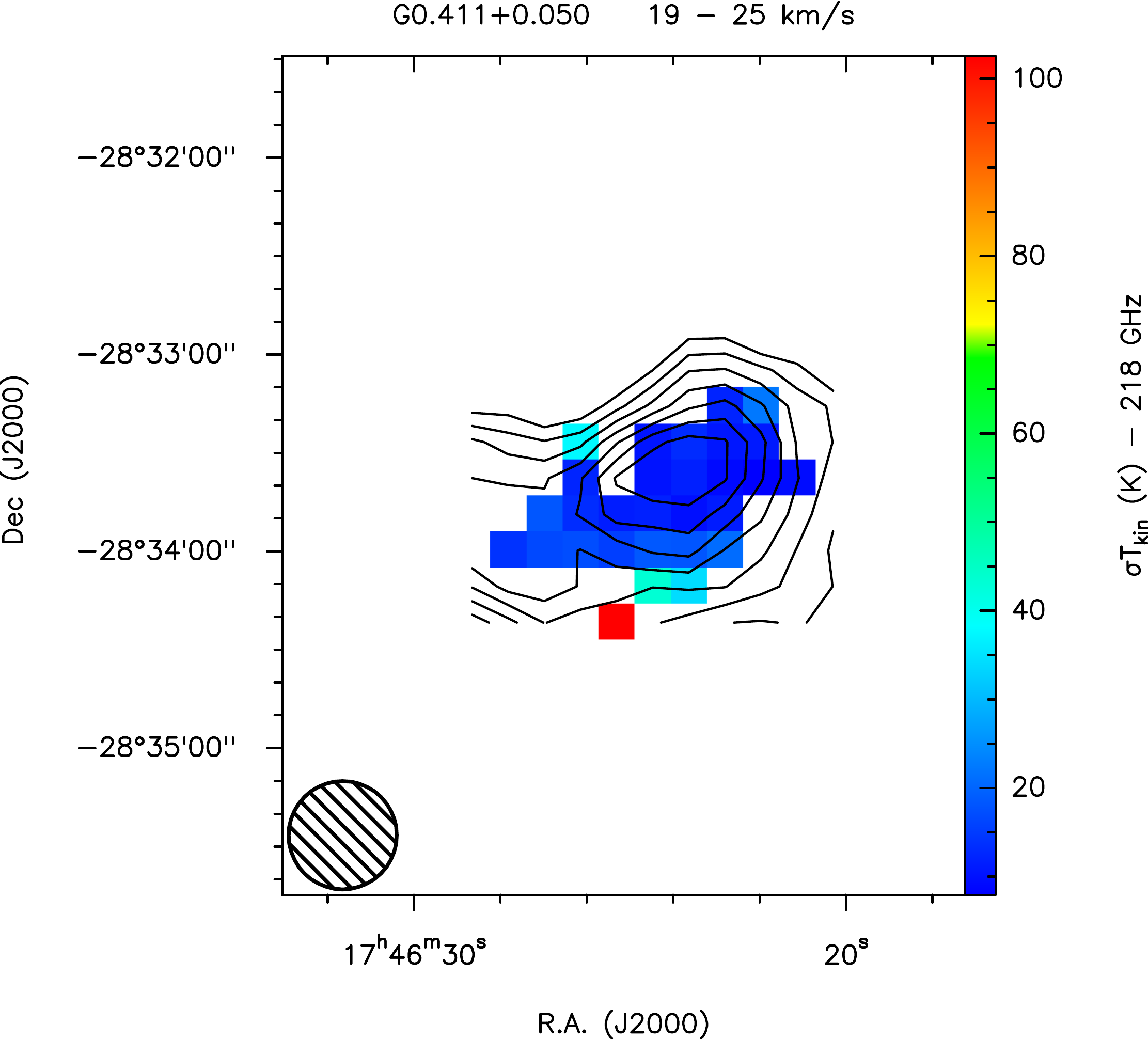}}\\ 
	\vspace{0.1cm}
	291 GHz temperatures \\
	\subfloat{\includegraphics[bb = 0 60 600 580, clip, height=5cm]{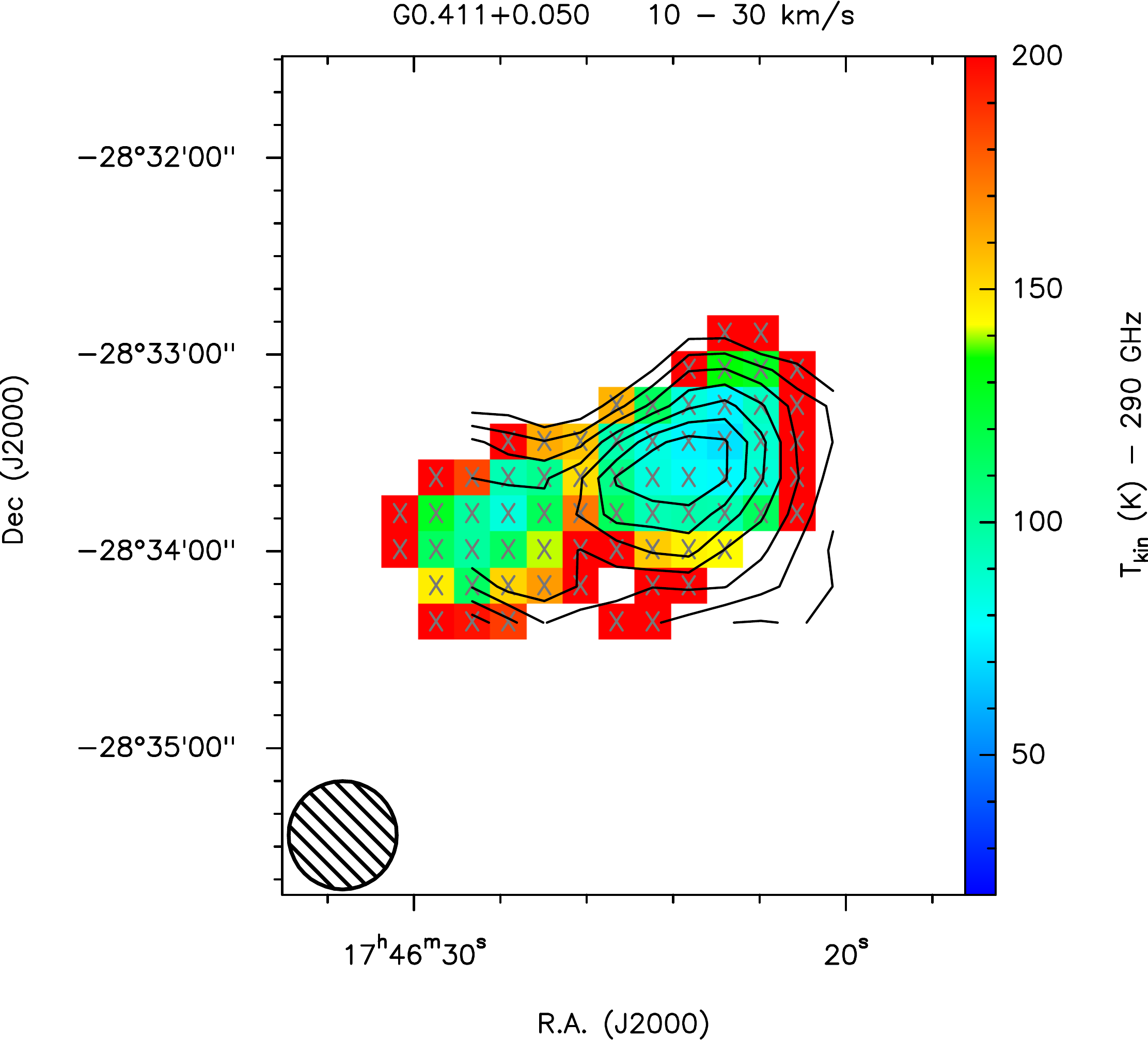}}
	\subfloat{\includegraphics[bb = 150 60 640 580, clip, height=5cm]{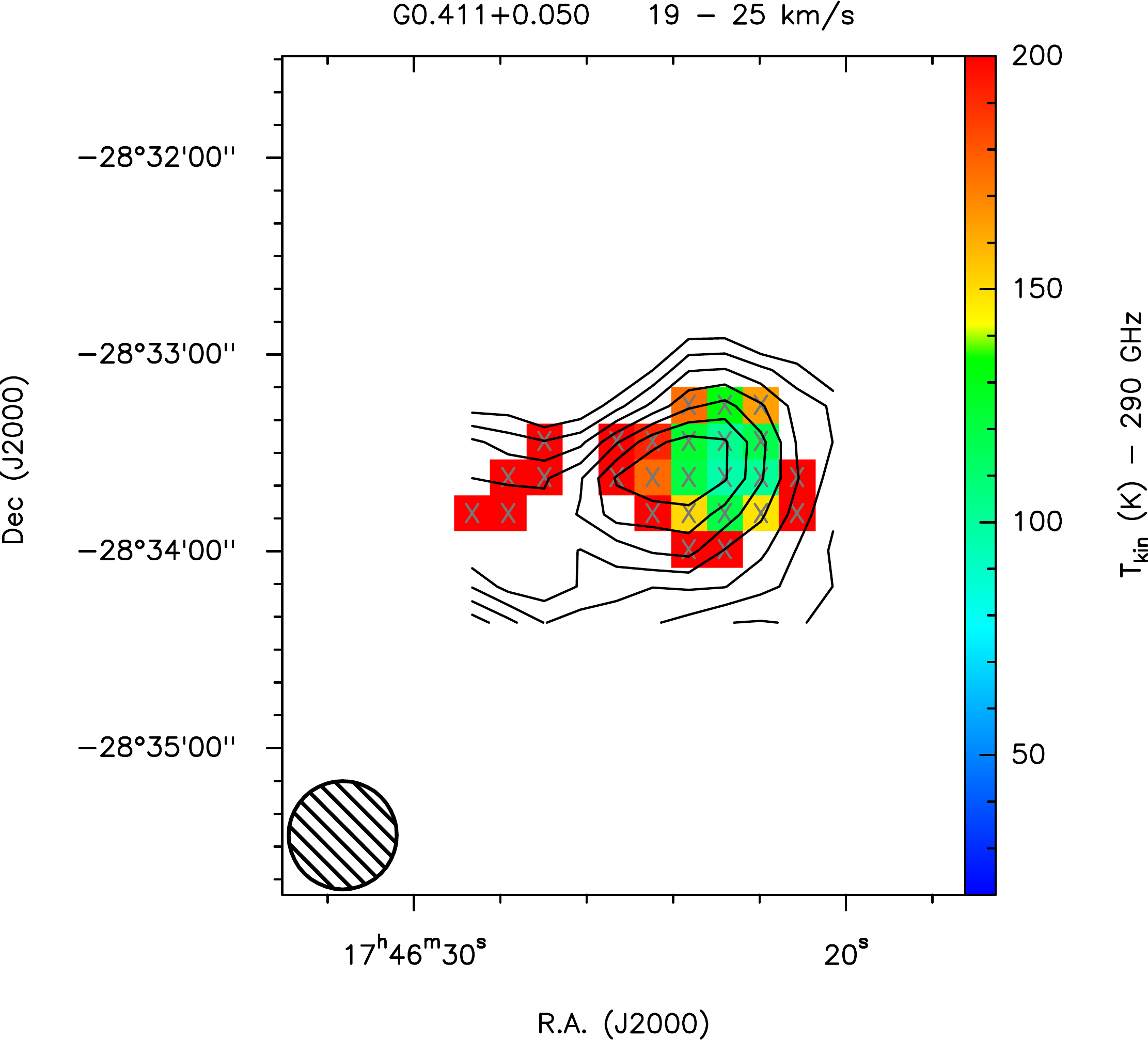}}\\
	\vspace{-0.5cm}
	\subfloat{\includegraphics[bb = 0 0 600 560, clip, height=5.39cm]{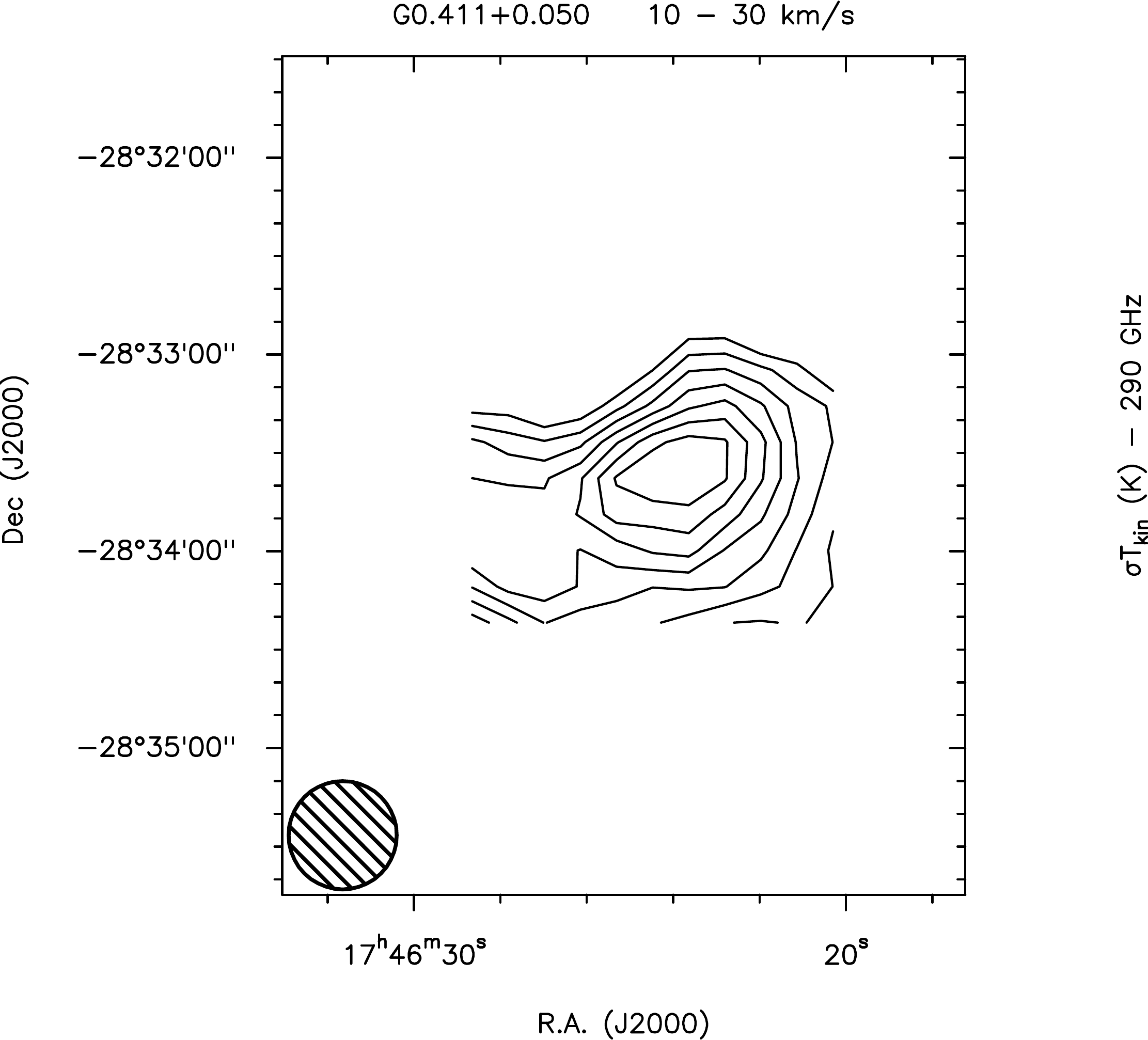}}
	\subfloat{\includegraphics[bb = 150 0 640 560, clip, height=5.39cm]{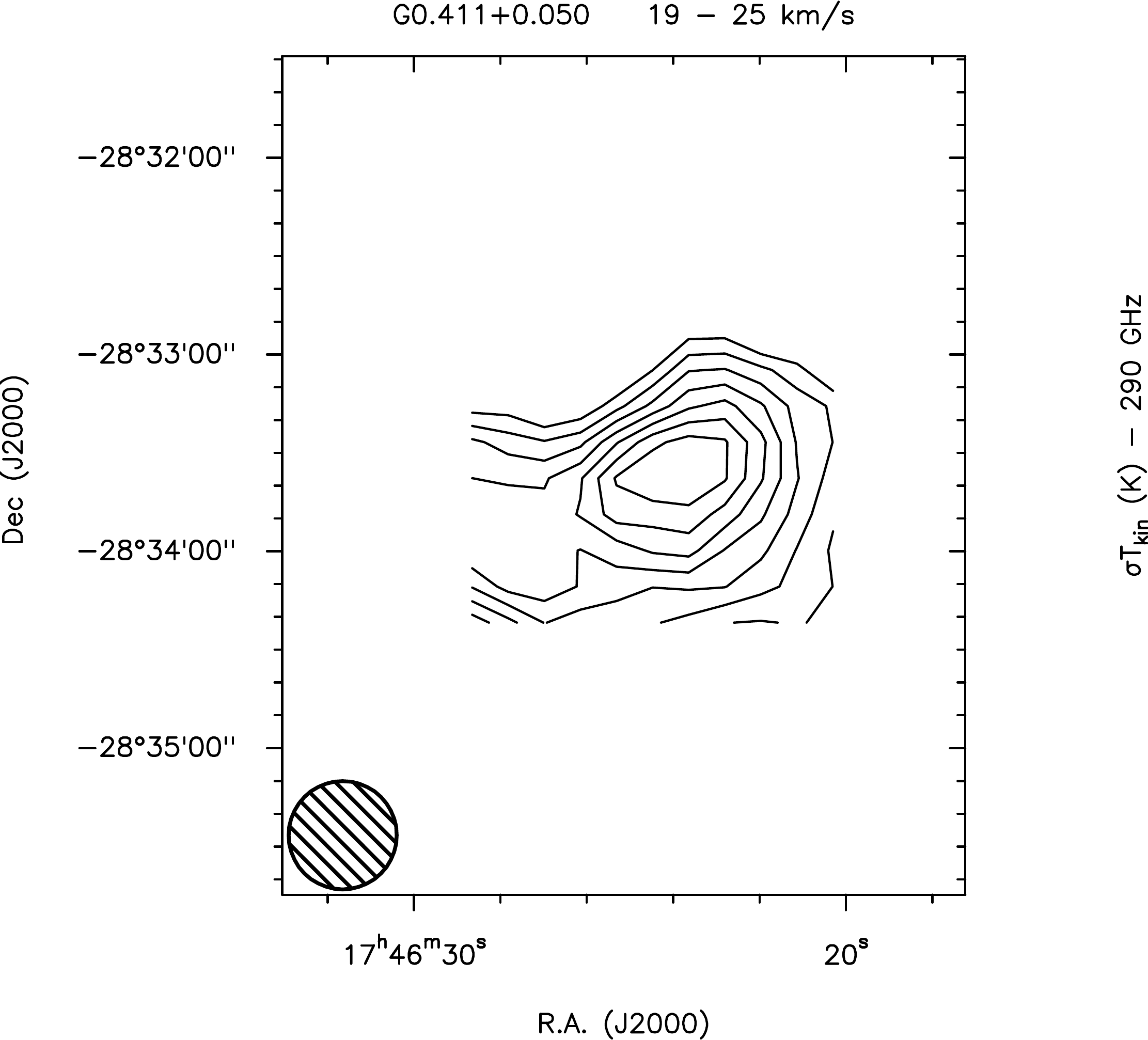}}\\ 
	\label{G0411-All-Temp-H2CO}
\end{figure*}

\begin{figure*}
	\caption{As Fig. \ref{20kms-All-Temp-H2CO}, for G0.480$-$0.006.}
	\centering
        218 GHz temperatures\\
	\subfloat{\includegraphics[bb = 0 60 600 580, clip, height=5cm]{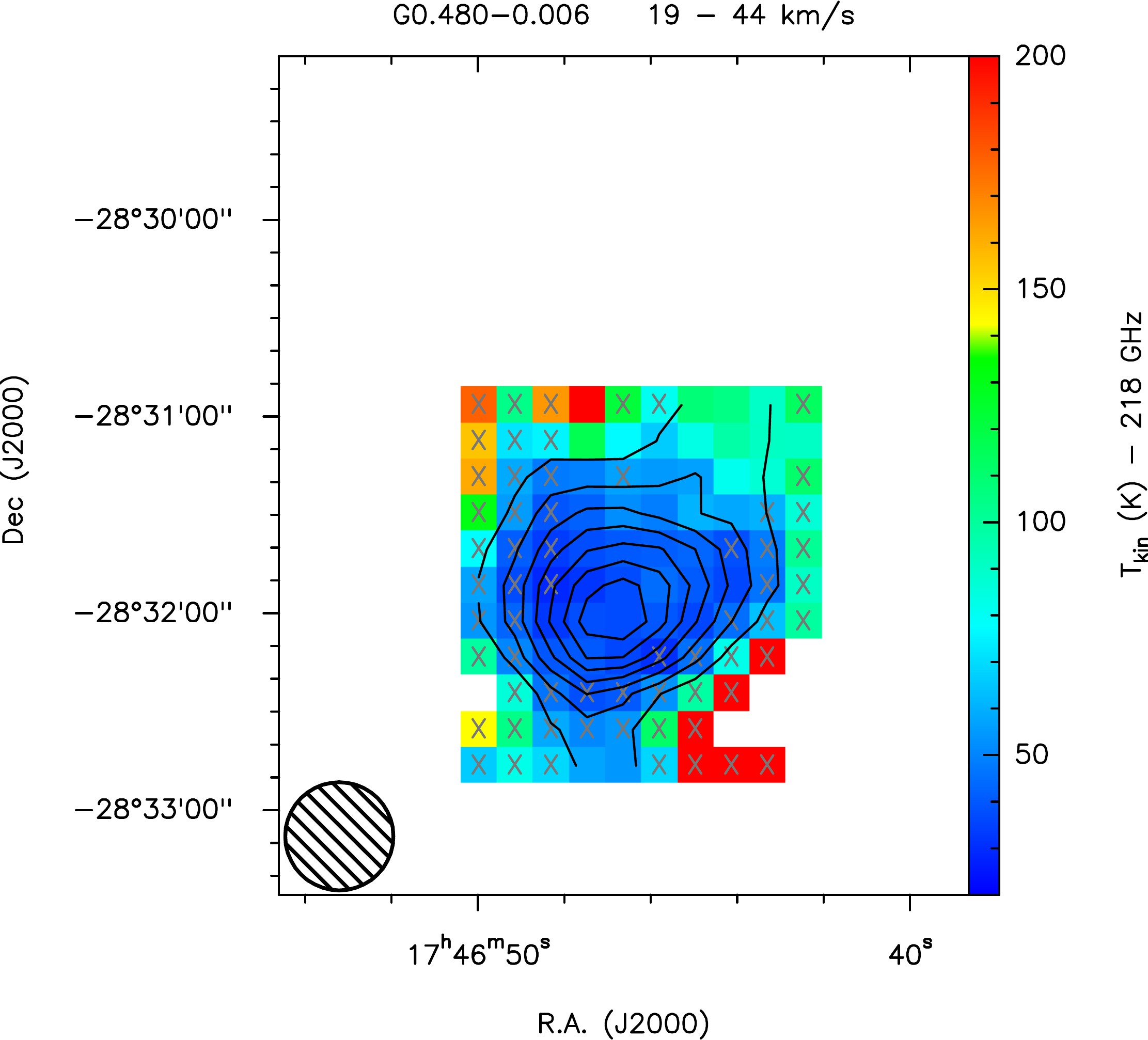}}
	\subfloat{\includegraphics[bb = 150 60 640 580, clip, height=5cm]{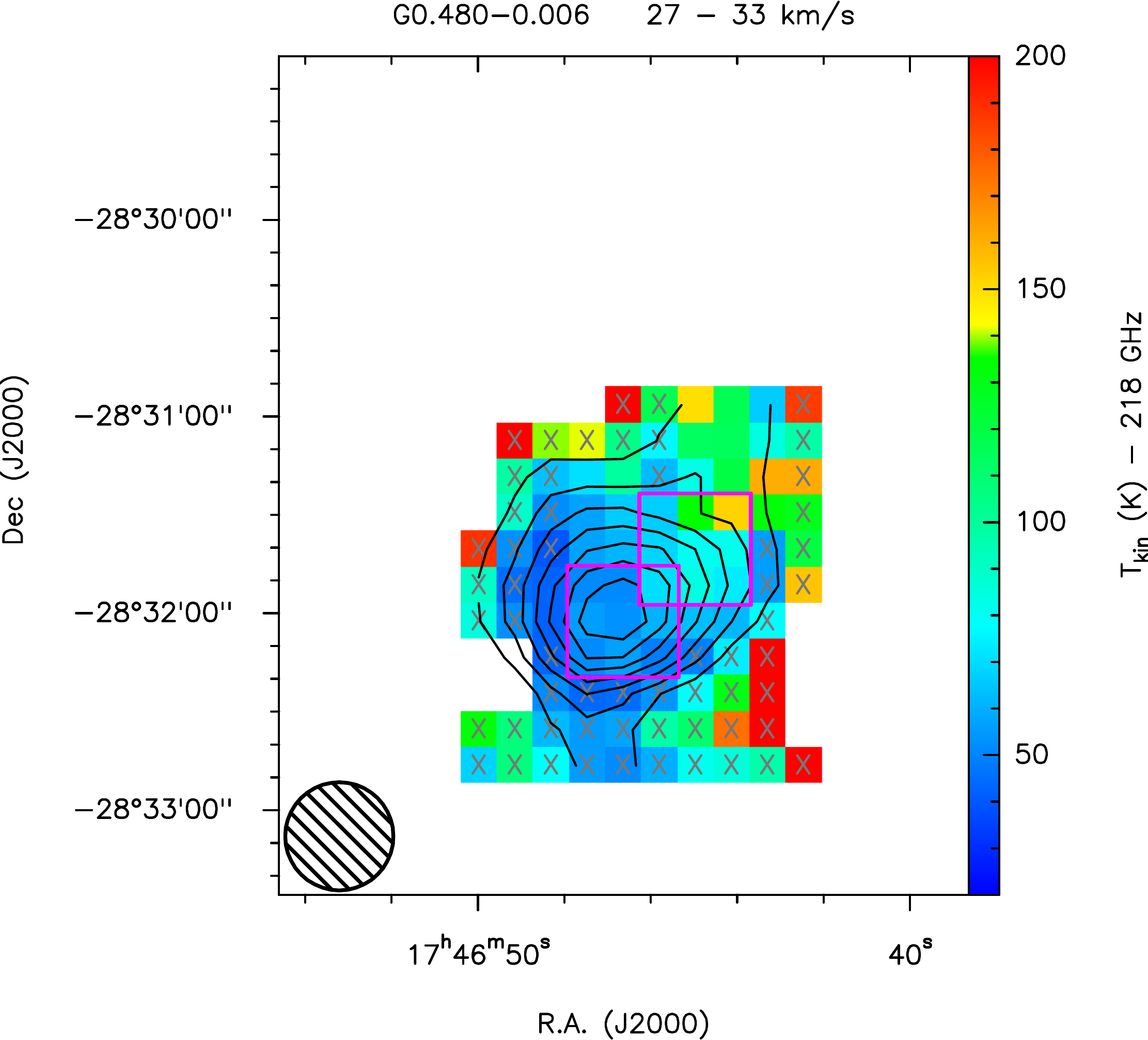}}\\
	\vspace{-0.5cm}
	\subfloat{\includegraphics[bb = 0 0 600 560, clip, height=5.39cm]{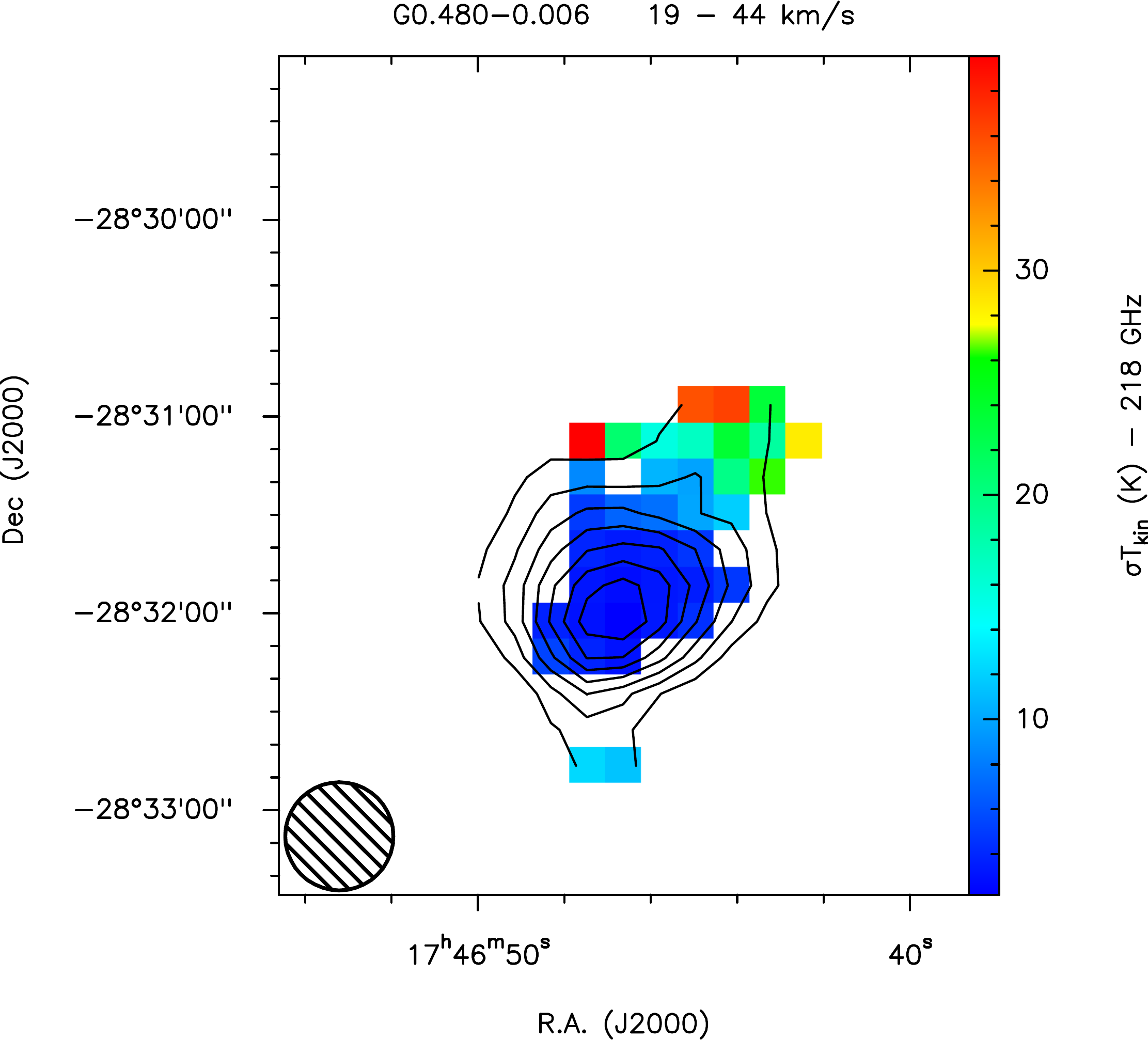}}
	\subfloat{\includegraphics[bb = 150 0 640 560, clip, height=5.39cm]{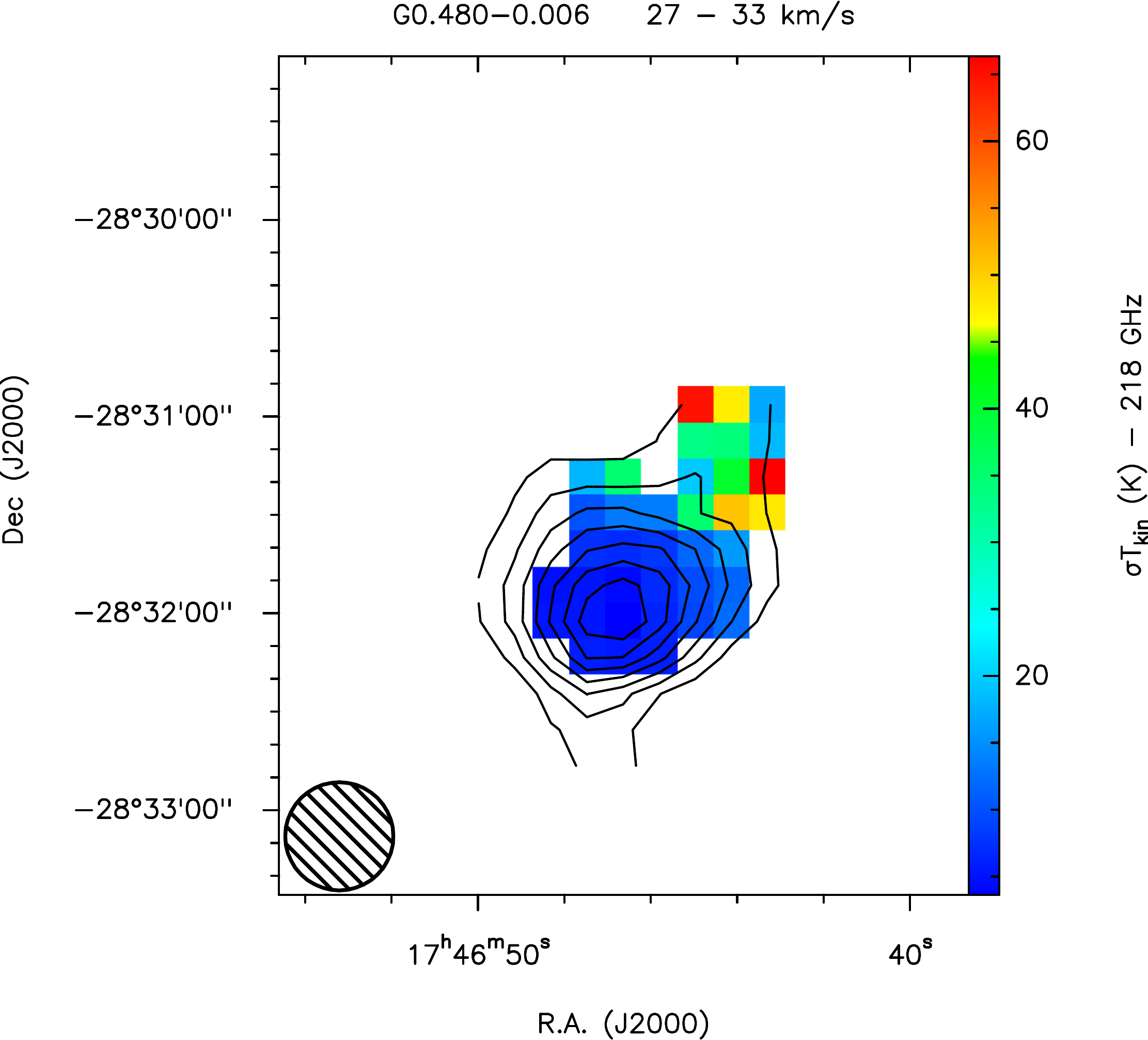}}\\ 
	\vspace{0.1cm}
	291 GHz temperatures \\
	\subfloat{\includegraphics[bb = 0 60 600 580, clip, height=5cm]{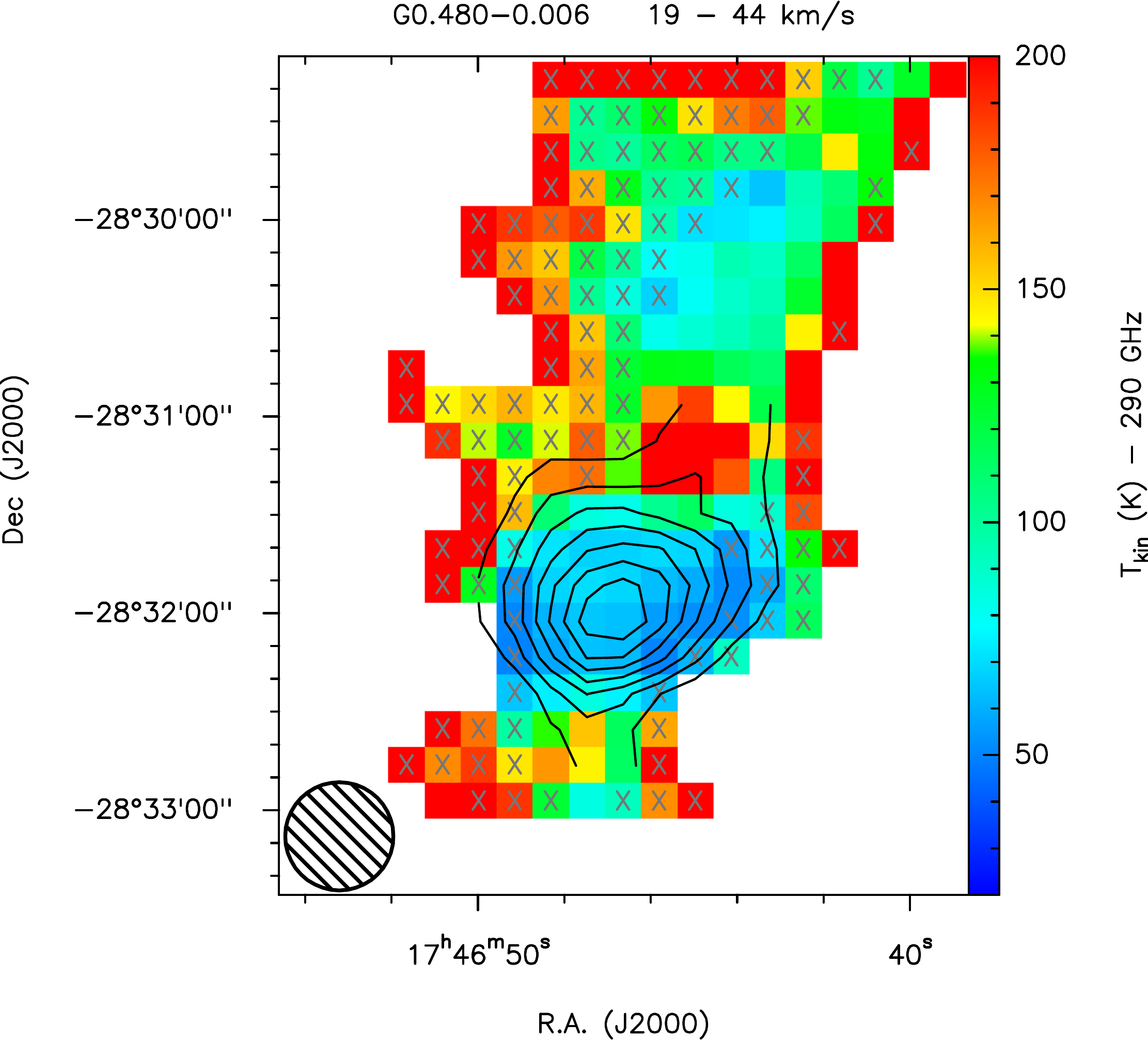}}
	\subfloat{\includegraphics[bb = 150 60 640 580, clip, height=5cm]{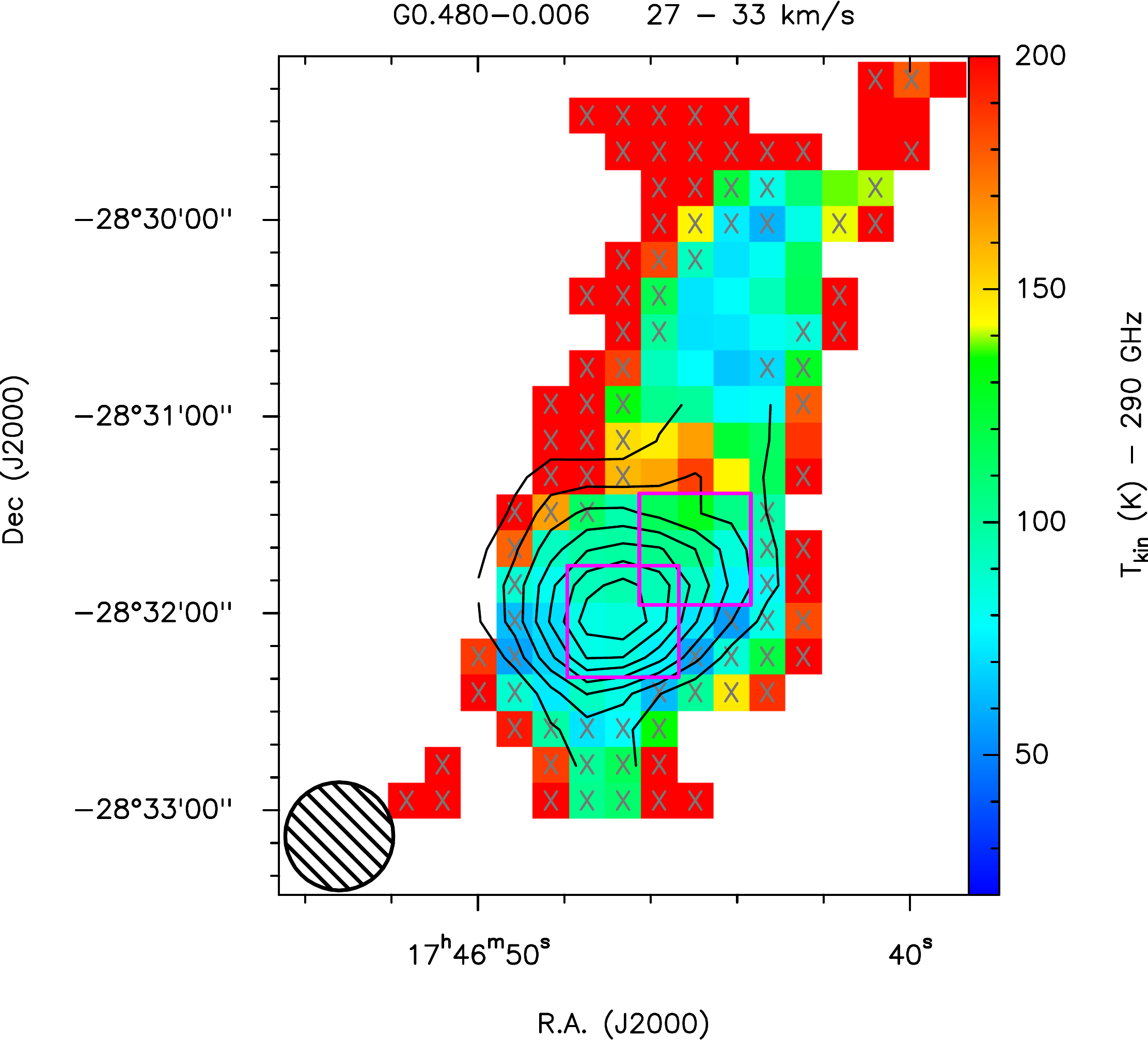}}\\
	\vspace{-0.5cm}
	\subfloat{\includegraphics[bb = 0 0 600 560, clip, height=5.39cm]{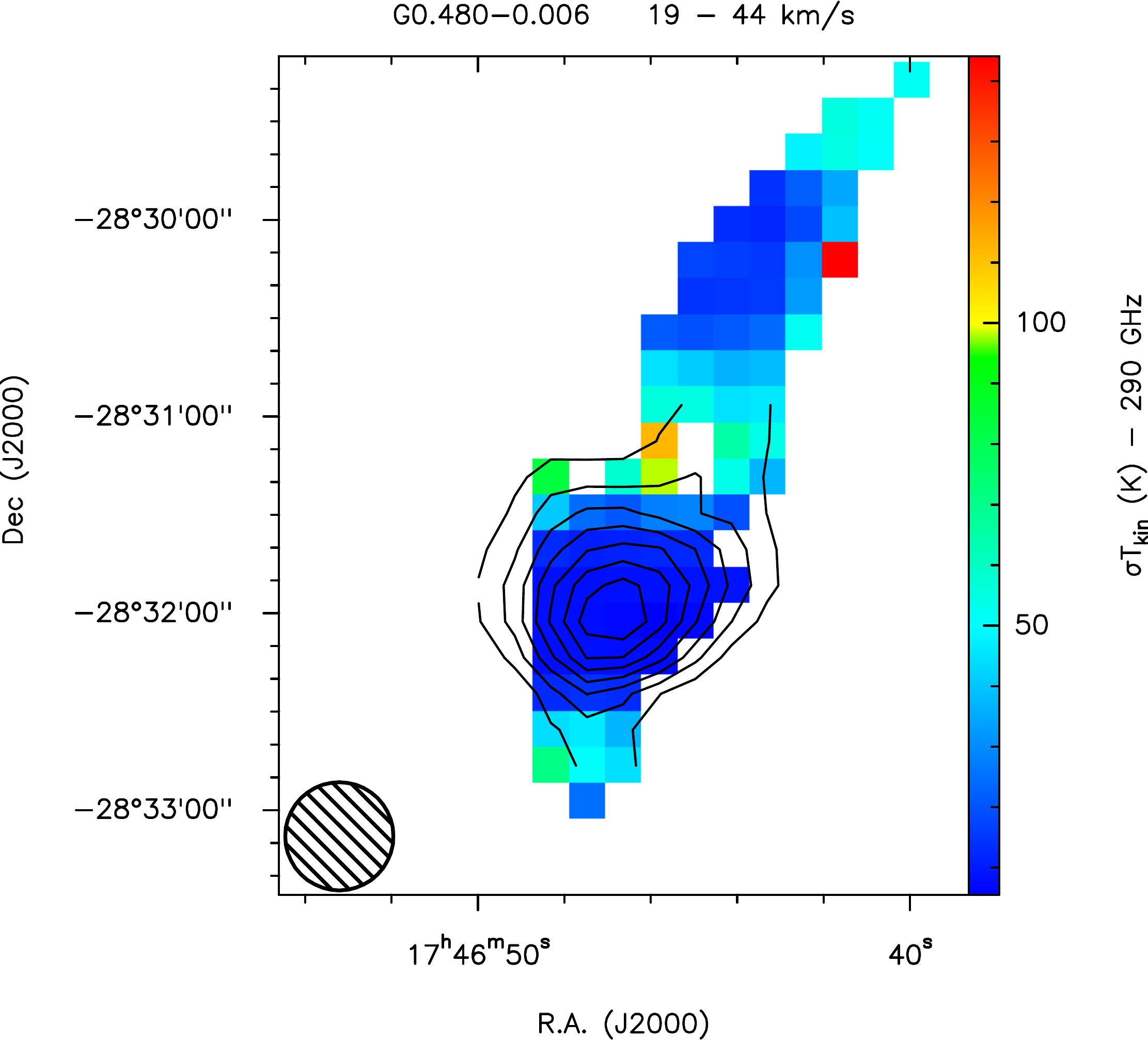}}
	\subfloat{\includegraphics[bb = 150 0 640 560, clip, height=5.39cm]{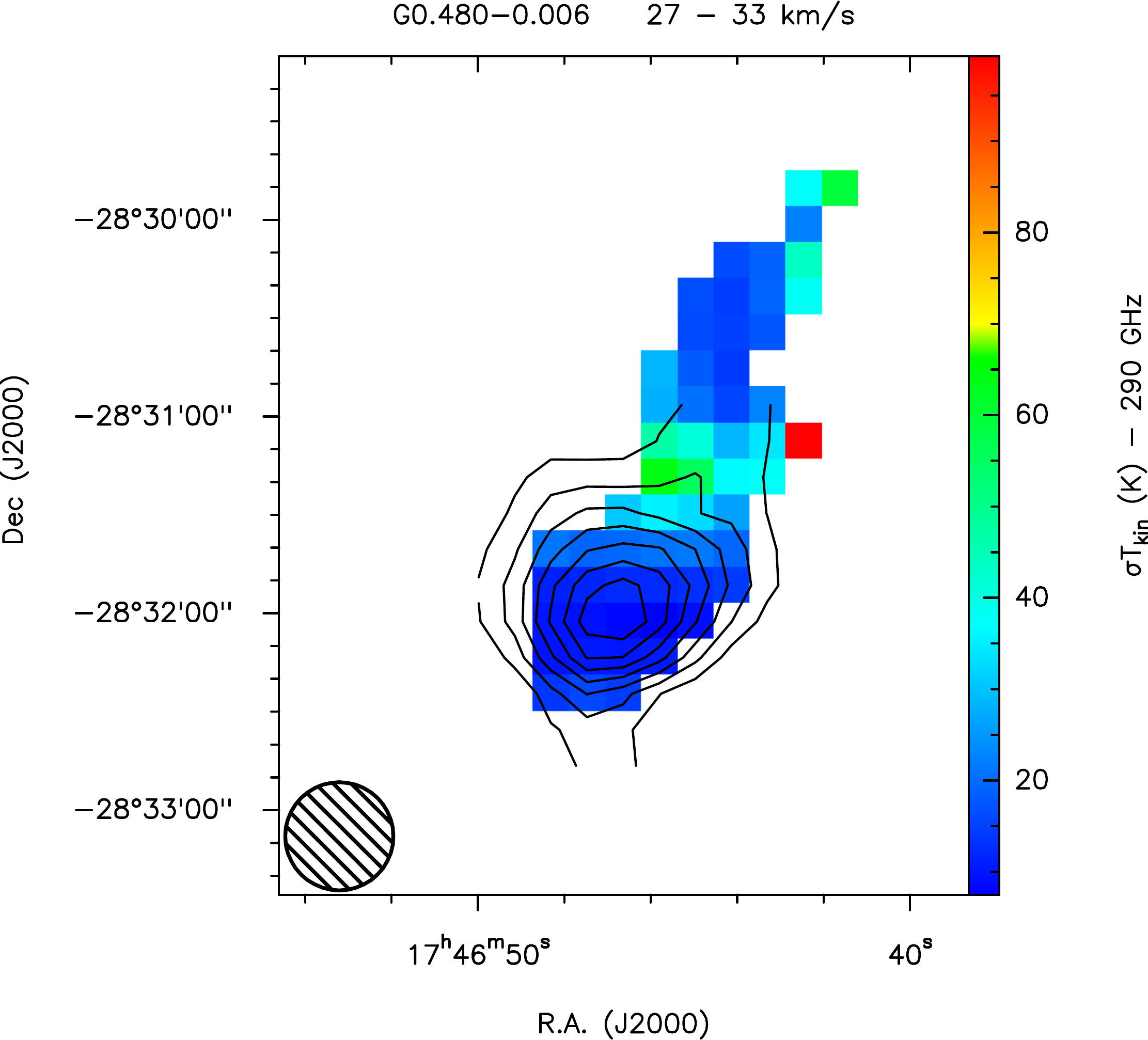}}\\ 
	\label{G0480-All-Temp-H2CO}
\end{figure*}

\clearpage

\begin{figure*}
	\caption{As Fig. \ref{20kms-All-Temp-H2CO}, for Sgr C.}
	\centering
	291 GHz temperatures \\
	\subfloat{\includegraphics[bb = 0 60 730 580, clip, height=5cm]{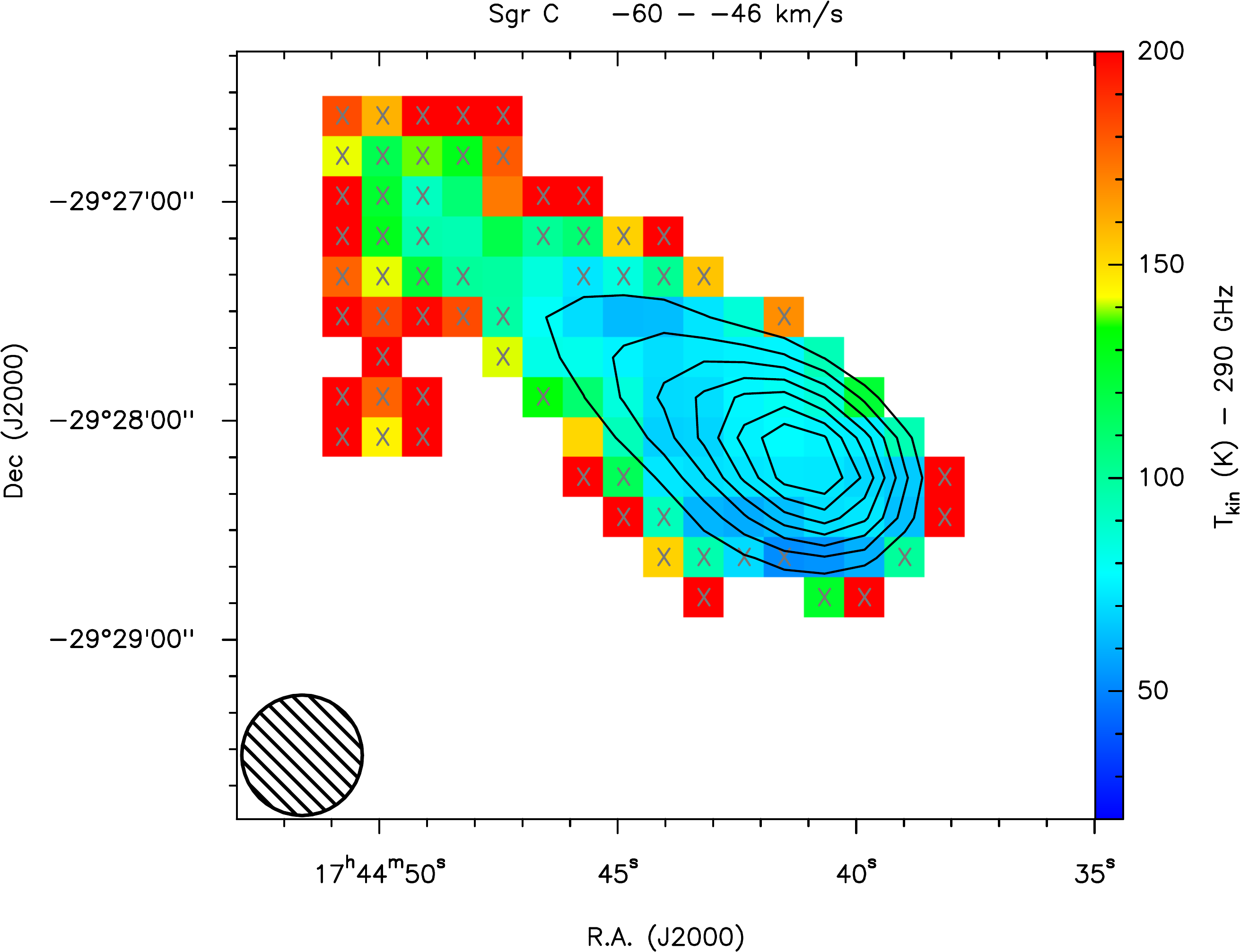}}
	\subfloat{\includegraphics[bb = 130 60 760 580, clip, height=5cm]{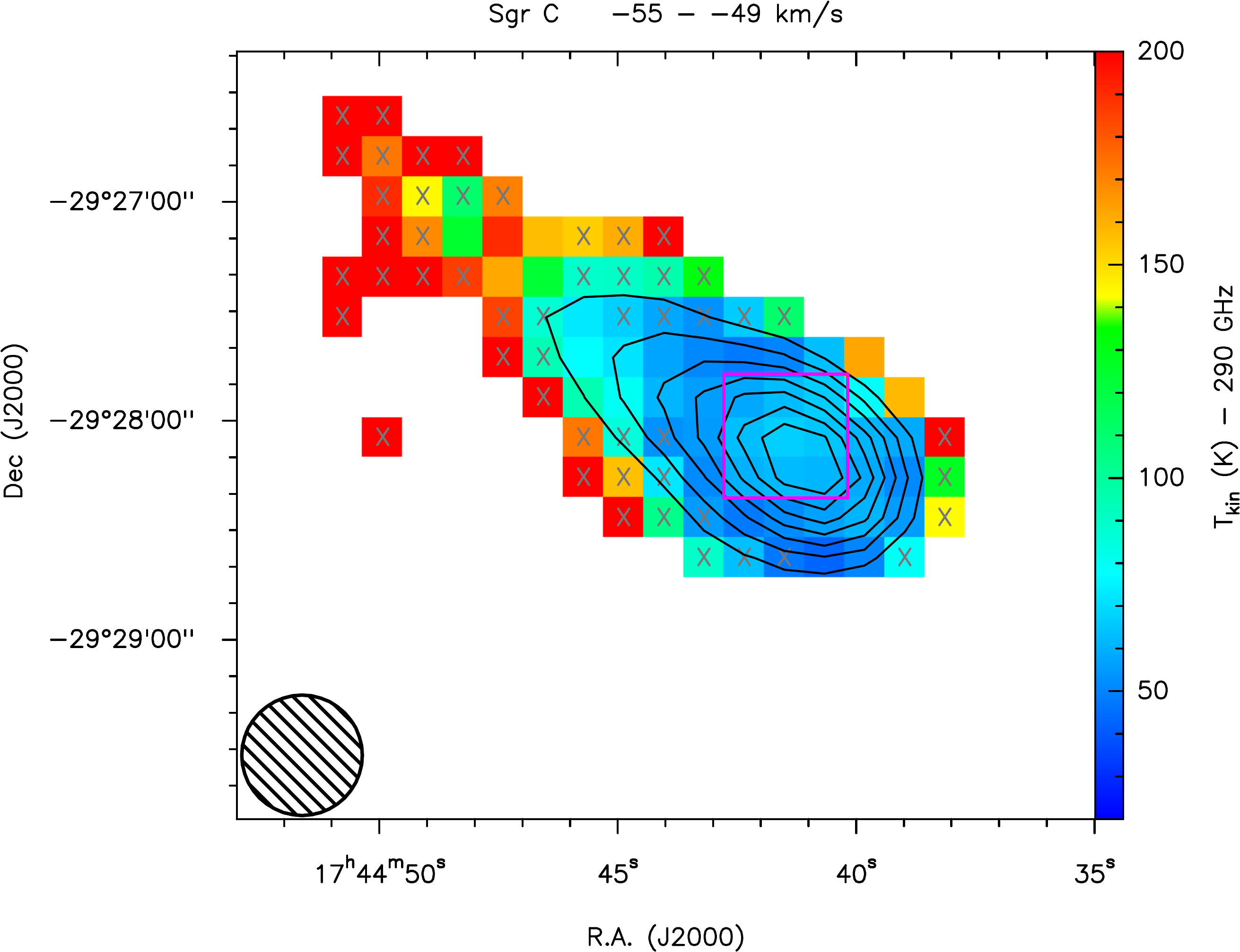}}\\
	\vspace{-0.5cm}
	\subfloat{\includegraphics[bb = 0 0 730 560, clip, height=5.39cm]{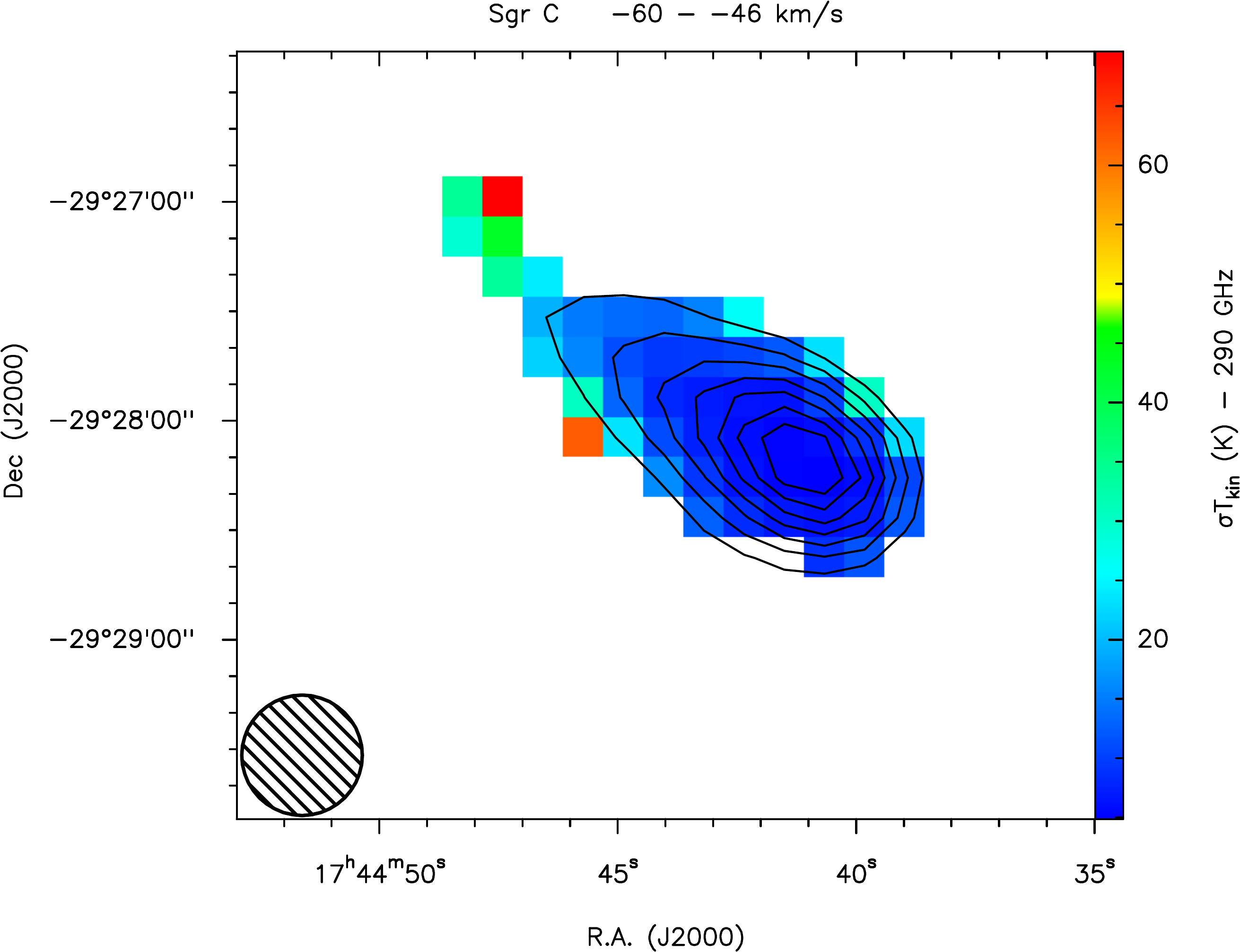}}
	\subfloat{\includegraphics[bb = 130 0 760 560, clip, height=5.39cm]{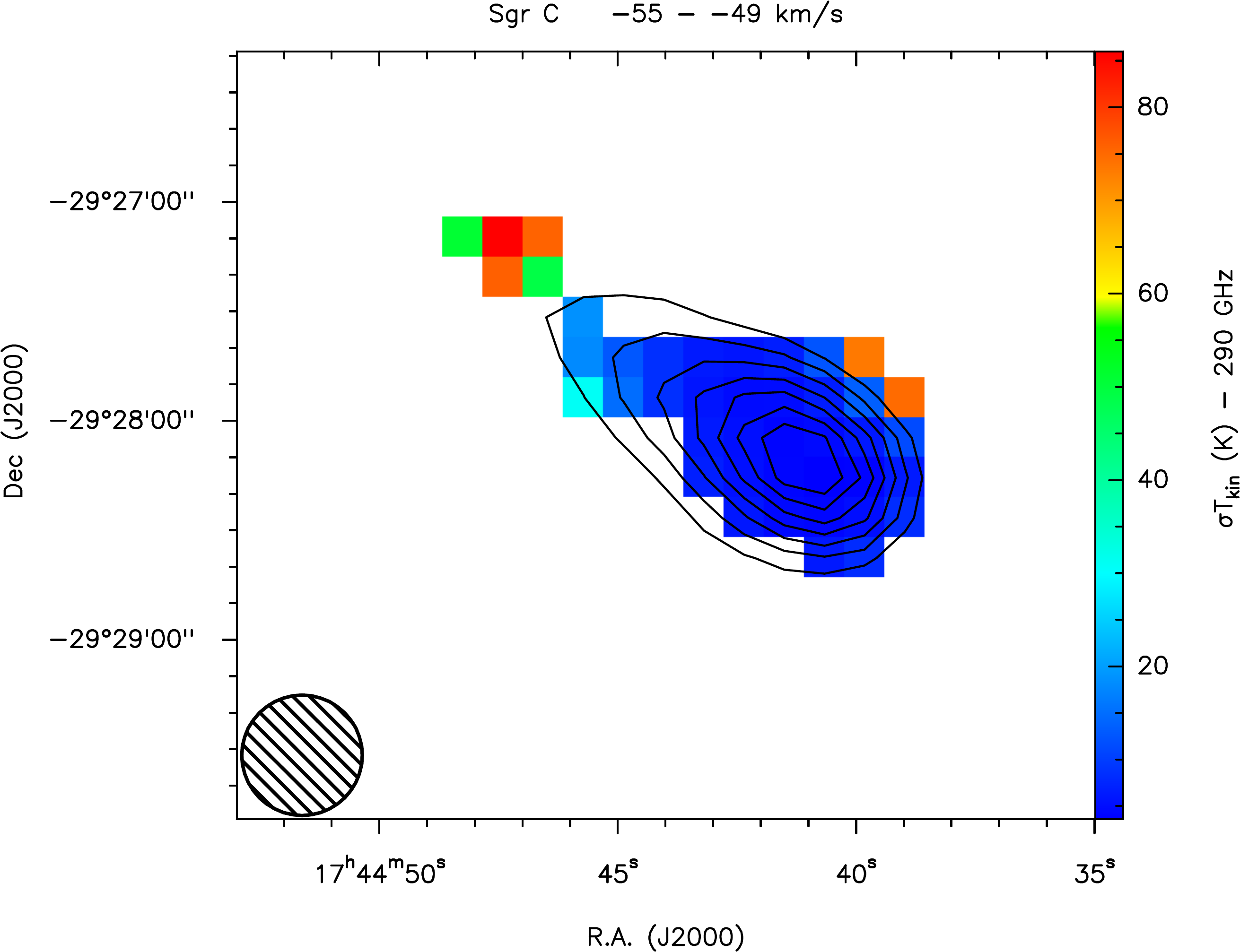}}\\ 
	\label{SGRC-All-Temp-H2CO}
\end{figure*}

\begin{figure*}
	\caption{As Fig. \ref{20kms-All-Temp-H2CO}, for Sgr D.}
	\centering
	291 GHz temperatures \\
	\subfloat{\includegraphics[bb = 0 60 710 580, clip, height=5cm]{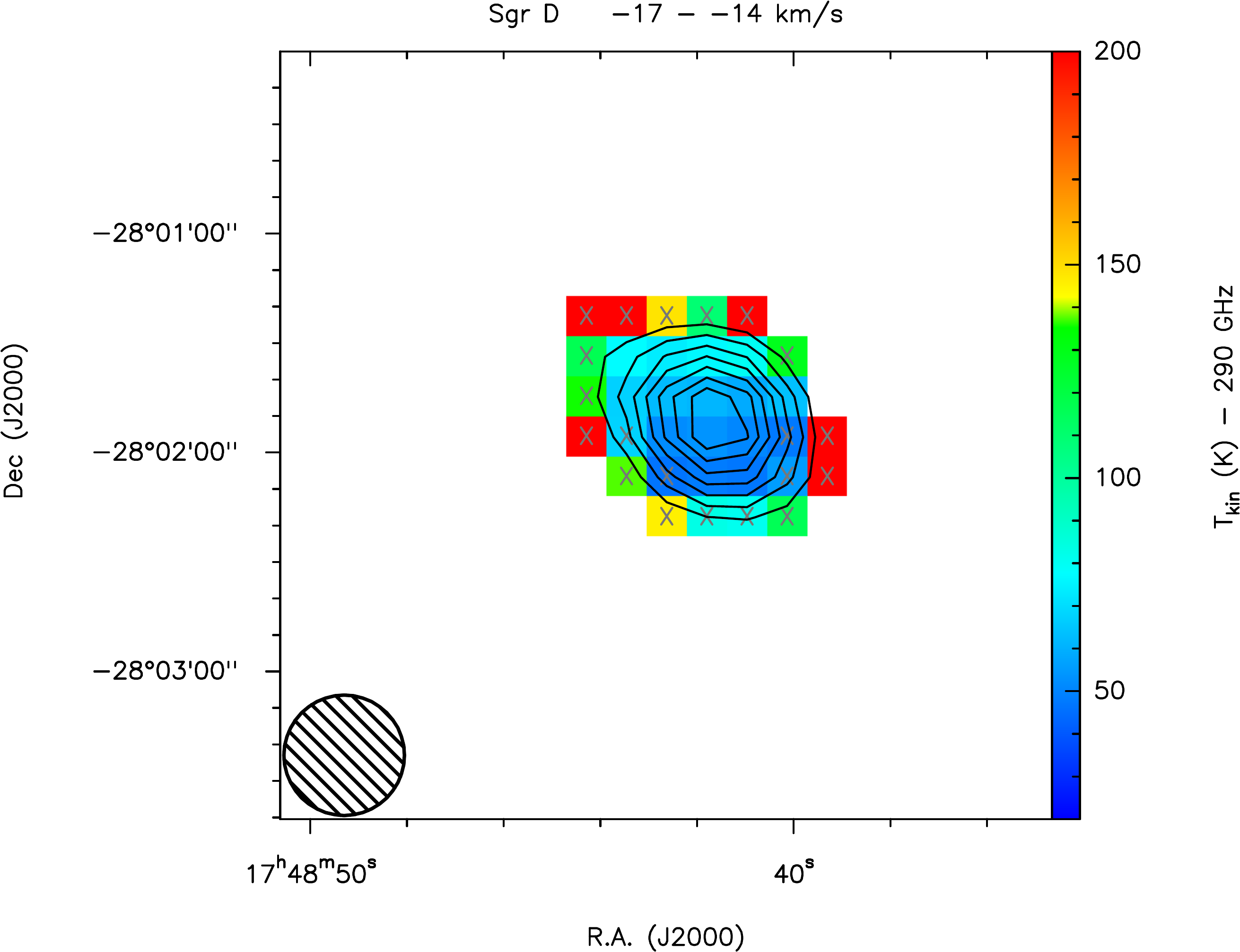}}
	\subfloat{\includegraphics[bb = 150 60 760 580, clip, height=5cm]{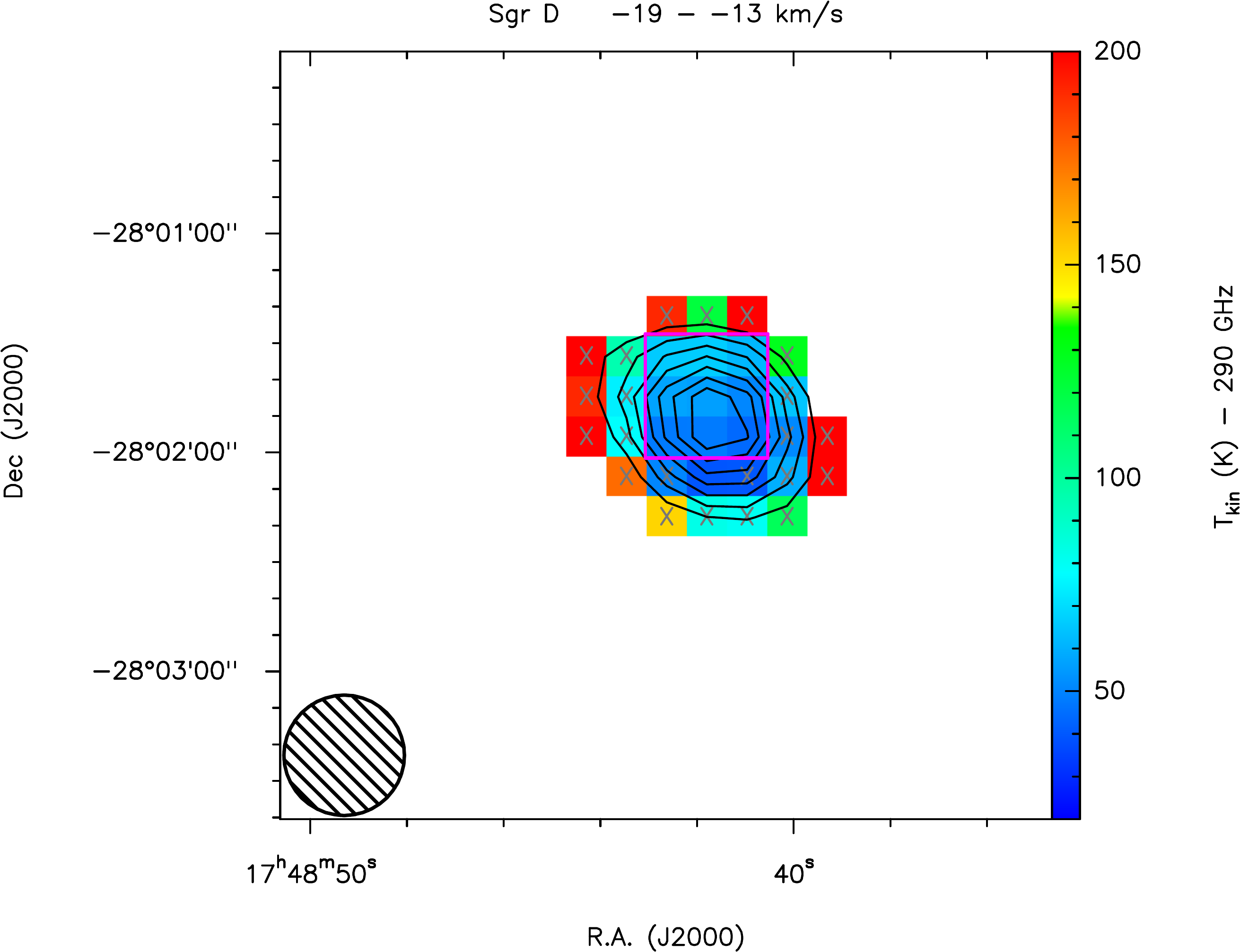}}\\
	\vspace{-0.5cm}
	\subfloat{\includegraphics[bb = 0 0 710 560, clip, height=5.39cm]{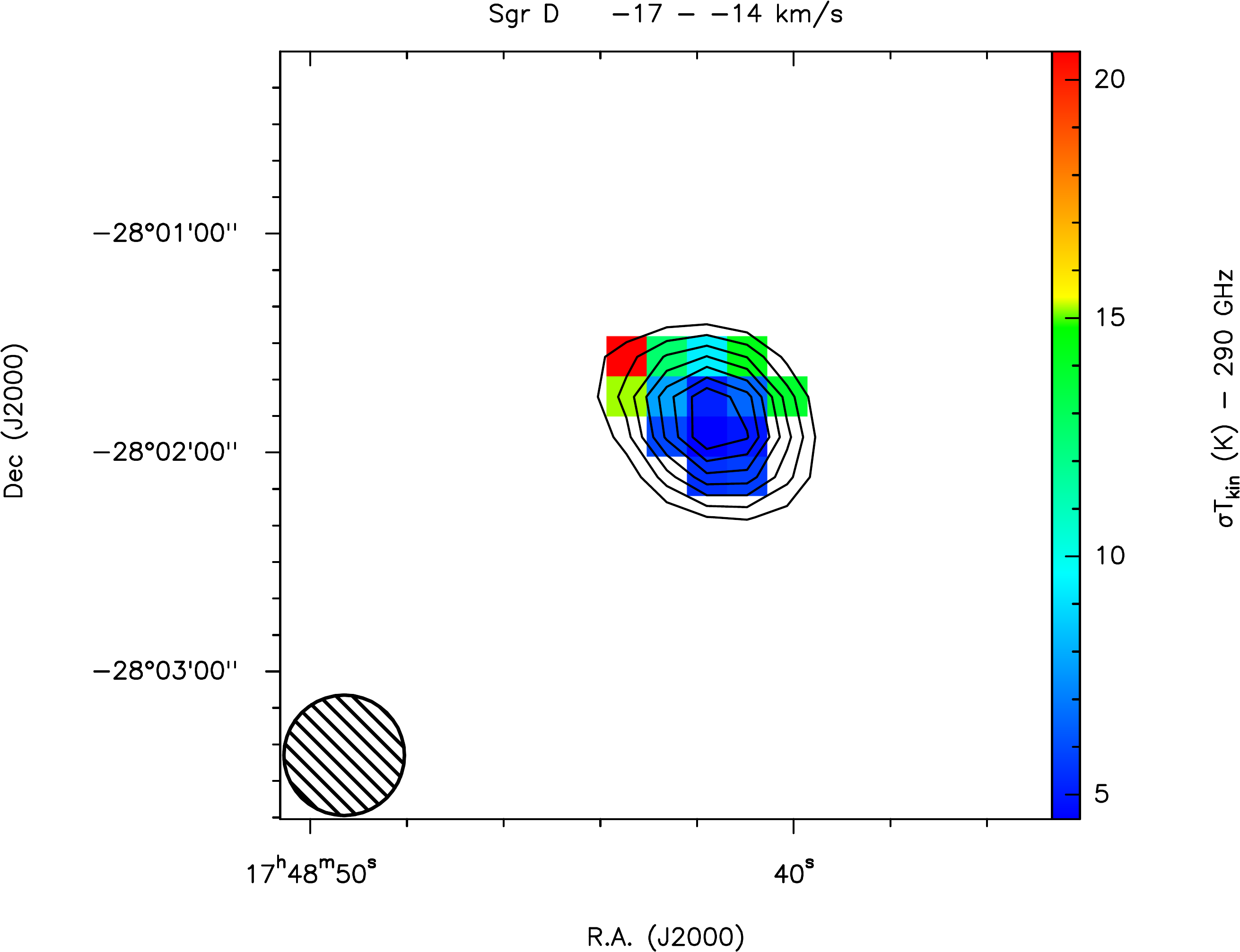}}
	\subfloat{\includegraphics[bb = 150 0 760 560, clip, height=5.39cm]{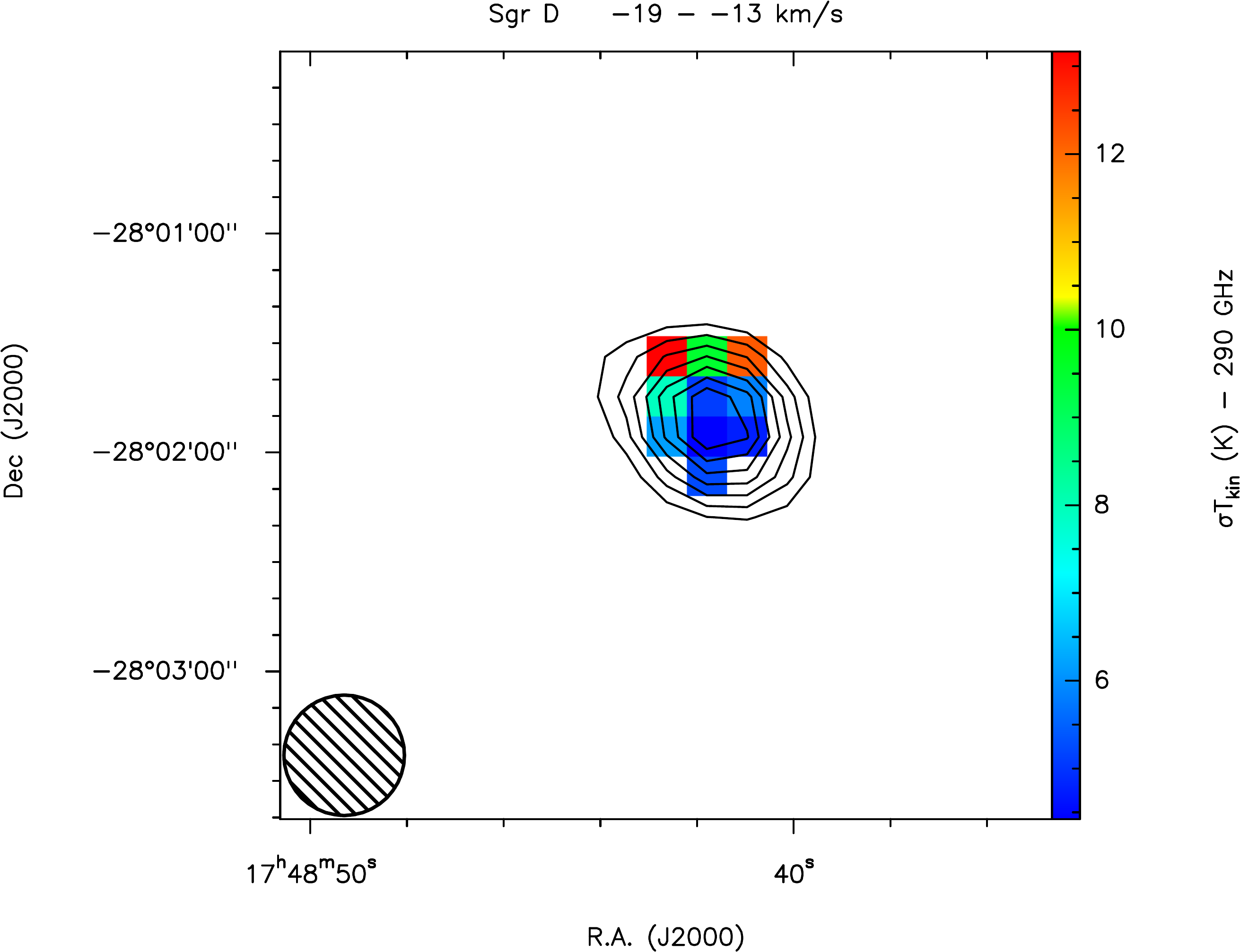}}\\ 
	\label{SGRD-All-Temp-H2CO}
\end{figure*}

\begin{figure*}
	\caption{The plots show the minimal, weighted average, and maximal temperatures of the temperature 
	maps of our seven sources. The vertical size of the boxes is given by the weighted average of the
	temperature $\pm$ 1$\sigma$ uncertainty. The values are taken from Table \ref{SourceTemp}. 
	Blue and yellow boxes show the results at 218 and 291 GHz, respectively.}
	\centering
	\subfloat{\includegraphics[height=5.5cm]{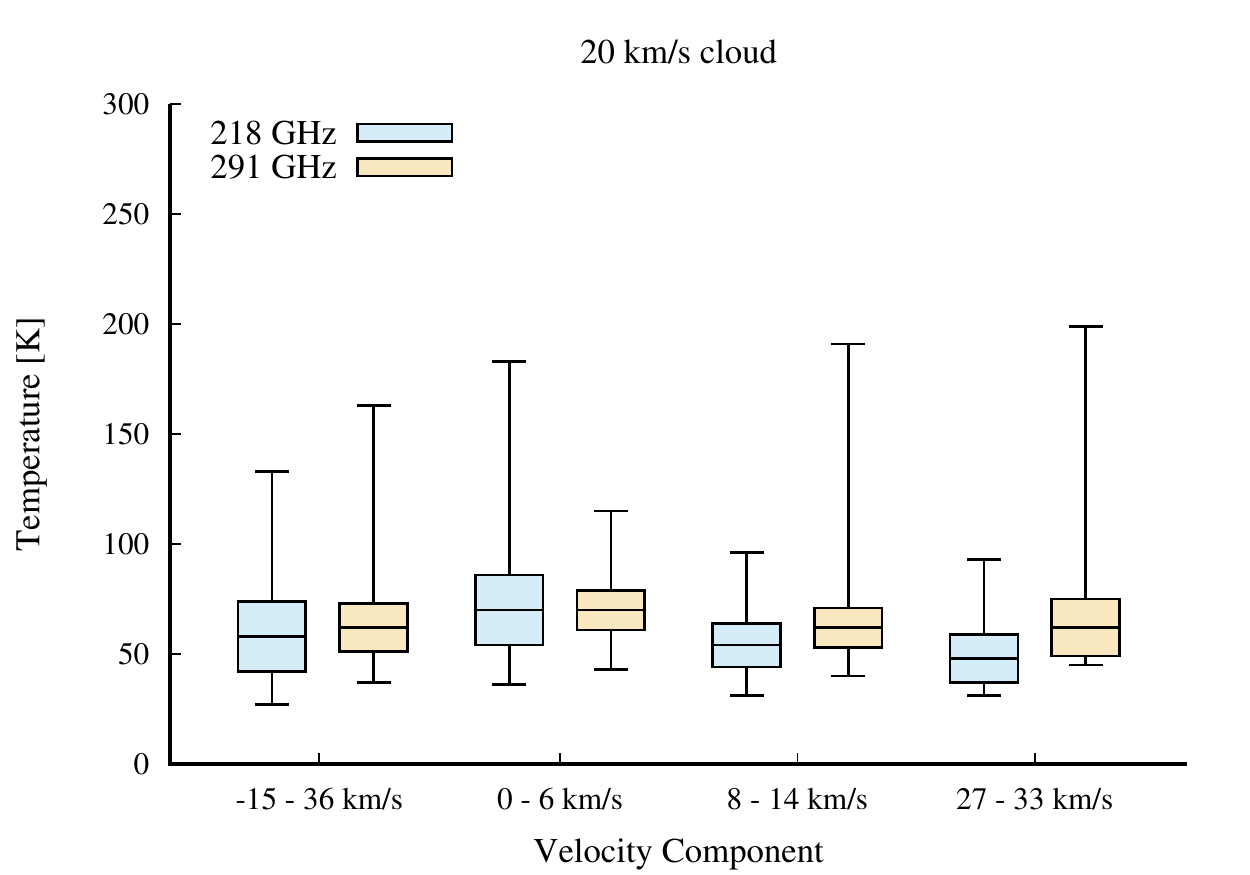}}
	\subfloat{\includegraphics[height=5.5cm]{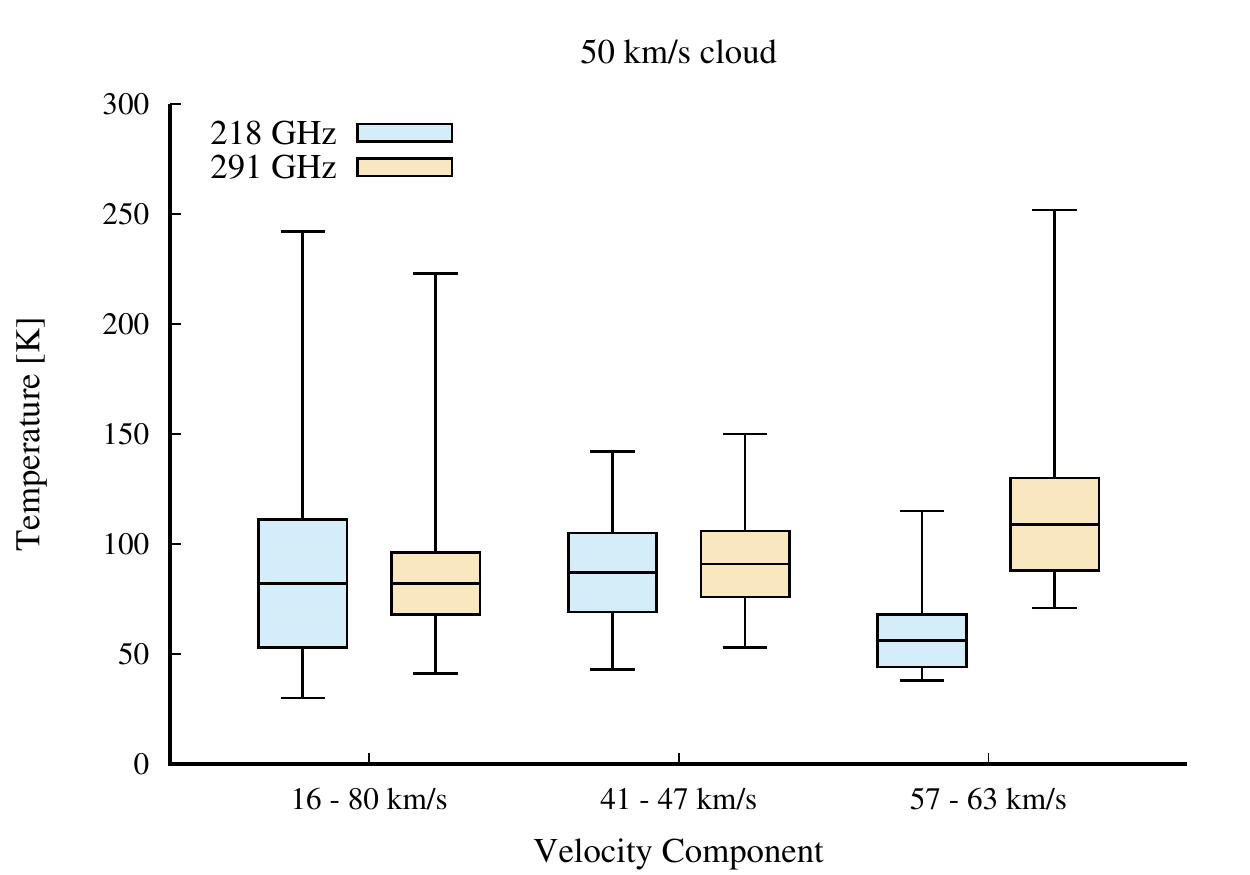}}\\
	\subfloat{\includegraphics[height=5.5cm]{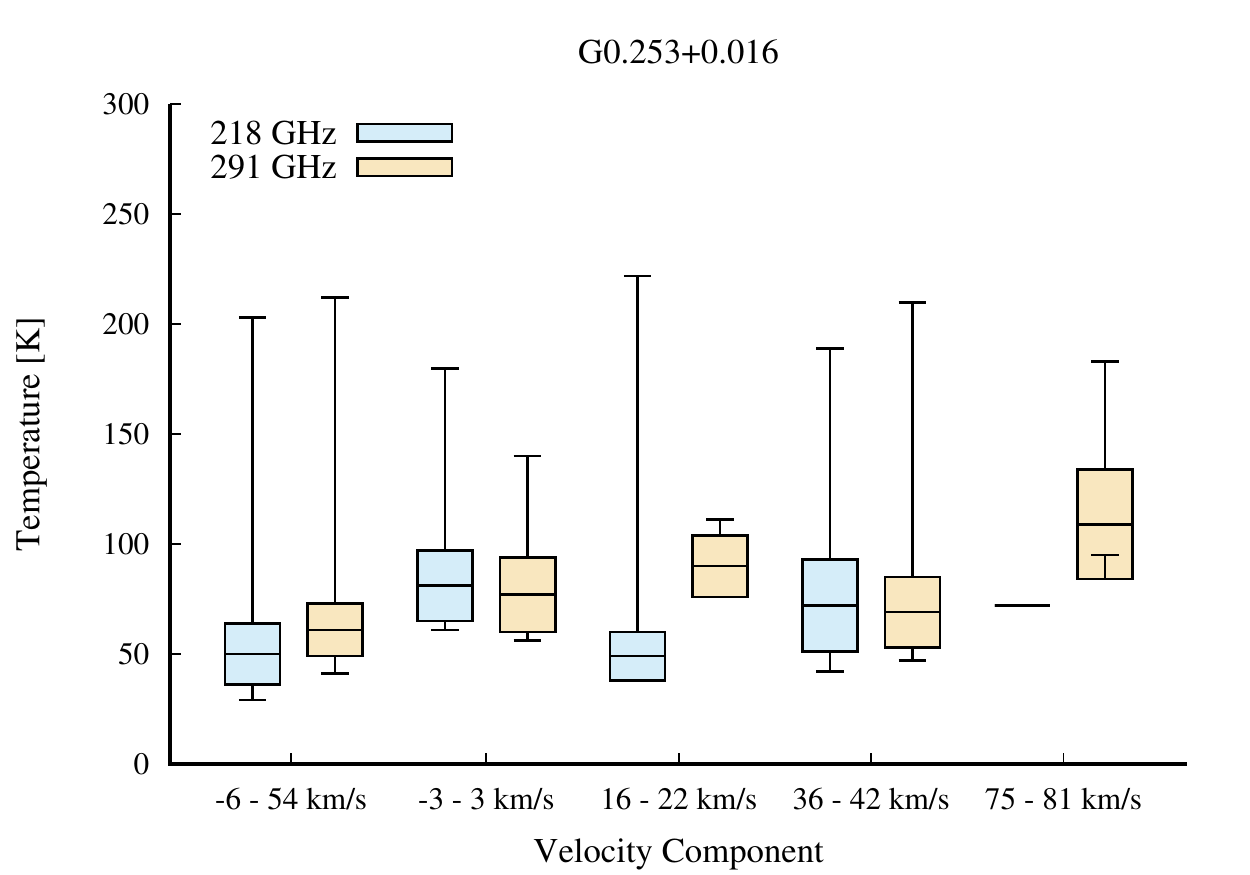}}
	\subfloat{\includegraphics[height=5.5cm]{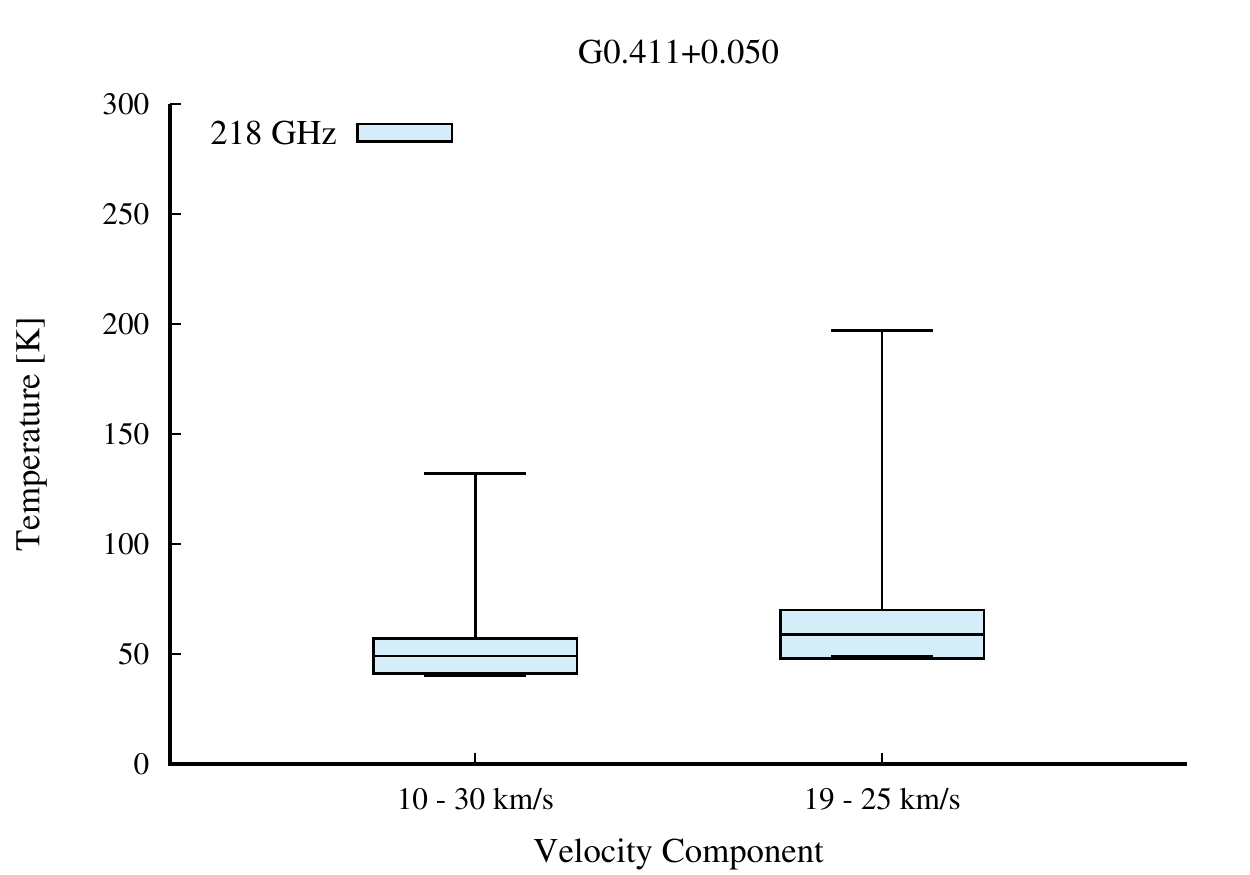}}\\
	\subfloat{\includegraphics[height=5.5cm]{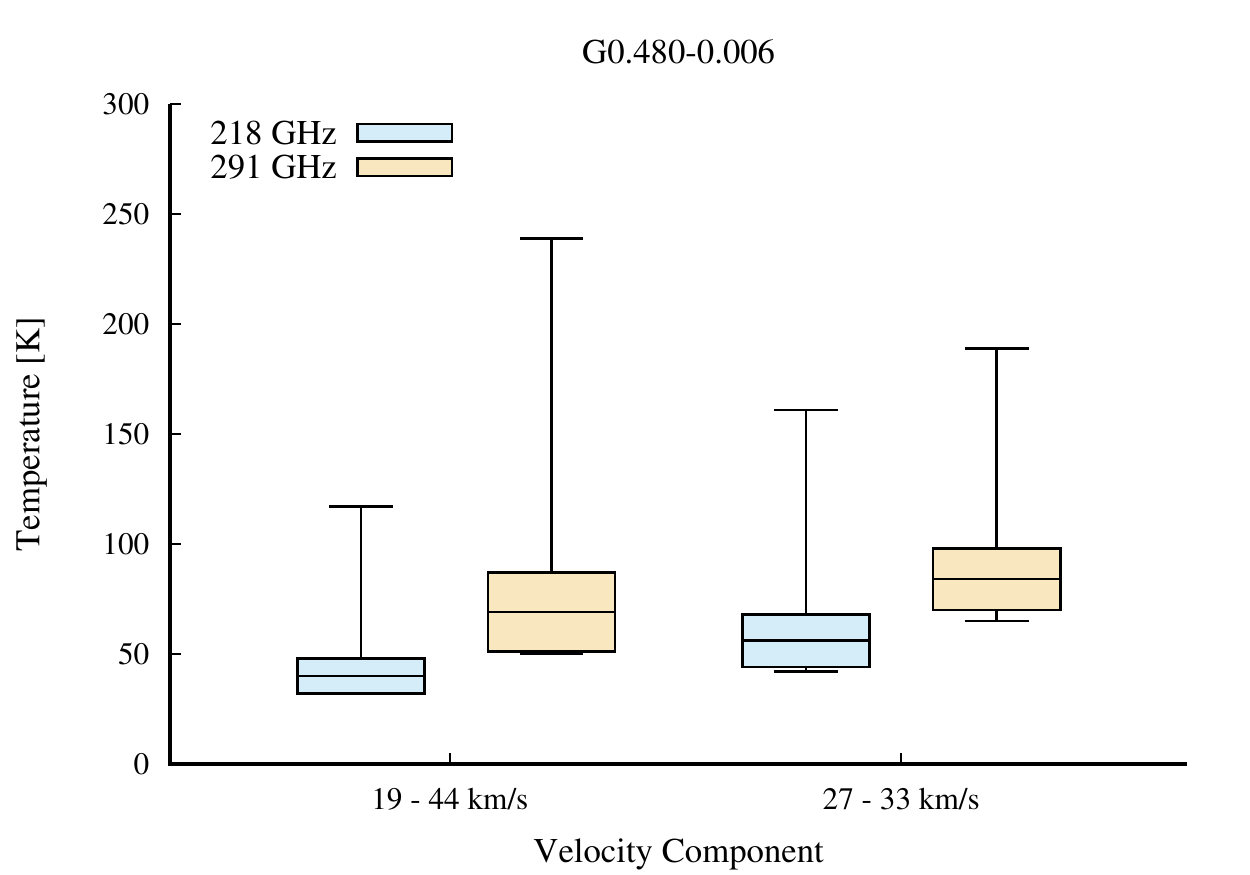}}
	\subfloat{\includegraphics[height=5.5cm]{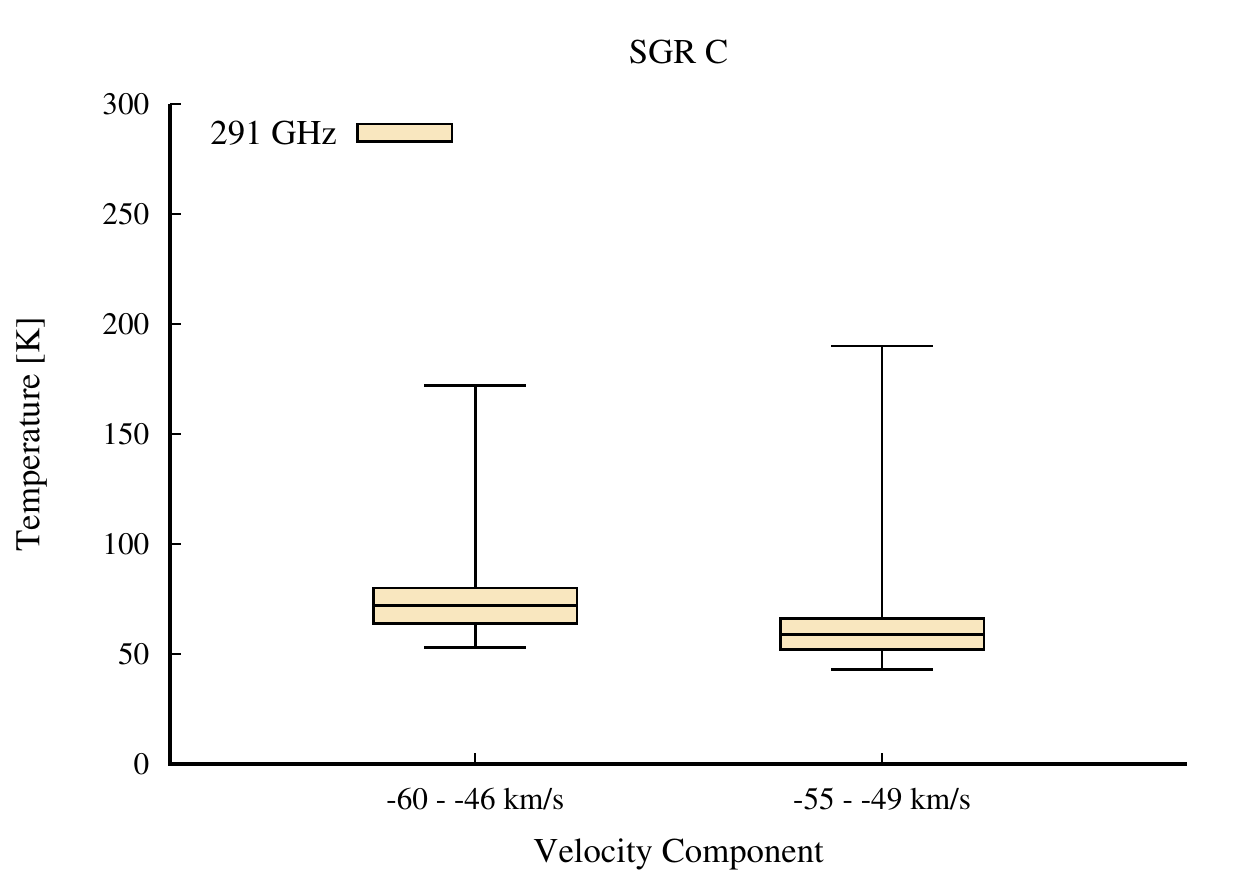}}\\
	\subfloat{\includegraphics[height=5.5cm]{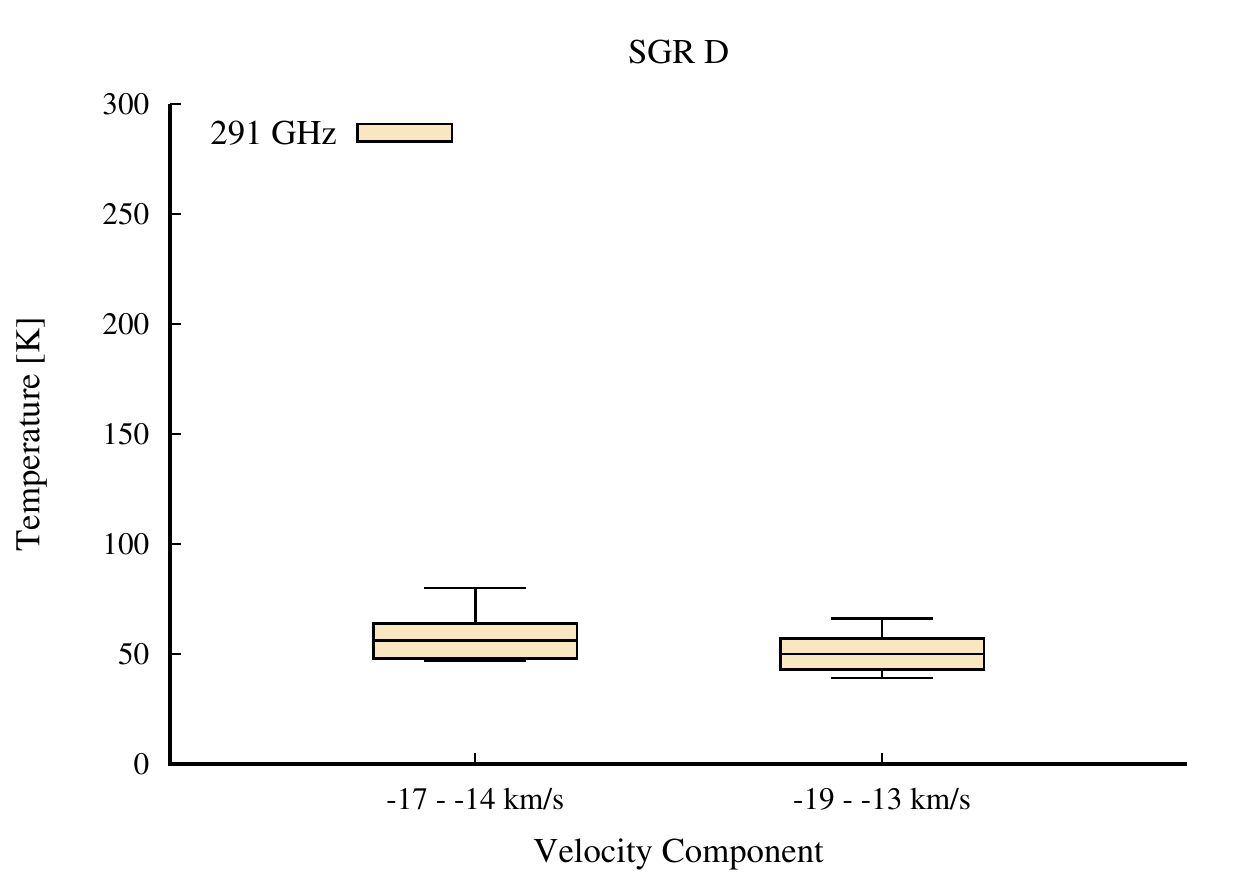}}
	\label{Average-Temp-Boxplot}
\end{figure*}

\clearpage

\section{Temperature vs Line width}

\begin{table*}
\centering
\caption{Line ratios, temperatures and line widths of Fig. \ref{TempvsLinewidth} at 218 and 291 GHz.}
\begin{tabular}{cccccccc}
Source & Velocity Component & \multicolumn{2}{c}{218 GHz} &\multicolumn{2}{c}{291 GHz}\\ 
& & R$_{321, \rm{}Average}$ & T$_{\rm{}Average}$ & Line width & R$_{422, \rm{}Average}$ & T$_{\rm{}Average}$ & Line width \\
& (km s$^{-1}$) & & (K) & (km s$^{-1}$) & & (K) & (km s$^{-1}$) \\ \hline
\multirow{4}{*}{50 km/s cloud} & \multirow{2}{*}{41 $-$ 47} & 0.409 & 104.9 & 24.6 & 0.464 & 113.4 & 20.9 \\
& & 0.346 & 82.0 & 26.4 & 0.38 & 82.5 & 23.1 \\
& \multirow{2}{*}{57 $-$ 63} & 0.279 & 61.7 & 10.1 & 0.440 & 103.2 & 20.5 \\
& & 0.208 & 45.5 & 14.1 & 0.489 & 126.4 & 17.8 \\ 
\multirow{6}{*}{G0.253+0.016} & $-$3 $-$ 3 & 0.355 & 83.0 & 18.4 & & &\\
& \multirow{2}{*}{16 $-$ 22} & 0.227 & 57.8 & 19.4 & & &\\
& & 0.198 & 43.7 & 11.6 & & &\\
& \multirow{3}{*}{36 $-$ 42} & 0.295 & 63.8 & 12.8 & 0.362 & 73.5 & 11.2 \\
& & 0.458 & 67.4 & 11.7 & 0.264 & 53.5 & 10.6\\
& & 0.460 & 131.6 & 19.2 & 0.392 & 87.1 & 17.9\\ 
G0.411+0.050 & 19 $-$ 25 & 0.255 & 55.6 & 7.3 & & \\ 
\multirow{2}{*}{G0.480$-$0.006} & \multirow{2}{*}{27 $-$ 33} & 0.333 & 74.9 & 11.6 & 0.416 & 92.9 & 12.3 \\ 
& & 0.252 & 54.4 & 9.3 & 0.385 & 85.0 & 9.7 \\ 
Sgr C & $-$55 $-$ $-$49 & & & & 0.299 & 62.9 & 8.2 \\ 
Sgr D & $-$19 $-$ $-$13 & & & & 0.250 & 51.2 & 4.0 \\
\end{tabular}
\label{TempLineWidth}
\end{table*}

\end{appendix}

\end{document}